\documentclass[oneside,11pt]{book}

\usepackage{coursenotes}

\renewcommand{\documentYear}{2025}

\title{Notes on Randomized Algorithms}

\author{James Aspnes}

\makeindex

\begin{document}

\frontmatter

\date{}

\hypersetup{pageanchor=false}
\maketitle
\hypersetup{pageanchor=true}

\pagebreak

\mbox{}

\vfill

Copyright © 2009–\documentYear{} by James Aspnes.  Distributed under a Creative
Commons Attribution-ShareAlike 4.0 International license:
\url{https://creativecommons.org/licenses/by-sa/4.0/}.

\clearpage
\phantomsection
\addcontentsline{toc}{chapter}{Table of contents}
\tableofcontents

\clearpage
\phantomsection
\addcontentsline{toc}{chapter}{List of figures}
\listoffigures

\clearpage
\phantomsection
\addcontentsline{toc}{chapter}{List of tables}
\listoftables

\clearpage
\phantomsection
\addcontentsline{toc}{chapter}{List of algorithms}
\listofalgorithms

\myChapter{Preface}{2025}{}

These are notes for the Yale
course CPSC 4690/5690 Randomized Algorithms.

Much of the structure of the course follows 
Mitzenmacher
and
Upfals's \emph{Probability and Computing}~\cite{MitzenmacherU2017},
with some
material from Motwani and Raghavan's
\emph{Randomized Algorithms}~\cite{MotwaniR1995}.  In most 
cases you'll find these textbooks contain much more detail than what is
presented here, so it is probably better to consider this document a supplement
to them than to treat it as your primary source of information.

The most recent version of these notes will be available at
\url{https://www.cs.yale.edu/homes/aspnes/classes/4690/notes.pdf}.
More stable archival versions may be found at 
\url{https://arxiv.org/abs/2003.01902}.

I would like to thank my many students and teaching fellows over the
years for their help in pointing out errors and omissions in
earlier drafts of these notes.

\mainmatter

\myChapter{Randomized algorithms}{2025}{}
\label{chapter-randomized-algorithms}

A randomized algorithm flips coins during its execution to determine
what to do next.  When considering a randomized algorithm, we usually
care about its \concept{expected worst-case} performance, which is the
average amount of time it takes on the worst input of a given size.
This average is computed over all the possible outcomes of the coin
flips during the execution of the algorithm.  We may also ask
for a \concept{high-probability bound}, 
showing that the algorithm doesn't consume too much resources most of the time.

In studying randomized algorithms, we consider pretty
much the same issues as for deterministic algorithms: how to design a
good randomized algorithm, and how to prove that it works within given
time or error bounds.  The main difference is that it is often easier
to design a randomized algorithm—randomness turns out to be a good
substitute for cleverness more often than one might expect—but
harder to analyze it.  So much of what one does is develop good
techniques for analyzing the often very complex random processes that
arise in the execution of an algorithm.  Fortunately, in doing so we
can often use techniques already developed by probabilists and
statisticians for analyzing less overtly algorithmic processes.

Formally, we think of a randomized algorithm as a machine $M$ that
computes $M(x,r)$, where $x$ is the problem input and $r$ is the
sequence of random bits.  
Our machine model is the usual \concept{random-access machine} or
\concept{RAM} model, where we have a memory space that is typically
polynomial in the size of the input $n$, and in constant time we can read
a memory location, write a memory location, or perform arithmetic
operations on integers of up to $O(\log n)$ bits.\footnote{This model
    is unrealistic in several ways: the assumption that we can perform
    arithmetic on $O(\log n)$-bit quantities in constant time omits at
    least a factor of $Ω(\log \log n)$ for addition and probably more
    for multiplication in any realistic implementation, while the
    assumption that we can address $n^c$ distinct locations in memory
    in anything less than $n^{c/3}$ time in the worst case requires
exceeding the speed of light.  But for reasonably small $n$, this
gives a pretty good approximation of the performance of real
computers, which do in fact perform arithmetic and access memory in a
fixed amount of time, although with fixed bounds on the size of both
arithmetic operands and memory.}
In this model, we may 
find it easier to think of the random bits as supplied as
needed by some subroutine, where generating a random integer of
size $O(\log n)$ takes constant time; the justification for this
assumption is that it takes constant time to read the next $O(\log
n)$-sized value from the random input.

Because the number of these various constant-time operations, and thus the running time for the
algorithm as a whole,
may depend on the
random bits, it is now a \concept{random variable}—a function on
points in some probability space.  The probability space $Ω$
consists of all possible sequences $r$, each of which is assigned a 
probability $\Prob{r}$ (typically $2^{-\card*{r}}$), and the running time for $M$ on some input $x$
is generally given as an \concept{expected value}\footnote{We'll see
    more details of these and other concepts from probability theory
    in Chapters~\ref{chapter-probability-theory}
    and~\ref{chapter-random-variables}.}
$\E_r[\Time(M(x,r))]$, where for any $X$,
\begin{align}
\E_r[X] &= ∑_{r ∈ Ω} X(r) \Prob{r}.
\end{align}
We can now quote the performance of $M$ in terms of this expected
value: where we would say that a deterministic algorithms runs in time
$O(f(n))$, where $n = \card*{x}$ is the size of the input, we instead
say that our randomized algorithm runs in \index{time!expected}\concept{expected time}
$O(f(n))$, which means that $\E_r[\Time(M(x,r))] = O(f(\card*{x}))$ for
all inputs $x$.

This is distinct from traditional \concept{worst-case analysis}, where there is
no $r$ and no expectation, and \concept{average-case analysis}, where
there is again no $r$ and the value reported is not a maximum but an
expectation over some distribution on $x$.  The following trivial example shows the distinction.

\section{Searching an array}
\label{section-searching-an-array}

Suppose we have an array $A$ with $n$ elements, and we are told that
one of the positions in the array contains a nonzero value (the
``prize'')
while every other position contains zero. How quickly
can we find the prize in the worst case?

Any \emph{deterministic} algorithm will probe the array positions in
some predictable order. An adversary can simulate the algorithm
running in an execution where it always sees $0$ until it has checked
every location. By putting the prize in the last place the
algorithm looks, we get a worst-case input that requires at least $n$
probes to find it.

In the average case, we might assume that instead of placing the
prize in the worst possible position, it is placed uniformly at random.
Now if we just scan the array from left to right, the expected number
of probes 
will be
\begin{align*}
    ∑_{i=1}^{n} i ⋅ \Prob{\text{algorithm does $i$ probes}}
    &= ∑_{i=1}^{n} i ⋅ \frac{1}{n}
    \\&= \frac{1}{n}⋅\frac{n(n+1)}{2}
    \\&= \frac{n+1}{2}.
\end{align*}
This is still $Θ(n)$, but we've improved the constant factor by almost
a factor of $2$. The cost is that we trust the adversary to place the
prize randomly—but there is no reason for the adversary to do this.

The trick to randomized algorithms is that we can
can obtain the same expected payoff even in the worst case by
supplying the randomness ourselves.
Let's suppose that instead of scanning the array predictably from left
to right, we flip a coin and either scan left-to-right or scan
right-to-left
with probability $1/2$ each. If the prize is in position $i$,
then scanning left-to-right finds it in $i$ probes, while scanning
right-to-left finds it in $n-i+1$ probes. The expected cost is thus
\begin{align*}
    \frac{1}{2}⋅i + \frac{1}{2}⋅(n-i+1) &= \frac{n+1}{2},
\end{align*}
the same as in the average case. But now we don't make any assumptions
about the input—by using just one bit of randomness we get this expected time no matter where the
adversary puts the prize.

A natural question is whether a more clever algorithm can do even
better. Usually it's pretty hard to prove lower bounds on algorithms,
but in this case we can use a classic result called \concept{Yao's
lemma}~\cite{Yao1977} to show a worst-case lower bound on the expected
cost of any randomized algorithm by constructing a probability
distribution on inputs that gives a bad average-case lower bound for
any deterministic algorithm. We'll give a proof of this in
§\ref{section-Yao-lemma}, but the intuition is that any randomized
algorithm acts like picking a deterministic algorithm at random, so if
there is an input distribution that is bad for all deterministic
algorithms, it is still bad even if we pick one of those algorithms
randomly.

In this case, the bad input distribution is simple: put the $1$ in
each position $A[i]$ with equal probability $1/n$. For a deterministic
algorithm, there will be some fixed sequence of positions $i_1,i_2,\dots$
that it examines as long as it only sees zeros. A smart deterministic
algorithm will not examine the same position twice, so the $1$ is
equally likely to be found $1,2,3,\dots,n$ probes. This gives the same
expected $\frac{n+1}{2}$ probes as for the simple randomized
algorithm, which shows that that algorithm is optimal.

We've been talking about searching an array, because that fits best in
our model of an input supplied to the algorithm, but essentially the
same analysis applies to brute-force inverting a black-box function.
Here we have a function $f$ and target output $y$, and we want to find
an input $x$ such that $f(x)=y$. The same analysis as for the array
case shows that this takes $\frac{n+1}{2}$ expected evaluations of $f$
assuming that exactly one $x$ works and we can't do anything clever.

Curiously, in this case it may be possible to improve this bound to
$O(√{n})$ evaluations if somehow we get our hands on a working
quantum computer. We'll come back to this when we discuss quantum
computing in general and Grover's algorithm in particular in
Chapter~\ref{chapter-quantum-computing}.

\section{Verifying polynomial identities}

This classic problem is described in
\cite[§1.1]{MitzenmacherU2017}.  Here we are given two
products of polynomials and we want to determine if they compute the
same function. For example, we might have
\begin{align*}
    p(x) &= {\left(x - 7\right)} {\left(x - 3\right)} {\left(x -
1\right)} {\left(x + 2\right)} {\left(2 \, x + 5\right)}
\\
    q(x) &= 2 \, x^{5} - 13 \, x^{4} - 21 \, x^{3} + 127 \, x^{2} +
121 \, x - 210
\end{align*}

These expressions both represent degree-5 polynomials, and it is not
obvious without multiplying out the factors of $p$ whether they are
equal or not.  Multiplying out all the factors of $p$ may take
as much as $O(d^2)$ time if we assume integer multiplication takes unit time and
do it the straightforward way.\footnote{It can be faster if we do
something sneaky like use fast Fourier
transforms~\cite{SchonhageS1971}.}  We can do better than this using
randomization.

The trick is that evaluating $p(x)$ and $q(x)$ takes only $O(d)$
integer operations, and we will find $p(x)=q(x)$ only if either (a)
$p(x)$ and $q(x)$ are really the same polynomial, or (b) $x$ is a
\concept{root} of $p(x)-q(x)$.  Since $p(x)-q(x)$ has degree at most
$d$, it can't have more than $d$ roots.  So if we choose $x$ uniformly
at random from some much larger space, it's likely that we will not
get a root.  Indeed, evaluating $p(11) = 112320$ and $q(11) = 120306$
quickly shows that $p$ and $q$ are not in fact the same.

This is an example of a
\concept{Monte Carlo algorithm}, which is an
algorithm that runs in a fixed amount of time but only gives the right
answer some of the time.  (In this case, with probability $1-d/r$, where
$r$ is the size of the range of random integers we choose $x$ from.)
Monte Carlo algorithms have the unnerving property of not indicating
when their results are incorrect, but we can make the probability of
error as small as we like by running the algorithm repeatedly.  For
this particular algorithm, 
the probability of error after $k$ trials is only $(d/r)^k$, which means
that for fixed $d/r$ we need $O(\log(1/ε))$ iterations to get
the error bound down to any given $ε$.  If we are really
paranoid, we could get the error down to $0$ by testing $d+1$ distinct values,
but now the cost is as high as multiplying out $p$
again.

The error for this algorithm is one-sided: if we find a 
\concept{witness} to the fact that $p≠q$, we are done, but if we
don't, then all we know is that we haven't found a witness yet.  We
also have the property that if we check enough possible witnesses, we
are guaranteed to find one.

A similar
property holds in the classic \index{primality
test!Miller-Rabin}\concept{Miller-Rabin primality test}, a randomized
algorithm for determining whether a large integer is prime or
not.

The idea of this algorithm is that Fermat's Little Theorem says
that Fermat's Little Theorem says that $a^{p-1} = 1 \pmod{p}$ when $p$
is prime Fermat's Little Theorem says that $a^{p-1} = 1 \pmod{p}$ when
$0<a<p$ and $p$ is prime, which we can compute by
first factoring $p-1$ as $2^k q$ where $q$ is odd, computing $a^q
\pmod{p}$, then repeatedly squaring this value to put back in all the
twos. It happens to be the case that when $p$ is prime, the last value
in this sequence that isn't $1$ will always be $-1$, but if $p$ is
composite, for some $a$ we will find that $a^{2^{j-1} q} ≠ ±1 \pmod{p}$
but $a^{2^j q} = 1 \pmod{p}$ anyway. Such a bad $a$ is a
witness to the fact that $p$ is composite.

The original version, due to
Gary Miller~\cite{Miller1976} showed that, as in
polynomial identity testing, it might be sufficient to choose a
sufficiently large set of candidate witnesses deterministically. Unfortunately,
this result depends on the truth of the
\index{Riemann hypothesis!extended}\index{hypothesis!extended Riemann}\concept{extended Riemann hypothesis},
a notoriously difficult open problem in number theory.
Michael Rabin~\cite{Rabin1980} 
demonstrated that choosing a random witness works with probability at
least $3/4$, so testing $t$ random witnesses fails
less than $(1/4)^t$ of the time.
So we don't need to prove the extended Riemann hypothesis 
as long as we were willing to accept a very small
probability of incorrectly identifying a composite number as prime.

For many years it was open whether it was possible to test primality
deterministically in polynomial time without unproven number-theoretic
assumptions, and the randomized Miller-Rabin algorithm was 
one of the most widely-used randomized algorithms for which no
good deterministic alternative was known.
Eventually, Agrawal~\etal~\cite{AgrawalKS2004} demonstrated how to test
primality deterministically using a different technique, although the
cost of their algorithm is high enough that Miller-Rabin is still used
in practice.

\section{Randomized QuickSort}
\label{section-quicksort}

The \concept{QuickSort} algorithm~\cite{Hoare1961quicksort} works as follows.  For simplicity, we assume that no two elements of the array being sorted are equal.

\begin{itemize}
 \item If the array has $>1$ elements,
\begin{itemize}
  \item Pick a \concept{pivot} $p$ uniformly at random from the elements of the array.
  \item Split the array into $A_{1}$ and $A_{2}$, where $A_{1}$
  contains all elements $< p$
  elements $> p$.
  \item Sort $A_{1}$ and $A_{2}$ recursively and return the sequence
  $A_{1}, p, A_{2}$.
\end{itemize}
\item Otherwise return the array.
\end{itemize}

The splitting step takes exactly $n-1$ comparisons, since we have to check each non-pivot against the pivot.  We assume all other costs are dominated by the cost of comparisons.  How many comparisons does randomized QuickSort do on average?

There are two ways to solve this problem: the dumb way and the smart
way.  We'll do it the dumb way first and save the smart way for
§\ref{section-quicksort-linearity-of-expectation}.

\subsection{Recurrence for QuickSort}
\label{section-quicksort-recurrence}

Let $T(n)$ be the expected number of comparisons done on an array of
$n$ elements.  We have $T(0) = T(1) = 0$ and for larger $n$,
\begin{align}
\label{eq-quicksort-recurrence}
    T(n) &= (n-1) + \frac{1}{n} ∑_{k=0}^{n-1} \left(T(k) + T(n-1-k)\right).
\end{align}
Why?  Because we do $(n-1)$ comparisons to split the piles, there are $n$ equally-likely choices for our pivot
(hence the $1/n$), and for each choice the expected cost of the
recursive sorts is $T(k) +
T(n-1-k)$, where $k$ is the number of elements that land in $A_{1}$.
Formally, we are using here the \concept{law of total probability},
which says that for any random variable $X$ and partition of the
probability space into events $B_{1}\dots{}B_{n}$, then 
\begin{align*}
    \Exp{X} &= ∑ \Prob{B_{i}} \ExpCond{X}{B_{i}},
\intertext{where}
\ExpCond{X}{B_{i}} &= ∑_{ω ∈ B_{i}} X(ω) \frac{\Prob{ω}}{\Prob{B_i}}
\end{align*}
is the \concept{conditional expectation} of $X$ conditioned on $B_i$,
which we can think of as just the average value of $X$ if we know that
$B_i$ occurred.  (See §\ref{section-law-of-total-probability}
for more details.)

So now we just have to solve this ugly recurrence.  We can reasonably
guess that when $n ≥ 1, T(n) ≤ an \log n$ for some constant
$a$.  Clearly this holds for $n = 1$.  Now apply induction on larger
$n$ to get
\begin{align*}
T(n)
&= (n-1) + \frac{1}{n} ∑_{k=0}^{n-1} \left(T(k) + T(n-1-k)\right) \\
&= (n-1) + \frac{2}{n} ∑_{k=0}^{n-1} T(k) \\
&= (n-1) + \frac{2}{n} ∑_{k=1}^{n-1} T(k) \\
&≤ (n-1) + \frac{2}{n} ∑_{k=1}^{n-1} a k \log k \\
&≤ (n-1) + \frac{2}{n} \int_{k=1}^{n} a k \log k \\
&= (n-1) + \frac{2a}{n} \left(\frac{n^2 \log n}{2} - \frac{n^2}{4} + \frac{1}{4}\right) \\
&= (n-1) + a n \log n - \frac{an}{2} + \frac{a}{2n}.
\end{align*}

If we squint carefully at this recurrence for a while we notice that
setting $a = 2$ makes this less than or equal to $a n \log n$, since
the remaining terms become $(n-1) - n + 1/n = 1/n - 1$, which is
negative for $n ≥ 1$.  We can thus confidently conclude that $T(n)
≤ 2n \log n$ (for $n ≥ 1$).

\section{Where does the randomness come from?}

Typically we assume in analyzing a randomized algorithm that we have
access to genuinely random bits $r$, and don't ask too carefully how
we can get these random bits. For practical applications, there are
basically three choices, from strongest (and most expensive) to
weakest (and cheapest):
\begin{enumerate}
    \item \index{physical
        randomness}\index{randomness!physical}\concept{Physical
        randomness.} Lock a cat in a box with a radioactive source and
        a Geiger counter attached to a solenoid aimed at a vial of
        prussic acid~\cite{Schroedinger1935}. Come back in an hour and check if the cat is
        still breathing. With an appropriately-tuned radioactive
        source, this generates one very unpredictable bit of
        randomness at the cost of one half of a cat on
        average.\footnote{Expected cost may be higher if the observer forgets their gas
        mask.}

        There are cheaper variants, mostly involving amplified quantum
        noise at the high end, or monitoring physical processes that
        are expected to be somewhat random (like intervals between
        keyboard presses or seek times of hard drive heads) at the low
        end. In each case you get a sequence of random bits that,
        under plausible physical assumptions, are effectively
        unpredictable.
        
        The \texttt{/dev/random} device in Linux systems gives you
        access to the cheap kind of physical randomness, and will
        block waiting for you to wave your mouse around if it runs
        out.
    \item \index{cryptographically secure pseudorandom number
        generator}\concept{Cryptographically-secure
        pseudorandomness.} Find some function that spits out a
        sequence of random-looking values given a \concept{seed}, such
        that if the seed is chosen uniformly at random (say using a
        physical random number generator), no polynomial-time program
        can distinguish the sequence from an actual random sequence
        with non-trivial probability. Usually expensive, but if your
        polynomial-time randomized algorithm fails using a CPRNG,
        you've succeeded in breaking its cryptographic assumptions.

        The \texttt{/dev/urandom} device in Linux systems gives you
        this, based on a seed derived from the same sources as \texttt{/dev/random}.
    \item \index{pseudorandom number
        generator}\concept{Statistical pseudorandomness.} As above, but choose a
        function that is not cryptographically secure but merely
        passes common statistical tests for things like $k$-wise
        independence of consecutive outputs. Which function to choose
        is largely a matter of convenience and fashion (although
        PRNGs in older standard libraries can be very, very bad).
        The advantage is that many PRNGs are very fast and will not
        slow your program down. The disadvantage is that you are
        relying on your program not doing anything that exposes the
        weakness of the PRNG.

        The \texttt{random} function in the standard C library is an
        example of this, although maybe not a good example. The cool
        kids mostly use Mersenne Twister~\cite{MatsumotoN1998}.
\end{enumerate}

One advantage of pseudorandom generators is that they allow for
debugging: if you run your program twice with the same key, it will do
the same thing. This was in fact an argument made by von Neumann
\emph{against} using physical randomness back in the old
days~\cite{VonNeumann1963}.

In practice, most people just use whatever PRNG is ready to hand, and
hope it works. As theorists, we will ignore this issue, and assume
that our random inputs are actually random.

\section{Classifying randomized algorithms}

Different random algorithms make different guarantees about the
likelihood of getting a correct output or even the possibility of
recognizing a correct output. These different guarantees have
established names in the randomized algorithms literature and are
correspond to various complexity classes in complexity theory.

\subsection{Las Vegas vs Monte Carlo}
\label{section-las-vegas-vs-monte-carlo}

One difference between QuickSort and polynomial equality testing
that QuickSort always succeeds, but may run for longer than you
expect; while the polynomial equality tester always runs in a fixed
amount of time, but may produce the wrong answer.
These are examples of two classes of randomized algorithms, which were
originally named by László Babai~\cite{Babai1979}:\footnote{To be
    more precise, Babai defined \emph{Monte Carlo algorithms} based on
    the properties of \index{simulation!Monte Carlo}
    \concept{Monte Carlo simulation}, a technique dating back to the
Manhattan project.  The term \emph{Las Vegas algorithm} was new.}
\begin{itemize}
\item A \concept{Las Vegas algorithm} fails with some probability,
but we can tell when it fails.  In particular, we can run it again
until it succeeds, which means that we can eventually succeed with
probability $1$ (but with a potentially unbounded running time).  
Alternatively, we can think of a Las Vegas algorithm as an algorithm
that runs for an unpredictable amount of time but always succeeds (we
can convert such an algorithm back into one that runs in bounded time
by declaring that it fails if it runs too long—a condition we can
detect).
QuickSort is an example of a Las Vegas algorithm.
\item A \concept{Monte Carlo algorithm} fails with some
probability, but we can't tell when it fails.  If the algorithm
produces a yes/no answer and the failure probability is significantly
less than $1/2$, we can reduce the probability of failure by running
it many times and taking a majority of the answers.  The polynomial
equality-testing algorithm in an example of a Monte Carlo algorithm.
\end{itemize}

The heuristic for remembering which class is which is that the names
were chosen to  
appeal to English speakers: in Las Vegas, the dealer can tell you whether you've won or lost, but in Monte Carlo, \emph{le croupier ne parle que Fran\c{c}ais}, so you have no idea what he's saying.

Generally, we prefer Las Vegas algorithms, because we like knowing
when we have succeeded.  But sometimes we have to settle for Monte
Carlo algorithms, which can still be useful if we can get the
probability of failure small enough.
For example, any time we try to estimate an average by
\concept{sampling} (say, inputs to a function we are trying to
integrate or political views of voters we are trying to win over) we
are running a Monte Carlo algorithm: there is always some possibility
that our sample is badly non-representative, but we can't tell if we
got a bad sample unless we already know the answer we are looking for.

\subsection{Machine model}

We won't spend a lot of time thinking about our machine model, but
we can avoid confusion later if we at least make some attempt to specify
a foundation for what we are doing. In a typical
algorithmic context we imagine we are working with a polynomial-time
\concept{random-access machine}\index{machine!random-access} or
\concept{RAM}. We will describe this model briefly and then discuss
what changes are needed to handle randomization.

Given an input of size $n$, a RAM provides a constant number of
registers of $O(\log n)$ bits each and a polynomial-sized array of
memory locations that also hold $O(\log n)$ bit each. The size of the
registers and locations are assumed to be large enough to hold an
index to a particular memory location.

There is a constant-size program that controls the RAM, that describes
a sequence of operations that are modeled on the instructions
available on a typical real-world CPU.  Examples of such operations
include loading memory locations to registers (with the address
obtained from some register); storing registers to memory; computing
arithmetic, comparison, shift, and bitwise logical operations on
values in registers; and performing jumps or conditional jumps to
other locations in the program. Each such operation is assumed to take
constant $(O(1))$ time even if it is working on values of $O(\log n)$
bits.

For a randomized algorithm, we need a source of randomness. The
natural assumption is that a probabilistic polynomial-time RAM
provides an additional operation that generates a random value. 
It is convenient to assume that this random value can be chosen
uniformly from a range whose size is provided at run time, at a cost
of $O(1)$ time as with other operations. This is a bit stronger than
the typical assumption in a probabilistic Turing machine of only
getting fair coin-flips, but it is consistent with the assumption that
other operations on $O(\log n)$-bit values take only constant time.

An odd consequence of assuming that we can generate uniform random
values is that bounded RAM computations—those that are guaranteed to
run in finite time—cannot be translated to bounded Turing machine
computations. This is because a Turing machine computation that
consumes $r$ random bits must choose between $2^{-r}$ outcomes, and
there is no way to combine $k$ of these outcomes to choose, for
example, an outcome that occurs with probability exactly $1/3$. In
practice this apparent disaster can be resolved by allowing for an
exponentially small error in the Turing machine computation or
relaxing the time bound to an expected time bound, neither of which is
likely to cause trouble in practical implementations.

\subsection{Randomized complexity classes}
\label{section-randomized-complexity-classes}

Las Vegas vs Monte Carlo is the typical distinction made by algorithm
designers, but complexity theorists have developed more elaborate classifications.
These include algorithms with ``one-sided'' failure properties.
For these algorithms, we never get a bogus ``yes'' answer but may get
a bogus ``no'' answer (or vice versa).  This gives us several
complexity classes that act like randomized versions of \classNP{},
co-\classNP{}, etc.:

\begin{itemize}
 \item The class \classR{} or \classRP{} (randomized \classP{}) consists of
 all languages $L$ for which a polynomial-time Turing machine $M$
 exists such that if $x∈L$, then $\Prob{M(x,r) = 1} ≥ 1/2$ and if
 $x\not∈L$, then $\Prob{M(x,r) = 1} = 0$.  In other words, we can
 find a witness that $x∈L$ with constant probability.  This is the
 randomized analog of \classNP{} (but it's much more practical, since
 with \classNP{} the probability of finding a winning witness may be exponentially small).
 \item The class co-\classR{} consists of all languages $L$ for which a
 poly-time Turing machine $M$ exists such that if $x\not∈L$, then
 $\Prob{M(x,r) = 1} ≥ 1/2$ and if $x∈L$, then $\Prob{M(x,r) = 1} =
 0$.  This is the randomized analog of co-\classNP{}.
 \item The class \classZPP{} (zero-error probabilistic $P$) is defined
 as $\classRP{}\cap{}\text{co-\classRP{}}$.  If we run both our \classRP{}
 and co-\classRP{} machines for polynomial time, we learn the correct
 classification of $x$ with probability at least $1/2$.  The rest of the time we learn only that we've failed (because both machines return 0, telling us nothing).  This is the class of (polynomial-time) Las Vegas algorithms.  The reason it is called ``zero-error'' is that we can equivalently define it as the problems solvable by machines that always output the correct answer eventually, but only run in \emph{expected} polynomial time.
 \item The class \classBPP{} (bounded-error probabilistic \classP{})
 consists of all languages $L$ for which a poly-time Turing machine
 exists such that if $x\not∈L$, then $\Prob{M(x,r) = 1} ≤ 1/3$,
 and if $x∈L$, then $\Prob{M(x,r) = 1} ≥ 2/3$.  These are the
 (polynomial-time) Monte Carlo algorithms: if our machine answers 0 or
 1, we can guess whether $x∈L$ or not, but we can't be sure.
 \item The class \classPP{} (probabilistic \classP{}) consists of all
 languages $L$ for which a poly-time Turing machine exists such that
 if $x\not∈L$, then $\Prob{M(x,r) = 1} ≥ 1/2$, and if $x∈L$,
 then $\Prob{M(x,r) = 1} < 1/2$.  Since there is only an exponentially
 small gap between the two probabilities, such algorithms are not
 really useful in practice; \classPP{} is mostly of interest to complexity theorists.
\end{itemize}

Assuming we have a source of random bits, any algorithm in \classRP{},
co-\classRP{}, \classZPP{}, or \classBPP{} is good enough for practical use.
We can usually even get away with using a pseudorandom number
generator, and there are plausible reasons to suspect that in fact every
one of these classes is equal to \classP{}.

\section{Classifying randomized algorithms by their methods}

We can also classify randomized algorithms by how they use their
randomness to solve a problem.  Some very broad
categories:\footnote{These are largely adapted from the introduction
to~\cite{MotwaniR1995}.}

\begin{itemize}
\item \indexConcept{avoiding worst-case inputs}{Avoiding worst-case
inputs}, by hiding the details of the algorithm from the adversary.
Typically we assume that an adversary supplies our input.
If the adversary can see what our algorithm is going to do (for
example, he knows which door we will open first), he can use this
information against us.  By using randomness, we can replace our
predictable deterministic algorithm by what is effectively a random
choice of many different deterministic algorithms.  Since the
adversary doesn't know which algorithm we are using, he can't (we
hope) pick an input that is bad for all of them.
\item \indexConcept{sampling}{Sampling}.  Here we use randomness to
find an example or examples of objects that are likely to be
typical of the population they are drawn from, either to estimate some
average value (pretty much the basis of all of statistics) or because
a typical element is useful in our algorithm (for example, when
picking the pivot in QuickSort).  Randomization means that the
adversary can't direct us to non-representative samples.
\item \indexConcept{hashing}{Hashing}.  Hashing is the process of
assigning a large object $x$ a small name $h(x)$ by feeding it to a
\concept{hash function} $h$.  Because the names are small, the
Pigeonhole Principle implies that many large objects hash to the same
name (a \concept{collision}).  If we have few objects that we actually
care about, we can avoid collisions by choosing a hash function that
happens to map them to different places.  Randomization helps here by
keeping the adversary from choosing the objects after seeing what our
hash function is.

Hashing techniques are used both in \concept{load balancing} (e.g.,
insuring that most cells in a \concept{hash table} hold only a few
objects) and in \concept{fingerprinting} (e.g., using a \concept{cryptographic hash
function} to record a \concept{fingerprint} of a file, so that we can
detect when it has been modified).
\item \indexConcept{random structure}{Building random structures}.
The \concept{probabilistic method} shows the existence of 
structures with some desired property 
(often graphs with interesting properties, but there are
other places where it can be used) by showing that a
randomly-generated structure in some class has a nonzero probability
of having the property we want.  If we can beef the probability up to
something substantial, we get a randomized algorithm for generating
these structures.
\item \indexConcept{symmetry breaking}{Symmetry breaking}.  In
\concept{distributed algorithms} involving multiple processes,
progress may be stymied by all the processes trying to do the same
thing at the same time (this is an obstacle, for example, in
\concept{leader election}, where we want only one process to declare
itself the leader).  Randomization can break these deadlocks.
\end{itemize}

\myChapter{Probability theory}{2025}{}
\label{chapter-probability-theory}

In this chapter, we summarize the parts of \concept{probability
theory} that we will use in the course.  This is not really a
substitute for reading an actual probability theory book like
Feller~\cite{Feller1968} or Grimmett and
Stirzaker~\cite{GrimmettS2001}, but the hope is that it's enough to
get by.

The basic idea of probability theory is that we want to model all
possible outcomes of whatever process we are studying simultaneously.
This gives the notion of a \concept{probability space}, which is the
set of all possible outcomes; for example, if we roll two dice, the
probability space would consist of all 36 possible combinations of
values.  Subsets of this space are called
\indexConcept{event}{events}; an example in the two-dice space would
be the event that the sum of the two dice is 11, given by the set
$A=\Set{\Tuple{5,6}, \Tuple{6,5}}$.  The probability of an event $A$ is
given by a
\index{measure!probability}
\concept{probability measure} $\Prob{A}$; for simple probability
spaces, this is just the sum of the probabilities of the individual
outcomes contains in $A$, while for more general spaces, we define the
measure on events first and the probabilities of individual outcomes
are derived from this.  Formal definitions of all of these concepts
are given later in this chapter.

When analyzing a randomized algorithm, the probability space describes
all choices of the random bits used by the algorithm, and we can think
of the possible executions of an algorithm as living within this
probability space.  More formally, the sequence of operations carried
out by an algorithm and the output it ultimately produces are examples
of \indexConcept{random variable}{random variables}—functions from a
probability space to some other set—which we will discuss in detail in
Chapter~\ref{chapter-random-variables}.

\section{Probability spaces and events}
\label{section-probability-spaces}

A 
\index{probability space!discrete}
\concept{discrete probability space} is a countable set $Ω$ of
\concept{points} or \concept{outcomes} $ω$.  Each $ω$ in
$Ω$ has a \concept{probability} $\Prob{ω}$, which is a real
value with $0≤ \Prob{ω} ≤ 1$.  It is required that
$∑_{ω ∈ Ω} = 1$.

An \concept{event} $A$ is a subset of $Ω$; its probability is
$\Prob{A} = ∑_{ω ∈ A} \Prob{ω}$.  We require that
$\Prob{Ω} = 1$, and it is immediate from the definition that
$\Prob{\emptyset} = 0$.

The \concept{complement} $\bar{A}$ or $\neg A$ of an event $A$ is
the event $Ω - A$.  It is always the case that $\Prob{\neg A} =
1-\Prob{A}$.

This fact is a special case of the general principle that if $A_1,
A_2, \dots$ forms a \concept{partition} of $Ω$—that is, if $A_i
\cap A_j = \emptyset$ when $i≠ j$ and $\bigcup A_i = Ω$—then
$∑ \Prob{A_i} = \Prob{Ω} = 1$.  It happens that $\neg A$ and $A$
form a partition of $Ω$ consisting of exactly two elements.

Even more generally, if $A_1, A_2, \ldots$ are
\concept{disjoint} events (that is, if $A_i \cap
A_j = \emptyset$ whenever $i≠ j$), then
$\Prob{\bigcup A_i} = ∑ \Prob{A_i}$.  This fact does not
hold in general for events that are not disjoint.

For discrete probability spaces, all of these facts can be proven
directly from the definition of probabilities for events.  For more
general probability spaces, it's no longer possible to express the
probability of an event as the sum of the probabilities of its
elements, and we adopt an axiomatic approach instead.

\subsection{General probability spaces}

More general probability spaces consist of a triple $(Ω,
ℱ, \Pr)$ where $Ω$ is a set of points, $ℱ$ is
a \concept{$σ$-algebra} (a family of subsets of $Ω$ that
contains $Ω$ and is closed under complement and countable
unions) of \concept{measurable sets}, and $\Pr$ is a function from
$ℱ$ to $[0,1]$ that gives $\Prob{Ω} = 1$ and satisfies
\concept{countable additivity}: when $A_1,\dots$ are disjoint,
$\Prob{\bigcup A_i} = ∑ \Prob{A_i}$.  This definition is needed for
uncountable spaces, because (under certain set-theoretic assumptions)
we may not be able to assign a meaningful probability to all subsets
of $Ω$.

Formally, this definition is often presented as three 
\concept{axioms of probability}, due to
Kolmogorov~\cite{Kolmogorov1933}:
\begin{enumerate}
    \item $\Prob{A} ≥ 0$ for all $A ∈ ℱ$.
    \item $\Prob{Ω} = 1$.
    \item For any countable collection of disjoint events $A_1, A_2, \dots$,
        \begin{align*}
            \Prob{\bigcup_i A_i} &= ∑_i \Prob{A_i}.
        \end{align*}
\end{enumerate}

It's not hard to see that the discrete probability spaces defined in
the preceding section satisfy these axioms.

General probability spaces arise in randomized algorithms when we have
an algorithm that might consume an unbounded number of random bits.
The problem now is that an outcome consists of countable sequence of
bits, and there are uncountably many such outcomes.
The solution is to consider as measurable events only those sets with
the property that membership in them can be determined after a finite
amount of time.  Formally, the probability space $Ω$ 
is the set $\Set{0,1}^{ℕ}$ of all countably infinite sequences
of 0 and 1 values indexed by the natural numbers, and the
measurable sets $ℱ$ are all sets that can be generated by
countable unions\footnote{As well as complements and countable
intersections.  However, it is not hard to show that that sets defined using these
operations can be reduced to countable unions of cylinder sets.}
of
\indexConcept{cylinder set}{cylinder sets}, where a cylinder set
consists of all extensions $xy$ of some finite prefix $x$.  
The probability measure itself is obtained by assigning the set of all
points that start with $x$ the probability $2^{-\card*{x}}$, and computing
the probabilities of other sets from the axioms.\footnote{This
turns out to give the same probabilities as if we consider each
outcome as a real number in the interval $[0,1]$ and use Lebesgue
measure to compute the probability of events.  For some applications,
thinking of our random values as real numbers (or even sequences of
real numbers) can make things easier: consider for example what
happens when we want to choose one of three outcomes with equal
probability.}

An oddity that arises in general probability spaces is it may be that
every particular outcome has probability zero but their union has
probability $1$.  For example, the probability of any particular
infinite string of bits is $0$, but the set containing all such
strings is the entire space and has probability $1$.  This is where
the fact that probabilities only add over \emph{countable} unions comes in.

Most randomized algorithms books gloss over general probability
spaces, with three good reasons.  The first is that 
if we truncate an algorithm after a
finite number of steps, we are usually get back to a discrete probability
space, which avoids a lot of worrying about measurability and
convergence.
The second is that we are often implicitly working in a probability
space that is either discrete or well-understood (like the space of
bit-vectors described above).
The last is that the 
\index{theorem!Kolmogorov extension}
\concept{Kolmogorov extension theorem}
says that if we specify 
$\Prob{A_1 \cap A_2 \cap \dots \cap A_k}$ consistently
for all
finite sets of events $\Set{A_1 \dots A_k}$, then there exists some
probability space that makes these probabilities work, even if we have
uncountably many such events.  So it's usually enough to specify how
the events we care about interact, without worrying about the details
of the underlying space.

\section{Boolean combinations of events}

Even though events are defined as sets, we often think of them as
representing propositions that we can combine using the usual Boolean
operations of NOT ($\neg$), AND ($∧$), and OR ($∨$).  In terms
of sets, these correspond to taking a complement $\bar{A} =
Ω ∖ A$, an 
intersection $A \cap B$, or a union $A \cup B$.

We can use the axioms of probability to calculate the probability of
$\bar{A}$:

\begin{lemma}
    \label{lemma-negation}
    \begin{align*}
        \Prob{\bar{A}} = 1-\Prob{A}.
    \end{align*}
\end{lemma}
\begin{proof}
    First note that $A \cap \bar{A} = \emptyset$, so $A \cup
    \bar{A} = Ω$ is a disjoint union of countably
    many\footnote{Countable need not be infinite, so 2 is countable.}
    events.  This gives $\Prob{A} + \Prob{\bar{A}} = \Prob{Ω} =
    1$.
\end{proof}

For example, if our probability space consists of the six outcomes of
a fair die roll, and $A = [\text{outcome is $3$}]$ with $\Prob{A} =
5/6$, then $\Prob{\text{outcome is not $3$}} = \Prob{\bar{A}} =
1-1/6 = 5/6$.  Though this example is trivial, using the formula does
save us from having to add up the five cases where we don't get $3$.

If we want to know the probability of $A \cap B$, we need to know more
about the relationship between $A$ and $B$.  For example, it could be
that $A$ and $B$ are both events representing a fair coin coming up
heads, with $\Prob{A} = \Prob{B} = 1/2$.  The probability of $A \cap B$
could be anywhere between $1/2$ and $0$:
\begin{itemize}
    \item For ordinary fair coins, we'd expect that half the time that
        $A$ happens, $B$ also happens.   
        This gives $\Prob{A \cap B} = (1/2)⋅ (1/2) = 1/4$.
        To make this formal, we might define our
        probability space $Ω$ as having four outcomes $\coinFlips{HH}$, $\coinFlips{HT}$,
        $\coinFlips{TH}$, and $\coinFlips{TT}$, each of which occurs with equal probability.
    \item But maybe $A$ and $B$ represent the same fair coin: then $A \cap
        B = A$ and $\Prob{A\cap B} = \Prob{A} = 1/2$.
    \item At the other extreme, maybe $A$ and $B$ represent two fair
        coins welded together so that if one comes up heads the other
        comes up tails.  Now $\Prob{A\cap B} = 0$.
    \item With a little bit of tinkering, we could also find probabilities
        for the outcomes in our four-outcome space to make 
        $\Prob{A} = \Prob{\coinFlips{HH}} + \Prob{\coinFlips{HT}} = 1/2$ and
        $\Prob{B} = \Prob{\coinFlips{HH}} + \Prob{\coinFlips{TH}} = 1/2$ while setting
        $\Prob{A \cap B} = \Prob{\coinFlips{HH}}$ to any value between $0$ and $1/2$.
\end{itemize}

The difference between the nice case where $\Prob{A \cap B}$ equals $1/4$ and
the other, more annoying cases where it doesn't is that in the first
case we have assumed that $A$ and $B$ are
\index{independence!of events}\index{independence}\index{independent events}\index{events!independent}\concept{independent}, which
is \emph{defined} to mean that $\Prob{A \cap B} = \Prob{A} \Prob{B}$.

In the real world, we expect events to be independent if they refer to
parts of the universe that are not causally related: if we flip
two coins
that aren't glued together somehow, then we assume that the
outcomes of the coins are independent.  But we can also get
independence from events that are not causally disconnected
in this way.  An example would be if we rolled a fair four-sided die
labeled $\coinFlips{HH}, \coinFlips{HT}, \coinFlips{TH}, \coinFlips{TT}$, where we take the first letter as
representing $A$ and the second as $B$.

There's no simple formula for $\Prob{A \cup B}$ when $A$ and $B$ are not
disjoint, even for independent
events, but we can compute the probability by splitting up into
smaller, disjoint events and using countable additivity:
\begin{align*}
    \Prob{A \cup B}
    &= \Prob{(A \cap B) \cup (A \cap \bar{B}) \cup (\bar{A} \cap B)} \\
    &= \Prob{A \cap B} + \Prob{A \cap \bar{B}} + \Prob{\bar{A} \cap B} \\
    &= \left(\Prob{A \cap B} + \Prob{A \cap \bar{B}}\right) 
        + \left(\Prob{\bar{A} \cap B} + \Prob{A \cap B}\right) - \Prob{A \cap B}\\
    &= \Prob{A} + \Prob{B} - \Prob{A \cap B}.
\end{align*}

The idea is that we can compute $\Prob{A\cup B}$ by adding up the
individual probabilities and then subtracting off the part where the
counted the event twice.

This is a special case of the general
\concept{inclusion-exclusion formula},
which says:
\begin{lemma}
\label{lemma-inclusion-exclusion}
For any finite sequence of events $A_1 \dots A_n$,
\begin{align}
\Prob{\bigcup_{i=1}^{n} A_i}
&=  ∑_{i}   \Prob{A_i}
   -∑_{i<j} \Prob{A_i \cap A_j}
   +∑_{i<j<k} \Prob{A_i \cap A_j \cap A_k}
   - \dots \nonumber \\
&= ∑_{S ⊆ \Set{1 \dots n}, S ≠ \emptyset}
   (-1)^{\card*{S}+1} \Prob{\bigcap_{i∈ S} A_i}.
    \label{eq-inclusion-exclusion}
\end{align}
\end{lemma}
\begin{proof}
Partition $Ω$ into $2^n$ disjoint events $B_T$,
where $B_T = \left(\bigcap_{i ∈ T} A_i\right) 
\cap \left(\bigcap_{i \notin T} \bar{A}_i\right)$
is the event that all $A_i$ occur for $i$ in $T$ and no
$A_i$ occurs for $i$ not in $T$.
Then $A_i$ is the union of all $B_T$ with $T \ni i$ and 
$\bigcup A_i$ is the union of all $B_T$ with
$T ≠ \emptyset$.

That the right-hand side gives the probability of this event is a
sneaky consequence of the binomial theorem, and in particular the
fact that $∑_{i=1}^{n} (-1)^i = ∑_{i=0}^{n} (-1)^i - 1
= (1-1)^n - 1$ is $-1$ if $n > 0$ and $0$ if $n=0$.
Using this fact after rewriting the right-hand side using the $B_T$
events gives
\begin{align*}
∑_{S ⊆ \Set{1 \dots n}, S ≠ \emptyset}
(-1)^{\card*{S}+1} \Prob{\bigcap_{i∈ S} A_i}
&= ∑_{S ⊆ \Set{1 \dots n}, S ≠ \emptyset}
(-1)^{\card*{S}+1} ∑_{T ⊇ S} \Prob{B_T} \\
    &= ∑_{T ⊆ \Set{1 \dots n}}
    \left(\Prob{B_T}
   ∑_{S ⊆ T, S ≠ \emptyset}
   (-1)^{\card*{S}+1} \right) \\
     &= ∑_{T ⊆ \Set{1 \dots n}}
     \left(
      - \Prob{B_T} 
      ∑_{i=1}^{n} (-1)^i \binom{\card*{T}}{i}
   \right) \\
   &= ∑_{T ⊆ \Set{1 \dots n}}
     \left(
      - \Prob{B_T} 
      ((1-1)^{\card*{T}} - 1)
   \right) \\
   &= ∑_{T ⊆ \Set{1 \dots n}}
   \Prob{B_T} \left((1-0^{\card*{T}}) \right) \\
     &= ∑_{T ⊆ \Set{1 \dots n}, T ≠ \emptyset} \Prob{B_T} \\
     &= \Prob{\bigcup_{i=1}^{n} A_i}.
 \end{align*}
 \end{proof}

\section{Conditional probability}
\label{section-conditional-probability}

The \index{probability!conditional}\index{conditional probability}\concept{probability of $A$
conditioned on $B$} or \concept{probability of $A$ given $B$}, written
$\ProbCond{A}{B}$, is defined by
\begin{align}
    \ProbCond{A}{B} &= \frac{\Prob{A\cap B}}{\Prob{B}},
    \label{eq-condition-probability-definition}
\end{align}
provided
$\Prob{B≠ 0}$.  If $\Prob{B} = 0$, we can't condition on
$B$.

Such conditional probabilities represent the effect of restricting our
probability space to just $B$, which can think of as computing the
probability of each event if we know that $B$ occurs.  The
intersection in the numerator limits $A$ to circumstances where $B$
occurs, while the denominator normalizes the probabilities so that,
for example, $\ProbCond{Ω}{B} = \ProbCond{B}{B} = 1$.

\subsection{Conditional probability and independence}

Rearranging \eqref{eq-condition-probability-definition} gives
$\Prob{A \cap B} = \Prob{B} \ProbCond{A}{B} = \Prob{A} \ProbCond{B}{A}$.
In many cases, knowing that $B$ occurs tells us nothing about whether
$A$ occurs; if so, we have $\ProbCond{A}{B} = \Prob{B}$, which implies that
$\Prob{A \cap B} = \ProbCond{A}{B} \Prob{B} = \Prob{A} \Prob{B}$—events $A$ and $B$
are \concept{independent}.
So $\ProbCond{A}{B} = \Prob{A}$ gives an alternative criterion for independence
when $\Prob{B}$ is nonzero.\footnote{If $\Prob{B}$ is zero, then $A$ and
$B$ are always independent.}

A \emph{set} of events $A_1, A_2, \dots$ is 
\index{independence!of sets of events}
independent if $A_i$ is
independent of $B$ 
when $B$ is any Boolean formula of the $A_j$ for $j≠ i$.  The idea
is that you can't predict $A_i$ by knowing anything about the rest of
the events.

A set of events $A_1, A_2, \dots$ is 
\index{independence!pairwise}
\concept{pairwise independent} if
each $A_i$ and $A_j$ are independent when $i≠j$.  It is possible for a
set of events to be pairwise independent but not independent; a simple
example is when $A_1$ and $A_2$ are the events that two independent
coins come up heads and $A_3$ is the event that both coins come up
with the same value.  The general version of pairwise independence is
\concept{$k$-wise independence}, which means that any subset of $k$
(or fewer) events are independent.

\subsection{Conditional probability and the law of total probability}
\label{section-law-of-total-probability}

The reason we like conditional probability in algorithm analysis is
that it gives us a natural way to model the kind of case analysis
that we are used to applying to deterministic algorithms.  Suppose we
are trying to prove that a randomized algorithm works (event $A$)
with a certain probability.  Most likely, the first random thing the algorithm does
is flip a coin, giving two possible outcomes $B$ and $\bar{B}$.
Countable additivity tells us that
$\Prob{A} = \Prob{A\cap B} + \Prob{A \cap \bar{B}}$, which we can rewrite
using conditional probability as
\begin{align}
    \Prob{A} &= \ProbCond{A}{B} \Prob{B} + \ProbCond{A}{\bar{B}}
\Prob{\bar{B}},
\end{align}
a special case of the \concept{law of total probability}.

What's nice about this expression is that we can often 
compute $\ProbCond{A}{B}$ and $\ProbCond{A}{\bar{B}}$ by looking at what the
algorithm does starting from the point where it has just gotten heads
($B$) or tails ($\bar{B}$), and use the formula to combine these
values to get the overall probability of success.

For example, if
\begin{align*}
\ProbCond{\text{class occurs}}{\text{snow}} &= 3/5, \\
\ProbCond{\text{class occurs}}{\text{no snow}} &= 99/100, \text{and} \\
\Prob{\text{snow}} &= 1/10, \\
\intertext{then}
\Prob{\text{class occurs}}
&= (3/5)⋅(1/10) + (99/100)⋅(1-1/10) &= 0.951.
\end{align*}

More generally, we can do the same computation for any partition of
$Ω$ into countably many disjoint events $B_i$:
\begin{align}
\Prob{A}
&= \Prob{\bigcup_i (A \cap B_i)}  \nonumber \\ \nonumber
&= ∑_i \Prob{A \cap B_i} \\
&= ∑_{i,\Prob{B_i} ≠ 0} \ProbCond{A}{B_i} \Prob{B_i},
\label{eq-law-of-total-probability}
\end{align}
which is the \concept{law of total probability}. Note that the last
step works for each term only if
$\ProbCond{A}{B_i}$ is well-defined, meaning
that $\Prob{B_i} ≠ 0$. But any case where $\Prob{B_i} = 0$ 
also has $\Prob{A∩B_i} = 0$,
so we get the correct answer 
if we simply omit these terms from both sums.

A special case arises when $\ProbCond{A}{\bar{B}} = 0$, which occurs, for
example, if $A ⊆ B$.  Then we just have
$\Prob{A} = \ProbCond{A}{B} \Prob{B}$.  If we consider an event $A = A_1 \cap A_2
\cap \dots \cap A_k$, then we can iterate this expansion to get
\begin{align}
\Prob{A_1 \cap A_2 \cap \dots \cap A_k}
&= \Prob{A_1 \cap \dots \cap A_{k-1}} \ProbCond{A_k}{A_1, \dots, A_{k-1}}
\nonumber \\
\nonumber
&= \Prob{A_1 \cap \dots \cap A_{k-2}} 
   \ProbCond{A_{k-1}}{A_1, \dots, A_{k-2}}
   \ProbCond{A_k}{A_1, \dots, A_{k-1}} \\ \nonumber
&= \dots \\ 
&= \prod_{i=1}^{k} \ProbCond{A_i}{A_1, \dots, A_i}.
\label{eq-conditional-probability-product}
\end{align}

Here $\ProbCond{A}{B,C,\dots}$ is short-hand for $\Prob{B\cap C\cap \dots}$,
the probability that $A$ occurs given that all of $B$, $C$, etc.,
occur.

\subsection{Examples}

Here we have some examples of applying conditional probability to
algorithm analysis.  Mostly we will be using some form of the law of
total probability.

\subsubsection{Racing coin-flips}
\label{section-racing-coin-flips}

Suppose that I flip coins and allocate a space for each heads that I
get before the coin comes up tails.  Suppose that you then supply me
with objects (each of which takes up one unit of space), one for each heads
that \emph{you} get before you get tails.  What are my chances of allocating
enough space?

Let's start by solving this directly using the law of total
probability.
Let $A_i$ be the event that I allocate $i$ spaces.  The event $A_i$ is
the intersection of $i$ independent events that I get heads in the
first $i$ positions and the event that I get tails in position $i+1$;
this multiplies out to $(1/2)^{i+1}$.  Let $B_i$ be the similar event
that you supply $i$ objects.  Let $W$ be the event that I win.
To make the $A_i$ partition the space, we must also add an extra event
$A_∞$ equal to the singleton set $\Set{ \coinFlips{HHHHHHH}\dots }$ consisting
of the all-$\coinFlips{H}$ sequence; this has probability $0$ (so it won't have
much of an effect), but we need to include it since $\coinFlips{HHHHHHH}\dots$ is
not contained in any of the other $A_i$.

We can compute
\begin{align}
\ProbCond{W}{A_i}
&= \ProbCond{B_0 \cap B_1 \cap \dots \cap B_i}{A_i} \nonumber \\ \nonumber
&= \Prob{B_0 \cap B_1 \cap \dots \cap B_i} \\ \nonumber
&= \Prob{B_0} + \Prob{B_1} + \dots + \Prob{B_i} \\ \nonumber
&= ∑_{j=0}^{i} (1/2)^i \\ \nonumber
&= (1/2) ⋅ \frac{1-(1/2)^{i+1}}{1-1/2} \\ 
&= 1-(1/2)^{i+1}.\label{eq-racing-coin-flips-conditional}
\end{align}

The clean form of this expression suggests strongly that there is a
better way to get it, and that this way involves taking the negation
of the intersection of $i+1$ independent events that occur with
probability $1/2$ each.  With a little reflection, we can see that the
probability that your objects \emph{don't} fit in my buffer is exactly
$(1/2)^{i+1}$

From the law of total probability \eqref{eq-law-of-total-probability},
\begin{align*}
\Prob{W}
&= ∑_{i=0}^{∞} (1-(1/2)^{i+1}) (1/2)^{i+1} \\
&= 1 - ∑_{i=0}^{∞} (1/4)^{i+1} \\
&= 1 - \frac{1}{4} ⋅ \frac{1}{1-1/4} \\
&= 2/3.
\end{align*}

This gives us our answer.  However, we again see an answer that is
suspiciously simple, which suggests looking for another way to find
it.  We can do this using conditional probability by defining new events
$C_i$, where $C_i$ contains all sequences of coin-flips for both
players where get $i$ heads in a row but at least
one gets tails on the $(i+1)$-th coin.  These events plus the
probability-zero event
$C_∞ = \Set{ \coinFlips{HHHHHHH}\dots, \coinFlips{TTTTTTT}\dots }$ partition the space, so
$\Prob{W} = ∑_{i=0}^∞ \ProbCond{W}{C_i} \Prob{C_i}$.

Now we ask, what is $\ProbCond{W}{C_i}$?  Here we only need to consider three
cases, depending on the outcomes of our $(i+1)$-th coin-flips.  The
cases $\Tuple{\coinFlips{H},\coinFlips{T}}$ and
$\Tuple{\coinFlips{T},\coinFlips{T}}$ cause me to win, while the case
$\Tuple{\coinFlips{T},\coinFlips{H}}$ causes me to
lose, and each case occurs with equal probability conditioned on $C_i$
(which excludes $\Tuple{\coinFlips{H},\coinFlips{H}}$).  So I win $2/3$ of the time conditioned on
$C_i$, and summing $\Prob{W} = ∑_{i=0}^∞ (2/3) \Prob{C_i} = 2/3$
since $\Prob{C_i}$ sums to $1$, because the union of these
events includes the entire space except for the probability-zero event $C_∞$.

Still another approach is to compute the probability that our runs
have exactly the same length ($∑_{i=1}^{∞} 2^{-i}⋅2^{-i} = 1/3$), and argue by symmetry that the
remaining $2/3$ probability is equally split between my run being
longer ($1/3$) and
your run being longer ($1/3$).  Since $W$ occurs if my run is just as
long or longer, $\Pr[W] = 1/3 + 1/3 = 2/3$.  A nice property of this
approach is that the only summation involved is over disjoint events, so 
we get to avoid using conditional probability entirely.

\subsubsection{Karger's min-cut algorithm}
\label{section-karger-min-cut}

\index{Karger's min-cut algorithm}
Here we'll give a simple algorithm for finding a global
\index{cut!minimum}
\index{minimum cut}
\concept{min-cut} in a
\concept{multigraph},\footnote{Unlike ordinary
graphs, multigraphs can have more than one edge between two vertices.}
due to David Karger~\cite{Karger1993}.

The idea is that we are given a multigraph $G$,
and we want to partition the vertices into nonempty sets $S$ and
$T$ such that the number of edges with one endpoint in $S$ and one
endpoint in $T$ is as small as possible.
There are many efficient ways to do
this, most of which are quite sophisticated.  There is also the algorithm
we will now present,
which
solves the problem with reasonable efficiency
using almost no sophistication at all (at least in
the algorithm itself).

The main idea is that given an edge $uv$, we can construct a new
multigraph $G_{1}$ by \indexConcept{contraction}{contracting} the
edge: in $G_{1}$, $u$ and $v$ are replaced by a single vertex, and any
edge that used to have either vertex as an endpoint now goes to the
combined vertex (edges with both endpoints in $\Set{u,v}$ are deleted).
Karger's algorithm is to contract edges chosen uniformly at random
until only two vertices remain.  All the vertices that got packed into
one of these become $S$, the others become $T$.  It turns out that 
this finds a 
minimum cut with probability at least $1/\binom{n}{2}$.

An example of the algorithm in action is given in
Figure~\ref{fig-karger-min-cut}.

\begin{figure}
    \centering
    \begin{tabular}{c}
    \begin{tikzpicture}
        \node (a) at (-0.7,0.5) {$a$};
        \node (b) at (-0.7,-0.5) {$b$};
        \node (c) at (0,0) {$c$};
        \node (d) at (1,0) {$d$};
        \node (e) at (1.7,0.5) {$e$};
        \node (f) at (1.7,-0.5) {$f$};
        \path
        (a) edge (b) edge (c)
        (b) edge (c)
        (c) edge (d)
        (d) edge (e) edge (f)
        (e) edge (f)
        ;
    \end{tikzpicture}
        \\
    \begin{tikzpicture}
        \node (ab) at (-1,0) {$ab$};
        \node (c) at (0,0) {$c$};
        \node (d) at (1,0) {$d$};
        \node (e) at (1.7,0.5) {$e$};
        \node (f) at (1.7,-0.5) {$f$};
        \path
        (ab) edge[bend left] (c)
        (ab) edge[bend right] (c)
        (c) edge (d)
        (d) edge (e) edge (f)
        (e) edge (f)
        ;
    \end{tikzpicture}
        \\
    \begin{tikzpicture}
        \node (ab) at (-1,0) {$ab$};
        \node (c) at (0,0) {$c$};
        \node (df) at (1,0) {$df$};
        \node (e) at (1.7,0.5) {$e$};
        \path
        (ab) edge[bend left] (c)
        (ab) edge[bend right] (c)
        (c) edge (df)
        (df) edge[bend left] (e)
        (df) edge[bend right] (e)
        ;
    \end{tikzpicture}
        \\
    \begin{tikzpicture}
        \node (ab) at (-1,0) {$ab$};
        \node (c) at (0,0) {$c$};
        \node (def) at (1,0) {$def$};
        \path
        (ab) edge[bend left] (c)
        (ab) edge[bend right] (c)
        (c) edge (def)
        ;
    \end{tikzpicture}
        \\
    \begin{tikzpicture}
        \node (abc) at (-1,0) {$abc$};
        \node (def) at (1,0) {$def$};
        \path
        (abc) edge (def)
        ;
    \end{tikzpicture}
    \end{tabular}
    \caption[Karger's min-cut algorithm]{Karger's min-cut algorithm.
    Initial graph (at top) has min cut $\Tuple{\Set{a,b,c},\Set{d,e,f}}$.  We
    find this cut by getting lucky and contracting edges $ab$, $df$, $de$,
    and $ac$ in that order.  The final graph (at bottom) gives the cut.}
    \label{fig-karger-min-cut}
\end{figure}

\begin{theorem}
\label{theorem-karger-min-cut}
Given any min cut $(S,T)$ of a graph $G$ on $n$ vertices,
Karger's algorithm outputs
$(S,T)$ with probability at least $1/\binom{n}{2}$.
\end{theorem}
\begin{proof}
Let $(S,T)$ be a min cut of size $k$.  Then the degree of each vertex
$v$ is at least $k$ (otherwise $(v,G-v)$ would be a smaller cut), and
$G$ contains at least $kn/2$ edges.  The probability that we contract
an $S$–$T$ edge is thus at most $k/(kn/2) = 2/n$, and the probability
that we don't contract one is $1-2/n = (n-2)/n$.  Assuming we missed
collapsing $(S,T)$ the first time, we now have a new graph $G_{1}$
with $n-1$ vertices in which the min cut is still of size $k$.  So now
the chance that we miss $(S,T)$ is $(n-3)/(n-1)$.  We stop when we
have two vertices left, so the last step succeeds with probability
$1/3$.  

We can compute the probability that the $S$–$T$ cut is never
contracted by applying \eqref{eq-conditional-probability-product},
which just tells us to multiply all the conditional probabilities
together:
\begin{align*}
\prod_{i=3}^{n} \frac{i-2}{i} &= \frac{2}{n(n-1)}.
\end{align*}
\end{proof}

This tells us what happens when we are considering a particular
min cut.
If the graph has more than one min cut, this only makes our life
easier.  Note that since each min cut turns up with probability at
least $1/\binom{n}{2}$, there can't be more than $\binom{n}{2}$ of
them.\footnote{The suspiciously combinatorial appearance of the
$1/\binom{n}{2}$ suggests that there should be some way of associating
minimum cuts with particular pairs of vertices, but I'm not aware of 
any natural way to do this. Sometimes the appearance of a simple
expression in a surprising context 
may just stem from the fact that there aren't very many
distinct simple expressions.}
But even if there is only one, we have a good chance of finding
it if we simply re-run the algorithm substantially more than $n^{2}$ times.

\myChapter{Random variables}{2025}{}
\label{chapter-random-variables}

Probabilities are fine when all we want to do is ask whether an
algorithm worked or not. But for many randomized algorithms
(particularly Las Vegas algorithms), we can structure the algorithm so
that it works eventually with probability $1$. This makes the
important question that of how long is eventually. To measure a
quantity like running time that depends on the random choices of the
algorithm, we need a \concept{random variable}, which is just a
function whose domain is some probability space $Ω$.\footnote{Technically, this only works
for discrete spaces.  In general, a random variable
is a 
\index{function!measurable}
\concept{measurable function} from a probability space
$(Ω,ℱ)$ to some other set $S$ equipped with its own
$σ$-algebra $ℱ'$.  What makes a function measurable in this
sense is that that for any set $A$ in $ℱ'$, the
inverse image $f^{-1}(A)$ must be in $ℱ$.
See §\ref{section-measurability} for more details.}

Even though random variables are just functions,
rather than writing a random variable
as $f(ω)$ everywhere, the convention is to write a random
variable as a capital letter ($X$, $Y$, $S$, etc.) and make the
argument implicit: $X$ is really $X(ω)$.
Variables that aren't random (or aren't variable)
are written in lowercase.

Most of the random variables we will consider will be 
\index{random variable!discrete}
\index{discrete random variable}
\conceptFormat{discrete random variables}.  A discrete random variable
takes on only countably many values, each with some nonzero
probability.

For example, consider the probability space corresponding to
rolling two independent fair six-sided dice.  There are 36 possible
outcomes in this space, corresponding to the $6×6$ pairs of
values $\Tuple{ x,y }$ we might see on the two dice.  
We could represent the value of each die as a random variable
$X$ or $Y$ given by $X(\Tuple{ x,y }) = x$ or
$Y(\Tuple{ x,y }) = y$,
but for many
applications, we don't care so much about the specific values on each
die.  Instead, we want to know the
sum $S = X+Y$
of the dice.  This value $S$ is also random variable; as a function on
$Ω$, it's defined by $S(\Tuple{ x,y }) = x+y$.

Random variables need not be real-valued.  
There's no reason why we can't think of the pair $\Tuple{ x,y }$ 
itself a
random variable, whose range is the set $[1\dots 6]× [1\dots 6]$.
Similarly, if we imagine choosing a
point uniformly at random in the unit square $[0,1]^2$, its
coordinates are a random variable.  
For a more exotic example, the \index{graph!random}\concept{random
graph} $G_{n,p}$ obtained by starting with $n$ vertices and including
each possible edge with independent probability $p$ is a random
variable whose range is the set of all graphs on $n$ vertices.

\section{Operations on random variables}

Random variables may be combined using standard arithmetic operators,
have functions applied to them, etc., to get new random variables.
For example, the random variable $X/Y$ is a function from $Ω$
that takes on the value $X(ω)/Y(ω)$ on each point $ω$.

\section{Random variables and events}

Any random variable $X$ allows us to define events based on its
possible values.  
Typically these are expressed by writing a predicate involving the
random variable in square brackets.
An example would be the probability that the sum of
two dice is exactly $11$: $[S = 11]$; or that the sum of the dice is
less than $5$: $[S < 5]$.  These are both sets of outcomes; we could
expand $[S = 11] = \Set{ \Tuple{ 5,6 }, \Tuple{ 6,5 } }$ or 
$[S < 5] = \Set{ \Tuple{ 1,1 }, \Tuple{ 1,2 }, \Tuple{ 1,3
}, \Tuple{ 2,1 }, \Tuple{ 2,2 }, \Tuple{ 3,1 } }$.  This allows us
to calculate the probability that a random variable has particular
properties: $\Prob{S = 11} = \frac{2}{36} = \frac{1}{18}$ and
$\Prob{S < 5} = \frac{6}{36} = \frac{1}{6}$.

Conversely, given any event $A$, we can define an
\index{random variable!indicator}
\concept{indicator random variable} $1_A$ that is $1$ when $A$ occurs
and $0$ when it doesn't.\footnote{Some people like 
writing $χ_A$ for these.

You may also see $[P]$ where $P$ is some
predicate, a convention known
as \concept{Iverson notation} 
or the \concept{Iverson bracket} that was invented by Iverson for the
programming language APL, appears in later languages like $C$ where
the convention is that true predicates evaluate to $1$ and false ones
to $0$, and ultimately popularized for use in mathematics—with the
specific choice of square brackets to set off the predicate—by 
Graham~\etal~\cite{GrahamKP1988}.

Out of these alternatives, I personally find $1_A$ to be the least
confusing.}
Formally, $1_A(ω) = 1$ for $ω$ in
$A$ and $1_A(ω) = 0$ for $ω$ not in $A$.

Indicator variables are mostly useful when combined with other random
variables.  For example, if you roll two dice and normally collect the sum of
the values but get nothing if it is $7$, we could write your payoff as 
$S⋅ 1_{[S ≠ 7]}$.

The \concept{probability mass function}
of a random variable gives $\Prob{X=x}$ for
each possible value $x$.
For example, our random variable $S$ has the probability mass function
show in Table~\ref{table-2d6-probability-mass-function}.
For a discrete random variable $X$, the probability mass function gives
enough information to calculate the probability of any event involving
$X$, since we can just sum up cases using countable additivity.  This
gives us another way to compute $\Prob{S<5} = \Prob{S=2} + \Prob{S=3} +
\Prob{S=4} = \frac{1+2+3}{36} = \frac{1}{6}$.

\begin{table}
    \renewcommand{\arraystretch}{1.5} \centering
    \begin{tabular}{cc}
        $S$ & Probability \\        
        \hline
2 & $1/36$ \\
3 & $2/36$ \\
4 & $3/36$ \\
5 & $4/36$ \\
6 & $5/36$ \\
7 & $6/36$ \\
8 & $5/36$ \\
9 & $4/36$ \\
10 & $3/36$ \\
11 & $2/36$ \\
12 & $1/36$ \\
    \end{tabular}
    \caption[Sum of two dice]{Probability mass function for the sum of two independent fair
six-sided dice}
\label{table-2d6-probability-mass-function}
\end{table}

For two random variables, the \concept{joint probability mass
function} gives $\Prob{X=x ∧ Y=y}$ for each pair of values $x$ and
$y$. This generalizes in the obvious way for more than two variables.

We will often refer to the probability mass function as giving the
\concept{distribution} or \concept{joint distribution} of a random
variable or collection of random variables, even though distribution
(for real-valued variables)
technically refers to the \concept{cumulative distribution function} 
$F(x) = \Prob{X ≤ x}$, which is 
generally not directly computable from the probability mass function for
\indexConcept{continuous random variable}{continuous random variables} that take on uncountably many
values.  To the extent that we can, we will try to avoid continuous
random variables, and the rather messy integration theory needed to handle
them.

Two or more random variables are \indexConcept{independent random
variables}{independent} if all sets of events involving different
random variables are independent.  In terms of probability mass
functions, $X$ and $Y$ are independent if 
$\Prob{X = x ∧ Y=y} = \Prob{X=x} ⋅ \Prob{Y=y}$ for any constants
$x$ and $y$.
In terms of cumulative distribution functions, $X$ and $Y$ are
independent if
$\Prob{X ≤ x ∧ Y≤y} = \Prob{X=x} ⋅ \Prob{Y=y}$ for any
constants $x$ and $y$.
As with events, we generally assume that random variables associated
with causally disconnected processes are independent, but this is not the
only way we might have independence.

It's not hard to see that the individual die values $X$ and $Y$ in our
two-dice example are independent, because every possible combination
of values $x$ and $y$ has the same probability $1/36 = \Prob{X=x}
\Prob{Y=y}$.  If we chose a different probability distribution on the
space, we might not have independence.

\section{Measurability}
\label{section-measurability}

For discrete probability spaces, any function on outcomes can
be a random variable.
The reason is that any event in a discrete probability space has a
well-defined probability.
For more general spaces,
in order to be useful, events involving a
random variable should have well-defined probabilities.  For
\index{random variable!discrete}
\indexConcept{discrete random variable}{discrete random variables}
that take on only countably many values (e.g., integers or rationals),
it's enough for the event $[X=x]$ 
(that is, the set $\Set{ω \mid X(ω) = x}$) 
to be in $ℱ$
for all $x$.  For real-valued random variables, we ask that the event
$[X ≤ x]$ be in $ℱ$.  In these cases, we say that $X$ is
\concept{measurable} with respect to $ℱ$, or just
\concept{measurable} $ℱ$.  More exotic random variables use
a definition of measurability that generalizes the real-valued
version, which we probably won't need.\footnote{The general version is
that if $X$ takes on values on another measure space $(Ω',
ℱ')$, then the inverse image 
$X^{-1}(A) = \Set{ω ∈ Ω \mid X(ω) ∈ A}$ of any set $A$ in
$ℱ'$ is in $ℱ$.  This means in particular that
$\Pr_{Ω}$ maps through $X$ to give a probability measure on
$Ω'$ by $\Pr_{Ω'}[A] = \Pr_{Ω}[X^{-1}(A)]$, and the
condition on $X^{-1}(A)$ being in $ℱ$ makes this work.}
Since we usually just assume that all of our random variables are
measurable unless we are doing something funny with $ℱ$ to
represent ignorance, this issue won't come up much.

\section{Expectation}
\label{section-expectation}

The \concept{expectation} or \concept{expected value} of a random
variable $X$ is given by $\Exp{X} = ∑_x x \Prob{X=x}$.  
This is essentially an average value of $X$ weighted by probability,
and it only makes sense
if $X$ takes on values that can be summed in this way (e.g., real or
complex values, or vectors in a real- or complex-valued vector space).
Even if the expectation makes sense, it may be that a particular
random variable $X$ doesn't have an expectation, because the sum fails
to converge.\footnote{Example: Let $X$ be the number of times you flip
a fair coin until it comes up heads.  We'll see later than $\Exp{X} =
1/(1/2) = 2$.  But $\Exp{2^X} = ∑_{n=1}^{∞} 2^n 2^{-n} = ∑_{n=1}^{∞}
1$, which diverges.  With some tinkering it is possible to come up
with even uglier cases, like an array that contains $1$ element on
average but requires infinite expected time to sort using a $Θ(n \log
n)$ algorithm.}

For an example that does work, if $X$ and $Y$ are independent fair six-sided dice, then
$\Exp{X} = \Exp{Y} = ∑_{k=1}^{6} k \left(\frac{1}{6}\right) =
\frac{21}{6} = \frac{7}{2}$, while $\Exp{X+Y}$ is the rather horrific
\begin{align*}
∑_{k=2}^{12} k ⋅ \Prob{X+Y = i} 
&= \frac{
\mbox{\tiny $
  2 ⋅ 1
+ 3 ⋅ 2
+ 4 ⋅ 3
+ 5 ⋅ 4
+ \dots
+ 9 ⋅ 4
+ 10 ⋅ 3
+ 11 ⋅ 2
+ 12 ⋅ 1
$}}{36} \\
&= \frac{252}{36} = 7.
\end{align*}

The fact that $7 = \frac{7}{2} + \frac{7}{2}$ here is not a
coincidence, but a consequence of \concept{linearity of expectation},
which is the subject of the next section.

\subsection{Linearity of expectation}
\label{section-linearity-of-expectation}

The main reason we like expressing the run times of algorithms in
terms of expectation is
\index{expectation!linearity of}
\concept{linearity of expectation}: $\Exp{aX+bY} = \Exp{aX} + \Exp{bY}$
for all random variables $X$ and $Y$ for which $\Exp{X}$ and $\Exp{Y}$ are
defined, and all constants $a$ and $b$.
This means that we can compute the running time for different parts of
our algorithm separately and then add them together, \emph{even if the
costs of different parts of the algorithm are not independent}.

The general version is $\Exp{∑ a_i X_i} = ∑ a_i
\Exp{X_i}$ for any \emph{finite}
collection of random variables $X_i$ and constants $a_i$,
which follows by applying induction to
the two-variable case.  A special case is $\Exp{cX} = c\Exp{X}$
when $c$ is constant.

For discrete random variables, linearity of expectation follows
immediately from the definition of expectation and the fact that the
event $[X=x]$ is the disjoint union of the events $[X=x, Y=y]$ for all
$y$:
\begin{align*}
\Exp{aX + bY}
&= ∑_{x,y} (ax+by) \Prob{X=x ∧ Y=y} \\
&= a ∑_{x,y} x \Prob{X=x,Y=y} + b ∑_{x,y} y \Prob{X=x,Y=y} \\
&= a ∑_x x ∑_y Pr[X=x,Y=y] + b ∑_y y ∑_x \Prob{X=x,Y=y} \\
&= a ∑_x x \Prob{X=x} + b ∑_y y \Prob{Y=y} \\
&= a \Exp{X} + b \Exp{Y}.
\end{align*}

A technical note: we are assuming that $\Exp{X}$, $\Exp{Y}$, and $\Exp{X+Y}$ all
exist.

This proof does \emph{not} require that $X$ and $Y$ be
independent.  The sum of two fair six-sided dice always has
expectation $\frac{7}{2} + \frac{7}{2} = 7$, whether they are
independent dice, the same die counted twice, or one die $X$ and its
complement $7-X$.

Linearity of expectation makes it easy to compute the expectations of
random variables that can be expressed as sums of other random variables.
One example that will come up a lot is a 
\index{random variable!binomial}
\concept{binomial random variable}, which is a sum
$S = ∑_{i=1}^{n} X_i$ of $n$ independent 
\index{random variable!Bernoulli}\index{Bernoulli random
variable}\conceptFormat{Bernoulli random variables},
each of which is $1$ with probability
$p$ and $0$ with probability $q=1-p$.
These are called binomial random variables because the probability
that $S$ is equal to $k$ is given by
\begin{equation}
    \Prob{S = k} = \binom{n}{k} p^k q^{n-k},
    \label{eq-binomial-random-variable-mass-function}
\end{equation}
which is the $k$-th term in the binomial expansion of $(p+q)^n$.
In this case each $X_i$ has $\Exp{X_i} = p$, so $\Exp{S}$ is
just $np$.  It is possible to calculate this fact directly from
\eqref{eq-binomial-random-variable-mass-function}, but it's
much more
work.\footnote{One way is to use the
\index{generating function!probability}
\concept{probability generating function}
$F(z) = ∑_{k=0}^{∞} \Prob{S = k} z^k = ∑_{k=0}^{∞} \binom{n}{k} p^k
q^{n-k} z^k = (pz+q)^n$.
Then take the derivative $F'(z) = ∑_{k=0}^{∞} \Prob{S=k} k z^{k-1}$ 
and observe $F'(1) = ∑_{k=0}^{∞} \Prob{S=k} k = \Exp{S}$.
Or we can write $F'(z)$ as $n (pz+q)^{n-1} p$, which gives
$F'(1) = n (p+q)^{n-1} p = np$.}

\subsubsection{Linearity of expectation for QuickSort}
\label{section-quicksort-linearity-of-expectation}

Previously (see §\ref{section-quicksort}), we showed that the QuickSort
algorithm used $≤ 2n log n$ comparisons on average by solving a
recurrence. Here we'll show that linearity of expectation lets us
compute the exact expected number of comparisons without quite so much
mess.

Imagine we use the following method for choosing pivots: we generate a
random permutation of all the elements in the array, and when asked to
sort some subarray $A'$, we use as pivot the first element of $A'$
that appears in our list.  Since each element is equally likely to be
first, this is equivalent to the actual algorithm.  Pretend that we
are always sorting the numbers $1\dots n$ and define for each pair of
elements $i < j$ the \concept{indicator variable} $X_{ij}$ to be $1$
if $i$ is compared to $j$ at some point during the execution of the
algorithm and $0$ otherwise.  Amazingly, we can actually compute the
probability of this event (and thus $\Exp{X_{ij}}$): the only time $i$
and $j$ are compared is if one of them is chosen as a pivot before
they are split up into different arrays.  How do they get split up
into different arrays?  This happens if some intermediate element $k$ is chosen as
pivot first, that is, if some $k$ with $i < k < j$ appears in the
permutation before both $i$ and $j$.  Occurrences of other elements
don't affect the outcome, so we can concentrate on the restriction of
the permutations to just the numbers $i$ through $j$, and we win if
this restricted permutation starts with either $i$ or $j$.  This event
occurs with probability $2/(j-i+1)$, so we have $\Exp{X_{ij}} =
2/(j-i+1)$.  Summing over all pairs $i < j$ gives:
\begin{align*}
\Exp{∑_{i<j} X_{ij}}
&= ∑_{i<j} \Exp{X_{ij}} \\
&= ∑_{i<j} \frac{2}{j-i+1} \\
&= ∑_{i=1}^{n-1} ∑_{k=2}^{n-i+1} \frac{2}{k} \\
&= ∑_{i=2}^{n} ∑_{k=2}^{i} \frac{2}{k} \\
&= ∑_{k=2}^{n} \frac{2(n-k+1)}{k} \\
&= ∑_{k=2}^{n} \left(\frac{2(n+1)}{k} - 2\right) \\
&= ∑_{k=2}^{n} \frac{2(n+1)}{k} - 2(n-1) \\
&= 2(n+1)(H_n - 1) - 2(n-1) \\
&= 2(n+1)H_n - 4n.
\end{align*}

Here $H_n = ∑_{i=1}^{n} \frac{1}{i}$ is the $n$-th
\concept{harmonic number}, equal to $\ln n + γ + O(n^{-1})$,
where $γ \approx 0.5772$ is the 
\concept{Euler-Mascheroni constant} (whose exact value is unknown!).
For asymptotic purposes we only need $H_n = Θ(\log n)$.

For the first step we are taking advantage of the fact that linearity
of expectation doesn't care about the variables not being independent.
The rest is just algebra.

This is pretty close to the bound of $2 n \log n$ we computed using
the recurrence 
in §\ref{section-quicksort-recurrence}.  Given that we now know the
exact answer, we could in principle go back and use it to solve the
recurrence exactly.\footnote{We won't.}

Which way is better?  Solving the recurrence requires less
probabilistic \concept{handwaving} (a more polite term might be
``insight'') but more grinding out inequalities, which is a pretty common trade-off.  Since I am personally not very clever I would try the brute-force approach first.  But it's worth knowing about better methods so you can try them in other situations.

\subsubsection{Linearity of expectation for infinite sequences}
\label{section-linearity-of-expectation-for-infinite-sequences}
\label{section-st-petersburg-paradox}

For infinite sequences of random
variables, linearity of expectation may break down.  
This is true even if the sequence is countable.
An example is the
\indexConcept{paradox!St.~Petersburg}{St.~Petersburg paradox}, in which a
gambler bets \$1 on a double-or-nothing game, then bets \$2 if they 
lose, then \$4, and so on, until they eventually wins and stops, up
\$1.  If we represent the gambler's gain or loss at stage $i$ as a random
variable $X_i$, it's easy to show that $\Exp{X_i} = 0$, because the
gambler either wins $\pm 2^i$ with equal probability, or doesn't play
at all.  So $∑_{i=0}^{∞} \Exp{X_i} = 0$.  
But $\Exp{∑_{i=0}^{∞} X_i} = 1$, because the probability that
the gambler doesn't eventually win is zero.\footnote{The trick here is
    that we are trading a probability-$1$ gain of $1$ against a
    probability-$0$ loss of $∞$.  So we could declare that
    $\Exp{∑_{i=0}^{∞} X_i}$ involves $0 ⋅
    (-∞)$ and is undefined.  But this would lose the useful
    property that expectation isn't affected by probability-$0$
    outcomes.  As often happens in mathematics, we are forced to
    choose between candidate definitions based on which bad consequences we most
want to avoid, with no way to avoid all of them.  So the standard
definition of expectation allows the St.~Petersburg paradox because
the alternatives are worse.}

Fortunately, these pathological cases don't come up often in algorithm
analysis, and with some additional side constraints we can apply
linearity of expectation even to infinite sums of random variables.
The simplest is when $X_i ≥ 0$ for all $i$; then
$\Exp{∑_{i=0}^{∞} X_i}$ exists and is equal to
$∑_{i=0}^{∞} \Exp{X_i}$ whenever the sum of the expectations
converges (this is a consequence of the monotone convergence theorem).
Another condition that works is if $\abs*{∑_{i=0}^{n} X_i} ≤ Y$
for all $n$, where $Y$ is a random variable with finite expectation;
the simplest version of this is when $Y$ is constant.
See~\cite[§5.6.12]{GrimmetS1992} or~\cite[§{}IV.2]{Feller1971}
for more details.

\subsection{Expectation and inequalities}

If $X ≤ Y$ (that is, if the event $[X ≤ Y]$ holds with probability
$1$), then $\Exp{X} ≤ \Exp{Y}$.  For finite discrete spaces 
the proof is trivial:
\begin{align*}
    \Exp{X}
    &= ∑_{ω∈Ω} \Prob{ω} X(ω)
    \\&≤ ∑_{ω∈Ω} \Prob{ω} Y(ω)
    \\&= \Exp{Y}.
\end{align*}

The claim continues to hold even in the general case, but the proof
is more work.

One special case of this that comes up often is that $X ≥ 0$ implies
$\Exp{X} ≥ 0$.

\subsection{Expectation of a product}

When two random variables $X$ and $Y$ are independent, it also holds
that $\Exp{XY} = \Exp{X}\Exp{Y}$.  The proof (at least for discrete random
variables) is straightforward:
\begin{align*}
    \Exp{XY} 
    &= ∑_{x} ∑_{y} xy \Prob{X = x ∧ Y= y} \\
    &= ∑_{x} ∑_{y} xy \Prob{X=x} \Prob{Y=y} \\
    &= \left(∑_x x \Prob{X=x} \right)\left(∑_y \Prob{Y=y}\right) \\
    &= \Exp{X} \Exp{Y}.
\end{align*}

For example, the expectation of the product of two independent
fair six-sided dice is $\left(\frac{7}{2}\right)^2 = \frac{49}{4}$.

This is not true for arbitrary random
variables.  If we compute the expectation of the product of a single
fair six-sided die with itself, we get 
$\frac{1⋅ 1 + 2 ⋅ 2 + 3 ⋅ 3 + 4 ⋅ 4 + 5 ⋅ 5 + 6
⋅ 6}{6} = \frac{91}{6}$, which is much larger.

One measure of the dependence between two random variables by
is the difference $\Exp{XY} - \Exp{X}⋅\Exp{Y}$. This is called the
\concept{covariance} of $X$ and $Y$, written $\Cov{X}{Y}$, and it is
$0$ when $X$ and $Y$ are independent and nonzero otherwise. 
Covariance will come back later when we
look at concentration bounds in
Chapter~\ref{chapter-concentration-bounds}.

\subsubsection{Wald's equation (simple version)}
\label{section-Walds-equation-simple}

Computing the expectation of a product does not often come up directly
in the analysis of a randomized algorithm.  
Where we might expect to do it is when we have a loop: one
random variable $N$ tells us the number of times we execute the 
loop, while another random variable $X$ tells us the cost of each
iteration.  The problem is that if each iteration is randomized, then
we really have a sequence of random variables $X_1, X_2, \dots$, and
what we want to calculate is
\begin{equation}
    \label{eq-walds-equation-sum}
    \Exp{∑_{i=1}^{N} X_i}.
\end{equation}
Here we can't use the sum formula directly, because $N$ is a random
variable, and we can't use the product formula, because the $X_i$ are
all different random variables.

If $N$ and the $X_i$ are all independent (which may or may not be the case
for the loop example),
and $N$ is bounded by some fixed maximum $n$, then we
can apply the product rule to get the value of
\eqref{eq-walds-equation-sum} by throwing in a few indicator variables.  The idea is
that the contribution of $X_i$ to the sum is given by $X_i 1_{[N≥i]}$,
and because we assume that $N$ is independent of the $X_i$, if we need
to compute $\Exp{X_i 1_{[N≥i]}}$, we can do so by computing
$\Exp{X_i}\Exp{1_{[N≥i]}}$.

So we get
\begin{align*}
    \Exp{∑_{i=1}^{N} X_i}
    &= \Exp{∑_{i=1}^{n} X_i 1_{[N≥i]}}
    \\&= ∑_{i=1}^{n} \Exp{X_i 1_{[N≥i]}}
    \\&= ∑_{i=1}^{n} \Exp{X_i} \Exp{1_{[N≥i]}}.
\end{align*}

For general $X_i$ we have to stop here. But if we also know that the
$X_i$ all have the same expectation $μ$, then $\Exp{X_i}$ doesn't depend on
$i$ and we can bring it out of the sum.  This gives
\begin{align}
    ∑_{i=1}^{n} \Exp{X_i} \Exp{1_{[N≥i]}}
    &= μ ∑_{i=1}^{n} \Exp{1_{[N≥i]}}
    \nonumber
    \\&= μ \Exp{N}.
    \label{eq-Walds-equation-simple}
\end{align}

This equation is a special case of
\index{equation!Wald's}
\concept{Wald's equation}, which we will see again in
§\ref{section-Walds-equation}.
The main difference between this version and the general version is
that here we had to assume that $N$ was independent of the $X_i$,
which may not be true if our loop is a \texttt{while} loop, and
termination after a particular iteration is correlated with the time
taken by that iteration.

But for simple cases, \eqref{eq-Walds-equation-simple} can still be useful.
For example, if we throw one six-sided die to get $N$, and then throw
$N$ six-sided dice and add them up, we get the same expected total
$\frac{7}{2}⋅\frac{7}{2} = \frac{49}{4}$ as if we just multiply two
six-sided dice.  This is true even though the actual distribution of
the values is very different in the two cases.

\section{Conditional expectation}
\label{section-conditional-expectation}

We can also define a notion of \index{expectation!conditional}\concept{conditional expectation},
analogous to conditional probability.  There are three versions of
this, depending on how fancy we want to get about specifying what
information we are conditioning on.

\subsection{Expectation conditioned on an event}
\label{section-expectation-conditioned-on-an-event}

The \index{expectation!conditioned on an event}
expectation of $X$ conditioned on an \emph{event} $A$ is
written $\ExpCond{X}{A}$ and defined by
\begin{equation}
    \label{eq-conditional-expectation-event}
    \ExpCond{X}{A} = ∑_x x \ProbCond{X=x}{A} = ∑_x x \frac{\Prob{X = x ∧ A}}{\Prob{A}}. 
\end{equation}
This is essentially the weighted average value of $X$
if we know that $A$ occurs.

Most of the properties that we see with ordinary expectations continue
to hold for conditional expectation.  For example, linearity
of expectation
\begin{equation}
    \label{eq-linearity-of-expectation-conditioned-on-an-event}
    \ExpCond{aX+bY}{A} = a \ExpCond{X}{A} + b \ExpCond{Y}{A}
\end{equation}
holds whenever $a$ and $b$ are constant on $A$.

Similarly if $\Prob{X ≥ Y}{A} = 1$, $\ExpCond{X}{A} ≥ \ExpCond{Y}{A}$.

Conditional expectation is handy because we can use it to compute
expectations by case analysis the same way we use conditional
probabilities using the law of total probability (see
§\ref{section-law-of-total-probability}).  If $A_1, A_2, \dots$ are a
countable partition of $Ω$, then
\begin{align}
    ∑_{i} \Prob{A_i} \ExpCond{X}{A_i}
    &= ∑_{i} \Prob{A_i} \parens*{ ∑_x x \ProbCond{X = x}{A_i} }
    \nonumber\\&= ∑_{i} \Prob{A_i} \parens*{ ∑_x x \frac{\Prob{X = x ∧ A_i}}{{\Prob A_i}} }
    \nonumber\\&= ∑_{i} ∑_x x \Prob{X = x ∧ A_i}
    \nonumber\\&= ∑_{x} x \parens*{ ∑_i \Prob{X = x ∧ A_i} }
    \nonumber\\&= ∑_{x} x \Prob{X = x}
    \nonumber\\&= \Exp{X}. \label{eq-law-of-total-expectation}
\end{align}

This is actually a special case of the law of iterated expectation, which we will see in the next section.

\subsection{Expectation conditioned on a random variable}
\label{section-expectation-conditioned-on-a-random-variable}

In the previous section, we considered computing $\Exp{X}$ by breaking
it up into disjoint cases $\ExpCond{X}{A_1}$, $\ExpCond{X}{A_2}$,
etc.  But keeping track of all the events in our partition of $Ω$ is a
lot of book-keeping.  Conditioning on a random variable lets us
combine all these conditional probabilities into a single expression
\begin{displaymath}
    \ExpCond{X}{Y},
\end{displaymath}
the
\index{expectation!conditioned on a random variable}\conceptFormat{expectation of $X$ conditioned on $Y$},
which is defined to have the value $\ExpCond{X}{Y=y}$ whenever $Y=y$.\footnote{If $Y$ is not discrete, the situation is more
complicated.  See~\cite[§§{}III.2 and V.9--V.11]{Feller1971}
or~\cite[§7.9]{GrimmetS1992}.}

Note that $\ExpCond{X}{Y}$ is generally a function of $Y$, unlike $\Exp{X}$
which is a constant.  This also means that $\ExpCond{X}{Y}$ is a
random variable, and its value can depend on which outcome $ω$ we
picked from our probability space $Ω$.
The intuition behind the definition is that $\ExpCond{X}{Y}$ is the
best estimate we can make of
$X$ given that we know the value of $Y$ but nothing else.

If we want to be formal about the definition, we can specific the
value of $\ExpCond{X}{Y}$ explicitly for each point $ω∈Ω$:
\begin{equation}
    \label{eq-conditional-expectation-random-variable-omega}
    \ExpCond{X}{Y}(ω) = \ExpCond{X}{Y=Y(ω)}.
\end{equation}
This is just another way of saying what we said already: if you want
to know what the expectation of $X$ is conditioned on $Y$ when you get
outcome $ω$, find the value of $Y$ at $ω$ and condition on seeing that.

Here is a simple example.  Suppose that $X$ and $Y$ are independent
fair coin-flips that take on the values $0$ and $1$ with equal
probability.  Then our probability space $Ω$ has four elements, and
looks like this:
\begin{equation*}
    \begin{array}{cc}
        \Tuple{0,0} & \Tuple{0,1} \\
        \Tuple{1,0} & \Tuple{1,1} \\
    \end{array}
\end{equation*}
where each tuple $\Tuple{x,y}$ gives the values of $X$ and $Y$.

We can also define the total number of heads as $Z=X+Y$.  If we label
all the points $ω$ in our probability space with $Z(ω)$, we get a
picture that looks like this:
\begin{equation*}
    \begin{array}{cc}
        0 & 1 \\
        1 & 2 \\
    \end{array}
\end{equation*}

This is what we see if we know the exact value of both coin-flips (or
at least the exact value of $Z$).

But now suppose we only know $X$, and want to compute $\ExpCond{Z}{X}$.
When $X=0$, $\ExpCond{Z}{X=0} = \frac{1}{2}⋅0 + \frac{1}{2}⋅1 =
\frac{1}{2}$; and when $X=1$, $\Exp{Z}{X=1} = \frac{1}{2}⋅1 +
\frac{1}{2}⋅2 = \frac{3}{2}$.  So drawing $\ExpCond{Z}{X}$ over our
probability space gives
\begin{equation*}
    \begin{array}{cc}
        \frac{1}{2} & \frac{1}{2} \\
        \frac{3}{2} & \frac{3}{2} \\
    \end{array}
\end{equation*}

We've averaged the value of $Z$ across each row, since each row
corresponds to one of the possible values of $X$.

If instead we compute $\ExpCond{Z}{Y}$, we get this picture instead:
\begin{equation*}
    \begin{array}{cc}
        \frac{1}{2} & \frac{3}{2} \\
        \frac{1}{2} & \frac{3}{2} \\
    \end{array}
\end{equation*}

Now instead of averaging across rows (values of $X$) we average across
columns (values of $Y$).  So the left column shows 
$\ExpCond{Z}{Y=0} = \frac{1}{2}$ and the right columns shows
$\ExpCond{Z}{Y=1} = \frac{3}{2}$, which is pretty much what we'd expect.

Nothing says that we can only condition on $X$ and $Y$.  What happens
if we condition on $Z$?

Now we are going to get fixed values for each possible value of $Z$.
If we compute $\ExpCond{X}{Z}$, then when $Z=0$ this will be $0$ (because
$Z=0$ implies $X=0$), and when $Z=2$ this will be $1$ (because $Z=2$
implies $X=1$).  The middle case is $\ExpCond{X}{Z=1} = \frac{1}{2}⋅0
+ \frac{1}{2}⋅1 = \frac{1}{2}$, because the two outcomes $\Tuple{0,1}$ and
$\Tuple{1,0}$ that give $Z=1$ are equally likely.  The picture is
\begin{equation*}
    \begin{array}{cc}
        0           & \frac{1}{2} \\
        \frac{1}{2} & 1           \\
    \end{array}
\end{equation*}

\subsubsection{Calculating conditional expectations}

Usually we will not try to $\ExpCond{X}{Y}$ individually for
each possible $ω$ or even each possible value $y$ of $Y$.  Instead, we
can use various basic facts to compute $\ExpCond{X}{Y}$ by applying
arithmetic to random variables.

The two basic facts to start with are:
\begin{enumerate}
    \item If $X$ is a function of $Y$, then $\ExpCond{X}{Y} = X$.
        Proof: Suppose $X = f(Y)$.  From
        \eqref{eq-conditional-expectation-random-variable-omega}, for
        each outcome $ω$, we
        have $\ExpCond{X}{Y}(ω) = \ExpCond{X}{Y=Y(ω)} =
        \ExpCond{f(Y(ω))}{Y=Y(ω)} = f(Y(ω)) = X(ω)$.
    \item If $X$ is independent of $Y$, then $\ExpCond{X}{Y} =
        \Exp{X}$.
        Proof: Now for each $ω$, we have $\ExpCond{X}{Y}(ω) =
        \ExpCond{X}{Y=Y(ω)} = \Exp{X}$.
\end{enumerate}

We also have a rather strong version of linearity of expectation.  If
$A$ and $B$ are both functions of $Z$, then
\begin{equation}
    \label{eq-linearity-of-expectation-conditioned-on-variable}
    \ExpCond{AX + BY}{Z} = A \ExpCond{X}{Z} + B \ExpCond{Y}{Z}.
\end{equation}

Here is a proof for discrete probability spaces.
For each value $z$ of $Z$, we have 
\begin{align*}
    \ExpCond{A(Z)X + B(Z)Y}{Z=z}
    &= \sum_{ω ∈ Z^{-1}(z)} \frac{\Prob{ω}}{\Prob{Z=z}} (A(z) X(ω) + B(z) Y(ω) 
    \\&= A(z) \sum_{ω ∈ Z^{-1}(z)} \frac{\Prob{ω}}{\Prob{Z=z}} X(ω)
    \\& \quad + B(z) \sum_{ω ∈ Z^{-1}(z)} \frac{\Prob{ω}}{\Prob{Z=z}} Y(ω)
    \\&= A(z) \ExpCond{X}{Z=z} + B(z) \ExpCond{Y}{Z=z},
\end{align*}
which is the value when $Z=z$ of $A \ExpCond{X}{Z} + B \ExpCond{Y}{Z}$.

This means that we can quickly simplify many conditional expectations.
If we go back to the example of the previous section, where $Z=X+Y$ is
the sum of two independent fair coin-flips $X$ and $Y$, then we can
compute
\begin{align*}
    \ExpCond{Z}{X}
    &= \ExpCond{X+Y}{X}
    \\&= \ExpCond{X}{X} + \ExpCond{Y}{X}
    \\&= X + \Exp{Y}
    \\&= X + \frac{1}{2}.
\end{align*}
Similarly, $\ExpCond{Z}{Y} = \ExpCond{X+Y}{Y} = \frac{1}{2} + Y$.

In some cases we have enough additional information to run this in
reverse.  If we know $Z$ and want to estimate $X$, we can use the fact
that $X$ and $Y$ are symmetric to argue that
$\ExpCond{X}{Z} = \ExpCond{Y}{Z}$.  But then $Z = \ExpCond{X+Y}{Z} = 2
\ExpCond{X}{Z}$, so $\ExpCond{X}{Z} = Z/2$.  Note that this works in
general only if the events $[X=a,Y=b]$ and
$[X=b,Y=a]$ have the same probabilities for all $a$ and $b$ even if we
condition on $Z$, which in
this case follows from the fact that
$X$ and $Y$ are independent and identically
distributed and that addition is commutative.  Other cases may be
messier.\footnote{An
example with $X$ and $Y$ identically distributed but not independent is
to imagine that we roll a six-sided die to get $X$, and let $Y=X+1$ if
$X < 6$ and $Y=1$ if $X=6$.  Now knowing $Z = X+Y = 3$ tells me that $X=1$
and $Y=2$ exactly, neither of which is $Z/2$.}

Other facts, like $X ≥ Y$ implies $\ExpCond{X}{Z} ≥ \ExpCond{Y}{Z}$,
can be proved using similar techniques.

\subsubsection{The law of iterated expectation}
\label{section-iterated-expectation}

The \index{expectation!iterated!law of}
\index{iterated expectation!law of}
\concept{law of iterated expectation}
says that
\begin{equation}
    \label{eq-law-of-iterated-expectation}
    \Exp{X} = \Exp{\ExpCond{X}{Y}}.
\end{equation}

When $Y$ is discrete, this is just
\eqref{eq-law-of-total-expectation} in disguise.  For each of the
(countably many) values of $Y$ that occurs with nonzero probability,
let $A_y$ be the event $[Y=y]$.  Then these events are a countable
partition of $Ω$, and
\begin{align*}
    \Exp{\ExpCond{X}{Y}}
    &= ∑_y \Prob{Y=y} \ExpCond{\ExpCond{X}{Y}}{Y=y}
    \\&= ∑_y \Prob{Y=y} \ExpCond{X}{Y=y}
    \\&= \Exp{X}.
\end{align*}
The trick here is that we use \eqref{eq-law-of-total-expectation} to
expand out the original expression in terms of the events $A_y$, then
notice that $\ExpCond{X}{Y}$ is equal to $\ExpCond{X}{Y=y}$ whenever
$Y=y$.

So as claimed, conditioning on a variable gives a way to write 
averaging over cases very compactly.

It's also not too hard to show that iterated
expectation works with partial conditioning:
\begin{equation}
    \label{eq-law-of-iterated-expectation-partial-conditioning}
    \ExpCond{\ExpCond{X}{Y,Z}}{Y} = \ExpCond{X}{Y}.
\end{equation}

\subsubsection{Conditional expectation as orthogonal projection}
\label{section-conditional-expectation-as-projection}

If you are comfortable with linear algebra, it may be helpful to think
about expectation conditioned on a random variable as a form of
projection onto a subspace.  In this section, we'll give a brief
description of how this works for a finite, discrete probability
space.  For a more general version, see for
example~\cite[§7.9]{GrimmetS1992}.

Consider the set of all real-valued random variables on the
probability space
$\Set{\coinFlips{TT}, \coinFlips{TH}, \coinFlips{HT}, \coinFlips{HH}}$
corresponding to flipping two independent fair coins.  We can think of
each such random variable as a vector in $ℝ^4$, where the four
coordinates give the value of the variable on the four possible
outcomes.  For example, the indicator variable for the event that the
first coin is heads would look like $X = \Tuple{0,0,1,1}$ and the
indicator variable for the event that the second coin is heads would
look like
$Y = \Tuple{0,1,0,1}$.

When we add two random variables together, we get a new random
variable.  This corresponds to vector addition: $X+Y = \Tuple{0,0,1,1}
+ \Tuple{0,1,0,1} = \Tuple{0,1,1,2}$.  Multiplying a random variable
by a constant looks like scalar multiplication $2X = 2⋅\Tuple{0,0,1,1}
= \Tuple{0,0,2,2}$.  Because random variables support both addition
and scalar multiplication, and because these operations obey the axioms
of a \concept{vector space}, we can treat the set of all real-valued
random variables defined on a given probability space as a vector
space, and apply all the usual tools from linear algebra to this
vector space.

One thing in particular we can look at is subspaces of this vector
space.  Consider the set of all random variables that are functions of
$X$.  These are vectors of the form $\Tuple{a,a,b,b}$, and adding any
two such vectors or multiplying any such vector by a scalar yields
another vector of this form.  So the functions of $X$ form a
two-dimensional subspace of the four-dimensional space of all random
variables.  An even lower-dimensional subspace is the one-dimensional
subspace of constants: vectors of the form $\Tuple{a,a,a,a}$.  As with
functions of $X$, this set is closed under addition and multiplication
by a constant.

When we take the expectation of $X$, we are looking for a constant
that gives us the average value of $X$.  In vector terms, this means
that $\Exp{\Tuple{0,0,1,1}} = \Tuple{1/2,1/2,1/2,1/2}$.  This
expectation vector is in fact the orthogonal projection of
$\Tuple{0,0,1,1}$ onto the subspace generated by $\mathbf{1} = \Tuple{1,1,1,1}$; we
can tell this because the dot-product of $X-\Exp{X}$ with $\mathbf{1}$
is $\Tuple{-1/2,-1/2,1/2,1/2} ⋅ \Tuple{1,1,1,1} = 0$.
If instead we take a conditional expectation, we are again doing an
orthogonal projection, but now onto a higher-dimensional subspace.  So
$\ExpCond{X+Y}{X} = \Tuple{1/2,1/2,3/2,3/2}$ is the orthogonal
projection of $X+Y = \Tuple{0,1,1,2}$ onto the space of all functions
of $X$, which is generated by the basis vectors $\Tuple{0,0,1,1}$ and
$\Tuple{1,1,0,0}$.  As in the simple expectation, the dot-product of $(X+Y) -
\ExpCond{X+Y}{X}$ with either of these basis vectors is $0$.

Many facts about conditional expectation translate in a
straightforward way to facts about projections.  Linearity of
expectation is equivalent to linearity of projection onto a subspace.
The law of iterated expectation $\Exp{\ExpCond{X}{Y}} = \Exp{X}$ says
that projecting onto the subspace of functions of $Y$ and then onto the
subspace of constants is equivalent to projection directly down to the
subspace of constants; this is true in general for projection
operations.  It's also possible to represent other features of
probability spaces in terms of expectations; for example, $\Exp{XY}$
acts like an inner product for random variables, $\Exp{X^2}$ acts like
the square of the Euclidean distance, and the fact that $\Exp{X}$ is an orthogonal
projection of $X$ means that $\Exp{X}$ is precisely the constant value $μ$ that
minimizes the distance $\Exp{(X-μ)^2}$.  We won't actually use any of
these facts in the following, but having another way to look at
conditional expectation may be helpful in understanding how it works.

\subsection{Expectation conditioned on a \texorpdfstring{$σ$}{sigma}-algebra}
\label{section-expectation-conditioned-on-a-sigma-algebra}

Expectation conditioned on a random variable
is actually a special case of the
\index{expectation!conditioned on a $σ$-algebra}expectation of
$X$ conditioned on a \emph{\index{$σ$-algebra}$σ$-algebra}
$ℱ$.  Recall
that a
$σ$-algebra is a family of subsets of $Ω$ that includes
$Ω$ and is closed under complement and countable union; for
discrete probability spaces, this turns out to be the set of all
unions of equivalence classes for some equivalence relation on
$Ω$,\footnote{Proof: 
Let $ℱ$ be a $σ$-algebra over a countable set $Ω$.
Let $ω \sim ω'$
if, for all $A$ in $ℱ$, $ω ∈ A$ if and only if
$ω' ∈ A$; this is an equivalence relation on $Ω$.
To show that the equivalence classes of $\sim$ are elements of
$ℱ$, for each $ω'' \not\sim ω$, let
$A_{ω''}$ be some element of $ℱ$ that contains $ω$
but not $ω''$.  Then $\bigcap_{ω''} A_{ω''}$ (a
countable intersection of elements of $ℱ$) contains
$ω$ and all points $ω' \sim ω$ but 
no points $ω'' \not\sim ω$; in other words, it's the
equivalence class of $ω$.  Since there are only countably many
such equivalence classes, we can construct all the elements of
$ℱ$ by taking all possible unions of them.}
and we think of $ℱ$ as representing knowledge of
which equivalence class we are in, but not which point in the
equivalence class we land on.  An example would be if $Ω$
consists of all values $(X_1, X_2)$ obtained from two die rolls, and
$ℱ$ consists of all sets $A$ such that whenever one point
$ω$ with $X_1(ω)+X_2(ω) = s$ is in $A$, so is every
other point $ω'$ with $X_1(ω') + X_2(ω') = s$.  (This
is the \index{generated by}\index{$σ$-algebra!generated by a random variable}
$σ$-algebra \conceptFormat{generated by} the random variable
$X_1+X_2$.)

A discrete random variable $X$ is \concept{measurable} with respect to
$ℱ$, or \concept{$ℱ$-measurable}, if every
event $[X=x]$ is contained in $ℱ$; in other words, knowing
only where we are in $ℱ$, we can compute exactly the value
of $X$.  This gives a formal way to define $σ(X)$: it is the
smallest $σ$-algebra $ℱ$ such that $X$ is
$ℱ$-measurable.

If $X$ is not $ℱ$-measurable, the best approximation we can
make to it given that we only know where we are in $ℱ$ is
$\ExpCond{X}{ℱ}$, which is defined as a random variable $Q$ that is (a)
$ℱ$-measurable; and (b) satisfies $\ExpCond{Q}{A} = \ExpCond{X}{A}$ for
any event $A ∈ ℱ$ with $\Prob{A} ≠ 0$.

For discrete probability spaces,
this just means that we replace $X$ with its average value across each
equivalence class: property (a) is satisfied because
$\ExpCond{X}{ℱ}$
is constant
across each equivalence class, meaning that $[\ExpCond{X}{ℱ}=x]$ is a union of
equivalence classes, and property (b) is satisfied because we define
$\ExpCond{\ExpCond{X}{ℱ}}{A} = \ExpCond{X}{A}$ for each equivalence class $A$, and
the same holds for unions of equivalence classes by a simple
calculation.

This gives the same result as $\ExpCond{X}{Y}$ if $ℱ$ is generated by
$Y$, or more generally as $\ExpCond{X}{Y_1,Y_2,\dots}$ if $ℱ$ is
generated by $Y_1,Y_2,\dots$.  In each case the intuition is that we
are getting the best estimate we can for $X$ given the information we
have.  It is also possible to define $\ExpCond{X}{ℱ}$ as a projection
onto the subspace of all random variables that are  $ℱ$-measurable,
analogously to the special case for $\ExpCond{X}{Y}$ described in
§\ref{section-conditional-expectation-as-projection}.

Sometimes it is convenient to use more than one $σ$-algebra to
represent increasing knowledge over time.  A \concept{filtration} is a
sequence of $σ$-algebras $ℱ_0 ⊆ ℱ_1 ⊆ ℱ_2 ⊆ \dots$, where each $ℱ_t$
represents the information we have available at time $t$.  That each
$ℱ_t$ is a subset of $ℱ_{t+1}$ means that any event we can determine
at time $t$ we can also determine at all future times $t' > t$: though
we may learn more information over time, we never forget what we
already know.  A common example of a filtration is when we have a
sequence of random variables $X_1, X_2, \dots$, and define $ℱ_t$ as
the $σ$-algebra $\Tuple{X_1,X_2,\dots,X_t}$ generated by
$X_1,X_2,\dots,X_t$.

When one $σ$-algebra is a subset of another, a version of the law of
iterated expectation applies: $ℱ ∈ ℱ'$ implies
$\ExpCond{\ExpCond{X}{ℱ'}}{ℱ} = \ExpCond{X}{ℱ}$.  One way to think about
this is that if we forget everything about $X$ we can't predict from
$ℱ'$ and then forget everything that's left that we can't predict from
$ℱ$, we get to the same place as if we just forget everything except
$ℱ$ to begin with.  The simplest version $\ExpCond{X}{ℱ'} = \Exp{X}$
is just what happens when $ℱ$ is
the trivial $σ$-algebra $\Set{∅,Ω}$, where all we know is that
something happened, but we don't know what.

\subsection{Examples}
\label{section-conditional-expectation-examples}

\begin{itemize}
    \item Let $X$ be the value of a six-sided die.  Let $A$ be the
        event that $X$ is even.  Then
        \begin{align*}
            \ExpCond{X}{A}
            &= ∑_x x \ProbCond{X=x}{A}
            \\&= (2+4+6)⋅\frac{1}{3}
            \\&= 4.
        \end{align*}
    \item Let $X$ and $Y$ be independent six-sided dice, and let
        $Z=X+Y$.  Then $\ExpCond{Z}{X}$ is a random variable whose
        value is $1+7/2$ when $X=1$, $2+7/2$ when $X=2$, etc.  We can
        write this succinctly by writing $\ExpCond{Z}{X} = X + 7/2$.
    \item Conversely, if $X$, $Y$, and $Z$ are as above, we can also
        compute $\Exp{X}{Z}$.  Here we are told what $Z$ is and must
        make an estimate of $X$.  
        
        For some values of $Z$, this nails down $X$ completely:
        $\ExpCond{X}{Z=2} = 1$ because $X$ you can only make $2$ in
        this model as $1+1$.  For other values, we don't know much
        about $X$, but can still compute the expectation.  For
        example, to compute $\ExpCond{X}{Z=5}$, we have to average
        $X$ over all the pairs $(X,Y)$ that sum to $5$.  This gives
        $\ExpCond{X}{Z=5} = \frac{1}{4}(1+2+3+4) = \frac{5}{2}$.
        (This is not terribly surprising, since by symmetry
        $\ExpCond{Y}{Z=5}$ should equal $\ExpCond{X}{Z=5}$, and since
        conditional expectations add just like regular expectations,
        we'd expect that the sum of these two expectations would be
        $5$.)

        The actual random variable $\ExpCond{X}{Z}$ summarizes these
        conditional expectations for all events of the form $[Z=z]$.
        Because of the symmetry argument above, we can write it
        succinctly as $\ExpCond{X}{Z} = \frac{Z}{2}$.  Or we could
        list its value for every $ω$ in our underlying probability
        space, as in done in 
        Table~\ref{table-conditional-expectations} for this and
        various other 
        conditional expectations on the 
        two-independent-dice space.
\end{itemize}

\begin{table}
{\small
    \begin{displaymath}
        \begin{array}{cccccccccc}
            ω & X & Y & Z=X+Y & \Exp{X} & \ExpCond{X}{Y=3} &
            \ExpCond{X}{Y} & \ExpCond{Z}{X} & \ExpCond{X}{Z} &
            \ExpCond{X}{X} \\
(1, 1)&1&1&2&7/2&7/2&7/2&1+7/2&2/2&1 \\
(1, 2)&1&2&3&7/2&7/2&7/2&1+7/2&3/2&1 \\
(1, 3)&1&3&4&7/2&7/2&7/2&1+7/2&4/2&1 \\
(1, 4)&1&4&5&7/2&7/2&7/2&1+7/2&5/2&1 \\
(1, 5)&1&5&6&7/2&7/2&7/2&1+7/2&6/2&1 \\
(1, 6)&1&6&7&7/2&7/2&7/2&1+7/2&7/2&1 \\
(2, 1)&2&1&3&7/2&7/2&7/2&2+7/2&3/2&2 \\
(2, 2)&2&2&4&7/2&7/2&7/2&2+7/2&4/2&2 \\
(2, 3)&2&3&5&7/2&7/2&7/2&2+7/2&5/2&2 \\
(2, 4)&2&4&6&7/2&7/2&7/2&2+7/2&6/2&2 \\
(2, 5)&2&5&7&7/2&7/2&7/2&2+7/2&7/2&2 \\
(2, 6)&2&6&8&7/2&7/2&7/2&2+7/2&8/2&2 \\
(3, 1)&3&1&4&7/2&7/2&7/2&3+7/2&4/2&3 \\
(3, 2)&3&2&5&7/2&7/2&7/2&3+7/2&5/2&3 \\
(3, 3)&3&3&6&7/2&7/2&7/2&3+7/2&6/2&3 \\
(3, 4)&3&4&7&7/2&7/2&7/2&3+7/2&7/2&3 \\
(3, 5)&3&5&8&7/2&7/2&7/2&3+7/2&8/2&3 \\
(3, 6)&3&6&9&7/2&7/2&7/2&3+7/2&9/2&3 \\
(4, 1)&4&1&5&7/2&7/2&7/2&4+7/2&5/2&4 \\
(4, 2)&4&2&6&7/2&7/2&7/2&4+7/2&6/2&4 \\
(4, 3)&4&3&7&7/2&7/2&7/2&4+7/2&7/2&4 \\
(4, 4)&4&4&8&7/2&7/2&7/2&4+7/2&8/2&4 \\
(4, 5)&4&5&9&7/2&7/2&7/2&4+7/2&9/2&4 \\
(4, 6)&4&6&10&7/2&7/2&7/2&4+7/2&10/2&4 \\
(5, 1)&5&1&6&7/2&7/2&7/2&5+7/2&6/2&5 \\
(5, 2)&5&2&7&7/2&7/2&7/2&5+7/2&7/2&5 \\
(5, 3)&5&3&8&7/2&7/2&7/2&5+7/2&8/2&5 \\
(5, 4)&5&4&9&7/2&7/2&7/2&5+7/2&9/2&5 \\
(5, 5)&5&5&10&7/2&7/2&7/2&5+7/2&10/2&5 \\
(5, 6)&5&6&11&7/2&7/2&7/2&5+7/2&11/2&5 \\
(6, 1)&6&1&7&7/2&7/2&7/2&6+7/2&7/2&6 \\
(6, 2)&6&2&8&7/2&7/2&7/2&6+7/2&8/2&6 \\
(6, 3)&6&3&9&7/2&7/2&7/2&6+7/2&9/2&6 \\
(6, 4)&6&4&10&7/2&7/2&7/2&6+7/2&10/2&6 \\
(6, 5)&6&5&11&7/2&7/2&7/2&6+7/2&11/2&6 \\
(6, 6)&6&6&12&7/2&7/2&7/2&6+7/2&12/2&6 \\
        \end{array}
    \end{displaymath}
}\caption{Various conditional expectations on two independent dice}
    \label{table-conditional-expectations}
\end{table}

\section{Applications}
\label{section-expectation-examples}

\subsection{Yao's lemma}
\label{section-Yao-lemma}

In Section~\ref{section-searching-an-array}, we considered a special case
of the unordered search problem, where we have an unordered array
$A[1..n]$ and want to find the location of a specific element $x$.
For deterministic algorithms, this requires probing $n$ array
locations in the worst case, because the adversary can place $x$ in
the last place we look.  Using a randomized algorithm, we can reduce
this to $(n+1)/2$ probes on average, either by probing according to a
uniform random permutation or just by probing from left-to-right or
right-to-left with equal probability.

Can we do better?  Proving lower bounds is a nuisance even for
deterministic algorithms, and for randomized algorithms we have even
more to keep track of.  But there is a sneaky trick that allows us to
reduce randomized lower bounds to deterministic lower bounds in many
cases.

The idea is that if we have a randomized algorithm that runs in time
$T(x,r)$ on input $x$ with random bits $r$, then for any fixed choice
of $r$ we have a deterministic algorithm.  So for each $n$, we find
some random $X$ with $\card{X} = n$ and show that, for any
deterministic algorithm that runs in time $T'(x)$, $\Exp{T'(X)} ≥
f(n)$.
But then $\Exp{T(X,R)} = \Exp{\ExpCond{T(X,R)}{R}} =
\Exp{\ExpCond{T_R(X)}{R}} ≥ f(n)$.

This gives us \index{lemma!Yao's}\concept{Yao's lemma}:
\begin{lemma}[Yao's lemma (informal version)\cite{Yao1977}]
    \label{lemma-yao}
    Fix some problem.
    Suppose there is a random distribution on inputs $X$ of size $n$ such that
    every deterministic algorithm for the problem has expected cost $T(n)$.

    Then the worst-case expected cost of any randomized algorithm 
    is at least $T(n)$.
\end{lemma}

For unordered search, putting $x$ in a uniform random array location
makes any deterministic algorithm take at least $(n+1)/2$ probes on
average.  So randomized algorithms take at least $(n+1)/2$ probes as
well.

\subsection{Geometric random variables}
\label{section-geometric-random-variables}

Suppose that we are running a Las Vegas algorithm that takes a fixed
amount of time $T$, but succeeds only with probability $p$ (which we
take to be independent of the outcome of any other run of the
algorithm).
If the algorithm fails, we run it again.  How long does it take on average to
get the algorithm to work?

We can reduce the problem to computing $\Exp{TX} = T \Exp{X}$, where $X$ is the
number of times the algorithm runs.  
The probability that $X=n$ is exactly $(1-p)^{n-1} p$, because we need
to get $n-1$ failures with probability $1-p$ each followed by a single
success with probability $p$, and by assumption all of these
probabilities are independent.
A variable with this kind of distribution is called a
\index{random variable!geometric}
\index{geometric distribution}
\index{distribution!geometric}
\concept{geometric random variable}.  We saw a special
case of this distribution earlier (§\ref{section-racing-coin-flips})
when we were looking at how many flips it would take on average to get
the first tails from a
fair coin
coin (in that case, $p$ was $1/2$).

Using conditional expectation, it's straightforward to compute
$\Exp{X}$.  Let $A$ be the event that the algorithm succeeds on the
first run, i.e., then event $[X=1]$.  Then
\begin{align*}
\Exp{X}
&= \ExpCond{X}{A} \Prob{A} + \ExpCond{X}{\bar{A}} \Prob{\bar{A}} \\
&= 1 ⋅ p + \ExpCond{X}{\bar{A}} ⋅ (1-p).
\end{align*}
The tricky part here is to evaluate $\ExpCond{X}{\bar{A}}$.  Intuitively, if
we don't succeed the first time, we've wasted one step and are back
where we started, so it should be the case that 
$\ExpCond{X}{\bar{A}} = 1 + \Exp{X}$.
If we want to be really careful, we can calculate this out formally
(no sensible person would ever do this):
\begin{align*}
\ExpCond{X}{\bar{A}}
&= ∑_{n=1}^{∞} n \ProbCond{X=n}{X≠ 1} \\
&= ∑_{n=2}^{∞} n \frac{\Prob{X=n}}{\Prob{X≠ 1}} \\
&= ∑_{n=2}^{∞} n \frac{(1-p)^{n-1} p}{1-p} \\
&= ∑_{n=2}^{∞} n (1-p)^{n-2} p \\
&= ∑_{n=1}^{∞} (n+1) (1-p)^{n-1} p \\
&= 1 + ∑_{n=1}^{∞} n (1-p)^{n-1} p \\
&= 1 + \Exp{X}.
\end{align*}

Since we know that $\Exp{X} = p + (1+\Exp{X})(1-p)$, a bit of algebra
gives $\Exp{X} = 1/p$, which is about what we'd expect.

There are more direct ways to get the same result.  If we don't have
conditional expectation to work with, we can try computing the sum
$\Exp{X} = ∑_{n=1}^{∞} n (1-p)^{n-1} p$ directly.  The easiest
way to do this is probably to use generating functions (see, for
example, \cite[Chapter 7]{GrahamKP1988} or \cite{Wilf2006}).
An alternative argument is given in \cite[§2.4]{MitzenmacherU2017};
this uses the fact that $\Exp{X} = ∑_{n=1}^{∞} \Prob{X ≥ n}$, which
holds when $X$ takes on only non-negative integer values.

\subsection{Coupon collector}
\label{section-coupon-collector}

In the \concept{coupon collector problem}, we throw balls uniformly
and independently into $n$ bins until every bin has at least one ball.
When this happens, how many balls have we used on
average?\footnote{The name comes from the problem of collecting
    coupons at random until you have all of them.  A typical
    algorithmic application is having a cluster of machines choose
jobs to finish at random from some list until all are done.  The
expected number of job executions to complete $n$ jobs is given
exactly by the solution to the coupon collector problem.}

Let $X_i$ be the number of balls needed to go from $i-1$ nonempty
bins to $i$ nonempty bins.  It's easy to see that $X_1 = 1$ always.
For larger $i$, each time we throw a ball, it lands in an empty bin
with probability $\frac{n-i+1}{n}$.  This means that $X_i$ has a
geometric distribution with probability $\frac{n-i+1}{n}$, giving
$\Exp{X_i} = \frac{n}{n-i+1}$ from the analysis in
§\ref{section-geometric-random-variables}.

To get the total expected number of balls, take the sum
\begin{align*}
\Exp{∑_{i=1}^{n} X_i}
&= ∑_{i=1}^{n} \Exp{X_i} \\
&= ∑_{i=1}^{n} \frac{n}{n-i+1} \\
&= n ∑_{i=1}^{n} \frac{1}{i} \\
&= n H_n.
\end{align*}

In asymptotic terms, this is $Θ(n \log n)$.

\subsection{Hoare's FIND}
\label{section-Hoares-find}
\label{section-QuickSelect}

\concept{Hoare's FIND}~\cite{Hoare1961QuickSelect}, 
often called \concept{QuickSelect}, 
is an algorithm for finding the $k$-th smallest
element of an unsorted array that works like QuickSort, only after
partitioning the array around a random pivot we throw away the part
that doesn't contain our target and recurse only on the surviving
piece.  As with QuickSort, we'd like to compute the expected number of
comparisons used by this algorithm, on the assumption that the cost of
the comparisons dominates the rest of the costs of the algorithm.

Here the indicator-variable trick gets painful fast.  It turns out to
be easier to get an upper bound by computing the expected number of
elements that are left after each split.

First, let's analyze the pivot step.
If the pivot is chosen uniformly, the number of elements $X$ smaller
than the pivot is uniformly distributed in the range $0$ to $n-1$.
The number of elements larger than the pivot will be $n-X-1$.
In the worst case, we find ourselves recursing on the large pile
always, giving a bound on the number of survivors $Y$ of 
$Y ≤ \max(X, n-X+1)$.

What is the expected value of $Y$?  
By considering both ways the
max can go, we get 
\begin{align*}
    \Exp{Y} &= \ExpCond{X}{X > n-X+1} \Prob{X > n-X+1}
    \\&\quad + \ExpCond{n-X+1}{n-X+1 ≥ X} \Prob{n-X+1 ≥ X}.
\end{align*}
For both conditional expectations, we are
choosing a value uniformly in either the range $\ceil{\frac{n-1}{2}}$
to $n-1$ or $\ceil{\frac{n-1}{2}}+1$ to $n-1$, and in either case the
expectation will be equal to the average of the two endpoints by
symmetry.  So we get
\begin{align*}
    \Exp{Y}
    &≤ \frac{n/2 + n-1}{2} \Prob{X > n-X+1}
       + \frac{n/2 + n}{2} \Prob{n-X+1 ≥ X} \\
       &= \left(\frac{3}{4} n - \frac{1}{2}\right) \Prob{X > n-X+1}
       + \frac{3}{4} n \Prob{n-X+1 ≥ X} \\
       &≤ \frac{3}{4} n.
\end{align*}

Now let $X_i$ be the number of survivors after $i$ pivot steps.  
Note that $\max(0, X_i-1)$ gives the number of comparisons at the following
pivot step, so that $∑_{i=0}^{∞} X_i$ is an upper bound on the
number of comparisons.

We
have $X_0 = n$, and from the preceding argument $\Exp{X_1} ≤ (3/4)n$.
But more generally, we can use the same argument to show that
$\ExpCond{X_{i+1}}{X_i} ≤ (3/4)X_i$, 
and by induction $\Exp{X_i} ≤ (3/4)^i n$.
We also have that $X_j = 0$ for all $j ≥ n$, because we lose at
least one element (the pivot) at each pivoting step.  This saves
us from having to deal with an infinite sum.

Using linearity of expectation,
\begin{align*}
\Exp{∑_{i=0}^{∞} X_i}
&= \Exp{∑_{i=0}^{n} X_i} \\
&= ∑_{i=0}^{n} \Exp{X_i}  \\
&≤ ∑_{i=0}^{n} (3/4)^i n  \\
&≤ 4n.
\end{align*}

\myChapter{Basic probabilistic inequalities}{2025}{}
\label{chapter-probabilistic-inequalities}

Here we're going to look at some inequalities useful for proving
properties of randomized algorithms.  These come in two flavors:
inequalities involving probabilities, which are useful for bounding
the probability that something bad happens, and inequalities involving
expectations, which are used to bound expected running times.  Later,
in Chapter~\ref{chapter-concentration-bounds}, we'll be doing both, by
looking at inequalities that show that a random variable is close to its
expectation with high probability.\footnote{Often, the phrase
\index{high probability}\index{probability!high}\concept{with high probability}
is used in algorithm analysis to mean specifically with probability at
least $1-n^{-c}$ for any fixed $c$. The idea is that if an algorithm
works with high probability in this sense, then the probability that
it fails each time you run it is at most $n^{-c}$, which means that if
you run it as a subroutine in a polynomial-time algorithm that calls
is at most $n^{c'}$ times, the total
probability of failure is at most $n^{c'} n^{-c} = n^{c'-c}$ by the
union bound. Assuming we can pick $c$ to be much larger than $c'$,
this makes the outer algorithm also work with high probability.}

\section{Markov's inequality}
\label{section-markovs-inequality}

This is the key tool for turning expectations of non-negative random
variables into (upper) bounds on
probabilities.  Used directly, it generally doesn't give very good
bounds, but it can work well if we apply it to $\Exp{f(X)}$ for a
fast-growing function $f$; for some examples, see Chebyshev's inequality
(§\ref{section-Chebyshevs-inequality}) or Chernoff bounds
(§\ref{section-chernoff-bounds}).

\index{inequality!Markov's}\index{Markov's inequality}
\index{inequality!Markov's}\concept{Markov's inequality} says that if
$X ≥ 0$ and $α > 0$, then 
\begin{align*}
    \Prob{X ≥ α} &≤ \frac{\Exp{X}}{α}.  
\end{align*}

The proof is immediate from the law of total probability
\eqref{eq-law-of-total-probability}.  We have
\begin{align*}
\Exp{X}
&= \ExpCond{X }{ X ≥ α} \Prob{X ≥ α} + \ExpCond{X }{ x < α} \Prob{X < α}
    \\&≥ α ⋅ \Prob{X ≥ α} + 0 ⋅ \Prob{X < α}
\\&= α ⋅ \Prob{X ≥ α};
\end{align*}
now solve for $\Prob{X ≥ α}$.

Markov's inequality doesn't work in reverse.  For example, consider
the following game: for each integer $k > 0$, with probability $2^{-k}$,
I give you $2^k$ dollars.  Letting $X$ be your payoff from the game,
we have $\Prob{X ≥ 2^k} = ∑_{j=k}^{∞} 2^{-k} = 2^{-k} ∑_{\ell=0}^{∞}
2^{-\ell} = \frac{2}{2^k}$.  The right-hand side here is exactly what we would get from
Markov's inequality if $\Exp{X} = 2$.  But in this case, $\Exp{X} ≠ 2$;
in fact, the expectation of $X$ is given by $∑_{k=1}^{∞} 2^k
2^{-k}$, which diverges.

\subsection{Applications}
\label{section-markovs-inequality-examples}

\subsubsection{Sum of fair coins}
 Flip $n$ independent fair coins, and let $S$ be the number
        of heads we get.  Since $\Exp{S} = n/2$, we get $\Prob{S = n} =
        1/2$.  This is much larger than the actual value $2^{-n}$, but
        it's the best we can hope for if we only know $\Exp{S}$: if we
        let $S$ be $0$ or $n$ with equal probability, we also get
        $\Exp{S} = n/2$.
\subsubsection{Randomized QuickSort}
 The expected running time for randomized QuickSort is $O(n \log
 n)$.  It follows that the probability that randomized QuickSort takes
 more than $f(n)$ time is $O(n \log n / f(n))$.  For example, the
 probability that it performs the maximum $\binom{n}{2} = O(n^{2})$
 comparisons is $O(\log n / n)$.  (It's possible to do much better than this.)
\subsubsection{Balls in bins}
 Suppose we toss $m$ balls in $n$ bins, uniformly and
 independently.  What is the probability that some particular bin
 contains at least $k$ balls?  The probability that a particular ball
 lands in a particular bin is $1/n$, so the expected number of balls
 in the bin is $m/n$.  This gives a bound of $m/nk$ that a particular bin
 contains $k$ or more balls. 
 Unfortunately this is not a very good bound.

\section{Union bound (Boole's inequality)}
\label{section-union-bound}

The \concept{union bound}\index{bound!union} or \concept{Boole's
inequality}\index{inequality!Boole's} says that for any countable collection of events $\Set{ A_{i} }$,
\begin{align}
\label{eq-union-bound}
\Prob{\bigcup A_i} ≤ ∑ \Prob{A_i}.
\end{align}

Combining Markov's inequality with linearity of expectation
        and indicator variables gives a succinct proof of the union
        bound:
        \begin{align*}
            \Prob{\bigcup A_i}
            &= \Prob{∑ 1_{A_i} ≥ 1} \\
            &≤ \Exp{∑ 1_{A_i}} \\
            &= ∑ \Exp{1_{A_i}} \\
            &= ∑ \Prob{A_i}.
        \end{align*}
        Note that for this to work for infinitely many events we need
        to use the fact that $1_{A_i}$ is non-negative.

If we prefer to avoid any issues with infinite sums of expectations,
the direct way to prove this is to replace $A_i$ with $B_i = A_i
∖ \bigcup_{j=1}^{i-1} A_i$.  Then $\bigcup A_i = \bigcup B_i$,
but since the $B_i$ are disjoint and each $B_i$ is a subset of the
corresponding $A_i$, we get 
$\Prob{\bigcup A_i} = \Prob{\bigcup B_i} = ∑
\Prob{B_i} ≤ ∑ \Prob{A_i}$.

The typical use of the union bound is to show that if an algorithm can
fail only if various improbable events occur, then the probability of
failure is no greater than the sum of the probabilities of these
events.  This reduces the problem of showing that an algorithm works
with probability $1-ε$ to constructing an
\concept{error budget} that divides the $ε$ probability of
failure among all the bad outcomes.

\subsection{Example: Balls in bins}

Suppose we toss $n$ balls uniformly and independently into $n$ bins.
What high-probability bound can we get
on the maximum number of balls in any one bin?\footnote{Algorithmic
    version: we insert $n$ elements into a hash table with $n$
    positions 
    using a random hash function.  What is the maximum number of
elements in any one position?}

Consider all $\binom{n}{k}$ sets $S$ of $k$ balls.  If we get at least $k$
balls in some bin, then one of these sets must all land in the same
bin.  Call the event that all balls in $S$ choose the same bin $A_S$.
The probability that $A_S$ occurs is exactly $n^{-k+1}$.  

Using the union bound, we get
\begin{align*}
    \Prob{\text{some bin gets at least $k$ balls}}
    &=
    \Prob{\bigcup_S A_S}
    \\&≤ ∑_S \Prob{A_S}
    \\&= \binom{n}{k} n^{-k+1}
    \\&≤ \frac{n^k}{k!} n^{-k+1}
    \\&= \frac{n}{k!}.
\end{align*}

If we want this probability to be low, we should
choose $k$ so that $k! \gg n$.  Stirling's formula says
that $k! ≥ √{2π k} (k/e)^k ≥ (k/e)^k$, which gives
$\ln (k!) ≥ k (\ln k - 1)$.
If we set $k = c \ln n / \ln \ln n$, we get
\begin{align*}
\ln (k!)
&≥ \frac{c \ln n}{\ln \ln n} \left(\ln c + \ln \ln n - \ln \ln \ln n
- 1\right)
\\&≥ c \ln n.
\end{align*}
when $n$ is sufficiently large.

It follows that the bound
$n/k!$ in this case is less than 
$n/\exp(c \ln n) = n ⋅ n^{-c} = n^{1-c}$.
For suitable choice of $c$ we get a high probability that every bin
gets at most $O(\log n / \log \log n)$ balls.

\section{Jensen's inequality}
\label{section-jensens-inequality}

This is mostly useful if we can calculate $\Exp{X}$ easily for some $X$,
but what we really care about is some other random variable $Y=f(X)$.

\concept{Jensen's inequality}\index{inequality!Jensen's} applies when $f$ is a 
\index{convex}
\concept{convex function}, which means that
for any $x$, $y$, and $0≤μ≤1$,
$f(λx+(1-λ)y) ≤ λf(x)+(1-λ)f(y)$.
Geometrically, this means that the line segment between any two points
on the graph of $f$ never goes below $f$; i.e., that the set of points
$\Set{ (x,y) \mid y ≥ f(x) }$ is convex.
If we want to show that a continuous function $f$ is convex, it's enough to show that 
that $f\left(\frac{x+y}{2}\right) ≤ \frac{f(x)+f(y)}{2}$ for all $x$
and $y$ (in effect, we only need to prove it for the case $λ =
1/2$).
If $f$ is twice-differentiable, an even easier way is to show that
$f''(x) ≥ 0$ for all $x$.

The inequality says that
if $X$ is a random variable and $f$ is convex then 
\begin{align}
\label{eq-jensens-inequality}
f(\Exp{X}) ≤ \Exp{f(X)}.
\end{align}

Alternatively, if $f$ is \concept{concave} (which means that
$f(λ x + (1-λ)y) ≥ λ f(x) + (1-λ) f(y)$, or
equivalently that $-f$ is
convex), the reverse inequality holds:
\begin{align}
\label{eq-jensens-inequality-concave}
f(\Exp{X}) ≥ \Exp{f(X)}.
\end{align}

In both cases, the direction of Jensen's inequality 
matches the direction of the inequality in the definition of convexity
or concavity.  This is not surprising because convexity or
or concavity is just Jensen's inequality for the random variable $X$
that equals $x$ with probability $λ$ and $y$ with probability
$1-λ$.  Jensen's inequality just says that this continues to work for any random
variable for which the expectations exist.

\subsection{Proof}

Here is a proof for the case that $f$ is convex and differentiable.  The idea is
that if $f$ is convex, then it lies above the tangent line at
$\Exp{X}$.  So we can define a linear function $g$ that represents
this tangent line, and get, for all $x$:
\begin{equation}
    \label{eq-jensens-inequality-tangent}
    f(x) ≥ g(x) = f(\Exp{X}) + (x - \Exp{X}) f'(\Exp{X}).
\end{equation}

But then
\begin{align*}
    \Exp{f(X)}
    &≥ \Exp{g(X)}
    \\&= \Exp{f(\Exp{X}) + (X - \Exp{X}) f'(\Exp{X})}
    \\&= f(\Exp{X}) + \Exp{(X - \Exp{X})} f'(\Exp{X})
    \\&= f(\Exp{X}) + \parens*{\Exp{X} - \Exp{X}} f'(\Exp{X})
    \\&= f(\Exp{X}).
\end{align*}

Figure~\ref{fig-jensens-inequality} shows what this looks like for a
particular convex $f$.

\begin{figure}
    \centering
\begin{tikzpicture}
    \draw (0,0) -- (8,2 );
    \draw [domain=0:8] plot (\x, {0.15*(\x - 4)*(\x - 4) + \x / 4});
    \draw [dashed] (4,0) -- (4,4.4);
    \node [below] at (4,0) {$\Exp{X}$};
    \node [below right] at (4,1) {$f(\Exp{X})$};
    \draw [fill=black] (4,1) circle [radius=0.05];
    \node [right] at (8,2) {$g(x)$};
    \node [right] at (8,4.4) {$f(x)$};
    \draw [fill=black] (4,2.2) circle [radius=0.05];
    \node [right] at (4,2.2) {$\Exp{f(X)}$};
\end{tikzpicture}
    \caption{Proof of Jensen's inequality}
    \label{fig-jensens-inequality}
\end{figure}

This is pretty much all linearity of expectation in action: $\Exp{X}$,
$f(\Exp{X})$, and $f'(\Exp{X})$ are all constants, so we can pull them
out of expectations whenever we see them.

The proof of the general case is similar, but for a non-differentiable
convex function it takes a bit more work
to show that the bounding linear function $g$ exists.

\subsection{Applications}

\subsubsection{Fair coins: lower bound}
 Suppose we flip $n$ fair coins, and we want to get a lower bound on
$\Exp{X^2}$, where $X$ is the number of heads.  The function $f:x\mapsto
x^2$ is convex (take its second derivative), so
\eqref{eq-jensens-inequality} gives $\Exp{X^2} ≥ (\Exp{X})^2 =
\frac{n^2}{4}$.

The actual value for $\Exp{X^2}$ is
$\frac{n^2}{4}+\frac{n}{4}$, which can
be found directly using generating functions\footnote{Here's how: The 
    \index{generating function!probability}
    \concept{probability generating function} for $X$ is $F(z) =
    \Exp{z^k} = ∑ z^k \Prob{X = k} = 2^{-n}
(1+z)^n$.  Then $z F'(z) = 2^{-n} n z (1+z)^{n-1} = ∑_k k z^k
\Prob{X=k}$.  Taking the derivative of this a second time gives
$2^{-n} n (1+z)^{n-1} + 2^{-n} n(n-1) z (1+z)^{n-2}
= ∑_k k^2 z^{k-1} \Prob{X=k}$.  Evaluate this monstrosity at $z=1$ to
get
$\Exp{X^2} 
= ∑_k k^2 \Prob{X=k}
= 2^{-n} n 2^{n-1} + 2^{-n} n(n-1) 2^{n-2}
= n/2 + n(n-1)/4
= \frac{2n + n^2 - n}{4}
= n^2/4 + n/4.$} or less directly using variance, which we will
encounter in §\ref{section-Chebyshevs-inequality}.
This is pretty close to the lower bound we got out of Jensen's
inequality, but we can't count on this happening in general.

\subsubsection{Fair coins: upper bound}
 For an upper bound, we can choose a concave $f$.  For example,
if $X$ is as in the previous example,
$\Exp{\lg X} ≤ \lg \Exp{X} = \lg \frac{n}{2} = \lg n - 1$.
This is probably pretty close to the exact value, as we will see
later that $X$ will almost always be within a factor of $1+o(1)$ of $n/2$.
It's not a terribly useful upper bound, because if we use it with
(say) Markov's inequality, the most we can prove is that $\Prob{X = n} =
\Prob{\lg X = \lg n} ≤ \frac{\lg n - 1}{\lg n} = 1 - \frac{1}{\lg n}$,
which is an even worse bound than the $1/2$ we can get from applying
Markov's inequality to $X$ directly.

\subsubsection{Sifters}

Here's an example of Jensen's inequality in action in the analysis of
an actual distributed algorithm.  For some problems in distributed
computing, it's useful to reduce coordinating a large number of
processes to coordinating a smaller number.  A
\concept{sifter}~\cite{AlistarhA2011} is a randomized mechanism for an
asynchronous
shared-memory system that sorts the processes into ``winners'' and
``losers,'' guaranteeing that there is at least one winner.  
The goal
is to make the expected number of winners as small as possible.
The problem is tricky, because processes can only communicate by
reading and writing shared variables, and an adversary gets to choose
which processes participate and fix the
schedule of when each of these processes perform their operations.

The current best known sifter is due to Giakkoupis and
Woelfel~\cite{GiakkoupisW2012}.  For $n$ processes, it uses an array $A$ of
$\ceil{\lg n}$ bits, each of which can be read or written by any of
the processes.  When a process executes the sifter, it chooses a
random index $r ∈ {1 \dots \ceil{\lg n}}$ with probability
$2^{-r-1}$ (this doesn't exactly sum to $1$, so the excess probability
gets added to $r=\ceil{\lg n}$).  The process then writes a $1$ to
$A[r]$ and reads $A[r+1]$.  If it sees a $0$ in its read (or chooses
$r=\ceil{\lg n}$), it wins;
otherwise it loses.

This works as a sifter, because no matter how many processes
participate, some process chooses a value of $r$ at least as large as
any other process's value, and this process wins.  To bound the
expected number of winners, take the sum over all $r$ over the random
variable $W_r$ representing the winners who chose this particular
value $r$.  A process that chooses $r$ wins if it carries out its read
operation before any process writes $r+1$.  If the adversary wants to
maximize the number of winners, it should let each process read as
soon as possible; this effectively means that a process that choose
$r$ wins if no process previously chooses $r+1$.  Since $r$ is twice
as likely to be chosen as $r+1$, conditioning on a process picking $r$
or $r+1$, there is only a $1/3$ chance that it chooses $r+1$.  So at
most $1/(1/3)-1 = 2 = O(1)$ process on average choose $r$ before some process
chooses $r+1$.  (A simpler argument shows that the expected number of
processes that win because they choose $r=\ceil{\lg n}$ is at most $2$
as well.)

Summing $W_r ≤ 2$ over all $r$ gives at most $2 \ceil{\lg n}$ winners on average.
Furthermore, if $k < n$ processes participate, essentially the same
analysis shows that only $2\ceil{\lg k}$ processes
win on average.
So this is a pretty effective tool for getting rid of excess
processes.

But it gets better.  Suppose that we take the winners of one sifter
and feed them into a second sifter.
Let $X_k$ be the number of processes left after $k$ sifters.  We
have that $X_0 = n$ and $\Exp{X_1} ≤ 2 \ceil{\lg n}$, but what can
we say about
$\Exp{X_2}$?  We can calculate $\Exp{X_2} =
\Exp{\ExpCond{X_2}{X_1}} ≤ 2 \ceil{\lg X_1}$.
Unfortunately, the ceiling means that
$2 \ceil{\lg x}$ is not a concave function, but
$f(x) = 2 (\lg x + 1) ≥ 2 \ceil{\lg x}$ is.
So $\Exp{X_2} ≤ f(f(n))$, and in general $\Exp{X_i} ≤ f^{(i)}(n)$,
where $f^{(i)}$ is the $i$-fold composition of $f$.
All the extra constants obscure what is going on a bit, but with
a little bit of algebra it is not too hard to show that
$f^{(i)}(n) = O(1)$ for $i = O(\log^* n)$.\footnote{The $\log^*$
    function counts how many times you need to hit $n$ with
    $\lg$ to reduce it to one or less.  So $\log^* 1 =
    0, \log^* 2 = 1, \log^* 4 = 2, \log^* 16 = 3, \log^* 65536 =
    4, \log^* 2^{65536} = 5$, and after that it starts getting
silly.}
So this gets rid of all but a constant number of processes very
quickly.

\myChapter{Concentration bounds}{2025}{}
\label{chapter-concentration-bounds}

If we really want to get tight bounds on a random variable $X$, the trick
will turn out to be picking some non-negative function $f(X)$ where (a) we can
calculate $\Exp{f(X)}$, and (b) $f$ grows fast enough that merely large values
of $X$ produce huge values of $f(X)$, allowing us to get small
probability bounds by applying Markov's inequality to $f(X)$.  This
approach is often used to show that $X$ lies close to $\Exp{X}$ with
reasonably high probability, what is known as a \concept{concentration
bound}.

Typically concentration bounds are applied to sums of random
variables, which may or may not be fully independent.  Which bound you
may want to use often depends on the structure of your sum.  A quick
summary of the bounds in this chapter is given in
Table~\ref{table-concentration-bounds}.  The rule of thumb is to use
Chernoff bounds (§\ref{section-chernoff-bounds}) if you have a
sum of independent $0$–$1$ random variables; the Azuma-Hoeffding
inequality (§\ref{section-Azuma-Hoeffding}) 
if you have bounded variables with a more complicated
distribution that may be less independent; and Chebyshev's
inequality (§\ref{section-Chebyshevs-inequality}) 
if nothing else works but you can somehow compute the
variance of your sum (e.g., if the $X_i$ are independent or have
easily computed covariance).  In the case of Chernoff bounds, you will
almost always end up using one of the weaker but cleaner versions in
§\ref{section-chernoff-bound-variants} rather than the general
version in §\ref{section-chernoff-bound-general}.

\begin{table}
    \begin{tabular}{ccr@{$\;≤\;$}l}
        Chernoff & $X_i ∈ \Set{0,1}$, independent & $\Prob{S ≥ (1+δ)\Exp{S}}$ & $\left(\frac{e^δ}{(1+δ)^{1+δ}}\right)^{\Exp{S}}$ \\[20pt]
        Azuma-Hoeffding & $\abs*{X_i} ≤ c_i$, martingale & $\Prob{S ≥ t}$ &  $\exp\left(-\frac{t^2}{2∑ c_i^2}\right)$ \\[20pt]
        Chebyshev & & $\Prob{\abs*{S - \Exp{S}} ≥ α}$ & $\frac{\Var{S}}{α^2}$ \\[20pt]
\end{tabular}
\caption[Concentration bounds]{Concentration bounds for $S = ∑ X_i$ (strongest to weakest)}
\label{table-concentration-bounds}
\end{table}

If none of these bounds work for your particular application, there are
many more out there. See for example the textbook by Dubhashi and
Panconesi~\cite{DubhashiP2009}.

\section{Chebyshev's inequality}
\label{section-Chebyshevs-inequality}

\index{inequality!Chebyshev's}\concept{Chebyshev's inequality} allows
us to show that a random variable is close to its mean, by applying
Markov's inequality to the
\concept{variance} of $X$, defined as
$\Var{X} 
= \Exp{(X-\Exp{X})^2}.$
It's a fairly weak concentration bound, that is most
useful when $X$ is a sum of random variables with limited
independence.  

Using Markov's inequality, calculate
\begin{align}
\label{eq-chebyshevs-inequality}
\Prob{\abs*{X - \Exp{X}} ≥ α} 
&= \Prob{(X-\Exp{X})^2 ≥ α^2} \nonumber\\
&≤ \frac{\Exp{(X-\Exp{X})^2}}{α^{2}} \nonumber\\
&= \frac{\Var{X}}{α^{2}}.
\end{align}

\subsection{Computing variance}

At this point it would be reasonable to ask why we are going through
$\Var{X} = \Exp{(X-\Exp{X})^2}$ rather than just using $\Exp{\abs*{X - \Exp{X}}}$.  The
reason is that $\Var{X}$ is usually easier to compute, especially if
$X$ is a sum.  In this section, we'll give some examples of computing
variance, including for various standard random variables that come up
often in randomized algorithms.

\subsubsection{Alternative formula}

The first step is to give an alternative formula for the variance that
is more convenient in some cases.

Expand
\begin{align}
    \Exp{\left(X - \Exp{X}\right)^2}
    &= \Exp{X^2 - 2X⋅\Exp{X} + \left(\Exp{X}\right)^2}
    \nonumber
    \\&= \Exp{X^2} - 2\Exp{X}⋅\Exp{X} + \left(\Exp{X}\right)^2
    \nonumber
    \\&= \Exp{X^2} - \left(\Exp{X}\right)^2.
    \label{eq-variance-alternative}
\end{align}

This formula is easier to use if you are estimating the variance from
a sequence of samples; by tracking $∑ x_i^2$ and $∑ x_i$, you can
estimate $\Exp{X^2}$ and $\Exp{X}$ in a single pass, without having
to estimate $\Exp{X}$ first and then go back for a second pass to calculate
$\left(x_i - \Exp{X}\right)^2$ for each sample.  We won't use this
particular application much, but this explains why the formula is
popular with statisticians.

\subsubsection{Variance of a Bernoulli random variable}

Recall that a Bernoulli random variable is $1$ with probability $p$
and $0$ with probability $q=1-p$; in particular, any indicator
variable is a Bernoulli random variable.

The variance of a Bernoulli random variable is easily calculated from
\eqref{eq-variance-alternative}:
\begin{align*}
    \Var{X}
    &= \Exp{X^2} - \left(\Exp{X}\right)^2
    \\&= p - p^2
    \\&= p (1-p)
    \\&= pq.
\end{align*}

\subsubsection{Variance of a sum}
\label{section-variance-of-a-sum}

If $S = ∑_i X_{i}$, then we can calculate
\begin{align*}
    \Var{S}
    &= \Exp{\left(∑_i X_i\right)^2} - \left(\Exp{∑_i X_i}\right)^2
    \\&= \Exp{∑_i ∑_j  X_i X_j} - ∑_i ∑_j \Exp{X_i} \Exp{X_j}
    \\&= ∑_i ∑_j \left(\Exp{X_i X_j} - \Exp{X_i} \Exp{X_j}\right).
\end{align*}

For any two random variables $X$ and $Y$, the quantity
$\Exp{XY} - \Exp{X}\Exp{Y}$ is called the \concept{covariance} of $X$
and $Y$, written $\Cov{X}{Y}$.  If we take the covariance of a
variable and itself, covariance becomes variance: $\Cov{X}{X} =
\Var{X}$.

We can use $\Cov{X}{Y}$ to rewrite the
above expansion as
\begin{align}
    \Var{∑_i X_i} &=∑_{i,j} \Cov{X_{i}}{X_{j}}
    \label{eq-variance-sum-covariance}
    \\&= ∑_i \Var{X_{i}} + ∑_{i≠j} \Cov{X_{i}}{X_{j}}
    \label{eq-variance-sum-symmetric}
    \\&= ∑_i \Var{X_{i}} + 2 ∑_{i < j} \Cov{X_{i}}{X_{j}}
    \label{eq-variance-sum-asymmetric}
\end{align}

Note that $\Cov{X}{Y} = 0$ when $X$ and $Y$ are independent; this makes
Chebyshev's inequality particularly useful for 
\indexConcept{pairwise independence}{pairwise-independent}
random variables, because then we can just sum up the variances of the
individual variables.

For non-independent random variables, covariance may be tricky to
compute. When cleverer approaches don't work,
a crude bound on the covariance can be obtained from the
\concept{covariance
inequality}\index{inequality!covariance}:\footnote{This is a disguised
version of the \concept{Cauchy-Schwarz
inequality}\index{inequality!Cauchy-Schwarz} for inner product spaces.
The idea is that we can treat the set of random variables on some
probability space as a vector space with the scalar multiplication
given by $aX$ and vector addition given by $X+Y$, and assign an
inner product by defining $\innerproduct{X}{Y} = \Exp{XY}$.
Then Cauchy-Schwarz says that
$\parens*{\Exp{XY}}^2
= \innerproduct{X}{Y}^2
≤ \innerproduct{X}{X}⋅\innerproduct{Y}{Y}
= \Exp{X^2}\Exp{Y^2}.$
The covariance inequality is what happens when we substitute in
the centered variables $X-\E X$ and $Y - \E Y$ for $X$ and $Y$.
}
\begin{align}
    \parens*{\Cov{X}{Y}}^2 &≤ \Var{X}\Var{Y},
    \label{eq-covariance-squared}
    \intertext{sometimes written as}
    \abs{\Cov{X}{Y}} &≤ √{\Var{X}\Var{Y}}.
    \label{eq-covariance-abs}
\end{align}

A typical application of covariance is when we have a sum $S = ∑ X_i$ of
non-negative random variables with small covariance; here applying
Chebyshev's inequality to $S$ can often be used to show that $S$ is
not likely to be much smaller than $\Exp{S}$, which can be handy if we
want to show that some lower bound holds on $S$ with some probability.
This complements Markov's inequality, which can only be used to get
upper bounds.  

For example, suppose $S = ∑_{i=1}^{n} X_i$, where the $X_i$ are
independent Bernoulli random variables with $\Exp{X_i} = p$ for all
$i$.  Then $\Exp{S} = np$, and $\Var{S} = ∑_i \Var{X_i} = npq$
(because the $X_i$ are independent).  Chebyshev's inequality then says
\begin{equation*}
    \Prob{\abs{S - \Exp{S}} ≥ α} ≤ \frac{npq}{α^2}.
\end{equation*}

The highest variance is when $p = 1/2$.  In this case, he probability that $S$ is
more than $β√{n}$ away from its expected value $n/2$ is bounded by
$\frac{1}{4β^2}$.  We'll see better bounds on this problem later, but
this may already be good enough for many purposes.

More generally, the approach of bounding $S$ from below
by estimating $\Exp{S}$ and either $\Exp{S^2}$ or $\Var{S}$ is known as
the \concept{second-moment method}.  In some cases, tighter bounds can
be obtained by more careful analysis.

\subsubsection{Variance of a geometric random variable}

Let $X$ be a geometric random variable with parameter $p$ as defined
in §\ref{section-geometric-random-variables}, so that $X$ takes on the
values $1, 2, \dots$ and $\Prob{X=n} = q^{n-1} p$, where $q=1-p$ as
usual.  What is $\Var{X}$?

We know that $\Exp{X} = 1/p$, so $\left(\Exp{X}\right)^2 = 1/p^2$.
Computing $\Exp{X^2}$ is trickier.  Rather than do this directly from
the definition of expectation, we can exploit the memorylessness of
geometric random variables to get it using conditional expectations,
just like we did for $\Exp{X}$ in
§\ref{section-geometric-random-variables}.

Conditioning on the two possible outcomes of the first trial, we have
\begin{equation}
    \Exp{X^2}
    = p + q \ExpCond{X^2}{X>1}.
    \label{eq-geometric-second-moment-split}
\end{equation}

We now argue that $\ExpCond{X^2}{X>1} = \Exp{(X+1)^2}$.  The intuition
is that once we have flipped one coin the wrong way, we are back where
we started, except that now we have to add that extra coin to $X$.
More formally, we have, for $n>1$, $\ProbCond{X^2 = n}{X>1} =
\frac{\Prob{X^2 = n}}{\Prob{X>1}} = \frac{q^{n-1}p}{q} = q^{n-2}p =
\Prob{X=n-1} = \Prob{X+1=n}$.  So we get the same probability mass
function for $X$ conditioned on $X>1$ as for $X+1$ with no
conditioning.

Applying this observation to the right-hand side of
\eqref{eq-geometric-second-moment-split} gives
\begin{align*}
    \Exp{X^2}
    &= p + q \Exp{(X+1)^2}
    \\&= p + q \left(\Exp{X^2} + 2 \Exp{X} + 1\right)
    \\&= p + q \Exp{X^2} + \frac{2q}{p} + q
    \\&= 1 + q \Exp{X^2} + \frac{2q}{p}.
\end{align*}

A bit of algebra turns this into
\begin{align*}
    \Exp{X^2}
    &= \frac{1+2q/p}{1-q}
    \\&= \frac{1+2q/p}{p}
    \\&= \frac{p+2q}{p^2}
    \\&= \frac{2-p}{p^2}.
\end{align*}

Now subtract $\left(\Exp{X}\right)^2 = \frac{1}{p^2}$ to get
\begin{equation}
    \label{eq-geometric-variance}
    \Var{X} = \frac{1-p}{p^2} = \frac{q}{p^2}.
\end{equation}

By itself, this doesn't give very good bounds on $X$.  For example, if
we want to bound the probability that $X=1$, we get
\begin{align*}
    \Prob{X=1}
    &= \Prob{X-\Exp{X} = 1-\frac{1}{p}}
    \\&≤ \Prob{\abs*{X-\Exp{X}} ≥ \frac{1}{p} - 1}
    \\&≤ \frac{\Var{X}}{\left(\frac{1}{p} - 1\right)^2}
    \\&= \frac{q/p^2}{\left(\frac{1}{p} - 1\right)^2}
    \\&= \frac{q}{\left(1- p\right)^2}
    \\&= \frac{1}{q}.
\end{align*}

Since $\frac{1}{q} ≥ 1$, we could have gotten this bound with much
less work.

The other direction is not much better.  We can easily calculate that
$\Prob{X≥n}$ is exactly $q^{n-1}$ (because this corresponds to
flipping $n$ coins the wrong way, no matter what happens with
subsequent coins).  Using Chebyshev's inequality gives
\begin{align*}
    \Prob{X≥n}
    &≤ \Prob{\abs*{X-\frac{1}{p}} ≥ n-\frac{1}{p}}
    \\&≤ \frac{q/p^2}{\left(n-\frac{1}{p}\right)^2}
    \\&= \frac{q}{\left(np - 1\right)^2}.
\end{align*}

This at least has the advantage of dropping below $1$ when $n$ gets
large enough, but it's only polynomial in $n$ while the true value is
exponential.

Where this might be useful is in analyzing the sum of a bunch of
geometric random variables, as occurs in the Coupon Collector problem
discussed in §\ref{section-coupon-collector}.\footnote{We are
following here a similar analysis in \cite[§3.3.1]{MitzenmacherU2017}.}
Letting $X_i$ be the number of balls to take us from $i-1$ to $i$
empty bins, we have previously argued that $X_i$ 
has a geometric distribution with $p=\frac{n-i-1}{n}$, so
\begin{align*}
    \Var{X_i} 
    &= \left.\frac{i-1}{n} \middle/ \left(\frac{n-i-1}{n}\right)^2 \right.
    \\&= n \frac{i-1}{\left(n-i-1\right)^2},
    \intertext{and}
    \Var{∑_{i=1}^{n} X_i}
    &= ∑_{i=1}^{n} \Var{X_i}
    \\&= ∑_{i=1}^{n} n \frac{i-1}{\left(n-i-1\right)^2}.
\end{align*}

Having the numerator go up while the denominator goes down makes this
a rather unpleasant sum to try to solve directly.  So we will follow
the lead of Mitzenmacher and Upfal and bound the numerator by $n$,
giving
\begin{align*}
    \Var{∑_{i=1}^{n} X_i} 
    &≤ ∑_{i=1}^{n} n \frac{n}{\left(n-i-1\right)^2}
    \\&= n^2 ∑_{i=1}^{n} \frac{1}{i^2}
    \\&≤ n^2 ∑_{i=1}^{∞} \frac{1}{i^2}
    \\&= n^2 \frac{π^2}{6}.
\end{align*}

The fact that $∑_{i=1}^{∞} \frac{1}{i^2}$ converges to $\frac{π^2}{6}$
is not trivial to prove, and was first shown by Leonhard Euler in 1735
some ninety years after the question was first
proposed.\footnote{\label{footnote-Basel-problem}See
    \wikipedia{Basel_Problem} for a history, or
    Euler's original paper~\cite{Euler1768}, available at
    \url{http://eulerarchive.maa.org/docs/originals/E352.pdf}, for the
    actual proof in the full glory of its original 18th-century 
typesetting.  Curiously, though Euler announced his result in 1735,
he didn't submit the journal version until 1749, and it didn't see
print until 1768.  Things moved more slowly in those
days.}  But it's easy to show that the series converges to something,
so even if we didn't have Euler's help, we'd know that the variance
is $O(n^2)$.

Since the expected value of the sum is $Θ(n \log n)$, this tells us
that we are likely to see a total waiting time reasonably close to this;
with at least a constant probability, it will be within $Θ(n)$ of
the expectation.
In fact, the distribution is much more sharply concentrated (see \cite[§5.4.1]{MitzenmacherU2017}
or \cite[§3.6.3]{MotwaniR1995}), but this bound at least gives us
something.

\subsection{More examples}
\label{section-Chebyshevs-inequality-examples}

Here are some more examples of Chebyshev's inequality in action.  Some
of these repeat examples for which we previously got crummier bounds
in §\ref{section-markovs-inequality-examples}.

\subsubsection{Flipping coins}
\label{section-pairwise-independent-coins}
\label{section-pairwise-independence-construction}

Let $X$ be the sum of $n$ independent fair coins.  Let $X_i$ be the
indicator variable for the event that the $i$-th coin comes up heads.
Then $\Var{X_i} = 1/4$ and $\Var{X} = ∑ \Var{X_i} = n/4$.
Chebyshev's inequality gives $\Prob{X = n} ≤ \Prob{\abs*{X-n/2} ≥ n/2}
≤ \frac{n/4}{(n/2)^2} = \frac{1}{n}$.
This is still not very good, but it's getting better.  It's also about
the best we can do given only the assumption of pairwise independence.

To see this,
let $n = 2^m-1$ for some $m$, and let $Y_1\dots Y_m$ be independent,
fair $0$–$1$ random variables.  For each non-empty subset $S$ of
$\Set{1\dots m}$, let $X_S$ be the exclusive OR of all $Y_i$ for $i∈
S$.
Then (a) the $X_S$ are pairwise independent; (b) each $X_S$ has
variance $1/4$; and thus (c) the same Chebyshev's inequality analysis
for independent coin flips above applies to $X = ∑_S X_S$,
giving $\Prob{\abs*{X-n/2}=n/2} ≤ \frac{\Var{S}}{(n/2)^2} =
\frac{n/4}{n^2/4} = 
\frac{1}{n}$.  In this case it is not actually possible for $X$ to
equal $n$, but we can have $X=0$ if all the $Y_i$ are $0$, which
occurs with probability $2^{-m} = \frac{1}{n+1}$.
So the Chebyshev's inequality is almost tight in this case.

\subsubsection{Balls in bins}
Let $X_{i}$ be the indicator that the $i$-th of $m$ balls lands in a
particular bin.  Then $\Exp{X_{i}} = 1/n$, giving $\Exp{∑ X_{i}} =
m/n$, and $\Var{X_{i}} = 1/n - 1/n^{2}$, giving $\Var{∑ X_{i}} =
m/n - m/n^{2}$.  So the probability that we get $k + m/n$ or more
balls in a particular bin is at most $(m/n - m/n^{2})/k^{2} <
m/nk^{2}$, and applying the union bound, the probability that we get
$k + m/n$ or more balls in any of the $n$ bins is less than $m/k^{2}$.
Setting this equal to $ε$ and solving for $k$ gives a
probability of at most $ε$ of getting more than $m/n +
√{m/ε}$ balls in any of the bins.  
This is not as good a bound as we will be able to prove later, but it's at least non-trivial.

\subsubsection{Lazy select}

This example comes from \cite[§{}3.3]{MotwaniR1995}; essentially the
same example, specialized to finding the median,
also appears in \cite[§3.5]{MitzenmacherU2017}.\footnote{
The history is that
Motwani and Raghavan adapted this algorithm from a similar algorithm
by Floyd and Rivest~\cite{FloydR1975}.  Mitzenmacher and Upfal give a
version that also includes the adaptations appearing Motwani and Raghavan,
although they don't say where they got it from, and it may be that
both textbook versions come from a common folklore source.
}

We want to find the $k$-th smallest element $S_{(k)}$ of a set $S$ of
size $n$.  (The parentheses around the index indicate that we are
considering the sorted version of the set $S_{(1)} < S_{(2)} \dots <
S_{(n)}$.)  The idea is to:
\begin{enumerate}
 \item Sample a multiset $R$ of $n^{3/4}$ elements of $S$ with
 replacement and sort them.  This takes $O(n^{3/4} \log n^{3/4}) =
 o(n)$ comparisons so far.
 \item Use our sample to find an interval that is likely to contain
 $S_{(k)}$.  The idea is to pick indices $\ell = (k-n^{3/4})n^{-1/4}$ and
 $r = (k+n^{3/4})n^{-1/4}$ and use $R_{(\ell)}$ and $R_{(r)}$ as
 endpoints (we are omitting some floors and maxes here to simplify the
 notation; for a more rigorous presentation see \cite{MotwaniR1995}).
        The hope is that the interval $P = [R_{(\ell)},R_{(r)}]$ in $S$ will both
 contain $S_{(k)}$, and be small, with $\card*{P} ≤ 4n^{3/4} + 2$.
 We can compute the elements of $P$ in $2n$ comparisons exactly by
 comparing every element with both $R_{(\ell)}$ and $R_{(r)}$.
 \item If both these conditions hold, sort $P$ ($o(n)$ comparisons)
 and return $S_{(k)}$.  If not, try again.
\end{enumerate}

We want to get a bound on how likely it is that $P$ either misses
$S_{(k)}$ or is too big.

For any fixed $k$, the probability that one sample in $R$ is less than
or equal to $S_{(k)}$ is exactly $k/n$, so the expected number $X$ of
samples $≤ S_{(k)}$ is exactly $kn^{-1/4}$.  The variance on $X$
can be computed by summing the variances of the indicator variables
that each sample is $≤ S_{(k)}$, which gives a bound $\Var{X} =
n^{3/4}((k/n)(1-k/n)) ≤ n^{3/4}/4$.  Applying Chebyshev's inequality
gives $\Prob{\abs*{X-kn^{-1/4}} ≥ √{n}} ≤ n^{3/4}/4n =
n^{-1/4}/4$.

Now let's look at the probability that $P$ misses $S_{(k)}$ because
$R_{(\ell)}$ is too big, where $\ell = kn^{-1/4}-√{n}$.  This is
\begin{align*}
\Prob{R_{(\ell)} > S_{(k)}} 
&= \Prob{X < kn^{-1/4}-√{n}} 
\\&≤ n^{-1/4}/4.
\end{align*}
(with the caveat that we are being sloppy about round-off errors).

Similarly,
\begin{align*}
\Prob{R_{(h)} < S_{(k)}} 
&= \Prob{X > kn^{-1/4}+√{n}}
\\&≤ n^{-1/4}/4.
\end{align*}

So the total probability that $P$ misses $S_{(k)}$ is at most
$n^{-1/4}/2$.

Now we want to show that $\card*{P}$ is small.  We will do so by
showing that it is likely that $R_{(\ell)} ≥ S_{(k-2n^{3/4})}$ and
$R_{(h)} ≤ S_{(k+2n^{3/4})}$.  Let $X_{\ell}$ be the number of
samples in $R$ that are $≤ S_{(k-2n^{3/4})}$ and $X_{r}$ be the
number of samples in $R$ that are $≤ S_{(k+2n^{3/4})}$.  Then we
have $\Exp{X_{\ell}} = kn^{-1/4}-2√{n}$ and $\Exp{X_{r}} =
kn^{-1/4}+2√{n}$, and $\Var{X_{l}}$ and $\Var{X_{r}}$ are both
bounded by $n^{3/4}/4$.

We can now compute
\begin{align*}
\Prob{R_{(l)} < S_{(k-2n^{3/4})}} = Pr[X_{l} > kn^{-1/4}-√{n}] < n^{-1/4}/4
\end{align*}
by the same Chebyshev's inequality argument as before, and get the
symmetric bound on the other side for $\Prob{R_{(r)} > S_{(k+2n^{3/4})}}$.  This gives a total bound of $n^{-1/4}/2$ that $P$
is too big, for a bound of $n^{-1/4} = o(n)$ that the algorithm fails on its first attempt.

The total expected number of comparisons is thus given by $T(n) = 2n +
o(n) + O(n^{-1/4} T(n)) = 2n + o(n)$.

\section{Chernoff bounds}
\label{section-chernoff-bounds}

To get really tight bounds, we apply Markov's inequality to $\exp(α S)$,
where $S = ∑_i X_{i}$.  This works best when the $X_i$ are
independent:
if this is the case, so are the
variables $\exp(αX_{i})$, and so we can easily calculate
$\Exp{\exp(αS)} = \Exp{\prod_i \exp(α X_i)} = \prod_i
\Exp{\exp(α X_{i})}$.

The quantity $\Exp{\exp(α S)}$, treated as a function of $α$,
is called the \concept{moment generating function} of $S$, because it
expands formally into $∑_{k=0}^{∞} \Exp{X^{k}}\frac{α^{k}}{k!}$, the
\index{generating function!exponential}\concept{exponential generating
function} for the series of $k$-th \indexConcept{moment}{moments}
$\Exp{X^{k}}$.  Note that it may not converge for all $S$ and
$α$;\footnote{For example, the moment generating function for
our earlier bad $X$ with $\Prob{X = 2^k} = 2^{-k}$ is equal to
$∑_k 2^{-k} e^{α k}$, which diverges unless $e^{α}/2 <
1$.}
we will be
careful to choose $α$ for which it does converge
and for which Markov's inequality gives us good bounds.

\subsection{The classic Chernoff bound}
\label{section-chernoff-bound-general}

The basic Chernoff bound applies to sums of independent 0--1 random
variables, which need not be identically distributed.  For
identically distributed random variables, the sum has a binomial
distribution, which we can either compute exactly or bound more
tightly using approximations specific to binomial tails; for sums of
bounded random variables that aren't necessarily 0--1, we can use Hoeffding's
inequality instead (see §\ref{section-Azuma-Hoeffding}).

Let each $X_{i}$ for $i=1\dots n$ be a 0--1 random variable with
expectation $p_{i}$, so that $\Exp{S} = μ = ∑_i p_{i}$.  
The plan is to show $\Prob{S ≥ (1+δ)μ}$ is small when $δ$ and $μ$ are
large, by applying Markov's inequality to $\Exp{e^{αS}}$, where $α$
will be chosen to make the bound as tight as possible for some
specific $δ$.
The first step is to get an upper bound on $\Exp{e^{αS}}$.

Compute
\begin{align*}
    \Exp{e^{αS}}
    &= \Exp{e^{α ∑ X_i}}
    \\&= ∏_i \Exp{e^{α X_i}}
    \\&= ∏_i \left(p_i e^α + (1-p_i)e^0\right)
    \\&= ∏_i \left(p_i e^α + 1-p_i\right)
    \\&= ∏_i \left(1 + \left(e^α-1\right)p_i\right)
    \\&≤ ∏_i e^{(e^α-1)p_i}
    \\&= e^{(e^α-1)∑_i p_i}
    \\&= e^{(e^α-1)μ}.
\end{align*}

The sneaky inequality step in the middle uses the fact that $(1+x) ≤
e^x$ for all $x$, which itself is one of the most useful
inequalities you can memorize.\footnote{For a proof of this
    inequality, observe that the function $f(x) = e^x - (1+x)$ has the 
    derivative
    $e^x - 1$, which is positive for $x > 0$ and negative for $x < 0$.
    It follows that $x=1$ is the unique minimum of $f$, at which
$f(1)=0$.}

What's nice about this derivation is that at the end, the $p_i$ have
vanished.
We don't care what random variables we started with or how many of
them there were, but only about their expected sum $μ$.

Now that we have an upper bound on $\Exp{e^{αS}}$, we can throw it
into Markov's inequality to get the bound we really want:
\begin{align*}
\Prob{S ≥ (1+δ)μ}
&=
\Prob{e^{α S} ≥ e^{α(1+δ)μ}}
\\
&≤
\frac{\Exp{e^{α S}}}{e^{α(1+δ)μ}}
\\
&≤
\frac{e^{\left(e^α - 1\right) μ}}{e^{α(1+δ)μ}}
\\
&=
\left(\frac{e^{e^α - 1}}{e^{α(1+δ)}}\right)^μ
\\
&=
\left(e^{e^α - 1 - α(1+δ)}\right)^μ.
\end{align*}

We now choose $α$ to minimize the base in the last expression, by
minimizing its exponent $e^{α}-1-α(1+δ)$.  Setting the
derivative of this expression with respect to $α$ to zero gives $e^{α} =
(1+δ)$ or $α = \ln (1+δ)$; luckily, this value of $α$ is
indeed greater than $0$ as we have been assuming.  Plugging this value in gives
\begin{align}
\Prob{S ≥ (1+δ)μ }
&≤
\left(e^{(1 + δ) - 1 - (1+δ)\ln (1+δ)}\right)^μ
\nonumber
\\
&=
\left(\frac{e^δ}{(1+δ)^{1+δ}}\right)^μ.
\label{eq-Chernoff-bound}
\end{align}

The base of this rather atrocious quantity is $e^{0}/1^{1} = 1$ at
$δ=0$, and its derivative is negative for $δ≥0$ (the
easiest way to show this is to substitute $δ=x-1$ first).  So
the bound is never greater than $1$ and is both decreasing and less than $1$ as soon as
$δ>0$.  We also have that the bound is exponential in $μ$
for any fixed $δ$.  

If we look at the shape of the base as a function of $δ$, we can
observe that 
when $δ$ is very large, we can replace $(1+δ)^{1+δ}$
with $δ^δ$ without changing the bound much (and to the
extent that we change it, it's an increase, so it still works as a
bound).  This turns the
base into $\frac{e^δ}{δ^δ} = (e/δ)^δ =
1/(δ/e)^δ$.  This is pretty close to Stirling's formula for
$1/δ!$ (there is a $√{2 π δ}$ factor missing).

For very small $δ$, we have that $1+δ \approx e^δ$, so
the base becomes approximately $\frac{e^δ}{e^{δ(1+δ)}} =
e^{-δ^2}$.  This approximation goes in the wrong direction (it's
smaller than the actual value) but with some fudging we can show
bounds of the form $e^{-μ δ^2 / c}$ for various constants $c$,
as long as $δ$ is not too big.

\subsection{Easier variants}
\label{section-chernoff-bound-variants}

The full Chernoff bound can be difficult to work with, especially
since it's hard to invert \eqref{eq-Chernoff-bound}
to find a good $δ$ that gives a particular
$ε$ bound.  Fortunately, there are approximate variants that 
substitute a weaker but less intimidating bound.  Some of the more useful are:

\begin{itemize}
 \item For $0 ≤ δ ≤ 1.81$,
 \begin{align}
   \Prob{S ≥ (1+δ)μ} &≤ e^{-μδ^{2}/3}.
   \label{eq-Chernoff-bound-one-third}
 \end{align}
 (The actual upper limit is slightly higher.)  Useful for small values of
 $δ$, especially because the bound can be inverted: if we want
 $\Prob{X ≥ (1+δ)μ} ≤ \exp(-μδ^{2}/3) ≤
 ε$, we can use any $δ$ with $√{3 \ln
 (1/ε) / μ} ≤ δ ≤ 1.81$. 

The proof of the approximate bound is to show that, in the given range,
 $e^{δ}/(1+δ)^{1+δ} ≤
 \exp(-δ^{2}/3)$.  This is easiest to do numerically; a
 somewhat more formal argument that the bound holds in the range
 $0≤δ≤1$ can be found in~\cite[Theorem 4.4]{MitzenmacherU2017}.
 \item For $0 ≤ δ ≤ 4.11$,
 \begin{align}
   \Prob{S ≥ (1+δ)μ} &≤ e^{-μδ^{2}/4}.
   \label{eq-Chernoff-bound-one-fourth}
 \end{align}
 This is a slightly weaker bound than the previous that holds over a larger
 range.  It gives $\Prob{X ≥ (1+δ)μ} ≤ ε$
 if $√{4 \ln (1/ε) / μ} ≤ δ ≤ 4.11$.
 Note that the version given on page 72 of~\cite{MotwaniR1995} is
 \emph{not correct}; it claims that the bound holds up to
 $δ=2e-1 \approx 4.44$, but it fails somewhat short of this value.
 \item For $R ≥ 2eμ$,
  \begin{align}
    \Prob{S ≥ R} &≤ 2^{-R}.
    \label{eq-Chernoff-bound-R}
  \end{align}
  Sometimes
 the assumption is replaced with the stronger $R≥ 6μ$ (this is
 the version given in~\cite[Theorem 4.4]{MitzenmacherU2017}, for
 example); one can also verify numerically that $R≥ 5μ$ (i.e.,
 $δ≥ 4)$ is enough.  The proof of the $2eμ$ bound is
 that $e^{δ}/(1+δ)^{(1+δ)} <
 e^{1+δ}/(1+δ)^{(1+δ)} =
 (e/(1+δ))^{1+δ} ≤ 2^{-(1+δ)}$ when
 $e/(1+δ) ≤ 1/2$ or $δ ≥ 2e-1$.  Raising this to
 $μ$ gives $\Prob{S ≥ (1+δ)μ} ≤
 2^{-(1+δ)μ}$ for $δ≥2e-1$.  Now substitute $R$
 for $(1+δ)μ$ (giving $R≥2eμ$) to get the full
 result.

Inverting this bound gives $\Prob{S ≥ R} ≤ ε$
when $R ≥ \min(2eμ, \lg(1/ε))$.
\end{itemize}

Figure~\ref{figure-chernoff-plot} shows the relation between the
various bounds, in the region where they cross each other.

\begin{figure}
    \centering
\begin{tikzpicture}[x=2.15cm,y=4cm]
    \newcommand{\logChernoff}[1]{(#1-(1+#1)*ln(1+#1))}
    \newcommand{\squaredOver}[2]{((-#1*#1/#2)-\logChernoff{#1})}
    \newcommand{\R}[1]{(-(1+#1)*ln(2)-\logChernoff{#1})}
    \draw [color=black] (0,0) -- (5,0);
    \node at (5,0) [right] {$\frac{e^δ}{(1+δ)^{(1+δ)}}$};
    \draw [color=blue,domain=0:3] plot (\x, {\squaredOver{\x}{3}});
    \node at (3, {\squaredOver{3}{3}}) [below,color=blue] {$e^{-δ^2/3}$};
    \node at ({1.81696/2},0) [color=blue,below] {$δ≤1.81+$};
    \draw [dashed,color=blue] (1.81696,-0.5) -- (1.81696,0.75);
    \draw [color=green,domain=0:5] plot (\x, {\squaredOver{\x}{4}});
    \node at (5, {\squaredOver{5}{4}}) [right,color=green] {$e^{-δ^2/4}$};
    \node at (2.5, {\squaredOver{2.5}{4}}) [color=green,above] {$δ≤4.11+$};
    \draw [dashed,color=green] (4.11566,-0.5) -- (4.11566,0.75);
    \draw [color=red,domain=3:4] plot (\x, {\R{\x}});
    \node at (4, {\R{4}}) [left,color=red] {$2^{-R/μ}$};
    \draw [dashed,color=red] (3.31107,-0.5) -- (3.31107,0.75);
    \node at ({3.31107-0.05},0) [color=red,below right] {\small{$R/μ≥4.32-$}};
\end{tikzpicture}
\caption[Comparison of Chernoff bound variants]{Comparison of Chernoff
    bound variants with exponent $μ$ omitted, plotted in logarithmic
    scale relative to the standard bound.  Each bound is valid only in
    in the region where it exceeds $e^δ/(1+δ)^{1+δ}$.}
\label{figure-chernoff-plot}
\end{figure}

\subsection{Lower bound version}

We can also use Chernoff bounds to show that a sum of independent
$0$–$1$ random variables isn't too small.  The essential idea is to repeat the
upper bound argument with a negative value of $α$, which makes
$e^{α(1-δ)μ}$ an increasing function in $δ$.  The resulting bound is:
\begin{align}
\label{eq-Chernoff-bound-negative}
\Prob{S ≤ (1-δ)μ }
&≤
\left(\frac{e^{-δ}}{(1-δ)^{1-δ}}\right)^μ.
\end{align}

A simpler but weaker version of this bound is
\begin{align}
\Prob{S ≤ (1-δ)μ }
&≤
e^{-μδ^{2}/2}.
\label{eq-Chernoff-bound-negative-one-half}
\end{align}
Both bounds hold for all $δ$ with
$0≤δ≤1$.

\subsection{Two-sided version}

If we combine~\eqref{eq-Chernoff-bound-one-third}
with~\eqref{eq-Chernoff-bound-negative-one-half}, we get
\begin{align}
\Prob{\abs*{S-μ} ≥ δ μ} &≤ 2e^{-μδ^2/3},
\label{eq-Chernoff-two-sided-one-third}
\end{align}
for $0 ≤ δ ≤ 1.81$.

Suppose that we want this bound to be less than $ε$.  Then we
need $2e^{-δ^2/3} ≤ ε$ or
$δ ≥ √{\frac{3 \ln (2/ε)}{μ}}$.  Setting $δ$ to exactly
this quantity, \eqref{eq-Chernoff-two-sided-one-third} becomes
\begin{equation}
\Prob{\abs*{S-μ} ≥ √{3 μ \ln (2/ε)}} ≤
ε,
\label{eq-Chernoff-two-sided}
\end{equation}
provided 
$ε ≥ 2 e^{-μ/3}$.

For asymptotic purposes, we can omit the constants,
giving
\begin{lemma}
\label{lemma-Chernoff-two-sided}
Let $S$ be a sum of independent $0$–$1$ variables with $\Exp{S} = μ$.
Then for any $0 < ε ≤ 2 e^{-μ/3}$, $S$ lies within
$O\left(√{μ \log (1/ε)}\right)$ of $μ$, with probability at least
$1-ε$.
\end{lemma}

\subsection{What if we only have a bound on $\Exp{S}?$}

For some applications, we may not know $\Exp{S} = ∑ \Exp{X_i}$ exactly,
but have
only an upper bound $\Exp{S} ≤ μ$. It is not immediately obvious that
we can use this upper bound in place of the actual value of $\Exp{S}$
when computing an upper tail bound on $S$, because $μ$ appears in the exponent
of all of our bounds, and it is not clear that the corresponding
reduction in $δ$ fully compensates
for this. However, there is a simple argument that shows that
all of these bounds hold even if we overestimate $\Exp{S}$.

Consider a sum $S$ of independent $0$-$1$ random variables with
$\Exp{S} = μ' ≤ μ$. We can turn this into a sum of independent
$0$-$1$ random variables with expectation exactly $μ$ by adding enough
new extra variables to make a second sum $T$ with $\Exp{T} = μ - μ'$.
Now $S+T$ satisfies the conditions for
\eqref{eq-Chernoff-bound}, and $S ≤ S+T$ always, so
\begin{align*}
\Prob{S ≥ (1+δ)μ }
&≤
\Prob{S+T ≥ (1+δ)μ }
\\&<=
\left(\frac{e^δ}{(1+δ)^{1+δ}}\right)^μ.
\end{align*}

This also works for any of the bounds in
§\ref{section-chernoff-bound-variants} that are derived from
\eqref{eq-Chernoff-bound}.

In the other direction, if we know $\Exp{S} ≥ μ$ and want to apply
the lower tail bound
\eqref{eq-Chernoff-bound-negative}, we can apply a slightly different
construction. Suppose that $\Exp{S} = μ' ≥ μ$. For each $X_i$,
construct a $0$-$1$ random variable $Y_i$ such that (a) all the $Y_i$
are independent of each other, (b) $Y_i ≤ X_i$ always, and (c)
$\Exp{Y_i} = \Exp{X_i} (μ/μ')$. The easiest way to do this is to
set $Y_i = X_i Z_i$ where each $Z_i$ is an independent biased coin
with expectation $μ/μ'$.

Let $T = ∑ Y_i$. Then $T≤S$ and $\Exp{T} = ∑
\Exp{Y_i} = ∑ \Exp{X_i} (μ/μ') = μ$. Since $T$ satisfies the
requirements of \eqref{eq-Chernoff-bound-negative}, we can argue
\begin{align*}
\label{eq-Chernoff-bound-negative}
\Prob{S ≤ (1-δ)μ}
&≤
\Prob{T ≤ (1-δ)μ}
    \\&≤
\left(\frac{e^{-δ}}{(1-δ)^{1-δ}}\right)^μ.
\end{align*}

As with the upper tail bound, this approach also works for simpified
versions of the lower tail bound like
\eqref{eq-Chernoff-bound-negative-one-half}.

For the two-sided variants, we are out of luck. The best we can do if
we know $a ≤ \Exp{S} ≤ b$ is to apply each of the one-sided bounds
separately.

\subsection{Almost-independent variables}

Chernoff bounds generally don't work very well for variables that are
not independent, and in most such cases we must use Chebyshev's
inequality (§\ref{section-Chebyshevs-inequality}) or the Azuma-Hoeffding
inequality (§\ref{section-Azuma-Hoeffding}) instead.  But there is one
special case that comes up occasionally where it helps to be able to
apply the Chernoff bound to variables that are \emph{almost}
independent in a particular technical sense.

\begin{lemma}
    \label{lemma-Chernoff-almost-independent}
    Let $X_1,\dots,X_n$ be 0–1 random variables with the property that
    $\ExpCond{X_i}{X_1,\dots,X_{i-1}} ≤ p_i ≤ 1$ for all $i$.  Let $μ =
    ∑_{i=1}^{n} p_i$ and $S = ∑_{i=1}^{n} X_i$.  Then \eqref{eq-Chernoff-bound} holds.

    Alternatively, let $\ExpCond{X_i}{X_1,\dots,X_{i-1}} ≥ p_i ≥ 0$ for
    all $i$, and let $μ = ∑_{i=1}^n p_i$ and $S = ∑_{i=1}^n X_i$ as
    before.  Then
    \eqref{eq-Chernoff-bound-negative} holds.
\end{lemma}
\begin{proof}
    Rather than repeat the argument for independent variables, we will
    employ a coupling, where we replace the $X_i$ with independent
    $Y_i$ so that $∑_{i=1}^{n} Y_i$ gives a bound an $∑_{i=1}^n X_i$.

    For the upper bound, let each $Y_i = 1$ with independent
    probability $p_i$.  Use the following process to generate a new
    $X'_i$ in increasing order of $i$: if $Y_i = 0$, set $X'_i = 0$.
    Otherwise set $X'_i = 1$ with
    probability $\ProbCond{X_i=1}{X_1=X'_1,\dots X'_{i-1}}/p_i$,
    Then $X'_i ≤ Y_i$, and 
    \begin{align*}
        \ProbCond{X'_i=1}{X'_1,\dots,X'_i}
        &= \parens*{\ProbCond{X_i=1}{X_1=X'_1,\dots,X_{i-1}=X'_{i-1}} / p_i} \Prob{Y_i = 1}
        \\&= \ProbCond{X_i=1}{X_1=X'_1,\dots,X_{i-1}=X'_{i-1}}.
    \end{align*}
     It follows that the $X'_i$ have the same joint distribution as
     the $X_i$, and so 
     \begin{align*}
         \Prob{∑_{i=1}^n X'_i ≥ μ(1+δ)}
         &= \Prob{∑_{i=1}^{n} X_i ≥ μ(1+δ)}
         \\&≤ \Prob{∑_{i=1}^n Y_i ≥ μ(1+δ)}
         \\&≤ \left(\frac{e^δ}{(1+δ)^{1+δ}}\right)^μ.
     \end{align*}
    
    For the other direction, generate the $X_i$ first and generate the
    $Y_i$ using the same rejection sampling trick.  Now the $Y_i$ are
    independent (because their joint distribution is) and each $Y_i$
    is a lower bound on the corresponding $X_i$.
\end{proof}

The lemma is stated for the general Chernoff bounds
\eqref{eq-Chernoff-bound} and \eqref{eq-Chernoff-bound-negative}, but
the easier versions follow from these, so they hold as well, as long as we are careful to
remember that the $μ$ in the upper bound is not necessarily the same $μ$
as in the lower bound.

\subsection{Other tail bounds for the binomial distribution}

The random graph literature can be a good source for bounds on the
binomial distribution.  See for example~\cite[§1.3]{Bollobas2001},
which uses normal approximation to get bounds that are slightly
tighter than Chernoff bounds in some cases, and \cite[Chapter
2]{JansonLR2000}, which describes several variants of Chernoff bounds as
well as tools for dealing with sums of random variables that aren't
fully independent.

\subsection{Applications}

\subsubsection{Flipping coins}
Suppose $S$ is the sum of $n$ independent fair coin-flips.
Then $\Exp{S} = n/2$ and  $\Prob{S = n} = \Prob{S ≥ 2 \Exp{S}}$ is bounded using
\eqref{eq-Chernoff-bound} by setting $μ = n/2$, $δ = 1$ to get
$\Prob{S = n} ≤ (e/4)^{n/2} = (2/√{e})^{-n}$.  This is not quite
as good as the real answer $2^{-n}$ (the quantity $2/√{e}$ is
about 1.213\dots), but it's at least exponentially small.

\subsubsection{Balls in bins again}
Let's try applying the Chernoff bound to the balls-in-bins problem.
Here we let $S = ∑_{i=1}^{m} X_{i}$ be the number of balls in a
particular bin, with $X_{i}$ the indicator that the $i$-th ball lands
in the bin, $\Exp{X_{i}} = p_{i} = 1/n$, and $\Exp{S} = μ = m/n$.  
To get a bound on $\Prob{S ≥ m/n + k}$, apply 
the Chernoff bound with $δ = kn/m$ to get
\begin{align*}
\Prob{S ≥ m/n + k}
&=
\Prob{S ≥ (m/n) (1+kn/m)}
\\
&≤
\frac{e^{k}}{(1+kn/m)^{1+kn/m}}.
\end{align*}

For $m=n$, this collapses to the somewhat nicer but still pretty
horrifying $e^{k}/(k+1)^{k+1}$.  

Staying with $m=n$, if we are bounding the probability of having large
bins, we can use the $2^{-R}$ variant to show that the probability
that any particular bin has more than $2 \lg n$ balls (for example),
is at most $n^{-2}$, giving the probability that there exists such a
bin of at most $1/n$.  This is not as strong as what we can get out of
the full Chernoff bound.  If we take the logarithm of
$e^{k}/(k+1)^{k+1}$, we get $k-(k+1) \ln (k+1)$; if we then substitute
$k = \frac{c \ln n}{\ln \ln n} - 1$, we get
\begin{align*}
&
\frac{c \ln n }{ \ln \ln n} - 1
- \frac{c \ln n}{ \ln \ln n} \ln \frac{c \ln n }{ \ln \ln n }
\\
&=
(\ln n) \left(
  \frac{c}{\ln \ln n}
  - \frac{1}{\ln n}
  - \frac{c}{\ln \ln n} \left(
     \ln c + \ln \ln n - \ln \ln \ln n
  \right)
\right)
\\
&=
(\ln n) \left(
  \frac{c}{\ln \ln n}
  - \frac{1}{\ln n}
  - \frac{c \ln c}{\ln \ln n}
  - c
  + \frac{c \ln \ln \ln n}{\ln \ln n}
\right)
\\
&=
(\ln n)(-c + o(1)).
\end{align*}

So the probability of getting more than $c \ln n / \ln \ln n$ balls in
any one bin is bounded by $\exp((\ln n)(-c + o(1))) = n^{-c+o(1)}$.
This gives a maximum bin size of $O(\log n / \log \log n)$ with any
fixed probability bound $n^{-a}$ for sufficiently large $n$.

\subsubsection{Flipping coins, central behavior}
Suppose we flip $n$ fair coins, and let $S$ be the number that come up
heads.  We expect $μ = n/2$ heads on average.  How many extra
heads can we get, if we want to stay within a probability bound of
$n^{-c}$?

Here we use the small-$δ$ approximation, which gives $\Prob{S ≥
(1+δ)(n/2)} ≤ \exp(-δ^{2}n/6)$.  Setting
$\exp(-δ^{2}n/6) = n^{-c}$ gives $δ = √{6 \ln
n^{c}/n} = √{6c \ln n / n}$.  The actual excess over the mean
is $δ(n/2) = (n/2) √{6c \ln n / n} =
√{\frac{3}{2} c n \ln n}$.  By symmetry, the same bound applies to extra tails.  So if we flip 1000 coins and see more than 676 heads (roughly the bound when c=3), we can reasonably conclude that either (a) our coin is biased, or (b) we just hit a rare one-in-a-billion jackpot.

In algorithm analysis, the $√{(3/2)c}$ part usually gets
absorbed into the asymptotic notation, and we just say that with
probability at least $1-n^{-c}$, the sum of $n$ random bits is within
$O(√{n \log n})$ of $n/2$.

\subsubsection{Permutation routing on a hypercube}

Here we use Chernoff bounds to show bounds on a classic
\index{routing!permutation}\indexConcept{permutation routing}{permutation-routing} algorithm for 
\indexConcept{hypercube network}{hypercube networks} due to
Valiant~\cite{Valiant1982}.  The presentation here is based on §4.2 of
\cite{MotwaniR1995}, which in turn is based on an improved version of
Valiant's original analysis that appeared in a follow-up paper with
Brebner~\cite{ValiantB1981}.  There's also a write-up of this in
\cite[§4.6.1]{MitzenmacherU2017}.

The basic idea of a hypercube architecture is that we have a collection of
$N=2^n$ processors, each with an $n$-bit address.  Two nodes are
adjacent if their addresses differ by one bit (see
Figure~\ref{fig-hypercube-network} for an example).  Though now mostly of
theoretical interest, these things were the
cat's pajamas back in the 1980s: see \wikipedia{Connection_Machine}.

\begin{figure}
\centering
\begin{tikzpicture}[node distance=3cm,auto,thick]
    \node[labeled] (000) {$000$};
    \node[labeled] (001) [right of=000] {$001$};
    \node[labeled] (010) [above of=000] {$010$};
    \node[labeled] (011) [above of=001] {$011$};
    \node[labeled] (100) [above right of=000] {$100$};
    \node[labeled] (101) [right of=100] {$101$};
    \node[labeled] (110) [above of=100] {$110$};
    \node[labeled] (111) [right of=110] {$111$};

    \path
        (000) edge (001) edge (010) edge (100)
        (001) edge (011) edge (101)
        (010) edge (011) edge (110)
        (011) edge (111)
        (100) edge (101) edge (110)
        (101) edge (111)
        (110) edge (111)
        ;
\end{tikzpicture}
\caption{Hypercube network with $n=3$}
\label{fig-hypercube-network}
\end{figure}

Suppose that at some point in a computation, each processor $i$ wants to
send a packet of data to some processor $π(i)$, where $π$ is a
permutation of the addresses.  But we can only send one packet per
time unit along each of the $n$ edges leaving a
processor.\footnote{Formally, we have a synchronous routing model with
unbounded buffers at each node, with a maximum capacity of one packet
per edge per round.}  How do we
route the packets so that all of them arrive in the minimum amount of time?

We could try to be smart about this, or we could use randomization.
Valiant's idea is to first route each process $i$'s packet to some random
intermediate destination 
$σ(i)$, then in the second phase, we route it
from $σ(i)$ to its ultimate destination $π(i)$.  
Unlike $π$, $σ$ is not necessarily a permutation; instead, $σ(i)$ is
chosen uniformly at random independently of all the other $σ(j)$.
This makes the choice of paths for different packets independent
of each other, which we will need later to apply Chernoff bounds..

Routing is done by a
\index{routing!bit-fixing}
\indexConcept{bit-fixing routing}{bit-fixing}:
if a packet is currently at node $x$ and heading for node $y$, find
the leftmost bit $j$ where $x_j ≠ y_j$ and fix it, by sending
the packet on to $x[x_j/y_j]$.
In the absence of contention, bit-fixing routes a packet to its
destination in at most $n$ steps.
The
hope is that the randomization will tend to spread the packets evenly across the
network, reducing the contention for edges enough that the actual time
will not be much more than this.

The first step is to argue that, during the first phase,
any particular packet is delayed at
most one time unit by any other packet whose path overlaps with it.
Suppose packet $i$ is delayed by contention on some edge $uv$.  Then
there must be some other packet $j$ that crosses $uv$ during this
phase.  From this point on, $j$ remains one step ahead of $i$ (until
its path diverges), so it can't block $i$ again unless both are
blocked by some third packet $k$ (in which case we charge $i$'s
further delay to $k$).  This means that we can bound the delays for
packet $i$ by counting how many other packets cross its
path.\footnote{A much more formal version of this argument is given
as~\cite[Lemma 4.5]{MotwaniR1995}.}
So now we just need a high-probability bound on the number of packets
that get in a particular packet's way.

Following the presentation in~\cite{MotwaniR1995}, define $H_{ij}$ to
be the indicator variable for the event that packets $i$ and $j$ cross
paths during the first phase.
Because each $j$ chooses its destination independently, once
we fix $i$'s path, the $H_{ij}$ are all independent.  So we can bound 
$S = ∑_{j≠i} H_{ij}$ using Chernoff bounds.  To do so, we must first
calculate an upper bound on $μ = \Exp{S}$.

The trick here is to observe that any path that crosses $i$'s path
must cross one of its edges, and we can bound the number of such paths
by bounding how many paths cross each edge.  For each edge $e$, let 
$T_e$ be the number of paths that
cross edge $e$, and for each $j$, let $X_j$ be the number of edges that path $j$
crosses.  Counting two ways, we have $∑_e T_e = ∑_j X_j$, and so
$\Exp{∑_e T_e} = \Exp{∑_j X_j} ≤ N(n/2)$.  By symmetry, all the
$T_e$ have the same expectation, so we get $\Exp{T_e} ≤
\frac{N(n/2)}{Nn} = 1/2$.

Now fix $σ(i)$.  This determines some path $e_1 e_2 \dots e_k$ for
packet $i$.  In general we do not expect $\ExpCond{T_{e_\ell}}{σ(i)}$
to equal $\Exp{T_{e_\ell}}$, because conditioning on $i$'s path
crossing $e_\ell$ guarantees that at least one path crosses this edge
that might not have.  However, if we let $T'_e$ be the number of
packets $j≠i$ that cross $e$, then we have $T'_e ≤ T_e$ always,
giving $\Exp{T'_e} ≤ \Exp{T_e}$, and because $T'_e$ does not depend on
$i$'s path, $\ExpCond{T'_e}{σ(i)} = \Exp{T'_e} ≤ \Exp{T_e} ≤ 1/2$.
Summing this bound over all $k≤n$ edges on $i$'s path gives 
$\ExpCond{∑_{j≠i} H_{ij}}{σ(i)} ≤ n/2$, which implies $\Exp{∑_{j≠i} H_{ij}}≤n/2$
after removing the conditioning on $σ(i)$.

Inequality \eqref{eq-Chernoff-bound-R} says that $\Prob{X ≥ R} ≤ 2^{-R}$
when $R ≥ 2eμ$.  Letting $X= ∑_{j≠i} H_{ij}$ and setting $R = 3n$ gives $R =
6(n/2) ≥ 6μ > 2eμ$, so
$\Prob{∑_j H_{ij} ≥ 3n} ≤ 2^{-3n} = N^{-3}$.  This says that any
one packet reaches its random destination with at most $3n$ added
delay (thus, in at most $4n$ time units) with probability at least
$1-N^{-3}$.  If we consider all $N$ packets, the total probability
that any of them fail to reach their random destinations in $4n$ time
units is at most $N⋅ N^{-3} = N^{-2}$.  Note that because we are
using the union bound, we don't need independence for this
step—which is good, because we don't have it.

What about the second phase?  Here, routing the packets from the
random destinations back to the real destinations is just the reverse
of routing them from the real destinations to the random destinations.
So the same bound applies, and with probability at most $N^{-2}$ some
packet takes more than $4n$ time units to get back (this assumes that
we hold all the packets before sending them back out, so there are no
collisions between packets from different phases).

Adding up the failure probabilities and costs for both stages gives a
probability of at most $2/N^2$ that any packet takes more than $8n$
time units to reach its destination.

The structure of this argument is pretty typical for applications of
Chernoff bounds: we get a very small bound on the
probability that something bad happens by applying Chernoff bounds to
a part of the problem where we have independence, then use the union
bound to extend this to the full problem where we don't.

\section{The Azuma-Hoeffding inequality}
\label{section-Azuma-Hoeffding}

The problem with Chernoff bounds is that they only work for sums of Bernoulli
random variables.
\index{inequality!Hoeffding's}\concept{Hoeffding's
inequality}~\cite{Hoeffding1963} is another concentration bound based on the
moment generating function that applies to any sum of bounded
independent random variables with mean $0$.\footnote{Note that the
    requirement that $\Exp{X_i} = 0$ can always be satisfied by
considering instead $Y_i = X_i - \Exp{X_i}$.}  It has the additional
useful feature that it generalizes nicely to some collections of
random variables that are not mutually independent, as we will
see in §\ref{section-Azumas-inequality}.  This more general version is
known as 
\index{inequality!Azuma's}\concept{Azuma's
inequality}~\cite{Azuma1967}
or the
\index{inequality!Azuma-Hoeffding}
\concept{Azuma-Hoeffding inequality}.\footnote{The history of
this is that Hoeffding~\cite{Hoeffding1963} proved it for independent random variables, and
observed that the proof was easily extended to martingales, while
Azuma~\cite{Azuma1967} actually went and did the work of proving it
for martingales.}

\subsection{Hoeffding's inequality}
\label{section-Hoeffdings-inequality}

This is the version for sums of bounded independent random variables.
We will consider the symmetric case, where each variable $X_i$
satisfies $\abs{X_i} ≤ c_i$ for some constant $c_i$.  Hoeffding's
original result considered bounds of the form $a_i ≤ X_i ≤ b_i$, and
is equivalent when $a_i = -b_i$.

The main tool is \concept{Hoeffding's lemma},
which states
\begin{lemma}
    \label{lemma-Hoeffding}
    Let $\Exp{X} = 0$ and $\abs{X} ≤ c$ with probability 1.
    Then
\begin{equation}
    \Exp{e^{α X}} ≤ e^{(α c)^2/2}.
    \label{eq-Hoeffding-lemma}
\end{equation}
\end{lemma}
\begin{proof}
The basic idea is that, for any $α$, $e^{α x}$ is a convex
function.  Since we want an upper bound, we can't use Jensen's
inequality \eqref{eq-jensens-inequality}, but we \emph{can} use the
fact that $X$ is bounded and we know its expectation.
Convexity of $e^{α x}$ means that, for any $x$ with $-c ≤ x ≤
c$, $e^{α x} ≤ λ e^{-α c} + (1-λ) e^{α
c}$, where $x = λ (-c) + (1-λ)c$.  Solving for $λ$
in terms of $x$
gives $λ = \frac{1}{2}\left(1-\frac{x}{c}\right)$ and 
$1-λ = \frac{1}{2}\left(1+\frac{x}{c}\right)$.
So
\begin{align*}
\Exp{e^{α X}}
&≤
  \Exp{
    \frac{1}{2}\left(1-\frac{X}{c}\right) e^{-α c}
    + \frac{1}{2}\left(1+\frac{X}{c}\right) e^{α c}
}
\\&= 
  \frac{e^{-α c} + e^{α c}}{2}
  - \frac{e^{-α c}}{2c} \Exp{X}
  + \frac{e^{α c}}{2c} \Exp{X}
\\&=
  \frac{e^{-α c} + e^{α c}}{2}
\\&= \cosh(α c).
\end{align*}

In other words, the worst possible $X$ is a fair choice between $\pm
c$, and in this case we get the hyperbolic cosine of $α c$ as its
moment generating function.

We don't like hyperbolic cosines much, because we are going to want to
take products of our bounds, and hyperbolic cosines don't multiply
very nicely.  
As before with $1+x$,
we'd be much happier if we could replace the $\cosh$ with a nice
exponential.
The Taylor series expansion of $\cosh x$ starts with
$1+x^2/2+\dots$, suggesting that we should approximate it with $\exp(x^2/2)$, and
indeed 
it is the case that for all $x$,
$\cosh x ≤ e^{x^2/2}$.  This can be shown by comparing the rest of
the Taylor series expansions:
\begin{align*}
\cosh x
&= \frac{e^{x} + e^{-x}}{2}
\\&= \frac{1}{2}
 \left(
  ∑_{n=0}^{∞} \frac{x^n}{n!}
 +∑_{n=0}^{∞} \frac{(-x)^n}{n!}
 \right)
\\&= ∑_{n=0}^{∞} \frac{x^{2n}}{(2n)!}
\\&≤ ∑_{n=0}^{∞} \frac{x^{2n}}{2^n n!}
\\&= ∑_{n=0}^{∞} \frac{(x^2/2)^n}{n!}
\\&= e^{x^2/2}.
\end{align*}
This gives the claimed bound
\begin{equation*}
    \Exp{e^{α X}} ≤ \cosh(αc) ≤ e^{(α c)^2/2}.
\end{equation*}
\end{proof}.

\begin{theorem}
\label{theorem-Hoeffding}
Let $X_1 \dots X_n$ be independent random variables with $\Exp{X_i} = 0$
and $\abs*{X_i} ≤ c_i$ for all $i$.  Then for all $t$,
\begin{align}
\label{eq-Hoeffdings-inequality}
\Prob{∑_{i=1}^{n} X_i ≥ t}
&≤ \exp\left(-\frac{t^2}{2∑_{i=1}^{n} c_i^2}\right).
\end{align}
\end{theorem}
\begin{proof}
Let $S=∑_{i=1}^{n} X_i$.
As with Chernoff bounds, we'll first calculate a bound on
the moment generating function
$\Exp{e^{α S}}$ 
and then apply Markov's inequality with a
carefully-chosen $α$.

    From \eqref{eq-Hoeffding-lemma}, we have $\Exp{e^{αX_i}} ≤
    e^{(αc_i)^2/2}$ for all $i$.
Using this bound and the independence of the $X_i$, we compute
\begin{align*}
\Exp{e^{α S}}
&= \Exp{\exp\left(α ∑_{i=1}^{n} X_i\right)}
\\&= \Exp{\prod_{i=1}^{n} e^{α X_i} }
\\&= \prod_{i=1}^{n} \Exp{e^{α X_i}}
\\&≤ \prod_{i=1}^{n} e^{(α c_i)^2 / 2}
\\&= \exp\left(∑_{i=1}^n \frac{α^2 c_i^2}{2}\right)
\\&= \exp\left(\frac{α^2}{2} ∑_{i=1}^n c_i^2\right).
\end{align*}

Applying Markov's inequality then gives (when $α > 0$):
\begin{align}
\nonumber
\Prob{S ≥ t}
&= \Prob{e^{α S} ≥ e^{α t}}
\\&≤ 
  \exp\left(
    \frac{α^2}{2} ∑_{i=1}^{n} c_i^2
   -α t
  \right).
\label{eq-Hoeffding-markov-step}
\end{align}

Now we do the same trick as in Chernoff bounds and choose $α$ to
minimize the bound.  
If we write $C$ for $∑_{i=1}^{n} c_i^2$,
this is done by minimizing the exponent
$\frac{α^2}{2} C - α t$,
which we do by taking the derivative
with respect to $α$ and setting it to zero:
$α C - t = 0$, or $α = t/C$.
At this point, the exponent becomes
$\frac{(t/C)^2}{2} C - (t/C) t = -\frac{t^2}{2C}$.

Plugging this into \eqref{eq-Hoeffding-markov-step} 
gives the bound \eqref{eq-Hoeffdings-inequality} claimed in the theorem.
\end{proof}

\subsubsection{Hoeffding vs Chernoff}

Let's see how good a bound this gets us for our usual test problem of
bounding $\Prob{S = n}$ where $S = ∑_{i=1}^{n} X_i$ is the sum of
$n$ independent fair coin-flips.  To make the problem fit the theorem,
we replace each $X_i$ by a rescaled version $Y_i = 2X_i - 1 = \pm 1$
with equal probability; this makes $\Exp{Y_i} = 0$ as needed, with
$\abs*{Y_i}
≤ c_i = 1$.  Hoeffding's inequality \eqref{eq-Hoeffdings-inequality}
then gives
\begin{align*}
\Prob{∑_{i=1}^{n} Y_i ≥ n}
&≤
\exp\left(-\frac{n^2}{2n}\right)
\\&= e^{-n/2} = (√{e})^{-n}.
\end{align*}
Since $√{e} \approx 1.649\dots$, this is actually slightly better
than the $(2/√{e})^{-n}$ bound we get using Chernoff bounds. 

On the other hand, Chernoff bounds work better if we have a more
skewed distribution on the $X_i$; for example, in the balls-in-bins
case, each $X_i$ is a Bernoulli random variable with $\Exp{X_i} = 1/n$.
Using Hoeffding's inequality, we get a bound $c_i$ on
$\abs*{X_i - \Exp{X_i}}$ of only $1-1/n$, which puts $∑_{i=1}^{n}
c_i^2$ very close to $n$, requiring $t = Ω(√{n})$ before we
get any non-trivial bound out of \eqref{eq-Hoeffdings-inequality},
pretty much the same as in the fair-coin case (which is not
surprising, since Hoeffding's inequality doesn't know anything about
the distribution of the $X_i$).
But we've already seen that Chernoff gives us that $∑ X_i = O(\log n
/ \log \log n)$ with high probability in this case.

\subsubsection{Asymmetric version}

The original version of Hoeffding's inequality~\cite{Hoeffding1963}
assumes
$a_i ≤ X_i ≤ b_i$, but $\Exp{X_i}$ is still zero for all
$X_i$.  In this version, the bound is
\begin{align}
\label{eq-Hoeffdings-inequality-asymmetric}
\Prob{∑_{i=1}^{n} X_i ≥ t}
 &≤ \exp\left(-\frac{2t^2}{∑_{i=1}^{n} (b_i - a_i)^2}\right).
\end{align}
This reduces to \eqref{eq-Hoeffdings-inequality} when $a_i = -c_i$ and
$b_i = c_i$.  The proof is essentially the same, but a little more
analytic sneakery is required to show that
$\Exp{e^{α X_i}} ≤ e^{α^2 (b_i-a_i)^2/8}$;
see~\cite{McDiarmid1989} for a proof of this that is a little more
approachable than Hoeffding's original paper.
For most applications, the only difference between the symmetric
version \eqref{eq-Hoeffdings-inequality} and the asymmetric version
\eqref{eq-Hoeffdings-inequality-asymmetric} is a small constant factor on
the resulting bound on $t$.

Hoeffding's inequality is not the tightest possible inequality that
can be obtained from the conditions under which it applies,
but is relatively simple and easy to work with.  For a particularly
strong version of Hoeffding's inequality and a discussion of some
variants, see~\cite{FanGL2012}.

\subsection{Azuma's inequality}
\label{section-Azumas-inequality}

A general rule of thumb is that most things that work for sums of
independent random variables also work for martingales, which are
sequences of random variables that have similar behavior but allow for more
dependence.

Formally, a \concept{martingale} is a sequence of random variables $S_0, S_1, S_2,
\dots$, where $\ExpCond{S_t }{ S_1, \dots, S_{t-1}} = S_{t-1}$.  In other words,
given everything you know up until time $t-1$, your best guess of the
expected value at time $t$ is just wherever you are now.

Another way to describe a martingale is to take the partial sums $S_t
= ∑_{i=1}^{t} X_t$ of a
\concept{martingale difference sequence}, which is a sequence of
random variables $X_1, X_2, \dots$ where 
$\ExpCond{X_t}{X_1 \dots X_{t-1}} =
0$.  So in this version, your expected change from time $t-1$ to $t$
averages out to zero, even if you try to predict it using all the
information you have at time $t-1$.

In some cases it makes sense to allow extra information to sneak in.
We can represent this using $σ$-algebras, in particular by using a
filtration of the form $ℱ_0 ⊆ ℱ_1 ⊆ ℱ_2 ⊆ \dots$, where each $ℱ_t$ is
a $σ$-algebra (see
§\ref{section-expectation-conditioned-on-a-sigma-algebra}).
A sequence $S_0, S_1, S_2, \dots$ is \concept{adapted} to a filtration
$ℱ_0 ⊆ ℱ_1 ⊆ ℱ_2 ⊆ \dots$ if each $S_t$ is $ℱ_t$-measurable.  This
means that at time $t$ the sum of our knowledge ($ℱ_t$) is enough to
predict exactly the value of $S_t$.  The subset relations also mean
that we remember everything there is to know about $S_{t'}$ for $t' <
t$.

The general definition of a martingale is a collection
$\SetWhere{(S_t,ℱ_t)}{t∈ℕ}$ where
\begin{enumerate}
    \item Each $S_t$ is $ℱ_t$-measurable; and
    \item $\ExpCond{S_{t+1}}{ℱ_t} = S_t$.
\end{enumerate}

This means that even if we include any extra information we might have
at time $t$, we still can't predict $S_{t+1}$ any better than by
guessing the current value $S_t$.  This alternative definition will be
important in some special cases, as when $S_t$ is a function of some
other collection of random variables that we use to define the $ℱ_t$.
Because $ℱ_t$ includes at least as much information as
$S_0,\dots,S_t$, it will always be the case that any sequence
$\Set{(S_t,ℱ_t)}$ that is a martingale in the general sense gives a
sequence $\Set{S_t}$ that is a martingale in the more specialized
$\ExpCond{S_{t+1}}{S_0,\dots,S_t} = S_t$ sense.

Martingales were invented to analyze fair gambling games, where your
return over some time interval is not independent of previous outcomes
(for example, you may change your bet or what game you are playing
depending on how things have been going for you), but it is always
zero on average given previous information.\footnote{Real
casinos give negative expected return, so your winnings in a real
casino form a 
\concept{supermartingale} 
with $S_{t} ≥ \ExpCond{S_{t+1}}{S_0 \dots S_{t}}$.  On the other hand,
the casino's take, in a well-run casino, is a \concept{submartingale}, a process with 
$S_t ≤ \ExpCond{S_{t+1}}{S_0 \dots S_{t}}$.  These definitions
also
generalize in the obvious way to the $\Set{(S_t,ℱ_t)}$ case.}
The nice thing about martingales is they allow for a bit of dependence
while still acting very much like sums of independent random
variables.

Where this comes up with Hoeffding's inequality is that we might have
a process that is reasonably well-behaved, but its increments are not
technically independent.  For example, suppose that a gambler plays a
game where they bet $x$ units $0 ≤ x ≤ 1$ at each round, and
receives $\pm x$ with equal probability.  
Suppose also that their bet at each round may depend on the outcome of
previous rounds (for example, they might stop betting entirely if they 
lose too much money).
If $X_i$ is their take at
round $i$, we have that $\ExpCond{X_i }{ X_1 \dots X_{i-1}} = 0$ and that
$\abs*{X_i} ≤ 1$.  This is enough to apply the martingale version of
Hoeffding's inequality, often called Azuma's inequality.

\index{inequality!Azuma's}\index{Azuma's inequality}
\begin{theorem}
\label{theorem-Azuma}
Let $\Set{S_k}$ be a martingale with $S_k = ∑_{i=1}^{k} X_i$ and
$\abs*{X_i} ≤ c_i$ for all $i$.
Then for all $n$ and all $t ≥ 0$:
\begin{align}
\label{eq-Azumas-inequality}
\Prob{S_n ≥ t} 
&≤ \exp\left(\frac{-t^2}{2∑_{i=1}^{n} c_i^2}\right).
\end{align}
\end{theorem}
\begin{proof}
Basically, we just show that 
$\Exp{e^{α S_n}} ≤ \exp\left(\frac{α^2}{2}
∑_{i=1}^{n} c_i^2\right)$—just like in the proof of
Theorem~\ref{theorem-Hoeffding}—and the rest follows using the same
argument.  The only tricky part is we can no longer use independence  
to transform $\Exp{\prod_{i=1}^{n} e^{α X_i}}$ into
$\prod_{i=1}^{n} \Exp{e^{α X_i}}$.

Instead, we use the martingale property.  For each $X_i$, we have
$\ExpCond{X_i}{X_1\dots X_{i-1}} = 0$ and $\abs*{X_i} ≤ c_i$ always.
Recall that $\ExpCond{e^{α X_i}}{X_1\dots X_{i-1}}$ 
is a random variable that takes
on the average value of $e^{α X_i}$ 
for each setting of $X_1\dots X_{i-1}$.
We can apply the same analysis as in the proof
of~\ref{theorem-Hoeffding} to show that this means that 
$\ExpCond{e^{α X_i}}{X_1 \dots X_{i-1}} ≤ e^{(α c_i)^2/2}$
always.

The trick is to use the fact that, for any random variables $X$ and
$Y$, $\Exp{XY} = \Exp{\ExpCond{XY}{X}} = \Exp{X\ExpCond{Y}{X}}$.

We argue by induction on $n$ that
$\Exp{\prod_{i=1}^{n} e^{α X_i}}
 ≤ \prod_{i=1}^{n} e^{(α c)^2/2}$.
The base case is when $n=0$.
For the induction step, compute
\begin{align*}
    \Exp{\prod_{i=1}^{n} e^{α X_i}}
    &=\Exp{\ExpCond{\prod_{i=1}^{n} e^{α X_i}}{X_1\dots X_{n-1}}}
\\&=
\Exp{
   \left(\prod_{i=1}^{n} e^{α X_i} \right)
   \ExpCond{e^{α X_n}}{X_1\dots X_{n-1}}}
\\&≤
\Exp{
   \left(\prod_{i=1}^{n-1} e^{α X_i} \right)
   e^{(α c_n)^2/2}
   }
\\&=
\Exp{
   \prod_{i=1}^{n-1} e^{α X_i} 
   }
   e^{(α c_n)^2/2}
\\&≤
 \left(\prod_{i=1}^{n-1} e^{(α c_i)^2/2}\right)
   e^{(α c_n)^2/2}
\\&=
 \prod_{i=1}^{n} e^{(α c_i)^2/2}
\\&=
 \exp\left(\frac{α^2}{2} ∑_{i=1}^{n} c_i^2\right).
\end{align*}

The rest of the proof goes through as before.
\end{proof}

Some extensions:
\begin{itemize}
    \item The asymmetric version of Hoeffding's inequality
        \eqref{eq-Hoeffdings-inequality-asymmetric} also holds for
        martingales.  So if each increment $X_i$ satisfies $a_i ≤ X_i
        ≤ b_i$ always,
\begin{align}
\label{eq-Azumas-inequality-asymmetric}
\Prob{∑_{i=1}^{n} X_i ≥ t}
 &≤ \exp\left(-\frac{2t^2}{∑_{i=1}^{n} (b_i - a_i)^2}\right).
\end{align}
\item 
The same bound works for bounded-difference
\indexConcept{supermartingale}{supermartingales}.  
A supermartingale is a process where
$\ExpCond{X_i}{X_1 \dots X_{i-1}} ≤ 0$; the idea is that my expected
gain at any step is non-positive, so my present wealth is always
superior to my future wealth.\footnote{The corresponding notion in the
    other direction is a \concept{submartingale}.  See
§\ref{section-submartingales-and-supermartingales}.}
If $\ExpCond{X_i}{X_1\dots X_{i-1}} ≤ 0$ and
$\abs*{X_i}≤ c_i$, then we can write $X_i = Y_i+Z_i$ where $Y_i =
\ExpCond{X_i}{X_1\dots X_{i-1}} ≤
0$ is predictable from $X_1\dots X_{i-1}$ and 
$\ExpCond{Z_i}{X_1 \dots X_{i-1}} = 0$.\footnote{This is known as a
\index{decomposition!Doob}\concept{Doob decomposition} and can be used
to extract a martingale $\Set{Z_i}$ from any stochastic process
$\Set{X_i}$.  For general
processes, $Y_i = X_i - Z_i$ will still be predictable, but may not
        satisfy $\ExpCond{Y_i}{X_1,\dots,X_{i-1}}
≤ 0$.}
Then we can bound $∑_{i=1}^{n} X_i$ by observing that it is no
greater than $∑_{i=1}^{n} Z_i$.

A complication is that we no longer have $\abs*{Z_i}
≤ c_i$; instead, $\abs*{Z_i} ≤ 2c_i$ (since leaving out $Y_i$ may
shift $Z_i$ up).
But with this revised bound,
\eqref{eq-Azumas-inequality} gives
\begin{align}
\nonumber
\Prob{∑_{i=1}^{n} X_i≥ t}
&≤ \Prob{∑_{i=1}^{n} Z_i≥ t}
\\
\label{eq-Azumas-inequality-supermartingale}
&≤ \exp\left(-\frac{t^2}{8∑_{i=1}^{n} c_i^2}\right).
\end{align}
\item Suppose that we stop the process after the first time $τ$
with $S_τ = ∑_{i=1}^{τ} X_i ≥ t$.  This is equivalent to
making a new variable $Y_i$ that is zero whenever $S_{i-1} ≥ t$ and
equal to $X_i$ otherwise.  This doesn't affect the
conditions $\ExpCond{Y_i}{Y_1\dots Y_{i-1}} = 0$ or $\abs*{Y_i} ≤ c_i$, 
but it makes it so $∑_{i=1}^{n} Y_i ≥ t$ if and only if 
$\max_{k ≤ n} ∑_{i=1}^{k} X_i ≥ t$.  Applying
\eqref{eq-Azumas-inequality} to $∑ Y_i$ then gives
\begin{align}
\label{eq-Azumas-inequality-max}
\Prob{\max_{k ≤ n} ∑_{i=1}^{k} X_i ≥ t}
 &≤ \exp\left(-\frac{t^2}{2∑_{i=1}^{n} c_i^2}\right).
\end{align}
\item Since the conditions on $X_i$ in Theorem~\ref{theorem-Azuma}
apply equally well to $-X_i$, we have
\begin{align}
\label{eq-Azumas-inequality-lower-bound}
\Prob{∑_{i=1}^{n} X_i≤ -t}
 &≤ \exp\left(-\frac{t^2}{2∑_{i=1}^{n} c_i^2}\right).
\intertext{which we can combine with \eqref{eq-Azumas-inequality} to
get the two-sided bound}
\label{eq-Azumas-inequality-two-sided}
\Prob{\abs*{∑_{i=1}^{n} X_i}≥ t}
 &≤ 2\exp\left(-\frac{t^2}{2∑_{i=1}^{n} c_i^2}\right).
\end{align}
\item The extension 
of Hoeffding's inequality to the case $a_i ≤ X_i ≤ b_i$ works
equally well for
Azuma's inequality, giving the same bound as in
\eqref{eq-Hoeffdings-inequality-asymmetric}.
\item Finally, one can replace the requirement that each $c_i$ be a
constant with a requirement that $c_i$ be predictable from $X_1\dots
X_{i-1}$ and that $∑_{i=1}^{n} c_i^2 ≤ C$ always and get
$\Prob{∑_{i=1}^{n} X_i ≥ t} ≤ e^{-t^2/2C}$.  This
generally doesn't come up unless you have an algorithm that explicitly
cuts off the process if $∑ c_i^2$ gets too big, but there is at
least one example of this in the literature~\cite{AspnesW1996}.

There are also cases where the asymmetric version works with
$a_i ≤ X_i ≤ b_i$ where a bound on $b_i-a_i$ is fixed but the
precise values of $a_i$ and $b_i$ may vary depending on
$X_1,\dots,X_{i-1}$. This shows up in the proof of McDiarmid's
inequality~\cite{McDiarmid1989}, which is described below in
§\ref{section-method-of-bounded-differences}.
\end{itemize}

\subsection{The method of bounded differences}
\label{section-method-of-bounded-differences}

To use Azuma's inequality, we need a bounded-difference martingale.
The easiest way to get such martingales is through the 
\concept{method of bounded differences}, which was popularized by a
survey paper by McDiarmid~\cite{McDiarmid1989}.  For this reason the
key result is often referred to as \index{inequality!McDiarmid's}\concept{McDiarmid's inequality}.

The basic idea of the method is to structure a problem so that we are
computing a
function $f(X_1,\dots, X_n)$ of a sequence of independent random
variables $X_1,\dots, X_n$.
To get our martingale, we'll imagine we reveal the $X_i$ one at a
time, and compute at each step the expectation of the final value of
$f$ based on just the inputs we've seen so far.

Formally, 
let $ℱ_t = \Tuple{X_1,\dots,X_t}$, the $σ$-algebra generated by $X_1$
through $X_t$.  This represents all the information we have at time
$t$.
Let $Y_t = \ExpCond{f}{ℱ_t}$, the expected value
of $f$ given the values of the first $t$ variables.  
Then $\Set{\Tuple{Y_t,ℱ_t}}$
forms a martingale, 
with $Y_0$ = $\Exp{f}$ and $Y_t = \ExpCond{f}{X_1,\dots,X_t} = f$.
So if we can find a bound $c_t$ on $\abs{Y_t - Y_{t-1}}$, we can apply
Azuma's inequality to get bounds on $Y_n - Y_0 = f - \Exp{f}$.

A sequence of random variables of the form $Y_t =
\ExpCond{Z}{ℱ_t}$, where $Z$ is some fixed random variable and $ℱ_0
⊆ ℱ_1 ⊆ ℱ_2 ⊆ \dots$ is a filtration, is a called 
a \concept{Doob martingale}, and this is one of the most common ways
to construct a martingale.
The proof
that a Doob martingale is in fact a martingale is immediate from the general version of the
law of iterated expectation
$\ExpCond{Y_{t+1}}{ℱ_t} = \ExpCond{\ExpCond{Z}{ℱ_{t+1}}}{ℱ_t} =
\ExpCond{Z}{ℱ_t} = Y_t$.
Not all martingales are Doob martingales: for example, the martingale
whose difference sequence consists of fair $±1$ coin-flips doesn't
converge to any random variable in the limit.\footnote{A 
question came up in class whether martingales that converge at some
finite time $n$ are
technically Doob martingales with respect to $ℱ_t =
\Tuple{Y_0, Y_1, \dots Y_t}$. This is sort of true, since we have that
$\ExpCond{Y_n}{ℱ_t} = Y_t$ for all $t$. However this doesn't help much for
constructing a particular martingale. If we let $Y_t = 0$ for all $t <
k$ and $Y_t = Y_k = ±1$ for all $t ≥ k$, then for any $n ≥ k$,
$(\Set{\Tuple{Y_t, \Tuple{Y_0,\dots,Y_t}}})$ is a Doob martingale
generated by $Y_n$. But this doesn't tell us anything about which
$Y_k$ does the coin-flip.}

To show that $Y_t$ meets the conditions for Azuma's inequality, we
require that $f$ has the 
\concept{bounded difference property}, which says that
there are bounds $c_t$ such
that for any
$x_1 \dots x_n$ and any $x'_t$, we have 
\begin{align}
\label{eq-bounded-difference}
\abs*{f(x_1\dots x_t \dots x_n) - f(x_1 \dots x'_t \dots x_n)} &≤ c_t.
\end{align}

We want to bound $\abs{Y_{t+1} - Y_t} = \abs{\ExpCond{f}{ℱ_{t+1}} -
\ExpCond{f}{ℱ_t}}$. We can do this showing $Y_{t+1} - Y_t ≤ c_{t+1}$.
Because $f$ can be replaced by $-f$ without changing the bounded
difference property, essentially
the same argument will show $Y_{t+1} - Y_t ≥ -c_{t+1}$.

Fix some possible value $x_{t+1}$ for $X_{t+1}$.  The bounded
difference property says that
\begin{align*}
    \lvert & f(X_1,\dots,X_t,x_{t+1},X_{t+2},\dots,X_n)
    \\ & -
    f(X_1,\dots,X_t,X_{t+1},X_{t+2},\dots,X_n) \rvert
    \\ ≤  & c_{t+1},
\end{align*}
so
\begin{align}
    \lvert &\ExpCond{f(X_1,\dots,X_t,x_{t+1},X_{t+2},\dots,X_n)}{X_1,\dots,X_t}
    \nonumber
    \\ & - 
    \ExpCond{f(X_1,\dots,X_t,X_{t+1},X_{t+2},\dots,X_n)}{X_1,\dots,X_t}\rvert
    \nonumber
    \\ ≤ &c_{t+1}.
    \label{eq-bounded-differences-ugly}
\end{align}

The second conditional expectation is just $Y_t$. What is the first one?
Recall
\begin{align*}
    Y_{t+1} 
    &= \ExpCond{f(X_1,\dots,X_n}{X_1,\dots,X_{t+1}}
    \\&= \Exp{f(x_1,\dots,x_{t+1},X_{t+2},\dots,X_{n}}
\end{align*}
when $X_i = x_i$ for all $i ≤ t+1$, since conditioning on
$X_1,\dots,X_{t+1}$ just replaces the random variables with their
actual values and then averages over the rest.  This is the same as
\begin{align*}
    &\ExpCond{f(X_1,\dots,X_t,x_{t+1},X_{t+2},\dots,X_n}{X_1,\dots,X_{t}}
    \\= &\Exp{f(x_1,\dots,x_{t+1},X_{t+2},\dots,X_{n})}
\end{align*}
when $x_{t+1}$ happens to be the value of $X_{t+1}$.  So the first
conditional expectation in
\eqref{eq-bounded-differences-ugly} is just $Y_{t+1}$, giving
\begin{align*}
    \abs*{Y_{t+1} - Y_t}& ≤ c_{t+1}.
\end{align*}

Now we can apply Azuma-Hoeffding to
get 
\begin{align*}
    \Prob{Y_n - Y_0 ≥ t} &≤ \exp\parens*{-\frac{t^2}{2∑_{i=1}^n
c_i^2}}.  
\end{align*}

This turns out to overestimate the possible range of $Y_{t+1}$.  With
a more sophisticated argument, it can be shown 
that for any fixed $x_1,\dots,x_t$, there exist
bounds $a_{t+1} ≤ Y_{t+1} - Y_t ≤ b_{t+1}$ such that $b_{t+1} -
a_{t+1} = c_{t+1}$.  We
would like to use this to apply the asymmetric version
of Azuma-Hoeffding given in
\eqref{eq-Azumas-inequality-asymmetric}.  A complication is that the specific
values of $a_{t+1}$ and $b_{t+1}$ may depend on the previous values
$x_1,\dots,x_t$, even if the bound $c_{t+1}$ on their maximum difference
does not. Fortunately, McDiarmid shows that the inequality works 
anyway, giving:
\begin{theorem}[\concept{McDiarmid's
    inequality}\index{inequality!McDiarmid's}~\cite{McDiarmid1989}]
\label{theorem-mcdiarmids-inequality}
Let $X_1,\dots,X_n$ be independent random variables and let
$f(X_1,\dots,X_n)$ have the bounded difference property with bounds $c_i$.
Then
\begin{align}
\label{eq-McDiarmids-inequality}
\Prob{f(X_1,\dots,X_n) - \Exp{f(X_1,\dots,X_n)} ≥ t}
 ≤ \exp\left(-\frac{2t^2}{∑_{i=1}^{n} c_i^2}\right).
\end{align}
\end{theorem}

The main difference between this and a direct application of
Azuma-Hoeffding is that the constant factor in the
exponent is better by a factor of $4$.

Since $-f$ satisfies the same constraints as $f$, the same bound holds
for $\Prob{f(X_1,\dots,X_n) - \Exp{f(X_1,\dots,X_n)} ≤ -t}$.  For some
applications it may make sense to apply the union bound to get a two-sided
version
\begin{align}
\label{eq-McDiarmids-inequality-two-sided}
    \Prob{\abs*{f(X_1,\dots,X_n) - \Exp{f(X_1,\dots,X_n)}} ≥ t}
 ≤ 2\exp\left(-\frac{2t^2}{∑_{i=1}^{n} c_i^2}\right).
\end{align}

\subsection{Applications}
\label{section-Azuma-Hoeffding-applications}

Here are some applications of the preceding inequalities.  Most of
these are examples of the method of bounded differences.

\subsubsection{Sprinkling points on a hypercube}

Suppose you live in a \index{hypercube network}hypercube, and
the local government has conveniently placed mailboxes on some
subset $A$ of the nodes.  If you start at a random location, how
likely is it that your distance to the nearest mailbox deviates
substantially from the average distance?

We can describe your position as a bit vector $X_1,\dots,X_n$, where
each $X_i$ is an independent random bit.  Let $f(X_1,\dots,X_n)$ be
the distance from $X_1,\dots,X_n$ to the nearest element of $A$.  Then
changing one of the bits changes this function by at most $1$.  So we
have $\Prob{\abs*{f - \Exp{f}} ≥ t} ≤ 2e^{-2t^2/n}$ by
\eqref{eq-McDiarmids-inequality}, giving a range of possible distances that is
$O(√{n \log n})$ with probability at least
$1-n^{-c}$ for any fixed $c>0$.\footnote{Proof: Let
    $t=√{\frac{1}{2}(c+1) n \ln n } = O(√{n \log n})$.
    Then $2e^{-2t^2/n} = 2e^{-(c + 1) \ln n} = 2 n^{-c-1} <
n^{-c}$ when $n$ is sufficiently large.}  Of course,
without knowing what $A$ is, we don't know what $E[f]$ is; but at
least we can be assured that (unless $A$ is very big)
the distance we have to walk to send our mail
will be pretty much the same pretty much wherever we start.

\subsubsection{Chromatic number of a random graph}
    
Consider a \index{graph!random}\concept{random graph} \index{$G(n,p)$}{$G(n,p)$}
consisting of $n$ vertices, where
each possible edge appears with independent probability $p$.
Let $χ$ be the \concept{chromatic number} of this graph, the
minimum number of colors necessary if we want to assign a color to
each vertex that is distinct for the colors of all of its neighbors.
The \index{martingale!vertex exposure}\concept{vertex exposure martingale} shows us the vertices of the
graph one at a time, along with all the edges between vertices that
have been exposed so far.  We define $X_t$ to be the expected value of
$χ$ given this information for vertices $1\dots t$.

If $Z_i$ is a random variable describing which edges are present
between $i$ and vertices less than $i$, then the $Z_i$ are all
independent, and we can write $χ = f(Z_1,\dots,Z_n)$ for some
function $f$ (this function may not be very easy to compute, but it
exists).  Then $X_t$ as defined above is just $\ExpCond{f}{Z_1,\dots,Z_t}$.

Now observe that $f$ has the bounded difference property with $c_t = 1$: if I change the
edges for some vertex $v_t$, I can't increase the number of colors I
need by more than $1$, since in the worst case I can always take
whatever coloring I previously had for all the other vertices and add
a new color for $v_t$. This implies that the difference between any
two graphs $G$ and $G'$ that differ only in the value of some $Z_t$ is at most one,
because going between them is an increase in one direction or the
other.

McDiarmid's inequality
\eqref{eq-McDiarmids-inequality}
then says that
$\Prob{\abs*{χ-\Exp{χ}} ≥ t} ≤ 2e^{-2t^2/n}$; in other words, the
chromatic number of a random graph is tightly concentrated around its
mean, even if we don't know what that mean is.

This proof is due to Shamir and Spencer~\cite{ShamirS1987}. Much
better bounds are known on the expected value and distribution of
$χ(G(n,p))$ for many values of $n$ and $p$ than are given by this
crude result. For a more recent paper that cites many of these
see~\cite{Heckel2021}.

\subsubsection{Balls in bins}

Suppose we toss $m$ balls into $n$ bins.  How many empty bins do
we get?  The probability that each bin individually is empty is
exactly $(1-1/n)^m$, which is approximately $e^{-m/n}$ when $n$ is
large.  So the expected number of empty bins is exactly $n(1-1/n)^m$.
If we let $X_i$ be the bin that ball $i$ gets tossed into, and let
$Y=f(X_1,\dots,X_m)$ be the number of empty bins, then changing a single
$X_i$ can change $f$ by at most $1$.  
So from~\eqref{eq-McDiarmids-inequality} we have
$\Prob{Y ≥ n(1-1/n)^m + t} ≤ e^{-2t^2/m}$.

\subsubsection{Probabilistic recurrence relations}
\label{section-Azuma-Hoeffding-recurrences}

Most \index{probabilistic recurrence}probabilistic recurrence
arguments (as in 
§\ref{section-probabilistic-recurrences})
can be interpreted as \index{supermartingale}supermartingales:
the current estimate of $T(n)$
is always greater than or equal to the expected estimate after doing one stage of the recurrence.
This fact can be used to get concentration bounds using
\eqref{eq-Azumas-inequality-supermartingale}.

For example, let's take the recurrence \eqref{eq-quicksort-recurrence}
for the expected number of comparisons for \index{QuickSort}QuickSort:
\begin{align*}
    T(n) &= (n-1) + \frac{1}{n} ∑_{k=0}^{n-1} \left(T(k) + T(n-1-k)\right).
\end{align*}

We showed in §\ref{section-quicksort-recurrence}
that the solution to this recurrence satisfies $T(n) ≤ 2n \ln n$.

To turn this into a supermartingale, imagine that we carry out a
process where we keep around at each step $t$ a set of unsorted blocks of
size $n^t_1,n^t_2,\dots,n^t_{k_t}$ for some $k_t$ (note that the
superscripts on $n_i^t$ are not exponents).  One step of the process
involves choosing one of the blocks (we can do this arbitrarily without
affecting the argument) and then splitting that block around a
uniformly-chosen pivot.  We will track a random variable $X_t$ equal
to $C_t + ∑_{i=1}^{k_t} 2n_i^t \ln n_i^t$, where $C_t$ is the number
of comparisons done so far and the summation gives an upper bound on
the expected number of comparisons remaining.

To show that this is in fact a supermartingale, observe that if we partition a
block of size $n$ we add $n-1$ to $C_t$ but replace the cost bound
$2n\ln n$ by an expected
\begin{align*}
2 ⋅ \frac{1}{n} ∑_{k=0}^{n-1} 2 k \ln k 
&≤ \frac{4}{n} \int_{2}^{n} n \ln n
\\
&= \frac{4}{n} \left(\frac{n^2 \ln n}{2} - \frac{n^2}{4} - \ln 2 + 1\right)
\\
&= 2 n \ln n - n - \ln 2 + 1
\\
&< 2 n \ln n - n.
\end{align*}

The net change is less than $-\ln 2$.
The fact that it's not zero suggests that we could improve the $2n
\ln n$ bound slightly, but since it's going down, we have a
supermartingale.

Let's try to get a bound on how much $X_{t}$ changes at
each step.  
The $C_t$ part goes up by at most $n-1$.
The summation can only go down; if we split a block of size $n_i$, the
biggest drop we get is if we split it evenly,\footnote{This can be
proven most easily using convexity of $n \ln n$.}
This gives a drop of 
\begin{align*}
2n \ln n - 2\left(2\frac{n-1}{2} \ln \frac{n-1}{2}\right)
&= 2n \ln n - 2(n-1) \ln \left(n \frac{n-1}{2n}\right)
\\
&= 2n \ln n - 2(n-1) (\ln n - \ln \frac{2n}{n-1})
\\
&= 2n \ln n - 2n \ln n + 2n \ln \frac{2n}{n-1} + 2 \ln n - 2 \ln \frac{2n}{n-1}
\\
&= 2n ⋅ O(1) + O(\log n)
\\
&= O(n).
\end{align*}
(with a constant tending to $2$ in the limit).

So we can apply \eqref{eq-Azumas-inequality-supermartingale} with $c_t
= O(n)$ to the at most $n$ steps of the algorithm, and get
\begin{align*}
\Prob{C_n - 2n \ln n ≥ t} ≤ e^{-t^2/O(n^3)}.
\end{align*}
This gives $C_n = O(n^{3/2})$ with constant probability or
$O(n^{3/2}√{\log n})$ with all but polynomial probability.
This is a rather terrible bound, but it's much better than $O(n^2)$.

Much tighter bounds are known: QuickSort in fact uses $Θ(n \log n)$
comparisons with high probability~\cite{McDiarmidH1992}.  If we aren't
too worried about constants, an easy way
to see the upper bound side of this is to adapt the analysis of
Hoare's FIND (§\ref{section-Hoares-find}).  For each element, the
number of elements in the same block is multiplied by a factor
of at most $3/4$ on average each time the element is compared, so the
chance that the element is not by itself is at most $(3/4)^k n$ after
$k$ comparisons.  Setting $k = \log_{4/3} (n^2/ε)$ gives that any
particular element is compared $k$ or times with probability at most
$ε/n$.  The union bound then gives a probability of at most $ε$ that
the most-compared element is compared $k$ or more times.  So the total
number of comparisons is $O(\log (n/ε))$ with probability $1-ε$, which
becomes $O(\log n)$ with probability $1-n^{-c}$ if we set $ε = n^{-c}$
for a fixed $c$.

We can formalize this argument itself using a supermartingale. Let
$C^t_i$ be the number of times $i$ has been compared as a non-pivot
in the first $t$ pivot steps and $N^t_i$ be the size of the block
containing $i$ after $t$ pivot steps. Let $Y^t_i = (4/3)^{C^t_i}
N^t_i$. Then if we pick $i$'s block at step $t+1$, the exponent goes
up by at most $1$ and $N^t$ drops by a factor of $3/4$, canceling out
the increase. If we don't pick $i$'s block, $Y^t_i$ is unchanged. In
either case we get $Y^t_i ≥ \ExpCond{Y^{t+1}_i}{ℱ_t}$ and $\Set{Y^t_i}$ is
a supermartingale.

Now let $Z^t = ∑_i Y^t_i$. Since this is greater than or equal to $∑_i
\ExpCond{Y^{t+1}_i}{ℱ_t} = Z^{t+1}$, $\Set{Z^t}$ is also a
supermartingale. For $t=0$, $Z^0 = n⋅(4/3)^0⋅n = n^2$. For $t = n$,
$Z^n = ∑_i (4/3)^{C^n_i}$. But then
\begin{align*}
    \Prob{\max_i C^n_i ≥ a \log_{4/3} n}
    &= \Prob{\max_i (4/3)^{C^n_i} ≥ n^a}
    \\&≤ \Prob{∑_i (4/3)^{C^n_i} ≥ n^a}
    \\&= \Prob{Z^n ≥ n^a}
    \\&≤ \frac{\Exp{Z^n}}{{n^a}}
    \\&≤ \frac{Z^0}{n^a}
    \\&= n^{a-2}.
\end{align*}

Choosing $a = c+2$ gives an $n^{-c}$ bound on $\Prob{\max_i C^n_i ≥
(c+2) \log_{4/3} n}$ and thus the same bound on $\Prob{\sum_i C^n_i ≥
(c+2) n \log_{4/3} n}$. This shows that the total number of
comparisons is $O(n \log n)$ with high probability.

\subsubsection{Multi-armed bandits}
\label{section-UCB1}

In the 
\index{bandit!multi-armed}
\concept{multi-armed bandit} problem, 
we must choose at each time step one of a fixed set of $k$
\indexConcept{arm}{arms} to pull.  Pulling arm $i$ at time $t$ yields
a return of $X^t_i$, a random payoff typically assumed to be between
$0$ and $1$.  Suppose that all the $X^t_i$ are independent, and that
for each fixed $i$, all $X^t_i$ have the same distribution, and thus
the same expected payoff.  Suppose also that we initially know nothing
about these distributions.  What strategy can we use to maximize our
expected payoff over a large number of pulls?  

More specifically,
we want to minimize our \concept{regret}, defined as
\begin{equation}
    \label{eq-ucb1-regret-definition}
    T_i ⋅ (μ^* - μ_i),
\end{equation}
where $T_i$ counts the number of times we pull arm $i$,
where $μ^*$ is the expected payoff of the best arm,
and $μ_i = \Exp{X_i}$ is the expected payoff of arm $i$.

The tricky part here is that when we pull an arm and get a bad return,
we don't know if we were just unlucky this time or it's actually a bad
arm.  So we have an incentive to try lots of different arms.  On the
other hand, the more we pull a genuinely inferior arm, the worse our
overall return.  We'd like to adopt a strategy that trades off between
exploration (trying new arms) and exploitation (collecting on the best
arm so far) to do as best we can in comparison to a strategy that
always pulls the best arm.

\paragraph{The UCB1 algorithm}

Fortunately, there is a simple algorithm due to
Auer~\etal~\cite{AuerCF2002} that solves this problem for
us.\footnote{This is not the only algorithm for solving multi-armed
    bandit problems, and it's not even the only algorithm in the
    Auer~\etal~paper.  But it has the advantage of being relatively
    straightforward to analyze.  For a more general survey of
    multi-armed bandit algorithms,
see~\cite{BubeckC2012} or~\cite{LattimoreS2020}.}
To start with, we pull each arm once.  For any subsequent pull,
suppose
that for each $i$, we have pulled the $i$-th arm $n_i$ times so far.
Let $n = ∑ n_i$, and let $\overline{x}_i$ be the average payoff from arm
$i$ so far.  Then the 
\index{algorithm!UCB1}
\index{UCB1 algorithm}
\conceptFormat{UCB1} algorithm pulls the arm that maximizes
\begin{equation}
    \label{eq-ucb1-criterion}
    \overline{x}_i + √{\frac{2 \ln n}{n_i}}.
\end{equation}

UCB stands for 
\concept{upper confidence bound}.  (The ``1'' is because it is
    the first, and simplest, of several algorithms of this general
    structure given in the paper.)  The idea is that we give arms that
we haven't tried very much the benefit of the doubt, and assume that their
actual average payout lies at the upper end of some plausible range of
values.\footnote{The ``bound'' part is because we don't attempt to
    compute the exact upper end of this confidence interval, which may
    be difficult, but
    instead use an upper bound derived from Hoeffding's inequality.
    This distinguishes the UCB1 algorithm of~\cite{AuerCF2002} from
    the \emph{upper confidence interval} approach of Lai and
    Robbins~\cite{LaiR1985} that it builds on.}

The quantity $√{\frac{2 \ln n}{n_i}}$ is a bit mysterious, 
but it arises in a fairly natural way from the asymmetric version of
Hoeffding's inequality.
With a small adjustment to deal with non-zero-mean variables,
\eqref{eq-Hoeffdings-inequality-asymmetric} says that, if
$S$ is a sum of $n$ random variables bounded between $a_i$ and $b_i$,
then
\begin{equation}
    \label{eq-Hoeffdings-inequality-asymmetric-nonzero-means}
    \Prob{∑_{i=1}^{n} \parens*{X_i - \Exp{X_i}} ≥ t}
    ≤ e^{-2t^2/∑_{i=1}^{n} (b_i-a_i)^2}.
\end{equation}
By applying \eqref{eq-Hoeffdings-inequality-asymmetric} to $-X_i$, we
also get
\begin{equation}
    \label{eq-Hoeffdings-inequality-asymmetric-nonzero-means-lower-bound}
    \Prob{∑_{i=1}^{n} \parens*{X_i - \Exp{X_i}} ≤ -t}
    ≤ e^{-2t^2/∑_{i=1}^{n} (b_i-a_i)^2}.
\end{equation}

Now consider $\overline{x}_i = \frac{1}{n_i} ∑_{j=1}^{n_i} X_j$ where each $X_j$
lies between $0$ and $1$.
This is equivalent to having 
$\overline{x}_i = ∑_{j=1}^{n_i} Y_j$ where $Y_j = X_j/n_i$ lies
between $0$ and $1/n_i$.
So
\eqref{eq-Hoeffdings-inequality-asymmetric-nonzero-means} says that
\begin{align}
    \Prob{\overline{x}_i - \Exp{\overline{x}_i} ≥ √{\frac{2 \ln n}{n_i}}}
    &≤ e^{-2(√{(2 \ln n)/n_i})^2/(n_i(1/n_i)^2)}
    \nonumber
    \\&= e^{-4 \ln n}
    \nonumber
    \\&= n^{-4}.
    \label{eq-ucb1-bonus-bound}
\end{align}
We also get a similar lower bound using
\eqref{eq-Hoeffdings-inequality-asymmetric-nonzero-means-lower-bound}.

So the bonus applied to $\overline{x}_i$ is really a high-probability
bound on how big the difference between the observed payoff and
expected payoff might be.  The $√{\ln n}$ part is there to make the
error probability be small as a function of $n$, since we will be
summing over a number of bad cases polynomial in $n$ and not a
particular $n_i$.
Applying the bonus to all arms make it likely
that the observed payoff of the best
arm stays above its actual payoff, so we won't forget to pull it.
The hope is that over time the same bonus applied to other arms will
not boost them up so much that we pull them any more than we have to.

However, in an infinite execution of the algorithm, even a bad arm
will be pulled infinitely often, as $\ln n$ rises enough to compensate
for the last increase in $n_i$.  This accounts for an $O(\log n)$ term
in the regret, as we will see below.  It also prevents us from getting
stuck refusing to give a good arm a second chance just because we had
an initial run of bad luck.

\paragraph{Analysis of UCB1}

The following theorem is a direct quote of \cite[Theorem 1]{AuerCF2002}:
\begin{theorem}[\cite{AuerCF2002}]
    \label{theorem-ucb1}
    For all $K > 1$, if policy UCB1 is run 
on $K$ machines having arbitrary reward
distributions $P_1,\dots, P_K$ with support in $[0, 1]$, then its expected regret after any number
$n$ of plays is at most
\begin{equation}
    \label{eq-ucb1-regret-bound}
    \left[8 ∑_{i:μ_i <μ^*} \frac{\ln n}{Δ_i}\right]
    + \left(1 + \frac{π^2}{3}\right) \left(∑_{j=1}^{K} Δ_j\right).
\end{equation}
\end{theorem}

Here $μ_i=\Exp{P_i}$ is the expected payoff for arm $i$, $μ^*$ as
before is $\max_i μ_i$, and $Δ_i = μ^* - μ_i$ is the regret for
pulling arm $i$.  The theorem states that our expected regret is
bounded by a term for each arm worse that $μ^*$ that grows
logarithmically in $n$ and is inversely proportional to how close the
arm is to optimal, plus a constant additional loss corresponding to
pulling every arm a little more than $4$ times on average.  The logarithmic regret in $n$ is a
bit of a nuisance, but an earlier lower bound of Lai and
Robbins~\cite{LaiR1985} shows that something like this is necessary in
the limit.

To prove Theorem~\ref{theorem-ucb1}, we need to get an upper bound on
the number of times each suboptimal arm is pulled during the first $n$
pulls.  Define
\begin{equation}
    \label{eq-ucb1-cts}
    c_{t,s} = √{\frac{2 \ln t}{s}},
\end{equation}
the bonus given to an arm that has been pulled $s$ times in the first
$t$ pulls.  

Fix some optimal arm.
Let $\overline{X}_{i,s}$ be the average return on arm $i$ after $s$
pulls and $\overline{X}^*_s$ be the average return on the optimal arm
after $s$ pulls.

If we pull arm $i$ after $t$ total pulls, when arm $i$ has
previously been pulled $s_i$ times and our optimal arm has been pulled
$s^*$ times, then we must have
\begin{equation}
    \label{eq-ucb1-bad-outcome}
    \overline{X}_{i,s_i} + c_{t,s_i} ≥ \overline{X}^*_{s^*} + C_{t,s^*}.
\end{equation}

This just says that arm $i$ with its bonus looks better than the
optimal arm with its bonus.

To show that this bad event doesn't happen, we need three
things:
\begin{enumerate}
    \item The value of $\overline{X}^*_{s^*} + c_{t,s^*}$ should be
        at least $μ^*$.
        Were it to be smaller,
        the observed value $\overline{X}^*_{s^*}$ would be more than
        $c_{t,s^*}$ away from its expectation.
        Hoeffding's inequality implies this doesn't happen too often.
    \item The value of $\overline{X}_{i,s_i} + c_{t,s_i}$ should not
        be too much bigger than $μ_i$.
        We'll again use Hoeffding's inequality to show that
        $\overline{X}_{i,s_i}$ is likely to be at most $μ_i +
        c_{t,s_i}$, making $\overline{X}_{i,s_i} + c_{t,s_i}$ at most
        $μ_i + 2c_{t,s_i}$.
    \item The bonus $c_{t,s_i}$ should be small enough that even
        adding $2c_{t,s_i}$ to $μ_i$ is not enough to beat $μ^*$.
        This means that we need to pick $s_i$ large enough that
        $2c_{t,s_i} ≤ Δ_i$.  For smaller values of $s_i$, we will just
        accept that we need to pull arm $i$ a few more times before we
        wise up.
\end{enumerate}

More formally, if none of the following events hold:
\begin{align}
    \overline{X}^*_{s^*} + c_{t,s^*} &≤ μ^*.  
    \label{eq-ucb1-optimal-too-small}
    \\
    \overline{X}_{i,s_i} &≥ μ_i + c_{t,s_i} 
    \label{eq-ucb1-other-too-big}
    \\
    \label{eq-ucb1-gap-too-small}
    μ^* - μ_i &< 2c_{t,s_i},
\end{align}
then $\overline{X}^*_{s^*} + c_{t,s^*} > μ^* > μ_i + 2c_{t,s_i} >
\overline{X}_{i,s_i} + c_{t,s_i}$, and we don't pull arm $i$ because the optimal
arm is better.  (We don't necessarily pull the optimal arm, but if we
don't, it's because we pull some other arm that still isn't arm $i$.)

For \eqref{eq-ucb1-optimal-too-small} and
\eqref{eq-ucb1-other-too-big}, we repeat the argument in
\eqref{eq-ucb1-bonus-bound}, plugging in $t$ for $n$ and $s_i$ or $s^*$
for $n_i$.  This gives a probability of at most
$2t^{-4}$ that either or both of these bad events occur.

For \eqref{eq-ucb1-gap-too-small}, we need to do something a little
sneaky, because the statement is not actually true when $s_i$ is
small.  So we will give $\ell_i$ free pulls to arm $i$, and only start
comparing arm $i$ to the optimal arm after we have done at least this
many pulls.  The value of $\ell_i$ is chosen so that, when $t ≤ n$ and
$s_i > \ell_i$,
\begin{align*}
    2c_{t,s_i} &≤ μ^* - μ_i,
    \intertext{which expands to,}
    2 √{\frac{2 \ln t}{s_i}} &≤ Δ_i,
    \intertext{giving}
    s_i &≥ \frac{8 \ln t}{Δ_i^2}.
\end{align*}

So we must set $\ell_i$ to be at least
\begin{displaymath}
    \frac{8 \ln n}{Δ_i^2} ≥ \frac{8 \ln t}{Δ_i^2}.
\end{displaymath}

Because $\ell_i$ must be an integer, we actually get
\begin{displaymath}
    \ell_i = \ceil*{ \frac{8 \ln n}{Δ_i^2} } ≤ 1 + \frac{8 \ln n}{Δ_i^2}.
\end{displaymath}

This explains (after multiplying by the regret $Δ_i$) the first term in
\eqref{eq-ucb1-regret-bound}.

For the other sources of bad pulls, 
apply the union bound to the $2t^{-4}$ error probabilities
we previously computed for all choices of $t≤n$, $s^* ≥ 1$, and $s_i >
\ell_i$.  This gives
\begin{align*}
    ∑_{t=1}^{n} ∑_{s^*=1}^{t-1} ∑_{s_i=\ell_i+1}^{t-1} 2t^{-4}
    &< 2 ∑_{t=1}^{∞} t^2 ⋅ t^{-4} 
    \\&= 2⋅\frac{π^2}{6}
    \\&= \frac{π^2}{3}.
\end{align*}
Again we have to multiply by the regret $Δ_i$ for pulling the $i$-th
arm, which gives the second term in \eqref{eq-ucb1-regret-bound}.

\section{Relation to limit theorems}
\label{section-CLT}

Since in many cases we are working with sums $S_n = ∑_{i=1}^n X_i$ of
independent, identically distributed random variables $X_i$, classical
limit theorems apply.  These relate the behavior of $S_n$ in the limit
to the common expectation $μ = \Exp{X_i}$ and variance $σ^2 =
\Var{X_i}$.\footnote{The quantity $σ = √{\Var{X}}$ is called the
\concept{standard deviation} of $X$, and informally gives a measure of the
typical distance of $X$ from its mean, measured in the same units as
$X$.}

These include the \concept{strong law of large
numbers}
\begin{align}
    \Pr{\lim_{n→∞} S_n/n = μ} = 1
    \label{eq-strong-law}
    \intertext{and the \concept{central limit theorem} (\concept{CLT})}
    \lim_{n→∞} \Prob{\frac{S_n-μn}{σ√{n}} ≤ t} = Φ(t),
\end{align}
where $Φ$ is the normal distribution function.

These can sometimes be useful in the analysis if randomized
algorithms, but often are not strong enough to get the results we
want.  The main problem is that the standard versions of both the
strong law and the central limit theorem say nothing about rate of
convergence.  So if we want (for example) to use the CLT to show that
$S_n$ is exponentially unlikely to be $nσ$ away
from the mean, we can't do it directly, because $nσ/σ$ is not a fixed
constant $t$, and for any fixed constant $t$, we don't know when the
limit behavior actually starts working.

But there are variants of these theorems that do bound rate of
convergence that can be useful in some cases.  An example is given in
§\ref{section-Berry-Esseen}.

\section{Anti-concentration bounds}
\label{section-anti-concentration-bounds}

It may be that for some problem you want to show that a sum of random
variables is far from its mean at least some of the time: this would
be an \concept{anti-concentration bound}.  Anti-concentration bounds
are much less well-understood than concentration bounds, but there are
known results that can help in some cases.

For variables where we know the distribution of the sum exactly (e.g.,
sums with binomial distributions, or sums we can attack with
generating functions), 
we don't need these.  But they may
be useful if computing the distribution of the sum directly is hard.

\subsection{The Berry-Esseen theorem}
\label{section-Berry-Esseen}

The 
\index{theorem!Berry-Esseen}
\concept{Berry-Esseen theorem}\footnote{Sometimes written
\emph{Berry-Esséen theorem} to help with the pronunciation of
Esseen's last name.} is an extension of the central limit theorem that 
characterizes how quickly a sum of 
independent identically-distributed
random variables converges to a normal distribution, as a
function of the \concept{third moment} of the random variables.  Its
simplest version says that if we have $n$ independent,
identically-distributed random variables $X_1 \dots X_n$, with
$\Exp{X_i} = 0$, $\Var{X_i} = \Exp{X_i^2} = σ^2$, and
$\Exp{\abs*{X_i}^3} ≤ ρ$,
then
\begin{align}
\label{eq-Berry-Esseen}
\sup_{-∞ < x < ∞}
\abs*{\Prob{\frac{1}{√{n}}∑_{i=1}^{n} X_i ≤ x}
 - Φ(x)}
    &≤ \frac{Cρ}{σ^3√{n}},
\end{align}
were $C$ is an absolute constant and $Φ$ is
the normal distribution function.
Note that the $σ^3$ in the denominator is really
$\Var{X_i}^{3/2}$.  Since the probability bound doesn't depend on $x$,
it's more useful toward the middle of the distribution than in the tails.

A classic proof of this result with $C=3$ can be found
in~\cite[§{}XVI.5]{Feller1971}.  
Shevtsova~\cite{Shevtsova2011} shows a stronger bound of
$C<0.4784$.

As with most results involving sums of random variables, there are
generalizations to martingales.  These are too involved to describe
here, but see \cite[§3.6]{HallH1980}.

\subsection{The Littlewood-Offord problem}

The \concept{Littlewood-Offord problem} asks, given a set
of $n$ complex numbers $x_1\dots x_n$ with $\abs*{x_i} ≥ 1$, for how
many assignments of $\pm 1$ to coefficients $ε_1 \dots
ε_n$ it holds that $\abs*{∑_{i=1}^{n} ε_i x_i} ≤ r$.  Paul
Erd\H{o}s showed~\cite{Erdos1945} that this quantity was at most
$cr2^n/√{n}$, where $c$ is a constant.  The quantity
$c2^n/√{n}$ here is really $\frac{1}{2}\binom{n}{\floor{n/2}}$: Erd\H{o}s's
proof shows that for each interval of length $2r$, the number of
assignments that give a sum in the interior of the interval is
bounded by at most the sum of the $r$ largest binomial
coefficients.

In random-variable terms, this means that if we are looking at
$∑_{i=1}^{n} ε x_i$, where the $x_i$ are constants with
$\abs*{x_i} ≥ 1$ and the $ε_i$ are independent $\pm 1$ fair
coin-flips, then 
$\Prob{\abs*{∑_{i=1}^{n} ε_i x_i} ≤ r}$ is
maximized by making all the $x_i$ equal to $1$.  This shows that any
distribution where the $x_i$ are all reasonably large will not be
any more concentrated than a binomial distribution.

There has been a lot of more recent work on variants of the
Littlewood-Offord problem, much of it by Terry Tao and Van Vu.
See
\url{http://terrytao.wordpress.com/2009/02/16/a-sharp-inverse-littlewood-offord-theorem/}
for a summary of much of this work.

\myChapter{Randomized search trees}{2025}{}
\label{chapter-randomized-search-trees}

These are data structures that are either trees or equivalent to
trees, and use randomization to maintain balance.  We'll start by
reviewing deterministic binary search trees and then add in the
randomization.

\section{Binary search trees}
\label{section-binary-search-trees}

A
\index{tree!binary search}
\index{search tree!binary}
\concept{binary search tree}
is a standard data structure for holding sorted data.  A 
\index{tree!binary}
\concept{binary tree}
is either empty, or it consists of a \concept{root} node containing a
\concept{key} and pointers to left and right
\indexConcept{subtree}{subtrees}.  What makes a
binary tree a binary search tree is the invariant that, both for the
tree as a whole and any subtree, all keys in the
left subtree are less than the key in the root, while all
keys in the right subtree are greater than the key in the
root.  This ordering property means that we can search for a
particular key by doing binary search: if the key is not at the root,
we can recurse into the left or right subtree depending on whether it
is smaller or bigger than the key at the root.

The efficiency of this operation depends on the tree being
\index{search tree!balanced}
\index{binary search tree!balanced}\concept{balanced}.
If each subtree always holds a constant fraction of the nodes in the
tree, then each recursive step throws away a constant fraction of the
remaining nodes.  So after $O(\log n)$ steps, we find the key we are
looking for (or find that the key is not in the tree).  But the
definition of a binary search tree does not by itself guarantee
balance, and in the worst case a binary search tree degenerates into a
linked list with $O(n)$ cost for all operations (see
Figure~\ref{figure-binary-search-tree}.  

\subsection{Rebalancing and rotations}

Deterministic binary search tree implementations include
sophisticated rebalancing mechanisms to adjust the structure of the
tree to preserve balance as nodes are inserted or
delete.  
Typically this is done using 
\index{tree!rotation}
\index{rotation!tree}\index{tree rotation}
\indexConcept{rotation}{rotations},
which are operations that change
the position of a parent and a child while preserving the
left-to-right ordering of keys
(see Figure~\ref{figure-tree-rotation}).

\begin{figure}
\centering
\begin{tikzpicture}
        \node (left) at (0,0) { 
            \begin{forest}
                [y [x,edge=red [A] [B]] [C]]
            \end{forest}
        };
        \node[red] (arrow) at (1.5,0.5) {$\leftrightarrow$};
        \node (right) at (3,0) {
            \begin{forest}
                [x [A] [y,edge=red [B] [C]]]
            \end{forest}
        };
\end{tikzpicture}
    \caption{Tree rotations}
    \label{figure-tree-rotation}
\end{figure}

Examples include 
\index{tree!AVL}\indexConcept{AVL tree}{AVL
trees}~\cite{AdelsonVelskiiL1962},
where the left and right subtrees of any node have heights that differ
by at most $1$;
\index{tree!red-black}\indexConcept{red-black tree}{red-black
trees}~\cite{GuibasS1978},
where a coloring scheme is used to maintain balance; and
\index{tree!scapegoat}\indexConcept{scapegoat tree}{scapegoat
trees}~\cite{GalperinR1993},
where no information is stored at a node but part of the tree is
rebuilt from scratch whenever an operation takes too long.
These all give $O(\log n)$ cost per operation (amortized in the case
of scapegoat trees), and vary in how much work is needed in
rebalancing.  Both AVL trees and red-black trees perform more
rotations than randomized rebalancing does on average.

\begin{figure}
\centering
\begin{tikzpicture}
    \node (balanced) at (0,0) [anchor=north] {
\begin{tikzpicture}
    \foreach \n/\x/\y in
    {4/0/0,2/-2/-1,6/2/-1,1/-3/-2,3/-1/-2,5/1/-2,7/3/-2}
    {\node (\n) at (\x,\y) {$\n$};}
    \path \foreach \u/\v in {4/2,4/6,2/1,2/3,6/5,6/7}
    { (\u) edge (\v) };
\end{tikzpicture}
};
\node (unbalanced) at (6,0) [anchor=north] {
\begin{tikzpicture}
    \foreach \n in {1,...,7} {\node (\n) at ($(0,0)+\n*(0.5,-1)$) {$\n$};}
    \path \foreach \u/\v in {1/2,2/3,3/4,4/5,5/6,6/7} {(\u) edge (\v) };
\end{tikzpicture}
};
\end{tikzpicture}
\caption{Balanced and unbalanced binary search trees}
\label{figure-binary-search-tree}
\end{figure}

\section{Random insertions}
\label{section-binary-search-tree-with-random-insertions}

Suppose we insert $n$ keys into an initially-empty binary search
tree in random order with no rebalancing.  This means that for each
insertion, we follow the same path that we would when searching for
the key, and when we reach an empty tree, we replace it with a tree
consisting solely of the key at the root.\footnote{This is not the
only way to generate a binary search tree at random. For example, we
could instead choose uniformly from the set of all $C_n$
binary search trees with $n$ nodes, where $C_n =
\frac{1}{n+1}\binom{2n}{n}$
is the $n$-th \index{number!Catalan}\concept{Catalan number}. For
$n≥3$, this gives a
different distribution that we don't care about.}

Since we chose a random order, each element is
equally likely to be the root, and all the elements less than the root
end up in the left subtree, while all the elements greater than the
root end up in the right subtree, where they are further partitioned
recursively.  This is exactly what happens in randomized QuickSort (see
§\ref{section-quicksort-recurrence}), so the structure of the tree
will exactly mirror the structure of an execution of
QuickSort. 
So,
for example, we can immediately observe from our previous analysis of
QuickSort that the \concept{total path length}—the sum of the depths
of the nodes—is $\Theta(n \log n)$, since the depth of each node is
equal to 1 plus the number of comparisons it participates in as a
non-pivot, and (using the same argument as for Hoare's FIND in 
§\ref{section-QuickSelect}) that the height of the tree is $O(\log n)$ with high
probability.\footnote{The argument for Hoare's FIND is that any node
    has at most $3/4$ of the descendants of its parent on average; 
    this gives for any node $x$
that $\Prob{\depth(x) > d} ≤
(3/4)^{d-1}n$, or a probability of at most $n^{-c}$ that $\depth(x) >
1 + (c+1) \log(n)/\log(4/3) \approx 1 + 6.952 \ln n$ for $c = 1$,
which we need to apply the union bound.  The right answer for the
actual height of a randomly-generated search tree in the limit
is $~4.31107 \ln n$~\cite{Devroye1988}
 so this bound is actually pretty close.
The real height is still nearly a factor of three worse than for a completely balanced
tree, which has max depth bounded by $1 + \lg n \approx 1 + 1.44269
\ln n$.}

When $n$ is small, randomized binary search trees can look pretty
scraggly.
Figure~\ref{figure-binary-search-tree-with-random-insertion} shows a
typical example.
\begin{figure}
    \centering
    \begin{tikzpicture}
        \foreach \n/\x/\y in
        {5/0/0,1/-2/-1,7/2/-1,3/-1/-2,6/1/-2,2/-1.5/-3,4/-0.5/-3}
        {\node (\n) at (\x,\y) {$\n$};}
        \path \foreach \u/\v in {5/1,5/7,1/3,7/6,3/2,3/4}
        { (\u) edge (\v) };
    \end{tikzpicture}
    \caption{Binary search tree after inserting 5 1 7 3 4 6 2}
    \label{figure-binary-search-tree-with-random-insertion}
\end{figure}

The problem with this approach in general is that we don't have any guarantees
that the input will be supplied in random order, and in the worst case
we end up with a linked list, giving $O(n)$ worst-case cost for all
operations.

\section{Treaps}
\label{section-treaps}

The solution to bad inputs is the same as for QuickSort: instead of assuming that
the input is permuted randomly, we assign random priorities to each
element and organize the tree so that elements with higher priorities
rise to the top.  The resulting structure is known as a
\concept{treap}~\cite{SeidelA1996}, because it satisfies the binary
search tree property with respect to keys and the \concept{heap}
property with respect to priorities.\footnote{The name ``treap'' for
    this data structure is now
    standard but the history is a little tricky.  According to Seidel
    and Aragon, essentially the same data structure (though with 
    non-random priorities) was previously called a 
\index{tree!cartesian}\concept{cartesian tree} by
Vuillemin~\cite{Vuillemin1980}, and
the word ``treap'' was initially applied by McCreight to a different
data structure—designed for storing two-dimensional data—that was
called a \index{tree!priority search}\index{search tree!priority}\concept{priority
search tree} in its published form.~\cite{McCreight1985}.}

There's an extensive
    page of information on treaps
    at
    \url{http://faculty.washington.edu/aragon/treaps.html}, maintained
by Cecilia Aragon, the co-inventor of treaps; they are also discussed
at length in~\cite[§8.2]{MotwaniR1995}.  We'll give a brief
description here.

To insert a new node in a treap, first walk down the tree according to
the key and insert the node as a new leaf.  Then go back up fixing the
heap property by rotating the new element up until it reaches an
ancestor with the same or higher priority.  (See
Figure~\ref{figure-treap-insertion} for an example.)
Deletion is the reverse of insertion: rotate a node until it has 0 or
1 children (by
swapping with its higher-priority child at each step), and then prune
it out, connecting its child, if any, directly to its parent.

\begin{figure}
    \begin{tikzpicture}
        \node (a0) at (0,0) {
            \begin{forest}
                [{$1,60$}
                    [,no edge]
                    [{$2,3$}
                        [,no edge]
                        [\textcolor{red}{$3,26$}]
                    ]
                ]
            \end{forest}
        };
        \node (a1) at (3,0) {
            \begin{forest}
                [{$1,60$}
                    [,no edge]
                    [\textcolor{red}{$3,26$}
                        [{$2,3$}]
                        [,no edge]
                    ]
                ]
            \end{forest}
        };
        \node (b0) at (0,-4) {
            \begin{forest}
                [{$1,60$}
                    [,no edge]
                    [{$3,26$}
                        [{$2,3$}]
                        [{$4,24$}
                            [,no edge]
                            [\textcolor{red}{$5,78$}]
                        ]
                    ]
                ]
            \end{forest}
        };
        \node (b1) at (3,-4) {
            \begin{forest}
                [{$1,60$}
                    [,no edge]
                    [{$3,26$}
                        [{$2,3$}]
                        [\textcolor{red}{$5,78$}
                            [{$4,24$}]
                            [,no edge]
                        ]
                    ]
                ]
            \end{forest}
        };
        \node (b2) at (6,-4) {
            \begin{forest}
                [{$1,60$}
                    [,no edge]
                    [\textcolor{red}{$5,78$}
                        [{$3,26$}
                            [{$2,3$}]
                            [{$4,24$}]
                        ]
                        [,no edge]
                    ]
                ]
            \end{forest}
        };
        \node (b3) at (9,-4) {
            \begin{forest}
                [\textcolor{red}{$5,78$}
                    [{$1,60$}
                        [,no edge]
                        [{$3,26$}
                            [{$2,3$}]
                            [{$4,24$}]
                        ]
                    ]
                    [,no edge]
                ]
            \end{forest}
        };
        \node (c0) at (0,-8.5) {
            \begin{forest}
                [{$5,78$}
                    [{$1,60$}
                        [,no edge]
                        [{$3,26$}
                            [{$2,3$}]
                            [{$4,24$}]
                        ]
                    ]
                    [{$6,18$}
                        [,no edge]
                        [\textcolor{red}{$7,41$}]
                    ]
                ]
            \end{forest}
        };
        \node (c1) at (4,-8.5) {
            \begin{forest}
                [{$5,78$}
                    [{$1,60$}
                        [,no edge]
                        [{$3,26$}
                            [{$2,3$}]
                            [{$4,24$}]
                        ]
                    ]
                    [\textcolor{red}{$7,41$}
                        [{$6,18$}]
                        [,no edge]
                    ]
                ]
            \end{forest}
        };
    \end{tikzpicture}
\caption[Inserting values into a treap]{Inserting values into a treap.
Each node is labeled with $k,p$ where $k$ is the key and $p$ the
priority.  Insertions of values not requiring rotations are not shown.}
    \label{figure-treap-insertion}
\end{figure}

Because of the heap property, the root of each subtree is always the
element in that subtree with the highest priority.  This means that
the structure of a treap is completely determined by
the priorities and the keys, no matter what order the elements 
arrive in.  We can imagine in retrospect that the treap is constructed
recursively by choosing the highest-priority element as the root, then
organizing all smaller-index and all larger-index nodes into the left
and right subtrees by the same rule.

If we assign the priorities independently and 
uniformly at random from a sufficiently
large set ($ω(n^2)$ is enough in the limit), then we get no
duplicates, and by symmetry all $n!$ orderings are equally likely.  So
the analysis of the depth of a treap with random priorities is
identical to the analysis of a binary search tree with random
insertion order.  It's not hard to see that the costs of search
insertion, and deletion operations are all linear in the depth of the
tree, so the expected cost of each of these operations is $O(\log n)$.

\subsection{Assumption of an oblivious adversary}
\label{section-treaps-oblivious}

One caveat is that this only works if the priorities of the elements
of the tree are in fact independent.  If operations on the tree are chosen 
by an \index{adversary!adaptive}\concept{adaptive adversary}, this
assumption may not work.  An adaptive adversary is one that can
observe the choice made by the algorithm and react to them: in this
case, a simple strategy would be to insert elements 1, 2, 3, 4, etc.,
in order, deleting each one and reinserting it until it has a lower 
priority value than all the smaller elements.  This might take a while
for the later elements, but the end result is the linked list again.
For this reason it is standard to assume in randomized data structures
that the adversary is
\index{adversary!oblivious}
\indexConcept{oblivious adversary}{oblivious},
meaning that it has to specify a sequence of operations without
knowing what choices are made by the algorithm.  Under this
assumption, whatever insert or delete operations the adversary
chooses, at the end of any particular sequence of operations we still
have independent priorities on all the remaining elements, and the
$O(\log n)$ analysis goes through.

\subsection{Analysis}
\label{section-treaps-analysis}

The analysis of treaps as carried out by Seidel and
Aragon~\cite{SeidelA1996} is a nice example of how to decompose a
messy process into simple variables, much like the
linearity-of-expectation argument for QuickSort
(§\ref{section-quicksort-linearity-of-expectation}).  The key
observation is that it's possible to bound both the expected depth of
any node and the number of rotations needed for an insert or delete
operation directly from information about the ancestor-descendant
relationship between nodes.

Define two classes of indicator variables.  For simplicity, we assume
that the elements have keys $1$ through $n$, which we also use as
indices.
\begin{enumerate}
    \item $A_{i,j}$ indicates the event that $i$ is an \concept{ancestor} of
        $j$, where $i$ is an ancestor of $j$ if it appears on the path
        from the root to $j$.  Note that every node is an ancestor of
        itself.
    \item $C_{i;\ell,m}$ indicates the event that $i$ is a
        \index{ancestor!common}\concept{common ancestor} of both
        $\ell$ and $m$; formally, $C_{i;\ell,m} = A_{i,\ell}A_{i,m}$.
\end{enumerate}

The nice thing about these indicator variables is that it's easy to
compute their expectations.  

For $A_{i,j}$, $i$ will be the ancestor
of $j$ if and only if $i$ has a higher priority than $j$ and there is
no $k$ between $i$ and $j$ that has an even higher
priority: in other words, if $i$ has the highest priority of all keys
in the interval $[\min(i,j), \max(i,j)]$.  
To see this, imagine that we are constructing the treap recursively,
by starting with all elements in a single interval and partitioning
each interval by its highest-priority element.  Consider the last
interval in this process that contains both $i$ and $j$, and suppose
$i < j$ (the $j>i$ case is symmetric).  If the
highest-priority element is some $k$ with $i < k < j$, then $i$ and
$j$ are separated into distinct intervals and neither is the ancestor
of the other.  If the highest-priority element is $j$, then $j$
becomes the ancestor of $i$.  The highest-priority element can't be
less than $i$ or greater than $j$, because then we get a smaller
interval that contains both $i$ and $j$.  So the only case where $i$
becomes an ancestor of $j$ is when $i$ has the highest priority.

It follows that $\Exp{A_{i,j}} = \frac{1}{\abs*{i-j}+1}$, where the
denominator is just the number of elements in the range
$[\min(i,j),\max(i,j)]$.

For $C_{i;\ell,m}$, $i$ is the common ancestor of both $\ell$ and $m$
if and only if it is has the highest priority in both
$[\min(i,\ell),\max(i,\ell)]$ and $[\min(i,m),\max(i,m)]$.  It turns
out that no matter what order $i$, $\ell$, and $m$ come in, these
intervals overlap so that $i$ must have the highest priority in
$[\min(i,\ell,m),\max(i,\ell,m)]$.  This gives
$\Exp{C_{i;\ell,m}} = \frac{1}{\max(i,\ell,m)-\min(i,\ell,m)+1}$.

\subsubsection{Searches}
\label{section-treap-analysis-searches}

From the $A_{i,j}$ variables we can compute $\depth(j) = ∑_i A_{i,j} -
1$.\footnote{We need the $-1$ because of the convention that the root
    has depth $0$, making the depth of a node one less than the number
    of its ancestors.  Equivalently, we could exclude $j$ from the
sum and count only proper ancestors.} So
\begin{align*}
    \Exp{\depth(j)} 
    &= \left(∑_{i=1}^{n} \frac{1}{\abs*{i-j}+1}\right) - 1 \\
    &= \left(∑_{i=1}^{j} \frac{1}{j-i+1}\right) + \left(∑_{i=j+1}^{n}
    \frac{1}{i-j+1}\right) - 1\\
    &= \left(∑_{k=1}^{j} \frac{1}{k}\right) +
    \left(∑_{k=2}^{n-j+1} \frac{1}{k}\right) - 1\\
    &= H_j + H_{n-j+1} - 2.
\end{align*}

This is maximized at $j=(n+1)/2$, giving $2H_{(n+1)/2} - 2 = 2 \ln n +
O(1)$.  So we get the same $2 \ln n + O(1)$ bound on the expected
depth of any one node that we got for QuickSort.  We can also sum over
all $j$ to get the exact value of the expected total path length (but
we won't).  These quantities bound the expected cost of
searches.

\subsubsection{Insertions and deletions}

For insertions and deletions, the question is how many rotations
we have to perform to float a new leaf up to its proper location
(after an insertion) or to float a deleted node down to a leaf (before
a deletion).  Since insertion is just the reverse of deletion, we can
get a bound on both by concentrating on deletion.  The trick is to
find some metric for each node that (a) bounds the number of rotations
needed to move a node to the bottom of the tree and (b) is easy to compute based on the $A$ and $C$ variables 

The \index{spine!left}\concept{left spine} of a subtree is the set of
all nodes obtained by starting at the root and following left
pointers; similarly the \index{spine!right}\concept{right spine} is
what we get if we follow the right pointers instead.  

When we rotate an element down, we are rotating either its left or
right child up.  This removes one element from either the right spine
of the left subtree or the left spine of the right subtree, but the
rest of the spines are left intact (see
Figure~\ref{figure-rotation-shortens-spines}).
Subsequent rotations will eventually remove all these elements by
rotating them above the target,
while other elements in the subtree will be
carried out from under the target without ever appearing as a child or parent
of the target.  Because each rotation removes exactly one element from
one or the other of the two spines, and we finish when both are empty,
the sum of the length of the spines gives the number of rotations.

\begin{figure}
    \begin{tikzpicture}
        \node[anchor=south] at (0,0) { Initial structure };
        \node[anchor=north] at (0,0) {
            \begin{forest}
                [\textbf{4}
                    [\textcolor{red}{2}
                        [1]
                        [\textcolor{red}{3},edge=red]
                    ]
                    [\textcolor{blue}{6}
                        [\textcolor{blue}{5},edge=blue]
                        [7]
                    ]
                ]
            \end{forest}
        };
        \node[anchor=south] at (3.5,0) { Rotated right };
        \node[anchor=north] at (3.5,0) {
            \begin{forest}
                [2 
                    [1]
                    [\textbf{4}
                        [\textcolor{red}{3}]
                        [\textcolor{blue}{6}
                            [\textcolor{blue}{5},edge=blue]
                            [7]
                        ]
                    ]
                ]
            \end{forest}
        };
        \node[anchor=south] at (6,0) { Rotated left };
        \node[anchor=north] at (6,0) {
            \begin{forest}
                [6
                    [\textbf{4}
                        [\textcolor{red}{2}
                            [1]
                            [\textcolor{red}{3},edge=red]
                        ]
                        [\textcolor{blue}{5}]
                    ]
                    [7]
                ]
            \end{forest}
        };
    \end{tikzpicture}
\caption[Tree rotation shortens spines]{Rotating $4$ right shortens
    the right spine of its left subtree by removing $2$.
    Rotating $4$ left shortens the left spine of the right subtree by
removing $6$.}
    \label{figure-rotation-shortens-spines}
\end{figure}

To calculate the length of the right spine of the left subtree of some
element $\ell$, start
with the predecessor $\ell-1$ of $\ell$.  Because there is no
element between them, either $\ell-1$ is a
descendant of $\ell$ or an ancestor of $\ell$.  In the former case
(for example, when $\ell$ is $4$ in
Figure~\ref{figure-rotation-shortens-spines}), we want to include all
ancestors of $\ell-1$ up to $\ell$ itself.  Starting with $∑_i
A_{i,\ell-1}$ gets all the ancestors of $\ell-1$, and subtracting off
$∑_i C_{i; \ell-1, \ell}$ removes any common ancestors of $\ell-1$
and $\ell$.  Alternatively, if $\ell-1$ is an ancestor of $\ell$,
every ancestor of $\ell-1$ is also an ancestor of $\ell$, so the
same expression $∑_i A_{i,\ell-1} - ∑_i C_{i;\ell-1,\ell}$
evaluates to zero.

It follows that the expected length of the right spine of the left
subtree is exactly
\begin{align*}
    & \Exp{∑_{i=1}^n A_{i,\ell-1} - ∑_{i=1}^n C_{i;\ell-1,\ell}}
    \\&= ∑_{i=1}^n \frac{1}{\abs*{i-(\ell-1)}+1}
    -∑_{i=1}^n \frac{1}{\max(i,\ell) - \min(i,\ell-1)+1}
    \\&= 
        ∑_{i=1}^{\ell-1} \frac{1}{\ell-i}
        + ∑_{i=\ell}^{n} \frac{1}{i-\ell+2}
        - ∑_{i=1}^{\ell-1} \frac{1}{\ell-i+1}
        - ∑_{i=\ell}^{n} \frac{1}{i-(\ell-1)+1}
    \\&=
        ∑_{j=1}^{\ell-1} \frac{1}{j}
        + ∑_{j=2}^{n-\ell+2} \frac{1}{j}
        - ∑_{j=2}^{\ell} \frac{1}{j}
        - ∑_{j=2}^{n-\ell+2} \frac{1}{j}
    \\&=
        1 - \frac{1}{\ell}.
\end{align*}

By symmetry, the expected length of the left spine of the right
subtree is $1 - \frac{1}{n-\ell+1}$.  So the total expected number of rotations
needed to delete the $\ell$-th element is
\begin{align*}
    2 - \frac{1}{\ell} - \frac{1}{n-\ell+1}
    &≤ 2.
\end{align*}

\subsubsection{Other operations}

Treaps support some other useful operations: for example, we can split
a treap into two treaps consisting of all elements less than and all
elements greater than a chosen pivot by rotating the pivot to the root
($O(\log n)$ rotations on average, equal to the pivot's expected depth
as calculated in §\ref{section-treap-analysis-searches}) and splitting
off the left and right subtrees.  Merging two treaps with
non-overlapping keys is the reverse of this and so it has the same expected complexity.

\section{Skip lists}
\label{section-skip-lists}

\begin{figure}
    \centering
    \includegraphics[scale=0.7]{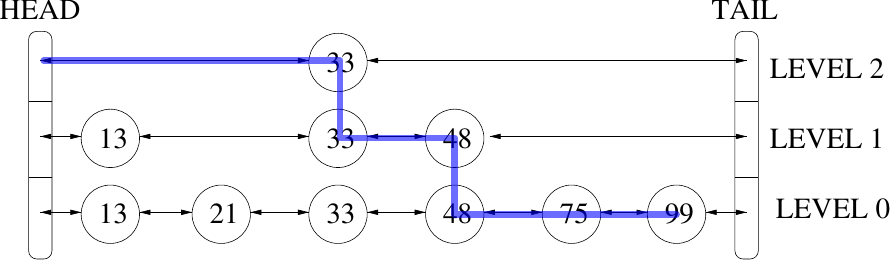}
    \caption[Skip list]{A skip list.  The blue search path for 99 is
        superimposed on an original image 
    from~\cite{AspnesS2007}.}
    \label{figure-skip-list-search-path}
\end{figure}

A skip list~\cite{Pugh1990}
is a randomized tree-like data structure based on linked
lists.  It consists of a level 0 list that is an ordinary sorted
linked list, together with higher-level lists that contain a random
sampling of the elements at lower levels.  When inserted into the
level $i$ list, an element flips a coin that tells it with probability
$p$ to insert itself in the level $i+1$ list as well.  
The result is that the element is represented by a tower of nodes, one
in each of the bottom $1+X$ many layers, where $X$ is a
geometrically-distributed random variable.
An example of a
small skip list is shown in Figure~\ref{figure-skip-list-search-path}.

Searches in a skip list are done by starting in the highest-level list
and searching forward for the last node whose key is smaller than
the target; the search then continues in the same way on the next
level down.  To bound the expected running time of a search, it helps
to look at this process backwards; the reversed search path starts at
level $0$ and continues going backwards until it reaches the first
element that is also in a higher level; it then jumps to the next
level up and repeats the process.  The nice thing about this reversed
process is that it has a simple recursive structure: if we restrict
a skip list to only those nodes to the left of and at the same level
or higher of a particular node, we again get a skip list.
Furthermore, the structure of this restricted skip list depends only
on coin-flips taken at nodes within it, so it's independent of
anything that happens elsewhere in the full skip list.

We can analyze this process by
tracking the number of nodes in the restricted skip list described
above, which is just the number of nodes in the current level that are earlier
than the current node.  If we move left, this drops by $1$; if up,
this drops to $p$ times its previous value on average.  So the
number of such nodes $X_k$ after $k$ steps satisfies the probabilistic recurrence 
\begin{align*}
\ExpCond{X_{k+1}}{X_k}
    &= (1-p)(X_k-1) + p(p X_k)
    \\&= (1-p+p^{2}) X_k - (1-p)
    \\&≤ (1-p+p^2) X_k,
\end{align*}
with $X_0 = n-1$ (since the last node is not included).
Wrapping expectations around both sides gives $\Exp{X_{k+1}} =
\Exp{\ExpCond{X_{k+1}}{X_k}}
≤ (1-p+p^2) \Exp{X_k}$,
and in general we get
$\Exp{X_k} ≤ (1-p+p^{2})^{k} X_0 < (1-p+p^2)^k n$. This is
minimized at $p = 1/2$, giving $\Exp{X_k} ≤ (3/4)^{k} n$, suspiciously
similar to the bound we computed before for random binary search
trees.

When $X_k = 0$, our search is done, so if $T$ is the time to
search, we have $\Prob{T ≥ k} = \Prob{X_k ≥ 1} ≤ (3/4)^k n$, by
Markov's inequality.  In particular, if we want to guarantee that we 
finish with
probability $1-ε$, we need to run for $\log_{4/3} (n/ε)$
steps.  This translates into an $O(\log n)$ bound on the expected search time,
and the constant is even the same as our (somewhat loose) bound for
treaps.

The space per element of a skip list also depends on $p$.  
Every
element needs one pointer for each level it appears in.
The number of levels each element appears in is a geometric random
variable where we are waiting for an event of probability $1-p$, so
the expected number of pointers is $\frac{1}{1-p}$.
For constant $p$ this is $O(1)$.
However, 
the space cost can reduced (at the cost of increasing search time) by
adjusting $p$.  For example, if space is at a premium, setting $p =
1/10$ produces $10/9$ pointers per node on average—not much more
than in a linked list—but still gives $O(\log n)$ search
time.

In general the trade-off is between
$n\left(\frac{1}{1-p}\right)$ total expected
pointers and $\log_{1/(1-p+p^2)} (n/ε)$ search time.  
For small $p$, the number of pointers scales as $1+O(p)$, while 
the constant factor in the search time is $\frac{1}{-\log(1-p+p^2)} =
O(1/p)$.

Skip lists are even easier to split or merge than treaps.  It's enough
to cut (or recreate) all the pointers crossing the boundary, without
changing the structure of the rest of the list.

\myChapter{Hashing}{2025}{}
\label{chapter-hashing}

The basic idea of hashing is that we have keys from a large set $U$,
and we'd like to pack them in a small set $M$ by passing them through
some function $h:U→M$, without getting too
many \indexConcept{collision}{collisions}, pairs of distinct keys $x$
and $y$ with $h(x) = h(y)$.  Where randomization comes in is that we
want this to be true even if the adversary picks the keys to hash.  At
one extreme, we could use a random function, but this will take a lot
of space to store.\footnote{An easy counting argument shows that
    almost all functions from $U$ to $M$ take $\card*{U} \log \card*{M}$
    bits to represent, no matter how clever you are about choosing
    your representation.  This forms the basis for
    \concept{algorithmic information theory}, which \emph{defines}
    an object as random if there is no way to reduce the number of
bits used to express it.}  So our goal will be to find functions with
succinct descriptions that are still random enough to do what we want.

The presentation in this chapter is based largely on
\cite[§§8.4-8.5]{MotwaniR1995} (which is in turn based on work of
Carter and Wegman~\cite{CarterW1977} on universal hashing and Fredman,
Koml\'{o}s, and Szemer\'{e}di~\cite{FredmanKS1984} on $O(1)$
worst-case hashing); on~\cite{PaghR2004} and~\cite{Pagh2006} for
cuckoo hashing; and on \cite[§5.5.3]{MitzenmacherU2017} for Bloom
filters.

\section{Hash tables}
\label{section-hash-tables}

Here we review \indexConcept{hash table}{hash
tables}, which are implementations of the \concept{dictionary} data
type mapping keys to values. The central idea is to use a
random-looking function to map keys to small numeric indices.
The first published version of this is due to Dumey~\cite{Dumey1956}.

Suppose we want to store $n$ elements from a universe $U$ of in a
table with \indexConcept{key}{keys} or \indexConcept{index}{indices}
drawn from an index space $M$ of size $m$.
Typically we assume $U = [\card*{U}] = \Set{0\dots \card*{U}-1}$ and $M =
[m] = \Set{0\dots m-1}$.

If $\card*{U} ≤ m$, we can just use an array.
Otherwise, we can map keys to positions in the array using a
\concept{hash function} $h:U\rightarrow M$.
This necessarily produces 
\indexConcept{collision}{collisions}: pairs $(x,y)$ with
$h(x) = h(y)$, and any design of a hash table must include some
mechanism for handling keys that hash to the same place.
Typically this is a secondary data structure in each bin, but we may
also place excess values in some other place.  Typical choices for
data structures are linked lists (\concept{separate chaining} or just
\concept{chaining}) or secondary
hash tables (see §\ref{section-FKS} below).  Alternatively, we
can push excess values into other positions in the same hash table
according to some deterministic rule
(\concept{open addressing} or \concept{probing}) or a second hash
function (see §\ref{section-cuckoo-hashing}).

For most of these techniques, the costs of insertions and searches 
will depend on how likely it is
that we get collisions.  An adversary that knows our hash function can
always choose keys with the same hash value, but we can avoid that by
choosing our hash function randomly.\footnote{In practice, hardly
anybody every does this. Hash functions are instead chosen based on
fashion and occasional experiments, often with additional goals like
cryptographic security or speed. For cryptographic security, the
\concept{SHA} family is standard. For speed, \concept{xxHash}
(see \url{https://cyan4973.github.io/xxHash/}) is probably the fastest
widely-used hash function with decent statistical properties.}
Our ultimate goal is to 
do each search in $O(1+n/m)$ expected time, which for $n
≤ m$ will be much better than the $Θ(\log n)$ time for
pointer-based data structures like balanced trees or skip lists.
The quantity $n/m$ is called the \concept{load factor} of the hash
table and is often abbreviated as $α$.

All of this only works if we are working in a RAM (random-access machine
model), where we can access arbitrary memory locations in time $O(1)$
and similarly compute arithmetic operations on $O(\log \card*{U})$-bit values
in time $O(1)$.  There is an argument that in reality any actual RAM
machine requires either $Ω{}(\log m)$ time to read one of $m$
memory locations (routing costs) or, if one is particularly pedantic,
$Ω{}(m^{1/3})$ time (speed of light + finite volume for each location).  We will ignore this argument.

We will try to be consistent in our use of variables to refer to the
different parameters of a hash table.
Table~\ref{table-hash-table-parameters} summarizes the meaning of
these variable names.

\begin{table}
\begin{tabular}{cl}
$U$ & Universe of all keys \\
$S⊆ U$ & Set of keys stored in the table \\
$n = \card*{S}$ & Number of keys stored in the table \\
$M$ & Set of table positions \\
$m = \card*{M}$ & Number of table positions \\
$α = n/m$ & Load factor
\end{tabular}
\caption{Hash table parameters}
\label{table-hash-table-parameters}
\end{table}

\section{Universal hash families}
\label{section-universal-hash-families}

A family of hash functions $H$ is \index{universal hash
family}\index{hash function!universal}\concept{2-universal} if for any
$x≠y$, $\Prob{h(x)=h(y)} ≤ 1/m$ for a uniform random $h∈H$.
It's 
\concept{strongly 2-universal} if for any
$x_{1}≠x_{2}∈U$, $y_{1},y_{2}∈M$, $\Prob{h(x_{1})=y_{1}
∧{} h(x_{2}) = y_{2}} = 1/m^{2}$ for a uniform random
$h∈H$.  Another way to describe strong $2$-universality is that
the values of the hash function are uniformly distributed and
pairwise-independent.

For $k > 2$,
\concept{$k$-universal} usually means \concept{strongly
$k$-universal}: Given distinct $x_{1}\dots{}x_{k}$, and any
$y_{1}\dots{}y_{k}$, $\Prob{∀i: h(x_{i})=y_{i}} =
m^{-k}$.  This is equivalent to the $h(x_i)$ values
for distinct $x_i$ and randomly-chosen $h$ having a uniform distribution and $k$-wise
independence.  It is possible to generalize the weak version of
2-universality to get a weak version of $k$-universality
($\Prob{\text{$h(x_{i})$ are all equal}} ≤ m^{-(k-1)}$), but this
generalization is not as useful as strong $k$-universality.

To analyze universal hash families, it is helpful to have some
notation for counting collisions.  We'll mostly be doing counting
rather than probabilities because it saves carrying around a lot of
denominators.  Since we are assuming uniform choices of $h$ we can
always get back probabilities by dividing by $\card*{H}$.
The particular notation we will be using follows
§8.4.1 of \cite{MotwaniR1995}; the original paper of Carter and
Wegman~\cite{CarterW1977} uses essentially the same notation with
slightly different formatting.

Let $δ(x,y,h) = 1$ if $x≠y$ and $h(x)=h(y)$, $0$ otherwise.
Abusing notation, we also define, for sets $X$, $Y$, and $H$,
$δ(X,Y,H) =
  ∑_{x∈X,y∈Y,h∈H} δ(x,y,h)$; and allow lowercase variables to stand in
  for singleton sets, as in
  $δ(x,Y,h) = δ(\Set{x},Y,\Set{h})$.
Now the statement that
$H$ is $2$-universal becomes $\forall x,y: δ(x,y,H) ≤
\card*{H}/m$; this says that only
$1/m$ of the functions in $H$ cause any particular 
distinct $x$ and $y$ to collide.

If $H$ includes all functions $U→M$, we get equality: a
random function gives $h(x) = h(y)$ with probability exactly $1/m$.
But we might do better if each $h$ tends to map distinct values to distinct places.
The following lemma shows we can't do too much better:
\begin{lemma}
\label{lemma-universal-hash-family-lower-bound}
For any family $H$, there exist $x,y$ such that
$δ(x,y,H) ≥
 \frac{\card*{H}}{m} \left(1-\frac{m-1}{\card*{U}-1}\right)$.
\end{lemma}
\begin{proof}
We'll count collisions in the inverse image of each element $z$.
Since all distinct pairs of elements of $h^{-1}(z)$ collide with each
other, we have
\begin{align*}
δ(h^{-1}(z), h^{-1}(z), h)
&= \card*{h^{-1}(z)}⋅\left(\card*{h^{-1}(z)}-1\right).
\end{align*}
Summing over all $z ∈ M$ gets all collisions, giving
\begin{align*}
δ(U,U,h) 
&= ∑_{z∈ M}
\left(\card*{h^{-1}(z)}⋅\left(\card*{h^{-1}(z)}-1\right)\right).
\end{align*}
Use convexity or Lagrange multipliers to argue that the right-hand
side is
minimized subject to $∑_{z} \card*{h^{-1}(z)} = \card*{U}$ when all
pre-images are the same size $\card*{U}/m$.
It follows that
\begin{align*}
δ(U,U,h) 
&≥ ∑_{z∈ M} \frac{\card*{U}}{m} \left(\frac{\card*{U}}{m} - 1\right)
\\
&= m \frac{\card*{U}}{m} \left(\frac{\card*{U}}{m} - 1\right)
\\
&= \frac{\card*{U}}{m} (\card*{U} - m).
\end{align*}

If we now sum over all $h$, we get
\begin{align*}
δ(U,U,H) 
&≥ \frac{\card*{H}}{m} \card*{U} (\card*{U} - m).
\end{align*}
There are exactly $\card*{U}(\card*{U}-1)$ ordered pairs $x,y$ for which
$δ(x,y,H)$ might not be zero; so
the Pigeonhole principle says some pair $x,y$ has
\begin{align*}
δ(x,y,H)
&≥ 
\frac{\card*{H}}{m}
\left(
\frac{\card*{U}(\card*{U}-m)}{\card*{U}(\card*{U}-1)}
\right)
\\
&= \frac{\card*{H}}{m} \left(1-\frac{m-1}{\card*{U}-1}\right).
\end{align*}
\end{proof}

Since $1-\frac{m-1}{\card*{U}-1}$ is likely to be very close to $1$, we
are happy 
if we get the 2-universal upper bound of $\card*{H}/m$.

Why we care about this:
With a $2$-universal hash family, chaining using linked lists costs
$O(1+n/m)$ expected time per operation.
The reason is that the expected cost of an operation on some key $x$
is proportional to the size of the linked list at $h(x)$ (plus $O(1)$
for the cost of hashing itself).  
But the expected size of this linked list is just the expected number
of keys $y$ in the dictionary that collide with $x$, which is
exactly $δ(x,S,H)/\card{H} ≤ n/m$.

\subsection{Linear congruential hashing}
\label{section-linear-congruential-hashing}
\label{section-example-of-a-2-universal-hash-family}

Universal hash families often look suspiciously like classic
pseudorandom number generators. Here is a $2$-universal hash family
based on taking remainders. It is assumed that the universe $U$ is a
subset of $ℤ_p$, the integers mod $p$ for some prime $p$. Effectively,
this just means that every element of $U$ can be represented by an
integer $x$ with $0 ≤ x ≤ p-1$.\footnote{An obvious question is where
we get $p$ from. We typically want the value of $p$ to be polynomial
in the size $n$ of the input; this will make the size of $p$ in bits
be $O(\log n)$, making operations in $ℤ_p$ take $O(1)$ time under the
typical unit-cost assumption. We also need $p$ to be large enough that
$p>m$.

A common approach is to pick some range of candidates
$\Set{q,\dots,2q}$ and argue that for sufficiently large $q$, the
Prime Number Theorem says that a randomly-chosen integer in this range
has probability $Θ(1/\log n)$ of being prime. So we will go through
$Θ(\log n)$ candidates on average before we find a good one.
If $q$ is, say, linear in $n$, we don't need to be particularly
clever about testing candidates: trial division up to $√{p}$ will
tell us if $p$ is prime or not in time $O(√{q}) = O(√{n})$,
giving a total expected cost of $O(√{n}⋅\log n)$ to find an actual
prime. If $q$ is larger, we might have to resort to the AKS primality
tester instead~\cite{AgrawalKS2004}, which has a running time
polynomial in $\log p$, the size of $p$ in bits, or use
Miller-Rabin~\cite{Miller1976,Rabin1980}, which has a better exponent
on the polynomial but which may give an incorrect answer with small probability.}

\begin{lemma}
\label{lemma-universal-hash-family-mod}
  Let $h_{ab}(x) = (ax + b \bmod p) \bmod m$, where
  $a∈ℤ_{p} ∖ \Set{0}, b∈ℤ_{p}$, and $p$ is a
  prime $≥ m$.
  Then $\Set{h_{ab}}$ is $2$-universal.
\end{lemma}
\begin{proof}
Again, we count collisions.
Split $h_{ab}(x)$ as $g(f_{ab}(x))$ where $f_{ab}(x) = ax+b
\bmod p$ and $g(x) = x \bmod m$.

    The intuition is that if we fix $x$ and $y$ and consider all
    $p(p-1)$ pairs
    $a,b$ with $a≠0$, all $p(p-1)$
    distinct pairs of values $r=f_{ab}(x)$ and $s=f_{ab}(y)$ are equally likely.  We then show
that feeding these values to $g$ produces no more collisions than
expected.

The formal statement of the intuition is that for any $0≤x,y≤p-1$ with
    $x≠y$, $δ(x,y,H) = δ(ℤ_{p},ℤ_{p},g)$.

To prove this, fix $x$ and $y$, and consider some pair
      $r≠s ∈ Z_{p}$. 
      Then the equations $ax+b = r$ and $ay+b = s$ have a unique
      solution for $a$ and $b$ mod
      $p$ (because $ℤ_{p}$ is a finite field).
      Furthermore this solution has $a≠0$ since otherwise $f_{ab}(x) =
      f_{ab}(y) = b$. So the function $q(a,b) =
      \Tuple{f_{ab}(x),f_{ab}(y)}$ is a bijection
      between pairs $a,b$ and
      pairs $r,s$. Any collisions will arise from applying $g$, giving
      $δ(x,y,H) = ∑_{a,b} δ(x,y,h_{ab})
      = ∑_{r≠s} δ(r,s,g) = δ(ℤ_{p},ℤ_{p},g)$.

      Now we just need to count how many distinct $r$ and $s$ collide.
      There are $p$ choices for $r$.
      For each $r$, the possible $s$ that map to the same
      remainder mod $m$ are in a set of the form $\Set{r', r'+m,
      r'+2m, \dots, r'+km}$ where $r' = r \bmod m$ and $r' + km ≤
      p-1$, which
      gives $k ≤ (p-1)/m$. There are $k+1$ elements of this set, but
      $s≠r$ means we can only use $k$ of them.
      This gives at most $k ≤ (p-1)/m$ choices for $s$ for each of the $p$
      choices for $r$. Multiplying out these bounds gives
      $δ(x,y,H) = δ(ℤ_p,ℤ_p,g) ≤ p(p-1)/m$.

      Since each choice of $a≠0$ and $b$ occurs with probability
      $\frac{1}{p(p-1)}$, this gives a probability of collision of
      at most $1/m$.
\end{proof}

A difficulty with this hash family is that it requires doing modular
arithmetic. A hash family that can be faster in practice is given by
Dietzfelbinger~\etal~\cite{DietzfelbingerHKP1997}, although it
requires a slight weakening of the notion of $2$-universality.  For
each $k$ and $\ell$ they define a class $H_{k,\ell}$ of functions from
$[2^k]$ to $[2^\ell]$ by defining
\begin{align*}
h_a(x) &= (ax \bmod 2^k) \operatorname{div} 2^{k-\ell},
\end{align*}
where $x \operatorname{div} y = \floor{x/y}$.  They prove~\cite[Lemma
2.4]{DietzfelbingerHKP1997} that if $a$
is a random \emph{odd} integer with $0 < a < 2^\ell$, and $x≠ y$,
$\Prob{h_a(x) = h_a(y)} ≤ 2^{-\ell + 1}$.  This increases by a factor
of $2$ the likelihood of a collision, but any extra costs from this
can often be justified in practice by the reduction in costs from working
with powers of $2$.

If we are willing to use more randomness (and more space), a method
called \conceptFormat{tabulation hashing}
(§\ref{section-tabulation-hashing}) gives a
simpler alternative that is $3$-universal.

\subsection{Tabulation hashing}
\label{section-tabulation-hashing}

\index{hash function!tabulation}
\indexConcept{tabulation hashing}{Tabulation
hashing}~\cite{CarterW1977} 
is a method for hashing fixed-length strings (or
things that can be represented as fixed-length strings) into
bit-vectors.  The description here follows Patrascu and
Thorup~\cite{PatrascuT2012}.

Let $c$ be the length of each string in characters, and let $s$ be the
size of the alphabet.  Initialize the hash function by constructing
tables $T_1 \dots T_c$ mapping
characters to independent random bit-vectors of size $\lg m$.  Define
\begin{align*}
    h(x) &= T_1[x_1] ⊕ T_2[x_2] ⊕ \dots T_c[x_c],
\end{align*}
where $⊕$ represents bitwise exclusive OR (what \verb:^: does in
C-like languages).\footnote{Letting $m$ be a power of $2$ and using
    exclusive OR is convenient on real computers.  If for some reason
    we don't like this approach, the same technique,
    with essentially the same analysis, works for arbitrary $m$ if we
replace bitwise XOR with addition mod $m$.}
This gives a family of hash functions that is $3$-wise independent
but not $4$-wise independent.

Like many hash algorithms, tabulation hashing was already in use
before it was formalized in general. \concept{Zobrist
hashing}\index{hashing!Zobrist}~\cite{Zobrist1970} is a special case of
tabulation hashing used to represent positions in board games like
Chess and Go, where $T_i[x_i]$ gives the contribution of having a
piece $x_i$ in position $i$. This is useful for games whose state
changes slowly because of a homomorphic property of tabulation
hashing: replacing $x_i$ by $x'_i$ while leaving all other $x_j$
unchanged does not require recomputing the entire hash function, since
$h[x'] = h[x] ⊕ T_i[x_i] ⊕ T'_i[x_i]$ in this case.

The intuition for why the hash values might be independent is that if we
have a collection of strings, and each string brings in an element of
$T$ that doesn't appear in the other strings, then that element is
independent of the hash values for the other strings and XORing it
with the rest of the hash value gives a random bit string that is
independent of the hash values of the other strings.  In fact, we
don't even need each string to include a unique value; it's enough if
we can order the strings so that each string gets a value that isn't
represented among its predecessors.

More formally, suppose we can order the strings $x^1,
x^2, \dots, x^n$ that we are hashing so that each has a position $i_j$ such
that $x^j_{i_j} ≠ x^{j'}_{i_j}$ for any $j' < j$, then we
have, for each value $v$, 
$\ProbCond{h(x^j) = v}{h(x^{j'}) = v_{j'}, \forall j' < j} = 1/m$.
It follows that the hash values are independent:
\begin{align*}
    \Prob{h(x^1) = v_1, h(x^2) = v_2, \dots, h(x^n) = v_n)}
&= \prod_{j=1}^{n} \ProbCond{h(x^j) = v_j}{h(x^1) = v_1 \dots
h(x^{j-1}) = v_{j-1}}
\\&= \frac{1}{m^n}
\\&= \prod_{j=1}^{n} \Prob{h(x^j) = v_j}.
\end{align*}

Now we want to show that when $n=3$, this actually works for all
possible distinct strings $x$, $y$, and $z$.  Let $S$ be the set of
indices $i$ such that $y_i ≠ x_i$, and similarly let $T$ be the set
of indices $i$ such that $z_i ≠ x_i$; note that both sets must be
non-empty, since $y ≠ x$ and $z ≠ x$.  If $S ∖ T$ is
nonempty, then (a) there is some index $i$ in $T$ where $z_i ≠ x_i$,
and (b) there is some index $j$ in $S ∖ T$ where $y_i ≠ x_i
= z_i$; in this case, ordering the strings as $x$, $z$, $y$ gives the
independence property above.  If $T ∖ S$ is nonempty, order
them as $x$, $y$, $z$ instead.  Alternatively, if $S = T$, then $y_i
≠ z_i$ for some $i$ in $S$ (otherwise $y=z$, since they both equal
$x$ on all positions outside $S$).  In this case, $x_i$, $y_i$, and
$z_i$ are all distinct.

For $n=4$, we can have strings \texttt{aa}, \texttt{ab}, \texttt{ba},
and \texttt{bb}.  If we take the bitwise exclusive OR of all four hash
values, we get zero, because each character is included exactly twice
in each position.  So the hash values are not independent, and we do
not get $4$-independence in general.

However, even though the outputs of tabulation hashing are not $4$-independent,
most reasonably small sets of inputs 
do give independence.  This can be used to show various miraculous
properties like working well for the cuckoo hashing algorithm
described in §\ref{section-cuckoo-hashing}.

\section{FKS hashing}
\label{section-FKS}

The FKS hash table, named for Fredman, Komlós, and
Szemerédi~\cite{FredmanKS1984}, is a method for storing
a static set $S$ so that we never pay more than
constant time for search (not just in expectation), while at the same
time not consuming too much space.  The assumption that $S$ is static
is critical, because FKS chooses hash functions based on the elements
of $S$.

If we were lucky in our choice of $S$, we might be able to do this
with
standard hashing.
 A \index{hash function!perfect}\concept{perfect hash function}
 for a set $S⊆ U$ is a hash function $h:U\rightarrow M$ that is
 injective on $S$ (that is, $x≠y → h(x)≠h(y)$ when
 $x,y∈ S$).
Unfortunately, we can only count on finding a perfect hash function if
$m$ is large:

\begin{lemma}
\label{lemma-perfect-hash-function}
    If $H$ is 2-universal and $\card*{S} = n$ with $\binom{n}{2} ≤ m$, then there is
 a perfect $h∈H$ for $S$.
\end{lemma}
\begin{proof}
    Let $h$ be chosen uniformly at random from $H$.  
    For each unordered pair $x≠y$ in $S$, let $X_{xy}$ be the indicator variable
    for the even that $h(x) = h(y)$, and let $C = ∑_{x≠y} X_{xy}$ be the
    total number of collisions in $S$.  Each $X_{xy}$ has expectation
    at most $1/m$, so $\Exp{C} ≤ \binom{n}{2} / m < 1$.  
    But we we can write $\Exp{C}$ as $\ExpCond{C}{C=0} \Prob{C = 0} +
    \ExpCond{C}{C≥1} \Prob{C≠0} ≥ \Prob{C≠0}$.  So $\Prob{C≠0} ≤
    \binom{n}{2} / m < 1$, giving $\Prob{C = 0} > 0$.  But if $C$ is
    zero with nonzero probability, there must be some $h$ that makes
    it $0$.  That $h$ is perfect for $S$.
\end{proof}

\begin{sloppypar}
Essentially the same argument shows that if $α \binom{n}{2} ≤ m$, then
$\Prob{\text{$h$ is perfect for $S$}} ≥ 1-α$.  This can be handy if
we want to find a perfect hash function and not just demonstrate that
it exists.
\end{sloppypar}

Using a perfect hash function, we get $O(1)$ search time using
$O(n^2)$ space.  But we can do better by using perfect hash functions
only at the second level of our data structure, which at top level
will just be an ordinary hash table.  This is the idea behind the
Fredman-Koml\'os-Szemer\'edi (FKS) hash table~\cite{FredmanKS1984}.

The short version is that we hash to $n = \card*{S}$ bins, then rehash perfectly
within each bin.  The top-level hash table stores a pointer to a
header for each bin, which gives the size of the bin and the hash
function used within it.  The $i$-th bin, containing $n_i$ elements,
uses $O(n_i^2)$ space to allow perfect hashing.  The total size is
$O(n)$ as long as we can show that $∑_{i=1}^{n} n_i^2 = O(n)$.
The time to do a search is $O(1)$ in the worst case: $O(1)$ for the
outer hash plus $O(1)$ for the inner hash.

\begin{theorem}
\label{theorem-FKS}
The FKS hash table uses $O(n)$ space.
\end{theorem}
\begin{proof}
Suppose we choose $h∈ H$ as the outer hash function, where $H$ is
some $2$-universal family of hash functions.  Compute:
\begin{align*}
∑_{i=1}^n n_i^2
&= ∑_{i=1}^n (n_i + n_i(n_i - 1))
\\
&= n + δ(S,S,h).
\end{align*}
The last equality holds because each ordered pair of distinct values
in $S$ that map
to the same bucket $i$ corresponds to exactly one collision in
$δ(S,S,h)$.

Since $H$ is 2-universal, we have $δ(S,S,H) ≤
    \card*{H} \frac{\card*{S} \parens*{\card*{S}-1}}{n} = \card*{H}
    \frac{n(n-1)}{n} = \card*{H} (n-1)$.
But then the Pigeonhole principle says there exists some $h ∈ H$
with
$δ(S,S,h) ≤ \frac{1}{\card*{H}} δ(S,S,H) = n-1$.
Choosing this $h$
gives $∑_{i=1}^{n} n_i^2 ≤ n + (n-1) = 2n-1 = O(n)$.
\end{proof}

If we want to find a good $h$ quickly, increasing the size of the
outer table to $n/α$ gives us a probability of at least
$1-α$ of getting a good one, using essentially the same
argument as for perfect hash functions. More generally, it's possible
to adapt FKS hashing to work with dynamic data sets by growing the
main table and each subtable as needed to keep the 2-lookup property
while maintaining $O(1)$ expected amortized cost per insertion.
This is know as
\concept{dynamic perfect hashing}\index{hashing!dynamic
perfect}\index{prefect hashing!dynamic} and was studied by
Dietzfelbinger~\etal~\cite{DietzfelbingerKMMRT1994}. In practice the
overhead of this approach is higher than other 2-lookup schemes, such
as the one described in the next section.

\section{Cuckoo hashing}
\label{section-cuckoo-hashing}

\index{hashing!cuckoo}\index{hash table!cuckoo}\index{cuckoo hash table}\indexConcept{cuckoo hashing}{Cuckoo hashing}~\cite{PaghR2004} 
is a hash table implementation that uses a single table and guarantees
at most $2$ probes per lookup, with $O(1)$ expected amortized cost per
insertion.

The name comes from the \concept{cuckoo}, a bird notorious
for stealing space for their own eggs in other birds' nests. In cuckoo
hashing, newly-inserted elements may steal slots from other elements,
forcing those elements to find an alternate nest.

The formal mechanism is to use two hash functions $h_1$ and $h_2$, and store
each element $x$ in one of the two positions $h_1(x)$ or $h_2(x)$.  This may require
moving other elements to their alternate locations to make room.  But the
payoff is that
each search takes only two reads, 
which can even be done in parallel.  This is optimal by a lower bound of
Pagh~\cite{Pagh2001}, which also shows a matching upper bound for
static dictionaries using a different technique. 

Cuckoo hashing was invented by Pagh and Rodler~\cite{PaghR2004}.  The
version described here is based on a simplified version from notes of
Pagh~\cite{Pagh2006}. The main difference is that it uses just one
table instead of the two tables—one for each hash
function—in~\cite{PaghR2004}.

\subsection{Structure}

We have a table $T$ of size $n$, with two
separate, independent hash functions $h_{1}$ and $h_{2}$.  These
functions are assumed to be $k$-universal for some sufficiently large
value $k$; as long as we never look at more than $k$ values at once,
this means we can treat them effectively as random functions.  In
practice, using crummy hash functions seems to work just fine, a
common property of hash tables.  There are also specific hash
functions that have been shown to work with particular variants of
cuckoo hashing~\cite{PaghR2004,PatrascuT2012}.  We will avoid these
issues by assuming that our hash functions are actually random.

Every key $x$ is stored either in $T[h_{1}(x)]$ or
$T[h_{2}(x)]$.  So the search procedure just looks at both of
these locations and returns whichever one contains $x$ (or fails if
neither contains $x$).

To insert a value $x_{1} = x$, we must put it in $T[h_{1}(x_{1})]$ or
$T[h_2(x_1)]$.  If one or both of these locations is empty, we put
it there.  Otherwise we have to kick out some value that is in the way
(this is the ``cuckoo'' part of cuckoo hashing, named after the bird
that leaves its eggs in other birds' nests).  
We do this by letting $x_2 = T[h_1(x_1)]$ and writing $x_1$ to
$T[h_1(x_1)]$.
We
now have a new ``nestless'' value $x_{2}$, which we swap with whatever
is in $T[h_{2}(x_{2})]$.  If that location was empty, we are done;
otherwise, we get a new value $x_{3}$ that we have to put in
$T[h_1(x_3)]$ and so on.  The procedure terminates when we find an
empty spot or if enough iterations have passed that we don't expect to
find an empty spot, in which case we rehash the entire table.  This
process can be implemented succinctly as shown in
Algorithm~\ref{algorithm-cuckoo-hashing-insert}.

\begin{algorithm}
    \caption[Insertion procedure for cuckoo hashing]{Insertion procedure for cuckoo hashing.  Adapted
from~\cite{Pagh2006}}
\label{algorithm-cuckoo-hashing-insert}
\newData{\Pos}{pos}
\Procedure{$\Insert(x)$}{
    \If{$T(h_1(x) = x$ \KwOr $T(h_2(x)) = x$}{
        \Return
    }
    $\Pos \leftarrow h_1(x)$\\
    \For{$i \leftarrow 1 \dots n$}{
        \If{$T[\Pos] = \bot$}{
            $T[\Pos] \leftarrow x$\\
            \Return
        }
        $x \rightleftarrows T[\Pos]$\\
        \eIf{$\Pos = h_1(x)$}{
            $\Pos \leftarrow h_2(x)$
        }{
            $\Pos \leftarrow h_1(x)$
        }
    }
    If we got here, rehash the table and reinsert $x$.
}
\end{algorithm}

A detail not included in the above code is that we always rehash (in
theory) after $m^{2}$ insertions; this avoids potential problems with
the hash functions used in the paper not being universal enough.  We
will avoid this issue by assuming that our hash functions are actually
random (instead of being approximately $n$-universal with reasonably
high probability).  For a more principled analysis of where the hash
functions come from, see~\cite{PaghR2004}.  An
alternative hash family 
that is known to work for a slightly different variant of
cuckoo hashing is tabulation hashing, as described in
§\ref{section-tabulation-hashing}; the proof that this works is found
in~\cite{PatrascuT2012}.

\subsection{Analysis}
The main question is how long it takes the insertion procedure to
terminate, assuming the table is not too full.  

First let's look at what happens during an insert if we have
many
nestless values.  We have a sequence of values $x_{1}, x_{2}, \dots$,
where each pair of values $x_{i}, x_{i+1}$ collides in $h_{1}$ or
$h_{2}$.  Assuming we don't reach the loop limit, there are three main
possibilities (the leaves of the tree of cases below):

\begin{enumerate}
 \item Eventually we reach an empty position without seeing the same key twice.
 \item Eventually we see the same key twice; there is some $i$ and
 $j>i$ such that $x_{j}=x_{i}$.  Since $x_{i}$ was already moved once,
 when we reach it the second time we will try to move it back,
 displacing $x_{i-1}$.  This process continues until we have restored
 $x_{2}$ to $T[h_{1}(x_{1})]$, displacing $x_{1}$ to
 $T[h_{2}(x_{1})]$ and possibly creating a new sequence of nestless values.  Two outcomes are now possible:
 \begin{enumerate}
  \item Some $x_{\ell}$ is moved to an empty location.  We win!
  \item Some $x_{\ell}$ is moved to a location we've already looked
      at.  We lose!  We find we are playing musical chairs with more players than chairs, and have to rehash.
 \end{enumerate}
\end{enumerate}

Let's look at the probability that we get the last
case, a \index{loop!closed}\concept{closed loop}.  Following the argument of Pagh and Rodler, we let $v$ be the number of distinct
nestless keys in the loop.  Since $v$ includes $x_1$ and at least one
other element blocking $x_1$ from being inserted at $T[h_1(x_1)]$, $v$
is at least $2$.  We can now count how many different ways
such a loop can form, and argue that in each case we include enough
information to reconstruct $h_1(u_i)$ and $h_2(u_i)$ for each of a
specific set of unique elements $u_1, \dots u_v$.

Formally, this means that we are expression the closed-loop case as a
union of many specific closed loops, and then bounding the probability
of each of these specific closed-loop events by the probability of the
event that $h_1$ and $h_2$ select the right values to make this
particular closed loop possible. Then we apply the union bound.

To describe each of the specific events, we'll provide this information:
\begin{itemize}
    \item The $v$ elements $u_1, \dots u_v$. Since we can fix $u_1 =
        x_1$, this leaves $v-1$ choices from $S$, giving $n^{(v-1)}$
        possibilities (we are overcounting by allowing duplicates, but
        that's not a problem for an upper bound). We'll require that the 
        other $u_i$ for $i>1$
        appear in the list in the same order they first appear in the
        sequence $x_1, x_2, \dots$.
    \item The $v-1$ locations we are trying to fit these $v$ elements
        into. There are at most $m^{(v-1)}$ choices for these. Again we order
        these locations by order of first appearance.
    \item The values of $i$, $j$, and $\ell$. These allow us to
        identify which segments of the sequence $x_1, x_2, \dots$
        correspond to new values $u_i$ and which are old values
        repeated (possibly in reverse order).
        
        There are at most $v$ choices for $i$ and $j$ (because we are
        still in the initial segment with no repeats), and at most
        $2v$ choices for $\ell$ if we count carefully (because we
        either land on either the initial no-duplicate sequence starting with
        $x_1$ or the second no-duplicate sequence starting with
        the second occurrence of $x_1$).

        All together, these give $2v^3$ choices.
    \item For each $i≠1$, whether the first occurrence of $u_i$
        appears in $h_1(u_i)$ or $h_2(u_i)$. This gives $2^{v-1}$
        choices, and allows us to correctly identify $h_1(u_i)$ or
        $h_2(u_1)$ from the value of $u_i$ and its first location and
        the other hash value for $u_i$ given the next location in the
        list.\footnote{The original analysis in
\cite{PaghR2004} avoids this by alternating between
two tables, so that we can
determine which of $h_1$ or $h_2$ is used at each step by parity.}
\end{itemize}

Multiplying everything out gives at most $2v^3 (2nm)^{(v-1)}$ choices
of closed loops with $v$ unique elements.
Since each particular loop allows us to determine both $h_1$ and $h_2$
for all $v$ of its elements, the probability that we get exactly these
hash values (so that the loop occurs) is $m^{-2v}$.
Summing over all closed loops with $v$ elements gives a total
probability of
\begin{align*}
2v^3 (2nm)^{v-1} ⋅ m^{-2v} 
&= 2v^3 (2n)^{v-1} m^{-v-1} 
\\&= 2v^3 (2n/m)^{v-1} m^{-2}.
\end{align*}

Now sum over all $v≥2$.  We get
\begin{align*}
    m^{-2} ∑_{v=2}^{n} 2v^3 (2n/m)^{v-1}
    &< m^{-2} ∑_{v=2}^{∞} 2v^3 (2n/m)^{v-1}.
\end{align*}
The series converges if $2n/m
< 1$, so for any fixed $α < 1/2$, the probability of any closed loop forming is
$O(m^{-2})$.

If we do hit a closed loop, then we pay $O(m)$ time to scan the
existing table and create a new empty table, and $O(n) = O(m)$ time on
average to reinsert
all the elements into the new table, assuming that this reinsertion
process doesn't generate any more closed loops and that the average
cost of an insertion that doesn't produce a closed loop is $O(1)$,
which we will show below.  But the rehashing
step only fails with probability $O(nm^{-2}) = O(m^{-1})$, so if it
does fail we can just try again until it works, and the expected total
cost is still $O(m)$.  Since we pay this $O(m)$ for each insertion with probability
$O(m^{-2})$, 
this adds only $O(m^{-1})$ to the expected cost of a single insertion.

Now we look at what happens if we don't get a closed loop.
This doesn't force us to rehash, but if the path is long enough, we
may still pay a lot to do an insertion.

It's a
little messy to analyze the behavior of keys that appear more than
once in the sequence, so the trick used in the paper is to observe
that for any sequence of nestless keys $x_{1}\dots{}x_{p}$, there is
a subsequence of size $p/3$ with no repetitions that starts with
$x_{1}$.  This will be either
the sequence $S_1$ given by $x_{1}\dots{}x_{j-1}$—the sequence starting with the first
place we try to insert $x_1$--or $S_2$ given by $x_{1}=x_{i+j-1}\dots{}x_{p}$,
the sequence starting with the second place we try to insert $x_1$.
Between these we have a third sequence $R$ where we revert some of the
moves made in $S_1$.  Because $\card*{S_1} + \card*{R} + \card*{S_2} ≥
p$, at least one of these three subsequences has size $p/3$.  But
$\card*{R}
≤ \card*{S_1}$, so it must be either $S_1$ or $S_2$.

We can then argue that the probability that we get a sequence of $v$
distinct keys in either $S_1$ or $S_2$
is at most $2(n/m)^{v-1}$.
The $(n/m)^{v-1}$ is because we need to hit a nonempty spot (which
happens with probability at most $n/m$) for the first $v-1$ elements
in the path, and since we assume that our hash functions are random,
the choices of these $v-1$ spots are all independent.  The $2$ is from
the union bound over $S_1$ and $S_2$.
If $T$ is length of the longer of $S_1$ or $S_2$, we get
$\Exp{T} =
∑_{v=1}^{∞} \Prob{T ≥ v} 
≤ ∑_{v=1}^{∞} 2 (n/m)^{v-1} = O(1)$,
assuming $n/m$ is bounded by a constant
less than $1$.  Since we already need $n/m ≤ 1/2$ to avoid the bad
closed-loop case, we can use this here as well.
We have to multiply $\Exp{T}$ by $3$ to get the bound on the actual
path, but this disappears into $O(1)$.

An annoyance with cuckoo hashing is that it has high space overhead
compared to more traditional hash tables: in order for the first part
of the analysis
above to work, the table must be at least half empty.  This can be
avoided at the cost of increasing the time complexity
by choosing between $d$ locations instead of
$2$.  This technique, due to Fotakis~\etal~\cite{FotakisPRS2003}, is known as 
\index{cuckoo hashing!$d$-ary}
\concept{$d$-ary cuckoo hashing}. For a suitable choice of $d$, it uses
$(1+ε)n$ space and guarantees that a lookup takes
$O(1/ε)$ probes, while insertion takes $(1/ε)^{O(\log
\log (1/ε))}$ steps in theory and appears to take
$O(1/ε)$ steps in practice according to experiments done by the authors.

\section{Practical issues}

Most hash functions used in practice do not have very good theoretical
guarantees, and indeed we have assumed in several places in this
chapter that we are using genuinely random hash functions when we
would expect our actual hash functions to be at most 2-universal.
There is some justification for doing this if there is enough entropy
in the set of keys $S$. A proof of this for many common applications
of hash functions is given by Chung~\etal~\cite{ChungMV2013}.

Even taking into account these results, 
hash tables that depend on strong properties of the hash function may
behave badly if the user supplies a crummy hash function.  For this
reason, many library implementations of hash tables are written
defensively, using algorithms that respond better in bad cases.  See
\url{https://svn.python.org/projects/python/trunk/Objects/dictobject.c} for
an example of a widely-used hash table implementation chosen
specifically because of its poor theoretical characteristics.

For large hash tables, local probing schemes are often faster than
chaining or cuckoo hashing, because it is
likely that all of the locations probed to find a particular value
will be on the same virtual memory page.  This means that a search for
a new value usually requires one cache miss instead of two.
\index{hashing!hopscotch}\indexConcept{hopscotch hashing}{Hopscotch
hashing}~\cite{HerlihyST2008} combines ideas from linear probing and
cuckoo hashing to get better performance than both in practice.

\section{Bloom filters}
\label{section-Bloom-filters}

See \cite[§5.5.3]{MitzenmacherU2017} for
basics and a formal analysis or \wikipedia{Bloom_filter} for many
variations and the collective wisdom of the masses, including
a pointer to a paper~\cite{DasguptaSSN2018} showing that fruit flies
use them. The
presentation here mostly
follows \cite{MitzenmacherU2017}.

\subsection{Construction}
\label{section-Bloom-filters-construction}

\indexConcept{Bloom filter}{Bloom filters} are a highly space-efficient 
randomized data structure invented by Burton H.
Bloom~\cite{Bloom1970} for storing sets of keys, with a small probability
for each key not in the set that it will be erroneously reported as being in
the set.

Suppose we have $k$ independent hash functions $h_1, h_2,
\dots, h_k$.  Our memory store $A$ is a vector of $m$ bits, all initially
zero.  To store a key $x$, set $A[h_i(x)] = 1$ for all $i$.  
To test
membership for $x$, see if $A[h_i(x)] = 1$ for all $i$.
The membership test always gives the right answer if $x$ is in fact in
the Bloom filter.  If not, we might decide that $x$ is in the Bloom
filter anyway, just because we got lucky.

\subsection{False positives}
\label{section-Bloom-filters-false-positives}

The probability of such 
\index{positive!false}\indexConcept{false positive}{false positives}
can be computed in two steps: first, we estimate how many of the bits
in the Bloom filter are set after inserting $n$ values, and then we
use this estimate to compute a probability that any fixed $x$ shows up
when it shouldn't.

If the $h_i$ are close to being independent random
functions,\footnote{We are going to sidestep the rather deep swamp of how
plausible this assumption is and what assumption we should be making
instead. However, it is known~\cite{KirschM2006} that starting
with two sufficiently random-looking hash functions 
$h$ and $h'$ and setting $h_i(x) = h(x)
+ ih'(x)$ works.} then with $n$ entries in the filter we have 
$\Prob{A[i] = 1} = 1-(1-1/m)^{kn}$, since each of the
$kn$ bits that we set while inserting the $n$ values has one chance in
$m$ of hitting position $i$.

We'd like to simplify this using the inequality $1+x ≤ e^x$, but it
goes in the wrong direction; instead, we'll use $1-x ≥ e^{-x-x^2}$,
which holds for $0 ≤ x ≤ 0.683803$ and in our application holds for $m
≥ 2$.
This gives
\begin{align*}
    \Prob{A[i] = 1} &≤ 1-(1-1/m)^{kn}
    \\
    &≤ 1-e^{-kn(1/m)(1+1/m)} \\
    &= 1 - e^{-kα(1+1/m)} \\
    &= 1 - e^{-kα'}
\end{align*}
where $α = n/m$ is the load factor and $α' = α(1+1/m)$
is the load factor fudged upward by a factor of $1+1/m$ to make the
inequality work.

Suppose now that we check to see if some value $x$ that we never
inserted in the Bloom filter appears to be present anyway.  This
occurs if $A[h_i(x)] = 1$ for all $i$.  
Since each $A[h_i(x)]$ is an independent sample from $A$,
the probability that they
all come up $1$ conditioned on $A$ is 
\begin{align}
    \left(\frac{∑ A[i]}{m}\right)^k.\label{eq-bloom-filter-false-positive-probability}
\end{align}
We have an upper bound 
$\Exp{∑ A[i]} ≤ m\left(1-e^{-kα'}\right)$,
and if we were born luckier, we might be able to get an upper bound on
the expectation of
\eqref{eq-bloom-filter-false-positive-probability} by applying
Jensen's inequality to the function $f(x) = x^k$.  But sadly this
inequality also goes in the wrong direction, because $f$ is convex for
$k > 1$.  So instead we will prove a concentration bound on $S=∑
A[i]$.

Because the $A[i]$ are not independent, we can't use off-the-shelf
Chernoff bounds. Instead, we
rely on McDiarmid's inequality. Our assumption is that the locations
of the
$kn$ ones that get written to $A$ are independent.  Furthermore,
changing the location of one of these writes changes $S$ by at most
$1$.  So McDiarmid's inequality \eqref{eq-McDiarmids-inequality} gives
$\Prob{S ≥ \Exp{S} + t} ≤ e^{-2t^2/kn}$,
which is bounded by $n^{-c}$ for $t ≥
√{\frac{1}{2}ckn \log n}$.  So as long as a reasonably large fraction of
the array is likely to be full, the relative
error from assuming $S = \Exp{S}$ is likely to be small.
Alternatively, if the array is mostly empty, 
then we don't
care about the relative error so much because the probability of
getting a false positive will already be exponentially small as a
function of $k$.

So let's assume for simplicity
that our false positive probability is exactly
$(1-e^{-kα'})^k$.
We can choose $k$ to minimize this quantity for
fixed $α'$ by
doing the usual trick of taking a derivative and
setting it to zero; to avoid weirdness with the $k$ in the exponent,
it helps to take the logarithm first (which doesn't affect the
location of the minimum), and it further helps to take the derivative
with respect to $x = e^{-α' k}$ instead of $k$ itself.  Note that when
we do this, $k = -\frac{1}{α'} \ln x$ still depends on $x$, and we will
deal with this by applying this substitution at an appropriate point.

Compute
\begin{align*}
\frac{d}{dx} \ln\left((1-x)^k\right)
&= \frac{d}{dx} k \ln (1-x) \\
&= \frac{d}{dx} \left(-\frac{1}{α'} \ln x\right) \ln (1-x) \\
&= -\frac{1}{α'} \left(\frac{\ln (1-x)}{x} - \frac{\ln 
x}{1-x}\right).
\end{align*}

Setting this to zero gives $(1-x) \ln(1-x) = x \ln x$, which by
symmetry has the unique solution $x=1/2$, giving $k = \frac{1}{α'}
\ln 2$.

In other words, to minimize the false positive rate for a known load
factor $α$, we want to choose $k = \frac{1}{α'} \ln 2 = \frac{1}{α(1+1/m)} \ln 2$, which
makes each bit one with probability approximately $1-e^{-\ln 2} = \frac{1}{2}$.
This makes intuitive sense, since having each bit be one or zero with
equal probability maximizes the entropy of the data.

The probability of a false positive for a given key is then $2^{-k} = 2^{-\ln
2/α'}$.  For a given maximum false positive rate $ε$, and
assuming optimal choice of $k$,
we need to keep $α' ≤ \frac{\ln^2 2}{\ln (1/ε)}$ 
or $α ≤ \frac{\ln^2 2}{(1+1/m)\ln(1/ε)}$.

Another way to look at this is that if we fix $ε$ and $n$, we need $m/(1+1/m) ≥ n ⋅
\frac{\ln (1/ε)}{\ln^2 2} \approx 1.442695 ⋅ n \lg(1/ε)$, which
works out to $m ≥ 1.442695 ⋅ n \lg(1/ε) + O(1)$.
This is very good for constant $ε$.

Note that for this choice
of $m$, we have $α = O(1/\ln(1/ε))$, giving $k = O(\log
(1/ε))$.  So for polynomial $ε$, we get $k = O(\log
n)$. This is closer to the complexity of tree lookups than hash table
lookups, so the main payoff for a sequential implementation is that we
don't have to store full keys.

\subsection{Comparison to optimal space}
\label{section-Bloom-filters-vs-optimal}

If we wanted to design a Bloom-filter-like data structure from scratch
and had no constraints on processing power, we'd be looking for
something that stored an index of size $\lg M$ into a family of
subsets $S_1, S_2, \dots S_M$ of our universe of keys $U$, where
$\card*{S_i} ≤ ε \card*{U}$ for each $i$ (giving the upper
bound on the false positive rate)\footnote{Technically, this gives a
weaker bound on false positives.  For standard Bloom filters, assuming
random hash functions, each key individually has at most an $ε$
probability of appearing as a false positive.  The hypothetical data
structure we are considering here—which is effectively deterministic—allows the set of false positives to
depend directly on the set of keys actually inserted in the data
structure, meaning that the adversary could arrange for a specific
key to appear as a false positive with probability $1$ by choosing
appropriate keys to insert.  So this argument may underestimate the
space needed to get make the false positives less predictable.  On the other
hand, we aren't charging the Bloom filter for the space needed to
store the hash functions, which could be quite a bit if they are
genuine random functions.}
and for any set $A⊆ U$ of
size $n$, $A ⊆ S_i$ for at least one $S_i$ (allowing us to
store $A$).

Let $N = \card*{U}$.  Then each set $S_i$ covers
$\binom{ε N}{n}$ of the $\binom{N}{n}$ subsets of size $n$.
If we could get them to overlap optimally (we can't), we'd still need
a minimum of 
$\binom{N}{n} \left/ \binom{ε{N}}{n} \right.
 = (N)_n / (ε N)_n \approx (1/ε)^n$
sets to cover everybody, where the approximation assumes $N \gg n$.
Taking the log gives $\lg M \approx n \lg (1/ε)$, meaning we
need about $\lg (1/ε)$ bits per key for the data structure.
Bloom filters use $1/\ln 2$ times this.

There are known data structures that approach this bound
asymptotically.  The first of these, due to
Pagh~\etal~\cite{PaghPR2005} also has
other desirable properties, like supporting deletions and faster
lookups if we can't look up bits in parallel.

More recently,
Fan~\etal~\cite{FanAKM2014} have described a variant of cuckoo hashing
(see §\ref{section-cuckoo-hashing}) called a 
\index{filter!cuckoo}
\concept{cuckoo filter}.  This is a cuckoo hash table that, instead of
storing full keys $x$, stores \indexConcept{fingerprint}{fingerprints}
$f(x)$, where $f$ is a hash function with $\ell$-bit outputs.  False
positives now arise if we happen to hash a value $x'$ with $f(x') =
f(x)$ to the same
location as $x$.  If $f$ is drawn from a $2$-universal family, this
occurs with probability at most $2^{-\ell}$.  So the idea is that by
accepting an $ε$ small rate of false positives, we can shrink the space
needed to store each key from the full key length to
$\lg (1/ε) = \ln(1/ε)/\ln 2$, the asymptotic minimum.

One complication is that, since we are throwing away the original key
$x$, when we displace a key from $h_1(x)$ to $h_2(x)$ or vice versa,
we can't recompute $h_1(x)$ and $h_2(x)$ for arbitrary $h_1$ and
$h_2$.  The solution proposed by Fan~\etal{} is to let
$h_2(x) = h_1(x) ⊕ g(f(x))$, where $g$ is a hash function that depends
only on the fingerprint.  This means that when looking at a
fingerprint $f(x)$ stored in position $i$, we don't need to know
whether $i$ is $h_1(x)$ or $h_2(x)$, since whichever it is, the other
location will be $i⊕g(f(x))$.  Unfortunately, this technique and some
other techniques used in the paper to crunch out excess empty space
break the standard analysis of cuckoo hashing, so the authors can only
point to experimental evidence that their data structure actually
works.  However, a variant of this data structure has been shown to
work by Eppstein~\cite{Eppstein2016}.

\subsection{Applications}
\label{section-Bloom-filters-applications}

Historically, Bloom filters were invented to act as a way of filtering
queries to a database table through fast but expensive\footnote{As
much as \$0.10/bit in 1970.} RAM before looking up the actual values on a slow but cheap tape
drive.  Nowadays the cost of RAM is low enough that this is less of an
issue in most cases, but Bloom filters are still popular in networking
and in distributed databases.

In networking, Bloom filters are useful in building network switches,
where incoming packets need to be matched against routing tables in
fractions of a nanosecond.
Bloom filters work particularly well for this when implemented in hardware,
since the $k$ hash functions can be computed in parallel.  False
positives, if infrequent enough, can be handled by some slower backup
mechanism.

In distributed databases, Bloom filters are used in the \concept{Bloomjoin} algorithm~\cite{MackertL1986}. 
Here we want to do a join on two
tables stored on different machines (a join is an operation where we
find all pairs of rows, one in each table, that match on some common
key).  A straightforward but expensive way to do this is to send the
list of keys from the smaller table across the network, then match
them against the corresponding keys from the larger table.  If there
are $n_s$ rows in the smaller table, $n_b$ rows in the larger table,
and $j$ matching rows in the larger table, this requires sending $n_s$
keys plus $j$ rows.  If instead we send a Bloom filter representing
the set of keys in the smaller table, we only need to send
$n \lg(1/ε) / \ln 2$ bits for the Bloom filter plus an extra
$ε n_b$ rows on average for the false positives.  This can be
cheaper than sending full keys across if the number of false positives
is reasonably small.

\subsection{Counting Bloom filters}
\label{section-counting-Bloom-filters}

It's not hard to modify a Bloom filter to support deletion.  The basic
trick is to replace each bit with a counter, so that whenever a value
$x$ is inserted, we increment $A[h_i(x)]$ for all $i$ and when it is
deleted, we decrement the same locations.  The search procedure now
returns $\min_i A[h_i(x)]$ (which means that it principle it can even
report back multiplicities, though with some probability of reporting
a value that is too high).  To avoid too much space overhead, each
array location is capped at some small maximum value $c$; once it
reaches this value, further increments have no effect.
The resulting
structure is called a \index{Bloom filter!counting}\concept{counting
Bloom filter}, due to Fan~\etal~\cite{FanCAB2000}.

We can only expect this to work if our chance of hitting the cap is
small.
Fan~\etal{} observe that the probability that the $m$ table
entries include one that is at least $c$ after $n$ insertions is bounded by 
\begin{align*}
    m \binom{nk}{c} \frac{1}{m^c} 
    &≤ m \left(\frac{enk}{c}\right)^c \frac{1}{m^c}
    \\ 
    &= m \left(\frac{enk}{cm}\right)^c
    \\
    &= m (ekα/c)^c.
\end{align*}
(This uses the bound $\binom{n}{k} ≤
\left(\frac{en}{k}\right)^k$, which follows from Stirling's
formula.)

For $k= \frac{1}{α}\ln 2$, this is
$m (e \ln 2 / c)^c$.
For the specific value of $c=16$ (corresponding to $4$ bits per entry),
they compute a bound of $1.37 × 10^{-15} m$, which they argue is
minuscule for all reasonable values of $m$ (it's a systems paper).

The possibility that a long chain of alternating insertions and
deletions might produce a false negative due to overflow is considered
in the paper, but the authors state that ``the probability of such a chain of events is so
low that it is much more likely that the proxy server would be
rebooted in the meantime and the entire structure reconstructed.''
An alternative way of dealing with this problem is to never decrement
a maxed-out register. This never produces a false negative, but may
cause the filter to slowly fill up with maxed-out registers, producing
a higher false-positive rate.

A fancier variant of this idea is the \index{Bloom
filter!spectral}\concept{spectral Bloom filter} of Cohen and Matias~\cite{CohenM2003},
which uses larger counters to track multiplicities of items.
The essential idea here is that
we can guess that the number of times a particular value
$x$ was inserted is equal to $\min_{i=1}^{k} A[h_i(x)]$), with some
extra tinkering to detect errors based on deviations from the typical
joint distribution of the $A[h_i(x)]$ values.
An even more sophisticated approach gives the count-min sketches of the
next section.

\section{Data stream computation}
\label{section-data-stream-computation}

In the \concept{data stream}\index{stream!data} model, we are given a
huge flood of data—far too big to store—in a single pass, and want to
incrementally build a small data structure, called a \concept{sketch},
that will allow us to answer statistical questions about the data
after we've processed it all.  The motivation is the existence of data
sets that are too large to store at all (network traffic statistics),
or too large to store in fast memory (very large database tables). By
building an appropriate small data structure using a single pass
through the data, we can still answer queries about with some loss of
accuracy. Examples we will consider include estimating the size of a
set presented over time with possible duplicate elements
(§\ref{section-cardinality-estimation}) or more general statistical
queries based on aggregate counts of some sort
(§\ref{section-count-min-sketches}).

In each of these cases, the answers we get will be approximate. We
will measure the quality of the approximation in terms of parameters
$(δ,ε)$, where we demand a relative error of at most $ε$ with
probability at least $1-δ$. We'd also like our data structure to have
size at most polylogarithmic in the number of samples $n$ and
polynomial in $δ$ and $ε$.

\subsection{Cardinality estimation}
\label{section-cardinality-estimation}

The \concept{cardinality estimation} or \concept{count-duplicates}
problem involves seeing a sequence of values $x_1, x_2, \dots, x_n$
and asking to compute the number of \emph{unique} values in this
sequence.

Without the uniqueness constraint, this is trivial: just keep a
counter. With the uniqueness constraint, exact counting is much
harder, since any data structure that lets us detect if we see a new
element also lets us test for membership. But if we are willing to
accept an approximation, we can get around this by using hashing and
then tracking statistical properties of the hashed values. To keep
the analysis simple, we will assume that our hashing function is a random
function, and not charge for storing its parameters.

Many algorithms of this type are based on a tool called a
\concept{Flajolet-Martin
sketch}\index{sketch!Flajolet-Martin}~\cite{FlajoletM1985}.
The simplest version of this is that we pick a random hash function
$h$, and use it to generate a geometric random variable $R_i$ for each $x_i$
by counting the number of trailing zeroes in the binary representation
of $h(x_i)$. We then track $R = \max_i R_i$ and estimate the number
$n$ of unique $x_i$ using $2^R$.

The intuition for why this counts unique $x_i$ is that sending in the
same $x_i$ twice produces the same $R_i = h(x_i)$ both times, so the
maximum $R$ is unaffected.

It's easy to show that this gives a reasonably good approximation to
$n$. For the upper bound, given $n$ samples, the expected number of
samples with $R_i ≥ k$ is $n⋅2^{-k}$, so for $k = \lg n + \ell$, the
probability that $2^R ≥ 2^k$ is at most $2^{-\ell}$ by Markov's
inequality. For the lower bound, we have $\Prob{R_i < k} = 1 - 2^{-k}$
so $\Prob{∀i: R_i < k} = (1-2^{-k})^n ≤ e^{-n 2^{-k}}$, which gives
$\Prob{\max R_i < \lg n - \ell} ≤ e^{-2^{\ell}}$.
The lower bound is a bit stronger than the upper bound, but
in both directions 
we get at least a constant probability of being within a (large)
constant factor of the correct answer. The expected value can also be
shown to converge to the correct value for large enough $n$, after
multiplying by a small correction factor $φ$ that compensates for
systematic round-off error caused by quantizing to powers of $2$.

To improve the constants, Flajolet and Martin proposed a technique
they called \concept{Probabilistic Counting with Stochastic Averaging}
or \concept{PCSA}. This splits the incoming stream into $m$ buckets
using a second hash function (or, in practice, the leading bits of the
first one). Each bucket gives its own estimate $\hat{n}_i = m φ 2^R$, then
these estimates are averaged to produce to final estimate. Flajolet
and Martin show that, with an appropriate multiplier, this estimate
has a typical error of $O(1/√{m})$. This analysis is pretty
involved so we will not repeat it here.

The original PCSA algorithm is not used much in practice. More popular
is a descendant called \concept{HyperLogLog}~\cite{FlajoletFGM2007} that replaces the
arithmetic mean $\frac{1}{m} \sum_i \hat{n}_i$ with a harmonic mean
\begin{displaymath}
    \frac{1}{\frac{1}{m}\sum_i \frac{1}{\hat{n}_i}}.
\end{displaymath}

As with PCSA, HyperLogLog requires using some carefully calculated
corrections to get an unbiased estimator. This can be avoided
using an auxiliary counter that is updated whenever the main data
structure changes, which also gives some improvement in the accuracy
of the estimate. This mechanism was originally proposed independently by
Cohen~\cite{Cohen2015} and Ting~\cite{Ting2014}, although the
description we give here is largely drawn from a more recent paper by
Pettie~\etal~\cite{PettieWY2020}, which refers to this approach as the
\concept{martingale transformation} of the original data structure.

What this transformation does is observe that for each state $s$ of the
HyperLogLog (or whatever) data structure, there is a probability $p_s$
that the next unique element will send it to a new state $s'$.
If we can calculate this probability, then we can update our estimated
count $\hat{λ}$ by increasing it by $1/p_s$ when the change occurs.
Because each new unique element gives an expected increase of exactly
$p_s ⋅ \frac{1}{p_s} = 1$, this makes the expected value in the
counter exactly equal to the actual count. (The connection between this
and martingales is that we just showed that $\hat{λ}_t - λ_t$ is a
martingale where $\hat{λ}_t$ is the value of the counter after $t$
steps and $λ_t$ is the actual number of unique elements.)

The only tricky part here is computing $p_s$. For HyperLogLog, there
is a $1/m$ chance that our new unique element lands in each of the $m$
buckets, and it lands in a bucket that currently stores $r_i$, the
probability that we increase $r_i$ is exactly $2^{-r_i-1}$. This
immediately gives
\begin{align*}
    p_s
    &= \frac{1}{m} \sum_i 2^{-r_i-1}
    \intertext{and thus}
    \frac{1}{p_s}
    &= \frac{1}{\frac{1}{m} \sum_i 2^{-r_i-1}},
\end{align*}
which looks suspiciously like the harmonic mean used on the final
estimates in HyperLogLog. As with the original HyperLogLog, it is
possible to show that the typical relative error for this sketch
is $O(1/√{m})$. See~\cite{PettieWY2020} for more details and some
further improvements.

If we don't care about practical engineering issues, there is a known
asymptotically optimal solution to the cardinality estimation
problem~\cite{KaneNW2010}, which doesn't even require assuming a
random oracle, but the constants give worse performance than the
systems that people actually use.

\subsection{Count-min sketches}
\label{section-count-min-sketches}

A \index{sketch!count-min}\concept{count-min sketch} is used for the case where
we are presented with a sequence of pairs $\Tuple{i_{t},c_{t}}$ where
$1≤i_{t}≤n$ is an \emph{index} and $c_{t}$ is a \emph{count},
and we want to construct a sketch that will allows us to
approximately answer statistical queries about the vector $a$ given by
$a_{i} = ∑_{t, i_t=i} c_{t}$.
These were developed by Cormode and Muthukrishnan~\cite{CormodeM2005},
although some of the presentation here is based
on~\cite[§15.4]{MitzenmacherU2017}. Structurally, they are related to
counting Bloom filters (§\ref{section-counting-Bloom-filters}).

Note that we are no longer interested in detecting unique values, but
instead want to avoid the cost of storing the entire vector $a$ when
most of its components are small or zero.
The goal is that the size of the sketch should be
polylogarithmic in the size of $a$ and the length of the stream, and
polynomial in the error bounds. We also want updating the sketch for
each new data point to be cheap.

The Cormode-Muthukrishnan count-min sketch is fairly versatile,
giving approximations of $a_{i}$, $∑_{i=\ell}^{r} a_{i}$, and $a⋅b$
(for any fixed $b$), and it can also be used for more complex tasks like
finding \indexConcept{heavy hitter}{heavy hitters}—indices with high
weight. The easiest case is approximating $a_{i}$ when all the
$c_{t}$ are non-negative, so we'll start with that.

\subsubsection{Initialization and updates}
\label{section-count-min-construction}

To construct a count-min sketch, build a two-dimensional
array $c$ with depth $d =
\ceil{\ln(1/δ)}$ and width $w =
\ceil{e/ε}$, where $ε$ is the
error bound and $δ$ is the probability of exceeding the error
bound.  Choose $d$ independent 
hash functions from some $2$-universal hash family; we'll use one of
these hash function for each row of the array.
Initialize $c$ to all zeros.

The update rule:
Given an update $(i_{t},c_{t})$, increment $c[j,h_{j}(i_{t})]$ by
$c_{t}$ for $j=1\dots d$.  (This is the \emph{count} part of count-min.)

\subsubsection{Queries}

Let's start with \index{query!point}\indexConcept{point query}{point queries}.
Here we want to estimate $a_{i}$ for some fixed $i$.  There are two
cases; the first handles non-negative increments only, while the
second handles arbitrary increments.
In both cases we will get an estimate whose error is
linear in both the error parameter $ε$ and the $\ell_1$-norm
$\norm*{a}_1 = ∑_i \abs*{a_i}$ of $a$.  It follows that the relative error will be low for heavy points, but we may get a large relative error for light points (and especially large for points that don't appear in the data set at all).

For the non-negative case, to estimate $a_i$, compute $\hat{a}_i =
\min_j c[j,h_j(i)]$.  (This is the \emph{min} part of coin-min.)
Then:
\begin{lemma}
\label{lemma-count-min-point-query-non-negative}
When all $c_t$ are non-negative,
for $\hat{a}_i$ as defined above:
\begin{align}
    \hat{a}_i &≥ a_i,
\label{eq-count-min-point-query-non-negative-lower-bound}
\intertext{and}
\Prob{\hat{a}_i ≤ a_i + ε \norm*{a}_1} &≥ 1-δ.
\end{align}
\end{lemma}
\begin{proof}
The lower bound is easy.  Since for each pair $(i,c_{t})$ we increment
each $c[j,h_{j}(i)]$ by $c_{t}$, we have an invariant that
$a_{i}≤ c[j,h_j(i)]$ for all $j$ throughout the computation, which
gives $a_i ≤ \hat{a}_i = \min_j c[j,h_j(i)]$.

For the upper bound, let $I_{ijk}$ be the indicator for the event that
$(i≠k) ∧ (h_{j}(i) = h_{j}(k))$, i.e., that we get a
collision between $i$ and $k$ using $h_{j}$.  The $2$-universality
property of
the $h_{j}$ gives $\Exp{I_{ijk}} ≤ 1/w ≤ ε/e$.

Now let $X_{ij} = ∑_{k=1}^{n} I_{ijk}a_{k}$.  Then $c[j,h_{j}(i)] =
a_{i} + X_{ij}$.  (The fact that $X_{ij}≥0$ gives an alternate
proof of the lower bound.)  Now use linearity of expectation to get
\begin{align*}
\Exp{X_{ij}} 
&= \Exp{∑_{k=1}^{n} I_{ijk}a_{k}} 
\\&= ∑_{k=1}^n a_{k}\Exp{I_{ijk}} 
\\&≤ ∑_{k=1}^{n} a_{k} (ε/e)
\\&= (ε/e)\norm*{a}_1.
\end{align*}
So $\Prob{c[j,h_{j}(i)] > a_{i} + ε\norm*{a}_{1}} = \Prob{X_{ij} >
e\Exp{X_{ij}}} < 1/e$, by Markov's inequality.  With $d$ choices for
$j$, and each $h_j$ chosen independently, the probability that
every count is too big is at most $(1/e)^{d} = e^{-d} ≤
\exp(-\ln(1/δ)) = δ$.
\end{proof}

Now let's consider the general case, where the increments $c_t$ might
be negative.  We still initialize and update the data structure as
described in §\ref{section-count-min-construction}, but now
when computing $\hat{a}_i$, we use the median count instead of the
minimum count: $\hat{a}_i = \median \Set{ c[j,h_j(i)] \mid j = 1 \dots n }$.
Now we get:
\begin{lemma}
\label{lemma-count-min-query-point-general}
For $\hat{a}_i$ as defined above,
\begin{align}
\label{eq-count-min-query-point-general}
\Prob{a_{i}-3ε\norm*{a}_{1} ≤ \hat{a}_{i} ≤
a_{i}+3ε\norm*{a}_{1}}
& > 1-δ^{1/4}.
\end{align}
\end{lemma}
\begin{proof}
    The basic idea is that for the median to be off by $t$, at least
    $d/2$ rows must give values that are off by $t$.  We'll show that
    for $t = 3ε\norm*{a}_{1}$, the expected number of rows that are
    off by $t$ is at most $d/8$.  Since the hash functions for the
    rows are chosen independently, we can use Chernoff bounds to show
    that with a mean of $d/8$, the chances of getting all the way to $d/2$ are
    small.

In detail, we again define the error term $X_{ij}$ as above, and observe that
\begin{align*}
\Exp{\abs*{X_{ij}}}
&= \Exp{\abs*{∑_{k} I_{ijk}a_{k}}}
\\&≤ ∑_{k=1}^n \abs*{a_{k}\Exp{I_{ijk}}}
\\&≤ ∑_{k=1}^n \abs*{a_{k}}(ε/e)
\\&= (ε/e)\norm*{a}_{1}.
\end{align*}
Using Markov's inequality, we get $\Prob{\abs*{X_{ij}}} >
3ε\norm*{a}_{1}] = \Prob{\abs*{X_{ij}} > 3e\Exp{X_{ij}}} < 1/3e <
1/8$.  
In order for the median to be off by more than
$3ε\norm*{a}_{1}$, we need $d/2$ of these low-probability
events to occur.  The expected number that occur is $μ = d/8$, so
applying the standard Chernoff bound \eqref{eq-Chernoff-bound} with $δ =
3$
we are looking at 
\begin{align*}
\Prob{S ≥ d/2}
&=
\Prob{S ≥ (1+3)μ}
\\
&≤ (e^{3}/4^{4})^{d/8}
\\&≤ (e^{3/8}/2)^{\ln (1/δ)}
\\&= δ^{\ln 2 - 3/8} 
\\&< δ^{1/4}.
\end{align*}
(The actual exponent is about 0.31, but 1/4 is easier to deal with).
This immediately gives \eqref{eq-count-min-query-point-general}.
\end{proof}

One way to think about this is that getting an estimate within
$ε\norm*{a}_{1}$ of the right value with probability at least
$1-δ$ requires 3 times the width and 4 times the depth—or 12 times the space and 4 times the time—when we aren't assuming increments are non-negative.

Next, we consider inner products.
Here we want to estimate $a⋅b$, where $a$ and $b$ are both
stored as count-min sketches using the same hash functions.  The paper concentrates on the case
where $a$ and $b$ are both non-negative, which has applications in
estimating the size of a join in a database.  The method is to
estimate $a⋅b$ as $\min_{j} ∑_{k=1}^w
c_{a}[j,k]⋅c_{b}[j,k]$.

For a single $j$, the sum consists of both good values and bad
collisions; we have $∑_{k=1}^w c_{a}[j,k]⋅c_{b}[j,k] =
∑_{k=1}^{n} a_{i}b_{i} + ∑_{p≠q, h_j(p)=h_j(q)}
a_{p}b_{q}$.
The second term has expectation 
\begin{align*}
∑_{p≠q} \Prob{h_{j}(p)=h_{j}(q)}a_{p}b_{q} 
&≤ ∑_{p≠q} (ε/e)a_{p}b_{q} 
\\&≤ ∑_{p,q}(ε/e)a_{p}b_{q} 
\\&≤ (ε/e)\norm*{a}_{1}\norm*{b}_{1}.
\end{align*}
As in the point-query case, we get probability at most $1/e$ that a
single $j$ gives a value that is too high by more than
$ε\norm*{a}_{1}\norm*{b}_{1}$, so the probability that
the minimum value is too high is at most $e^{-d} ≤ δ$.

\subsubsection{Finding heavy hitters}

Here we want to find the heaviest elements in the set: those indices
$i$ for which $a_{i}$ exceeds $φ\norm*{a}_{1}$ for some constant
threshold $0 < φ ≤ 1$. We assume that all $c_t$ are non-negative.
Because $\norm*{a}_1 = ∑_i a_i$, we know that there will be at most
$1/φ$ heavy hitters.  But the tricky part is figuring out which
elements they are.

The output at any stage will be approximate in the following sense: it is
guaranteed that any $i$ such that $a_i ≥ φ \norm*{a}_1$ is included, and
each $i$ with $a_i < (φ-ε)$ that previously appeared in the stream is included
with probability at most $1-δ$. This is similar to what we would get
if we just ran a point query on all possible $i$, but (a) there are
many possible $i$ and (b) we won't ever output an $i$ we've never
seen.

The trick is to extend the data structure and update procedure to track all the
heavy elements found so far (stored in a heap, with the minimum
estimate at the top), as well as
$\norm*{a}_{1} = ∑ c_{t}$.  When a new increment $(i,c)$ comes in,
we first update the count-min structure and then do a point query on
$a_{i}$; if $\hat{a}_{i} ≥ φ\norm*{a}_{1}$, we insert $i$ into
the heap. We also delete any elements at the top of the heap that 
have a point-query estimate below threshold.

Because $\hat{a}_i ≥ a_i$, every heavy hitter is correctly identified.
However, it's possible that an index stops being a heavy hitter at some
point (because the threshold $φ\norm{a}_1$ rose since we included it).
In this case it may get removed from the heap, but if it becomes a
heavy hitter again, we'll put it back.

\section{Locality-sensitive hashing}
\label{section-locality-sensitive-hashing}

\index{hashing!locality-sensitive}\indexConcept{locality-sensitive
hashing}{Locality-sensitive hashing} was invented by Indyk and
Motwani~\cite{IndykM1998} to solve the problem of designing a data
structure that finds approximate nearest neighbors to query points in
high dimension. We'll mostly be following this paper in this section,
concentrating on the hashing parts.

Because approximate nearest neighbors has many applications, there has
been quite a bit of work since the Indyk and Motwani paper on
locality-sensitive hashing in particular and approximate nearest
neighbor data structures in general. (A relatively recent survey is
that of Andoni~\etal~\cite{AndoniIR2018}.) So this section should
be read more as an example of how the idea of hashing can be adapted
to contexts where we want to preserve some information in the input
data than as a guide to solving nearest neighbors.

\subsection{Approximate nearest neighbor search}
\label{section-approximate-nearest-neighbor-search}

In the \concept{nearest neighbor search} problem
(\concept{NNS} for short), we are given a set of $n$ points $P$ in a
metric space with distance function $d$, 
and we want to construct a data
structure that allows us to quickly find the closet point $p$ in $P$ to
any given query point $q$.  We could always compute the distance
between $q$ and each possible $p$, but this takes time $O(n)$,
and we'd like to get lookups to be sublinear in $n$.

Indyk and Motwani were particularly interested
in what happens in $ℝ^d$ for high dimension $d$ under various
natural metrics.  Because the volume of a ball in a
high-dimensional space grows exponentially with the dimension, this
problem suffers from the \concept{curse of
dimensionality}~\cite{Bellman1957}: simple
techniques based on, for example, assigning points in $P$ to nearby
locations in a grid
may require searching exponentially many grid locations.
Indyk and Motwani deal with this through a combination of
randomization and solving the weaker problem of \concept{$ε$-nearest
neighbor search} (\concept{$ε$-NNS}), where it's acceptable to return a
different 
point $p'$ as long as $d(q,p') ≤ (1+ε) \min_{p ∈ P} d(q,p)$.

This problem can be solved by reduction to a simpler problem called
\index{point location in equal balls}\concept{$ε$-point
location in equal balls} or
\index{PLEB}\concept{$ε$-PLEB}.  In this problem, we
are given $n$ radius-$r$ balls centered on points $c$ in a set $C$, and we
want a data structure that returns a point $c' ∈ C$ with
$d(q,c') ≤ (1+ε)r$ if there is at least one point $c$
with $d(q,c) ≤ r$.  If there is no such point, the data
structure may or may not return a point (it might say no, or
it might just return a point that is too far away, which we
can discard).  The difference between an $ε$-PLEB and NNS is
that an $ε$-PLEB isn't picky about returning the closest point to $q$
if there are multiple points that are all good enough.  Still, we can
reduce NNS to $ε$-PLEB.

The easy reduction is to use binary search.
Let
$R = \frac{\max_{x,y∈ P} d(x,y)}{\min_{x,y∈ P, x≠ y} d(x,y)}$.
Given a point $q$, look for the minimum
$\ell ∈ \Set{(1+ε)^0, (1+ε)^1, \dots, R}$
for which an $ε$-PLEB data
structure with radius $\ell$ and centers $P$ returns a point $p$ with
$d(q,p) ≤ (1+ε)\ell$; then return this point as the
approximate nearest neighbor.

This requires $O(\log_{1+ε} R)$ instances of the $ε$-PLEB data
structure and $O(\log \log_{1+ε} R)$ queries.  The blowup as a function of
$R$ can be avoided using a more sophisticated data structure called a
\index{tree!ring-cover}\concept{ring-cover tree}, defined in the
paper.  We won't talk about ring-cover trees because they are (a)
complicated and (b) not randomized.  Instead, we'll move directly to
the question of how we solve $ε$-PLEB.

\subsection{Locality-sensitive hash functions}
\label{section-locality-sensitive-hash-functions}

\begin{definition}[\cite{IndykM1998}]
A family of hash functions $H$ is
\concept{$(r_1,r_2,p_1,p_2)$-sensitive} for $d$ if, for any
points $p$ and $q$, if $h$ is chosen uniformly from $H$,
\begin{enumerate}
\item If $d(p,q) ≤ r_1$, then $\Prob{h(p) = h(q)} ≥ p_1$, and
\item If $d(p,q) > r_2$, then $\Prob{h(p) = h(q)} ≤ p_2$.
\end{enumerate}
\end{definition}

These are useful if $p_1 > p_2$ and $r_1 < r_2$; that is, we are more
likely to hash inputs together if they are closer.  Ideally, we can
choose $r_1$ and $r_2$ to build $ε$-PLEB data structures
for a range of
radii sufficient to do binary search as described above (or build a
ring-cover tree if we are doing it right).  For the moment, we will
aim for an $(r_1,r_2)$-PLEB data structure, which 
returns a point within $r_1$ with high probability if one exists,
and never returns a point farther away than $r_2$.

There is some similarity between locality-sensitive hashing and a more
general dimension-reduction technique known as the
\index{lemma!Johnson-Lindenstrauss}\concept{Johnson-Lindenstrauss
lemma}~\cite{JohnsonL1984}; this says that projecting $n$ points in
a high-dimensional space to $O(ε^{-2} \log n)$ dimensions using
an appropriate random matrix preserves $\ell_2$ distances between the
points to within relative error $ε$ (in fact, even a 
random matrix with entries in $\Set{-1,0,+1}$ is enough~\cite{Achlioptas2003}).  Unfortunately,
dimension reduction by itself is not enough to solve approximate
nearest neighbors in sublinear time, because we may still need to
search a number of boxes exponential in $O(ε^{-2} \log n)$,
which will be polynomial in $n$. But we'll look at the
Johnson-Lindenstrauss lemma and its many other applications more
closely in Chapter~\ref{chapter-dimension-reduction}.

\subsection{Constructing an \texorpdfstring{$(r_1,r_2)$}{(r1,r2)}-PLEB}

The first trick is to amplify the difference between $p_1$ and $p_2$
so that we can find a point within $r_1$ of our query point $q$ if one
exists.  This is done in three stages: First, we concatenate multiple hash
functions to drive the probability that distant points hash together
down until we get few collisions: the idea here is that we are taking
the AND of the events that we get collisions in the original hash
function.  Second, we hash our query point and
target points multiple times to bring the probability that nearby
points hash together up: this is an OR.  Finally, we iterate the procedure to drive
down any remaining probability of failure below a target probability
$δ$: another AND.

For the first stage, 
let $k = \log_{1/p_2} n$ and define a composite hash
function $g(p) = (h_1(p)\dots h_k(p))$.
If $d(p,q) > r_2$, $\Prob{g(p) = g(q)} ≤ p_2^k = p_2^{\log_{1/p_2}
n} = 1/n$.  Adding this up over all $n$ points in our data structure
gives us one false match for $q$ on average.

However, we may 
not be able to find the correct match for $q$, since $p_1$ may not be
all that much larger than $p_2$.  For this, we do a second round of
amplification, where now we are taking the OR of events we want
instead of the AND of events we don't want.

Let $\ell = n^ρ$, where $ρ = \frac{\log(1/p_1)}{\log(1/p_2)} =
\frac{\log p_1}{\log p_2} < 1$,
and choose hash functions $g_1 \dots g_\ell$
independently as above.  To store a point $p$, put it in a bucket for
$g_j(p)$ for each $j$; these buckets are themselves stored in a hash
table (by hashing the value of $g_j(p)$ down further) so that they fit
in $O(n)$ space.  Suppose now that $d(p,q) ≤ r_1$ for some $p$.
Then 
\begin{align*}
\Prob{g_j(p) = g_j(q)} 
&≥ p_1^k 
\\&= p_1^{\log_{1/p_2} n} 
\\&= n^{-\frac{\log 1/p_1}{\log 1/p_2}} 
\\&= n^{-ρ}
\\&= 1/\ell.
\end{align*}
So by searching through $\ell$ independent buckets we find $p$ with
probability at least $1-(1-1/\ell)^\ell = 1-1/e + o(1)$.
We'd like to guarantee that we only have to look at $O(n^ρ)$ points
(most of which we may reject) during this process; but we can do this by stopping if we see more
than $2\ell$ points.  Since we only expect to see $\ell$ bad points in
all $\ell$ buckets, this event only happens with probability
$1/2$.  So even adding it to the probability of failure from the hash
functions not working right we still have only a constant probability
of failure $1/e + 1/2 + o(1)$.

Iterating the entire process $O(\log (1/δ))$ times then gives the
desired bound $δ$ on the probability that this process fails to
find a good point if one exists.  

Multiplying out all the costs gives a cost of a query of $O(k\ell
\log(1/δ)) = O\left(n^ρ (\log_{1/p_2} n) \log (1/δ)\right)$ hash
function evaluations and $O(n^ρ \log(1/δ))$ distance
computations.  The cost to insert a point is just $O(k\ell
\log(1/δ)) = O\left(n^ρ (\log_{1/p_2} n) \log
(1/δ)\right)$ hash function evaluations, the same number as for a
query.

\subsection{Hash functions for Hamming distance}
\label{section-locality-sensitive-hashing-for-hamming-distance}

Suppose that our points are $d$-bit vectors and that we use Hamming
distance for our metric.  In this case, using the family of one-bit
projections $\SetWhere{ h_i }{ h_i(x) = x_i }$ gives a locality-sensitive hash
family~\cite{AnderssonBRT1996}.

Specifically, we can show this family is
$\parens*{r,r(1+ε),1-\frac{r}{d},1-\frac{r(1+ε)}{d}}$-sensitive.
The argument is trivial: if two points $p$ and $q$ are at distance $r$
or less, they differ in at most $r$ places, so the probability that
they hash together is just the probability that we don't pick one of
these places, which is at least $1-\frac{r}{d}$.  Essentially the same
argument works when $p$ and $q$ are far away.

These are not particularly clever hash functions, so the heavy lifting
will be done by the $(r_1,r_2)$-PLEB construction.  Our goal is to
build an $ε$-PLEB for any fixed $r$, which will correspond to
an $(r,r(1+ε))$-PLEB.  The main thing we need to do,
following~\cite{IndykM1998} as always, is compute
a reasonable bound on 
$ρ = \frac{\log p_1}{\log p_2} =
\frac{\ln(1-r/d)}{\ln(1-(1+ε)r/d)}$.  This is essentially just
a matter of hitting it with enough inequalities, although there are a
couple of tricks in the middle.

Compute
\begin{align}
ρ
&= \frac{\ln(1-r/d)}{\ln(1-(1+ε)r/d)}
\nonumber\\&= \frac{(d/r) \ln(1-r/d)}{(d/r) \ln (1-(1+ε)r/d)}
\nonumber\\&= \frac{\ln((1-r/d)^{d/r})}{\ln ((1-(1+ε)r/d)^{d/r})}
\nonumber\\&≤ \frac{\ln(e^{-1}(1-r/d))}{\ln e^{-(1+ε)}}
\nonumber\\&= \frac{-1 + \ln(1-r/d)}{-(1+ε)}
\nonumber\\&= \frac{1}{1+ε} - \frac{\ln(1-r/d)}{1+ε}.
\label{eq-Indyk-Motwani-rho-bound}
\end{align}

Note that we used the fact that $1+x ≤ e^x$ for all $x$
in the denominator and $(1-x)^{1/x} ≥ e^{-1}(1-x)$ for $x ∈ [0,1]$
in the numerator.  The first fact is our usual favorite inequality.

The second can be proved in a number of ways.  The most visually 
intuitive is
that $(1-x)^{1/x}$ and $e^{-1}(1-x)$
are equal at $x=1$ and equal in the limit as $x$ goes to $0$, while 
$(1-x)^{1/x}$ is concave in between $0$ and $1$
and $e^{-1}(1-x)$ is linear.  Unfortunately it is rather painful to
show that $(1-x)^{1/x}$ is in fact concave.  An alternative is to
rewrite the inequality $(1-x)^{1-x} ≥ e^{-1}(1-x)$ as $(1-x)^{1/x-1}
≥ e^{-1}$, apply a change of variables $y = 1/x$ to get
$(1-1/y)^{y-1} ≥ e^{-1}$ for $y ∈ [1,∞)$, and then argue that
(a) equality holds in the limit as $y$ goes to infinity, and (b)
the left-hand-side is a nonincreasing function, since
\begin{align*}
    \frac{d}{dy}
\ln\left((1-1/y)^{y-1}\right) 
&= \frac{d}{dy} \left[(y-1)\left(\ln(y-1) - \ln
y\right) \right]
\\&= \ln(1-1/y) + (y-1)\left(\frac{1}{y-1} - \frac{1}{y}\right)
\\&= \ln(1-1/y) + 1 - (1-1/y)
\\&= \ln(1-1/y) + 1/y
\\&≤ -1/y + 1/y
\\&= 0.
\end{align*}

We now return to \eqref{eq-Indyk-Motwani-rho-bound}.
We'd really like the second term to be small enough that we can just
write $n^ρ$ as $n^{1/(1+ε)}$.  (Note that even though it
looks negative, it isn't, because $\ln(1-r/d)$ is negative.)
So we
pull a rabbit out of a hat by assuming that $r/d < 1/\ln
n$.\footnote{Indyk and Motwani pull this rabbit out of a hat a few
steps earlier, but it's pretty much the same rabbit either way.}
This assumption can be justified by modifying the algorithm so that
$d$ is padded out with up to $d \ln n$ unused junk bits if necessary.
Using this assumption, we get
\begin{align*}
n^ρ
&< n^{1/(1+ε)} n^{-\ln(1-1/\ln n)/(1+ε)}
\\&= n^{1/(1+ε)} (1-1/\ln n)^{-\ln n}
\\&≤ e n^{1/(1+ε)}.
\end{align*}

Plugging into the formula for $(r_1,r_2)$-PLEB gives
$O(n^{1/(1+ε)} \log n \log (1/δ))$ hash function evaluations
per query, each of which costs $O(1)$ time, plus $O(n^{1/(1+ε)}
\log(1/δ))$ distance computations, which will take $O(d)$ time
each.  If we add in the cost of the binary search, we have to multiply
this by $O(\log \log_{1+ε} R\, \log \log \log_{1+ε} R)$, where the log-log-log comes
from having to adjust $δ$ so that the error doesn't accumulate
too much over all $O(\log \log R)$ steps.  The end result is that we
can do approximate nearest-neighbor queries in
\begin{align*}
    O\left(n^{1/(1+ε)} \log (1/δ) (\log n + d) \log
    \log_{1+ε} R
    \log \log \log_{1+ε} R\right)
\end{align*}
time.  For $ε$ reasonably large, this is much better than
naively testing against all points in our database, which takes
$O(nd)$ time (although it does produce an exact result).

\subsection{Hash functions for \texorpdfstring{$\ell_1$}{l1} distance}

Essentially the same approach works for (bounded) $\ell_1$ distance,
using \concept{discretization}, where we replace a continuous variable
over some range with a discrete variable.  Suppose we are working in
$[0,1]^d$ with the $\ell_1$ metric.  Represent each coordinate $x_i$
as a sequence of $d/ε$ values $x_{ij}$ in \concept{unary}, for
$j=1\dots ε d$, 
with $x_{ij} = 1$ if $ε j/d < x_i$.  Then the Hamming distance
between the bit-vectors representing $x$ and $y$ is proportional to
the $\ell_1$ distance between the original vectors, plus an error term
that is bounded by $ε$.  We can then use the hash functions for
Hamming distance to get a locality-sensitive hash family.

A nice bit about this construction is that we don't actually have to
build the bit-vectors; instead, we can specify a coordinate $x_i$ and a
threshold $c$ and get the same effect by recording whether $x_i > c$
or not.  

Note that this does increase the cost slightly: we are
converting $d$-dimensional vectors into $(d/ε)$-long bit
vectors, so the $\log (n+d)$ term becomes $\log(n+d/ε)$.  When
$n$ is small, this effectively multiples the cost of a query
by an extra $\log (1/ε)$.  More significant is that we have to
cut
$ε$ in half to obtain the same error bounds, because we now
pay $ε$ error for the data structure itself and an additional
$ε$ error for the discretization.  So our revised
cost for the $\ell_1$ case is
\begin{align*}
    O\left(n^{1/(1+ε/2)} \log (1/δ) (\log n + d/ε) \log
    \log_{1+ε/2} R
    \log \log \log_{1+ε/2} R\right).
\end{align*}

\myChapter{Dimension reduction}{2025}{}
\label{chapter-dimension-reduction}

In this chapter, we will discuss how randomization can be used to
reduce the dimension of a set of points in a way that approximately
preserves the distance between points. The main tool for doing this is
a family of closely-related results that collectively are known as the
Johnson-Lindenstrauss lemma.

We can't really do full justice to the Johnson-Lindenstrauss lemma and its
applications here, although we will try to hit the high points. There
is an excellent survey by Freksen~\cite{Freksen2021} if you'd like to learn
more.

\section{The Johnson-Lindenstrauss lemma}
\label{section-Johnson-Lindenstrauss-lemma}

The \index{lemma!Johnson-Lindenstrauss}\concept{Johnson-Lindenstrauss
lemma}~\cite{JohnsonL1984} says that it is possible to project a set of $n$ vectors in a
space of arbitrarily high dimension onto an $O(\log n)$-dimensional
subspace, such that the distances between the vectors are
approximately preserved.  There are several versions of the lemma,
depending on how the projection is done, but in each case we want to
show that for sufficiently large $k$ as a function of $n$ there is
some $k×d$ matrix $A$ such that for any two $d$-dimensional vectors
$u$ and $v$ in our set,
$(1-ε)\norm{u-v}^2 ≤ \norm{Au-Av}^2 ≤ (1+ε)\norm{u-v}^2$.  Typical
choices for $A$ are:
\begin{itemize}
    \item A projection matrix for a uniform random $k$-dimensional
        subspace. (Original Johnson-Lindenstrauss
        paper~\cite{JohnsonL1984}, also used in the
        Dasgupta-Gupta~\cite{DasguptaG2003} proof described below in
        §\ref{section-JLT-proof}.)
    \item A matrix whose elements are i.i.d.~normal random variables.
        (Indyk-Motwani~\cite{IndykM1998}.)
    \item A matrix whose elements are i.i.d.~variables with a
        particular distribution on $\Set{-1,0,+1}$.
        (Achlioptas~\cite{Achlioptas2003}.)
    \item A matrix whose elements are $±1$
        coin-flips~\cite{Achlioptas2003}. This is perhaps
        the easiest to keep track of, but the constants in the error
        bounds are a bit worse than the $\Set{-1,0,+1}$ version.
\end{itemize}

In each case, we multiply the matrix by an appropriate fixed scale
factor $c$ so that $\Exp{\norm{cAu}^2} = \norm{u}^2$. The resulting
linear projection is known as a
\concept{Johnson-Lindenstrauss
transformation}\index{transformation!Johnson-Lindenstrauss}
(\concept{JLT} for short).

\subsection{Reduction to single-vector case}

Most proofs of the theorem reduce to the case of a single vector, by
finding a value of $k$ such that
$\Prob{(1-ε)\norm{u}^2 ≤ \norm{Au}^2 ≤ (1+ε)\norm{u}^2} ≥ 1-2/n^2$,
and then using the union bound to show that this same property holds
for all $\binom{n}{2}$ vectors $u-v$ with nonzero probability.
This shows the existence of a good matrix, and we can generate matrices and test them until we find one that actually works.

\subsection{A relatively simple proof of the lemma}
\label{section-JLT-proof}

This proof is due to Dasgupta and Gupta~\cite{DasguptaG2003}, and is
somewhat simpler than many proofs of the theorem. The basic idea is
that if $A$ projects onto a uniform random $k$-dimensional subspace,
then its effect on the length of an arbitrary nonzero vector $u$ is
the same as its effect on the length of a unit vector $u/\norm{u}$,
which is the same as the effect of some specific fixed $k×d$ matrix $B$
applied to a random unit vector $Y$ drawn uniformly from the surface
of the $d$-dimensional sphere $S^d$. An easy choice for $B$ is just
the matrix that extracts the first $k$ coordinates from $Y$.

So how do we get a random unit vector $Y$? The usual trick is to use
the fact that the multivariate normal distribution is radially
symmetric: if we generate $d$ independent normally-distributed random
variables $X_1,\dots,X_d$, then the vector $Y = X/\norm{X}$ is
uniformly distributed over $S^d$.\footnote{Radial symmetry is
immediate from the density $\frac{1}{√{2π}} e^{-x^2/2}$ of
the univariate normal distribution. If we consider a vector
$\Tuple{X_1, \dots, X_d}$ of independent $N(0,1)$ random variables,
then the joint density is given by the product $∏_{i=1}^{d}
\frac{1}{2π} e^{-x_i^2/2} = (2π)^{-d/2} e^{-∑ x_i^2 / 2}$. But $∑
x_i^2 = r^2$ where $r$ is the distance from the origin, meaning this
distribution has the same density at all points at the same distance.}

This gives
$Z = BY = (X_1,\dots,X_k)/\norm{X}$, and 
$\norm{Z}^2 = (X_1^2 + X_2^2 + \dots + X_k^2) / (X_1^2 + \dots + X_d^2)$.
If each value $X_i^2$ were exactly equal to its expectation $1$ (the
variance of a standard $N(0,1)$ normal random variable), then 
$\norm{Z}^2$ would be exactly equal to $k/d$, and we could
adjust the length of $Z$ to equal $1$
by multiplying $Z$ by a correction factor of $√{d/k}$.
But we generally won't have each $X_i^2$ equal to $1$;
instead, we will use a Chernoff bound argument to show that
$\norm{Z}^2$ lies between $(1-ε)(k/d)$ and $(1+ε)(k/d)$,
for suitable $ε$ and $k$,
with high probability.

Following the approach in~\cite{DasguptaG2003}, we'll write $β$ for
$(1-ε)$ or $(1+ε)$ as appropriate.
Starting with the lower bound, we want to show that for $β=1-ε$, it is unlikely that
\begin{align*}
    \norm{Z}^2 &≤ β(k/d)
\intertext{which expands to}
    \frac{∑_{i=1}^k X_i^2}{∑_{i=1}^{d} X_i^2} &≤ β(k/d).
\end{align*}

Having a ratio between two sums is a nuisance, but we can multiply out
the denominators to turn it into something we can apply a
Chernoff-style argument to.

\begin{align*}
    \Prob{\frac{∑_{i=1}^k X_i^2}{∑_{i=1}^{d} X_i^2} ≤ β(k/d)}
    &= \Prob{d ∑_{i=1}^k X_i^2 ≤ β k ∑_{i=1}^{d} X_i^2}
    \\&= \Prob{\beta k (X_1^2+\dots+X_d^2) - d(X_1^2+\dots X_k^2) ≥ 0}
\\
&=
    \Prob{\exp\parens*{t\parens*{\beta k (X_1^2+\dots+X_d^2) - d(X_1^2+\dots X_k^2)}} ≥ 1}
\\
&≤
    \Exp{\exp(t(\beta k (X_1^2+\dots+X_d^2) - d(X_1^2+\dots X_k^2)))}
\\
&=
    \Exp{\exp(t \beta k X^2)}^{d-k} \Exp{\exp(t(\beta k - d)X^2)}^k,
\end{align*}
where $t$ can be any value greater than $0$, the shift from
probability to expectation uses Markov's inequality, 
and in the last step we replace each independent occurrence of $X_i$
with a standard normal random variable $X$.

Now we just need to be able to compute the moment generating function
$\Exp{e^{sX^2}}$ for $X^2$. The quick way to do this is to notice
that $X^2$ has a chi-squared distribution with one degree of freedom
(since the chi-squared distribution with $k$ degrees of freedom
is just the distribution of the sum of squares of $k$ independent
normal random variables),
and look up its m.g.f.~$(1-2s)^{-1/2}$ (for $s < 1/2$).

We can substitute this m.g.f.~in the formula above to get
\begin{align*}
    \Prob{(X_1^2+\dots+X_k^2)/(X_1^2+\dots+X_d^2) ≤ \beta(k/d)}
&≤
    \parens*{\frac{1}{1-2\beta k t}}^{(d-k)/2}
    \parens*{\frac{1}{2t(\beta k - d)}}^{k/2}.
\end{align*}

The rest is just the usual trick of finding the minimum value of this
expression over all values $t$ (that don't produce negative denominators in
the m.g.f.!) by differentiating and setting to $0$.  This turns out to
be easiest
if we maximize $1/g(t)^2$ where $g(t)$ is the above expression; see
the Dasgupta-Gupta paper for details. For $β=1-ε<1$, they show that
the optimal $t$ is $\frac{1-β}{2β(d-kβ)}$, which gives, after a few
intermediate steps,
\begin{align*}
    \Prob{\norm{Z}^2 ≤ \beta(k/d)}
&≤ e^{(k/2)(1-\beta+\ln \beta)}.
\end{align*}

The same argument applied to $g(-t)$ gives essentially the same bound
for $β=1+ε>1$:
\begin{align*}
    \Prob{\norm{Z}^2 ≥ β(k/d)}
&≤
    e^{(k/2)(1-β+\ln β)}.
\end{align*}

The $1-β+\ln β$ factors are very close to $0$. For $β = 1-ε$, we have
$1-β+\ln β = ε + \ln (1-ε) = Θ(-ε^2)$, and for $β = 1+ε$, we similarly
have $1-β+\ln β = -ε + \ln(1+ε) = Θ(-ε^2)$. So in either case we need
$k = Θ(ε^{-2} \ln n)$ to make the probability bound polynomially small
in $n$.

A more precise calculation (see~\cite{DasguptaG2003}) includes the
next term in the Taylor series expansion of $1±ε$ to get an exact
inequality. This gives a
good map at $k ≥ 4(ε^2/2 - ε^3/3)^{-1}\ln n$.
Note that both our crummy asymptotic bound and this more precise bound
depend on $n$ but not $d$.

Constructing a projection matrix for a random $k$-dimensional space is
mildly painful. The easiest way to do it may be to generate a random
$k×d$ matrix of independent $N(0,1)$ variables and then apply
Gram-Schmidt orthogonalization, which takes $O(k^2 d)$ time.  A faster
approach that gives similar results
is to use a matrix of independent random variables that are
$±1$ with probability $1/6$ each and $0$ the rest of the time.
This doesn't produce exactly the same distribution on projections, but
it can be shown (with
much more effort) to still work pretty well~\cite{Achlioptas2003}.

If we leave out enough details, we can summarize all of these results
as a single lemma:
\begin{lemma}[\cite{JohnsonL1984}]
    \label{lemma-Johnson-Lindenstrauss}
    For every set of $n$ points $X$ in $ℝ^d$, and any $ε$ with $0 < ε
    < 1$, there is a linear
    projection $f:ℝ^d→ℝ^k$ with $k = O(ε^{-2} \log n)$ such that for any
    $u$ and $v$ in $X$,
    \begin{equation*}
        (1-ε) \norm{u-v}^2 ≤ \norm{f(u)-f(v)} ≤ (1+ε) \norm{u-v}.
    \end{equation*}
\end{lemma}

\subsection{Distributional version}

It's worth nothing that the argument for
Lemma~\ref{lemma-Johnson-Lindenstrauss} doesn't use any property of $X$
except its size. This means that the same argument works (with nonzero
probability) without needing to know $X$. This gives the
``distributional'' version of the lemma, which says

\begin{lemma}[\cite{JohnsonL1984}]
    \label{lemma-Johnson-Lindenstrauss-distributional}
    For every $d$, $0 < ε < 1$, and $0 < δ < 1$, there exists a
    distribution over linear functions $f:ℝ^d → R^k$ with $k = O(ε^{-2}
    \log (1/δ))$ such that for every $x ∈ R^d$,
    \begin{equation*}
        \Prob{(1-ε)\norm{x}^2 ≤ \norm{f(x)}^2 ≤ (1+ε)\norm{x}^2} ≥ 1-δ.
    \end{equation*}
\end{lemma}

This can be handy for applications where we don't know the vectors we
will be working with in advance.

\section{Applications}

The intuition is that the Johnson-Lindenstrauss lemma lets us reduce
the dimension of some problem involving distances between $n$ points,
where we are willing to tolerate a small constant relative error, from
some arbitrary $d$ to $O(\log n)$ (or $O(\log (1/δ))$ if we just care
about the error probability per pair of points). So applications tend
to fall into one of two categories:
\begin{enumerate}
    \item We want to run some algorithm on a set of points in a
        high-dimensional space, but the
        cost of the algorithm is high (perhaps exponential!) as a
        function of the dimension.
    \item We want to run some algorithm on a set of points in a
        high-dimensional space, but we don't want to pay the
        linear-in-the-dimension space costs.
\end{enumerate}

If we are lucky, we'll get both payoffs, winning on both time and space.

For example, suppose we have a set of $n$ points $x_1, x_2, \dots,
x_n$ representing the centers of various clusters in a $d$-dimensional
space, and we want to rapidly classify incoming points $y$ into one of
these clusters by finding which $x_i$ $y$ is closest to. If we do this
naively, this is a $Θ(nd)$ operation, since it takes $Θ(d)$ time to
compute each distance between $y$ and some $x_i$. If instead we are
willing to accept the inaccuracy associated with
Johnson-Lindenstrauss, we can fix a matrix $A$ in advance, replace
each $x_i$ with $Ax_i$, and find the $x_i$ that is (approximately)
closest to $y$ using $O(d \log n)$ time to reduce $y$ to $Ay$ and $O(n
\log n)$ time to compute the distance between $Ay$ and each $Ax_i$. In
this case we are reducing both the time complexity of our
classification algorithm (at least if we don't count the
pre-processing time to generate the $Ax_i$) and the amount of data we
need to store.

An example of space savings is the use of the Johnson-Lindenstrauss
transform  in
streaming algorithms (see
§\ref{section-data-stream-computation}).
Freksen~\cite{Freksen2021} gives a simple example of estimating
$\norm{x}^2$ where $x$ is a vector of counts of items from a set of
size $n$ presented one at
a time. If we don't charge for the space to store the JLT function $f$, we
can simply add the $i$-th column of the matrix to our running total whenever we see item $i$,
and we need only store $O\parens*{ε^{-2} \log (1/δ)}$ distinct
numerical values of an appropriate precision to estimate $\norm{x}^2$
to within $ε$ relative error with probability at least $1-δ$. The
problem is that in reality we do need to represent $f$ somehow, and
even for a $±1$ matrix this will take $Θ(n ε^{-2} \log (1/δ))$ space. 
Fortunately it can be shown that generating $f$ using a
$4$-independent hash function reduces the space for $f$ to $O(\log
n)$, giving the \concept{Tug-of-War sketch}\index{sketch!Tug-of-War}
of Alon~\etal~\cite{AlonMS1996}, one of the first compact streaming
data structures.
Though this is a nice application of the JLT,
it's worth mentioning Cormode and Muthukrishnan~\cite{CormodeM2005} 
observe that this is still significantly more costly for most queries than their
own count-min sketch.

\myChapter{Martingales and stopping times}{2025}{}
\label{chapter-stopping-times}

In §\ref{section-Azumas-inequality}, we used martingales 
to show that the outcome of some process was tightly concentrated.
Here we will show how martingales interact with 
\index{stopping time}\index{time!stopping}\conceptFormat{stopping times}, which
are random variables that control when we stop carrying out some task.
This will
require a few new definitions.

\section{Definitions}
\label{section-stopping-times-definitions}

The general form of a \index{martingale}martingale
$\Set{X_t,ℱ_t}$ consists of:
\begin{itemize}
\item A sequence of random variables $X_0, X_1, X_2, \dots$; and
\item A \concept{filtration} $ℱ_0 ⊆ ℱ_1
⊆ ℱ_2 \dots$, where each $σ$-algebra
$ℱ_t$ represents our knowledge at time $t$;
\end{itemize}
subject to the requirements that:
\begin{enumerate}
\item The sequence of random variables is \indexConcept{adapted
sequence of random variables}{adapted} to the filtration, which just
means that each $X_t$ is $ℱ_t$-measurable or equivalently
that $ℱ_t$ (and thus all subsequent $ℱ_{t'}$ for
$t' ≥ t$) includes all knowledge of $X_t$; and
\item The \concept{martingale property} 
    \begin{align}
        \ExpCond{X_{t+1}}{ℱ_t}
        &= X_t
        \label{eq-martingale-property-general}
    \end{align}
holds for all $t$.
\end{enumerate}

We will also
also need the following definition of a \index{time!stopping}\concept{stopping time}.
Given a filtration $\Set{ℱ_t}$,
a random variable $τ$ is a
stopping time for $\Set{ℱ_{t}}$ if $τ ∈ ℕ \cup
\Set{∞}$ and the
event $[τ≤t]$ is $ℱ_{t}$-measurable for all
$t∈ℕ$.\footnote{Different authors impose
    different conditions on the range of $τ$; for example,
    Mitzenmacher and Upfal~\cite{MitzenmacherU2017} exclude the case $τ = ∞$. 
    We allow $τ = ∞$ to represent
    the outcome where we never stop.
    This can be handy for modeling processes where this outcome is
    possible, although in practice
    we will typically insist
that it occurs only with probability zero.} In
simple terms, $τ$ is a stopping time if you know at time $t$
whether to stop there or not.

What we like about martingales is that iterating the martingale
property shows that $\Exp{X_t} = \Exp{X_0}$ for all fixed $t$.  We will
show that, under reasonable conditions, the same holds for $X_τ$ when
$τ$ is a stopping time.  (The random variable $X_τ$ is defined in
the obvious way, as a random variable that takes on the value of $X_t$
when $τ = t$.)

\section{Submartingales and supermartingales}
\label{section-submartingales-and-supermartingales}

In some cases we have a process where instead of getting equality in 
\eqref{eq-martingale-property-general}, we get an inequality instead.
A \concept{submartingale} replaces
\eqref{eq-martingale-property-general} with
\begin{align}
    X_t &≤ \ExpCond{X_{t+1}}{ℱ_t}
    \label{eq-submartingale-property-general}
    \intertext{while a \concept{supermartingale} satisfies}
    X_t &≥ \ExpCond{X_{t+1}}{ℱ_t}.
    \label{eq-supermartingale-property-general}
\end{align}

In each case, what is ``sub'' or ``super'' is the value at the
current time compared to the expected value at the next time.
Intuitively, a submartingale corresponds to a process where you
win on average, while a supermartingale is a process where you
lose on average.  Casino games (in profitable casinos) 
are submartingales for the house
and supermartingales for the player.

Sub- and supermartingales can be reduced to martingales by subtracting
off the expected change at each step.  For example, if
$\Set{X_t}$ is a submartingale with respect to $\Set{ℱ_t}$,
then the process $\Set{Y_t}$ defined recursively by
\begin{align*}
    Y_0 &= X_0 \\
Y_{t+1} &= Y_t + X_{t+1} - \ExpCond{X_{t+1}}{ℱ_t}
\end{align*}
is a martingale, since
\begin{align*}
    \ExpCond{Y_{t+1}}{ℱ_t}
    &= \ExpCond{Y_t + X_{t+1} -
\ExpCond{X_{t+1}}{ℱ_t}}{ℱ_t}
\\&= Y_t + \ExpCond{X_{t+1}}{ℱ_t} - \ExpCond{X_{t+1}}{ℱ_t} 
\\&= Y_t.
\end{align*}

One way to think of this is that $Y_t = X_t + Δ_t$, where
$Δ_t$ is a predictable, non-decreasing 
\concept{drift process} that starts at $0$.  For supermartingales, the same result
holds, but now $Δ_t$ is non-increasing.
This ability to decompose an adapted stochastic process into the sum of a
martingale and a predictable drift process is known as the
\index{theorem!Doob decomposition}\concept{Doob decomposition
theorem}.

\section{The optional stopping theorem}
\label{section-optional-stopping-theorem}

If $(X_t, ℱ_t)$ is a martingale, then applying induction to
the martingale property shows that $\Exp{X_t} = \Exp{X_0}$ for any
fixed time $t$.
The
\index{theorem!optional stopping}
\concept{optional stopping theorem}
shows that this also happens for $X_τ$ when $τ$ is a stopping time,
under various choices of additional conditions:
\begin{theorem}
\label{theorem-optional-stopping}
Let $(X_{t},ℱ_{t})$ be a martingale and $τ$ a stopping
time for $\Set{ℱ_{t}}$.  Then $\Exp{X_{τ}} = E[X_{0}]$ if at
least one of the following conditions holds:
\begin{enumerate}
    \item \textbf{Bounded time.}  There is a fixed $n$ such that $τ≤n$
        always.
    \item \textbf{Finite time and bounded range.}  
        $\Prob{τ<∞} = 1$, and there is a fixed $M$ such that for all $t ≤ τ$, 
        $\abs*{X_t} ≤ M$.
    \item \textbf{Finite expected time and bounded increments.}
        $\Exp{τ} < ∞$, and
        there is a fixed $c$ such that $\abs*{X_{t+1}-X_t} ≤ c$
        for all $t < τ$.
    \item \textbf{General case.}  All three of the following
        conditions hold:
        \begin{enumerate}
          \item\label{item-optional-stopping-finite-time} $\Prob{τ<∞} = 1$,
          \item\label{item-optional-stopping-finite-expectation}
              $\Exp{\abs*{X_{τ}}} < ∞$, and
          \item\label{item-optional-stopping-funky-condition}
              $\lim_{t→∞} \Exp{ X_{t}⋅1_{[τ >
              t]} } = 0$.
        \end{enumerate}
\end{enumerate}
\end{theorem}

It would be nice if we could show $\Exp{X_{τ}} = E[X_{0}]$ without the
side conditions, but in general this isn't true.  For example, the
double-after-losing martingale strategy in the St.~Petersburg paradox
(see §\ref{section-st-petersburg-paradox}) eventually yields $+1$ with
probability $1$, so if $τ$ is the time we stop playing, we have
$\Prob{τ<∞} = 1$, $\Exp{\abs*{X_{τ}}} < ∞$, but $\Exp{X_{τ}} =
1 ≠ \Exp{X_{0}} = 0$.  
To make this happen, we have to violate all of bounded
time ($τ$ is not bounded), bounded range ($\abs*{X_t}$ roughly doubles every step
until we stop), bounded increments ($\abs*{X_{t+1}-X_t}$ doubles
every step as well), and at least one of the three conditions of the general
case (the last one: $\lim_{t→∞} \Exp{X_t}⋅1_{[τ>t]} = -1 ≠ 0$).

To prove Theorem~\ref{theorem-optional-stopping}, we'll use a
truncation argument.
The intuition is that for any fixed $n$, we can
\index{truncation}\conceptFormat{truncate} $X_τ$ to $X_{\min(τ,n)}$
and show that $\Exp{X_{\min(τ,n)}} = \Exp{X_0}$.  This will immediately
give the bounded time case.  For the other cases, the argument is
that $\lim_{n→∞} \Exp{X_{\min(τ,n)}}$ converges to $\Exp{X_τ}$
provided the missing part $\Exp{X_τ - X_{\min(τ,n)}}$ converges to zero.
How we do this depends on which assumptions we are making.

We'll start by formalizing the core truncation argument:
\begin{lemma}
\label{lemma-optional-stopping-easy}
Let $(X_{t},ℱ_{t})$ be a martingale and $τ$ a stopping
time for $\Set{ℱ_{t}}$.  Then for any $n∈ℕ$,
$\Exp{X_{\min(τ,n)}} = E[X_{0}]$.
\end{lemma}
\begin{proof}
    Define $Y_{t} = X_{0} + ∑_{i=1}^{t} (X_{t}-X_{t-1}) 1_{[τ > t-1]}$.  
Then $(Y_{t},ℱ_{t})$ is a martingale, because we can
calculate
$\ExpCond{Y_{t+1}}{ℱ_t} = \ExpCond{Y_t + (X_{t+1}-X_t)1_{[τ >
t]} }{
ℱ_t} = Y_t + 1_{[τ > t]} ⋅ \ExpCond{X_{t+1}-X_t}{
ℱ_t} = Y_t$; effectively, we are treating
$1_{[τ≤t-1]}$ as a sequence of bets, and we know that adjusting our
bets doesn't change the martingale property.  But then $\Exp{X_{\min(τ,n)}} =
\Exp{Y_{n}} = \Exp{Y_{0}} = \Exp{X_{0}}$.
\end{proof}

As claimed, this gives us the bounded-time variant for free. 
If $τ≤n$ always, then $X_{τ} = X_{\min(τ,n)}$, and
$\Exp{X_τ} = \Exp{X_{\min(τ,n)}} = \Exp{X_0}$.

For each of the unbounded-time variants, we will apply some version of the
following strategy:
\begin{enumerate}
    \item Observe that since $\Exp{X_{\min(τ,n)}} = \Exp{X_0}$ is a
        constant for any fixed $n$, $\lim_{n→∞} \Exp{X_{\min(τ,n)}}$
        converges to $\Exp{X_0}$.
    \item Argue using whatever assumptions we are making that
        $\lim_{n→∞} \Exp{X_{\min(τ,n)}}$ also converges to $\Exp{X_τ}$.
    \item Conclude that $\Exp{X_0} = \Exp{X_τ}$, since they are both
        limits of the same sequence.
\end{enumerate}

For the middle step, start with
\begin{equation*}
    \label{eq-optional-stopping-tail}
    X_τ = X_{\min(τ,n)} + 1_{[τ>n]}(X_τ-X_n).
\end{equation*}
This holds because either
$τ≤n$, and we just get $X_{τ}$, or $τ>n$, and we get
$X_{n}+(X_{τ}-X_{n}) = X_{τ}$.

Taking the expectation of both sides gives
\begin{align*}
    \Exp{X_τ} 
    &= \Exp{X_{\min(τ,n)}} + \Exp{1_{[τ>n]}(X_τ-X_n)}
    \\&= \Exp{X_0} + \Exp{1_{[τ>n]}(X_τ-X_n)}.
\end{align*}
So if we can show that the right-hand term goes to zero in the limit,
we are done.

For the bounded-range case, we have $\abs*{X_τ - X_n} ≤ 2M$, so
$\abs*{\Exp{1_{[τ>n]}}(X_τ-X_n)} ≤ 2M ⋅ \Prob{τ>n}$.  Since in this
case we assume $\Prob{τ<∞} = 1$, $\lim_{n→∞} \Prob{τ>n} = 0$, and the
theorem holds.

For bounded increments, we have
\begin{align*}
    \abs*{\Exp{(X_{τ}-X_{n})1_{[τ>n]}}}
    &= \abs*{\Exp{∑_{t≥n} (X_{t+1}-X_{t}) 1_{[τ>t]}}} 
  \\&≤ \Exp{∑_{t≥n} \abs*{(X_{t+1}-X_{t})} ⋅ 1_{[τ>t]}}
  \\&≤ \Exp{∑_{t≥n} c ⋅ 1_{[τ>t]}}
  \\&≤ c \Exp{∑_{t≥n} 1_{[τ>t]}}.
\end{align*}
But $\Exp{τ} = ∑_{t=0}^{∞} 1_{[τ>t]}$.  Under the assumption that this
sequence converges, its tail goes to zero, and again the theorem
holds.

For the general case,
we can expand
\begin{align*}
\Exp{X_{τ}}
    &= \Exp{X_{\min(τ,n)}} + \Exp{1_{[τ>n]}X_{τ}} - \Exp{1_{[τ>n]}X_{n}}
    \intertext{which implies}
    \lim_{n→∞} \Exp{X_{τ}}
    &= \lim_{n→∞}\Exp{X_{\min(τ,n)}} + \lim_{n→∞}\Exp{1_{[τ>n]}X_{τ}}
    - \lim_{n→∞}\Exp{1_{[τ>n]}X_{n}},
\end{align*}
assuming all these limits exist and are finite.
We've already established that the first limit is $\Exp{X_0}$,
which is exactly what we want.
So we just need to show that the other
two limits both converge to zero.
For the last limit, 
we just use condition
(\ref{item-optional-stopping-funky-condition}), which 
gives $\lim_{n→∞}
\Exp{1_{[τ>n]}X_{n}}=0$; no further argument is needed.  But we still
need to show that the middle limit also vanishes.

Here we use condition
(\ref{item-optional-stopping-finite-expectation}).  Observe that
$\Exp{1_{[τ>n]} X_{τ}} = ∑_{t=n+1}^{∞}
\Exp{1_{[τ=t]} X_{t}}$.  Compare
this with $\Exp{X_{τ}} = ∑_{t=0}^{∞} \Exp{1_{[τ=t]} X_{t}}$;
this is an absolutely
convergent series (this is why we need
condition~(\ref{item-optional-stopping-finite-expectation})), so in the limit the sum of the terms for $i=0\dots
n$ converges to $\Exp{X_{τ}}$.  But this means that the sum of the
remaining terms for $i=n+1\dots ∞$ converges to zero.  So the
middle term goes to zero as $n$ goes to infinity.
This completes the proof.

\section{Applications}
\label{section-optional-stopping-applications}

Here we give some example of the Optional Stopping Theorem in action.
In each case, the trick is to find an appropriate martingale and
stopping time, and let the theorem do all the work.

\subsection{Random walks}
\label{section-stopping-times-and-random-walks}

Let $X_{t}$ be an \index{random walk!unbiased}\indexConcept{unbiased random walk}{unbiased}
$±1$ \concept{random walk} that starts at $0$, adds $\pm 1$ to its
current position with equal probability at each step, and
stops if it reaches $-a$ or $+b$.\footnote{This is called a
\indexConcept{random walk!with two absorbing barriers}{random walk with two absorbing barriers}.}  We'd like to calculate the
probability of reaching $+b$ before $-a$.
Let $τ$ be the time at which the
process stops.

We can easily show that $\Prob{τ<∞} = 1$ and
$\Exp{τ} < ∞$ by observing that from any state of the random walk,
there is a probability of at least $2^{-(a+b)}$ that it stops within
$a+b$ steps (by flipping heads $a+b$ times in a row), so that if we
consider a sequence of intervals of length $a+b$, the expected number
of such intervals we can have before we stop is at most $2^{a+b}$,
giving $\Exp{τ} ≤ (a+b) 2^{a+b}$ (we can do better than this).

We also
have bounded increments by the definition of the process (bounded
range also works, at least up until time $τ$).  So $\Exp{X_{τ}} = E[X_{0}] = 0$ and the probability
$p$ of landing on $+b$ instead of $-a$ must satisfy $p b - (1-p)a = 0$,
giving $p = \frac{a}{a+b}$.

Now suppose we want to find $\Exp{τ}$.
Let $Y_{t} = X_{t}^{2}-t$.
Then $Y_{t+1} = (X_t \pm 1)^2 - (t+1)
= X_t^2 \pm 2X_t + 1 - (t+1)
= (X_t^2 - t) \pm 2X_t = Y_t \pm 2X_t$.
Since the plus and minus cases are equally likely,
they cancel out in expectation and
$\ExpCond{Y_{t+1}}{ℱ_t} = Y_t$: we just showed $Y_t$ is a
martingale.\footnote{This construction generalizes in a nice way to
arbitrary martingales.
Suppose $\Set{X_t}$ is a martingale with respect to $\Set{ℱ_t}$.
Let $Δ_t = X_t - X_{t-1}$,
and let $V_t = \VarCond{Δ_t}{ℱ_{t-1}}$ be the conditional variance of
the $t$-th increment (note that this is a random variable that may depend on
previous outcomes).  We can easily show that $Y_t
= X_t^2 - ∑_{i=1}^{t} V_i$ is a martingale.  The proof is that
\begin{align*}
\ExpCond{Y_t}{ℱ_{t-1}} 
&= \ExpCond{X_t^2 - ∑_{i=1}^{t} V_t }{ ℱ_{t-1}}
\\
&= \ExpCond{(X_{t-1} + Δ_t)^2}{ℱ_{t-1}} - ∑_{i=1}^{t} V_i
\\
&= \ExpCond{X_{t-1}^2 + 2X_{t-1}Δ_t + Δ_t^2}{ℱ_{t-1}} - ∑_{i=1}^{t} V_i
\\
&=
 X_{t-1}^2 
 + 2X_{t-1}\ExpCond{Δ_t}{ℱ_{t-1}}
 + \ExpCond{Δ_t^2}{ℱ_{t-1}}
 - ∑_{i=1}^{t} V_i
\\
&=
 X_{t-1}^2 
 + 0
 + V_t
 - ∑_{i=1}^{t} V_i
\\
&=
 X_{t-1}^2 
 - ∑_{i=1}^{t-1} V_i
\\
&= Y_{t-1}.
\end{align*}
For the $\pm 1$ random walk case, we have $V_t = 1$ always, giving
$∑_{i=1}^t V_i = t$ and $\Exp{X_τ^2} = \Exp{X_0^2} + \Exp{τ}$ when
$τ$ is a stopping time satisfying the conditions of the Optional
Stopping Theorem.  
For the general case, the same argument gives $\Exp{X_τ^2} = \Exp{X_0^2} +
\Exp{∑_{t=1}^{τ} V_t}$ instead: the expected square
position of $X_t$ is incremented by the conditional variance at each
step.}
We can also show it has bounded increments (at least up until
time $τ$), because $\abs*{Y_{t+1}-Y_t} = 2\abs*{X_t} ≤ \max(a,b)$.

From Theorem~\ref{theorem-optional-stopping},
$\Exp{Y_{τ}} = 0$, which gives $\Exp{τ} =
\Exp{X_{τ}^{2}}$.  But we can calculate $\Exp{X_{τ}^{2}}$: it is $a^{2}
\Prob{X_{τ} = -a} + b^{2} \Prob{X_{t} = b} = a^{2}(b/(a+b)) +
b^{2}(a/(a+b)) = (a^{2}b+b^{2}a)/(a+b) = ab$.

\subsubsection{Random walk with one absorbing barrier}

If we have a random walk that only stops at $+b$,\footnote{This would
be a \indexConcept{random walk!with one absorbing barrier}{random walk with one absorbing barrier}.} then if $τ$ is
the first time at which $X_τ = b$, $τ$ is a stopping time.
However, in this case $\Exp{X_τ} = b ≠ \Exp{X_0} = 0$.  So the
optional stopping theorem doesn't apply in this case.  But we have
bounded increments, so
Theorem~\ref{theorem-optional-stopping}
would apply if $\Exp{τ} < ∞$.  It follows that the expected time until
we reach $b$ is unbounded, either because sometimes we never reach
$b$, or because we always reach $b$ but sometimes it takes a very long
time.  \footnote{In fact, we always reach $b$.  An easy way to see this
is to imagine a sequence of intervals of length $n_1, n_2, \dots,$
where $n_{i+1} = \left(b + ∑_{j=1}^{i} n_j\right)^2$.  At the end of
the $i$-th interval, we are no lower than $-∑_{j=0}^{i} n_j$, so we
only need to go up $√{n_{i+1}}$ positions to reach $a$ by the end of
the $(i+1)$-th interval.  Since this is just one standard deviation,
it occurs with constant probability, so after a finite expected number
of intervals, we will reach $+a$.  Since there are infinitely many
intervals, we reach $+a$ with probability $1$.}

\subsubsection{Random walk with one absorbing and one reflecting
barrier}

Another variant is a process that stops at $+b$, but never drops below
$0$; any step from $0$ goes to $1$. This is called a
\indexConcept{random walk!with one absorbing and one reflecting
barrier}{random walk with one absorbing and one reflecting barrier}.
The reflecting barrier means that this process is no longer a
martingale. However, it is still possible to analyze this random walk
by reducing its behavior to a random walk with absorbing barriers at
$±b$ using the \concept{reflection principle}.

Let $Z_t$ be a random walk with absorbing barriers at $±b$. Then $X_t
= \abs{Z_t}$ is a random walk with an absorbing barrier at $+b$ and a
reflecting barrier at $0$. Starting at $Z_t = 0$, the $Z$ process
reaches $±b$ in $b^2$ steps on average. So this also gives the
expected time for the $X$ process to reach $+b$.

\subsubsection{Biased random walks}
\label{section-martingales-for-biased-random-walks}

We can also consider a \index{random walk!biased}\concept{biased
random walk} where $+1$ occurs with probability $p$ and $-1$
with probability $q = 1-p$.  If $X_t$ is the position of the random
walk at time $t$, and $ℱ_t$ is the associated $σ$-algebra, then $X_t$
isn't a martingale with respect to $ℱ_t$. But there are at least two
ways to turn it into one:
\begin{enumerate}
    \item Define $Y_t = X_t - (p-q)t$. Then 
        \begin{align*}
            \ExpCond{Y_{t+1}}{ℱ_t}
            &= \ExpCond{X_{t+1} - (p-q)(t+1)}{ℱ_t} 
            \\&= p\parens*{X_t + 1}{ℱ_t} + q\parens*{X_t - 1}{ℱ_t} -
            (p-q)(t+1)
            \\&= (p+q)X_t + (p-q) - (p-q)(t+1)
            \\&= X_t - (p-q) t
            \\&= Y_t,
        \end{align*}
        and $Y_t$ is a martingale with respect to $ℱ_t$.
    \item Define $Z_t = (q/p)^{X_t}$. Then
        \begin{align*}
            \ExpCond{Z_{t+1}}{ℱ_t}
            &= p (q/p)^{X_t+1} + q (q/p)^{X_t-1}
            \\&= (q/p)^{X_t} \parens*{p (q/p) + q (p/q)}
            \\&= (q/p)^{X_t} (q + p)
            \\&= (q/p)^{X_t}
            \\&= Z_t.
        \end{align*}
        Again we have a martingale with respect to $ℱ_t$.
\end{enumerate}

Now let's see what we can do with the Optional Stopping Theorem.

\begin{enumerate}
    \item Suppose we start with $X_0 = 0$ and we want to know the
        probability $p_b$ that we will reach $+b$ before we reach
        $-a$.

        Let $τ$ be the first time at which $X_τ ∈ \Set{-a,b}$.
        We can use the same argument as in the unbiased case to show 
        that $\Prob{τ < ∞} = 1$ and $\Exp{τ} < ∞$, because from any
        position $X_t$ there is at least a $p^{a+b} > 0$ chance that
        the next $a+b$ steps will all be $+1$ and we will reach $b$.
        Since we can flip this $p^{a+b}$-probability coin independently
        every $a+b$ steps, eventually we reach $b$ if we haven't
        already reached $-a$. The expected time is bounded by $(a+b)
        / p^{a+b}$.

        We also have that $0 < Z_t ≤ \max\parens*{(q/p)^a,(q/p)^b}$ for all
        $t≤τ$. This gives us bounded range, so the
        finite-time/bounded-range case of OST applies.

        We thus have $\Exp{Z_τ} = p_b (q/p)^b + (1-p_b) (q/p)^{-a} =
        \Exp{Z_0} = (q/p)^0 = 1$. Solving for $p_b$ gives
        \begin{align}
            p_b
            &= \frac{1-(q/p)^{-a}}{(q/p)^b - (q/p)^{-a}}.
            \label{eq-biased-random-walk-with-two-absorbing-barriers}
        \end{align}

        (Note that this only makes sense if $q ≠ p$.)

        As a test, if $-a = -1$ and $+b = +1$, then we get
        \begin{align*}
            p_b 
            &= \frac{1-p/q}{q/p - p/q}
            \\&= \frac{pq-p^2}{q-p}
            \\&= \frac{p(q-p)}{q-p}
            \\&= p,
        \end{align*}
        which is what we would expect since we hit $-a$ or $+b$ on the
        first step.

        A more interesting case is if we set $p<1/2$. Then $q/p > 1$,
        and
        \begin{align}
            p_b
            &= \frac{1-(q/p)^{-a}}{(q/p)^b - (q/p)^{-a}}
            \nonumber
            \\&= (q/p)^{-b} ⋅ \frac{1 - (q/p)^{-a}}{1-(q/p)^{-a-b}}
            \nonumber
            \\&< (q/p)^{-b}.
            \label{eq-biased-random-walk-with-two-absorbing-barriers-clean-bound}
        \end{align}

        Any walk
        that is biased against us will be an exponentially improbable hill
        to climb.

    \item Now suppose we want to know $\Exp{τ}$, the average time at
        which we first hit $a$ or $b$. We already argued $\Exp{τ}$ is
        finite, and it's easy to see that
        $\Set{Y_t}$ has bounded increments, so we
        can use the finite-expected-time/bounded-increment case of OST
        to get $\Exp{Y_τ} = \Exp{Y_0} = 0$, or $\Exp{X_τ - (p-q)τ} =
        0$. It follows that $\Exp{τ} = \Exp{X_τ} / (p-q)$.

        But we can compute $\Exp{X_τ}$, since it is just $p_b b -
        (1-p_b)a$. So $\Exp{τ} = \frac{p_b b - (1-p_b) a}{p-q}$. If $p
        > q$, and $a$ and $b$ are both large enough to make $p_b$ very close
        to $1$, this will be approximately $b/(p-q)$, the time to
        climb to $b$ using our average return of $p-q$ per step.
\end{enumerate}

\subsection{Wald's equation}
\label{section-Walds-equation}

Suppose we run a Las Vegas algorithm until it succeeds, and the $i$-th
attempt costs $X_i$, where all the $X_i$ are independent,
satisfy $0 ≤ X_i ≤ c$ for some $c$, and have a common mean $\Exp{X_i}
= μ$.

Let
$N$ be the number of times we run the algorithm.  Since we can tell
when we are done, $N$ is a stopping time with respect to some
filtration $\Set{ℱ_i}$ to which the $X_i$ are
adapted.\footnote{A stochastic process $\Set{X_t}$ is \concept{adapted}
to a filtration $\Set{ℱ_t}$ if each $X_t$ is $ℱ_t$-measurable.}
Suppose also that $\Exp{N}$ exists.  What
is $\Exp{∑_{i=1}^{N} X_i}$?

If $N$ were not a stopping time, this might be a very messy problem
indeed.  But when $N$ is a stopping time, we can apply it to the
martingale $Y_t = ∑_{i=1}^{t} (X_i - μ)$.  This has bounded
increments ($0 ≤ X_i ≤ c$, so $-c ≤ X_i - \Exp{X_i} ≤ c$), and
we've already said $\Exp{N}$ is finite (which implies $\Prob{N<∞} =
1$), so Theorem~\ref{theorem-optional-stopping}
applies.  We thus have
\begin{align*}
0 &= \Exp{Y_N}
\\
&= \Exp{∑_{i=1}^{N} (X_i - μ)}
\\
&= \Exp{∑_{i=1}^{N} X_i} - \Exp{∑_{i=1}^{N} μ}
\\
&= \Exp{∑_{i=1}^{N} X_i} - \Exp{N} μ.
\end{align*}

Rearranging this gives \concept{Wald's equation}:
\begin{align}
\label{eq-walds-equation}
\Exp{∑_{i=1}^{N} X_i}
&= \Exp{N} μ.
\end{align}

This is the same formula as in §\ref{section-Walds-equation-simple},
but we've eliminated the bound on $N$ and allowed for much more
dependence between $N$ and the $X_i$.\footnote{In fact, looking
closely at the proof reveals that we don't even need the $X_i$ to be
independent of each other.  We just need that $\ExpCond{X_{i+1}}{ℱ_i}
= μ$ for all $i$ to make $(Y_t,ℱ_t)$ a martingale.  But if we don't
carry any information from one iteration of our Las Vegas algorithm to
the next, we'll get independence anyway.  So the big payoff is not
having to worry about whether $N$ has some devious dependence on the
$X_i$.}

\subsection{Maximal inequalities}
\label{section-maximal-inequalities-for-martingales}

Suppose we have a martingale $\Set{X_i}$ with $X_i ≥ 0$ always, and we
want to bound $\max_{i≤n} X_i$. We can do this using the Optional
Stopping Theorem:
\begin{lemma}
    \label{lemma-easy-doobs-inequality}
    Let $\Set{X_i}$ be a martingale with $X_i ≥ 0$. Then for any
    fixed $n$, 
    \begin{equation}
        \label{eq-easy-doobs-inequality}
        \Prob{\max_{i≤n} X_i ≥ α} ≤ \frac{\Exp{X_0}}{α}.
    \end{equation}
\end{lemma}
\begin{proof}
    The idea is to pick a stopping time $τ$ such that $\max_{i≤n} X_i
    ≥ α$ if and only if $X_τ ≥ α$.

    Let $τ$ be the first time such that $X_τ ≥ α$ or $τ ≥ n$. Then $τ$ is
    a stopping time for $\Set{X_i}$, since we can determine from
    $X_0,\dots,X_t$ whether $τ≤t$ or not. We also have that $τ≤n$
    always, which is equivalent to $τ = \min(τ,n)$.
    Finally, $X_τ ≥ α$ means that $\max_{i≤n} X_i ≥ X_τ ≥ α$, and
    conversely if there is some $t≤n$ with $X_t = \max_{i≤n} X_i ≥ α$,
    then $τ$ is the first such $t$, giving $X_τ ≥ α$.

    Lemma~\ref{lemma-optional-stopping-easy} says $\Exp{X_τ} =
    \Exp{X_0}$. So Markov's inequality gives
    \begin{displaymath}
        \Prob{\max_{i≤n} X_i ≥ α}
        = \Prob{X_τ ≥ α}
        ≤ \frac{\Exp{X_τ}}{α}
        = \frac{\Exp{X_0}}{α},
    \end{displaymath}
    as claimed.
\end{proof}

Lemma~\ref{lemma-easy-doobs-inequality} is a special case of
\index{inequality!Doob's martingale}
\concept{Doob's martingale inequality}, which says that for a
non-negative \emph{submartingale} $\Set{X_i}$, 
\begin{equation}
    \label{eq-doobs-inequality}
\Prob{\max_{i≤n} X_i ≥ α} ≤ \frac{\Exp{X_n}}{α}.
\end{equation}

The proof is similar, but requires showing
first that $\Exp{X_τ} ≤ \Exp{X_n}$ when $τ≤n$ is a stopping time and
$\Set{X_i}$ is a submartingale. 

Doob's martingale inequality is what you get if you generalize
Markov's inequality to martingales. The analogous generalization of
Chebyshev's inequality is 
\index{inequality!Kolmogorov's}\concept{Kolmogorov's inequality}, which says:
\begin{lemma}
    \label{lemma-Kolmogorov-inequality}
    For sums $S_i =
    ∑_{j=1}^{i} X_j$ of independent random variables $X_1,X_2,\dots,X_n$
    with $\Exp{X_i} = 0$,
\begin{equation}
    \label{eq-Kolmogorov-inequality}
    \Prob{\max_{i≤n} \abs*{S_i} ≥ α} ≤ \frac{\Var{S}}{α^2}.
\end{equation}
\end{lemma}
\begin{proof}
    Let $Y_i = S_i^2 - \Var{S_i}$. Then $\Set{Y_i}$ is a martingale.
    This implies that $\Exp{Y_n} = \Exp{S_n^2} - \Var{S} = Y_0 = 0$ and
    thus that $\Exp{S_n^2} = \Var{S}$. 
    It's easy to see that $\Set{S_i^2}$ is a submartingale since
    partial sums can only increase over time.
    Now apply \eqref{eq-doobs-inequality}.
\end{proof}

In general, because we can always stop updating a martingale once we
hit a particular threshold, other martingale concentration bounds like
Azuma-Hoeffding will also apply to the maximum or minimum of a
martingale over a given interval.

\subsection{Waiting times for patterns}
\label{section-waiting-times-for-patterns}

Let's suppose we flip coins until we see some pattern 
appear: for example, we might flip coins until we see \coinFlips{HTHH}.  What is
the expected number of coin-flips until this happens?

A very clever trick due to Li~\cite{Li1980} solves this problem
exactly using the Optional Stopping Theorem.  Suppose our pattern is
$x_1 x_2 \dots x_k$.  We imagine an army of
gamblers, one of which shows up before each coin-flip.  Each gambler
starts by betting \$1 that next coin-flip will be
$x_1$.  If they win, they bets \$2 
that the next coin-flip will be $x_2$,
continuing to play double-or-nothing until either they lose (and is
down \$1) or wins her last bet on $x_k$ (and is up $2^{k}-1$).  Because
each gambler's winnings form a martingale, so does their sum, and so
the expected total return of all gamblers up to the stopping time
$τ$ at which our pattern first occurs is $0$.

We can now use this fact to compute $\Exp{τ}$.  When we stop at time
$τ$, we have one gambler who has won $2^{k}-1$.  We may also have
other gamblers who are still in play.  For each $i$ with $x_1 \dots
x_i = x_{k-i+1} \dots x_k$, there will be a gambler with net winnings
$∑_{j=1}^{i} 2^{j-1} = 2^{i}-1$.  The remaining gamblers will all be at $-1$.

Let $χ_i = 1$ if $x_1 \dots x_i = x_{k-i+1} \dots x_k$, and $0$
otherwise.  Then the number of losers is given by $\tau - ∑_{i=1}^{k}
χ_i$ and the total expected payoff is
\begin{align*}
\Exp{X_τ}
&= \Exp{-(τ-∑_{i=1}^{k} χ_i) + ∑_{i=1}^{k} χ_i \left(2^{i}-1\right)}
\\&= \Exp{-τ + ∑_{i=1}^{k} χ_i \left(2^{i}\right)}
\\&= 0.
\end{align*}
It follows that $\Exp{τ} = ∑_{i=1}^{k} χ_i 2^i$.

As a quick test, the pattern \text{H} has $\Exp{τ} = 2^{1} = 2$.
This is consistent with what we know about geometric distributions.

For a longer example, the pattern \coinFlips{HTHH} only overlaps with its prefix
\coinFlips{H}, so
in this case we have $\Exp{τ} = ∑ χ_i 2^{i} = 16 + 2 = 18$.  But
\coinFlips{HHHH} overlaps with all of its prefixes, giving $\Exp{τ} =
16 + 8 + 4 + 2 = 30$.
At the other extreme, \coinFlips{THHH} has no overlap at all and gives
$\Exp{τ} = 16$.  

In general, for a pattern of length $k$, we expect a
waiting time somewhere between $2^k$ and $2^{k+1} - 2$—almost a factor
of $2$ difference depending on how much overlap we get.

This analysis generalizes in the obvious way to biased coins and
larger alphabets. See the paper~\cite{Li1980} for details.

\myChapter{Markov chains}{2025}{}
\label{chapter-Markov-chains}

A (discrete time) \concept{Markov chain} is a sequence of random
variables $X_0,X_1,X_2,\dots$, which we think of as the position of
some particle at increasing times in $ℕ$, where the distribution of
$X_{t+1}$ depends only on the value of $X_t$.
A typical example of a Markov chain is a random walk on a graph: each
$X_t$ is a node in the graph, and a step moves to one of the neighbors
of the current node chosen at random, each with equal probability.

Markov chains come up
in randomized algorithms both because the execution of any randomized
algorithm is, in effect, a Markov chain (the random variables are the
states of the algorithm); and because we can often sample from
distributions that are difficult to sample from directly by designing
a Markov chain that converges to the distribution we want.  Algorithms
that use this latter technique are known as
\concept{Markov chain Monte Carlo} algorithms, and rely on the
fundamental fact that a Markov chain that satisfies a few
straightforward conditions will always converge in the limit to a
\concept{stationary distribution}, no matter what state it starts in.

An example of this technique that predates randomized algorithms is
card shuffling: each permutation of the deck is a state, and the
shuffling operation sends the deck to a new state each time it is
applied.  Assuming the shuffling operation is not too deterministic,
it is possible to show that enough shuffling will eventually produce a
state that is close to being a uniform random permutation.  The big
algorithmic question for this and similar Markov chains is how quickly
this happens: what is the \concept{mixing time} of the Markov chain, a
measure of how long we have to run it to get close to its limit
distribution (this notion is defined formally in
§\ref{section-mixing-time}).  
Many of the techniques in this chapter will be aimed at finding bounds
on the mixing time for particular Markov processes.

If you want to learn more about Markov chains than presented here,
they are usually
covered in general probability textbooks (for example,
in~\cite{Feller1968} or~\cite{GrimmettS2001}), mentioned in many
linear algebra textbooks~\cite{Strang2003}, covered in some detail in
stochastic processes textbooks~\cite{KarlinT1975}, and covered in
exquisite detail in many books dedicated specifically to the
subject~\cite{KemenyS1976,KemenySK1976}.
Good sources for mixing times for Markov chains are the textbook of
Levin, Peres, and Wilmer~\cite{LevinPW2009} and the survey paper by
Montenegro and Tetali~\cite{MontenegroT2005}.
An early reference on the mixing times for random walks on graphs that
helped inspire much subsequent work is
the Aldous-Fill manuscript~\cite{AldousF2001}, which can be found
on-line at \url{http://www.stat.berkeley.edu/~aldous/RWG/book.html}.

\section{Basic definitions and properties}
\label{section-Markov-basics}

A \concept{Markov chain} or \concept{Markov process} is a stochastic
process\footnote{A \concept{stochastic process} is just a sequence of
random variables $\Set{S_t}$, where we usually think of $t$ as
representing time and the sequence as representing the evolution of
some system over time.  Here we are considering discrete-time
processes, where $t$ will typically be a non-negative integer.} 
where the distribution of $X_{t+1}$ depends only on the value
of $X_{t}$ and not any previous history.  
Formally, this means that
\begin{equation}
    \label{eq-Markov-memoryless}
    \ProbCond{X_{t+1} = j}{X_t=i_t,X_{t-1}=i_{t-1},\dots,X_0=i_0}
    = \ProbCond{X_{t+1}=j}{X_t=i_t}.
\end{equation}
A stochastic process with this property is called
\concept{memoryless}: at any time, you know where you are, and you can
figure out where you are going, but you don't know where you were
before.

The \concept{state space} of the chain is just the set of all values
that each $X_{t}$ can have.  A Markov chain is \indexConcept{finite
Markov chain}{finite} or \indexConcept{countable Markov
chain}{countable} if it has a finite or countable state space,
respectively.  We'll mostly be interested in finite Markov
chains (since we have to be able to fit them inside our computer), but
countable Markov chains will come up in some contexts.\footnote{If the state space is not countable, we run into the
    same measure-theoretic issues as with continuous random variables,
    and have to replace \eqref{eq-Markov-memoryless} with the more
    general condition that
    \begin{equation*}
        \ExpCond{1_{[X_t ∈ A]}}{X_t, X_{t-1}, \dots, X_0}
        = \ExpCond{1_{[X_t ∈ A]}}{X_t},
    \end{equation*}
    provided $A$ is measurable with respect to some appropriate
$σ$-algebra.  We don't really want to deal with this, and for the most
part we don't have to, so we won't.}

We'll also assume that our Markov chains are \concept{homogeneous},
which means that $\ProbCond{X_{t+1} = j}{X_{t} = i}$ doesn't depend on
$t$.

For a homogeneous countable Markov chain, we can describe its behavior completely
by giving the state space and the one-step \concept{transition
probabilities} $p_{ij} = \ProbCond{X_{t+1} = j}{X_{t} = i}$.  Given
$p_{ij}$, we can calculate two-step transition probabilities
\begin{align*}
p^{(2)}_{ij}
&= \ProbCond{X_{t+2} = j}{X_{t} = i}
\\&=
∑_{k} \ProbCond{X_{t+2} = j}{X_{t+1} = k} \ProbCond{X_{t+1} = k }{ X_{t} = i}
\\&= ∑_{k} p_{ik}
p_{kj}.
\end{align*}

This is identical to the formula for matrix multiplication.
For a Markov chain with $n$ states, we can specify the transition
probabilities $p_{ij}$ using an $n×n$
\concept{transition matrix}
\index{matrix!transition}
$P$ with $P_{ij} = p_{ij}$, and the two-step transition probabilities
are given by $p^{(2)})_{ij} = P^2_{ij}$.
More generally, the $t$-step transition probabilities are given by $p^{(t)}_{ij} =
(P^{t})_{ij}$.

Conversely, given any matrix with non-negative entries where the rows
sum to $1$ ($∑_{j} P_{ij} = 1$, or $P\mathbf{1} = \mathbf{1}$,
where $\mathbf{1}$ in the
second equation stands for the
all-ones vector), there is a
corresponding Markov chain given by $p_{ij} = P_{ij}$.  Such a matrix
is called a \concept{stochastic matrix}; and for every stochastic
matrix there is a corresponding finite Markov chain and vice versa.

The general formula for $(s+t)$-step transition probabilities is that
$p^{(s+t)}_{ij} = ∑_{k} p^{(s)}_{ik}p^{(t)}_{kj}$.  This is known
as the \concept{Chapman-Kolmogorov equation} and is equivalent to the
matrix identity $P^{s+t} = P^{s}P^{t}$. It is also similar to the
formula for counting paths in a directed graph, and we can use
the correspondence between matrices and (labeled) directed graphs to
give visual depictions of particular Markov chains, as shown in
Figure~\ref{fig-Markov-chain-picture}.

\begin{figure}
    \begin{tikzpicture}
        \foreach \x in {1,...,4} {
            \node[draw,circle] (\x) at (\x*1.5,0) { \x };
        }
        \foreach \x/\xx in {1/2,2/3,3/4} {
            \draw (\x) edge[->,out=60,in=120] node[midway,above] {$p_{\x\xx}$} (\xx);
            \draw (\xx) edge[->,out=210,in=330] node[midway,below] {$p_{\xx\x}$} (\x);
        }
        \draw (1) edge[->,loop left] node[left] {$p_{11}$} (1);
        \draw (4) edge[->,loop right] node[right] {$p_{44}$} (4);
        \node (matrix) at (10,0) {$
            \begin{bmatrix}
            p_{11} & p_{12} & 0 & 0 \\
            p_{21} & 0 & p_{23} & 0 \\
            0 & p_{32} & 0 & p_{34} \\
            0 & 0 & p_{43} & p_{44}
            \end{bmatrix}
        $};
    \end{tikzpicture}
    \caption[Drawing a Markov chain as a digraph]{Drawing a Markov
    chain as a directed graph. Nodes represent states. Edges, labeled
    with probabilities, represent possible transitions.
    Zero-probability transitions are usually omitted. The
    corresponding transition matrix is shown on the right.}
    \label{fig-Markov-chain-picture}
\end{figure}

A distribution over states of a finite Markov chain at some time $t$ can be
given by a row vector $x$, where $x_{i} = Pr[X_{t} = i]$.  To compute
the distribution at time $t+1$, we use the law of total probability:
$\Prob{X_{t+1} = j} = ∑_{i} \Prob{X_{t} = i} \ProbCond{X_{t+1} = j
}{X_{t} =
i} = ∑_{i} x_{i} p_{ij}$.  Again we have the formula for matrix
multiplication (where we treat $x$ as a $1×n$ matrix); so the
distribution vector at time $t+1$ is just $xP$, and at time $t+n$ is
$xP^{n}$.

We like Markov chains for two reasons:
\begin{enumerate}
 \item They describe what happens in a randomized algorithm; the state space is just the set of all states of the algorithm, and the Markov property holds because the algorithm can't remember anything that isn't part of its state.  So if we want to analyze randomized algorithms, we will need to get good at analyzing Markov chains.
 \item They can be used to do sampling over interesting distributions.
 Under appropriate conditions (see below), the state of a Markov chain
 converges to a \concept{stationary distribution}.  If we build the right Markov chain, we can control what this stationary distribution looks like, run the chain for a while, and get a sample close to the stationary distribution.
\end{enumerate}

In both cases we want to have a bound on how long it takes the Markov chain to converge, either because it tells us when our algorithm terminates, or because it tells us how long to mix it up before looking at the current state.

\subsection{Examples}
\label{section-markov-chain-examples}

\begin{itemize}
 \item A fair $±1$ random walk.  The state space is $ℤ$,
 the transition probabilities are $p_{ij} = 1/2$ if $\abs*{i-j} = 1$,
 $0$ otherwise.  This is an example of a Markov chain that is also a martingale.
 \item A fair $±1$ random walk on a cycle.  As above, but now the
 state space is $ℤ/m$, the integers mod $m$.  This is a finite Markov
        chain. It is also in some sense a martingale, although we
        usually don't define martingales over finite groups.
 \item Random walks with absorbing and/or reflecting barriers.
 \item Random walk on a graph $G = (V,E)$.  The state space is $V$,
 the transition probabilities are $p_{uv} = 1/d(u)$ if $uv ∈ E$.
 
 One can also have more general transition probabilities, where the
 probability of traversing a particular edge is a property of the edge
 and not the degree of its source.  In principle we can represent any
 Markov chain as a random walk on graph in this way: the states become
 vertices, and the transitions become edges, each labeled with its
 transition probability.  
 It's conventional in this representation to
 exclude edges with probability $0$ and include self-loops for any
 transitions $i→i$.  

 If the resulting graph is small enough or has a nice structure, this
 can be a convenient way to draw a Markov chain.
 \item The Markov chain given by $X_{t+1} = X_{t}+1$ with probability
 $1/2$, and $0$ with probability $1/2$.  The state space is
 $ℕ$.
 \item A finite-state machine running on a random input.  The sequence
     of states acts as a Markov chain, assuming each input symbol is
     independent of the rest.
 \item A classic randomized algorithm for 2-SAT, due to
     Papadimitriou~\cite{Papadimitriou1991}.  Each state is a truth-assignment.  The
 transitional probabilities are messy but arise from the following
 process: pick an unsatisfied clause, pick one of its two variables
 uniformly at random, and invert it.  Then there is an absorbing state
 at any satisfying assignment.  With a bit of work, it can be shown
 that the Hamming distance between the current assignment and some
 satisfying assignment follows a random walk that is either unbiased
        or biased toward $0$, giving
        a satisfying assignment after $O(n^2)$ steps on
        average.\footnote{The proof of this is not too hard: given an
        unsatisfying assignment $x$ and a satisfying assignment $y$,
        any clause that is not satisfied by $x$ includes at least one
        variable that is satisfied by $y$. If we get lucky and flip
        this variable, we reduce the distance by one. If not, we
        increase it by at most 1 (depending on whether $y$ satisfies
        only one or both variables). So we get a process bounded by an
        unbiased random walk with a reflecting barrier at $n$ and an
        absorbing barrier at $0$, assuming we don't hit some other
        satisfying assignment $y'$ first.}
    This algorithm is not necessarily all that good, because there is a clever deterministic algorithm
        that solves 2-SAT in time linear in the size of the formula~\cite{AspvallPT1979}, but it
        has the nice property of not being so clever.
 \item A similar process works for
 2-colorability, 3-SAT, 3-colorability, etc., although for 
 \classNP-hard problems, it may take a while to reach an absorbing
 state.
 The constructive Lovász Local Lemma proof from
 §\ref{section-constructive-lll} also follows this pattern.
\end{itemize}

\section{Convergence of Markov chains}
\label{section-Markov-chain-convergence-methods}

We want to use Markov chains for sampling.  Typically this means that
we have some subset $S$ of the state space, and we want to know what
proportion of the states are in $S$.  If we can't sample states from
from the state space a priori, we may be able to get a good
approximation by running the Markov chain for a while and hoping that
it converges to something predictable.

To show this, we will proceed though several steps:
\begin{enumerate}
    \item We will define a class of distributions, known as
        \conceptFormat{stationary distributions}, that we hope to
        converge to (§\ref{section-stationary-distribution}
    \item We will define a distance between distributions, the
        \conceptFormat{total variation distance}
        (§\ref{section-total-variation-distance}).
    \item We will define the \conceptFormat{mixing time} of a Markov
        chain as the minimum time for the distribution of the position
        of a particle to get within a given total variation distance
        of the stationary distribution (§\ref{section-mixing-time}).
    \item We will describe a technique called \conceptFormat{coupling}
        that can be used to bound total variation distance
        (§\ref{section-coupling}), in terms of the probability that
        two dependent variables $X$ and $Y$ with the distributions we
        are looking at are or are not equal to each other.
    \item We will define \concept{reducible} and \concept{periodic}
        Markov chains, which have structural properties that 
        prevent convergence to a unique stationary distribution
        (§\ref{section-irreducible-and-aperiodic}).
    \item We will use a coupling between two copies of a
        Markov chain to show
        that \emph{any} Markov chain that does not have these
        properties does converge in
        total variation distance to a unique stationary distribution
        (§\ref{section-Markov-chain-convergence}).
    \item Finally, we will show that if we can construct a
        coupling between two copies of a particular chain that 
        causes both copies to reach the same state quickly on average, then we
        can use this expected \conceptFormat{coupling time} to bound
        the mixing time of the chain that we defined previously.
        This will give us a practical tool for showing that many
        Markov chains not only converge eventually but converge at a
        predictable rate, so we can tell when it is safe to stop
        running the chain and take a sample
        (§\ref{section-coupling-examples}).
\end{enumerate}

Much of this section follows the approach of Chapter 4 of
Levin~\etal~\cite{LevinPW2009}.

\subsection{Stationary distributions}
\label{section-stationary-distribution}

For a finite Markov chain, there is a transition matrix $P$ in which
each row sums to $1$.  We can write this fact compactly as
$P\mathbf{1} = \mathbf{1}$, where $\mathbf{1}$ is the all-ones column
vector.  This means that $\mathbf{1}$ is a right eigenvector of $P$
with eigenvalue $1$.\footnote{Given a square matrix $A$, a vector $x$
is a right eigenvector of $A$ with eigenvalue $λ$ is $Ax = λx$.
Similarly, a vector $y$ is a left eigenvalue of $A$ with eigenvalue
$λ$ if $yA = λA$.}  Because the left eigenvalues of a matrix are equal
to the right eigenvalues, this means that there will be at least one
left eigenvector $π$ such that $πP = π$, and in fact it is possible to
show that there is at least one such $π$ that represents a probability
distribution in that each $π_i≥0$ and $∑ π_i = 1$.  Such a
distribution is called a 
\index{distribution!stationary}\concept{stationary
distribution}\footnote{Spelling is important here.  
A \emph{stationery distribution} would involve handing out office supplies.}
of the
Markov chain, and if $π$ is unique, the probability $π_i$ is called
the \index{probability!stationary}\concept{stationary probability} for $i$.

Every finite Markov chain has at least one stationary distribution,
but it may not be unique.  For example, if the transition matrix is
the identity matrix (meaning that the particle never moves), then all
distributions are stationary.

If a Markov chain does have a unique stationary distribution, we can
calculate it from the transition matrix, by observing that the
equations
\begin{align*}
    πP &= P
    \intertext{and}
    π\mathbf{1} &= 1
\end{align*}
together give $n+1$ equations in $n$ unknowns, which we can solve for
$π$. (We need an extra equation because the stochastic property of $P$
means that it has rank at most $n-1$.)

Often this will be impractical, especially if our state space is
large enough or messy enough that we can't write down the entire
matrix.  In these cases we may be able to take advantage of a special
property of some Markov chains, called \concept{reversibility}. We'll
discuss this in §\ref{section-reversible-chains}. For the moment we
will content ourselves with showing that we do in fact converge to
some unique stationary distribution if our Markov chain has the right
properties.

\subsection{Total variation distance}
\label{section-total-variation-distance}

\begin{definition}
    \label{def-total-variation-distance}
Let $X$ and $Y$ be random variables defined on the same probability
space.  Then the \concept{total variation distance}
\index{distance!total variation} between $X$ and $Y$, written
$d_{TV}(X,Y)$ is given by
\begin{equation}
    \label{eq-total-variation-distance}
    d_{TV}(X,Y) = \sup_A \left(\Prob{X ∈ A} - \Prob{Y ∈ A}\right),
\end{equation}
where the supremum is taken over all sets $A$ for which $\Prob{X∈A}$
and $\Prob{Y∈A}$ are both defined.\footnote{For discrete random
    variables, this just means all $A$, since we can write $\Prob{X∈A}$ as $∑_{x∈A}
    \Prob{X=x}$; we can also replace $\sup$ with $\max$ for this case.
    For continuous random variables, we want that
    $X^{-1}(A)$ and $Y^{-1}(A)$ are both measurable.  If our $X$ and
    $Y$ range over the states of a 
    countable Markov chain, we will be
    working with discrete random variables, so we can just consider
    all $A$.}  
\end{definition}

An equivalent definition is
\begin{equation*}
    d_{TV}(X,Y) = \sup_A \abs*{\Prob{X ∈ A} - \Prob{Y ∈ A}}.
\end{equation*}
The reason this is equivalent is that if $\Prob{X∈A} - \Prob{Y∈A}$ is
negative, we can replace $A$ by its complement.

Less formally, given any test set $A$, $X$ and $Y$ show up in $A$ with
probabilities that differ by at most $d_{TV}(X,Y)$.  This is usually
what we want for sampling, since this says that if we are testing some
property (represented by $A$) of the states we are sampling, the
answer ($\Prob{X∈A}$) that we get for how likely this property is to occur is close to the
correct answer ($\Prob{Y∈A}$).

Total variation distance is a property of distributions, and is not
affected by dependence between $X$ and $Y$.  For finite Markov chains,
we can define the total variation distance between two distributions
$x$ and $y$ as
\begin{equation*}
    d_{TV}(x,y) = \max_A ∑_{i∈A} (x_i - y_i) = \max_A \abs*{∑_{i∈A} (x_i - y_i)}.
\end{equation*}
A useful fact is that $d_{TV}(x,y)$ is directly connected to the $\ell_1$
distance between $x$ and $y$.  If we let $B = \SetWhere{i}{x_i ≥ y_i}$,
then
\begin{align*}
    d_{TV}(x,y) 
    &= \max_A ∑_{i∈A} (x_i - y_i)
    \\&≤ ∑_{i∈B} (x_i - y_i),
\end{align*}
because if $A$ leaves out an element of $B$, or includes and element
of $\overline{B}$, this can only reduce the sum.  But if we consider
$\overline{B}$ instead, we get
\begin{align*}
    d_{TV}(y,x) 
    &= \max_A ∑_{i∈A} (y_i - x_i)
    \\&≤ ∑_{i∈B} (y_i - x_i).
\end{align*}
Now observe that
\begin{align*}
    \norm{x-y}_1
    &= ∑_i \abs{x_i-y_i}
    \\&= ∑_{i∈B} (x_i-y_i) + ∑_{i∈\overline{B}} (y_i - x_i)
    \\&= d_{TV}(x,y) + d_{TV}(y,x)
    \\&= 2 d_{TV}(x,y).
\end{align*}
So $d_{TV}(x,y) = \frac{1}{2} \norm{x-y}_1$.

\subsubsection{Total variation distance and expectation}

Sometimes it's useful to translate a bound on total variation to a
bound on the error when getting the expectation of a random variable.
The following lemma may be handy:
\begin{lemma}
    \label{lemma-total-variation-expectation}
Let $x$ and $y$ be two distributions of some discrete random variable $Z$.
    Let $\E_x(Z)$ and $\E_y(Z)$ be the expectations of $Z$ with
    respect to each of these distributions.
    Suppose that $\abs{Z} ≤ M$ always.
    Then 
    \begin{equation}
        \abs*{\E_x(Z) - \E_y(Z)} ≤ 2 M ⋅ d_{TV}(x,y).
        \label{eq-total-variation-expectation}
    \end{equation}
\end{lemma}
\begin{proof}
    Compute
\begin{align*}
    \abs*{\E_x(Z) - \E_y(Z)}
    &= \abs*{∑_z z \parens*{\Pr_x(Z = z) - \Pr_y(Z=z)}}
    \\&≤ ∑_z \abs*{z} ⋅ \abs*{\Pr_x(Z = z) - \Pr_y(Z=z)}
    \\&≤ M \, ∑_z \abs*{\Pr_x(Z = z) - \Pr_y(Z=z)}
    \\&≤ M \, \norm{x-y}_1
    \\&= 2 M ⋅ d_{TV}(x,y).
\end{align*}
\end{proof}

\subsection{Mixing time}
\label{section-mixing-time}

We are going to show that well-behaved finite Markov chains eventually
converge to some stationary distribution $π$ in total variation distance.
This means that for any $ε > 0$, there is a 
\index{$t_{\mix}$}
\index{time!mixing}\concept{mixing time}
$t_{\mix}(ε)$ such that for any initial distribution $x$ and any $t ≥ t_{\mix}(ε)$,
\begin{equation*}
    d_{TV}(xP^t,π) ≤ ε.
\end{equation*}

It is common to standardize $ε$ as $1/4$: if we write just
$t_{\mix}$, this means $t_{\mix}(1/4)$.  The choice of
$1/4$ is somewhat arbitrary, but has some nice technical properties
that we will see below.

\subsection{Coupling of Markov chains}
\label{section-coupling}

A \concept{coupling} of two random variables $X$ and $Y$ is a joint
distribution on $\Tuple{X,Y}$ that gives the correct marginal
distribution for each of $X$ and $Y$ while creating a dependence
between them with some desirable property (for example, minimizing
total variation distance or maximize $\Prob{X=Y}$).

We will use
couplings between Markov chains to prove convergence.
Here we take two copies
of the chain, once of which starts in an arbitrary distribution, and
one of which starts in the stationary distribution, and show that we
can force them to converge to each other by carefully correlating
their transitions.  Since the second chain is always in the stationary
distribution, this will show that the first chain converges to the
stationary distribution as well.

The tool that makes this work is the
\index{lemma!coupling}\concept{Coupling Lemma}:\footnote{It turns out that the bound in the Coupling Lemma is tight in the
following sense: for any given distributions on $X$ and $Y$, there
exists a joint distribution giving these distributions such that
$d_{TV}(X,Y)$ is exactly equal to $\Prob{X≠Y}$ when $X$ and $Y$ are
sampled from the joint distribution. For discrete distributions, the
easiest way to construct the joint distribution is first to let
to let $Y=X=i$ for each $i$ with probability
$\min(\Prob{X=i},\Prob{Y=i})$,
and then distribute the remaining probability for $X$ over all the cases where
$\Prob{X=i} > \Prob{Y=i}$ and similarly for $Y$ over all the cases
where $\Prob{Y=i} > \Prob{X=i}$.
Looking at the unmatched values for $X$ gives 
$\Prob{X≠Y} ≤ ∑_{\SetWhere{x}{\Pr{X=i} > \Pr{Y=i}}} \parens*{\Prob{X=i} -
\Prob{Y=i}} ≤ d_{TV}(X,Y)$. So in this case $\Prob{X≠Y} =
d_{TV}(X,Y)$.

Unfortunately, the fact that there always exists a perfect coupling in
this sense does not mean that we can express it in any convenient way,
or that even if we could, it would arise from the kind of causal, step-by-step
construction that we will use for couplings between Markov processes.}
\begin{lemma}
    \label{lemma-coupling}
For any discrete random variables $X$ and $Y$,
\begin{align*}
    d_{TV}(X,Y) &≤ \Prob{X ≠ Y}.
\end{align*}
\end{lemma}
\begin{proof}
    Let $A$ be any set for which $\Prob{X∈A}$ and $\Prob{Y∈A}$ are
    defined.  Then
    \begin{align*}
        \Prob{X∈A} &= \Prob{X∈A ∧ Y∈A} + \Prob{X∈A ∧ Y∉A}, \\
        \Prob{Y∈A} &= \Prob{X∈A ∧ Y∈A} + \Prob{X∉A ∧ Y∈A}, \\
        \intertext{and thus}
        \Prob{X∈A} - \Prob{Y∈A} &= \Prob{X∈A ∧ Y∉A} - \Prob{X∉A ∧ Y∈A}
        \\&≤ \Prob{X∈A ∧ Y∉A}
        \\&≤ \Prob{X ≠ Y}.
    \end{align*}

    Since this holds for any particular set $A$, it also holds when we
    take the maximum over all $A$ to get $d_{TV}(X,Y)$.
\end{proof}

For Markov chains, our goal will be to find a useful coupling between a
sequence of random variables $X_0, X_1, X_2, \dots$ corresponding to the Markov
chain starting in an arbitrary distribution with a second sequence
$Y_0, Y_1, Y_2, \dots$ corresponding to the same chain starting in a
stationary distribution.  What will make a coupling useful is if
$\Prob{X_t ≠ Y_t}$ is small for reasonably large $t$: since $Y_t$ has
the stationary distribution, this will show that $d_{TV}(xP^t, π)$ is
also small.

Our first use of this technique will be to show, using a rather generic
coupling, that Markov chains
with certain nice properties converge to their stationary distribution
in the limit.  Later we will construct specialized couplings for
particular Markov chains to show that they converge quickly.
But first we will consider what properties a Markov chain must have to
converge at all.

\subsection{Irreducible and aperiodic chains}
\label{section-irreducible-and-aperiodic}

Not all chains are guaranteed to converge to their stationary
distribution.  If some states are not reachable from other states, it
may be that starting in one part of the chain will keep us from ever
reaching another part of the chain.  Even if the chain is not
disconnected in this way, we might still not converge if the
distribution oscillates back and forth due to some periodicity in the
structure of the chain.  But if these conditions do not occur, then we
will be able to show convergence.

Let $p_{ij}^t$ be $\ProbCond{X_t = j}{X_0 = i}$.

A Markov chain is \concept{irreducible} if, for all states $i$ and $j$, there
exists some $t$ such that $p_{ij}^t ≠ 0$.  This says that we can reach
any state from any other state if we wait long enough.  If we think of
a directed graph of the Markov chain where the states are vertices and
each edge represents a transition that occurs with nonzero
probability, the Markov chain is irreducible if its graph is strongly
connected.

The \concept{period} of a state $i$ of a Markov chain is
$\gcd\parens*{\SetWhere{t>0}{p_{ii}^t ≠ 0}}$.
If the period of $i$ is $m$, then starting from $i$ we can only return to $i$ at times
that are multiples of $m$.  If $m=1$, state $i$ is said to be
\concept{aperiodic}.  A Markov chain as a whole is aperiodic if all of
its states are aperiodic.  In graph-theoretic terms, this means that
the graph of the chain is not $k$-partite for any $k>1$.
Reversible chains are also an interesting special case: if a chain is
reversible, it can't have a period greater than $2$, since we can
always step off a node and step back.

If our Markov chain is not aperiodic, we can make it aperiodic by
flipping a coin at each step to decide whether to move or not.  This
gives a \index{Markov chain!lazy}\concept{lazy Markov chain} whose
transition probabilities are given by $\frac{1}{2} p_{ij}$ when $i≠j$
and $\frac{1}{2} + \frac{1}{2} p_{ij}$ when $i=j$.  This doesn't
affect the stationary distribution: if we replace our transition
matrix $P$ with a new transition matrix $\frac{P+I}{2}$, and $πP = π$,
then $π\parens*{\frac{P+I}{2}} = \frac{1}{2} πP + \frac{1}{2} πI =
\frac{1}{2} π + \frac{1}{2} π = π$.

Unfortunately there is no quick fix for reducible Markov chains.  But
since we will often be designing the Markov chains we will be working
with, we can just take care to make sure they are not reducible.

We will later need the following lemma about aperiodic Markov chains, which is related to the
\concept{Frobenius problem} of finding the minimum value that cannot
be constructed using coins of given denominations:
\begin{lemma}
    \label{lemma-aperiodic}
    Let $i$ be an aperiodic state of some Markov chain.
    Then there is a time $t_0$ such that $p_{ii}^t ≠ 0$ for all $t
    ≥ t_0$.
\end{lemma}
\begin{proof}
    Let $S = \SetWhere{t}{p_{ii}(t) ≠ 0}$.  Since $\gcd(S) = 1$, there
    is a finite subset $S'$ of $S$ such that $\gcd{S'} = 1$.  Write the
    elements of $S'$ as $m_1,m_2,\dots,m_k$ and let $M = ∏_{j=1}^{k}
    m_j$.  From the extended Euclidean algorithm, there exist integer
    coefficients $a_j$ with $\abs{a_j} ≤ M/m_j$ such that $∑_{j=1}^{k}
    a_j m_j = 1$.  We would like to use each $a_j$ as the number of
    times to go around the length $m_j$ loop from $i$ to $i$.
    Unfortunately many of these $a_j$ will be negative.

    To solve this problem, we replace $a_j$ with $b_j = a_j + M/m_j$.
    This makes all the coefficients non-negative, and gives
    $∑_{j=1}^{k} b_j m_j = kM + 1$.  This implies that there is a
    sequence of loops that gets us from $i$ back to $i$ in $kM+1$
    steps, or in other words that $p_{ii}^{kM+1} ≠ 0$.  By repeating
    this sequence $\ell$ times, we can similarly show that $p_{ii}^{\ell
    k M + \ell} ≠ 0$ for any $\ell$.
    
    We can also pad any of these
    sequences out by as many copies of $M$ as we like.
    In particular, given $\ell k M + \ell$, where $\ell ∈
    \Set{0,\dots,M-1}$, we can add $(kM-\ell)M$ to 
    it to get $(kM)^2 + \ell$.  This means that we can express any $t
    ∈ \Set{(kM)^2,\dots,(kM)^2+M-1}$ as a sum of elements of $S$, or
    equivalently that $p_{ii}^t ≠ 0$ for any such $t$.  But for larger
    $t$, we can just add in more copies of $M$.  So in fact $p_{ii}^t
    ≠ 0$ for all $t ≥ t0 = (kM)^2$.
\end{proof}

\subsection{Convergence of finite irreducible aperiodic Markov chains}
\label{section-Markov-chain-convergence}

We can now show:
\begin{theorem}
    \label{theorem-Markov-chain-convergence}
    Any finite irreducible aperiodic Markov chain converges to a
    unique stationary distribution in the limit.
\end{theorem}
\begin{proof}
    Consider two copies of the chain $\Set{X_t}$ and $\Set{Y_t}$,
    where $X_0$ starts in some arbitrary distribution $x$ and $Y_0$ starts
    in a stationary distribution $π$.  Define a coupling between
    $\Set{X_t}$ and $\Set{Y_t}$ by the rule: (a) if $X_t ≠ Y_t$, then
    $\ProbCond{X_{t+1} = j ∧ Y_{t+1} = j'}{X_t = i ∧ Y_t = i'} = p_{ij}
    p_{i'j'}$; and (b) if $X_t = Y_t$, then $\ProbCond{X_{t+1} =
    Y_{t+1} = j}{X_t = Y_t = i} = p_{ij}$.  Intuitively, we let both
    chains run independently until they collide, after which we run
    them together.  Since each chain individually moves from state $i$
    to state $j$ with probability $p_{ij}$ in either case, we have
    that $X_t$ evolves normally and $Y_t$ remains in the stationary
    distribution.

    Now let us show that $d_{TV}(xP^t,π) ≤ \Prob{X_t ≠ Y_t}$ goes to
    zero in the limit.
    Pick some state $i$.  Let $r$ be the maximum over all states $j$
    of the \concept{first
    passage time} $f_{ji}$ where $f_{ji}$ is the minimum time $t$ such
    that $p_{ji}^t ≠ 0$.  Let $s$ be a time such that $p_{ii}^t ≠ 0$
    for all $t ≥ s$ (the existence of such an $s$ is given by
    Lemma~\ref{lemma-aperiodic}).

    Suppose that at time $\ell(r+s)$, where $\ell ∈ ℕ$, $X_{\ell(r+s)}
    = j ≠ j' = Y_{\ell(r+s)}$.  Then there are times $\ell(r+s)+u$ and
    $\ell(r+s)+u'$, where $u,u' ≤ r$, such that $X$ reaches $i$ at
    time $\ell(r+s)+u$ and $Y$ reaches $i$ at time $\ell(r+s)+u'$ with
    nonzero probability.  Since $(r+s-u) ≤ s$, then having reached $i$
    at these times, $X$ and $Y$ both return to $i$ at time
    $\ell(r+s)+(r+s) = (\ell+1)(r+s)$ with nonzero probability.
    Let $ε > 0$ be the product of these nonzero probabilities; then
    $\Prob{X_{(\ell+1)(r+s)} ≠ Y_{(\ell+1)(r+s)}} ≤ (1-ε)
    \Prob{X_{\ell(r+s)} = Y_{\ell(r+s)}}$, and in general we have
    $\Prob{X_t ≠ Y_t} ≤ (1-ε)^{\floor{t/(r+s)}}$, which goes to zero in
    the limit.  This implies that $d_{TV}(xP^t,π)$ also goes to zero
    in the limit (using the Coupling Lemma), and since any initial
    distribution (including a stationary distribution) converges to
    $π$, $π$ is the unique stationary distribution as claimed.
\end{proof}

This argument requires that the chain be finite, because otherwise we
cannot take the maximum over all the first passage times.  For
infinite Markov chains, it is not always enough to be irreducible and
aperiodic to converge to a stationary distribution (or even to have a
stationary distribution at all).  However, with some additional
conditions a similar result can be shown: see for example
\cite[§6.4]{GrimmettS2001}.

\section{Reversible chains}
\label{section-reversible-chains}

A Markov chain with transition probabilities $p_{ij}$
is \concept{reversible} if there is a distribution $π$ such that, for
all $i$ and $j$,
\begin{equation}
\label{eq-detailed-balance}
π_{i}p_{ij} = π_{j}p_{ji}.
\end{equation}
These are called the \concept{detailed balance
equations}—they say
that in the stationary distribution, the probability of seeing a
transition from $i$ to $j$ is equal to the probability of seeing a
transition from $j$ to $i$).  If this is the case, then $∑_{i}
π_{i}p_{ij} = ∑_{i} π_{j}p_{ji} = π_{j}$, which means
that $π$ is stationary.

It's worth noting that this works for countable chains even if they
are not finite, because the sums always converge since each term is
non-negative and $∑_i π_i p_{ij}$ is dominated by $∑_i π_i = 1$.
However, it may not be the case for any particular $p$ that there
exists a corresponding stationary distribution $π$. If this happens,
the chain is not reversible.

\subsection{Stationary distributions}
\label{section-reversible-chains-stationary-distribution}

The detailed balance equations often give a very quick way to compute the stationary
distribution, since if we know $π_{i}$, and $p_{ij}≠0$, then
$π_{j} = π_{i}p_{ij}/p_{ji}$.  If the transition probabilities
are reasonably well-behaved (for example, if $p_{ij} = p_{ij}$ for all
$i,j$), we may even be able to characterize the stationary
distribution up to a constant multiple even if we have no way to
efficiently enumerate all the states of the process.

What \eqref{eq-detailed-balance} says is that if we start in the stationary
distribution and observe either a forward transition $\Tuple{X_0,X_1}$ or
a backward transition $\Tuple{X_1,X_0}$, we can't tell which is which;
$\Prob{X_0 = i ∧ X_1 = j} = \Prob{X_1 = i ∧ X_0 = j}$.  This extends
to longer sequences.  The probability that $X_0 = i$, $X_1 = j$, and
$X_2 = k$ is given by $π_i p_{ij} p_{jk} = p_{ji} π_j p_{jk} = p_{ji}
p_{kj} π_k$, which is the probability that $X_0 = k$, $X_1 = j$, and
$X_0 = i$.  (A similar argument works for finite sequences of any
length.)  So a reversible Markov chain is one with no arrow of time
in the stationary distribution.

A typical reversible Markov chain is a random walk on a graph, where
a step starting from a vertex $u$ goes to one of its neighbors $v$,
which each neighbor chosen with probability $\frac{1}{d(u)}$.  This
has a stationary distribution
\begin{align*}
    π_u 
    &= \frac{d(u)}{∑_u d(u)}
    \\&= \frac{d(u)}{2\card{E}},
    \intertext{which satisfies}
    π_u p_{uv}
    &= \frac{d(u}{2\card{E}} ⋅ \frac{1}{d(u)}
    \\&= \frac{d(v}{2\card{E}} ⋅ \frac{1}{d(v)}
    \\&= π_v p_{vu}.
\end{align*}

If we don't know $π$ in advance, we can often guess it by observing that
\begin{align}
    π_i p_{ij} &= π_j p_{ji}
    \nonumber
    \intertext{implies}
    π_j &= π_i \frac{p_{ij}}{p_{ji}},
    \label{eq-reversible-calculation}
\end{align}
provided $p_{ij} ≠ 0$.  This gives us the ability to calculate
$π_k$ starting from any initial state $i$ as long as there is
some chain of transitions $i = i_0 → i_1 → i_2 → \dots i_\ell = k$
where each step $i_{m} → i_{m+1}$ has $p_{i_m,i_{m+1}} ≠ 0$.  For a
random walk on a graph, this implies that $π$ is unique as long as the
graph is connected.  This of course only works for reversible chains;
if we try to do this with a non-reversible chain, we are likely
to get a contradiction.

For example, if we consider a biased random walk on the $n$-cycle,
which moves from $i$ to $(i+1) \bmod n$ with probability $p$ and in the
other direction with probability $q = 1-p$, then applying 
\eqref{eq-reversible-calculation} repeatedly would give
$π_i = π_0 \parens*{\frac{p}{q}}^i$.
This is not a problem when $p=q=1/2$, since we get $π_i = π_0$ for all
$i$ and can deduce that $π_i = 1/n$ is the unique stationary
distribution.  But if we try it for $p = 2/3$, then we get $π_i = π_0
2^i$, which is fine up until we hit $π_0 = π_n = π_0 2^n$.  So for
$p≠q$, this process is not reversible, which is not surprising if we
realize that the $n=60, p=1$ case describes precisely the movement of 
the second hand on a clock.\footnote{It happens to be the case that
$π_i = 1/n$ is a stationary distribution for any value of $p$, we just
can't prove this using \eqref{eq-detailed-balance}.}

Reversible chains have a special role in Markov chain theory, because
some techniques for proving convergence only apply to reversible
chains (see §\ref{section-Cheegers-inequality}).  They are also handy
for sampling from large, irregular state spaces, because by tuning the
transition probabilities locally we can often adjust the relative
likelihoods of states in the stationary distribution to be closer to
what we want (see §\ref{section-Metropolis-Hastings}).

\subsection{Examples}
\label{section-reversible-chain-examples}

\begin{description}
 \item[Random walk on a weighted graph] Here each edge has a weight
 $w_{uv}$ where $0 < w_{uv} = w_{vu} < ∞$, with self-loops
 permitted.  A step of the random walk goes from $u$ to $v$ with
 probability $w_{uv}/∑_{v'} w_{uv'}$.  It is easy to show that
 this random walk has stationary distribution $π_{u} = ∑_{u}
 w_{uv} / ∑_{u}∑_{v} w_{uv}$, generalizing the previous case, and that the resulting Markov chain satisfies the detailed balance equations.
 \item[Random walk with uniform stationary distribution]  Now let $Δ$
 be the maximum degree of the graph, and traverse each each with
 probability $1/Δ$, staying put on each vertex $u$ with probability
 $1-d(u)/Δ$.  The stationary distribution is uniform, since for
 each pair of vertices $u$ and $v$ we have $p_{uv} = p_{vu} = 1/Δ$ if $u$ and
 $v$ are adjacent and $0$ otherwise.
        This is a special case of the Metropolis-Hastings algorithm
        (see §\ref{section-Metropolis-Hastings}).
\end{description}

\subsection{Time-reversed chains}
\label{section-time-reversed-chains}

Another way to get a reversible chain is to take an arbitrary chain
with a stationary distribution and rearrange it so that it can run
both forwards and backwards in time. This is not necessarily useful,
but it illustrates the connection between reversible and irreversible
chains.

Given a finite Markov chain with transition matrix $P$ and stationary
distribution $π$, define the corresponding
\concept{time-reversed chain} with
matrix $P^{*}$ where $π_{i}p_{ij} = π_{j}p^{*}_{ji}$.

To make sure that this actually works, we need to verify that:
\begin{enumerate}
    \item The matrix $P^{*}$ is stochastic:
        \begin{align*}
            ∑_{j} p^{*}_{ij} 
            &= ∑_{j} p_{ji}π_{j}/π_{i} 
            \\&= π_{i}/π_{i}
            \\&= 1.
        \end{align*}
    \item The reversed chain has the same stationary distribution as
        the original chain:
        \begin{align*}
            ∑_{j} π_j p^{*}_{ji}
            &= ∑_{j} π_i p_{ij}
          \\&= π_i.
        \end{align*}
    \item And that in
general $P^{*}$'s paths starting from the stationary distribution are
a reverse of $P$'s paths starting from the same distribution.  
For length-$1$ paths, this is just $π_j p^{*}_{ji} = π_i p_{ij}$.
For longer paths, this follows from an argument similar to that for reversible chains.
\end{enumerate}

This gives
an alternate definition of a reversible chain as a chain for which $P
= P^{*}$.

We can also use time-reversal to generate reversible chains from arbitrary
chains.  The chain with transition matrix $(P+P^{*})/2$ (corresponding
to moving $1$ step forward or back with equal probability at each step) is always a reversible chain.

Examples:
\begin{itemize}
 \item Given a biased random walk on a cycle that moves right with
 probability $p$ and left with probability $q$, its time-reversal is
 the walk that moves left with probability $p$ and right with
 probability $q$.  (Here the fact that the stationary distribution is uniform makes things simple.)
 The average of this chain with its time-reversal is an unbiased
 random walk.
 \item Given the random walk defined by $X_{t+1} = X_{t}+1$ with
 probability $1/2$ and $0$ with probability $1/2$, we have $π_{i}
 = 2^{-i-1}$.  This is not reversible (there is a transition from $1$
 to $2$ but none from $2$ to $1$), but we can reverse it by setting
 $p^{*}_{ij} = 1$ for $i = j+1$ and $p^{*}_{0i} = 2^{-i-1}$.  (Check:
 $π_{i}p_{i i+1} = 2^{-i-1}(1/2) = π_{i+1}p^{*}_{i+1 i} =
 2^{-i-2}(1); π_{i}p_{i0} = 2^{-i-1}(1/2) = π_{0}p^{*}_{0i} =
 (1/2) 2^{-i-1}$.)
\end{itemize}

These examples work because the original chains are simple and have
clean stationary distributions.
Reversed versions of chains with messier stationary distributions are
usually messier.
In practice, building reversible chains using time-reversal is often
painful precisely because we don't have a good characterization of the
stationary distribution of the original non-reversible chain. So we
will often design our chains to be reversible from the start rather
than relying on after-the-fact flipping.

\subsection{Adjusting stationary distributions with the Metropolis-Hastings algorithm}
\label{section-metropolis}
\label{section-Metropolis-Hastings}

Sometimes we have a reversible Markov chain, but it doesn't have the
stationary distribution we want.
The \concept{Metropolis-Hastings
algorithm}\cite{MetropolisRRTT1953,Hastings1970} (sometimes just
called \concept{Metropolis}) lets us start with a reversible Markov
chain $P$ with a known stationary distribution $π$ and convert it to
a chain $Q$ on the same states with a different stationary distribution $μ$, where $μ_{i} = f(i) /
∑_{j} f(j)$ is proportional to some function $f≥ 0$ on states
that we can compute easily.

A typical application is that we want to sample according to
$\ProbCond{i}{A}$, but $A$ is highly improbable (so we can't just use
\index{sampling!rejection}\concept{rejection sampling}, where we sample random points from the
original distribution until we find one for which $A$ holds), and
$\ProbCond{i}{A}$ is easy to compute for any fixed $i$ but tricky to compute
for arbitrary events (so we can't use divide-and-conquer).  If we let
$f(i) \propto{} \ProbCond{i}{A}$, then Metropolis-Hastings will do exactly what we want, assuming it converges in a reasonable amount of time.

Let $q$ be the transition probability for
$Q$.  Define, for $i≠ j$,
\begin{align*}
q_{ij} 
&= p_{ij} \min\left(1, \frac{π_{i}f(j)}{π_{j}f(i)}\right) 
\\&= p_{ij} \min\left(1,
\frac{π_{i}μ_{j}}{π_{j}μ_{i}}\right)
\end{align*}
and let $q_{ii}$ be whatever
probability is left over.  Now consider two states $i$ and $j$, and
suppose that $π_{i}f(j) ≥ π_{j}f(i)$.  Then
\begin{align*}
q_{ij} &= p_{ij}
\intertext{which gives}
μ_{i}q_{ij} &= μ_{i}p_{ij},
\intertext{while}
μ_{j}q_{ji} 
&=
μ_{j}p_{ji}(π_{j}μ_{i}/π_{i}μ_{j}) 
\\&=
p_{ji}(π_{j}μ_{i}/π_{i}) 
\\&= μ_{i} (p_{ji}π_{j}) /
π_{i}
\\&= μ_{i} (p_{ij}π_{i}) / π_{i} 
\\&= μ_{i}p_{ij}
\end{align*}
(note the use of reversibility of $P$ in the second-to-last step).
So we have $μ_j q_{ji} = μ_i p_{ij} = μ_i q_{ij}$
and $Q$ is a reversible Markov chain with stationary
distribution $μ$.

We can simplify this when our underlying chain $P$ has a uniform
stationary distribution (for example, when it's the
random walk on a graph with maximum degree $Δ$, where we 
traverse each edge with
probability $1/Δ$).  Then we have $π_i
= π_j$ for all $i, j$, so the new transition probabilities $q_{ij}$
are just $\frac{1}{Δ} \min(1, f(j)/f(i))$.  Most of our
examples of reversible chains will be instances of this case.

\section{The coupling method}
\label{section-coupling-method}
\label{section-coupling-examples}

In order to use the stationary distribution of a Markov chain to do sampling, we need to have a bound on the rate of convergence to tell us when it is safe to take a sample.  There are two standard techniques for doing this: coupling, where we show that a copy of the process starting in an arbitrary state can be made to converge to a copy starting in the stationary distribution; and spectral methods, where we bound the rate of convergence by looking at the second-largest eigenvalue of the transition matrix.  We'll start with coupling because it requires less development.

(See also \cite{Guruswami2000} for a survey of the relationship between the various methods.)

Note: these notes will be somewhat sketchy.  If you want to read more
about coupling, a good place to start might be Chapter 12 of
\cite{MitzenmacherU2017}; Chapter 4-3
(\url{http://www.stat.berkeley.edu/~aldous/RWG/Chap4-3.pdf}) of the
unpublished but nonetheless famous \concept{Aldous-Fill manuscript}
(\url{http://www.stat.berkeley.edu/~aldous/RWG/book.html},
\cite{AldousF2001}), which is a
good place to learn about Markov chains and Markov chain Monte Carlo
methods in general; or even an entire book~\cite{Lindvall1992}.
We'll mostly be using examples from the Aldous-Fill text.

\subsection{Random walk on a cycle}
\label{section-random-walk-on-a-cycle}

Let's suppose we do a random walk on $ℤ_{m}$, where to avoid
periodicity at each step we stay put with probability $1/2$, move
counterclockwise with probability $1/4$, and move clockwise with
probability $1/4$: in other words, we are doing a lazy unbiased random
walk.  What's a good choice for a coupling to show this process converges quickly?

Specifically, we need to create a joint process $(X_t, Y_t)$,
where each of the marginal processes $X$ and $Y$ looks like a lazy
random walk on 
$ℤ_{m}$,
$X_0$ has whatever distribution our real process starts with, and $Y_0$ has the stationary
distribution.  Our goal is to structure the combined process so that
$X_t = Y_t$ as soon as possible.

Let $Z_t = X_t - Y_t \pmod{m}$.  If $Z_t = 0$, then $X_t$ and $Y_t$
have collided and we will move both together.  If $Z_t ≠ 0$, then
flip a coin to decide whether to move $X_t$ or $Y_t$; whichever one
moves then moves up or down with equal probability.  It's not hard to
see that this gives a probability of exactly $1/2$ that $X_{t+1} =
X_t$, $1/4$ that $X_{t+1} = X_t + 1$, and $1/4$ that $X_{t+1} =
X_t-1$, and similarly for $Y_t$.  So the transition functions for $X$
and $Y$ individually are the same as for the original process.

Whichever way the first flip goes, we get $Z_{t+1} = Z_t \pm 1$ with
equal probability.  So $Z$ acts as an unbiased random walk on
$ℤ_m$ with an 
absorbing barriers at $0$; this is equivalent to a random walk on
$0\dots m$ with absorbing barriers at both endpoints.  
The expected time for this random walk to reach a barrier starting
from an arbitrary initial state is at most $m^2/4$, so if $τ$ is
the first time at which $X_τ = Y_τ$, we have $\Exp{τ} ≤
m^2/4$.\footnote{If we know that $Y_0$ is uniform, then $Z_0$ is also
uniform, and we can use this fact to get a slightly smaller bound on
$\Exp{τ}$, around $m^2/6$.  But this will cause problems if we want
to re-run the coupling starting from a state where $X_t$ and $Y_t$
have not yet converged.}

Using Markov's inequality, 
after $t = 2(m^{2}/4) = m^2/2$ steps we have
$\Prob{X_{t}≠Y_{t}} = \Prob{τ>m^2/2} ≤ \frac{\Exp{τ}}{m^2/2} ≤ 1/2$.
We can also iterate the whole argument, starting over in whatever
state we are in at time $t$ if we don't converge.  This gives at most
a $1/2$ chance of not converging for each interval of $m^2/2$ steps.
So after 
$α m^{2}/2$ steps we will have $\Prob{X_{t}≠Y_{t}} ≤
2^{-α}$.  This gives
$t_{\mix}(ε) ≤ \frac{1}{2} m^{2} \ceil{\lg(1/ε)}$, where as before
$t_{\mix}(ε)$ is the time needed to make $d_{TV}(X_t,π) ≤ ε$
(see §\ref{section-mixing-time}).

The choice of $2$ for the constant in Markov's inequality could be
improved.  The following lemma gives an optimized version of this
argument:
\begin{lemma}
\label{lemma-coupling-time}
Let the expected \index{time!coupling}\concept{coupling time}, at which two coupled
processes $\Set{X_t}$ and $\Set{Y_t}$ starting from
an arbitrary state are first
equal, be $T$.  Then $d_{TV}\left(X_{T_ε},
Y_{T_ε}\right) ≤
    ε$ for $T_ε ≥ Te \ceil{\ln(1/ε)}$.
\end{lemma}
\begin{proof}
Essentially the same argument as above, but replacing $2$ with a
constant $c$ to be determined.
Suppose we restart the process every $cT$
steps.  Then at time $t$ we have a total variation bounded by 
$c^{-\floor{t/cT}}$.  The expression $c^{-t/cT}$ is minimized by
minimizing $c^{-1/c}$ or equivalently $-\ln c / c$, which occurs at
$c=e$.  
    This gives $t_{\mix}(ε) ≤ Te \ceil{\ln(1/ε)}$.
\end{proof}

It's worth noting that the random walk example was very carefully rigged to make
the coupling argument clean.  A similar argument still 
works (perhaps with a 
change in the bound) for other irreducible aperiodic walks on the ring,
but the details are messier.

\subsection{Random walk on a hypercube}
\label{section-random-walk-on-a-hypercube-via-coupling}

Start with a bit-vector of length $n$.  At each step, choose an index
uniformly at random, and set the value of the bit-vector at that index
to $0$ or $1$ with equal probability.  How long until we get a
nearly-uniform distribution over all $2^{n}$ possible bit-vectors?

Here we apply the same transformation to both the $X$ and $Y$ vectors.
It's easy to see that the two vectors will be equal once every index
has been selected once.  The waiting time for this to occur is just
the waiting time $nH_{n}$ for the coupon collector problem.
We can
either use this expected time directly to show that the process mixes
in time $O(n \log n \log (1/ε))$ as above, 
or apply a sharp concentration bound to the coupon collector process.
It is known that 
\index{concentration bound!coupon collector problem}
\index{coupon collector problem!concentration bounds}(see
\cite[§5.4.1]{MitzenmacherU2017}
or \cite[§3.6.3]{MotwaniR1995}), 
$\lim_{n→∞} \Prob{T ≥ n(\ln n + c)} =
1-\exp(-\exp(-c))$), so in the limit $n \ln n + n \ln \ln
(1/(1-ε)) = n \ln n + O(n \log(1/ε))$ would seem to be enough.
But
this is a little tricky: we don't know from this
bound alone how fast the probability converges as a function of $n$,
so to do this right we need to look into the bound in more detail.

Fortunately, we are looking at large enough values of $c$ that we can get a bound that is just as good
using a simple argument.  We have
\begin{align*}
    \Prob{\text{there is an empty bin after $t$ insertions}}
    &≤ n\parens*{1-\frac{1}{n}}^t
    \\& ≤ n e^{-t/n},
\end{align*}
and setting $t= n \ln n + cn$ gives a bound of $e^{-c}$.
We can then set $c = \ln(1/ε)$ to get a 
$\Prob{X_t ≠ Y_t} ≤ ε$ at time $n \ln n + n \ln(1/ε)$.

We can improve the bound slightly by observing that, on average, half
the bits in $X_{0}$ and $Y_{0}$ are already equal; doing this right
involves summing over many cases, so we won't do it.

This is an example of a Markov chain with the 
\index{mixing!rapid}\concept{rapid mixing} property: the mixing time
is polylogarithmic in the number of states ($2^n$ in this case) and
$1/ε$.  For comparison, the random walk on the ring is not
rapid mixing, because the coupling time is polynomial in $n=m$ rather
than $\log n$.

\subsection{Various shuffling algorithms}
\label{section-shuffling-algorithms}

Here we have a deck of $n$ cards, and we repeatedly apply some random
transformation to the deck to converge to a stationary distribution
that is uniform over all permutations of the cards (usually this is
obvious by symmetry, so we won't bother proving it).  Our goal is to
show that the expected \index{time!coupling}\concept{coupling time} at which our deck ends
up in the same permutation as an initially-stationary deck is small.
Typically we do this by linking identical cards on each side of the coupling, so
that each linked pair moves through the same positions with each step.
When all $n$ cards are linked, we will have achieved $X = Y$.

Lest somebody try implementing one of these shuffling algorithms, it's
probably worth mentioning that they are all terrible. If you actually
want to shuffle an array of values, the usual approach is to do a
\index{shuffle!Fisher-Yates}\concept{Fisher-Yates shuffle}~\cite{FisherY1948,Durstenfeld1964}: swap an
element chosen uniformly at random into the first position, then
shuffle the remaining $n-1$ positions recursively. This gives a
uniform shuffle in $O(n)$ steps. Another classical shuffling algorithm
is the \index{shuffle!Rao-Sandelius}\concept{Rao-Sandelius shuffle}~\cite{Rao1961,Sandelius1962}, where we
flip a coin for each element to divide the original array into two
random arrays, then shuffle each pile recursively. This has worse
expected running time in theory than Fisher-Yates ($O(n \log n)$
instead of $O(n)$), but it parallelizes well and uses only random bits
instead of random values from a larger range. The more recent
\index{shuffle!Bacher~\etal}\concept{MergeShuffle} algorithm of
Bacher~\etal~\cite{BacherBHL2015} is even faster in practice; this
paper also gives a more complete history of shuffling algorithms than
given here.

\subsubsection{Move-to-top}
This is a variant of card shuffling that is interesting mostly because
it gives about the easiest possible coupling argument.  At each step,
we choose one of the cards uniformly at random (including the top
card) and move it to the top of the deck.  How long until the deck is
fully shuffled, i.e., until the total variation distance between the
actual distribution and the stationary distribution is bounded by
$ε$?

Here the trick is that when we choose a card to move to the top in the
$X$ process, we choose the same card in the $Y$ process.  It's not
hard to see that this links the two cards together so that they are
always in the same position in the deck in all future states.  So to
keep track of how well the coupling is working, we just keep track of
how many cards are linked in this way, and observe that as soon as
$n-1$ are, the two decks are identical.

Note: Unlike some of the examples below, we don't consider two cards to be linked just because they are in the same position.  We are only considering cards that have gone through the top position in the deck (which corresponds to some initial segment of the deck, viewed from above).  The reason is that these cards never become unlinked: if we pick two cards from the initial segment, the cards above them move down together.  But deeper cards that happen to match might become separated if we pull a card from one deck that is above the matched pair while its counterpart in the other deck is below the matched pair.

Given $k$ cards linked in this way, the
probability that the next step links another pair of cards is exactly
$(n-k)/n$.  So the expected time until we get $k+1$ cards is
$n/(n-k)$, and if we sum these waiting times for $k=0\dots n-1$, we
get $nH_{n}$, the waiting time for the coupon collector problem.  So the bound on the mixing time is the same as for the random walk on a hypercube.

\subsubsection{Random exchange of arbitrary cards}
Here we pick two cards uniformly and independently at random and swap
them.  (Note there is a $1/n$ chance they are the same card; if we
exclude this case, the Markov chain has period $2$.)  To get a
coupling, we reformulate this process as picking a random card and a
random location, and swapping the chosen card with whatever is in the
chosen location in both the $X$ and $Y$ processes.

First let's observe that the number of linked cards never decreases.
Let $x_{i}$, $y_{i}$ be the position of card $i$ in each process, and
suppose $x_{i}=y_{i}$.  If neither card $i$ nor position $x_{i}$ is
picked, $i$ doesn't move, and so it stays linked.  If card $i$ is
picked, then both copies are moved to the same location; it stays
linked.  If position $x_{i}$ is picked, then it may be that $i$
becomes unlinked; but this only happens if the card $j$ that is picked
has $x_{j}≠y_{j}$.  In this case $j$ becomes linked, and the number of linked cards doesn't drop.

Now we need to know how likely it is that we go from $k$ to $k+1$
linked cards.  We've already seen a case where the number of linked
cards increases; we pick two cards that aren't linked and a location
that contains cards that aren't linked.  The probability of doing this
is $((n-k)/n)^{2}$, so our total expected waiting time is $n^{2}
∑ (n-k)^{-2} = n^{2} ∑ k^{-2} ≤ n^2 π^{2}/6$. The final bound is $O(n^{2}
\log(1/ε))$.

This bound is much worse that the bound for move-to-top, which is
surprising.  In fact, the real bound is $O(n \log n)$ with high
probability, although the proof uses very different methods (see
\url{http://www.stat.berkeley.edu/~aldous/RWG/Chap7.pdf}).  
This shows that the coupling method doesn't always give tight bounds
(or perhaps we need a better coupling?).

\subsubsection{Random exchange of adjacent cards}
\label{section-shuffling-with-adjacent-swaps}

Suppose now that we only swap adjacent cards.  Specifically, we choose
one of the $n$ positions $i$ in the deck uniformly at random, and then
swap the cards a positions $i$ and $i+1 \pmod{n}$ with probability
$1/2$.  (The $1/2$ is there for the usual reason of avoiding periodicity.)

So now we want a coupling between the $X$ and $Y$ processes where each
possible swap occurs with probability $\frac{1}{2n}$ on both sides,
but somehow we correlate things so that like cards are pushed together
but never pulled apart.  The trick is that we will use the same
position $i$ on both sides, but be sneaky about when we swap.
In particular, we will aim to arrange things so that once some card is
in the same position in both decks, both copies move together, but
otherwise one copy changes its position by $\pm 1$ relative to the
other with a fixed probability $\frac{1}{2n}$.

The coupled process works like this.  Let $D$ be the set of indices
$i$ where the same card appears in both decks at position $i$ or at
position $i+1$.  Then we do:

\begin{enumerate}
 \item For $i∈D$, swap $(i,i+1)$ in both decks with probability $\frac{1}{2n}$.
 \item For $i\not∈D$, swap $(i,i+1)$ in the $X$ deck only with probability $\frac{1}{2n}$.
 \item For $i\not∈D$, swap $(i,i+1)$ in the $Y$ deck only with probability $\frac{1}{2n}$.
 \item Do nothing with probability $\frac{\card*{D}}{2n}$.
\end{enumerate}

It's worth checking that the total probability of all these events is
$\card*{D}/2n + 2(n-\card*{D})/2n + \card*{D}/2n = 1$.  More important is that if we
consider only one of the decks, the probability of doing a swap at
$(i,i+1)$ is exactly $\frac{1}{2n}$ (since we catch either case 1 or 2
for the $X$ deck or 1 or 3 for the $Y$ deck).

Now suppose that some card $c$ is at position $x$ in $X$ and $y$ in
$Y$.  If $x=y$, then both $x$ and $x-1$ are in $D$, so the only way
the card can move is if it moves in both decks: linked cards stay
linked.  If $x≠y$, then $c$ moves in deck $X$ or deck $Y$, but not
both.  (The only way it can move in both is in case 1, where $i=x$ and
$i+1=y$ or vice versa; but in this case $i$ can't be in $D$ since the
copy of $c$ at position $x$ doesn't match whatever is in deck $Y$, and
the copy at position $y$ doesn't match what's in deck $X$.)  In this
case the distance $x-y$ goes up or down by $1$ with equal probability
$\frac{1}{2n}$.  Considering $x-y \pmod{n}$, we have a ``lazy'' random walk
that moves with probability $1/n$, with absorbing barriers at $0$ and
$n$.  The worst-case expected time to converge is $n(n/2)^{2} =
n^{3}/4$, giving $\Prob{\text{time for $c$ to become linked} ≥
αn^{3}/2} ≤ 2^{-α}$ using the usual argument.  Now
apply the union bound to get $\Prob{\text{time for every $c$ to become
linked} ≥ αn^{3}/2} ≤ n2^{-α}$ to get an
expected coupling time of $O(n^{3} \log n)$.

To simplify the argument, we assumed that the deck wraps around, so
that we can swap the first and last card in addition to swapping
physically adjacent cards.  If we restrict to swapping $i$ and $i+1$
for $i ∈ \Set{0,\dots,n-2}$, we get a slightly different process that
converges in essentially the same time $O(n^3 \log n)$.
A result of David Bruce Wilson~\cite{Wilson2004})
shows both this upper bound holds
and that the bound is optimal up to a constant factor.

\subsubsection{Real-world shuffling}
In real life, the handful of people who still use physical playing
cards tend to use a \concept{dovetail shuffle}, which is closely
approximated by the reverse of a process where each card in a deck is
independently assigned to a left or right pile and the left pile is
place on top of the right pile.  Coupling doesn't really help much
here.  Instead, the process can be analyzed using more sophisticated
techniques due to Bayer and Diaconis~\cite{BayerD1992}.
The short version of the result is that $\Theta(\log n)$ shuffles are
needed to randomize a deck of size $n$.

\subsection{Path coupling}
\label{section-path-coupling}

If the states of our Markov process are the vertices of a graph, we
may be able to construct a coupling by considering a path between two
vertices and showing how to shorten this path on average at each step.
This technique is known as \index{coupling!path}{path
coupling}~\cite{BubleyD1997}.
Typically, the graph we use will be the graph of possible transitions
of the underlying Markov chain (possibly after making all edges
undirected).

There are two ideas at work here.  The first is that the expected distance
$\Exp{d(X_t,Y_t)}$ between $X_t$ and $Y_t$ in the graph gives an upper
bound on $\Prob{X_t≠Y_t}$ (by Markov's inequality, since if $X_t≠Y_t$
then $d(X_t,Y_t)≥1$).  The second is that to show that
$\ExpCond{d(X_{t+1},Y_{t+1})}{ℱ_t} ≤ α ⋅ d(X_t,Y_t)$ for
some $α<1$, it is enough to show how to contract a single edge, that
is, to show that 
$\ExpCond{d(X_{t+1},Y_{t+1}}{d(X_t,Y_t)=1} ≤ α$.  The reason is that if we have a
coupling that contracts one edge, we can apply this inductively along
each edge in the path to get a coupling between all the vertices in
the path that still leaves each pair with expected distance at most
$α$. The result for the whole path then follows from linearity of
expectation.

Formally,
instead of just looking at $X^{t}$ and $Y^{t}$, consider a path of
intermediate states 
$X^{t} = Z^t_0Z^t_1Z^t_2\dots{}Z^t_m = Y^{t}$,
where $d(Z_{i,t},Z_{i+1,t}) = 1$ for each $i$ (the vertices
are adjacent in the graph).
We now construct a coupling only
for adjacent nodes that reduces their distance on average.  The idea
is that $d(X^{t},Y^{t}) ≤ ∑ d\parens*{Z^t_i,Z^t_{i+1}}$, so if the
distance between each adjacent pair shrinks on average, so does the
total length of the path.

The coupling on each edge gives a joint conditional probability
\begin{displaymath}
    \ProbCond{Z^{t+1}_{i}=z'_i,Z^{t+1}_{i+1}=z'_{i+1}}{Z^t_{i}=z_i,Z^{t+1}_{i+1}=z_{i+1}}.
\end{displaymath}
We can extract from this a conditional distribution on $Z^{t+1}_{i+1}$
given the other three variables:
\begin{displaymath}
    \ProbCond{Z^{t+1}_{i+1}=z'_{i+1}}{Z^{t+1}_{i}=z'_{i},Z^t_{i}=z_i,Z^{t}_{i+1}=z_{i+1}}.
\end{displaymath}
Multiplying these conditional probabilities together lets us compute
a joint distribution on $X^{t+1},Y^{t+1}$ conditioned on $X^t,Y^t$,
which is the ordinary coupling we really want.

It's worth noting that the path is entirely notional, and we don't
actually keep it around
around between steps of the coupled Markov chains.  The only purpose
of $Z_0, Z_1, \dots, Z_m$ is to show that $X$ and $Y$ move closer
together.  Even though we could imagine that we are coalescing
these nodes together to create a new path at each step (or throwing in
a few extra nodes if some $Z_i,Z_{i+1}$ move away from each other), 
we are really computing a fresh path between $X$ and $Y$ at each step and
throwing it away as soon as it has done its job.

\subsubsection{Random walk on a hypercube}
\label{section-path-coupling-random-walk}

As a warm-up,
let's redo the argument about lazy random walks on a hypercube from
§\ref{section-random-walk-on-a-hypercube-via-coupling} using path
coupling. Each state $X^t$ or $Y^t$ is an $n$-bit vector, and with
probability $1/2$ a step flips one of the bits chosen uniformly at
random. The obvious metric $d$ is Hamming distance: $d(x,y)$ is the
number of indices $i$ for which $x_i ≠ y_i$.

For path coupling, we only need to push adjacent $Z^t_i$ and
$Z^t_{i+1}$. Adjacency means that there is exactly one index $j$ at
which these two bit-vectors differ. We apply the following coupling
(which looks suspiciously like the more generic coupling in
§\ref{section-random-walk-on-a-hypercube-via-coupling}):
\begin{enumerate}
    \item Pick a random index $r$.
    \item If $r ≠ j$, which occurs with probability $1-1/n$, flip
        position $r$ in both $Z^t_i$ and $Z^t_{i+1}$ with probability
        $1/2$. Whether we flip or not, we get
        $d(Z^{t+1}_i,Z^{t+1}_{i+1}) = 1$.
    \item If $r = j$, pick a new bit value $b$ uniformly at random
        and set position $j$ in
        both $Z^{t+1}_i$ and $Z^{t+1}_{i+1}$ to $b$. 
        From either side alone, this looks like flipping bit $j$ with
        probability $1/2$, exactly as in the original process. But now
        $d(Z^{t+1}_i, Z^{t+1}_{i+1}) = 0$.
\end{enumerate}

Averaging over both cases gives $\ExpCond{d(Z^{t+1}_{i},Z^{t+1}_{i+1})}{ℱ_t} = 1-1/n$. It
follows that $\ExpCond{d(X^t,Y^t)}{ℱ_0} ≤ (1-1/n)^t d(X_0,Y_0) ≤ n
e^{-n/t}$. Since $d(X^t,Y^t)$ is always at least $1$ if it's not zero,
this gives $\Prob{X^t ≠ Y^t} ≤ n e^{-n/t}$ by Markov's inequality, so
$d_{TV}(X^t,Y^t) ≤ ε$ after $O\parens*{n \log\frac{n}{ε}}$ steps.

\subsubsection{Sampling graph colorings}
\label{section-coupling-graph-colorings}

For this example, we'll look at sampling $k$-colorings of a 
graph with maximum degree $Δ$.  We will assume that
$k≥2Δ+1$.\footnote{Jerrum~\cite{Jerrum1995} provided the random walk
we'll be using, and Bubley and Dyer~\cite{BubleyD1997} used it as one
of their examples.
    Mitzenmacher and Upfal~\cite[§12.5]{MitzenmacherU2017} give an analysis based on a standard coupling for a
    different version of the Markov chain that works for the same
    bound on $k$.
More sophisticated results and a history of the
problem can be found in \cite{DyerGM2002}.}
For smaller values of $k$, it might still be the case that the chain
converges reasonably quickly for some graphs, but our analysis will not show this.

Unlike the hypercube case, the path coupling we will construct might
allow the distance between $X^t$ and $Y^t$ to rise with some
probability.  But the expected distance will always decrease, which is
enough.

Consider the following chain on proper $k$-colorings of a graph with
maximum
degree $Δ$.  At each step,
we choose one of the $n$ nodes $v$, 
compute the set $S$ of colors not
found on any of $v$'s neighbors, and recolor $v$ with a color chosen
uniformly from $S$ (which may be its original color).

Suppose $p_{ij} ≠ 0$.  Then $i$ and $j$ differ in at most one place
$v$, and so the set $S$ of permitted colors for each process—those not found on $v$'s
neighbors—are the
same.  This gives $p_{ij} = p_{ji} = \frac{1}{n⋅\card*{S}}$, and
the detailed
balance equations \eqref{eq-detailed-balance} hold when $π_i$ is
constant for all $i$.   So we have a reversible Markov chain with a
uniform stationary distribution.  Now we will apply a path coupling to
show that we converge to this stationary distribution reasonably
quickly when $k$ is large enough.

We'll think of colorings as vectors.  Given two colorings $x$ and $y$,
let $d(x,y)$ be the Hamming distance between them, which is the number
of nodes $u$ for which $x_u ≠ y_u$.  To show
convergence, we will construct a coupling that shows that
$d(X^{t},Y^{t})$ converges to $0$ over time starting from arbitrary
initial points $X^{0}$ and $Y^{0}$.  

A complication is that it's not immediately evident
that the length of the shortest path from $X^t$ to $Y^t$ in the
transition graph of our Markov chain is
$d(X^{t},Y^{t})$.  The problem is that it may not be possible to
transform $X^{t}$ into $Y^{t}$ one node at a time without producing
improper colorings.  With enough colors, we can explicitly construct a
short path between $X^{t}$ and $Y^{t}$ that uses only proper
colorings; but for this particular process it is easier to simply
extend the Markov chain to allow improper colorings, and show that our
coupling works anyway.  This also allows us to start with an improper
coloring for $X^{0}$ if we are particularly lazy.
The stationary distribution is not affected, because if $i$ is a
proper coloring and $j$ is an
improper coloring that differs from $i$ in exactly one place, we have
$p_{ij} = 0$ and $p_{ji} ≠ 0$, so the detailed balance equations hold
with $π_j = 0$.

The natural coupling to consider given
adjacent $X^{t}$ and $Y^{t}$ is to pick the same node and the same new
color for both, provided we can do so.  If we pick the one node $v$ on which they differ, and
choose a color that is not used by any neighbor (which will be the
same for both copies of the process, since all the neighbors have the
same colors), then we get $X^{t+1} = Y^{t+1}$; this event occurs with
probability at least $1/n$.  If we pick a node that is neither
$v$ nor adjacent to it, then the distance between $X$
and $Y$ doesn't change; either both get a new identical color or both
don't.  

Things get a little messier when we pick some node $u$ adjacent to
$v$, an event that occurs with probability at most $Δ/n$.
Let $c$ be the color of $v$ in $X^t$, $c'$ the color of $v$ in $Y^t$,
and $T$ the set of colors that do not appear among the other neighbors
of $u$.  Let $\ell = \card{T} ≥ k - (Δ-1)$.

Conditioning on choosing $u$ to recolor, $X^{t+1}$ picks a
color uniformly from $T ∖ \Set{c}$ and $Y^{t+1}$ picks a color
uniformly from $T ∖ \Set{c'}$.  We'd like these colors to be the same
if possible, but these are not the same sets, and they aren't even
necessarily the same size.

There are three cases:
\begin{enumerate}
    \item Neither $c$ nor $c'$ are in $T$.  Then $X^{t+1}$ and
        $Y^{t+1}$ are choosing a new color from the same set, and we
        can make both choose the same color: the distance between $X$
        and $Y$ is unchanged.
    \item Exactly one of $c$ and $c'$ is in $T$.  Suppose that it's
        $c$.  Then $\card*{T ∖ \Set{c}} = \ell-1$
        and $\card*{T ∖ \Set{c'}} = \card*{T} = \ell$.  Let $X^{t+1}$
        choose a new color $c''$ first.  Then let $Y^{t+1}_u =
        c''$ with probability $\frac{\ell-1}{\ell}$ (this gives a
        probability of $\frac{1}{\ell}$ of picking each color in $T ∖
        \Set{c}$, which is what we want), and let $Y^{t+1}_u = c$ with
        probability $\frac{1}{\ell}$.  Now the distance between $X$
        and $Y$ increases with probability $\frac{1}{\ell}$.
    \item Both $c$ and $c'$ are in $T$.  For each $c''$ in $T ∖
        \Set{c,c'}$, let $X^{t+1}_u = Y^{t+1}_u = c''$ with probability
        $\frac{1}{\ell-1}$; since there are $\ell-2$ such $c''$, this
        accounts for $\frac{\ell-2}{\ell-1}$ of the probability.
        Assign the remaining $\frac{1}{\ell-1}$ to $X^{t+1}_u = c',
        Y^{t+1}_u = c$.
        In this case the distance between $X$ and $Y$ increases with
        probability $\frac{1}{\ell-1}$, making this the worst case.
\end{enumerate}

Putting everything together, we have a $1/n$ chance of picking a node
that guarantees to reduce $d(X,Y)$ by $1$, and at most a $Δ/n$ chance
of picking a node that may increase $d(X,Y)$ by at most
$\frac{1}{\ell-1}$ on average, where $\ell ≥ k - Δ + 1$, giving a
maximum expected increase of $\frac{Δ}{n} ⋅ \frac{1}{k-Δ}$.  
So
\begin{align*}
    \ExpCond{d(X^{t+1},Y^{t+1})-d(X^t,Y^t)}{d(X^t,Y^t)=1}
    &≤ \frac{-1}{n} + \frac{Δ}{n} ⋅ \frac{1}{k-Δ}
    \\&= \frac{1}{n} \parens*{-1 + \frac{Δ}{k-Δ}}
    \\&= \frac{1}{n} \parens*{\frac{-(k-Δ)+Δ}{k-Δ}}
    \\&= - \frac{1}{n} \parens*{\frac{k-2Δ}{k-Δ}}.
\end{align*}

So we get 
\begin{align*}
    d_{TV}(X^t,Y^t)
    &≤ \Prob{X^t≠Y^t}
    \\&≤ \parens*{1-\frac{1}{n}⋅\frac{k-2Δ}{k-Δ}}^t ⋅ \Exp{d(X^0,Y^0)}
    \\&≤ \exp\parens*{-\frac{t}{n}⋅\frac{k-2Δ}{k-Δ}} ⋅ n.
\end{align*}

For fixed $k$ and $Δ$ with $k > 2Δ$, this is $e^{-Θ(t/n)} n$, which will be less than
$ε$ for $t = Ω\parens*{n \parens*{\log n + \log (1/ε)}}$.

\subsubsection{Sampling independent sets}
\label{section-sampling-independent-sets}

For a more complicated example of path coupling, let's try 
sampling independent sets of vertices on a graph $G = (V,E)$ with $n$
vertices and $m$ edges.
If we can bias in favor of larger sets, we might even get a good
independent set approximation!  The fact that this is hard will
console us when we find that it doesn't work.

A natural way to set up
the random walk is to represent each potentially independent set as a
bit vector, where $1$ indicates membership in the set, and at each
step we pick one of the $n$ bits uniformly at random and set it to $0$
or $1$ with probability $1/2$ each, provided that the resulting set is
independent.  If the resulting set is not independent, we stay
put.\footnote{In statistical physics, this process of making a local
    change with probability proportional to how much we like the
    result goes by the name of
\index{dynamics!Glauber}\concept{Glauber dynamics}.}

It's easy to see that $d(x,y) = \norm*{x-y}_1$ 
is a bound on the length of the minimum
number of transitions to get from $x$ to $y$, since we can always
remove all the extra ones from $x$ and put back the extra ones in $y$
while preserving independence throughout the process.  
(This argument also shows that the Markov chain is irreducible.)
It may be hard to find the exact minimum path
length, so 
we'll use this
distance instead for our path coupling.

We can easily show that the stationary distribution of this process is
uniform.  The essential idea is that if we can transform one
independent set $S$ into another $S'$ by flipping a bit, then we can go back by
flipping the bit the other ways.  Since each transition happens with
the same probability $1/n$, we get $π_S⋅(1/n) = π_{S'}⋅(1/n)$ and $π_S
= π_{S'}$. Since we can apply this equation along a path between any
two states, all states must have the same probability in the unique
stationary distribution.

To prove convergence, it's tempting to start with
the obvious coupling, even though it doesn't actually work.  Pick the same
position and value for both copies of the chain.  If $x$ and $y$ are
adjacent, then they coalesce with probability $1/n$ (both probability
$1/2n$ transitions are feasible for both copies, since the neighboring
nodes always have the same state).  What is the probability that they
diverge?  We can only be prevented from picking a value if the value
is $1$ and some neighbor is $1$.  So the bad case is when $x_{i} = 1$,
$y_{i} = 0$, and we attempt to set some neighbor of $i$ to $1$; in the
worst case, this happens $Δ/2n$ of the time, which is at least $1/n$
when $Δ ≥ 2$.  No coalescence here!

This could be a sign that our random walk is no good, or it could be a
sign that our coupling is no good.
But we can avoid figuring out which is the case by using
a sneakier random walk.
We'll adopt the approach used in~\cite[§12.6]{MitzenmacherU2017}.

Here the idea is that we pick a random edge $uv$, and then try to do one of the following operations, all with equal probability:
\begin{enumerate}
 \item Set $u=v=0$.
 \item Set $u=0$ and $v=1$.
 \item Set $u=1$ and $v=0$.
\end{enumerate}
In each case, if the result would be a non-independent set, we instead
do nothing.

Verifying that this has a uniform stationary distribution is mildly
painful if we are not careful, since there may be several different
transitions that move from some state $x$ to the same state $y$.  But
for each transition (occurring with probability $\frac{1}{3m}$), we
can see that there is a reverse transition that occurs with equal
probability; so the detailed balance equations
\eqref{eq-detailed-balance} hold with uniform
probabilities.  Note that we can argue this even though we don't know
what the actual stationary probabilities are, since we don't know how
many independent sets our graph has.

So now what happens if we run two coupled copies of this process,
where the copies differ on exactly one vertex $i$?

First, every neighbor of $i$ is $0$ in both processes.  A transition
that doesn't involve any neighbors of $i$ will have the same effect on
both processes.  So we need to consider all choices of edges where one
of the endpoints is either $i$ or a neighbor $j$ of $i$.  In the case
where the other endpoint isn't $i$, we'll call it $k$; there may be
several such $k$.

If we choose $ij$ and don't try to set $j$ to one, we always coalesce the
states.  This occurs with probability $\frac{2}{3m}$.  If we try to
set $i$ to zero and $j$ to one, we may fail in both processes, because $j$ may have a
neighbor $k$ that is already one; this will preserve the distance
between the two processes.
Similarly, if we try to set $j$ to
one as part of a change to some $jk$, we will also get a divergence
between the two processes: in this case, the distance will actually
increase.  This can only happen if $j$ has at most one neighbor $k$
(other than $i$) that is already in the independent set; if there are
two such $k$, then we can't set $j$ to one no matter what the state of
$i$ is.

This argument suggests that we need to consider three cases for each
$j$, depending on the number $s$ of nodes $k ≠ i$ that are adjacent to $j$ 
and have $x_k = y_k = 1$.  In each case, we assume $x_i=0$ and
$y_i=1$, and that all other nodes have the same value in both $x$ and
$y$.  (Note that these assumptions mean that any such $k$ can't be
adjacent to $i$, because we have $y_k = y_i = 1$.)
\begin{itemize}
\item $s=0$.  Then if we choose $ij$, we can always set $i$ and $j$
however we like, giving a net $-\frac{1}{m}$ expected change to the
distance.  However, this is compensated for by up to $d-1$ attempts to
set $j=1$ and $k=0$ for some $k$, all of which fail in one copy of the
process but succeed in the other.  Since $k$ doesn't change, each of
these failures adds only $1$ to the distance, which becomes at most
$\frac{d-1}{3m}$ total.  So our total expected change in this case is
at most $\frac{d-4}{3m}$.
\item $s=1$.  Here attempts to set $i=0$ and $j=1$ fail in both
processes, giving only a $-\frac{2}{3m}$ expected change after picking
$ij$.  Any change to $jk$ fails only if we set $j=1$, which we can
only do in the $x$ process and only if we also set $k=0$ for the
unique $k$ that is currently one.  This produces an increase in the
distance of $2$ with probability $\frac{1}{3m}$, exactly canceling out
the decrease from picking $ij$.  Total expected change is $0$.
\item $s=2$.  Now we can never set $j=1$.  So we drop $-\frac{2}{3m}$
from changes to $ij$ and have no change in distance from updates to
$jk$ for any $k≠ i$.
\end{itemize}

Considering all three cases, in the worst case we
have $\ExpCond{d(X_{t+1},Y_{t+1}}{X_t,Y_t} = d(X_t,Y_t) + \frac{Δ-4}{3m}$.
For $Δ≤3$ (a pretty restrictive case), this is $-1/3m$, giving a
decent enough expected coupling time of $O(m \log n)$.

Here, we've considered the case where all independent sets have the
same probability.  One can also bias the random walk in favor of
larger independent sets by accepting increases with higher probability
than decreases (as in Metropolis-Hastings); this samples independent
sets of size $s$ with probability proportional to $λ^s$.
Some early examples of this approach are given
in~\cite{LubyV1997,LubyV1999,DyerG2000}.  The question of exactly
which values of $λ$ give polynomial convergence times is still
open; see~\cite{MosselWW2007} for some more recent bounds.

\subsubsection{Simulated annealing}
\label{section-metropolis-convergence}
\label{section-simulated-annealing}

Recall that the
\index{Metropolis-Hastings!convergence}Metropolis-Hastings algorithm
constructs a reversible Markov chain with a desired stationary
distribution from any reversible Markov chain on the same states (see
§\ref{section-metropolis} for details.)

A variant, which generally involves tinkering with the chain
while it's running, is the global optimization heuristic known as
\concept{simulated annealing}~\cite{KirkpatrickGV1983}.  Here we have some function $g$ that we
are trying to minimize.  So we set $f(i) = \exp(-αg(i))$ for
some $α>0$.  Running Metropolis-Hastings gives a stationary
distribution that is exponentially weighted to small values of $g$; if
$i$ is the global minimum and $j$ is some state with high $g(j)$, then
$π(i) = π(j) \exp(α(g(j)-g(i))$, which for large enough
$α$ goes a long way towards compensating for the fact that in
most problems there are likely to be exponentially more bad $j$'s than
good $i$'s.  The problem is that the same analysis applies if $i$ is a
local minimum and $j$ is on the boundary of some depression around
$i$; large $α$ means that it is exponentially unlikely that we escape this depression and find the global minimum.

The simulated annealing hack is to vary $α$ over time;
initially, we set $α$ small, so that we get mixing time close
to that of the original Markov chain.  This gives us a sample that is
roughly uniform, with a small bias towards states with smaller $g(i)$.
After some time we increase $α$ to force the process into
better states.  The hope is that by increasing $α$ slowly, by
the time we are stuck in some depression, it's a deep one—optimal or
close to it.  If it doesn't work, we can randomly restart and/or
decrease $α$ repeatedly to jog the chain out of whatever
depression it is stuck in.  How to do this effectively is deep voodoo
that depends on the structure of the underlying chain and the shape of
$g(i)$, so most of the time people who use simulated annealing just
try it out with some generic \concept{annealing schedule} and hope it gives some useful result.  (This is what makes it a heuristic rather than an algorithm.  Its continued survival is a sign that it does work at least sometimes.)

Here are toy examples of simulated annealing with provable
convergence times.

\paragraph{Single peak} Let's suppose $x$ is a random walk on an $n$-dimensional hypercube
(i.e., $n$-bit vectors where we set $1$ bit at a time), $g(x) =
\card*{x}$, and we want to maximize $g$.  Now a transition that increase
$\card*{x}$ is accepted always and a transition that decreases
$\card*{x}$ is accepted only with probability $e^{-α}$.  For large
enough $α$, this puts a constant fraction of $π$ on the
single peak at $x=\mathbf{1}$; the observation is that that there are
only $\binom{n}{k} ≤ n^{k}$ points with $k$ zeros, so the total
weight of all points is at most $π(\mathbf{1}) ∑_{k≥0}
n^{k} \exp(-αk) = π(\mathbf{1}) ∑ \exp(\ln n -
α)^{k} = π(\mathbf{1}) / (1 - \exp(\ln n - α)) =
π(\mathbf{1}) ⋅ O(1)$ when $α > \ln n$, giving
$π(\mathbf{1}) = Ω{}(1)$ in this case.

So what happens with convergence?  Let $p = \exp(-α)$.  Let's
try doing a path coupling 
between two adjacent copies $x$ and $y$ of the Metropolis-Hastings
process, where we first pick a bit to change, then pick a value to
assign to it, accepting the change in both processes if we can.  The
expected change in $\abs*{x-y}$ is then $(1/2n) (-1-p)$, since if we
pick the bit where $x$ and $y$ differ, we have probability $1/2n$ of
setting both to $1$ and probability $p/2n$ of setting both to $0$, and
if we pick any other bit, we get the same distribution of outcomes in
both processes.  This gives a general bound of
$\ExpCond{\abs*{X_{t+1}-Y_{t+1}} }{ \abs*{X_{t}-Y_{t}}} ≤ (1-(1+p)/2n)
\abs*{X_{t}-Y_{t}}$, from which we have $\Exp{\abs*{X_{t}-Y_{t}}} ≤
\exp(-t(1+p)/2n) \Exp{\abs*{X_{0}-Y_{0}}} ≤ n \exp(-t(1+p)/2n)$.  So
after $t = 2n/(1+p) \ln(n/ε)$ steps, we expect to converge to
within $ε$ of the stationary distribution in total variation
distance.
This gives an $O(n \log n)$ algorithm for finding the peak.

This is kind of a silly example, but if we suppose that $g$ is better
disguised (for example, $g(x)$ could be $\abs*{x ⊕ r}$ where $r$
is a random bit vector), then we wouldn't really expect to do much
better than $O(n)$.  So $O(n \log n)$ is not bad for an algorithm with
no brains at all.

\paragraph{Somewhat smooth functions}
Now we'll let $g:2^{n}→ℕ$ be some arbitrary
\index{Lipschitz function}Lipschitz function, that is, a function with
the property that $\abs*{g(x)-g(y)} ≤ \abs*{x-y}$, and ask
for what values of $p = e^{-α}$ the Metropolis-Hastings walk with
$f(i) = e^{-αg(i)}$ can be shown to converge quickly.  Given
adjacent states $x$ and $y$, with $x_{i}≠y_{i}$ but $x_{j}=y_{j}$
for all $j≠i$, we still have a probability of at least $(1+p)/2n$
of coalescing the states by setting $x_{i}=y_{i}$.  But now there is a
possibility that if we try to move to $(x[j/b], y[j/b])$ for some $j$
and $b$, that $x$ rejects while $y$ does not or vice versa (note if
$x_{j}=y_{j}=b$, we don't move in either copy of the process).
Conditioning on $j$ and $b$, this occurs with probability $1-p$
precisely when $x[j/b] < x$ and $y[j/b] ≥ y$ or vice versa, giving
an expected increase in $\abs*{x-y}$ of $(1-p)/2n$.  We still get an
expected net change of $-2p/2n = -p/n$ provided there is only one
choice of $j$ and $b$ for which this occurs.  So we converge in time
$τ(ε) ≤ (n/p) \log(n/ε)$ in this
case.\footnote{You might reasonably ask if such functions $g$ exist.
One example is $g(x) = (x_{1}⊕{}x_{2}) + ∑_{i>2} x_{i}$.}

One way to think of this is that the shape of the neighborhoods of
nearby points is similar.  If I go up in a particular direction from
point $x$, it's very likely that I go up in the same direction from
some neighbor $y$ of $x$.

If there are more bad choices for $j$ and $b$, then we need a much
larger value of $p$: the expected net change is now $(k(1-p)-(1+p))/2n =
(k-1-(k+1)p)/2n$, which is only negative if $p > (k-1)/(k+1)$.  This
gives much weaker pressure towards large values of $g$, which still
tends to put us in high neighborhoods but creates the temptation to
fiddle with $α$ to try to push us even higher once we think we are close to a peak.

\section{Spectral methods for reversible chains}
\label{section-spectral-convergence-bounds}

(See also \url{http://www.stat.berkeley.edu/~aldous/RWG/Chap3.pdf}, from which many of the details in the notes below are taken.)

The problem with coupling is that (a) it requires cleverness to come
up with a good coupling; and (b) in many cases, even that doesn't
work—there are problems for which no coupling that only depends on
current and past transitions coalesces in a reasonable amount of
time.\footnote{A coupling that doesn't require looking into the future
is called a \index{coupling!causal}\concept{causal
coupling}. An example of a Markov chain for which causal couplings are
known not to work is the one used for sampling perfect matchings in bipartite
graphs as described in
§\ref{section-congestion-perfect-matchings}~\cite{KumarR1999}.}
When we run into these problems, we may be able to show convergence
instead using a linear-algebraic approach, where we look at the
eigenvalues of the transition matrix of our Markov chain.
This approach works best for reversible Markov chains, where $π_i
p_{ij} = π_j p_{ij}$ for all states $i$ and $j$ and some
distribution $π$.

\subsection{Spectral properties of a reversible chain}
\label{sections-spectral-properties-of-reversible-chains}

Suppose that $P$ is the transition matrix of an irreducible,
reversible Markov chain.  Then it has a unique
stationary distribution $π$ that is a left \concept{eigenvector} corresponding
to the \concept{eigenvalue} $1$, which just means that $πP = 1π$.  
For irreducible aperiodic chains, $P$ will have a total of $n$ real eigenvalues
$λ_1 > λ_2 ≥ λ_3 ≥ \dots ≥ λ_n$, where $λ_1 = 1$ and $λ_n > -1$.  (This
follows for symmetric chains from the
\index{theorem!Perron-Frobenius}\concept{Perron-Frobenius theorem},
which we will not attempt to prove here; for general irreducible
aperiodic reversible chains, we will show below that we can reduce to the symmetric case.)
Each eigenvalue $λ_i$ has a corresponding eigenvector $u^i$, a nonzero
vector that
satisfies $u^i P = λ_i u^i$.  In Markov chain terms, these
eigenvectors correspond to deviations from the stationary distribution
that shrink over time.

For example, the transition matrix
\begin{align*}
S =
\begin{bmatrix}
p & q \\
q & p 
\end{bmatrix}
\end{align*}
corresponding to a Markov chain on two states that stays in the same
state with probability $p$ and switches to the other state with
probability $q = 1-p$ has eigenvectors $u_1 = \rowvector{1& 1}$ and
$u_2 = \rowvector{1 &-1}$ with corresponding eigenvalues $λ_1 = 1$ and
$λ_2 = p-q$, as shown by computing
\begin{align*}
\rowvector{1&1}
\begin{bmatrix}
p & q \\
q & p 
\end{bmatrix}
&= \rowvector{p+q & q+p} = 1 ⋅ \rowvector{1 & 1}
\intertext{and}
\rowvector{1&-1}
\begin{bmatrix}
p & q \\
q & p 
\end{bmatrix}
&= \rowvector{p-q& q-p} = (p-q) ⋅ \rowvector{1 & -1}.
\end{align*}

\subsection{Analysis of symmetric chains}

To make our life easier, we will assume that in addition to being
reversible, our Markov chain has a uniform stationary distribution.
Then $π_i = π_j$ for all $i$ and $j$, and so reversibility implies
$p_{ij} = p_{ji}$ for all $i$ and $j$ as well, meaning that the
transition matrix $P$ is symmetric.  Symmetric matrices have the very
nice property that their eigenvectors are all orthogonal (this is
the \index{theorem!spectral}\concept{spectral theorem}), which allows
for a very straightforward decomposition of the distribution on the
initial state (represented as a vector $x_0$) as a linear combination
of the eigenvectors.  For example, if initially we put all our weight on state $1$,
we get $x_0 = \rowvector{1& 0} = \frac{1}{2} u^1 + \frac{1}{2} u^2$.

If we now take a step in the chain, we multiply $x$ by $P$.  We can
express this as
\begin{align*}
xP
&= \parens*{\frac{1}{2} u^1 + \frac{1}{2} u^2} P
\\&= \frac{1}{2} u^1 P + \frac{1}{2} u^2 P
\\&= \frac{1}{2} λ_1 u^1 + \frac{1}{2} λ_2 u^2.
\end{align*}

This uses the defining property of eigenvectors, that $u^i P = λ_i
u^i$.

In general, if $x = ∑_i a_i u^i$, then $xP = ∑_i a_i λ_i u^i$ and
$xP^t = ∑ a_i λ_i^t u^i$.  For any eigenvalue $λ_i$ with 
$\abs*{λ_i} < 1$, $λ_i^t$ goes to zero in the limit.  So only those
eigenvectors with $λ_i=1$ survive.  For irreducible aperiodic chains, these
consist only of the stationary distribution $π = u^i / \norm*{u^i}_1$.
For reducible or periodic chains, there may be some additional eigenvalues with
$\abs*{λ_i}=1$, but the only possibility that arises for an
irreducible reversible
chain is $λ_n = -1$, corresponding to a chain with period $2$.\footnote{Chains with periods greater than $2$ (which are never
reversible) have pairs of
complex-valued eigenvalues that are roots of unity, which happen to
cancel out to only produce real probabilities in $vP^{t}$.  Chains
that aren't irreducible will have one eigenvector with eigenvalue $1$
for each final component; the stationary distributions of these chains
are linear combinations of these eigenvectors (which are just the
stationary distributions on each component).}

Assuming
that $\abs*{λ_{2}} ≥ \abs*{λ_{n}}$, as $t$ grows
large $λ_{2}^{t}$ will dominate the other smaller eigenvalues,
and so the size of $λ_{2}$ will control the rate of
convergence of the underlying Markov process.

This assumption is
always true for lazy walks that stay put with probability $1/2$,
because all eigenvalues of a lazy walk are non-negative. The reason is
that any such walk has transition matrix
$\frac{1}{2}(P+I)$, where $I$ is the identity matrix and $P$ is the
transition matrix of the unlazy version of the walk.  If $xP = λx$
for some $x$ and $λ$, then $x\parens*{\frac{1}{2}(P+I)} =
\parens*{\frac{1}{2}(λ+1)}x$.  This means that $x$ is still an
eigenvector of the lazy walk, and its corresponding eigenvalue is
$\frac{1}{2}(λ+1) ≥ \frac{1}{2}((-1)+1) ≥ 0$.

But what does having small $λ_2$ mean for total variation distance?  If $x^t$
is a vector representing the distribution of our position at time $t$,
then $d_{TV} = \frac{1}{2} ∑_{i=1}^{n} \abs*{x^t_i - π_i} =
\frac{1}{2} \norm*{x^t - π}_1$.  But we know that 
$x^0 = π + ∑_{i=2}^{n} c_i u^i$ for some coefficients $c_i$, and
$x^t = π + ∑_{i=2}^{n} λ_i^t c_i u^i$.  So we are looking for a bound
on $\norm*{∑_{i=2}^{n} λ_i^t c_i u^i}_1$.

It turns out that it is easier to get a bound on the $\ell_2$ norm
$\norm*{∑_{i=2}^{n} λ_i^t c_i u^i}_2$.
Here we use the fact that the eigenvectors are orthogonal.  This means
that the Pythagorean theorem holds and $\norm*{∑_{i=2}^{n} λ_i^t c_i
u^i}_2^2 = ∑_{i=2}^n λ_i^{2t} c_i^2 \norm{u^i}^2_2 = ∑_{i=2}^{n}
λ_i^{2t} c_i^2$
if we normalize each $u^i$ so that $\norm{u^i}_2^2 = 1$.  But then
\begin{align*}
    \norm*{x^t - π}_2^2
    &= ∑_{i=2}^{n} λ_i^{2t} c_i^2
    \\&≤ ∑_{i=2}^{n} λ_2^{2t} c_i^2
    \\&= λ_2^{2t} ∑_{i=2}^{n} c_i^2
    \\&= λ_2^{2t} \norm{x^0 - π}_2^2.
    \intertext{Now take the square root of both sides to get}
    \norm*{x^t-π}_2
    &=< λ_2^t \norm*{x^0 - π}.
\end{align*}

To translate this back into $d_{TV}$, we use the inequality
\begin{equation*}
    \norm{x}_2 ≤ \norm{x}_1 ≤ √{n} ⋅ \norm{x}_2,
\end{equation*}
which holds for any $n$-dimensional vector $x$.
Because $\norm{x^0}_1 = \norm{π}_1 = 1$, $\norm{x^0-π}_1 ≤ 2$ by the
triangle inequality, which also gives $\norm{x^0-π}_2 ≤ 2$.  So
\begin{align*}
    d_{TV}(x^t,π)
    &= \frac{1}{2} \norm{x^t-π}_1
    \\&≤ \frac{√{n}}{2} λ_2^t \norm{x^0-π}_2
    \\&≤ λ_2^t √{n}.
\end{align*}
to get a bound on $\norm{x-π}_1$.

If we want to get $d_TV(x^t,n) ≤ ε$, we will need $t \ln(λ_2) + \ln
√{n}
≤ \ln ε$ or $t ≥ \frac{1}{\ln (1/λ_2)}\parens*{\frac{1}{2} \ln n + \ln(1/ε)}$.

The factor $\frac{1}{\ln (1/λ_2)} = \frac{1}{-\ln λ_2}$ can be
approximated by $τ_2 = \frac{1}{1-λ_2}$, which is called the 
\concept{mixing rate} or
\index{time!relaxation}\concept{relaxation time} 
\index{$τ_2$}
of the Markov chain.  Indeed, our old friend $1+x ≤ e^x$ implies $\ln
(1+x) ≤ x$, which gives $\ln λ_2 ≤ λ_2 - 1$ and thus $\frac{1}{-\ln
λ_2} ≤ \frac{1}{1 - λ_2} = τ_2$.  So $τ_2 \parens*{\frac{1}{2}\ln n + \ln (1/ε)}$ gives a
conservative estimate on the time needed to achieve $d_{TV}(x^t, π) ≤ ε$
starting from an arbitrary distribution.

It's worth comparing this to the mixing time $t_{\mix} =
\argmin_t d_{TV}(x^t,π) ≤ 1/4$ that we used with coupling. With $τ_2$,
we have to throw in an extra factor of $\frac{1}{2} \ln n$ to get the
bound down to $1$, but after that the bound continues to drop by a
factor of $e$ every $τ_2$ steps.  With $t_{\mix}$, we have to go through
another $t_{\mix}$ steps for each factor of $2$ improvement in
total variation distance, even if the initial drop avoids the log
factor.  Since we can't always tell whether coupling arguments or spectral
arguments will do better for a particular chain, and since convergence
bounds are often hard to obtain no matter how many techniques we throw
at them, we will generally be happy with any reasonable bound on
either $τ_2$ or $t_{\mix}$.

\subsection{Analysis of asymmetric chains}
\label{section-spectral-convergence-asymmetric}

If the stationary distribution is not uniform, then in general the
transition matrix will not be symmetric.  We can make it symmetric by
scaling the probability of being in the $i$-th state by $π_i^{-1/2}$.
The idea is to decompose our transition matrix $P$ as $ΠAΠ^{-1}$,
where $Π$ is a diagonal matrix with $Π_{ii} = π_i^{-1/2}$, and 
$A_{ij} = √{\frac{π_i}{π_j}} P_{ij}$.  Then $A$ is symmetric,
because
\begin{align*}
    A_{ij} 
    &= √{\frac{π_i}{π_j}} P_{ij}
    \\&= √{\frac{π_i}{π_j}} \frac{π_j}{π_i} P_{ji}
    \\&= √{\frac{π_j}{π_i}} P_{ji}
    \\&= A_{ji}.
\end{align*}

This means in particular that $A$ has orthogonal eigenvectors
$u^1,u^2,\dots,u^n$ with corresponding eigenvalues
$λ_1≥λ_2≥\dots≥λ_n$.  These eigenvalues carry over to $P = ΠAΠ^{-1}$,
since $(u^i Π^{-1}) P = u^i Π^{-1} Π A Π^{-1} = u^i A Π^{-1} = λ_i
u^i Π^{-1}$.

So now given an initial distribution $x_0$, we can first pass it
through $Π$ and then apply the same
reasoning as before to show that $\norm{πΠ-xΠ}_2$ shrinks by $λ_2$
with every step of the Markov chain.  The difference now is that the
initial distance is affected by the scaling we did in $Π^{-1}$; so instead
of getting $d_{TV}(x^t,π) ≤ λ_2^t √{n}$, we get $d_{TV}(x^t,π) ≤
λ_2^t (π_{\min})^{-1/2}$, where $π_{\min}$ is the smallest probability of
any single node in the stationary distribution $π$.  The bad initial
$x_0$ in this case is the one that puts all its weight on this node,
since this maximizes its distance from $π$ after scaling.

(A uniform $π$ is a special case of this, since when $π_i = 1/n$ for
all $i$, $π_{\min}^{-1/2} = (1/n)^{-1/2} = √{n}$.)

For more details on this, see \cite[§3.4]{AldousF2001}.

So now we just need a tool for bounding $λ_{2}$.
For a small chain with a known transition matrix, we can just feed it
to our favorite linear algebra library, but most of the time we will
not be able to construct the matrix explicitly.  So we need a way to
bound $λ_2$ indirectly, in terms of other structural properties of our
Markov chain.

\section{Conductance}
\label{section-conductance}
\label{section-Cheegers-inequality}

The \concept{conductance} or \concept{Cheeger constant}
\index{$Φ$}
$Φ(S)$ of a
set $S$ of states in a Markov chain is
\begin{align}
\label{eq-conductance-set}
Φ(S) &= \frac{∑_{i∈ S, j ∉ S} π_i p_{ij}}{π(S)}.
\end{align}
This is the probability of leaving $S$ on the next step starting from
the stationary distribution conditioned on being in $S$. The
conductance is a measure of how easy it is to escape from a set.
It can also be thought of as a weighted version of edge expansion.

The conductance of a Markov chain as a whole is obtained by taking the
minimum of $Φ(S)$ over all $S$ that occur with probability at most
$1/2$:
\begin{align}
\label{eq-conductance}
Φ &= \min_{0 < π(S) ≤ 1/2} Φ(S).
\end{align}

The usefulness of conductance is that it
bounds $λ_{2}$:
\begin{theorem}
\label{theorem-conductance}
In a reversible Markov chain, 
\begin{equation}
\label{eq-conductance-bound}
1 - 2Φ ≤ λ_{2} ≤ 1 - Φ^{2}/2.
\end{equation}
\end{theorem}

The bound \eqref{eq-conductance-bound} is known as the 
\index{inequality!Cheeger}\concept{Cheeger inequality}.
We won't attempt to prove it here, but the intuition is that in order
for a reversible chain not to mix, it has to get stuck in some subset
of the states. Having high conductance prevents this disaster, while having
low conductance causes it.\footnote{For a slightly more detailed
version of this intuition, any
eigenvector that is not the stationary distribution is going to be
orthogonal to the stationary distribution after applying an
appropriate scaling to both. This means we can split its
weights into positive weights and negative weights, and reducing
its length means shifting weight from the positive weights to the
negative weights. If we can't shift very much of this weight, this
means that one of these two sets of weights has low conductance.
Conversely, if we have an $S$ with minimum 
conductance, putting positive weight on $S$ and negative weight on its
complement in the right way can give us a vector that doesn't shrink
very quickly.

A real version of this argument, along with several other proofs of
Cheeger's inequality, can be found in~\cite{Chung2007}.}

For lazy walks we always have $λ_{2} = λ_{\max}$, and so we can
convert \eqref{eq-conductance-bound} to a bound on the relaxation time:
\begin{corollary}
\label{corollary-conductance-relaxation-time}
\begin{align}
\label{eq-conductance-relaxation-time}
\frac{1}{2Φ} &≤ τ_{2} ≤ \frac{2}{Φ^{2}}.
\end{align}
\end{corollary}

In other words, high conductance implies low relaxation time and vice versa, up to squaring.

\subsection{Easy cases for conductance}
\label{section-conductance-easy-cases}

For very simple Markov chains we can compute the conductance directly.
Consider a lazy random walk on a cycle.  Any proper subset $S$
has at least two outgoing edges, each of which carries a flow of
$1/4n$, giving $Φ_{S} ≥ (1/2n)/π(S)$.  If we now take the
minimum of $Φ_{S}$ over all $S$ with $π(S) ≤ 1/2$, we get
$φ ≥ 1/n$, which gives $τ_{2} ≤ 2n^{2}$.  This is essentially the same bound as we got from coupling.

Here's a slightly more complicated chain.  Take two copies of $K_{n}$,
where $n$ is odd,
and join them by a path with $n$ edges.  Now consider $Φ_{S}$ for
a lazy random walk on this graph where $S$ consists of half the graph,
split at the edge in the middle of the path.  There is a single outgoing edge
$uv$, with $π(u) = d(u)/2\card*{E} = 2/(2n(n-1)n/2 + n) = 2n^{-2}$ and
$p_{uv} = 1/4$, for $π(u) p_{uv} = n^{-2}/2$.  By symmetry, we
have $π(S) → 1/2$ as $n → ∞$,
giving $Φ_{S} → n^{-2}(1/2)/(1/2) = n^{-2}$.  So we
have $n^{2}/2 ≤ τ_{2} ≤ 2n^{4}$.

How does this compare to the actual mixing time?  In the stationary
distribution, we have a constant probability of being in each of the
copies of $K_{n}$.  Suppose we start in the left copy.  At each step
there is a $1/n$ chance that we are sitting on the path endpoint.  We
then step onto the path with probability $1/n$, and reach the other
end before coming back with probability $1/n$.  So (assuming we can
make this extremely sloppy handwaving argument rigorous) it takes at
least $n^{3}$ steps on average before we reach the other copy of
$K_{n}$, which gives us a rough estimate of the mixing time of
$Θ(n^{3})$.  In this case the exponent is exactly in the middle of the bounds derived from conductance.

Curiously, it's not hard to argue that this is in fact the worst
possible conductance we can get from a lazy random walk. Consider an
arbitrary connected graph $G$ and any subset $S$ of $V(G)$. Then $S$
has at least one outgoing edge $uv$, and we take it with probability
$π_u p_{uv} = \frac{d(u)}{∑_w d(w)} ⋅ \frac{1}{2d(u)} = \frac{1}{∑_w
d(w)} = \frac{1}{2\card{E}} ≥ \frac{1}{n(n-1)}$. Assuming $π(S) ≤
1/2$, this gives $Φ(S) ≥ \frac{2}{n(n-1)} = Θ(1/n^2)$. This
immediately gives that a lazy random walk on \emph{any} graph
converges in time $\widetilde{O}(n^4)$.

\subsection{Edge expansion using canonical paths}
\label{section-edge-expansion-using-canonical-paths}

(Here and below we are mostly following the presentation of
Guruswami~\cite{Guruswami2000}, but with slightly different examples.)

For more complicated Markov chains, it is helpful to have a tool for
bounding conductance that doesn't depend on intuiting what sets have
the smallest boundary.  The \concept{canonical paths}~\cite{JerrumS1989} technique does this
by assigning a unique path $γ_{xy}$ from each state $x$ to each
state $y$ in a way that doesn't send too many paths across any one
edge.  So if we have a partition of the state space into sets $S$ and
$T$, then there are $\card*{S}⋅\card*{T}$ paths from states in $S$
to states in $T$, and since (a) every one of these paths crosses an
$S$–$T$ edge, and (b) each $S$–$T$ edge carries at most $ρ$
paths, there must be at least $\card*{S}⋅\card*{T}/ρ$ edges
from $S$ to $T$.  Note that because there is no guarantee we chose
good canonical paths, this is only useful for getting lower
bounds on conductance—and thus upper bounds on mixing time—but
this is usually what we want.

Let's start with a small example.  Let $G =
C_{n}\mathbin{\square}C_{m}$, the $n×m$ torus.  A lazy random
walk on this graph moves north, east, south, or west with probability
1/8 each, wrapping around when the coordinates reach $n$ or $m$.  Since this is a random walk on a regular graph, the stationary distribution is uniform.  What is the relaxation time?

Intuitively, we expect it to be $O(\max(n,m)^{2})$, because we can
think of this two-dimensional random walk as consisting of two
one-dimensional random walks, one in the horizontal direction and one
in the vertical direction, and we know that a random walk on a cycle
mixes in $O(n^{2})$ time.  Unfortunately, the two random walks are not
independent: every time I take a horizontal step is a time I am not
taking a vertical step.  We \emph{can} show that the expected coupling
time is $O(n^{2}+m^{2})$ by running two sequential instances of the coupling argument for the cycle, where we first link the two copies in the horizontal direction and then in the vertical direction.  So this gives us one bound on the mixing time.  But what happens if we try to use conductance?

Here it is probably easiest to start with just a cycle.  Given points
$x$ and $y$ on $C_{n}$, let the canonical path $γ_{xy}$ be a
shortest path between them, breaking ties so that half the paths go
one way and half the other.  Then each each is crossed by exactly $k$
paths of length $k$ for each $k=1\dots(n/2-1)$, and at most $n/4$
paths of length $n/2$ ($0$ if $n$ is odd), giving a total of $ρ
≤ \binom{n/2}{2} + \frac{n}{4} = \frac{(n/2-1)(n/2)}{2} + \frac{n}{4} = n^{2}/8$ paths across the edge.

If we now take an $S$–$T$ partition where $\card*{S} = m$, we get at
least $m(n-m)/ρ = 8m(n-m)/n^{2}$ $S$–$T$ edges.  This peaks at
$m=n/2$, where we get $2$ edges—exactly the right number—and in
general when $m≤n/2$ we get at least $8m(n/2)/n^{2} = 4m/n$
outgoing edges, giving a conductance $Φ_{S} ≥
(1/4n)(4m/n)/(m/n) = 1/n$.  

This is essentially what we got before,
except we have to divide by $2$ because we are doing a lazy walk.
Note that for small $m$, the bound is a gross underestimate, since we
know that every nonempty proper subset has at least $2$ outgoing edges.

Now let's go back to the torus $C_{n}\mathbin{\square}C_{m}$.  Given
$x$ and $y$, define $γ_{xy}$ to be the L-shaped path that first
changes $x_{1}$ to $y_{1}$ by moving the shortest distance vertically,
then changes $x_{2}$ to $y_{2}$ by moving the shortest distance
horizontally.  For a vertical edge, the number of such paths that
cross it is bounded by $n^{2}m/8$, since we get at most $n^{2}/8$
possible vertical path segments and for each such vertical path
segment there are $m$ possible horizontal destinations to attach to
it.  For a horizontal edge, $n$ and $m$ are reversed, giving
$m^{2}n/8$ paths.  To make our life easier, let's assume $n≥m$,
giving a maximum of $ρ = n^{2}m/8$ paths over any edge.

For $\card*{S} ≤ nm/2$, this gives at least
\begin{displaymath}
    \frac{\card*{S}⋅\card*{G∖ S}}{ρ} ≥
    \frac{\card*{S}(nm/2)}{n^{2}m/8} = \frac{4\card*{S}}{n}
\end{displaymath}
outgoing edges.  We thus have
\begin{displaymath}
    Φ(S) ≥ \frac{1}{8nm}⋅\frac{4\card*{S}/n}{\card*{S}/nm} =
    \frac{1}{2n}.
\end{displaymath}
This
gives $τ_{2} ≤ 2/Φ^{2} ≤ 8n^{2}$.
Given that we assumed $n ≥ m$, this is essentially
the same $O(n^2 + m^2)$ bound that we would get from coupling.

\subsection{Congestion}
\label{section-congestion}

For less symmetric chains, we weight paths by the probabilities of
their endpoints when counting how many cross each edge, and treat the
flow across the edge as a capacity.  This gives the
\concept{congestion} of a collection of canonical paths $Γ =
\Set{γ_{xy}}$, which is computed as
\begin{align*}
ρ(Γ) &= \max_{uv ∈ E} \frac{1}{π_{u}p_{uv}} ∑_{γ_{xy} \ni uv} π_{x} π_{y},
\end{align*}
and we define $ρ = \min_{Γ} ρ(Γ)$.

The intuition here is that the congestion bounds the ratio between the
canonical path flow across an edge and the Markov chain flow across
the edge.  If the congestion is not too big, this means that any cut
that has a lot of canonical path flow across it also has a lot of
Markov chain flow.  When $π(T) ≥ \frac{1}{2}$, the total canonical path flow
$π(S) π(T)$ is at least $\frac{1}{2} π(S)$. This means that when
$π(S)$ is large but less than $\frac{1}{2}$, the Markov chain flow leaving $S$ is also large.
This gives a lower bound on conductance.

Formally, we have the following lemma:
\begin{lemma}
\label{lemma-congestion}
For any reversible aperiodic irreducible Markov chain,
\begin{align}
Φ &≥ \frac{1}{2ρ}
\intertext{from which it follows that}
τ_2 &≤ 8ρ^2.
\end{align}
\end{lemma}
\begin{proof}
Pick some set of canonical paths $Γ$ with $ρ(Γ) =
ρ$.  Now pick some $S$–$T$ partition with 
    $π(S)≤1/2$ and $φ(S) = φ$.
Consider a flow where we route $π(x)π(y)$ units of flow along
each path $γ_{xy}$ with $x∈S$ and $y∈T$.  This gives a
total flow of $π(S)π(T) ≥ π(S)/2$.  We are going to show that we need a lot of capacity across the $S$–$T$ cut to carry this flow, which will give the lower bound on conductance.

For any $S$–$T$ edge $uv$, we have
 \begin{align*}
\frac{1}{π_{u}p_{uv}} ∑_{γ_{xy} \ni uv} π_{x} π_{y}
&≤ ρ
\intertext{or}
∑_{γ_{xy} \ni uv} π_{x} π_{y}
&≤
ρ ⋅ π_{u} p_{uv}.
\end{align*}

Since each $S$–$T$ path crosses at least one $S$–$T$ edge, we have
\begin{align*}
π(S) π(T)
&=
∑_{x ∈ S, y ∈ T} π_{x} π_{y}
\\
&≤
∑_{u ∈ S, v ∈ T, uv ∈ E} ∑_{γ_{xy} \ni uv} π_{x} π_{y}
\\
&≤
∑_{u ∈ S, v ∈ T, uv ∈ E} ρ π_{u} p_{uv}.
\\
&=
ρ ∑_{u ∈ S, v ∈ T, uv ∈ E} π_{u} p_{uv}.
\end{align*}

But then
\begin{align*}
Φ(S) 
&= 
\frac{∑_{u ∈ S, t ∈ T, uv ∈ E} π_{u} p_{uv}}{π_{S}}
\\
&≥ \frac{π(S) π(T)/ρ}{π(S)}
\\
&= \frac{π(T)}{ρ}
\\
&≥ \frac{1}{2ρ}.
\end{align*}

To get the bound on $τ_2$, use
\eqref{eq-conductance-relaxation-time} to compute
$τ_{2} ≤ 2/Φ^{2} ≤ 8ρ^{2}$.
\end{proof}

\subsection{Examples}
\label{section-spectral-convergence-bounds-examples}

Here are some more examples of applying canonical paths.

\subsubsection{Lazy random walk on a line}
\label{section-congestion-line}
Consider a lazy random walk on a line with reflecting barriers
at $0$ and $n-1$.  Here $π_x = \frac{1}{n}$ for all $x$.  There
aren't a whole lot of choices for canonical paths; the obvious choice
for $γ_{xy}$ with $x < y$ is $x, x+1, x+2, \dots, y$.  This puts
$(n/2)^2$ paths across the middle edge (which is the most heavily
loaded), each of which has weight
$π_x π_y = n^{-2}$.  So the congestion is
$\frac{1}{(1/n)(1/4)} (n/2)^2 n^{-2} = n$, giving a mixing time of at
most $8n^2$.  This is a pretty good estimate.

\subsubsection{Random walk on a hypercube}
\label{section-congestion-hypercube}
Let's try a more complicated example: the random walk on a hypercube
from §\ref{section-random-walk-on-a-hypercube-via-coupling}.
Here at each step we pick some coordinate uniformly at random and set
it to $0$ or $1$ with equal probability; this gives a transition
probability $p_{uv} = \frac{1}{2n}$ whenever $i$ and $j$ differ by
exactly one bit.  A plausible choice for canonical paths is to let
$γ_{xy}$ use \index{routing!bit-fixing}\concept{bit-fixing routing}, where
we change bits in $x$ to the corresponding bits in $y$ from left to
right.  To compute congestion, pick some edge $uv$, and let $k$ be the
bit position in which $u$ and $v$ differ.  A path $γ_{xy}$ will
cross $uv$ if $u_{k}\dots u_n = x_{k}\dots x_n$ (because when we are
at $u$ we haven't fixed those bits yet) and $v_1 \dots v_k = y_1 \dots
y_k$ (because at $v$ we have fixed all of those bits).  There are
$2^{k-1}$ choices of $x_1 \dots x_{k-1}$ consistent with the first
condition and $2^{n-k}$ choices of $y_{k+1}\dots y_n$ consistent with
the second, giving exactly $2^{n-1}$ total paths $γ_{xy}$
crossing $uv$.  Since each path occurs with weight $π_x π_y =
2^{-2n}$, and the flow across $uv$ is $π_u p_{uv} = 2^{-n}
\frac{1}{2n}$, we can calculate the congestion
\begin{align*}
ρ(Γ) &= \max_{uv ∈ E} \frac{1}{π_{u}p_{uv}} ∑_{γ_{xy} \ni uv} π_{x} π_{y}
\\
&= \frac{1}{2^{-n}/2n} ⋅ 2^{n-1} ⋅ 2^{-2n}
\\
&= n.
\end{align*}
This gives a relaxation time $τ_2 ≤ 8ρ^2 = 8n^2$, which when we
account for the large state space gives $t_{\mix}(ε) ≤ 8n^2
\parens*{\frac{1}{2} \ln 2^n + \ln (1/ε)} = O(n^3)$.  In this
case the bound is substantially worse than what we previously proved
using coupling.

The fact that the number of canonical paths that cross a particular
edge is exactly one half the number of nodes in the hypercube is not an
accident: if we look at what information we need to fill in to compute
$x$ and $y$ from $u$ and $v$, we need (a) the part of $x$ we've
already gotten rid of, plus (b) the part of $y$ we haven't filled in
yet.  If we stitch these two pieces together, we get all but one of the
$n$ bits we need to specify a node in the hypercube, the missing bit
being the bit we flip to get from $u$ to $v$.  This sort of thing
shows up often in conductance arguments where we build our canonical
paths by fixing a structure one piece at a time.

\subsubsection{Matchings in a graph}
\label{section-congestion-matchings}

A \concept{matching} in a graph $G = (V,E)$ is a subset of the edges with no
two elements adjacent to each other; equivalently, it's a subgraph
that includes all the vertices in which each vertex has degree at most
$1$.  We can use a random walk to
sample matchings from an arbitrary graph uniformly.

Here is the random walk.  Let $S_t$ be the matching at time $t$.  
At each step, we choose an edge $e ∈ E$
uniformly at random, and flip a coin to decide whether to include it
or not.  If the coin comes up tails, set $S_{t+1} = S_t ∖ \Set{
e }$ (this may leave $S_{t+1} = S_t$ if $S_t$ already omitted $e$);
otherwise, set $S_{t+1} = S_t \cup \Set{e}$ unless one of the endpoints
of $e$ is
already incident to an edge in $S_t$, in which case set $S_{t+1} =
S_t$.

Because this chain contains many self-loops, it's aperiodic.  It's
also straightforward to show that any transition between two adjacent
matchings occurs with probability exactly $\frac{1}{2m}$, where
$m=\card*{E}$, and
thus that the chain is reversible with a uniform stationary
distribution.  We'd like to bound the congestion of the chain to show
that it converges in a reasonable amount of time.

Let $N$ be the number of matchings in $G$, and let $n = \card*{V}$ and
$m = \card*{E}$ as usual.  Then $π_S = 1/N$ for all
$S$ and $π_{S} p_{ST} = \frac{1}{2Nm}$.  Our congestion for
any transition $ST$ will then be $2Nm N^{-2} = 2m/N$
times the number of paths that cross $ST$; ideally this number of
paths will be
at most $N$ times some small polynomial in $n$ and/or $m$.

Suppose we are trying to get from some matching $X$ to another
matching $Y$.  The graph $X \cup Y$ has maximum degree $2$, so each of
its connected components is either a path or a cycle; in addition, we
know that the edges in each of these paths or cycles alternate
whether they come from $X$ or $Y$, which among other things implies
that the cycles are all even cycles.

We can
transform $X$ to $Y$ by processing these components in a methodical
way: first order the components (say, by the increasing identity of
the smallest vertex in each), then for each component replace the $X$
edges with $Y$ edges.  If we do this cleverly enough, we can guarantee
that for any transition $ST$, the set of edges $(X∪Y)∖(S∪T)$ always
consists of a matching plus at most two extra edges, and that $S$, $T$,
$(X∪Y)∖(S∪T)$ are enough to reconstruct $X$ and $Y$.  Since there are
at most $Nm^2$ choices of $(X∪Y)∖(S∪T)$, this will give at most $Nm^2$
canonical paths across each transition.\footnote{We could improve the
constant a bit by using $N\binom{m}{2}$, but we won't.}

Here is how we do the replacement within a component.
If $C$ is a cycle with $k$ vertices, order the vertices $v_0,\dots,v_{k-1}$
such that $v_0$ has the smallest index in $C$ and $v_0 v_1 ∈ X$.  Then
$X∩C$ consists of the even-numbered edges $e_{2i} = v_{2i} v_{2+1}$
and $Y∩C$ consists of the odd-numbered edges $e_{2i+1} = v_{2i+1}
v_{2i+2}$, where we take $v_k = v_0$.
We now delete $e_0$, then alternate between deleting $e_{2i}$ and
adding $e_{2i-1}$ for $i = 1 \dots k/2-1$.  Finally we add $e_{k-1}$.

The number of edges in $C$ at each step in this process is always one
of $k/2$, $k/2-1$, or $k/2-2$, and since we always add or delete an
edge, the number of edges in $C∩(S∪T)$ for any transition will be at least
$k/2-1$, which means that the total degree of the vertices in
$C∩(S∪T)$ is at least $k-2$.  We also know that $SUT$ is a matching, since it's either
equal to $S$ or equal to $T$.  So there are at least $k-2$ vertices in
$C∩(S∪T)$ with degree $1$, and none with degree $2$, leaving at most
$2$ vertices with degree $0$.  In the complement $C∖(S∪T)$, this
becomes at most two vertices with degree $2$, with the rest having
degree $1$.  If these vertices are adjacent, removing the edge between
them leaves a matching; if not, removing one edge adjacent to each
does so.  So two extra edges are enough.

A similar argument works for paths, which we will leave as an exercise
for the reader.

Now suppose we know $S$, $T$, and $(XUY)∖(S∪T)$.  We can compute
$X∪Y = ((X∪Y)∖(S∪T))∪(S∪T)$; this means we can reconstruct the graph
$X∪Y$, identify its components, and so on.  We know which component
$C$ we
are working on because we know which edge changes between $S$ and $T$.
For any earlier component $C'$, we have $C'∩S = C'∩Y$ (since we
finished already), and similarly $C'∖S = C'∩X$.  For components $C'$ we
haven't reached yet, the reverse holds, $C'∩S = C∩X$ and $C'∖S = C∩Y$.
This determines the edges in both $X$ and $Y$ for all edges not in
$C$.

For $C$, we know from $C∩(X∪Y)$ which vertex is $v_0$.  In a cycle, whenever we
remove an edge, its lower-numbered endpoint is always an even distance
from $v_0$, and similarly when we add an edge, its lower-numbered
endpoint is always an odd distance from $v_0$.  So we can orient the
cycle and tell which edges in $S∪T$ have already been added (and are
thus part of $Y$) and which have been removed (and are thus part of
$X$).  Since we know that $C∩Y = C∖X$, this is enough to reconstruct
both $C∩X$ and $C∩Y$.  (The same general idea works for paths,
although for paths we do not need to be as careful to figure out
orientation since we can only leave $v_0$ in one direction.)

The result is that given $S$ and $T$, there are at most $Nm^2$ choices
of $X$ and $Y$ such that the canonical path $γ_{XY}$ crosses $ST$.
This gives $ρ ≤ \frac{Nm^2 N^{-2}}{N^{-1} \frac{1}{2m}} = 2m^3$, 
which gives $τ_2 ≤ 32m^6$.  This is not great but it is at least
polynomial in the size of the graph.
For actual sampling, this translates to $O\parens*{m^6\parens*{\log N + \log
\frac{1}{ε}}} = O\parens*{m^6\parens*{m + \log \frac{1}{ε}}}$
steps to get a total variation distance of $ε$.

This may or may not give rapid mixing, depending on how big $N$ is
relative to $m$. For many graphs, the number of matchings $N$ will be
exponential in $m$, and so the $m^6$ term will be polylogarithmic in
$N$.
But $N$ can be much smaller in dense graphs.
For example, on a clique we have
$N=m+1$, since no matching can contain more than one edge, making the
bound $τ_2 ≤ 32m^6 = Θ(N^6)$ polynomial in $N$ but \emph{not}
polynomial in $\log N$, which is what we want.
The actual mixing time in this case is $Θ(m)$ (for the upper bound, wait
to delete the only edge, and do so on both sides of the coupling; for
the lower bound, observe that until we delete the only edge we are
still in our initial state). This is much better than the upper bound
from canonical paths, but it still doesn't give rapid maxing.

\subsubsection{Perfect matchings in dense bipartite graphs}
\label{section-congestion-perfect-matchings}

(Basically doing \cite[§11.3]{MotwaniR1995}.)

A similar chain can be used to sample perfect matchings in dense
bipartite graphs, as shown by Jerrum and Sinclair~\cite{JerrumS1989}
based on an algorithm by Broder~\cite{Broder1986} that turned out to
have a bug in the analysis~\cite{Broder1988}.

A \concept{perfect matching} of a graph $G$ is a subgraph that
includes all the vertices and gives each of them exactly one incident
edge.  We'll be looking for perfect matchings in a
\index{graph!bipartite}\concept{bipartite graph}
consisting of $n$ left vertices $u_1,\dots,u_n$ and $n$ right vertices
$v_1,\dots,v_n$, where every edge goes between a left vertex and a
right vertex.  We'll also assume that the graph is \concept{dense},
which in this case means that every vertex has at least $n/2$
neighbors.  This density assumption was used in the original
Broder and Jerrum-Sinclair papers but removed in a later paper by Jerrum,
Sinclair, and Vigoda~\cite{JerrumSV2004}.

The random walk is similar to the random walk from the previous
section restricted to the set of matchings with either $n-1$ or $n$
edges.  At each step, we first flip a coin (the usual
lazy walk trick); if it comes up heads, we choose an edge $uv$
uniformly at random and apply one of the following transformations
depending on how $uv$ fits in the current matching $m_t$:
\begin{enumerate}
\item If $uv ∈ m_t$, and $\card*{m_t} = n$, set $m_{t+1} = m_t
    ∖ \Set{uv}$.
\item If $u$ and $v$ are both unmatched in $m_t$, and $\card*{m_t} =
n-1$, set $m_{t+1} = m_t \cup \Set{uv}$.
\item If exactly one of $u$ and $v$ is matched to some other node $w$,
and $\card*{m_t} = n-1$, perform a rotation that deletes the $w$ edge
and adds $uv$.
\item If none of these conditions hold, set $m_{t+1} = m_t$.
\end{enumerate}

The walk can be started from any perfect matching, which can be found
in $O(n^{5/2})$ time using a classic algorithm of Hopcroft and
Karp~\cite{HopcroftK1973}. This algorithm repeatedly searches for an
\concept{augmenting path}, which is a path in $G$ between two unmatched
vertices that alternates between edges not in the matching and edges
in the matching. When it finds an augmenting path, it switches the out
edges for the in edges, increasing the size of the matching.

We can show that the walk
converges in polynomial time using a canonical path argument.  This is
done in two stages: first, we define canonical paths between all
perfect matchings.  Next, we define a short path from any matching of
size $n-1$ to some nearby perfect matching, and build paths between
arbitrary matchings by pasting one of these short paths on one or both
ends of a long path between perfect matchings. This gives a collection
of canonical paths that we can show to have low congestion.

To go between perfect matchings $X$ and $Y$, consider $X \cup Y$
as a collection of paths and even cycles as in
§\ref{section-congestion-matchings}.
In some standard order, fix each path cycle by first deleting one
edge to make room, then using rotate operations to move the rest of the edges, then
putting the last edge back in.  For any transition $S$–$T$ along the way, we
can use the same argument that we can compute $X∪Y$ from $S∪T$
by supplying the missing edges, which will consist of a matching
of size $n$ plus at most one extra edge.  So if $N$ is the number of matchings of
size $n$ or $n-1$, the same argument used previously shows that at
most $Nm$ of these long paths cross any $S$–$T$ transition.

For the short paths, we must use the density property.  The idea is
that for any matching that is not perfect, we can find an augmenting path of
length at most $3$ between two unmatched nodes on either side.
Pick some unmatched nodes $u$ and $v$.
Each of these nodes is adjacent to at least $n/2$ neighbors; if any of
these neighbors are unmatched, we just found an augmenting path of
length $1$.  Otherwise the $n/2$ neighbors of $u$ and the $n/2$ nodes
matched to neighbors of $v$ overlap (because $v$ is unmatched, leaving
at most $n-1$ matched nodes and thus at most $n/2-1$ nodes that are
matched to something that's not a neighbor of $v$).  So for each
matching of size $n-1$, in at most two steps (rotate and then add an
edge) we can reach some specific perfect matching.  There are at most
$m^2$ ways that we can undo this, so each perfect matching is
associated with at most $m^2$ smaller matchings.  This blows up the
number of canonical paths crossing any transition by roughly $m^4$; by
counting carefully we can thus show congestion that is $O(m^6)$
($O(m^4)$
from the blow-up, $m$ from the $m$ in $Nm$, and $m$ from $1/p_{ST}$).

It follows that for this process, $τ_2 = O(m^{12})$.  (I think a
better analysis is possible.)

As noted earlier, this is an example of a process for which causal
coupling doesn't work in less than exponential time~\cite{KumarR1999},
a common problem with Markov chains that don't have much symmetry.
So it's not surprising that stronger techniques were developed
specifically to attack this problem.

An issue we need to deal with is that when we sample a matching with
this process, we don't necessarily get a perfect matching, since many
states are going to have only $n-1$ edges. But we have already shown
that the perfect matchings make up at least an $m^{-2}$ fraction of
the total. So if we don't get a perfect matching after a polynomial
number of steps, we can simply start over and try again. This adds a
polynomial factor to the already-bad running time, but it is still
polynomial. Alternatively we can argue that the Markov chain
sampled only at times when it yields a perfect matching is (a) still a
Markov chain, and (b) has a total variation distance from its
stationary distribution that we can bound using the bounds on the
unsampled chain. The details of this are a little messy but it also
works.

\myChapter{Approximate counting}{2025}{}
\label{chapter-approximate-counting}

(See also \cite[Chapter 11]{MotwaniR1995}.)

The basic idea: we have some class of objects, and we want to know how many of
them there are.  Ideally we can build an algorithm that just prints
out the exact number, but for many problems this is hard.

A
\index{approximation scheme!fully polynomial-time randomized}
\concept{fully polynomial-time randomized approximation scheme} or
\concept{FPRAS} for a numerical problem outputs a number that is
between $1-ε$ and $1+ε$ times the correct answer,
with probability at least $3/4$ (or some constant bounded away from
$1/2$—we can amplify to improve it), in time polynomial in the
input size $n$ and $1/ε$.  In this chapter, we'll be hunting
for FPRASs.  But first we will discuss briefly why we can't just count
directly in many cases.

\section{Exact counting}
\label{section-exact-counting}

A typical application is to a problem in the complexity class
\classSharpP, problems that involve counting the number of accepting
computation paths in a nondeterministic polynomial-time Turing
machine.  Equivalently, these are problems that involve counting for
some input $x$ the number of values $r$ such that $M(x,r) = 1$, where
$M$ is some machine in \classP.  
An example would be the problem \concept{\#SAT} of counting the number
of satisfying assignments of some CNF formula.

The class \index{\#P}$\classSharpP$ (which is
usually pronounced \concept{sharp P} or \concept{number P}) was defined
by Leslie Valiant in a classic paper~\cite{Valiant1979}.
The central result in this paper 
is
\index{theorem!Valiant's}
\concept{Valiant's theorem}.  This shows that any problem in
$\classSharpP$
can be \indexConcept{reduction}{reduced} (by \indexConcept{Cook
reduction}{Cook reductions}, meaning that we are
allowed to use the target problem as a subroutine instead of just
calling it once) to the problem of computing the \concept{permanent}
of a square $0$–$1$ matrix $A$, where the permanent is given by the formula
$∑_{π} \prod_{i} A_{i,π(i)}$, where the sum ranges over all
$n!$ permutations $π$ of the indices of the matrix.  An equivalent
problem is counting the number of \concept{perfect matchings}
(subgraphs including all vertices in which every vertex has degree
exactly $1$) of a bipartite graph.  Other examples of
$\classSharpP$-complete problems are \#SAT (defined above) and
\concept{\#DNF} (like \#SAT, but the input is in DNF form; \#SAT
reduces to \#DNF by negating the formula and then subtracting the
result from $2^n$).

Exact counting of \classSharpP-hard problems is likely to be very
difficult: 
\index{theorem!Toda's}
\concept{Toda's theorem}~\cite{Toda1991} says that being
able to make even a single query to a \classSharpP-oracle is enough to
solve any problem in the \concept{polynomial-time hierarchy}, which 
contains most of the complexity classes you have probably heard of.
Nonetheless, it is often possible to obtain good approximations to
such problems.

\section{Counting by sampling}
\label{section-counting-by-sampling}

If many of the things we are looking for are in the target set, we can
count by sampling; this is what poll-takers do for a living.  Let $U$
be the universe we are sampling from and $G$ be the ``good'' set of
points we want to count.  Let $ρ = \card*{G}/\card*{U}$.  If we can
sample uniformly from $U$, then we can estimate $ρ$ by taking $N$
independent samples and dividing the number of samples in $G$ by $N$.
This will give us an answer $\hat{ρ}$ whose expectation is $ρ$,
but the accuracy may be off.  Since the variance of each sample is
$ρ(1-ρ) \approx ρ$ (when $ρ$ is small), we get
$\Var{∑ X_i} \approx mρ$, giving a standard deviation
of $√{m ρ}$.  For this to be less than our allowable error
$ε m ρ$, we need $1 ≤ ε √{m ρ}$ or $N ≥
\frac{1}{ε^2 ρ}$.  This gets bad if $ρ$ is exponentially
small.  So if we are to use sampling, we need to make sure that we
only use it when $ρ$ is large.

On the positive side, if $ρ$ is large enough we can easily compute
how many samples we need using Chernoff bounds.
The following lemma gives a convenient estimate; it is based on
\cite[Theorem 11.1]{MotwaniR1995} with a slight improvement on the
constant:
\begin{lemma}
\label{lemma-sampling}
Sampling $N$ times gives relative error $ε$ with probability at
least $1-δ$ provided $ε ≤ 1.81$ and 
\begin{align}
\label{eq-sampling}
N &≥ \frac{3}{ε^2 ρ} \ln \frac{2}{δ}.
\end{align}
\end{lemma}
\begin{proof}
Suppose we take $N$ samples, and let $X$ be the total count for these
samples.  Then $\Exp{X} = ρ N$, and 
\eqref{eq-Chernoff-two-sided-one-third} gives (for $ε ≤ 1.81$):
\begin{align*}
\Prob{\abs*{X - ρ N} ≥ ε ρ N}
&≤ 2 e^{-ρ N ε^2 / 3}.
\end{align*}
Now set $2 e^{-ρ N ε^2 / 3} ≤ δ$ and solve for $N$ to
get \eqref{eq-sampling}.
\end{proof}

\subsection{Generating samples}

An issue that sometimes comes up in this process is that it is not
always obvious how to generate a sample from a given universe $U$.
What we want is a function that takes random bits as input and
produces an element of $U$ as output.

In the simplest case, we know
$\card{U}$ and have a function $f$ from $\Set{0,\dots,\card{U}-1}$ to
$U$ (the application of such a function is called
\concept{unranking}).
Here we can do \index{sampling!rejection}\concept{rejection sampling}:
we choose a bit-vector of length $\ceil{\lg \card{U}}$, interpret it
as an integer $x$ expressed in binary, and try again until $x <
\card{U}$.  This requires less than $2 \ceil{\lg \card{U}}$ bits on average in
the worst case, since we get a good value at least half the
time.

Unranking (and the converse operation of ranking)
can be a non-trivial problem in combinatorics, and there are
entire textbooks on the subject~\cite{StantonW2013}.  But some general
principles apply.  If $U$ is a union of disjoint sets $U_1 ∪ U_2 ∪ \dots ∪
U_n$, we can compute the sizes of each $\card{U_i}$, and we
can unrank each $U_i$, then
we can map some $x∈\Set{0\dots \card{U}-1}$ to a specific
element of $U$ by choosing the maximum $i$ such that
$∑_{j=1}^{i-1} \card{U_j} ≤ x$ and choosing the $k$-th element
of $U_i$, where $k = x - ∑_{j=1}^{i-1} \card{U_j}$.  This is
essentially what happens if we want to compute the $x$-th day
of the year.  The first step of this process requires computing the
sizes of $O(n)$ sets if we generate the sum one element at a time,
although if we have a fast way of computing $∑_{i=1}^j \card{U_i}$ for
each $i$, we can cut this time to $O(\log n)$ such computations down using binary search.
The second step depends on whatever mechanism we use to unrank within
$U_i$.

An example would be generating one of the $\binom{n}{k}$ subsets of a
set of size $k$ uniformly at random.  If we let $S$ be the set of all
such $k$-subsets, we can express $S$ as $\bigcup_{i=1}^{n} S_i$ where $i$
is the set of all $k$-subsets that have the $i$-th element as their
smallest element in some fixed ordering.  But then we can easily
compute $\card{S_i} = \binom{n-i}{k-1}$ for each $i$, apply the
above technique to pick a particular $S_i$ to start with, and then
recurse within $S_i$ to get the rest of the elements.

A more general case is when we can't easily sample from $U$ but we can sample
from some $T ⊇ U$.  Here rejection sampling comes to our rescue as
long as $\card{U}/\card{T}$ is large enough.  
For example, if we want to generate a $k$-subset of $n$ elements when
$k \ll √{n}$, we can choose a list of $k$ elements with replacement and
discard any lists that contain duplicates.  But this doesn't work so well
for larger $k$.  In this particular case, there is an easy out: we can
sample $k$ elements without replacement and forget their order.  This
corresponds to sampling from a universe $T$ of ordered $k$-subsets,
then mapping down to $U$.  As long as the inverse image of each
element of $U$ has the same size in $T$, this will give us a uniform
sample from $U$.

When rejection sampling fails, we may need to
come up with a more clever approach to concentrate on the particular
values we want.  One approach that can work well if we can express $U$
as a union of sets that are not necessarily disjoint is the Karp-Luby
technique~\cite{KarpL1985}, discussed in
§\ref{section-approximating-sharp-dnf} below.

All of this assumes that we are interested in getting uniform samples.
For non-uniform samples, we replace the assumption that we are trying
to generate a uniform $X ∈ \Set{0 \dots m-1}$ with some $X ∈
\Set{0\dots m-1} $ for which we can efficiently calculate
$\Prob{X ≤ i}$ for each $i$.  In this case, we can (in principle)
generate a continuous random variable $Y ∈ [0,1]$ and choose the
maximum $i$ such that $\Prob{X ≤ i} ≤ Y$, which we can find using
binary search.

The complication is that we can't actually generate $Y$ in finite time.
Instead, we generate $Y$ one bit at a time; this gives a sequence of
rational values $Y_t$ where each $Y_t$ is of the form $k/2^t$.  We can
stop when every value in interval $[Y_t,Y_t+2^{-t})$ lies with the
range corresponding to some particular value of $X$.  This technique
is closely related to \concept{arithmetic coding}~\cite{WittenNC1987},
and requires generating $H(X) + O(1)$ bits on average, where $H(X) =
- ∑_{i=0}^{m-1} \Prob{X=i} \lg \Prob{X=i}$ is the (base 2)
\concept{entropy} of $X$.

\section{Approximating \#KNAPSACK}
\label{section-approximating-sharp-knapsack}

Here is an algorithm for approximating the number of solutions to a
\concept{KNAPSACK} problem, due to Dyer~\cite{Dyer2003}.  We'll
concentrate on the simplest version, $0$–$1$ KNAPSACK, following
the analysis in Section 2.1 of~\cite{Dyer2003}.

For the $0$–$1$ KNAPSACK problem, we are given a set of $n$ objects of weight
$0 ≤ a_1 ≤ a_2 ≤ \dots a_n ≤ b$, and we want to find a $0$–$1$
assignment $x_1, x_2, \dots, x_n$ such that $∑_{i=1}^{n} a_i x_i ≤ b$
(usually while optimizing some property that prevents us from setting
all the $x_i$ to zero).
We'll assume that the $a_i$ and $b$ are all integers.

For \#KNAPSACK, we want to compute $\card*{S}$, where $S$ is the set of
all assignments to the $x_i$ that make $∑_{i=1}^{n} a_i x_i ≤ b$.

There is a well-known 
\index{approximation scheme!fully polynomial-time}
\concept{fully polynomial-time approximation scheme} 
for optimizing KNAPSACK, based on dynamic programming.  The
idea is that a maximum-weight solution can be found exactly in time
polynomial in $b$, and if $b$ is too large, we can reduce it by
rescaling all the $a_i$ and $b$ at the cost of a small amount of
error.  A similar idea is used in Dyer's algorithm: the KNAPSACK
problem is rescaled so that size of the solution set $S'$
of the rescaled version can be computed in polynomial time.
Sampling is then used to determine what proportion of the solutions in
$S'$ correspond to solutions of the original problem.

Scaling step: Let $a'_i = \floor{n^2 a_i/b}$.  Then $0 ≤ a'_i
≤ n^2$ for all $i$.
Taking the floor creates some error: if we try to reconstruct $a_i$
from $a'_i$, the best we can do is argue that
$a'_i ≤ n^2 a_i / b < a'_i +1$
implies
$(b/n^2) a'_i ≤ a_i < (b/n^2) a'_i + (b/n^2)$.
The reason for using $n^2$ as our rescaled bound is
that the total error in the
upper bound on $a_i$, summed over all
$i$, is bounded by $n(b/n^2) = b/n$, a fact that will become important
soon.

Let $S' = \SetWhere{ \vec{x} }{ ∑_{i=1}^n a'_i x_i ≤
n^2 }$ be the set of solutions to the rescaled knapsack problem,
where we substitute $a'_i$ for $a_i$ and $n^2 = (n^2/b)b$ for $b$.  
Claim: $S ⊆ S'$.  Proof: $\vec{x} ∈ S$ if and only if
$∑_{i=1}^n a_i x_i ≤ b$.  But then 
\begin{align*}
∑_{i=1}^n a'_i x_i 
&= ∑_{i=1}^n \floor{n^2 a_i/b} x_i
\\
&≤ ∑_{i=1}^n (n^2/b) a_i x_i
\\
&= (n^2/b) ∑_{i=1}^n a_i x_i
\\
&≤ n^2,
\end{align*}
which shows $\vec{x} ∈ S'$.

The converse does not hold.  However, we can argue that any $\vec{x}
∈ S'$ can be shoehorned into $S$ by setting at most one of the $x_i$
to $0$.  Consider the set of all positions $i$ such that $x_i = 1$ and
$a_i > b/n$.  If this set is empty, then $∑_{i=1}^n a_i x_i ≤
∑_{i=1}^n b/n = b$, and $\vec{x}$ is already in $S$.
Otherwise, pick any position $i$ with $x_i = 1$ and $a_i > b/n$, and
let $y_j = 0$ when $j=i$ and $y_j = x_j$ otherwise.
If we recall that the total error from the floors in our approximation was at most
$(b/n^2)n = b/n$, the intuition is that deleting $y_j$ removes it. Formally, we
can write
\begin{align*}
∑_{j=1}^n a_j y_j
&= ∑_{j=1}^n a_j x_j - a_i
\\
&< ∑_{j=1}^n ((b/n^2) a'_j + b/n^2) x_j - b/n
\\
&≤ (b/n^2) ∑_{j=1}^n a'_j x_j + b/n - b/n
\\
&≤ (b/n^2) n^2
\\
&=b.
\end{align*}
Applying this mapping to all elements $\vec{x}$ of $S$ maps at most
$n+1$ of them to each $\vec{y}$ in $S'$; it follows that $\card*{S'}
≤ (n+1) \card*{S}$, which means that if we can sample elements of
$S'$ uniformly, each sample will hit $S$ with probability at least
$1/(n+1)$.

To compute $\card*{S'}$, let
$C(k,m) = \card*{\SetWhere{ \vec{x} }{ ∑_{i=1}^k a'_i x_i ≤ m}}$ be the
number of subsets of $\Set{a'_1,\dots,a'_k}$ that sum to $m$ or less.
Then $C(k,m)$ satisfies the recurrence
\begin{align*}
C(k,m) &= C(k-1,m-a'_k) + C(k-1,m) \\
C(0,m) &= 1
\end{align*}
where $k$ ranges from $0$ to $n$ and $m$ ranges from $0$ to $n^2$,
and we treat $C(k-1,m-a'_k) = 0$ if $m-a'_k < 0$.
The idea is that $C(k-1,m-a'_k)$ counts all the ways to make $m$ if we
include $a'_k$, and $C(k-1,m)$ counts all the ways to make $m$ if we
exclude it.  The base case corresponds to the empty set (which sums to
$≤ m$ no matter what $m$ is).

We can compute a table of all values of $C(k,m)$ by iterating through
$m$ in increasing order; this takes $O(n^3)$ time.  At the end of this
process, we can read off $\card*{S'} = C(n,n^2)$.

But we can do more than this: we can also use the table of counts to 
sample uniformly from $S'$.
The probability that $x'_n = 1$ for a uniform random element of $S'$
is exactly $C(n-1, n^2-a'_n)/C(n,n^2)$; having chosen $x'_n = 1$
(say), the probability that $x'_{n-1} = 1$ is then $C(n-2,
n^2-a'_n-a'_{n-1})/C(n-2,n^2-a'_n)$, and so on.  So after making
$O(n)$ random choices (with $O(1)$ arithmetic operations for each
choice to compute the probabilities) 
we get a uniform element of $S'$, which we can
test for membership in $S$ in an additional $O(n)$ operations.

We've already established that $\card*{S}/\card*{S'} ≥ 1/(n+1)$, so we
can apply Lemma~\ref{lemma-sampling} to get $ε$ relative error
with probability at least $1-δ$ using
$\frac{3(n+1)}{ε^2}\ln \frac{2}{δ}$ samples.
This gives a cost of $O(n^2 \log(1/δ) / ε^2)$ for the
sampling step, or a total cost of $O(n^3 + n^2
\log(1/δ)/ε^2)$ after including the cost of building the
table (in practice, the second term will dominate unless we are
willing to accept $ε = ω(1/√{n})$).

It is possible to improve this bound.  Dyer~\cite{Dyer2003} shows that using
\concept{randomized rounding}
on the $a'_i$ instead of just truncating them gives a
FPRAS that runs in $O(n^{5/2} √{\log(1/ε)} + n^2/ε^2)$ time.
We won't do this here, but we will see randomized rounding used in
a different context in §\ref{section-max-sat}.

\section{Approximating \#DNF}
\label{section-approximating-sharp-dnf}
\label{section-Karp-Luby}

A classical algorithm of Karp and Luby~\cite{KarpL1985} gives an FPRAS
for \#DNF.  We'll describe how this works, mostly following the
presentation in~\cite[§11.2]{MotwaniR1995}.  The key idea of the
Karp-Luby technique is to express the set $S$ whose size
we want to know, as
a union of a polynomial number of simpler sets $S_1,\dots,S_k$.
If for each $i$ we can sample uniformly from $S_i$, compute
$\card*{S_i}$, and determine membership in $S_i$, then some clever
sampling will let us approximate $\card*{S}$ by sampling from a
\emph{disjoint} union of the $S_i$ to determine how much $∑
\card*{S_i}$ overestimates $\card*{S}$.

To make this concrete, let's look at the specific problem studied by
Karp and Luby of approximating the number of satisfying assignment to
a DNF formula.
A \concept{DNF formula} is a formula that is in \index{normal
form!disjunctive}\concept{disjunctive
normal form}: it is an OR of zero or more \indexConcept{clause}{clauses},
each of which is an AND of variables or their negations.  An example
would be $(x_1 ∧ x_2 ∧ x_3) ∨ (\neg x_1 ∧ x_4) ∨
x_2)$.  The \concept{\#DNF} problem is to count the number of
satisfying assignments of a formula presented in disjunctive normal
form.

Solving \#DNF exactly is \classSharpP-complete, so we don't expect to
be able to do it.  Instead, we'll get an FPRAS by cleverly sampling
solutions.  The need for cleverness arises because just sampling
solutions directly by generating one of the $2^n$ possible assignments
to the $n$ variables may find no satisfying assignments at all, since
the size of any individual clause might be big enough that getting a
satisfying assignment for that clause at random is exponentially
unlikely.

So instead we will sample pairs $(x,i)$, where $x$ 
is an
assignment that satisfies clause $C_i$; these are easier to find,
because if we know which clause $C_i$ we are trying to satisfy, we can
read off the satisfying assignment from its variables.
Let $S$ be the set of such pairs.
For each pair $(x,i)$, define $f(x,i) = 1$ if and only if $C_j(x) =
0$ for all $j < i$.  Then $∑_{(x,i) ∈ S} f(x,i)$ 
counts every satisfying
assignment $x$, because (a) there exists some $i$ such that $x$
satisfies $C_i$, and (b) only the smallest such $i$ will have $f(x,i)
= 1$.  
In effect, $f$ is picking out a single canonical satisfied clause from
each satisfying assignment.  Note that we can compute $f$ efficiently
by testing $x$ against all clauses $C_j$ with $j < i$.

Our goal is to estimate the proportion $ρ$ of ``good'' pairs with
$f(x,i) = 1$ out of all pairs in $S'=\biguplus S_i$, and then use this to estimate
$\card{S} = ∑_{(x,i) ∈ S'} f(x,i) = ρ \card*{S'}$.
If we can sample from $S$ uniformly,
the proportion $ρ$ of ``good'' pairs with $f(x,i) = 1$ is at least
$1/m$, because every satisfying assignment $x$ contributes at most $m$
pairs total, and one of them is good.

The only tricky part is figuring out how to sample pairs $(x,i)$ with
$C_i(x) = 1$ so that all pairs occur with the same probability.  Let
$S_i = \Set{ (x,i) \mid C_i(x) = 1}$.  Then we can compute $\card*{S_i} =
2^{n-k_i}$ where $k_i$ is the number of literals in $C_i$.  Using this
information, we can
sample $i$ first with probability $\card*{S_i} / ∑_j
\card*{S_j}$, then sample $x$ from $S_i$ just by picking independent
uniform random values for the
$n-k$ variables not fixed by $C_i$.

With $N ≥ \frac{4}{ε^2 (1/m)} \ln \frac{2}{δ} =
\frac{4m}{ε^2} \ln \frac{2}{δ}$, we obtain an estimate
$\hat{ρ}$ for the proportion of pairs $(x,i)$ with $f(x,i) = 1$
that is within $ε$ relative error of $ρ$ with probability at
least $1-δ$.  Multiplying this by $∑ \card*{S_i}$ then gives the
desired count of satisfying assignments.

It's worth noting that there's nothing special about DNF formulas in
this method.  Essentially the same trick will work for estimating the
size of the union of any collection of sets $S_i$ where we can (a)
compute the size of each $S_i$; (b) sample from each $S_i$
individually; and (c) test membership of our sample $x$ in $S_j$ for
$j < i$.

\section{Approximating exponentially improbable events}
\label{section-exponentially-improbable-events}

For \#KNAPSACK and \#DNF, we saw how restricting our sampling to a
cleverly chosen sample space could boost the hit ratio $ρ$ to
something that gave a reasonable number of samples using
Lemma~\ref{lemma-sampling}.  For other problems, it is often not clear
how to do this directly, and the best sample spaces we can come up
with make our target points an exponentially small fraction of the
whole.

In these cases, it is sometimes possible to approximate this
exponentially small fraction as a product of many more reasonable
ratios.  The idea is to express our target set as the last of a
sequence of sets $S_0, S_1, \dots, S_k$, where we can compute the size
of $S_0$ and can estimate $\card*{S_{i+1}}/\card*{S_i}$ accurately for
each $i$.  This gives $\card*{S_k} = \card*{S_0} ⋅ \prod_{i=1}^{k}
\frac{\card*{S_{i+1}}}{\card*{S_i}}$, with a relative error that grows
roughly linearly with $k$.  Specifically, if we can approximate each
$\card*{S_{i+1}}/\card*{S_i}$ ratio to between $1-ε$ and
$1+ε$ of the correct value, then the product of these ratios
will be between $(1-ε)^k$ and $(1+ε)^k$ of the correct
value; these bounds approach $1-ε k$ and $1 + ε k$ in
the limit as $ε k$ goes to zero, using the binomial theorem,
although to get a real bound we will need to do more careful error
analysis.

\subsection{Matchings}

We saw in §\ref{section-congestion-matchings} that a random
walk on matchings on a graph with $m$ edges has mixing time $τ_2
≤ 32m^6$, where the walk is defined by selecting an edge uniformly
at random and flipping whether it is in the matching or not,
while rejecting any steps that produce a non-matching.  This allows us
to sample matchings of a graph with $δ$ total variation
distance from the uniform distribution in $O\parens*{m^6 \parens*{\log
N + \log
\frac{1}{δ}}}$ time, where $N$ is the number of matchings. Since every
matching is a subset of the edges, we can crudely bound $N ≤ 2^m$
which lets us rewrite the sampling time as $O\parens*{m^6 \parens*{m +
\log \frac{1}{δ}}}$.

Suppose now that we want to count matchings instead of sampling them.
It's easy to show that for any particular edge $uv ∈ G$, at least
half of all matchings in $G$ don't include $uv$: the reason is that
if $M$ is a matching in $G$, then $M' = M ∖ \Set{uv}$ is also a
matching, and at most two matchings $M'$ and $M' \cup \Set{uv}$ are
mapped to any one $M'$ by this mapping.  

Order the edges of
$G$ arbitrarily as $e_1, e_2, \dots, e_m$.  Let $S_i$ be the set of
matchings in $G ∖ \Set{ e_1 \dots e_i }$.  Then $S_0$ is the set
of all matchings, and we've just argued that
$ρ_{i+1} = \card*{S_{i+1}} / \card*{S_i} ≥ 1/2$.
We also know that
$\card*{S_m}$ counts the number of matchings in a graph with no edges,
so it's exactly one.  So we can use the
product-of-ratios trick to compute $S_0 = \prod_{i=1}^{m}
\frac{\card*{S_i}}{\card*{S_{i+1}}}$.

A random walk of length $O\parens*{m^6 \parens*{m + \log
\frac{1}{η}}}$ can sample matchings from $S_i$ with a probability
$ρ'$ of
getting a matching in $S_{i+1}$ that is between $(1-η) ρ_{i+1}$ and
$(1+η) ρ_{i+1}$.  From Lemma~\ref{lemma-sampling}, we can
estimate $ρ'$ within relative error $γ$ with probability at least
$1-ζ$ using
$O\left(\frac{1}{γ^2 ρ'} \log \frac{1}{ζ}\right) =
O\left(\frac{1}{γ} \log \frac{1}{ζ}\right)$ samples.
Combined with the error on $ρ'$, this gives relative error at most
$γ + η + γη$ in
$O\parens*{m^6 \parens*{m +\log \frac{1}{η}} \log \frac{1}{γ} \log
\frac{1}{ζ}}$ operations.\footnote{This is the point where
    sensible people start hauling out the $\widetilde{O}$ notation, where
    a function is $\widetilde{O}(f(n))$ if it $O(f(n) g)$ where $g$ is
    polylogarithmic in $n$ and any other parameters that may be
running around ($\frac{1}{ε}$, $\frac{1}{η}$, etc.).}
If we then multiply out all the estimates for
$\card*{S_{i}}/\card*{S_{i+1}}$, we get an estimate of $S_0$ that is 
    at most
    $(1+γ+η+γη)^m$ times the correct value with
    probability at least $1-mζ$ (with a similar bound on the other
    side), in
    total time
$O\parens*{m^7 \parens*{m \log \frac{1}{η}} \log \frac{1}{γ} \log
\frac{1}{ζ}}$.

To turn this into a fully polynomial-time approximation scheme, given
$ε$, $δ$, and $m$, we need to
select $η$, $γ$, and $ζ$ to get relative error $ε$
with probability at least $1-δ$. 
Letting $ζ = δ/m$ gets the $δ$ part.
For $ε$, we need
$(1+γ+η+γη)^m ≤ 1+ε$.
Suppose that $ε < 1$ and let $γ = η = ε/6m$.
Then
\begin{align*}
(1+γ+η+γη)^m
    &≤ \left(1+\frac{ε}{2m}\right)^m
  \\&≤ e^{ε/2}
  \\&≤ 1+ε.
\end{align*}

Plugging these values into our cost formula gives
$O(m^8)$ times a bunch of factors that are polynomial in $\log m$ and
$\frac{1}{ε}$, which we can abbreviate as
$\widetilde{O}(m^8)$.

\subsection{Other problems}

Similar methods work for other problems that self-reduce by
restricting particular features of the solution.  Examples include
colorings (fix the color of some vertex), independent sets (remove a
vertex), and approximating the volume of a convex body (take the
intersection with a sphere of appropriate radius;
see~\cite[§11.4]{MotwaniR1995} for a slightly less sketchy
description).  We will not go into details on any of these
applications here.

\myChapter{Hitting times}{2025}{}
\label{chapter-hitting-times}

In addition to using Markov chains for sampling, we can also use
Markov chains to model processes that we hope will eventually
reach some terminating state. These Markov chains will generally
not be irreducible, since often the terminating states are
inescapable, and may or may not have the other desirable properties we
needed for convergence analysis. So instead of looking at convergence
to a stationary distribution that might not even exist, we will be
mostly interested in the \concept{hitting time}\index{time!hitting}
for some subset of $A$ of the states, defined as the minimum $τ$ such
that $X^τ ∈ A$ starting from some given initial distribution on
$X^0$. As suggested by the notation, hitting times will a special case
of stopping times (see Chapter~\ref{chapter-stopping-times}), and if
we are very lucky we may be able to use the optional stopping theorem
to bound them. If we are less lucky, we will use whatever tools we
can.

\section{Waiting times}
\label{section-waiting-times}

The simplest case is when our Markov chain consists of two states: one
where something hasn't happened yet, and one where it has. If the
something happens with probability $p$ at each step, then $τ$ has a
geometric distribution, and we can calculate $\Exp{τ}$ using the usual
conditional-probability argument for a geometric random variable (see
§\ref{section-geometric-random-variables}):
\begin{align*}
    \Exp{τ}
    &= p⋅1 + (1-p)⋅(1 + \Exp{τ})
    \\&= 1 + (1-p)\Exp{τ}
    \intertext{which has the solution}
    \Exp{τ}
    &= 1/p.
\end{align*}

A more interesting case is when we have a Markov chain that we can
describe using a sequence of waiting times. For example, an
\concept{epidemic} process involves a population of $n$ agents of
which initially only one is infected, and at each step we pair two
agents chosen uniformly at random, and if one is infected and the
other not, the uninfected agent becomes infected. 
Such processes are used as the basis for \index{gossip
algorithm}\index{algorith!gossip}\conceptFormat{gossip} or
\index{epidemic
algorithm}\index{algorithm!epidemic}\conceptFormat{epidemic
algorithms}~\cite{DemersGHIL1987},
where the infection corresponds to knowing some piece of
information we want to broadcast through the population.

If we let $X^t$ be the number of agents infected after $t$ steps, we
have a Markov process where 
\begin{align*}
    p_{k,k+1} 
    &= \frac{k(n-k)}{\binom{n}{2}}
    \\
    p_{k,k}
    &= 1 - p_{k,k+1},
\end{align*}
and all other transition probabilities are $0$. So any trajectory of
this process involves starting in $1$, moving to $2$ after some time,
then $3$, and so on until all $n$ agents are infected.

The hitting time $τ_n$ for $n$ will be the sum of the waiting times for each
of these steps. This gives
\begin{align*}
    \Exp{τ_n}
    &= ∑_{k=1}^{n-1} \frac{1}{p_{k,k+1}}
    \\&= ∑_{k=1}^{n-1} \frac{\binom{n}{2}}{k(n-k)}
    \\&= \binom{n}{2}
        ∑_{k=1}^{n-1} \parens*{\frac{1/n}{k} + \frac{1/n}{n-k}}
    \\&= \frac{(n-1)}{2}
        ∑_{k=1}^{n-1} \parens*{\frac{1}{k} + \frac{1}{n-k}}
    \\&= (n-1) H_{n-1}
    \\&= Θ(n \log n).
\end{align*}

A similar process in the same model solves
\concept{leader election}\index{election!leader}, by
starting with $n$ leaders and having one drop out whenever two leaders
meet~\cite{AngluinADFP2006}. The end state has exactly one leader.

As with an epidemic, we can reduce the state down to a
single number: we'll track only how many leaders remain in each
state. This number drops by one when two leaders meet, so we have  
\begin{align*}
    p_{k,k-1} &= \frac{k(k-1)}{n(n-1)}.
\end{align*}

Summing the waiting times for each of these transitions gives
\begin{align*}
    \Exp{τ}
    &= ∑_{k=2}^{n} \frac{n(n-1)}{k(k-1)}
    \\&= n(n-1) ∑_{k=2}^{n} \parens*{\frac{1}{k-1}-\frac{1}{k}}
    \\&= n(n-1) \parens*{1 - \frac{1}{n}}
    \\&= (n-1)^2.
\end{align*}

Usually it will be difficult get exact expectations like this, but
it's nice when it happens.

\section{Probabilistic recurrences for waiting times}
\label{section-probabilistic-recurrences}

We can generalize waiting time analysis a bit using a probabilistic
recurrence relation of the form
\begin{equation}
    T(n) = 1 + T(n-X_{n}),
    \label{eq-probabilistic-recurrence-waiting}
\end{equation}
where $X_{n}$ is a random variable with $0 ≤ X_{n} ≤ n$
and $T(0) = 0$. 

The idea is that we can characterize progress with a single integer
parameter as in the previous section, but instead of always going from
$n$ to $n-1$ after some expected waiting time we may jump from $n$ to
any value $n-X_n$ in the range $0$ to $n$ with some random distribution
depending on $n$.

We'll start by giving some examples of processes that fit within this
framework, then show (using a result of Karp~\etal~\cite{KarpUW1988}) 
how to bound these recurrences using only a lower
bound on $\Exp{X_n}$ for each $n$ that satisfies a simple
monotonicity property.

\subsection{Examples}
\label{section-probabilistic-recurrences-examples}

Here are some processes that fit within this framework:
\begin{itemize}
    \item How long does it take to get our first heads if we repeatedly
        flip a coin that comes up heads with probability $p$?
        Even though we probably already know the answer to this,
        We can solve it by solving the recurrence $T(1) = 1 + T(1-X_1), T(0)
        = 0$,
        where $\Exp{X_1} = p$.
    \item 
        In §\ref{section-QuickSelect}, we bounded the running time of 
        \concept{Hoare's FIND}~\cite{Hoare1961QuickSelect}, 
        often called \concept{QuickSelect}, an algorithm for finding
        the $k$-th smallest element of an unsorted array.
        Recall that this algorithm works in a series of rounds, where
        in each round we remove some of the remaining elements.

        If we consider the expected number rounds this takes, 
        we get a recurrence
        \eqref{eq-probabilistic-recurrence-waiting} where $X_n$ is the
        number of elements we remove.
        Computing $\Exp{X_{n}}$ exactly is tricky,
        since after splitting our array of size $n$ into piles
        of size $n'$ and $n-n'-1$, we have to pick one or the other (or
        possibly just the pivot alone) based on the value of $k$.
        But we showed that $\Exp{X_{n}} ≥ n/4$.
    \item Suppose we start with $n$ biased coins that each come up heads
        with probability $p$.  In each round, we flip all the coins and throw
        away the ones that come up tails.  How many rounds does it take to
        get rid of all of the coins? This tells us how tall a
        skip list (\cite{Pugh1990}, or see §\ref{section-skip-lists} can get.
        Here we have $\Exp{X_{n}} = (1-p)n$.
    \item In §\ref{section-coupon-collector}, we looked at
        the \concept{coupon collector problem}, where we sample from
        $1\dots n$ with replacement until we see ever value at least once.
        We can model this by a recurrence in which $T(k)$ is the time to get
        all the coupons given there are $k$ left that we haven't seen. 

        A mild complication is that we want to keep $n$ for the total
        number of coupons, so we have to rewrite
        \eqref{eq-probabilistic-recurrence-waiting} to use $k$ instead
        of $n$.
        With this adjustment,
        $X_{k}$ is $1$ with probability $k/n$ and $0$ with probability
        $(n-k)/n$, giving $\Exp{X_{k}} = k/n$.
    \item Let's play \concept{Chutes and Ladders} without the chutes and
        ladders. We start at location $n$, and whenever it's our turn, we
        roll a fair six-sided die $X$ and move to $n-X$ unless this value is
        negative, in which case we stay put until the next turn.  How many
        turns does it take to get to $0$?
\end{itemize}

\subsection{The Karp-Upfal-Wigderson bound}
\label{section-KUW}

Here we describe a mechanical solution to
\eqref{eq-probabilistic-recurrence-waiting} that works for many
distributions on $X_n$.

It was original described in a paper by Karp, Upfal, and Wigderson on
analyzing parallel search algorithms~\cite{KarpUW1988}.
The
bound applies when $\Exp{X_n}$ is bounded below by a non-decreasing
function $μ(n)$.

\index{Karp-Upfal-Wigderson bound}
\index{bound!Karp-Upfal-Wigderson}
\begin{lemma}
\label{lemma-karp-upfal-wigderson}
Let $a$ be a constant, let $T(n) = 1 + T(n-X_n)$, where for each $n$, $X_n$ is an
integer-valued random variable satisfying $0 ≤ X_n ≤ n-a$ and let
$T(a) = 0$.  Let $\Exp{X_n} ≥ μ(n)$ for all $n > a$, where $μ$ is a
positive non-decreasing function of $n$.  Then
\begin{align}
\label{eq-karp-upfal-wigderson}
\Exp{T(n)} &≤ \int_a^n \frac{1}{μ(t)} \,dt.
\end{align}
\end{lemma}

To get an intuition for why this works, imagine that 
$X_n$ is the speed at which we drop from $n$, expressed in units per round.  
Traveling at this speed, it takes $1/X_n$ rounds to cross from $k+1$
to $k$ for any such interval we pass.  From the point of view of the
interval $[k,k+1]$, we don't know which $n$ we are going to start from
before we cross it, but we do know that for any $n ≥ k+1$ we start
from, our speed will be at least $μ(n) ≥ μ(k+1)$ on average.  So
the time it takes will be at most
$\int_{k}^{k+1} \frac{1}{μ(t)} \,dt$ on average, and the total time
is obtained by summing all of these intervals.

Of course, this intuition is not even close to a real proof (among
other things, there may be a very dangerous confusion in there
between $1/\Exp{X_n}$ and $\Exp{1/X_n}$), so we will give a real proof as
well.

\begin{proof}[Proof of Lemma~\ref{lemma-karp-upfal-wigderson}]
This is essentially the same proof as in Motwani and
Raghavan~\cite{MotwaniR1995}, but we add some extra detail to allow for
the possibility that $X_n=0$.  

Let $p = \Prob{X_n=0}$, $q = 1-p = \Prob{X_n≠
0}$.  Note we have $q>0$ because otherwise $\Exp{X_n} = 0 < μ(n)$.  Then we have
\begin{align*}
\Exp{T(n)} 
&= 1 + \Exp{T(n-X_n)} \\
&= 1 + p \ExpCond{T(n-X_n)}{X_n=0} + q \ExpCond{T(n-X_n)}{X_n≠ 0} \\
&= 1 + p \Exp{T(n)} + q \ExpCond{T(n-X_n)}{X_n≠ 0}.
\end{align*}

Now we have $\Exp{T(n)}$ on both sides, which we don't like very much.  So we collect it on the left-hand side:
\begin{align*}
(1-p) \Exp{T(n)} &= 1 + q \ExpCond{T(n-X_n)}{X_n ≠ 0},
\end{align*}
divide both sides by $q = 1-p$, and apply the induction hypothesis:
\newcommand{\lemmaKUWugly}[1]{\int_a^{#1} \frac{1}{μ(t)}\,dt}
\begin{align*}
\Exp{T(n)} &= 1/q + \ExpCond{T(n-X_n)}{X_n ≠ 0} \\
 &= 1/q + \ExpCond{\ExpCond{T(n-X_n)}{X_n}}{X_n ≠ 0} \\
 &≤ 1/q + \ExpCond{\lemmaKUWugly{n-X_n}}{X_n ≠ 0} \\
&= 1/q + \ExpCond{\lemmaKUWugly{n} - \int_{n-X_n}^{n}
\frac{1}{μ(t)}\,dt}{X_n ≠ 0} \\
&≤ 1/q + \lemmaKUWugly{n} - \ExpCond{\frac{X_n}{μ(n)}}{X_n ≠ 0} \\
&≤ 1/q + \lemmaKUWugly{n} - \frac{\ExpCond{X_n}{X_n ≠ 0}}{μ(n)}.
\end{align*}

The second-to-last step uses the fact that $μ(t)≤ μ(n)$ for
$t≤ n$.

It may seem like we don't know what $\ExpCond{X_n}{X_n≠ 0}$ is.  But we know
that $X_n≥ 0$, so we have $\Exp{X_n} = p \ExpCond{X_n}{X_n=0} + q \ExpCond{X_n}{X_n≠ 0} = q
\ExpCond{X_n}{X_n≠ 0}$.  So we can solve for $\ExpCond{X_n}{X_n≠ 0} = E[X_n]/q$.  So let's plug this in:
\begin{align*}
\Exp{T(n)}
&≤ 1/q + \lemmaKUWugly{n} - \frac{\Exp{X_n}/q}{μ(n)} \\
&≤ 1/q + \lemmaKUWugly{n} - 1/q \\
&= \lemmaKUWugly{n}.
\end{align*}

This concludes the proof.
\end{proof}

Now we just need to find some applications.

\subsubsection{Applications}

Let's go through the list of examples we had in
§\ref{section-probabilistic-recurrences-examples}.

\subsubsection{Waiting for heads}

For the recurrence $T(1) = 1 + T(1-X_1)$ with $\Exp{X_1} = p$, we set
$μ(n) = p$ and get $\Exp{T(1)} ≤ \int_0^1 \frac{1}{p} \,dt =
\frac{1}{p}$, which happens to be exactly the right answer.

\subsubsection{QuickSelect}

In QuickSelect, we pick a random pivot and split the original array of
size $n$ into three piles of size $m$ (less than the pivot), $1$ (the
pivot itself), and $n-m-1$ (greater than the pivot).  We then figure
out which of the three piles contains the $k$-th smallest element
(depend on how $k$ compares to $m-1$) and recurse, stopping when we
hit a pile with $1$ element.  It's easiest to analyze this by assuming
that we recurse in the largest of the three piles, i.e., that our
recurrence is $T(n) = 1 + \max(T(m), T(n-m-1))$, where $m$ is uniform
in $0\dots n-1$.  The exact value of $\Exp{\max(m, n-m-1)}$ is a little
messy to compute (among other things, it depends on whether $n$ is odd
or even), but it's not hard to see that it's always less than
$(3/4)n$.  So letting $μ(n) = n/4$, we get
\begin{align*}
\Exp{T(n)} ≤ \int_1^{n} \frac{1}{t/4} \,dt = 4 \ln n.
\end{align*}

\subsubsection{Tossing coins}

Here we have $\Exp{X_{n}} = (1-p)n$.  If we let $μ(n) = (1-p)n$ and plug into the formula without thinking about it too much, we get
\begin{align*}
\Exp{T(n)} &≤ \int_0^n \frac{1}{(1-p)t} \,dt = \frac{1}{1-p}(\ln n - \ln 0).
\end{align*}

That $\ln 0$ is trouble.  We can fix it by making $μ(n) =
(1-p)\ceil{n}$, to get
\begin{align*}
\Exp{T(n)} &≤ \int_{0+}^n \frac{1}{(1-p)\lceil t \rceil} \,dt \\
&= \frac{1}{1-p} ∑_{k=1}^{n} \frac{1}{k} \\
&= \frac{H_n}{1-p}.
\end{align*}

\subsubsection{Coupon collector}
Now that we know how to avoid dividing by zero, this is easy and fun.
Let $μ(x) = \ceil{x}/n$, then we have
\begin{align*}
\Exp{T(n)} & ≤ \int_{0+}^{n} \frac{n}{\lceil t \rceil}\,dt \\
&= ∑_{k=1}^{n} \frac{n}{k} \\
&= n H_n.
\end{align*}

As it happens, this is the exact answer for this case.  This will
happen whenever $X$ is always a 0--1 variable and we define $μ(x) =
\ExpCond{X}{n=\ceil{x}}$, which can be seen by spending far too much
time thinking about the precise sources of error in the inequalities
in the proof.

\subsubsection{Chutes and ladders}
Let $μ(n)$ be the expected drop from position $n$.  We have to be a
little bit careful about small $n$, but we can compute that in general
$μ(n) = \frac{1}{6} ∑_{i=0}^{\min(n,6)} i$.  For
fractional values $x$ we will set $μ(x) = μ(\ceil{x})$ as before.  Then we have
\begin{align*}
\Exp{T(n)} & ≤ \int_{0+}^{n} \frac{1}{μ(t)}\,dt \\
&= ∑_{k=1}^{n} \frac{1}{μ(k)} \\
\end{align*}

We can summarize the values in the following table:

\begin{align*}
\begin{array}{llll}
n&μ(n)&1/μ(n)&∑ 1/μ(k)\\
1&1/6&6&6\\
2&1/2&2&8\\
3&1&1&9\\
4&5/3&3/5&48/5\\
5&5/2&2/5&10\\
≥6&7/2&2/7&10+(2/7)(n-5)=(2/7)n + 65/7\\
\end{array}
\end{align*}

This is a slight overestimate; for example, we can calculate by hand
that the expected waiting time for $n=2$ is $6$ and for $n=3$ that it is $20/3
= 6 + 2/3$.

We can also consider the generalized version of the game where we
start at $n$ and drop by $1\dots n$ each turn as long as the drop
wouldn't take us below 0.  Now the expected drop from position $k$ is
$k(k+1)/2n$, and so we can apply the formula to get
\begin{align*}
\Exp{T(n)} &≤ ∑_{k=1}^{n} \frac{2n}{k(k+1)}.
\end{align*}

The sum of $\frac{1}{k(k+1)}$ when $k$ goes from $1$ to $n$ happens to
have a very nice value; it's exactly $\frac{n}{n+1} = 1 +
\frac{1}{n+1}$.\footnote{Proof: Trivially true for $n=0$; for larger
$n$, compute 
$∑_{k=1}^{n} \frac{1}{k(k+1)} 
∑_{k=1}^{n-1} \frac{1}{k(k+1)} + \frac{1}{n(n+1)}
= \frac{n-1}{n} + \frac{1}{n(n+1)}
= \frac{(n-1)(n+1) - 1}{n(n+1)}
= \frac{n^{2}}{n(n+1)} 
= n/(n+1)$.}
So in this case we can rewrite the bound as $2n⋅ \frac{n}{n+1} =
\frac{2n^{2}}{n+1}$.

\subsection{High-probability bounds}
\label{section-probabilistic-recurrences-high-probability}

So far we have only considered bounds on the expected value of $T(n)$.
Suppose we want to show that $T(n)$ is in fact small with high
probability, i.e., a statement of the form $\Prob{T(n) ≥ t} ≤
ε$.  There are two natural ways to do this: we can repeatedly apply Markov's inequality to the expectation bound, or we can attempt to analyze the recurrence in more detail.  The first method tends to give weaker bounds but it's easier.

\subsubsection{High-probability bounds from expectation bounds}

Given $\Exp{T(n)} ≤ m$, we have $\Prob{T(n) ≥ α m} ≤
α^{-1}$.  This does not give a very good bound on probability; if
we want to show $\Prob{T(n) ≥ t} ≤ n^{-c}$ for some constant $c$ (a
typical high-probability bound), we need $t ≥ n^{c}m$.  But we can
get a better bound if $m$ bounds the expected time starting from any reachable state, as is the case for the class of problems we have been considering.

The idea is that if $T(n)$ exceeds $αm$, we restart the analysis
and argue that $\ProbCond{T(n) ≥ 2α m}{T(n) ≥ α m}
≤ α^{-1}$, from which it follows that $\Prob{T(n) ≥
2αm} ≤ α^{-2}$.  In general, for any non-negative
integer $k$, we have $\Prob{T(n) ≥ kαm} ≤ α^{-k}$.
Now we just need to figure out how to pick $α$ to minimize this
quantity for fixed $t$.

Let $t = kαm$.  Then $k = t/αm$ and we seek to
minimize $α^{-t/αm}$.  Taking the logarithm gives
$-(t/m)(\ln α)/α$.  The $t/m$ factor is irrelevant to
the minimization problem, so we are left with minimizing $-(\ln
α)/α$.  Taking the derivative gives $-α^{-2} +
α^{-2} \ln α$; this is zero when $\ln α = 1$ or $α
= e$.  (This popular constant shows up often in problems like this.)
So we get $\Prob{T(n) ≥ kem} ≤ e^{-k}$, or, letting $k =
\ln(1/ε)$, $\Prob{T(n) ≥ em \ln(1/ε)} ≤ ε$.

So, for example, we can get an $n^{-c}$ bound on the probability of
running too long by setting our time bound to $em \ln(n^{c}) = cem \ln
n = O(m \log n)$.  We can't hope to do better than $O(m)$, so this bound is tight up to a log factor.

\subsubsection{Detailed analysis of the recurrence}
\label{section-probabilistic-recurrences-detailed-analysis}

As Lance Fortnow has
explained,\footnote{\url{http://weblog.fortnow.com/2009/01/soda-and-me.html}}
getting rid of log factors is what theoretical computer science is all about.
So we'd like to do better than an $O(m \log n)$ bound if we can.  In some cases this is not too hard.

Suppose for each $n$, $T(n) = 1 + T(X)$, where $\Exp{X} ≤ αn$
for a fixed constant $α$.  Let $X_{0} = n$, and let $X_{1},
X_{2}$, etc., be the sequence of sizes of the remaining problem at
time 1, 2, etc.  Then we have $\Exp{X_{1}} ≤ αn$ from our
assumption.  But we also have $\Exp{X_{2}} = \Exp{\ExpCond{X_{2}}{X_{1}}} ≤
\Exp{αX_{1}} = α\Exp{X_{1}} ≤ α^{2}n$, and by
induction we can show that $\Exp{X_{k}} ≤ α^{k}n$ for all
$k$.  Since $X_{k}$ is integer-valued, $\Exp{X_{k}}$ is an upper bound
on $\Prob{X_{k} > 0}$; we thus get $\Prob{T(n) ≥ k} = \Prob{X_{k} > 0}
≤ α^{k}n$.  We can solve for the value of $k$ that makes
this less than $ε$: $k = -\log(n/ε) / \log α
= \log_{1/α} n + \log_{1/α} (1/ε)$.

For comparison, the bound on the expectation of $T(n)$ from
Lemma~\ref{lemma-karp-upfal-wigderson}
is $H(n)/(1-α)$.  This is actually pretty close to
$\log_{1/α} n$ when $α$ is close to 1, and is not too bad
even for smaller $α$.  But the difference is that the dependence
on $\log(1/ε)$ is additive with the tighter analysis, so for
fixed $c$ we get $\Prob{T(n) ≥ t} ≤ n^{-c}$ at $t = O(\log n +
\log n^{c}) = O(\log n)$ instead of $O(\log n \log n^{c}) = O(\log^{2}
n)$.

\subsection{More general recurrences}

The Karp-Upfal-Wigderson bound is pretty powerful, but it only applies
to a very narrow class of recurrences, that fortunately correspond
naturally to Markov chain hitting times. For more general recurrences,
a collection of tools that may be helpful can be found in a paper by 
Roura~\cite{Roura2001}. But these are outside the scope of this chapter.

\section{Lyapunov functions}
\label{section-Lyapunov-functions}

Waiting time analysis works well for processes like the ones in
§\ref{section-waiting-times} that only go in one
direction. But what if sometimes we go backwards, or if it's not even
possible to arrange the states of our Markov chain neatly in one
dimension? Here it can be helpful to construct a
\concept{Lyapunov function}\index{function!Lyapunov},
which is often called a
\concept{potential function}\index{function!potential}
in the computer science literature. This is a function that reduces
each state of a complex process to a single value in $ℝ$, ideally
producing some sort of semimartingale that drops in a predictable way
that so that we can use it to get a bound on the time to reach our
target set.

An example of a Lyapunov function is the integral in
\eqref{eq-karp-upfal-wigderson}. This doesn't necessarily compute the
time for $T$ to reach $0$ exactly, but it drops by $1$ on average and
so gives an upper bound on the hitting time.

We've also seen examples of Lyapunov functions
§\ref{section-stopping-times-and-random-walks}.
For an unbiased random walk, we saw that the function $X_t^2$ drops by
$1$ on average, making $X_t^2 - t$ a martingale and $X_t^2$ a Lyapunov
function. A similar result
holds with $X_t - (p-q)t$ for biased random walks, making $X_t$ its
own Lyapunov function (with a rate of descent that is not necessarily
$1$).

Let's try this for a more complicated process. The
\concept{rock-paper-scissors} process is similar to the epidemic
process, except that instead of having two states infected and
uninfected, where infected overwrites uninfected when agents in these
states meet, we have three states with a cyclic overwriting relation.
This means that when a rock agent meets a paper agent, we get two
paper agents (``paper covers rock''), and similarly paper gets
replaced by scissors (``scissors cuts paper'') and scissors gets
replaced by rock (``rock breaks scissors''). Meetings between agents
in the same state have no effect. As in the epidemic case,
we choose pairs of agents to interact uniformly at random from some
population of size $n$.\footnote{This is also a special case of the
\concept{Lotka-Volterra model} in population dynamics. The translation
is to convert rocks to wolves, paper to grass, and scissors to sheep,
and adopt the rules ``sheep eats grass'', ``wolf eats sheep'', and
``grass starves wolf.''}

Processes like this have been extensively studied in a variety of
models; see~\cite{SzolnokiMJSRP2014} for a survey. The rock-paper-scissors 
process in particular is often modeled in its continuous limit as a
system of differential equations:
\begin{align*}
    r' &= rs - pr \\
    p' &= pr - sp \\
    s' &= sp - rs,
\end{align*}
where $r$, $p$, and $s$ represent the densities of rocks, papers, and
scissors in the population at some time and primes are used to indicate
derivatives with respect to time.

This system of differential equations does
not have simple closed-form solutions, but it does have some
interesting invariants.
If we compute $(r+p+s)' = r' + p' + s' = (rs - pr) + (pr - sp) + (sp - rs) = 0$,
we see that the total concentration of agents $r+p+s$ doesn't change over time.
This is not especially surprising.
But we can also compute
\begin{align*}
    (rps)'
    &= r'ps + rp's + rps'
    \\&= (rs - pr)ps + rs(pr - sp) + rp(sp - rs)
    \\&= rps(s - p) + rps(r - s) + rps(p - r)
    \\&= 0.
\end{align*}
This shows that the product of the three concentrations also doesn't
change over time, giving a family of closed-loop trajectories in the
continuous case. This makes the rock-paper-scissors process a useful
way to build an oscillator in a dynamical system.

In the discrete case, we don't have infinitesimal rocks breaking
infinitesimal scissors over infinitesimal time intervals. Instead, we
have a Markov chain. The states of this Markov chain are triples of
counts $\Tuple{R_t, P_t, S_t}$, and at each step we see transitions
\begin{align*}
    \Tuple{R,P,S} &→ \Tuple{R-1,P+1,S}
    & \text{with probability $RP/\binom{n}{2}$}
    \\ \Tuple{R,P,S} &→ \Tuple{R,P-1,S+1}
    & \text{with probability $PS/\binom{n}{2}$}
    \\ \Tuple{R,P,S} &→ \Tuple{R+1,P,S-1}
    & \text{with probability $SR /\binom{n}{2}$},
\end{align*}
with the remaining probability causing no change.

This is a finite Markov chain that is aperiodic because of the no-op
transitions, but it's not irreducible. The three states
$\Tuple{n,0,0}$, $\Tuple{0,n,0}$, and $\Tuple{0,0,n}$ all have no
outgoing transitions: once we're all rocks, we all stay rocks forever.
Furthermore there is a path from every configuration with $n$ agents
that isn't one of these three terminal configurations to one of these
configurations. This means that eventually this process will converge
to a uniform configuration, which is trouble if we want to use the
process as an oscillator but maybe not so bad if we are trying to
achieve agreement. But whatever our goal is, we can ask how long will this take?

The $rps$ invariant for the continuous process suggests looking for a
similar invariant or near-invariant for the discrete process. Let's
define a Lyapunov function
\begin{align*}
    Φ &= RPS,
\end{align*}
and look at the expected change in this function at each step.

Suppose paper covers rock, which occurs with probability
$PR/\binom{n}{2}$. Then the change in $Φ$ is
\begin{align*}
    ∆Φ
    &= (R-1)(P+1)S - RPS
    \\&= RPS + RS - PS - S - RPS
    \\&= (R-P-1)S.
\end{align*}
If we multiply this by the probability that the event occurs, we get a
contribution to $\Exp{∆Φ}$ of $(R-P-1)⋅RPS/\binom{n}{2}$.
By symmetry, the other two transitions yield contributions of
$(P-S-1)⋅RPS/\binom{n}{2}$ and $(S-R-1)⋅RPS/\binom{n}{2}$. Adding
these up gives
\begin{align*}
    \Exp{∆Φ}
    &= \frac{RPS}{\binom{n}{2}} \parens*{(R-P-1)+(P-S-1)+(S-R-1)},
    \intertext{which very nicely cancels down to}
    \\&= RPS⋅\frac{-3}{\binom{n}{2}}.
\end{align*}

What this tells us is that $Φ$ drops, on average, by a $Θ(n^{-2})$
fraction of its current value. We can express this as
\begin{align}
    \ExpCond{Φ_{t+1}}{Φ_t}
    &= Φ_t \parens*{1 - \frac{3}{\binom{n}{2}}},
    \nonumber
    \intertext{which, iterated, gives for any $t$}
    \Exp{Φ_t}
    &= Φ_0 \parens*{1- \frac{3}{\binom{n}{2}}}^t.
    \label{eq-rps-expectation}
\end{align}

Alternatively we can use this to argue that 
\begin{align}
    Z_t 
    &= Φ_t \parens*{1 - \frac{3}{\binom{n}{2}}}^{-t}
    \label{eq-rps-martingale}
\end{align}
is a martingale.

For proving an upper bound on the hitting time, the first formulation
\eqref{eq-rps-expectation}
is more useful. Using our old friend $1+x ≤ e^x$, we can rewrite the
bound as $\Exp{Φ_t} ≤ e^{-6t/n(n-1)}$. Now observe that $Φ$ is
never greater than $(n/3)^3$ and never less than $n-2$ as long as it
is not zero. So we can compute
\begin{align*}
    \Prob{Φ_t ≠ 0}
    &= \Prob{Φ_t ≥ n-2}
    \\&≤ \frac{\Exp{Φ_t}}{n-2}
    \\&≤ \frac{(n/3)^3 e^{-6t/n(n-1)}}{n-2}
    \\&= O\parens*{n^2 \exp\parens*{-Θ(t/n^2)}}.
\end{align*}
We can immediately see that there is some value $t = O(n^2 \log n)$ that
knocks $\Prob{Φ_t ≠ 0}$ down to at most $1/2$.  As when converging to a stationary
distribution, if we lose this coin-flip, we can restart the argument
and try again. This gives an expected waiting time of $O(n^2 \log n)$ to reach
$Φ=0$.

We are not quite done. Having $Φ=0$ only means that one of our three
species has disappeared; for full convergence, we need to lose two.
But once we are down to two remaining species (say rock and paper), we
have an epidemic process. From our previous analysis, we know that
only $O(n \log n)$ additional steps are needed to get down to one.

It's worth noting that hitting times are generally stopping times. So
perhaps there is a cleaner argument using \eqref{eq-rps-martingale}
and the optional stopping theorem (see
Theorem~\ref{theorem-optional-stopping}). Let $τ$ be the first time at which
$Φ_τ = 0$. We've already shown that $\Exp{τ}$ is finite, so the
bounded-increments case of Theorem~\ref{theorem-optional-stopping} is
tempting. But if we get unlucky and $Φ_{t+1} = Φ_t ≠ 0$ for some large
$t$, $Z_t$ can increase by an arbitrarily large amount.
The bounded time and bounded range cases also don't apply.
It's tempting to see if the general case works, but since we know that
$Z_τ = 0 ≠ Z_0$ and OST implies $Z_τ = Z_0$, we can rule out applying
the theorem to this particular martingale and stopping time no matter
how clever we get. What we \emph{can} do is use a fixed time $t$ and compute
$\Exp{Z_t} = Z_0 = Φ_0$, but at this point we are just reinventing
\eqref{eq-rps-expectation}.

To salve our disappointment, let's get a lower bound on the expected
fixation time. Use $1-x ≥ e^{-x/2}$, which holds for $0≤x≤2$ at least,
to get
\begin{align*}
    \Exp{Φ_t} 
    &= Φ_0 \parens*{1-Θ(1/n^2)}^t
    \\&≥ Φ_0 \exp\parens*{-Θ(t/n^2)}.
\end{align*}

If we start with $Φ_0 = Θ(n^3)$, there is some $t = Θ(n^2)$
such that $\Exp{Φ_t}$ is still $Θ(n^3)$, just with a smaller constant.
But then we can argue
\begin{align*}
    Θ(n^3)
    &= \Exp{Φ_t}
    \\&= \Exp{Φ_t} \Prob{Φ_t ≠ 0}
    \\&≤ O(n^3) \Prob{Φ_t ≠ 0},
\end{align*}
from which $\Prob{Φ_t ≠ 0} = Ω(1)$. So there is a constant probability
of not reaching $Φ_t = 0$ after $Θ(n^2)$ steps, which gives an
expected hitting time for $Φ = 0$ of $Ω(n^2)$. This bound is missing a
log factor compared to our upper bound. I suspect that this
could be improved with a better analysis.

\section{Drift analysis}

The Lyapunov/potential function approach has two main steps: (a)
finding a good Lyapunov function and (b) using it to get a bound on
hitting time. The first step can be frustrating because we know that a
good Lyapunov function always exists, since we can in principle just
take the expected hitting time from the current state. But finding a
good, clean Lyapunov function that we can write down and reason about
is tougher. The second step can involve a lot of awkward futzing with
supermartingales and such, and we can avoid some of this awkwardness
but letting other people package it up in a few convenient theorems.
For this purpose, it's helpful to look at the literature on
\concept{drift analysis}, which mostly shows up in the context of
\index{algorithm!evolutionary}\indexConcept{evolutionary algorithm}{evolutionary algorithms}, a
class of optimization algorithms modeled on natural evolutionary
processes. In this section, we'll describe some of the results in this
area, largely following a survey by Lengler~\cite{Lengler2020}.
Many of these results are at their core repackagings of the optional
stopping theorem, but it can be convenient not to have to go through
the details of the OST every time we want to use them.

As before, the idea is that we will come up with some one-dimensional
stochastic process $\Set{X_t}$ on the non-negative reals $ℝ^+_0$ that
is related to the process we are trying to bound.
Our hope is that $X_t$ either drops by a consistent additive amount or
a consistent multiplicative factor on average at each step.

For \index{drift!additive}\concept{additive drift}, we have the following:
\begin{theorem}[{\cite[Theorem 2.3.1]{Lengler2020}}, adapted
    from~\cite{HeY2004}]
    \label{theorem-drift-additive}
    Let $(X_t)_{t≥0}$ b e sequence of non-negative random variables
    with a finite state space $S ⊆ ℝ^+_0$ such that $0 ∈ S$. Let $T =
    \inf\SetWhere{t ≥ 0}{X_t = 0}$.
    \begin{enumerate}
        \item If there exists $δ>0$ such that for all $s ∈ S ∖
            \Set{0}$ and for all $t ≥ 0$,
            \begin{align*}
                ∆_t(s) &= \ExpCond{X_t - X_{t+1}}{X_t = s} ≥ δ,
            \end{align*}
            then
            \begin{align*}
                \Exp{T} &≤ \frac{\Exp{X_0}}{δ}.
            \end{align*}
        \item If there exists $δ > 0$ such that for all $s ∈ S ∖
            \Set{0}$ and for all $t ≥ 0$,
            \begin{align*}
                ∆_t(s) &= \ExpCond{X_t - X_{t+1}}{X_t = s} ≤ δ,
            \end{align*}
            then
            \begin{align*}
                \Exp{T} &≥ \frac{\Exp{X_0}}{δ}.
            \end{align*}
    \end{enumerate}
\end{theorem}

The proof of Theorem~\ref{theorem-drift-additive} is a fairly simple
application of the optional stopping theorem. Depending on which case
we are in, $X_t + tδ$ is either a supermartingale or submartingale;
we have a bounded range because of the assumption that $S$ is
finite; and we have finite time in the supermartingale case
because the downward drift implies and finite state space implies that
there is always a path to $0$ that we can take with nonzero
probability, and in the submartingale case because if we don't 
$\Exp{T}$ is infinite. So in either case we are looking at 
arguments that we've done before.

Where the theorem is helpful is that it saves us from having to repeat
these arguments. For example, in the case of a biased random walk with
reflecting barriers at $0$ and $n$ and a probability of $p > 1/2$ of
dropping at each step, we can compute $\ExpCond{X_t - X_{t+1}}{X_t =
s} ≥ p - q$ (note that the convention here is that we are tracking
expected drop, which is the negative of the expected change) and
instantly get $\Exp{T} ≤ n/(p-q)$ for any starting point $X_0 ≤ n$.
In the other direction we are not so lucky: the reflecting barrier at $n$ means
that $\ExpCond{X_t - X_{t+1}}{X_t = n} = 1$, so the best bound we get
starting at $X_0 = n$ is $\Exp{T} ≥ n$, which we don't really need the theorem to get.

In some cases, it can be hard to find a Lyapunov function with
constant additive drift. Rescaling may help in this case: the idea is
to stretch the Lyapunov function locally by multiplying by the inverse
of the expected drop, so that the new stretched function has expected
drop at least 1. This trick has been reinvented several times in
different contexts, dating back at least to the
\concept{probabilistic recurrence relations} of
Karp~\etal~\cite{KarpUW1988}. We'll quote the version from the
evolutionary analysis literature given by Lengler~\cite{Lengler2020},
which is known as the 
\index{drift!variable}\concept{variable drift} theorem:

\begin{theorem}[{\cite[Theorem 2.3.3]{Lengler2020}}, adapted from \cite{Johannsen2010,RoweS2014}]
    \label{theorem-drift-variable}
    Let $(X_t)_{t≥0}$ b e sequence of non-negative random variables
    with a finite state space $S ⊆ ℝ^+_0$ such that $0 ∈ S$. Let
    $s_{\min} = \min(S∖\Set{0})$, let $T =
    \inf\SetWhere{t ≥ 0}{X_t = 0}$, and for $t≥0$ and $s∈S$ let
    $∆_t(s) = \ExpCond{X_t-X_{t+1}}{X_t = s}$. If there is an
    increasing function $h:ℝ^+ → ℝ^+$ such that for all $s ∈
    S∖\Set{0}$ and all $t ≥ 0$,
    \begin{equation*}
        ∆_t(s) ≥ h(s),
    \end{equation*}
    then
    \begin{equation*}
        \Exp{T} ≤ \frac{s_{\min}}{h(s_{\min})} + \Exp{∫_{s_{\min}}^{X_0}
        \frac{1}{h(σ)} dσ},
    \end{equation*}
    where the expectation in the latter term is over the random choice
    of $X_0$.
\end{theorem}

For the full proof, see~\cite[Theorem 2.3.3]{Lengler2020} or the original
\cite[Theorem 4.6]{Johannsen2010}. The intuition (using the notation
of~\cite{Lengler2020})
is that we can apply Theorem~\ref{theorem-drift-additive} to a
function
\begin{align*}
    g(s) &= 
    \begin{cases}
        \frac{s_{\min}}{h(s_{\min})} 
        + ∫_{s_{\min}}^s \frac{1}{h(σ)} dσ & \text{when $s ≥
        s_{\min}$}, \\
        \frac{s}{s_{\min}} & \text{when $s ≤ s_{\min}$}.
    \end{cases}
\end{align*}

The slope of this function for large $s$ is $1/h(s)$, which means that
the expected change in $g(s)$ when we drop by $h(s)$ is roughly
$h(s)⋅(1/h(s)) = 1$. The exact bound requires observing that the
integral is concave and applying Jensen's inequality. For $s$ close to
$s_{\min}$, the linear term covers any drop that goes below $s_{\min}$
(where we don't care about $h(s)$, because we are never going to
land there).

The actual proof does a case analysis on all possible changes in $s$
depending on whether they involve going up, going down but staying
about $s_{\min}$, or going down and crossing to $0$. Adding up all of
these cases shows an expected drift for $g(s)$ of at least $1$,
reducing to Theorem~\ref{theorem-drift-additive}. The somewhat messy 
bound is just what we get when we run the bound from the theorem back through $g$.

A special case of variable drift is 
\index{drift!multiplicative}\concept{multiplicative drift}, where
$∆_t(s) ≥ δs$ for some constant $δ>0$. In this case the bound in
Theorem~\ref{theorem-drift-variable} simplifies to
\begin{align}
    \Exp{T}
    &≤ \frac{1+\Exp{\ln(X_0/s_{\min})}}{δ}.
    \label{eq-drift-multiplicative}
\end{align}

For example, we showed in §\ref{section-Lyapunov-functions} that the
function $RPS$ for the rock-paper-scissors process has multiplicative
drift with $δ = 3/\binom{n}{2}$, $s_{\min} = n-2$, and $X_0 ≤
(n/3)^3$. Plugging these values in \eqref{eq-drift-multiplicative}
gives a bound on the time to reach $RPS = 0$ of 
$\frac{1+\ln((n/3)^3/(n-2))}{\binom{n}{2}} ≈ \frac{3}{2} n^2 \ln n$.
So we can get the same bound as before without having to go through
all the intermediate steps.

There are many drift theorem results in the literature, including lower
bounds and tail inequalities. The Lengler survey~\cite{Lengler2020} summarizes many of
these results. Another good source for Lyapunov-style bounds in
general is the textbook of Menshikov~\etal~\cite{MenshikovPW2016}.

\myChapter{The probabilistic method}{2025}{}
\label{chapter-probabilistic-method}

The \concept{probabilistic method} is a
tool for proving the existence of objects with particular
combinatorial properties, by showing that some process generates these
objects with nonzero probability.

The relevance of this to randomized
algorithms is that in some cases we can make the probability large
enough that we can actual produce such objects.

We'll mostly be following Chapter 5 of~\cite{MotwaniR1995} with some
updates for more recent results.  If you'd like to read more about
these technique, a classic reference on the
probabilistic method in combinatorics is the text of Alon and
Spencer~\cite{AlonS1992}.

\section{Randomized constructions and existence proofs}
\label{section-randomized-constructions}

Suppose we want to show that some object exists, but we don't know how
to construct it explicitly.  One way to do this is to devise some
random process for generating objects, and show that the probability
that it generates the object we want is greater than zero.  This
implies that something we want exists, because otherwise it would be
impossible to generate; and it works even if the nonzero probability
is very, very small.  The systematic development of the method is
generally attributed to the notoriously productive mathematician Paul
Erdős and his frequent collaborator Alfréd Rényi.

From an algorithmic perspective, the probabilistic method is useful
mainly when we can make the nonzero probability substantially larger
than zero—and especially if we can recognize when we've won.  But
sometimes just demonstrating the existence of an object is a start.

We give a couple of example of the probabilistic method in action
below.  In each case, the probability that we get a good outcome is
actually pretty high, so we could in principle generate a good outcome
by retrying our random process until it works.  There are some more
complicated examples of the method for which this doesn't work, either
because the probability of success is vanishingly small, or because we
can't efficiently test whether what we did succeeded (the last example
below may fall into this category).  This means that we often end up
with objects whose existence we can demonstrate even though we can't
actually point to any examples of them.  For example, it is known that
there exist \index{network!sorting}\indexConcept{sorting network}{sorting networks} (a
special class of circuits for sorting numbers in parallel) that sort
in time $O(\log n)$, where $n$ is the number of values being
sorted~\cite{AjtaiKS1983}, and these these can be generated randomly with
nonzero probability.  But the best explicit constructions of
such networks take time $\Theta(\log^2 n)$, and the question of how to
find an \emph{explicit} network that achieves $O(\log n)$ time has
been open for decades despite many efforts to solve it. 

\subsection{Set balancing}
\label{section-set-balancing-randomized}

Here we have a collection of vectors $v_1, v_2, \dots, v_n$ in
$\Set{0,1}^m$.  We'd like to find $\pm 1$ coefficients $ε_1,
ε_2, \dots ε_n$ that minimize the \concept{discrepancy}
$\max_j \abs*{X_j}$ where $X_j = ∑_{i=1}^{n} ε_i v_{ij}$.

This is called \concept{set balancing}\index{balancing!set} because we
are trying to balance attributes (represented by the vectors) between
two sets (represented by the coefficients). We can also take the
transpose and imagine that we have a collection of sets (represented
by the columns of the matrix made up of the $v_i$) that we are trying
to divide in half as evenly as possible by assigning their common
elements to the $-1$ and $+1$ piles.

If we choose the $ε_i$ randomly, Hoeffding's inequality
\eqref{eq-Hoeffdings-inequality} says for each fixed $j$ that
$\Prob{\abs*{X_j} > t}
< 2\exp(-t^2/2n)$ (since there are at most $n$ non-zero values
$v_{ij}$).  Setting $2\exp(-t^2/2n) ≤ 1/m$ gives $t ≥ √{2n \ln
2m}$.
So by the union bound, we have $\Prob{\max_j \abs*{X_j} > √{2n \ln
2m}} < 1$: a solution exists.

Since we only proved that the probability of winning in nonzero, this
doesn't necessary give us a good solution. In
§\ref{section-set-balancing-deranomdized} we will revisit this problem
and show how to remove the randomness while still getting the same
bound.

\subsection{Ramsey numbers}
\label{section-ramsey-numbers}

Consider a collection of $n$ schoolchildren, and imagine that each
pair of schoolchildren either like each other or dislike each other.  We
assume that these preferences are symmetric: if $x$ likes $y$, then
$y$ likes $x$, and similarly if $x$ dislikes $y$, $y$ dislikes $x$.  Let
$R(k,h)$ be the smallest value for $n$ that ensures that among any
group of $n$ schoolchildren, there is either a subset of $k$ children
that all like each other or a subset of $h$ children that all dislike
each other.\footnote{In terms of graphs, any graph $G$ with at least
$R(k,h)$ nodes contains either a \concept{clique} of size $k$ or an
\concept{independent set} of size $h$.}

It is not hard to show that $R(k,h)$ is finite for all $k$ and
$h$.\footnote{A simple proof, due to Erdős and
    Szekeres~\cite{ErdosS1935}, proceeds by showing
    that $R(k,h) ≤ R(k-1,h) + R(k,h-1)$.  Given a graph $G$ of size 
    at least $R(k-1,h) + R(k,h-1)$, choose a vertex $v$ and partition the
    graph into two induced subgraphs $G_1$ and $G_2$, where $G_1$
    contains all the vertices adjacent to $v$ and $G_2$ contains all
    the vertices not adjacent to $v$.  
    Either $\card*{G_1} ≥ R(k-1,h)$ or $\card*{G_2} ≥ R(k,h-1)$.
    If $\card*{G_1} ≥ R(k-1,h)$, then $G_1$
    contains either a clique of size $k-1$ (which makes a clique of size $k$
    in $G$ when we add $v$ to it) or an independent set of size $h$
    (which is also in $G$).  Alternatively, if $\card*{G_2} ≥
    R(k,h-1)$, then $G_2$ either gives us a clique of size $k$ by
itself or an independent set of size $h$ after adding $v$.  Together
with the fact that $R(1,h)=R(k,1)=1$, this recurrence gives $R(k,h)
≤ \binom{(k-1)+(h-1)}{k-1}$.}
The exact value of $R(k,h)$ is
known only for small values of $k$ and $h$.\footnote{There is a fairly
current table at \wikipedia{Ramsey's_Theorem}.  Some noteworthy values
are $R(3,3)=6$, $R(4,4)=18$, and $43 ≤ R(5,5) ≤ 46$.  One problem
with computing exact values
is that as $R(k,h)$ grows, the number of graphs one needs to consider
gets very big.  There are $2^{n \choose 2}$ graphs with $n$ vertices,
and even detecting the presence or absence of a moderately-large
clique or independent set in such a graph can be expensive.  This
pretty much rules out any sort of brute-force approach based on simply
enumerating candidate graphs. But even smarter approaches have yielded
only slow progress: for example, it took almost three decades for the
best known upper bound on 
$R(5,5)$ to drop
from $49$~\cite{McKayR1997} to 
$48$~\cite{AngeltveitM2018} to
the current value of
$46$~\cite{AngeltveitM2025}.}

But we can use the
probabilistic method to show that for $k=h$, it is reasonably large.
The following theorem is due to Erdős, and was the first known
lower bound on $R(k,k)$.
\begin{theorem}[\cite{Erdos1947}]
\label{theorem-ramsey-numbers}
If $k ≥ 3$, then $R(k,k) > 2^{k/2}$.
\end{theorem}
\begin{proof}
Suppose each pair of schoolchildren flip a fair coin to decide whether
they like each other or not.  Then the probability that any particular
set of $k$ schoolchildren all like each other is $2^{-\binom{k}{2}}$
and the probability that they all dislike each other is the same.
Summing over both possibilities and all subsets gives a bound of
$\binom{n}{k}2^{1-\binom{k}{2}}$ on the probability that there is at
least one subset in which everybody likes everybody or everybody dislikes
everybody.  For $n = 2^{k/2}$, we have 
\begin{align*}
\binom{n}{k} 2^{1-\binom{k}{2}}
&≤ \frac{n^k}{k!} 2^{1-\binom{k}{2}} \\
&= \frac{2^{k^2/2 + 1 - k(k-1)/2}}{k!} \\
&= \frac{2^{k^2/2 + 1 - k^2/2 + k/2}}{k!} \\
&= \frac{2^{1 + k/2}}{k!} \\
&< 1.
\end{align*}

Because the probability that there is an all-liking or all-hating
subset is less than $1$, there must be some chance that we get a
collection that doesn't have one.  So such a collection exists.  It
follows that $R(k,k) > 2^{k/2}$, because we have shown that not all
collections at $n = 2^{k/2}$ have the Ramsey property.
\end{proof}

The last step in the proof uses the fact that $2^{1+k/2} < k!$ for
$k≥3$, which can be tested explicitly for $k=3$ and proved by
induction for larger $k$.  The resulting bound is a little bit weaker
than just saying that $n$ must be large enough that 
$\binom{n}{k}2^{1-\binom{k}{2}} ≥ 1$, but it's easier to use.

The proof can be generalized to the case where $k≠ h$ by tweaking the
bounds and probabilities appropriately.
Note that even though this process generates a graph with no large
cliques or independent sets with reasonably high probability, we don't have
any good way of testing the result, since testing for the existence of
a clique is \classNP-hard.

\section{Approximation algorithms}
\label{section-randomized-approximation}

One use of a randomized construction is to approximate the solution to
an otherwise difficult problem.  In this section, we start with a
trivial approximation algorithm for the largest cut in a
graph, and then show a more powerful randomized approximation
algorithm, due to Goemans and Williamson~\cite{GoemansW1994}, that
gets a better approximation ratio for a much larger class of problems.

\subsection{MAX CUT}
\label{section-max-cut}

We've previously seen (§\ref{section-karger-min-cut}) a randomized algorithm
for finding small cuts in a graph.  What if we want to find a large
cut?

Here is a particularly brainless algorithm that finds a large cut.  For each
vertex, flip a coin: if the coin comes up heads, put the vertex in
$S$, otherwise, put in it $T$.  For each edge, there is a probability
of exactly $1/2$ that it is included in the $S$–$T$ cut.  It follows
that the expected size of the cut is exactly $m/2$.

One consequence of this is that every graph has a cut that
includes at least half the edges.  Another is that this algorithm
finds such a cut, with probability at least
$\frac{1}{m+1}$.  To prove
this, let $X$ be the random variable representing the number of edges
in the cut, and let $p$ be the probability that $X ≥ m/2$.
Then
\begin{align*}
    m/2 &= \Exp{X}
      \\&= (1-p) \ExpCond{X}{X < m/2}
 + p \ExpCond{X}{X ≥ m/2}
 \\&≤ (1-p) \frac{m-1}{2} + p m.
 \end{align*}
 Solving this for $p$ gives
 the claimed bound.\footnote{This is tight for $m=1$, but I suspect it's
an underestimate for larger $m$.  The main source of slop in the
analysis seems to be the step $\ExpCond{X}{X ≥ m/2} ≤ m$; using a
concentration bound, we should be able to show a much stronger
inequality here and thus a much larger lower bound on $p$.}

By running this enough times to get a good cut, we get a
polynomial-time randomized 
algorithm for approximating the \concept{maximum cut} within a factor
of $2$, which is pretty good considering that MAX CUT is
\classNP-hard.

There exist better approximation algorithms.  Goemans and
Williamson~\cite{GoemansW1995} give a 0.87856-approximation algorithm
for MAX CUT based on randomized rounding of a semidefinite
program.\footnote{Semidefinite programs are like linear programs
except that the variables are vectors instead of scalars, and the
objective function and constraints apply to linear combinations of
dot-products of these variables. The Goemans-Williamson MAX CUT
algorithm is based on a relaxation of the integer optimization problem of maximizing $∑ x_i x_j$
where each $x_i ∈ \Set{-1,+1}$ encodes membership in $S$ or $T$.
They instead allow $x_i$ to be any unit vector in an $n$-dimensional
space, and then take the sign of the dot-product with a random unit
vector $r$ to map each optimized $x_i$ to one side of the cut or the
other.}
The analysis of this algorithm is a little involved, so we won't attempt to show this here, but we will describe (in
§\ref{section-max-sat})
an earlier
result, also due to Goemans and Williamson~\cite{GoemansW1994}, that gives a
$\frac{3}{4}$-approximation to MAX SAT using a similar technique.

\subsection{MAX SAT}
\label{section-max-sat}

Like MAX CUT, MAX SAT is an \classNP-hard optimization problem that
sounds like a very short story about Max.  We are given a
\concept{satisfiability problem} in 
\index{normal form!conjunctive}\concept{conjunctive normal
form}: as a conjunction (AND) of $m$
\indexConcept{clause}{clauses},
each of which is the OR of a bunch of variables or
their negations. We want to choose values for the $n$ variables that
\concept{satisfy} as many of the clauses as possible, where a clause
is satisfied if it contains a true variable or the negation of a false
variable.\footnote{The presentation here follows 
    \cite[§5.2]{MotwaniR1995}, which in turn is mostly based on a classic paper of
Goemans and Williamson~\cite{GoemansW1994}.}

We can instantly satisfy at least $m/2$ clauses on average by
assigning values to the variables independently and uniformly at
random; the analysis is the same as in
§\ref{section-max-cut} for large cuts, since a random
assignment makes the first variable in a clause true with
probability $1/2$.  Since this approach doesn't require thinking and
doesn't use the fact that many of our clauses may have more than one
variable, we can probably do better.  Except we can't do better in the
worst case, because it might be that our clauses consist entirely of
$x$ and $\neg x$ for some variable $x$; clearly, we can only satisfy
half of these.  We could do better if we knew all of our clauses
consisted of at least $k$ distinct literals (satisfied with
probability $1-2^{-k}$), but we can't count on this.  We also can't
count on clauses not being duplicated, so it may turn out that
skipping a few hard-to-satisfy small clauses hurts us if they occur
many times.

Our goal will be to get a good \concept{approximation ratio},
defined as the ratio between the number of clauses we manage to
satisfy and the actual maximum that can be satisfied.  The tricky part
in designing
\indexConcept{approximation algorithm}{approximation algorithms} is
showing that the denominator here won't be too big.  We can do this
using a standard trick, of expressing our original problem as an
\concept{integer program} and then \indexConcept{relaxation}{relaxing}
it to a \concept{linear program}\footnote{A \concept{linear program}
is an optimization problem where we want to maximize (or minimize)
some linear \concept{objective function} of the variables subject to linear-inequality
constraints.  A simple example would be to maximize $x+y$ subject to
$2x+y ≤ 1$ and $x + 3y ≤ 1$; here the \concept{optimal solution}
is the assignment $x=\frac{2}{5}, y=\frac{1}{5}$, which sets the
objective function to its maximum possible value $\frac{3}{5}$.  An
\concept{integer program} is a linear program where some of the
variables are restricted to be integers.  
Determining if an integer program even has a solution is
\classNP-complete; in contrast, linear programs can be solved in
polynomial time.  We can \concept{relax} an integer program to a
linear program by dropping the requirements that variables be
integers; this will let us find a \concept{fractional solution} that
is at least as good as the best integer solution, but might be
undesirable because it tells us to do something that is ludicrous in
the context of our original problem, like only putting half a passenger on the
next plane.

Linear programming has an interesting history.  The basic ideas were
developed independently by Leonid Kantorovich in the Soviet Union and
George Dantzig in the United States around the start of the Second
World War.  Kantorovich's work had direct relevance to Soviet planning
problems, but wasn't pursued seriously because it threatened the
political status of the planners, required computational resources
that weren't available at the time, and looked suspiciously like
trying to sneak a capitalist-style price system into the planning
process; for a fictionalized account of this tragedy, see~\cite{Spufford2012}.
Dantzig's work, which included the development of the
\concept{simplex method} for solving linear programs,
had a higher impact, although its publication
was delayed until 1947 by wartime secrecy.}
whose solution doesn't have to
consist of integers.  We then convert the fractional solution back to
an integer solution by rounding some of the variables to integer
values randomly in a way that preserves their expectations, a
technique known as \concept{randomized rounding}.\footnote{
Randomized rounding was invented by Raghavan and
Thompson~\cite{RaghavanT1987}; the particular application here is due
to 
Goemans and
Williamson~\cite{GoemansW1994}.}

Here is the integer program (taken from \cite[§5.2]{MotwaniR1995}).
We let $z_j ∈ \Set{0,1}$ represent whether clause $C_j$ is satisfied,
and let $y_i ∈ \Set{0,1}$ be the value of variable $x_i$.
We also let $C_j^+$ and $C_j^-$ be the set of variables that appear in
$C_j$ with and without negation.
The problem is to maximize
\begin{displaymath}
∑_{j=1}^{m} z_j
\end{displaymath}
subject to
\begin{displaymath}
∑_{i ∈ C_j^+} y_i + ∑_{i ∈ C_j^-} (1-y_i) ≥ z_j
\end{displaymath}
for all $j$.

The main trick here is to encode OR in the constraints; there is no
requirement that $z_j$ is the OR of the $y_i$ and $(1-y_i)$ values,
but we maximize the objective function by setting it that way.

Sadly, solving integer programs like the above is \classNP-hard (which
is not surprising, since if we could solve this particular one, we
could solve SAT).  But if we drop the requirements that $y_i, z_j ∈
\Set{0,1}$ and replace them with $0 ≤ y_i ≤ 1$ and $0 ≤ z_j ≤ 1$,
we get a linear program—solvable in polynomial time—with an
optimal value at least as good as the value for the integer program,
for the simple reason that any solution to the integer program is also
a solution to the linear program.

The problem now is that the solution to the linear program is likely
to be \indexConcept{fractional solution}{fractional}: instead of
getting useful $0$–$1$ values, we might find out we are supposed to make
$x_i$ only $2/3$ true.  So we need one more trick to turn the
fractional values back into integers.  This is 
the randomized rounding step:
given a fractional assignment $\hat{y}_i$, we set $x_i$ to
true with probability $\hat{y}_i$.

So what does randomized rounding do to clauses?  In our fractional
solution, a clause might have value $\hat{z}_j$, obtained by summing
up bits and pieces of partially-true variables.  We'd like to argue
that the rounded version gives a similar probability that $C_j$ is
satisfied.

Suppose $C_j$ has $k$ variables; to make things simpler, we'll pretend
that $C_j$ is exactly $x_1 ∨ x_2 ∨ \dots x_k$.  Then the
probability that $C_j$ is satisfied is exactly $1-\prod_{i=1}^{k}
(1-\hat{y}_i)$.  This quantity is minimized subject to $∑_{i=1}^{k}
\hat{y}_i ≥ \hat{z}_j$ by setting all $\hat{y}_i$ equal to
$\hat{z}_j/k$ (easy application of Lagrange multipliers, or can be
shown using a convexity argument).  Writing $z$ for $\hat{z}_j$, this
gives
\begin{align*}
\Prob{\text{$C_j$ is satisfied}}
&= 1 - \prod_{i=1}^{k} (1-\hat{y}_i)
\\
&≥ 1 - \prod_{i=1}^{k} (1-z / k)
\\
&= 1 - (1-z/k)^k
\\
&≥ z (1 - (1-1/k)^k).
\\
&≥ z (1-1/e).
\end{align*}
The second-to-last step looks like a typo, but it actually works.  The idea is
to observe that the function $f(z) = 1-(1-z/k)^k$ is concave (Proof:
$\frac{d^2}{dz^2} f(z) = -\frac{k-1}{k}(1-z/k)^{k-2} < 0$), while
$g(z) = z(1-(1-1/k)^k)$ is linear, so since $f(0) = 0 = g(0)$ and
$f(1) = 1-(1-1/k)^k = g(1)$, any point in between must have $f(z) ≥
g(z)$.  An example of this argument for $k=3$ is depicted in
Figure~\ref{fig-max-sat-magic-trick}.

\begin{figure}
    \centering
    \begin{tikzpicture}[scale=8]
    \draw (1,0) -- (0,0) -- (0,{1-(1-1.0/3)^3)});  \draw[green] (0,0) -- (1, {1-(1-1.0/3)^3)});
    \draw[domain=0:1, blue] plot (\x, {1-(1-\x/3)^3});
\end{tikzpicture}
    \caption[Tricky step in MAX SAT argument]{Tricky step in MAX
    SAT argument, showing $(1-(1-z/k)^k) ≥ z(1-(1-1/k)^k)$ when $0 ≤ z
    ≤ 1$.  The case $k=3$ is depicted.}
    \label{fig-max-sat-magic-trick}
\end{figure}

Since each clause is satisfied with probability at least $\hat{z}_j
(1-1/e)$, the expected number of satisfied clauses is at least
$(1-1/e) ∑_j \hat{z}_j$, which is at least $(1-1/e)$ times the
optimum.  This gives an approximation ratio of slightly more than
0.632, which is better than $1/2$, but still kind of weak.

So now we apply the second trick from~\cite{GoemansW1994}: we'll
observe that, on a per-clause basis, we have a randomized rounding
algorithm that is good at satisfying small clauses (the coefficient
$(1-(1-1/k)^k)$ goes all the way up to $1$ when $k=1$), and our
earlier dumb algorithm that is good at satisfying big clauses.  We
can't combine these directly (the two algorithms demand different
assignments), but we can run both in parallel and take whichever gives
a better result.

To show that this works, let $X_j$ be indicator for the event that
clause $j$ is satisfied by the randomized-rounding algorithm and $Y_j$
the indicator for the even that it is satisfied by the simple
algorithm.  Then if $C_j$ has $k$ literals,
\begin{align*}
    \Exp{X_j} + \Exp{Y_j}
    &≥ (1-2^{-k}) + (1-(1-1/k)^k) \hat{z}_j
    \\
    &≥ \parens*{(1-2^{-k}) + (1-(1-1/k)^k)}) \hat{z}_j
    \\
    &= \parens*{2-2^{-k}-(1-1/k)^k} \hat{z}_j.
\end{align*}
The coefficient here is exactly $3/2$ when $k=1$ or $k=2$, and rises
thereafter, so for integer $k$ we have $\Exp{X_j} + \Exp{Y_j} ≥
(3/2)\hat{z}_j$.  Summing over all $j$ then gives $\Exp{∑_j
X_j} + \Exp{∑_j Y_j} ≥ (3/2) ∑_j \hat{z}_j$.
But then one of the two expected sums must beat $(3/4) ∑_j
\hat{z}_j$, giving us a $(3/4)$-approximation algorithm.

\section{The Lovász Local Lemma}
\label{section-lovasz-local-lemma}

Suppose we have a finite set of bad events $\mathcal{A}$, and we
want to show that with nonzero probability, none of these events
occur.  Formally, we want to show
$\Prob{\bigcap_{A∈\mathcal{A}} \bar{A}} > 0$.

Our usual trick so far has been to use the union bound
\eqref{eq-union-bound} to show that $∑_{A ∈ \mathcal{A}} \Prob{A} < 1$.
But this only works if the events are actually improbable.  If the
union bound doesn't work, we might be lucky enough to have the events
be independent; in this case, 
$\Prob{\bigcap_{A∈\mathcal{A}} \bar{A}} 
= \prod_{A∈\mathcal{A}} \Prob{\bar{A}} > 0$, as long as each event
$\bar{A}$ occurs with positive probability.  But most of the
time, we'll find that the events we care about aren't independent, so
this won't work either.

The 
\index{lemma!Lovász local}
\concept{Lovász Local Lemma}~\cite{ErdosL1975} handles a situation intermediate
between these two extremes, where events are generally not independent
of each other, but each collection of events that are not independent
of some particular event $A$ has low total probability.  In the original
version, it's non-constructive: the lemma shows a nonzero probability
that none of the events occur, but this probability may be very small
if we try to sample the events at random, and there is no guidance for
how to find a particular outcome that makes all the events false.

Subsequent
work~\cite{Beck1991,Alon1991,MolloyR1998,CzumajS2000,Srinivasan2008,Moser2009,MoserT2010}
showed how, when the events $A$ are determined by some underlying set
of independent variables and independence between two events is 
detected by having non-overlapping sets of underlying variables, an
actual solution could be found in polynomial expected time.  The final
result in this series, due to Moser and Tardos~\cite{MoserT2010}, gives
the same bounds as in the original non-constructive lemma, using the
simplest algorithm imaginable: whenever some bad event $A$ occurs, try
to get rid of
it by resampling all of its variables, and continue until no bad
events are left. We'll describe this algorithm in
§\ref{section-Moser-Tardos}.

\subsection{General version}
\label{section-Lovasz-local-lemma-general}

A formal statement of the general lemma is:\footnote{This version is
adapted from~\cite{MoserT2010}.}
\begin{lemma}
\label{lemma-lovasz-local-lemma}
Let $\mathcal{A} = A_1,\dots,A_m$ be a finite collection of events on some
probability space, and 
for each $A∈\mathcal{A}$, let $Γ(A)$ be a set of events such
that $A$ is independent of all events not in $Γ^+(A) =
\Set{A} \cup Γ(A)$.
If there exist real numbers $0 < x_A < 1$ such that,
for all events $A ∈ \mathcal{A}$
\begin{align}
\label{eq-lll-condition}
\Prob{A} &≤ x_A \prod_{B ∈ Γ(A)} (1-x_B),
\intertext{then}
\label{eq-lll-result}
\Prob{\bigcap_{A∈\mathcal{A}} \bar{A}}
& ≥ \prod_{A∈ \mathcal{A}} (1-x_A).
\end{align}
\end{lemma}
In particular, this means that the probability that none of the $A_i$
occur is not zero, since we have assumed $x_{A_i} < 1$ holds for all $i$.

The role of $x_A$ in the original proof is to act as an upper bound on the
probability that $A$ occurs given that some collection of other
events doesn't occur.   For the constructive proof, the $x_A$ are
used to show a bound on the number of resampling steps needed until
none of the $A$ occur.

\subsection{Symmetric version}
\label{section-Lovasz-local-lemma-symmetric}

For many applications, having to come up with the $x_A$ values can be
awkward.  The following symmetric version is often used instead:
\begin{corollary}
\label{corollary-symmetric-lovasz-local-lemma}
Let $\mathcal{A}$ and $Γ$ be as in
Lemma~\ref{lemma-lovasz-local-lemma}.
Suppose that there are constants $p$ and $d$, such that 
for all $A∈ \mathcal{A}$, we have
$\Prob{A} ≤ p$ and $\card*{Γ(A)} ≤ d$.
Then if $ep(d+1) < 1$, 
$\Prob{\bigcap_{A∈ \mathcal{A}} \bar{A}} ≠ 0$.
\end{corollary}
\begin{proof}
Basically, we are going to pick a single value $x$ such that $x_A =
x$ for all $A$ in $\mathcal{A}$, and \eqref{eq-lll-condition} is
satisfied.  This works as long as
$p ≤ x (1-x)^d$, as in this case we have, for all $A$,
$\Prob{A} ≤ p ≤ x (1-x)^d ≤ x (1-x)^{\card*{Γ(A)}}
= x_A \left(\prod_{B ∈ Γ(A)} (1-x_B)\right)$.

For fixed $d$, $x (1-x)^d$ is maximized using the usual trick:
$\frac{d}{dx} x (1-x)^d = (1-x)^d - x d (1-x)^{d-1} = 0$ gives
$(1-x) - x d = 0$ or $x = \frac{1}{d+1}$.
So now we need $p ≤ \frac{1}{d+1} \left(1-\frac{1}{d+1}\right)^d$.
It is possible to show that $1/e < \left(1-\frac{1}{d+1}\right)^d$ for all
    $d≥ 0$.\footnote{Observe that $\lim_{d → ∞}
    \parens*{1-\frac{1}{d+1}}^d = e^{-1}$ and that 
    $f(d) = \parens*{1-\frac{1}{d+1}}^d = e^{-1}$ is a decreasing function.
    The last part can be shown by taking the derivative of $\log
    f(d)$. See the solution to
Problem~\ref{problem-all-must-have-candy} for the gory details.
    }
      So $ep(d+1) ≤ 1$
implies $p ≤ \frac{1}{e(d+1)} ≤ \left(1-\frac{1}{d+1}\right)^d
\frac{1}{d+1} ≤ x (1-x)^{\card*{Γ(A)}}$ as required by
Lemma~\ref{lemma-lovasz-local-lemma}.
\end{proof}

\subsection{Applications}
\label{section-Lovasz-local-lemma-applications}

Here we give some simple applications of the lemma.

\subsubsection{Graph coloring}

Let's start with a problem where we know
what the right answer should be.  Suppose we want to color the
vertices of a cycle with $c$ colors, so that no edge has two endpoints
with the same color.  How many colors do we need?

Using brains, we can quickly figure out that $c=3$ is enough.  Without
brains, we could try coloring the vertices randomly: but in a cycle
with $n$ vertices and $n$ edges, on average $n/c$ of the edges will be
monochromatic, since each edge is monochromatic with probability
$1/c$.  If these bad events were independent, we could argue that
there was a $(1-1/c)^n > 0$ probability that none of them occurred,
but they aren't, so we can't.  Instead, we'll use the local lemma.

The set of bad events $\mathcal{A}$ is just the set of events $A_i =
[\text{edge $i$ is monochromatic}]$.
We've already computed $p=1/c$.  To get $d$, notice that each edge
only shares a vertex with two other edges, so $\card*{Γ(A_i)} ≤ 2$.
Corollary~\ref{corollary-symmetric-lovasz-local-lemma} then says that
there is a good coloring as long as $ep(d+1) = 3e/c ≤ 1$, which
holds as long as $c ≥ 9$.  We've just shown we can $9$-color a
cycle.  If use the asymmetric version,
we can set all $x_A$ to $1/3$ and show
that $p ≤ \frac{1}{3} \left(1-\frac{1}{3}\right)^2 = \frac{4}{27}$
would also work; with this we can $7$-color a cycle. This is still not as
good as what we can do if we are paying attention, but not bad for a
procedure that doesn't use the structure of the problem much.

\subsubsection{Satisfiability of \texorpdfstring{$k$}{k}-CNF formulas}

A more sophisticated application is demonstrating satisfiability for
$k$-CNF formulas where each variable appears in a bounded number of
clauses.  Recall that a \concept{$k$-CNF formula} consists of $m$
\indexConcept{clause}{clauses}, each of which consists of exactly $k$
variables or their negations (collectively called
\indexConcept{literal}{literals}.  It is
\indexConcept{satisfiability}{satisfied} if at least one literal in
every clause is assigned the value true.

Suppose that each variable appears in at most $\ell$ clauses.  Let
$\mathcal{A}$ consist of all the events $A_i = [\text{clause $i$ is
not satisfied}]$.  Then, for all $i$, $\Prob{A_i} = 2^{-k}$ exactly, and
$\card*{Γ(A_i)} ≤ d = k(\ell-1)$ since each variable in $A_i$ is
shared with at most $\ell-1$ other clauses, and $A_i$ will be
independent of all events $A_j$ with which it doesn't share a
variable.  So if $ep(d+1) = e 2^{-k} (k (\ell - 1) + 1) < 1$,
which holds if $\ell < \frac{2^k+e(k-1)}{ek}$,
Corollary~\ref{corollary-symmetric-lovasz-local-lemma} tells us that a
satisfying assignment exists.\footnote{To avoid over-selling this
claim, it's worth noting that the bound on $\ell$ only reaches $2$ at
$k=4$, although it starts growing pretty fast after that.}

Corollary~\ref{corollary-symmetric-lovasz-local-lemma} doesn't let us
actually find a satisfying assignment, but it turns out we can do that
too. We'll return to this when we talk about the constructive version
of the local lemma in §\ref{section-constructive-lll}. 

\subsubsection{Hypergraph $2$-colorability}

This was the original motivation for the lemma. A 
\index{hypergraph}
\index{hypergraph!$k$-uniform}
\concept{$k$-uniform hypergraph}
$G = \Tuple{V,E}$ has a set of \indexConcept{hyperedge}{hyperedges}
$E$, each of which contains exactly $k$ vertices from $V$.
A \index{coloring!hypergrah}\concept{$2$-coloring} of the hypergraph
assigns one of two colors to each vertex in $v$, so that no edge is
monochromatic.

Suppose we color vertices at random. Pick some edge $S$, and let $A_S$ be
the probability that it is monochromatic. Then $\Prob{A_S} = 2^{1-k}$
exactly. We also have that $A_S$ depends only on events $A_T$ where
$S∩T≠∅$. Suppose that there is a bound $d$ on the number of hyperedges
$T$ that overlap with any hyperedge $S$. Then
Corollary~\ref{corollary-symmetric-lovasz-local-lemma} says that there exists a good coloring as long as $e 2^{1-k} (d+1) < 1$.
This holds for any hypergraph with $d < 2^{k-1} / e - 1$.

Unlike graph coloring, it's not so obvious how to solve hypergraph
coloring with a simple greedy algorithm. So here the local lemma seems
to buy us something we didn't already have.

\subsection{Non-constructive proof}
\label{section-non-constructive-lll}

This is essentially the same argument presented in
\cite[§5.5]{MotwaniR1995}, but we adapt the notation to match the
statement in terms of neighborhoods $Γ(A)$ instead of edges in a
\concept{dependency graph}, where an event is independent of all
events that are not its successors in the graph.  The two formulations are 
identical, since we can always represent the neighborhoods $Γ(A)$
by creating an edge from $A$ to $B$ when $B$ is in $Γ(A)$; and
conversely we can convert a dependency graph into a neighborhood
structure by making $Γ(A)$ the set of successors of $A$ in the
dependency graph.

The proof is a bit easier than the constructive
version, while also being a bit more general because the neighborhood
structure doesn't depend on identifying a collection of underlying
independent variables.

The idea is to order the events in $\mathcal{A}$ as $A_1, A_2, \dots,
A_n$ and expand $\Prob{\bigcap_{i=1}^{n} \bar{A}_i}$ as
$∏_{i=1}^{n} \ProbCond{\bar{A}_i}{\bigcap_{j=1}^{i-1} \bar{A}_j}$.
If we can show that every factor in the product is nonzero, we are done.

To do this, we will show by induction on $\card*{S}$ the more general
statement that for any $A$ and \emph{any} $S ⊆
\mathcal{A}$ with $A ∉ S$,
\begin{align}
\label{eq-lll-outer-induction}
\ProbCond{A}{\bigcap_{B ∈ S} \bar{B}}
&≤ x_A.
\end{align}

When $\card*{S} = 0$, this just says $\Prob{A} ≤ x_A$, which follows
immediately from \eqref{eq-lll-condition}.

For larger $S$, split $S$ into 
$S_1 = S \cap Γ(A)$, the events in $S$ that might not be
independent of $A$; 
and $S_2 = S ∖ Γ(A)$,
the events in $S$ that we know to be independent of $A$.
If $S_2 = S$, then $A$ is independent of all events in $S$, and
\eqref{eq-lll-outer-induction} follows immediately
from $\ProbCond{A}{\bigcap_{B ∈ S} \bar{B}} = \Prob{A} ≤ x_A
\prod_{B ∈ Γ(A)} (1-x_B) ≤ x_A$.  Otherwise $\card*{S_2} <
\card*{S}$, which means that we can assume the induction hypothesis
holds for $S_2$.

Write $C_1$ for the event $\bigcap_{B ∈ S_1} \bar{B}$ and
$C_2$ for the event $\bigcap_{B ∈ S_2} \bar{B}$. Then $\bigcap_{B ∈ S}
\bar{B} = C_1 ∩ C_2$ and we can expand
\begin{align}
    \ProbCond{A}{\bigcap_{B∈ S} \bar{B}}
&= \frac{\Prob{A \cap C_1 \cap C_2}}{\Prob{C_1 \cap C_2}}
\nonumber\\
&= \frac{\ProbCond{A \cap C_1 }{ C_2} \Prob{C_2}}{\ProbCond{C_1 }{ C_2} \Prob{C_2}}
\nonumber\\
\label{eq-lll-fraction}
&= \frac{\ProbCond{A \cap C_1 }{ C_2}}{\ProbCond{C_1 }{ C_2}}.
\end{align}

We don't need to do anything particularly clever with the numerator:
\begin{align}
\ProbCond{A \cap C_1 }{ C_2}
&≤ \ProbCond{A }{ C_2}
\nonumber\\
&= \Prob{A}
\nonumber\\
&≤ 
\label{eq-lll-numerator}
x_A \prod_{B ∈ Γ(A)} (1-x_B),
\end{align}
from \eqref{eq-lll-condition} and the fact that $A$ is independent of
all $B$ in $S_2$ and thus also independent of $C_2$.

For the denominator, we expand $C_1$ back out to $\bigcap_{B ∈ S_1}
\bar{B}$ and break out the induction hypothesis.  To bound
$\ProbCond{\bigcap_{B ∈ S_1} \bar{B}}{C_2}$, we order
$S_1$ arbitrarily as $\Set{B_1,\dots,B_r}$, for some $r$, and show by
induction on $\ell$ as $\ell$ goes from $1$ to $r$ that
\begin{align}
\label{eq-lll-inner-induction}
\ProbCond{\bigcap_{i=1}^{\ell} \bar{B}_i }{ C_2}
&≥ \prod_{i=1}^{\ell} (1-x_{B_i}).
\end{align}

The proof is that, for $\ell=1$,
\begin{align*}
    \ProbCond{\bar{B}_1 }{ C_2}
    &= 1-\ProbCond{B_1 }{ C_2}
\\
&≥ 1 - x_{B_1}
\end{align*}
using the outer induction hypothesis \eqref{eq-lll-outer-induction},
and for larger $\ell$, we can compute
\begin{align*}
    \ProbCond{\bigcap_{i=1}^{\ell} \bar{B}_i }{ C_2}
    &= \ProbCond{ \bar{B}_\ell}{
\left(\bigcap_{i=1}^{\ell-1}\bar{B}_i\right)
\cap C_2 }
⋅ \ProbCond{\bigcap_{i=1}^{\ell-1} \bar{B}_i }{ C_2}
\\
&≥ (1-x_{B_\ell}) \prod_{i=1}^{\ell-1} (1-x_{B_i})
\\
&= \prod_{i=1}^{\ell} (1-x_{B_i}),
\end{align*}
where the second-to-last step uses the outer induction hypothesis \eqref{eq-lll-outer-induction} for
the first term and the inner induction hypothesis
\eqref{eq-lll-inner-induction} for the rest.
This completes the proof of the inner induction.

When $\ell = r$, we get
\begin{align}
\nonumber
\ProbCond{C_1}{C_2}
&=
\ProbCond{\bigcap_{i=1}^{r} \bar{B}_i }{ C_2}
\\
\label{eq-lll-denominator}
&≥ \prod_{B ∈ S_1} (1-x_B).
\end{align}

Substituting \eqref{eq-lll-numerator} and \eqref{eq-lll-denominator}
into \eqref{eq-lll-fraction} gives
\begin{align*}
    \ProbCond{A }{\bigcap_{B∈ S} \bar{B}}
&≤ 
\frac{
    x_A \left(\prod_{B ∈ Γ(A)} (1-x_B)\right),
}{
    \prod_{B ∈ S_1} (1-x_B)
}
\\
&=
x_A \left(\prod_{B ∈ Γ(A) ∖ S_1} (1-x_B)\right)
\\
&≤
x_A.
\end{align*}
This completes the proof of the outer induction.

To get the bound \eqref{eq-lll-result}, we reach back inside the proof
and repeat the argument for \eqref{eq-lll-denominator} with
$\bigcap_{A∈\mathcal{A}} \bar{A}$ in place of $C_1$ and
without the conditioning on $C_2$.  We order $\mathcal{A}$ arbitrarily
as $\Set{ A_1, A_2, \dots, A_m }$ and show by induction on $k$ that
\begin{align}
\label{eq-lll-final-induction}
\Prob{\bigcap_{i=1}^{k} \bar{A}_i}
&≥ \prod_{i=1}^k (1-x_{A_k}).
\end{align}
For the base case we have $k=0$ and $\Prob{Ω} ≥ 1$, using the
usual conventions on empty products.  For larger $k$, we have
\begin{align*}
    \Prob{\bigcap_{i=1}^{k} \bar{A}_i}
&= 
\ProbCond{\bar{A}_k 
}{ \bigcap_{i=1}^{k-1} \bar{A}_i}
    \Prob{\bigcap_{i=1}^{k-1} \bar{A}_i}
\\
&≥ (1-x_{A_k}) \prod_{i=1}^{k-1} (1-x_{A_i})
\\
&≥ \prod_{i=1}^k (1-x_{A_k}),
\end{align*}
where in the second-to-last step we use \eqref{eq-lll-outer-induction}
for the first term and the induction hypothesis \eqref{eq-lll-final-induction} for the big
product.

Setting $k=m$ finishes the proof.

\subsection{Constructive proof}
\label{section-constructive-lll}
\label{section-Moser-Tardos}

We now describe he constructive proof of the Lovász local lemma due
to Moser and Tardos~\cite{MoserT2010}, which is based on a slightly
more specialized construction of Moser alone~\cite{Moser2009}.  This
version applies when our set of bad events $\mathcal{A}$ is defined in
terms of a set of \emph{independent}
variables $\mathcal{P}$, where each $A∈
\mathcal{A}$ is determined by some set of 
variables $\vbl(A) ⊆
\mathcal{P}$, and $Γ(A)$ is defined to be the set of all events
$B ≠ A$ that share at least one variable with $A$; i.e.,
$Γ(A) =
\SetWhere{B ∈ \mathcal{A} ∖ \Set{A} }{ \vbl(B) \cap \vbl(A) ≠ \emptyset}$.

In this case, we can attempt to find an assignment to the variables
that makes none of the $A$ occur using the obvious algorithm of
sampling an initial state randomly, then resampling all variables in
$\vbl(A)$ whenever we see some bad $A$ occur.  Astonishingly, this
actually works in a reasonable amount of time, without even any
cleverness in deciding which $A$ to resample at each step, if
the conditions for Lemma~\ref{lemma-lovasz-local-lemma} hold for $x$
that are not too large.
In particular, we will show that a good assignment is found after
each $A$ is resampled at most
$\frac{x_A}{1-x_A}$ times on average.\footnote{Even though there is a
lot of resampling going on here, Moser-Rardos is not actually a
sampling algorithm, in the sense that some solutions to the original
problem may be much more likely to be produced than others. There has
been some more recent work on the
\concept{sampling Lovász local lemma}\index{Lovász local
lemma!sampling}, where the
goal is to obtain a close-to-uniform sample from the space of possible
solutions. This is much harder than just finding one solution and
usually requires stronger constraints on the problem. Some recent
examples of this work can be found in~\cite{HeWY2022,WangY2024}.}

\begin{lemma}
\label{lemma-lovasz-local-lemma-constructive}
Under the conditions of Lemma~\ref{lemma-lovasz-local-lemma}, the
Moser-Tardos procedure does at most
\begin{align*}
∑_A \frac{x_A}{1-x_A}
\end{align*}
resampling steps on average.
\end{lemma}

We defer the proof of
Lemma~\ref{lemma-lovasz-local-lemma-constructive} for the moment.  For
most applications, the following symmetric version is easier to work
with:
\begin{corollary}
\label{corollary-symmetric-lovasz-local-lemma-constructive}
Under the conditions of
Corollary~\ref{corollary-symmetric-lovasz-local-lemma},
the Moser-Tardos procedure does at most
$m/d$ resampling steps on average.
\end{corollary}
\begin{proof}
Follows from Lemma~\ref{lemma-lovasz-local-lemma-constructive} and
the choice of $x_A = \frac{1}{d+1}$ in the proof of
Corollary~\ref{corollary-symmetric-lovasz-local-lemma}.
\end{proof}

How this expected $m/d$ bound translates into actual time depends on the cost
of each resampling step.  The expensive part at each step is likely to
be the cost of finding an $A$ that occurs and thus needs to be
resampled.

Intuitively, we might expect the resampling to work because if each
$A∈\mathcal{A}$ has a small enough neighborhood $Γ(A)$ and a
low enough probability, then whenever we resample $A$'s variables,
it's likely that we fix $A$ and unlikely that we break too many $B$
events in
$A$'s neighborhood.  It turns out to be tricky to quantify how this
process propagates outward,
so the actual proof uses a different idea that
essentially looks at this process in reverse, looking for each
resampled event $A$ at a set of previous events whose resampling we can
blame for making $A$ occur, and then showing that this tree (which
will include every resampling operation as one of its vertices) can't be
too big.

The first step is to fix some strategy for choosing which event $A$ to
resample at each step.  We don't care what this strategy is; we just
want it to be the case that the sequence of events depends only on the
random choices made by the algorithm in its initial sampling and each
resampling.  We can then define an \concept{execution log} $C$ that
lists the sequence of events $C_1, C_2, C_3, \dots$ that are resampled
at each step of the execution.

From $C$ we construct a \index{tree!witness}\concept{witness tree} $T_t$ for each
resampling step $t$ whose nodes are
labeled with events, with the property that the children of a node $v$
labeled with event $A_v$ are labeled with events in $Γ^+(A_v) =
\Set{A_v} ∪ Γ(A_v)$.

The root of $T_t$ is labeled with $C_t$.
To construct the rest of the tree, we work backwards through $C_{t-1},
C_{t-2}, \dots, C_1$, and for each event $C_i$ we encounter, we attach
$C_i$ as a child of the deepest $v$ we can find with $C_i ∈
Γ^+(A_v)$, choosing arbitrarily if there is more than one such
$v$, and discarding $C_i$ if there is no such $v$.

Now we can ask what the probability is that we see some particular
witness tree $T$ in the execution log.  Each vertex of $T$
corresponds to some event $A_v$ that we resample because it
occurs; in order for it to occur, the previous assignments of each
variable in $\vbl(A_v)$ must have made $A_v$ true, which occurs with
probability $\Prob{A_v}$.  But since we resample all the variables in
$A_v$, any subsequent assignments to these variables are independent
of the ones that contributed to $v$; with sufficient
\index{handwaving}handwaving (or a rather detailed coupling argument
as found in~\cite{MoserT2010}) this gives that each event $A_v$ occurs
with independent probability $\Prob{A_v}$, 
giving $\Prob{T} = \prod_{v ∈ T} \Prob{A_v}$.

Why do we care about this?  Because every event we resample is the
root of some witness tree, and we can argue that every event we
resample is the root of a \emph{distinct} witness tree.  The proof is
that since we only discard events $B$ that have $\vbl(B)$ disjoint
from all nodes already in the tree, once we put $A$ at the root, any
other instance of $A$ gets included.  So the witness tree rooted at
the $i$-th occurrence of $A$ in $C$ will include exactly $i$ copies of
$A$, unlike the witness tree rooted at the $j$-th copy for $j ≠ i$.

Now comes the sneaky trick: we'll bound how many distinct witness
trees $T$ we expect to see rooted at $A$, given that each occurs with
probability $\prod_{v ∈ T} \Prob{A_v}$.  This is done by
constructing a \concept{branching process} using the $x_B$ values
from Lemma~\ref{lemma-lovasz-local-lemma} as
probabilities of a node with label $A$ having a child labeled $B$ for
each $B$ in $Γ^+(A)$,
and doing algebraic
manipulations on the resulting probabilities until $\prod_{v ∈ T}
\Prob{A_v}$ shows up. We can then sum over the expected number of
copies of trees to get a bound on the expected number of events in the
execution log (since each such even is the root of some tree), which
is equal to the expected number of resamplings.

Consider the process where we start
with a root labeled $A$, and for each vertex $v$ with label $A_v$,
give it a child labeled $B$ for each $B∈Γ^+(A_v)$ with
independent probability $x_B$.
We'll now calculate the probability $p_T$ that this process
generates a particular tree $T$ in the set $\mathcal{T}_A$ of trees
with root $A$.

Let $x'_{B} = x_B \prod_{C ∈ Γ(B)} (1-x_{C})$.  Note that 
\eqref{eq-lll-condition} says precisely that $\Prob{B} ≤ x'_{B}$.

For each vertex $v$ in $T$, let $W_v ⊆ Γ^+(A_v)$ be
the set of events $B ∈ Γ^+(A_v)$ that \emph{don't} occur as
labels of children of $v$.  
The probability of getting $T$ is equal
to the product of the probabilities
at each $v$ of getting all of its children and none
of its non-children.
The non-children of $v$ collectively contribute
$\prod_{B ∈ W_v} (1-x_B)$ to the product, and 
$v$ itself contributes $x_{A_v}$ (via the product for its parent),
unless $v$ is the root node.  So we can express the giant product as
\begin{align*}
p_T
&= \frac{1}{x_A} 
 \prod_{v ∈ T}
   \left(
      x_{A_v} \prod_{B ∈ W_v} (1-x_B)
   \right).
\end{align*}
We don't like the $W_v$ very much, so we get rid of them by taking the
product of \emph{B} in $Γ^+(A)$, then dividing out the ones that
aren't in $W_v$.
This gives
\begin{align*}
p_T
&= \frac{1}{x_A} 
 \prod_{v ∈ T}
   \left(
      x_{A_v} \prod_{B ∈ W_v} (1-x_B)
   \right).
    \\&= \frac{1}{x_A} ∏_{v∈T}
    \parens*{x_{A_v} \parens*{∏_{B ∈ Γ^+(A_v)} (1-x_B)} \parens*{∏_{B ∈ Γ^+(A_v) ∖ W_v}
    \frac{1}{1-x_B}}}.
\end{align*}
This seems like we exchanged one annoying index set for another, but
each element of $Γ^+(A_V) ∖ W_v$ is $A_{v'}$ for some child of $v$ in
$T$. So we can push these factors down to the children, and since we
are multiplying over all vertices in $T$, they will each show up
exactly once except at the root. To keep the products clean, we'll
throw in $\frac{1}{1-x_A}$ for the root as well, but compensate for
this by multiplying by $1-x_A$ on the outside of the product. This
gives
\begin{align*}
p_T
&= \frac{1-x_A}{x_A}
  \prod_{v ∈ T} 
   \left(
     \frac{x_{A_v}}{1-x_{A_v}}
     \prod_{B ∈ Γ^+(A_v)} (1-x_B)
   \right)
\\
&= \frac{1-x_A}{x_A}
  \prod_{v ∈ T} 
   \left(
     x_{A_v}
     \prod_{B ∈ Γ(A_v)} (1-x_B)
   \right)
\\
&= \frac{1-x_A}{x_A}
 \prod_{v ∈ T} x'_{A_v}.
\end{align*}

Now we can bound the expected number of trees rooted at $A$ that
appear in $C$, assuming \eqref{eq-lll-condition} holds.
Letting $\mathcal{T}_A$ as before be the set of all such trees
and $N_A$ the number that appear in $C$, we have
\begin{align*}
\Exp{N_A}
&= ∑_{T ∈ \mathcal{T}_A} \Prob{\text{$T$ appears in $C$}}
\\
&≤ ∑_{T ∈ \mathcal{T}_A} \prod_{v ∈ T} \Prob{A(v)}
\\
&≤ ∑_{T ∈ \mathcal{T}_A} \prod_{v ∈ T} x'_{A_v}
\\
&= ∑_{T ∈ \mathcal{T}_A} \frac{x_A}{1-x_A} p_T
\\
&= \frac{x_A}{1-x_A} ∑_{T ∈ \mathcal{T}_A} p_T
\\
&≤ \frac{x_A}{1-x_A}.
\end{align*}

The last sum is bounded by one because occurrences of particular trees
$T$ in $\mathcal{T}_A$ are all disjoint events.

Now sum over all $A$, and we're done.

\myChapter{Derandomization}{2025}{}
\label{chapter-derandomization}

\indexConcept{derandomization}{Derandomization} is the process of
taking a randomized algorithm and turning it into a deterministic
algorithm.  This is useful both for practical reasons (deterministic
algorithms are more predictable, which makes them easier to debug and
gives hard guarantees on running time) and theoretical reasons (if we
can derandomize any randomized algorithm we could show results like
$\classP=\classRP$,\footnote{The class $\classRP$ consists of all
    languages $L$ for which there is a polynomial-time randomized
    algorithm that correctly outputs ``yes'' given an input $x$ in
    $L$ with probability at least $1/2$, and never answers ``yes''
    given an input $x$ not in $L$.  See
    §\ref{section-randomized-complexity-classes} for a more extended
    description of $\classRP$ and other randomized complexity classes.} which would reduce the number of complexity
classes that complexity theorists otherwise have to deal with).  It
may also be the case that derandomizing a randomized algorithm
can be used for \concept{probability amplification}, where we replace
a low probability of success with a higher probability, in this case
$1$.

There are basically two approaches to derandomization:
\begin{enumerate}
\item Reduce the number of random bits used down to $O(\log n)$, and
then search through all choices of random bits exhaustively.
For example, if we only need pairwise independence, we could use the XOR
technique from
§\ref{section-pairwise-independence-construction}
to replace a large
collection of variables with a small collection of random bits.

Except for the exhaustive search part, this is how randomized
algorithms are implemented in practice: rather than burning random
bits continuously, a \concept{pseudorandom generator} is initialized
from a \concept{seed} consisting of a small number of random bits.
For pretty much all of the randomized algorithms we know about, we
don't even need to use a particularly strong pseudorandom generator.
This is largely
because current popular generators are the products of a process of evolution: 
pseudorandom generators that cause
wonky behavior or fail to pass tests that approximate the assumptions
made about them by typical randomized algorithms are abandoned in
favor of better generators.\footnote{Having cheaper computers helps
here as well.  Nobody would have been willing to spend 2496 bytes on the state
vector for Mersenne Twister~\cite{MatsumotoN1998} back in 1975, but by
1998
this amount of memory was trivial for pretty much any computing device
except the tiniest microcontrollers.}

From a theoretical perspective, pseudorandom generators offer the
possibility of eliminating randomization from all randomized
algorithms, except there is a complication.  While (under reasonable
cryptographic assumptions) there exist
\index{pseudorandom generator!cryptographically secure}\indexConcept{cryptographically secure pseudorandom generator}{cryptographically secure pseudorandom generators} whose
output is indistinguishable from a genuinely random source by
polynomial-time algorithms (including algorithms originally intended
for other purposes), such generators are inherently incapable of
reducing the number of random bits down to the $O(\log n)$ needed for
exhaustive search.  The reason is that any pseudorandom generator with
only polynomially-many seeds can't be cryptographically secure,
because we can distinguish it from a random source by just checking
its output against the output for all possible seeds.  Whether there
is some other method for transforming an arbitrary algorithm in $\classRP$ or
$\classBPP$ into a deterministic algorithm remains an open problem in
complexity theory (and beyond the scope of this course).

\item Start with
a specific randomized protocol and analyze its behavior enough that we
can replace the random bits it uses with specific,
deterministically-chosen bits we can compute.  This is the main
approach we will describe below.  A non-constructive variant of this
shows that we can always replace the random bits used by all inputs of
a given size with a few carefully-selected fixed sequences (Adleman's
Theorem, described in §\ref{section-adlemans-theorem}).  More
practical is the \conceptFormat{method of conditional probabilities},
which chooses random bits sequentially based on which value is more
likely to give a good outcome
(see §\ref{section-method-of-conditional-probabilities}).
\end{enumerate}

\section{Deterministic vs.\ randomized algorithms}
\label{section-deterministic-vs-randomized-algorithms}

In thinking about derandomization, it can be helpful to have more than
one way to look at a randomized algorithm.  So far, we've describe
randomized algorithms as random choices of deterministic algorithms
($M_r(x)$)
or, equivalently, as deterministic algorithms that happen to have
random inputs ($M(r,x)$).  This gives a very static view of how randomness is
used in the algorithm.  A more dynamic view is obtained by thinking of
the computation of a randomized algorithm as a \concept{computation
tree}, where each \index{computation path}path through the tree
corresponds to a computation with a fixed set of random bits and a
branch in the tree corresponds to a random decision.  In either case
we want an execution to give us the right answer with reasonably high
probability, whether that probability measures our chance of getting a
good deterministic machine for our particular input or landing on a
good computation path.

\section{Adleman's theorem}
\label{section-adlemans-theorem}

The idea of picking a good deterministic machine is the basis for
\index{theorem!Adleman's}
\concept{Adleman's theorem}\cite{Adleman1978}, 
a classic result in complexity
theory.
Adleman's theorem says that we can always replace randomness by an
oracle that presents us with a fixed string of \concept{advice} $p_n$
that depends only on the size of the input $n$ and has size polynomial
in $n$.  The formal statement
of the theorem relates the class $\classRP$, which
is the class of problems for which there exists a
polynomial-time Turing machine $M(x,r)$ that outputs
$1$ at least half the time when $x∈L$ and never when
$x\not∈L$; and the class $\classPpoly$, which is the class of problems for which
there is a polynomial-sized string 
$p_{n}$ for each input size $n$ and a 
polynomial-time Turing machine
$M'$ such that $M'(x,p_{\card*{x}}$) outputs $1$ if and only if
$x∈L$.
\begin{theorem}
\label{theorem-adlemans-theorem}
$\classRP ⊆ \classPpoly$.
\end{theorem}
\begin{proof}
The intuition is that if any one random string has a constant
probability of making $M$ happy, then by choosing enough random
strings we can make the probability that $M$ fails using on every
random string for any given input so small that even after we sum over
all inputs of a particular size, the probability of failure is still
small using 
the union bound \eqref{eq-union-bound}.
This is an example of 
\concept{probability amplification}, where we
repeat a randomized algorithm many times to reduce its failure
probability.

Formally,
consider any fixed input
$x$ of size $n$, and imagine running $M$ repeatedly on this input with
$n+1$ independent sequences of random bits $r_{1}, r_{2}, \dots{},
r_{n+1}$.  If $x\not∈L$, then $M(x,r_i)$ never outputs $1$.  If
$x∈L$, then for each $r_i$, there is an independent probability of
at least $1/2$ that $M(x,r_i) = 1$.  So $\Prob{M(x,r_i) = 0} ≤ 1/2$,
and $\Prob{\forall i M(x,r_i) = 0} ≤ 2^{-(n+1)}$.  If we sum this
probability of failure for each individual $x∈L$ of length $n$ over the at most
$2^{n}$ such elements, we get a probability that any of them fail of
at most $2^{n}2^{-(n+1)} = 1/2$.  Turning this upside down, any
sequence of $n+1$ random inputs includes a \concept{witness} that
$x∈L$ for \emph{all} inputs $x$ with probability at least $1/2$.
It follows that a good sequence $r_1, \dots, r_{n+1}$, exists.

Our advice $p_{n}$ is now some good sequence
$p_n = \Tuple{r_{1}\dots{}r_{n+1}}$, and the deterministic
advice-taking algorithm that uses it runs $M(x,r_i)$ for each $r_i$
and returns true if and only if at least one of these executions returns true.
\end{proof}

The classic version of this theorem shows that anything you can do
with a polynomial-size randomized circuit (a circuit made up of AND,
OR, and NOT gates where some of the inputs are random bits,
corresponding to the $r$ input to $M$) can be done with a
polynomial-size deterministic circuit (where now the $p_{n}$ input is
baked into the circuit, since we need a different circuit for each
size $n$ anyway).  

A limitation of this result is that 
ordinary algorithms seem to be better
described by \concept{uniform} families of circuits,
where there exists a polynomial-time algorithm that, given input $n$,
outputs the circuit $C_{n}$ for processing size-$n$ inputs.  In
contrast, the class
of circuits generated by Adleman's theorem is most likely
\concept{non-uniform}: the process of finding the good witnesses
$r_{i}$ is not something we know how to do in polynomial time
(with the usual caveat that we can't prove much about what we can't do in polynomial time).

\section{Limited independence}

For some algorithms, it may be that full independence is not needed
for all of the random bits.  If the amount of independence needed is
small enough, this may allow us to reduce the actual number of random
bits consumed down to $O(\log n)$, at which point we can try all
possible sequences of random bits in polynomial time.

Variants of this technique have been used heavily in the cryptography
and complexity; see~\cite{LubyW2005} for a survey of some early work
in this area.  We'll do a quick example of the method
before before moving onto more direct
approaches.

\subsection{MAX CUT}

Let's look at the randomized MAX CUT algorithm from
§\ref{section-max-cut}.  In the original algorithm, we use $n$
independent random bits to assign the $n$ vertices to $S$ or $T$.
The idea is that by assigning each vertex independently at
random to one side of the cut or the other, each edge appears in
the cut with probability $1/2$, giving a total of $m/2$ edges in the
cut in expectation. 

Suppose that we replace these $n$ independent random bits with $n$
pairwise-independent bits generated by taking XORs of subsets of
$\ceil{\lg (n+1)}$ independent random bits as described in
§\ref{section-pairwise-independence-construction}.  Because the bits
are pairwise-independent, the probability that the two endpoints of an
edge are assigned to different sides of the cut is still exactly
$1/2$.  So on average we get $m/2$ edges in the cut as before, and
there is at least one sequence of random bits that guarantees a cut at
least this big.

But with only $\ceil{\lg (n+1)}$ random bits, there are only
$2^{\ceil{\log (n+1)}} < 2(n+1)$
possible sequences of random bits.  If we try all of them, then we
find a cut of size $m/2$ always.  The total cost is $O(n(n+m))$ if we
include the $O(n+m)$ cost of testing each cut.  Note that this
algorithm does not generate all $2^n$ possible cuts, but among those
it \emph{does}
generate, there must be a large one.

In this particular case, we'll see below how to get the same result at
a much lower cost, using more knowledge of the problem.
So we have the typical trade-off between algorithm
performance and algorithm designer effort.

\section{The method of conditional probabilities}
\label{section-method-of-conditional-probabilities}

The \concept{method of conditional probabilities}~\cite{Raghavan1988} 
follows an execution of the randomized algorithm, but at
each point where we would otherwise make a random decision, it makes a
decision that minimizes the conditional probability of losing.

Structurally, this is similar to the method of bounded differences
(see §\ref{section-method-of-bounded-differences}).  
Suppose our randomized algorithm generates $m$ random values $X_1,
X_2, \dots, X_m$.  Let $f(X_1,\dots,X_m)$ be the indicator variable
for our randomized algorithm failing (more generally, we can make it
an expected cost or some other performance measure).  Extend $f$ to
shorter sequences of values by defining 
$f(x_1,\dots,x_k) = \Exp{f(x_1,\dots,x_k,X_{k+1},\dots,X_m)}$.  Then
$Y_k = f(X_1,\dots,X_k)$ is a \index{martingale!Doob}\index{Doob martingale}{Doob
martingale}, just as in the method of bounded differences.
This implies that, for any partial sequence of values $x_1,\dots,x_k$,
there exists some next value $x_{k+1}$ such that 
$f(x_1,\dots,x_k) ≥ f(x_1,\dots,x_k,x_{k+1})$.
If we can find this value, we can follow a path on which $f$ always
decreases, and obtain an outcome of the algorithm $f(x_1,\dots,x_m)$
less than or equal to the initial value $f(\Tuple{})$.  If our
outcomes are $0$–$1$ (as in failure probabilities), and our initial
value for $f$ is less than $1$, this means that
we reach an outcome with $f=0$.

The tricky part here is that it may be hard to compute
$f(x_1,\dots,x_k$).  (It's always possible to do so in principle by
enumerating all assignments of the remaining variables, but if we have
time to do this, we can just search for a winning assignment
directly.)  What makes the method of conditional probabilities
practical in many cases is that it's not necessary for $f$ to compute the
actual probability of failure, as long as (a) $f$ gives an upper bound on
the real probability of failure, at least in terminal configurations,
and (b) $f$ has the property used in the argument that for any
partial sequence $x_1,\dots,x_k$ there exists an extension
$x_1,\dots,x_k,x_{k+1}$ with $f(x_1,\dots,x_k) ≥
f(x_1,\dots,x_k,x_{k+1})$.  
Such an $f$ is called a \concept{pessimistic estimator}.
If we can find a pessimistic estimator that is easy to
compute and starts out less than $1$, then we can just follow it down the
tree to a leaf that doesn't fail.

\subsection{MAX CUT using conditional probabilities}
\label{section-max-cut-via-conditional-probabilities}

Again we consider the algorithm from §\ref{section-max-cut} that
assigns vertices to $S$ and $T$ at random.
To derandomize this algorithm, at each step we pick a vertex and
assign it to the side of the cut that maximizes the conditional
expectation of the number of edges that cross the cut.  We can compute
this conditional expectation as follows:
\begin{enumerate}
    \item For any edge that already has both endpoints assigned to $S$
        or $T$, it's either in the cut or it isn't: add $0$ or $1$ as
        appropriate to the conditional expectation.
    \item For any edge with only one endpoint assigned, there's a
        $1/2$ probability that the other endpoint gets assigned to the
        other side (in the original randomized algorithm).  Add $1/2$
        to the conditional expectation for these edges.
    \item For any edge with neither endpoint assigned, we again have a
        $1/2$ probability that it crosses the cut.  Add $1/2$ for
        these as well.
\end{enumerate}

So now let us ask how assigning a particular previously unassigned
vertex $v$ to $S$ or $T$
affects the conditional probability.  For any neighbor $w$ of $v$ that is
not already assigned, adding $v$ to $S$ or $T$ doesn't affect the
$1/2$ contribution of $vw$.  So we can ignore these.  The only effects
we see are that if some neighbor $w$ is in $S$, assigning $v$ to $S$
decreases the conditional expectation by $1/2$ and assigning $v$ to
$T$ increases the expectation by $1/2$.  So to maximize the
conditional expectation, we should assign $v$ to whichever side
currently holds fewer of $v$'s neighbors—the obvious greedy
algorithm, which runs in $O(n+m)$ time if we are reasonably clever
about keeping track of how many neighbors each unassigned node has in
$S$ and $T$.  The advantage of the method of conditional probabilities
here is that we immediately get that the greedy algorithm achieves a
cut of size $m/2$, which might require actual intelligence to prove
otherwise.

\subsection{Deterministic construction of Ramsey graphs}
\label{section-deterministic-construction-of-ramsey-graphs}

Here is an example that is slightly less trivial.  For this example we
let $f$ be a count of bad events rather than a failure probability,
but the same method applies.

Recall from §\ref{section-ramsey-numbers} that if $k ≥ 3$,
for $n ≤ 2^{k/2}$ there exists a graph with $n$ nodes and no cliques
or independent sets of size $k$.  The proof of this fact is to observe
that each subset of $k$ vertices is bad with probability $2^{-k+1}$,
and when $\binom{n}{k}2^{-k+1} < 1$, the expected number of bad
subsets in $G_{n,1/2}$ is less than $1$, showing that some good graph
exists.

We can turn this into a deterministic $n^{O(\log n)}$ algorithm for
finding a $k$-Ramsey graph in the worst case when $n = 2^{k/2}$.
The trick is to set the edges to be present or absent one at a time,
and for each edge, take the value that minimizes the expected number
of bad subsets conditioned on the choices so far.  We can easily
calculate the conditional probability that a subset is bad in $O(k)$
time: if it already has both a present and missing edge, it's not bad.
Otherwise, if we've already set $\ell$ of its edges to the same value,
it's bad with probability exactly $2^{-k+\ell}$.  Summing this over all
$O(n^k)$ subsets takes $O(n^k k)$ time per edge, and we have $O(n^2)$
edges, giving $O(n^{k+2} k) = O\left(n^{(2 \lg n + 2 + \lg \lg
n)}\right) = n^{O(\log n)}$ time total.

It's worth mentioning that there are faster deterministic
constructions in the literature. The best construction that I
am aware of is given by 
Cohen~\cite{Cohen2019},
which constructs a graph with
$n$ vertices and no clique or independent set of size $2^{(\log \log
n)^c}$ for a particular constant $c$, in polylogarithnmic time per
edge.

\subsection{Derandomized set balancing}
\label{section-set-balancing-deranomdized}

Recall the set balancing problem from
§\ref{section-set-balancing-randomized}:
given vectors $v_1, v_2, \dots, v_n$ in
$\Set{0,1}^m$, we'd like to find $\pm 1$ coefficients $ε_1,
ε_2, \dots ε_n$ that minimize
$\max_j \abs*{X_j}$ where $X_j = ∑_{i=1}^{n} ε_i v_{ij}$.
Hoeffding tells us that choosing the $ε_i$ independently at random gives a nonzero 
probability that this quantity is at most $√{2n \ln 2m}$.

Because there may be very complicated dependence between the $X_j$, it
is difficult to calculate the probability of the event
$\bigcup_j [\abs*{X_j} ≥ t]$, whether conditioned on some of the $ε_i$
or not.
However,
we can calculate the probability of the individual events 
$[\abs*{X_j} ≥ t]$ exactly.
Conditioning on $ε_1, \dots, ε_k$, the expected value of
$X_j$ is just $\ell = ∑_{i=1}^{k} ε_i v_{ij}$, and the distribution
of $Y=X_j - \Exp{X_j}$ is the sum $r ≤ n-k$ independent $\pm 1$ random
variables.  
So
\begin{align*}
    \ProbCond{\abs*{X_j} > t}{ε_1, \dots, ε_k}
    &= 1 - \Prob{-t-\ell ≤ Y ≤ t-\ell}
    \\&= 1 -
    ∑_{i=\ceil{\frac{-t-\ell+r}{2}}}^{\floor{\frac{t-\ell+r}{2}}}
    \binom{r}{i} 2^{-r}.
\end{align*}
This last expression involves a linear number of terms, each of which
we can calculate using a linear number of operations on rational
numbers that fit in a linear number of bits, so we
can calculate the probability exactly in polynomial time by just
adding them up.

For our pessimistic estimator, we take
\begin{displaymath}
    U(ε_1,\dots,ε_k) = ∑_{i=j}^{n} 
\ProbCond{\abs*{X_j} > √{2n \ln 2m}}{ε_1,\dots,ε_k}.  
\end{displaymath}
Since each term in the sum is a Doob martingale, the sum
is a martingale as well, so 
$\ExpCond{U(ε_1,\dots,ε_{k+1})}{ε_1,\dots,ε_k}
= U(ε_1,\dots,ε_k)$.  It follows that 
for any choice of $ε_1,\dots,ε_k$ there exists some
$ε_{k+1}$ such that $U(ε_1,\dots,ε_k) ≥
U(ε_1,\dots,ε_{k+1})$, and
we can determine this winning
$ε_{k+1}$ explicitly.
Our previous argument
shows that $U(\Tuple{}) < 1$, which implies that our final value
$U(ε_1,\dots,ε_n)$ will also be less than $1$.  Since
$U$ is an integer, this means it must be
$0$, and we find an assignment in which $\abs*{X_j} < √{2n \ln 2m}$ for all $j$.

\myChapter{Probabilistically-checkable proofs}{2025}{}
\label{chapter-PCP}
\chaptermark{PCP and hardness of approximation}

In this chapter, we discuss the
\concept{PCP theorem}\index{theorem!PCP}~\cite{AroraLMSS1998}, which
is an alternative characterization of \classNP{} that can be used
to show hardness for various approximation problems assuming 
$\classP ≠ \classNP$. What connects the PCP theorem to randomized
algorithms is that class of
\conceptFormat{probabilistically-checkable
proofs}\index{proof!probabilistically-checkable}\index{probabilistically-checkable
proof} that it describes are what you get when you replace the
deterministic verifier that checks a witness
in the usual definition of $\classNP$ with a
randomized algorithm. This allows for very efficient verification,
since a randomized verified can demand a proof that includes
sufficient internal consistency checks that a spot-check of only a few
randomly-chosen bits of the proof is enough to detect an incorrect
proof with constant probability.
Formally, this means that we consider a verifier that is a 
probabilistic Turing machine that uses $r$
random bits to select $q$ bits
of the proof to look at, and accepts bad proofs
with less than some constant probability $ρ$ while accepting all good
proofs.

This turns out to have strong consequences for approximation
algorithms.  We can think of a probabilistically-checkable proof as a kind
of constraint satisfaction problem, where the constraints apply to the
tuples of $q$ bits that the verifier might look at, the number of
constraints is bounded by the number of possible random choices $2^r$,
and each constraint enforces some
condition on those $q$ bits corresponding to the verifier's response
to seeing them. If we can satisfy at least $ρ⋅2^r$ of the constraints,
we've constructed a proof that is not bad. This means that there is a
winning certificate for our original $\classNP$ machine, and that
whatever input $x$ we started with is in the language accepted by that
machine. So a polynomial-time approximation algorithm for this
constraint satisfaction problem will let us generate
probabilistically-checkable proofs for the corresponding problem in \classNP, 
which means there can't be such an approximation algorithm unless 
$\classP = \classNP$.

The above is only a sketchy, high-level overview of where we are going.
Below we fill in some of the details. We are mostly following the
approach of~\cite[§§18.1–18.4]{AroraB2007}, with much of the actual
text of this chapter adapted from~\cite{Aspnes2020complexity}.

\section{Probabilistically-checkable proofs}
\label{section-PCP}

A \concept{$\Tuple{r(n),q(n),ρ}$-\classPCP{} verifier} for a language $L$
consists of two polynomial-time computable functions $f$ and $g$,
where:
\begin{itemize}
    \item 
        $f(x,r)$ takes an input $x$ of length $n$ and a string $r$ of length
        $r(n)$
        and outputs a sequence $i$ of $q(n)$ indices $i_1,i_2,\dots,i_{q(n)}$,
        each in the range $0\dots q(n)⋅2^{r(n)}$; and
    \item 
        $g(x,π_i)$ takes as input the same $x$ as $f$ and a sequence
        $π_i = π_{i_1} π_{i_2} \dots π_{i_{(q)n}}$ and outputs either
        $1$ (for accept) or $0$ for (reject); and
    \item if $x∈L$, then there exists a sequence $π$ that causes
        $g(x,π_{f(x,r)})$ to output $1$ always
        (\concept{completeness}); and
    \item if $x∉L$, then for any sequence $π$, $g(x,π_{f(x,r)})$
        outputs $1$ with probability at most $ρ$
        (\concept{soundness}).
\end{itemize}

We call the string $π$ a \concept{probabilistically-checkable proof}
or \concept{PCP} for short.

Typically, $ρ$ is set to $1/2$, and we just write
$\Tuple{r(n),q(n)}$-\classPCP{} verifier for
$\Tuple{r(n),q(n),1/2}$-\classPCP{} verifier.

The class $\classPCP(r(n),q(n))$ is the class of languages $L$ for which
there exists an $\Tuple{O(r(n)),O(q(n)),1/2}$-\classPCP{} verifier.

The \index{PCP theorem}\index{theorem!PCP}\concept{\classPCP{} theorem}
says that $\classNP = \classPCP(\log n, 1)$.  That is, any language in
$\classNP$ can be recognized by a $\classPCP$-verifier that is allowed to
look at only a constant number of bits selected using $O(\log n)$
random bits from a proof of polynomial length, which is fooled at most
half the time by bad proofs.  In fact, $3$ bits is enough
\cite{Hastad2001}.  We won't actually prove this here, but we will
describe some consequences of the theorem, and give
some hints about how the proof works.

\section{GRAPH NON-ISOMORPHISM}

The \concept{GRAPH NON-ISOMORPHISM} (\concept{GNI}) problem takes as
input two graphs $G$ and $H$, and asks if $G$ is not isomorphic to
$H$ (written $G \not≃H$).  Recall that $G$ \emph{is} isomorphic to $H$
(written $G≃H$) if there is a permutation of the
vertices of $G$ that makes it equal to $H$.  
An \classNP-machine can easily solve the GRAPH ISOMORPHISM problem by
guessing the right permutation, which puts the complementary GRAPH
NON-ISOMORPHISM problem in \classcoNP.\footnote{Neither problem is
known not to be in \classP, even under the assumption that $\classP ≠
\classNP$. The exact complexity of GRAPH ISOMORPHISM is still open.}
But in general we don't know if problems in \classcoNP problems have
\classNP-style proofs, and there is no obvious general method
that a prover could use to
convince a deterministic verifier that two graphs are \emph{not
} isomorphic.

Instead, we'll describe two protocols that allow an omniscient prover
to demonstrate that two graphs are not isomorphic to a suspicious,
randomized verifier. The first, described in
§\ref{section-graph-non-isomorphism-with-private-coins}, is an
\concept{interactive proof}\index{proof!interactive}, meaning
that the verifier can ask questions that the prover must answer. The
second, described in §\ref{section-graph-non-isomorphism-PCP}, is not
interactive: instead, the prover supplies a (very large) proof that
the verifier can check by looking at a single randomly-chosen bit.

\subsection{GRAPH NON-ISOMORPHISM with private coins}
\label{section-graph-non-isomorphism-with-private-coins}

Here is how the prover can convince a verifier that two graphs are not
isomorphic, assuming that the verifier can flip coins that are not
visible to the prover.

Given graphs $G$ and $H$, the verifier picks one or the other with equal
probability, then randomly permutes its vertices to get a test graph
$T$. It then asks the
prover which of $G$ or $H$ it picked.

If the graphs are not isomorphic, the prover can distinguish $T≃G$
from $T≃H$
and answer the
question correctly with probability $1$. If they are isomorphic, then
the prover can only guess: $T$ gives no information at all about which
of $G$ or $H$ was used to generate it. So in this case it answers
correctly only with probability $1/2$. This gives us the completeness
and soundness properties needed for a probabilistically-checkable
proof. Unfortunately we need to do more work to make this
non-interactive.

\subsection{A probabilistically-checkable proof for GRAPH NON-ISOMORPHISM}
\label{section-graph-non-isomorphism-PCP}

To turn the interactive proof 
into a probabilistically-checkable proof, have the prover
build a bit-vector $π$ indexed by
every possible graph on $n$ vertices,
writing a $1$ for each graph that is isomorphic
to $H$. Now the verifier can use $Θ(n^2)$ random bits
to construct $T$ as
above, look it up in this gigantic table, and accept if and only if
(a) it chose $G$ and $π[T] = 0$, or (b) it chose $H$ and $π[T] = 1$.
If $G$ and $H$ are non-isomorphic, the verifier accepts every time.
But if they are isomorphic, no matter what proof $π'$ is supplied,
there is at least a $1/2$ chance that $π'[T]$ is wrong.
This puts GRAPH NON-ISOMORPHISM in $\classPCP(n^2, 1)$.

(A minor technical issue here is that the verifier only needs $O(n
\log n)$ bits to construct $T$. But we give it $Θ(n^2)$ bits, so that
the table can be big enough to hold all graphs and still fit in the
maximum proof size $q(n) 2^{r(n)}$.)

\section{$\classNP ⊆ \classPCP(\poly(n), 1)$}
\label{section-PCP-poly}

Here we give a weak version of the PCP theorem, adapted
from~\cite[§18.4]{AroraB2007}, showing that any
problem in $\classNP$ has a probabilistically-checkable proof where
the verifier uses polynomially-many random bits but only needs to look
at a constant number of bits of the proof, which we can state
succinctly as $\classNP ⊆
\classPCP(\poly(n),1)$.\footnote{This is a rather weak
result, since (a) the full PCP theorem gives $\classNP$ using only
$O(\log n)$ random bits, and (b) $\classPCP(\poly(n),1)$ is known to
be equal to $\classNEXP$~\cite{BabaiFL1991}. But the construction is still useful
for illustrating many of the ideas behind probabilistically-checkable
proofs.} The proof itself will be
exponentially long.  

The central step is to construct a $\Tuple{\poly(n),1)}$-PCP for a particular
$\classNP$-complete problem; we can then take any other problem in
$\classNP$, reduce it to this problem, and use the construction to get
a PCP for that problem as well.

\subsection{QUADEQ}

The particular problem we will look
at is QUADEQ, the language of systems of quadratic equations over
$ℤ_2$ that have solutions.  

This is in $\classNP$ because we can
guess and verify a solution; it's $\classNP$-hard because we can use
quadratic equations over $ℤ_2$ to encode instances of SAT, using the
representation $0$ for false, $1$ for true, $1-x$ for $¬x$, $xy$ for
$x∧y$, and $1-(1-x)(1-y) = x + y + xy$ for $x∨y$. We may also need to introduce
auxiliary variables to keep the degree from going up: for example, to
encode the clause $x∨y∨z$, we introduce an auxiliary variable 
$q$ representing $x∨y$ and enforce the constraints
$q = x∨y$ and $1 = q∨z = x∨y∨z$ using two equations
\begin{align*}
    x + y + xy &= q, \\
    q + z + qz & = 1.
\end{align*}

It will
be helpful later to rewrite these in a standard form with only zeros
on the right:
\begin{align*}
    q + x + y + xy &= 0 \\
    q + z + qz + 1 & = 0.
\end{align*}
This works because we can move summands freely from one side of an
equation to the other since all addition is mod $2$.

\subsection{The Walsh-Hadamard Code}

An $\classNP$ proof for QUADEQ just gives an assignment to the
variables that makes all the equations true.  Unfortunately, this
requires looking at the entire proof to check it.  To turn this
into a $\classPCP(\poly(n),1)$ proof, we will make heavy use of a rather
magical 
\index{code!error-correcting}
\concept{error-correcting code} called the 
\index{code!Walsh-Hadamard}
\concept{Walsh-Hadamard code}.

This code expands an $n$-bit string $x$ into a $2^n$-bit
\concept{codeword} $H(x)$, where $H(x)_i = x⋅i$ when $x$ and the index
$i$ are both interpreted as $n$-dimensional vectors over $ℤ_2$ and
$⋅$ is the usual vector dot-product $∑_{i=1}^{n} x_j i_j$.  This
encoding has several very nice properties, all of which we will need:
\begin{enumerate}
    \item It is a linear code: $H(x+y) = H(x) + H(y)$ when all strings
        are interpreted as vectors of the appropriate dimension over
        $ℤ_2$.
    \item It is an error-correcting code with distance $2^{n-1}$.  If
        $x≠y$, then exactly half of all $i$ will give $x⋅i ≠ y⋅i$.
        This follows from the \concept{subset sum principle}, which
        says that a random subset of a non-empty set $S$ is equally
        likely to have an even or odd number of elements.\footnote{Proof:
        If $S$ is non-empty, pick some element $x$, and then the
        function $f$ that removes $x$ if present and adds $x$ if
        missing is a bijection between even and odd subsets.
        If you like generating functions, an alternative proof
        uses the Binomial Theorem to show $∑_{\text{even $i$}} \binom{n}{i} -
        ∑_{\text{odd $i$}} \binom{n}{i} = ∑_{i=0}^{n} (-1)^n
        \binom{n}{i} = (1 + (-1))^n = 0^n = 0$ when $n≠0$.}
        So for any particular nonzero $x$, exactly half of the $x⋅i$ values
        will be $1$, since $i$ includes each one in $x$ with
        independent probability $1/2$.  This makes $d(H(0),H(x)) =
        2^{n-1}$. But then the linearity of $H$ gives $d(H(x),H(y)) = d(H(0),d(H(x+y))) =
        2^{n-1}$ whenever $x≠y$.
    \item It is 
        \index{testable!locally}
        \index{test!local}
        \concept{locally testable}: Given an alleged codeword $w$, we
        can check if $w$ is close to being a legitimate codeword
        $H(x)$ by sampling a constant number of bits from $w$.  (We do
        not need to know what $x$ is to do this.)

        Our test is: Pick two indices $i$ and $j$ uniformly at random,
        and check if $w_i + w_j = w_{i+j}$.  A legitimate codeword
        will pass this test always.  It is also possible to show using
        Fourier analysis (see \cite[Theorem 19.9]{AroraB2007}) that if
        $w$ passes this test with probability $ρ ≥ 1/2$, then there is
        some $x$ such that $\Pr_i[H(x)_i = w_i] ≥ ρ$ (equivalently,
        $d(H(x),w) ≤ 2^n(1-ρ)$, in which case we say $w$ is
        \concept{$ρ$-close} to $H(x)$.
    \item It is
        \index{decoding!local}
        \index{decodable!locally}
        \concept{locally decodable}: If $w$ is $ρ$-close to $H(x)$,
        then we can compute $H(x)_i$ by choosing a random index $r$
        and computing $w_r + w_{i+r}$.  This will be equal to
        $H(x)_i$ if both bits are correct (by linearity of $H$).  The
        probability that both bits are correct is at least $1-2δ$ if
        $ρ = 1-δ$.
    \item It allows us to check an unlimited number of linear
        equations in $x$ by looking up a single bit of $H(x)$.  This again uses the
        subset sum principle.  Give a system of linear equations
        $x⋅y_1 = 0, x⋅y_2 = 0, \dots x⋅y_m =0$, choose a random subset
        $S$ of $\Set{1,\dots,m}$, let $i = ∑_{j∈S} y_j$,
        and query $H(x)_i = x ⋅ ∑_{j∈S} y_j$.
        This will be $0$ always if the equations hold and $1$
        with probability $1/2$ if at least one is violated.

        This gets a little more complicated if we have any ones on the
        right-hand side.  But we can handle an equation of the form
        $x⋅y=1$ by rewriting it as $x⋅y+1 = 0$, and then extending $x$
        to include an extra constant $1$ bit (which we can test is
        really one by looking up $H(x)_i$ for an appropriate index $i$).
\end{enumerate}

\subsection{A PCP for QUADEQ}

So now to construct a PCP for QUADEQ, we build:
\begin{enumerate}
    \item An $n$-bit solution $u$ to the system of quadratic equations, which
        we think of as a function $f: \Set{0,1}^n → \Set{0,1}$ and
        encode as $f = H(u)$.
    \item An $n^2$-bit vector $w = u⊗ u$ where $(u ⊗ u)_{ij} =
        u_i u_j$, which we encode as $g = H(u ⊗ u)$.
\end{enumerate}

To simplify our life, we will assume that one of the equations is $x_1
= 1$ so that we can use this constant $1$ later (note that we can
trivially reduce the unrestricted version of QUADEQ to the version
that includes this assumption by adding an extra variable and
equation).  A different approach
that does not require this assumption is given in \cite[§18.4.2, Step 3]{AroraB2007}.

To test this PCP, the verifier checks:
\begin{enumerate}
    \item That $f$ and $g$ are $(1-δ)$-close to real codewords for some suitably small $δ$.
    \item That for some random $r, s$, $f(r)f(s) = g(r⊗ s)$.
        This may let us know if $w$ is inconsistent with $u$.  Define $W$
        as the $n×n$ matrix with $W_{ij} = w_{ij}$ and $U$ as the
        $n×n$ matrix $U = u⊗ u$ (so $U_{ij} = u_i u_j)$.  Then
        $g(r⊗ s) = w⋅(r ⊗ s) = ∑_{ij} w_{ij} r_i s_j =
        r W s$ and $f(r) f(s) = (u⋅r)(u⋅s) = \parens*{∑_i u_i r_i}
        \parens*{∑_j u_j s_j} = ∑_{ij} r_i U_{ij} r_j = r U s$, where
        we are treating $r$ as a row vector and $s$ as a column vector.  Now
        apply the random subset principle to argue that if $U≠W$, then
        $rU ≠ rW$ at least half the time, and if $rU≠rW$, then $rUs ≠
        rWs$ at least half the time.  This gives a probability of at
        least $1/4$ that we catch $U≠W$, and we can repeat the test a
        few times to amplify this to whatever constant we want.
    \item That our extra constant-$1$ variable is in fact $1$ (lookup
        on $u$).
    \item That $w$ encodes a satisfying assignment for the original
        problem.  This just involves checking a system of linear
        equations using $w$.
\end{enumerate}

Since we can make each step fail with only a small constant
probability, we can make the entire process fail with the sum of these
probabilities, also a small constant.

\section{\classPCP{} and approximability}
\label{section-PCP-approximation}

Suppose we want to use the full $\classPCP$ theorem $\classNP =
\classPCP(\log n, 1)$ to actually decide some language $L$ in $\classNP$.
What do we need to do?

\subsection{Approximating the number of satisfied verifier queries}

If we can somehow find a PCP for $x∈L$ and verify it, then we know
$x∈L$.  So the obvious thing is to try to build an algorithm for
generating PCPs.  But actually generating a PCP may be hard.
Fortunately, even getting an good approximation will be enough.
We illustrate the basic idea using MAX SAT, the problem of finding an
assignment that maximizes the number of satisfied clauses in a 3CNF
formula.

Suppose we have some language $L$ with a PCP verifier $V$.  If
$x∈L$, there exists a proof of polynomial length such that every
choice of $q = O(\log n)$ bits by $V$ from the proof will be accepted by $V$.  We
can encode this verification step as a Boolean formula: for each
sequence of bits $S = \Tuple{π_{i_1},\dots,π_{i_q}}$, write a polynomial-size formula $φ_S$
with variables in $π$ that checks if $V$ will accept
$π_{i_1},\dots,π_{i_q}$ for our given input $x$.
Then we can test if
$x∈L$ by testing if $φ = \bigwedge_S φ_S$ is satisfiable or not.

But we can do better than this.  Suppose that we can approximate the
number of $φ_S$ that can be satisfied to within a factor of $2-ε$.
Then if $φ$ has an assignment that makes all the $φ_S$ true (which
followed from completeness if $x∈L$), our
approximation algorithm will give us an assignment that makes at least
a $\frac{1}{2-ε} > \frac{1}{2}$ fraction of the $φ_S$ true.
But we can never make more than $\frac{1}{2}$ of the $φ_S$ true if
$x∉L$.  So we can run our hypothetical approximation algorithm, and if
it gives us an assignment that satisfies more than half of the $φ_S$,
we know $x∈L$.  If the approximation runs in $\classP$, we just solved
SAT in $\classP$ and showed $\classP=\classNP$.

\subsection{Gap-preserving reduction to MAX SAT}

Maximizing the number of subformulas $φ_S$ that are satisfied is a
strange problem, and we'd like to state this result in terms of a more
traditional problem like MAX SAT. We can do this by converting each
$φ_S$ into a 3CNF formula (which makes $φ$ also 3CNF), but the cost is
that we reduce the \concept{gap} between negative instances $x∉L$ and
negative instances $x∈L$.

The $\classPCP$ theorem gives us a gap of $(1/2,1)$ between negative
and positive instances.
If each $φ_S$ is represented by $k$ 3CNF clauses,
then it may be that violating a single $φ_S$ only maps to violating
one of those $k$ clauses.  So where previously we either satisfied at most
$1/2$ of the $φ_S$ or all of them, now we might have
a negative instance where we can still satisfy a $1-\frac{1}{2k}$
fraction of the
clauses.  So we only get $\classP=\classNP$ if we are given a
poly-time approximation algorithm for MAX SAT that is at least this
good; or, conversely, we only show that $\classP≠\classNP$ implies
that there is no MAX SAT approximation that gets more than
$1-\frac{1}{2k}$ of optimal.

This suggests that we want to find a version of the $\classPCP$
theorem that makes $k$ as small as possible.  Fortunately,
Håstad~\cite{Hastad2001} showed that it is possible to construct a
PCP-verifier for 3SAT with the miraculous property that (a) $q$ is
only $3$, and (b) the verification step involves testing only if
$π_{i_1} ⊕ π_{i_2} ⊕ π_{i_3} = b$, where $i_1,i_2,i_3,$ and $b$ are
generated from the random bits.

There is a slight cost: the
completeness parameter of this verifier is only $1-ε$ for any fixed $ε>0$,
meaning that it doesn't always recognize a valid proof, and the
soundness parameter is $1/2+ε$.  But checking 
$π_{i_1} ⊕ π_{i_2} ⊕ π_{i_3} = b$ requires a 3CNF formula of only 4
clauses.  So this means that there is no approximation algorithm for
MAX SAT that does better than $7/8+δ$ of optimal in all cases, unless
$\classP = \classNP$.  This matches the $7/8$ upper bound given by
just picking an assignment at random.\footnote{It's common in the approximation-algorithm
literature to quote approximation ratios for maximization problems as
the fraction of the best solution that we can achieve, as in a
$7/8$-approximation for MAX SAT satisfying $7/8$ of the maximum
possible clauses.  This leads to rather odd statements when we start
talking about lower bounds (``you can't do better than $7/8+δ$'') and
upper bounds (``you can get at least $7/8$''), since the naming of the
bounds is reversed from what they actually say.  For this reason
complexity theorists have generally standardized on always treating
approximation ratios as greater than $1$, which for maximization
problems means reporting the inverse ratio, $8/7-ε$ in this case.  I
like $7/8$ better than $8/7$, and there is no real possibility of
confusion, so I will stick with $7/8$.}

This is an example of a reduction argument, since we reduced 3SAT
first to a problem of finding a proof that would make a particular
PCP-verifier happy and then to MAX SAT.  The second reduction is an
example of a 
\index{reduction!gap-preserving}
\concept{gap-preserving reduction},
in that it takes an instance of a  problem with a non-trivial gap $(1/2+ε,1-ε)$ and
turns it into an instance of a problem with a non-trivial gap
$(7/8+ε,1-ε)$.  Note that do be gap-preserving, a reduction doesn't
have to preserve the value of the gap, it just has to preserve the
existence of a gap.  So a 
\index{reduction!gap-reducing}
\concept{gap-reducing reduction} like this
is still gap-preserving.  We can also consider
\index{reduction!gap-amplifying}
\index{gap-amplifying reduction}
\conceptFormat{gap-amplifying reductions}: in a sense, Håstad's
verifier gives a reduction from 3SAT to 3SAT that amplifies the reduction from the
trivial $(1-1/m,1)$ that follows from only being able to satisfy $m-1$
of the $m$ clauses in a negative instance
to the much more useful $(1/2+ε,1-ε)$.

\subsection{Other inapproximable problems}

Using inapproximability of MAX SAT, we can find similar
inapproximability results for other \classNP-complete optimization
problems by looking for gap-preserving reductions from MAX SAT.  In
many cases, we can just use whatever reduction we already had for
showing that the target problem was \classNP-hard.  This gives
constant-factor inapproximability bounds for problems like GRAPH
3-COLORABILITY (where the value of a solution is the proportion of
two-colored edges) and MAXIMUM INDEPENDENT SET (where the value of a
solution is the size of the independent set).  In each case we observe
that a partial solution to the target problem maps back to a partial
solution to the original SAT problem.

In some cases we can do better, by applying a gap-amplification step.
For example, suppose that no polynomial-time algorithm for INDEPENDENT
SET can guarantee an approximation ratio better than $ρ$, assuming
$\classP ≠ \classNP$.  Given a graph $G$, construct the graph $G^k$ on
$\binom{n}{k}$ vertices where each vertex in $G^k$ represents a set of
$k$ vertices in $G$, and $ST$ is an edge in $G^k$ if $S∪T$ is
\emph{not} an independent set in $G$.  
Let $I$ be an independent set for $G$.
Then the set $I^k$ of all $k$-subsets of $I$ is an independent set in $G^k$
($S∪T ⊆ I$ is an independent set for any $S$ and $T$ in $I^k$).
Conversely, given any independent set $J ⊆ G^k$, its union $\bigcup J$
is an independent set in $G$ (because otherwise there is an edge either
within some element of $J$ or between two elements of $J$).  So any
maximum independent set in $G^k$ will be $I^k$ for some maximum
independent set in $G$.

This amplifies approximation ratios: given an independent set $I$ such
that $\card{I}/\card{OPT} = ρ$, then $\card{I^k}/\card{OPT^k} =
\binom{\card{I}}{k}/\binom{\card{OPT}}{k} ≈ ρ^k$.  If $k$ is constant,
we can compute $G^k$ in polynomial time.  If we can then compute a
$ρ^k$-approximation to the maximum independent set in $G^k$, we can
take the union of its elements to get a $ρ$-approximation to the
maximum independent set in $G$.  By making $k$ sufficiently large,
this shows that approximating the maximum independent set to within
any constant $ε>0$ is \classNP-hard.

There is a stronger version of this argument that uses expander graphs
to get better amplification, which shows that $n^{-δ}$ approximations
are also \classNP-hard for any $δ < 1$.  See \cite[§18.3]{AroraB2007}
for a description of this argument.

\section{Dinur's proof of the \classPCP{} theorem}
\label{section-PCP-Dinur}

Here we give a very brief outline of Dinur's proof of the \classPCP{}
theorem~\cite{Dinur2007}.  This is currently the simplest known proof
of the theorem, although it is still too involved to present in detail
here.  For a more complete description, see §18.5 of
\cite{AroraB2007}, or Dinur's paper, which is pretty accessible.

A \concept{constraint graph} is a graph $G=(V,E)$,
where the vertices in $V$ are interpreted as variables, and each edge
$uv$ in $E$ carries a \concept{constraint} $c_{uv} ⊆ Σ^2$ that
specifies what assignments of values in some \concept{alphabet} $Σ$
are permitted for the endpoints of $uv$.  A \concept{constraint
satisfaction problem} asks for an assignment $σ:V→Σ$ that minimizes
$\UNSAT_σ(G) = \Pr_{uv∈E}[\Tuple{σ(u),σ(v)}∉c_{uv}]$, the
probability that a randomly-chosen constraint is unsatisfied.  The
quantity $\UNSAT(G)$ is defined as minimum value of $\UNSAT_σ(G)$:
this is the smallest proportion of constraints that we must leave
unsatisfied.  In the other direction, the \concept{value} $\val(G)$ of
a constraint satisfaction problem is $1-\UNSAT(G)$: this is the
largest proportion of constraints that we can satisfy.
\footnote{Though Dinur's proof doesn't need this,
we can also consider a \concept{constraint hypergraph},
where each $q$-\concept{hyperedge} is a $q$-tuple $e = v_1 v_2 \dots v_q$
relating $q$ vertices, and a constraint $c_e$ is a subset of $Σ^q$
describing what assignments are permitted to the vertices in $e$.  As
in the graph case, our goal is to minimize $\UNSAT_σ(G)$, which is now
the proportion of hyperedges whose constraints are violated, or
equivalently to maximize $\val_σ(G) = 1- \UNSAT_σ(G)$.
This gives us the $q$-CSP problem: given a $q$-ary constraint
hypergraph, find an assignment $σ$ that maximizes $\val_σ(G)$.
An example of a constraint hypergraph arises from 3SAT.
Given a formula $φ$, construct a graph $G_φ$ in which each vertices
represents a variable, and the constraint on each 
$3$-hyperedge enforces
least one of the literals in some clause is true.
The MAX 3SAT problem asks to find an assignment $σ$ that maximizes the
proportion of satisfied clauses, or $\val_σ(G_φ)$.}

An example of a constraint satisfaction problem is GRAPH
3-COLORABILITY: here $Σ = \Set{r,g,b}$, and the constraints just
require that each edge on the graph we are trying to color (which will
be the same as the constraint graph $G$!) has two different colors on
its endpoints.  If a graph $G$ with $m$ edges has a 3-coloring, then
$\UNSAT(G) = 0$ and $\val(G) = 1$; if $G$ does not, then $\UNSAT(G) ≥
1/m$ and $\val(G) ≤ 1-1/m$.  This gives a $(1-1/m,1)$ gap between
the best value we can get for non-$3$-colorable vs.~$3$-colorable
graphs.  Dinur's proof works by amplifying this gap.

Here is the basic idea:
\begin{enumerate}
    \item We first assume that our input graph is $k$-regular (all
        vertices have the same degree $k$) and an expander (every subset
        $S$ with $\card{S} ≤ m/2$ has $δ\card{S}$ external neighbors
        for some constant $δ>0$).  Dinur shows that even when
        restricted to graphs satisfying these
        assumptions,
        GRAPH 3-COLORABILITY is still \classNP-hard.
    \item We then observe that coloring our original graph $G$ has a gap
        of $(1-1/m,1)$, or that $\UNSAT(G) ≥ 1/m$.  This follows
        immediately from the fact that a bad coloring must include at
        least one monochromatic edge.
    \item To amplify this gap, we apply a two-stage process.
        
        First, we
        construct a new constraint graph $G'$ (that is no longer a graph coloring
        problem) with $n$ vertices, where the constraint graph has an
        edge between any two vertices at distance $2d+1$ or less in $G$,
        the label on each vertex $v$ is a ``neighborhood map'' assigning
        a color of
        every vertex within distance $d$ of $v$, and the constraint on
        each edge $uv$ says that the maps for $u$ and $v$ (a) assign
        the same color to each vertex in the overlap between the two
        neighborhoods, and (b) assigns colors to the endpoints
        of any edge in either neighborhood that are permitted by 
        the constraint on that edge.  Intuitively, this means
        that a bad edge in a coloring of $G$ will turn into
        many bad edges in $G'$, and the expander assumption means that
        many bad edges in $G$ will also turn into many bad edges in
        $G'$.
        In particular, Dinur shows that with
        appropriate tuning this process amplifies the $\UNSAT$ value
        of $G$ by a constant.  Unfortunately, we also blow up the size
        of the alphabet by $Θ(k^d)$.

        So the second part of the amplification knocks the size of the
        alphabet back down to $2$.  This requires replacing each node
        in $G'$ with a set of nodes in a new constraint graph $G''$,
        where the state of the nodes in the set encodes the state of
        the original node, and some coding-theory magic is used to
        preserve the increased gap from the first stage (we lose a
        little bit, but not as much as we gained).

        The net effect of both stages is to take a constraint graph $G$ of size
        $n$ with $\UNSAT(G) ≥ ε$ and turn it into a constraint graph $G''$ of
        size $cn$, for some constant $c$, with $\UNSAT(G'') ≥ 2ε$.
    \item Finally, we repeat this process $Θ(\log m) = Θ(\log n)$ times
        to construct a constraint graph with size $c^{O(\log n)} n = \poly(n)$
        and gap $(1/2,1)$.  Any solution to this constraint graph gives a PCP for GRAPH
        3-COLORABILITY for the original graph $G$.
\end{enumerate}

\section{The Unique Games Conjecture}

The \classPCP{} theorem, assuming $\classP≠\classNP$,
gives us fairly strong inapproximability
results for many classic optimization problems, but in many cases
these are not tight: there is a still a gap between the lower bound
and the best known upper bound.  The 
\index{conjecture!Unique Games}
\concept{Unique Games Conjecture} of
Khot~\cite{Khot2002}, if true, makes many of these bounds tight.

The Unique Games Conjecture was originally formulated in terms of an
interactive proof game with two provers.  In this game, the verifier
$V$ picks a query $q_1$ to ask of prover $P_1$ and a query $q_2$ to
ask of prover $P_2$, and then checks consistency of the prover's
answers $a_1$ and $a_2$.  (The provers cannot communicate, and so answer the queries
independently, although they can coordinate their strategy before
seeing the queries.)  This gives a \index{game!unique}\concept{unique game}
if, for each answer $a_1$ there is exactly one answer $a_2$ that will
cause $V$ to accept.

Equivalently, we can model a unique game as a restricted $2$-CSP: the labels on
the vertices are the answers, and the consistency condition on each
edge is a bijection between the possible labels on each endpoint.
This corresponds to the two-prover game the same way PCPs correspond
to single-prover games, in that a labeling just encodes the answers
given by each prover.

A nice feature of unique games is that they are easy to solve in
polynomial time: pick a label for some vertex, and then use the unique
constraints to deduce the labels for all the other vertices.  So the
problem becomes interesting mostly when we have games for which there
is no exact solution.

For the Unique Games Conjecture, we consider the set of all unique
games with gap $(δ,1-ε)$; that is, the set consisting of the union of
all unique games with approximations with ratio $1-ε$ or better and
all unique games with no approximations with ratio better than $δ$.
The conjecture states that for any $δ$ and $ε$, there exists some
alphabet size $k$ such that it is \classNP-hard to determine of these
two piles a unique game $G$ with this alphabet size lands in.\footnote{Note
that this is not a decision problem, in that the machine $M$ considering
$G$ does not need to do anything sensible if $G$ is in the gap;
instead, it is an example of a \index{problem!promise}\concept{promise
problem} where we have two sets $L_0$ and $L_1$, $M(x)$ must output $i$
when $x∈L_i$, but $L_0∪L_1$ does not necessarily cover all of
$\Set{0,1}^*$.}

Unfortunately, even though the Unique Games Conjecture has many
consequences that are easy to state (for example, the usual
$2$-approximation to MINIMUM VERTEX COVER is optimal, as is the
$0.878$-approximation to MAX CUT of Goemans and
Williamson~\cite{GoemansW1995}), actually proving these consequences
requires fairly sophisticated arguments.  So we won't attempt to do
any of them here, and instead will point the interested reader to
Khot's 2010 survey paper~\cite{Khot2010}, which gives a table of 
bounds known at that time and citations to where they came from.

There is no particular consensus among complexity theorists as to
whether the Unique Games Conjecture is true or not, but it would be
nifty if it were.

\myChapter{Quantum computing}{2025}{}
\label{chapter-quantum-computing}

\indexConcept{quantum computing}{Quantum computing} is a currently
almost-entirely-theoretical branch of randomized algorithms that
attempts to exploit the fact that probabilities at a microscopic scale
arise in a mysterious way from more fundamental
\index{amplitude!probability}\indexConcept{probability
amplitude}{probability amplitudes}, which are complex-valued and can cancel
each other out where probabilities can only add.  In a quantum
computation, we replace random bits with
\indexConcept{quantum bit}{quantum
bits}—\indexConcept{qubit}{qubits} for short—and replace random
updates to the bits with quantum operations.

To explain how this works, we'll start by re-casting our usual
model of a randomized computation to make it look more like the
standard
\index{circuit!quantum}\indexConcept{quantum circuit}{quantum circuits} of
Deutsch~\cite{Deutsch1989}.  We'll then get quantum computation
by replacing all the real-valued probabilities with complex-valued
amplitudes.

\section{Random circuits}
\label{section-random-circuits}

Let's consider a very simple randomized computer whose memory consists
of two bits.  We can describe our knowledge of the state of this
machine using a vector of length $4$, with the coordinates in the
vector giving the probability of states $00$, $01$, $10$, and $11$.
For example, if we know that the initial state is always $00$, we
would have the (column) vector
\begin{align*}
    \begin{bmatrix} 1 \\ 0 \\ 0 \\ 0 \end{bmatrix}.
\end{align*}

Any such \index{vector!state}\concept{state vector} $x$ for the system
must consist of non-negative real values that sum to $1$; this is just
the usual requirement for a discrete probability space.  Operations on
the state consist of taking the old values of one or both bits and
replacing them with new values, possibly involving both randomness and
dependence on the old values.  
The law of total probability applies here, so we can calculate the new
state vector $x'$ from the old state vector $x$ by the rule
\begin{align*}
    x'_{b'_1 b'_2} &= ∑_{b_1 b_2} x_{b_1 b_2} \ProbCond{X_{t+1} = b'_1
b'_2}{X_t = b_1 b_2}.
\end{align*}
These are linear functions of the previous state vector, so we can
summarize the effect of our operation using a transition matrix $A$,
where $x' = Ax$.\footnote{Note that this is the reverse of the convention we
adopted for Markov chains in Chapter~\ref{chapter-Markov-chains}.
There it was convenient to have $P_{ij} = p_{ij} = \ProbCond{X_{t+1} =
j}{X_t = i}$.  Here we defer to the physicists and make the update
operator come in front of its argument, like any other function.}

We imagine that these operations are carried out by feeding the
initial state into some circuit that generates the new state. This
justifies the calling this model of computation a 
\index{circuit!random}\concept{random circuit}. But the actual
implementation might be an ordinary computer than is just flipping
coins. If we can interpret each step of the computation as applying a
transition matrix, the actual implementation doesn't matter.

For example, if we negate the second bit $2/3$ of the time while
leaving the first bit alone, we get the matrix
\begin{align*}
    A &=
    \begin{bmatrix}
        1/3 & 2/3 & 0 & 0 \\
        2/3 & 1/3 & 0 & 0 \\
          0 & 0 & 1/3 & 2/3 \\
          0 & 0 & 2/3 & 1/3
    \end{bmatrix}.
\end{align*}

One way to derive this matrix other than computing each entry directly
is that it is the \index{product!tensor}\concept{tensor product} of
the matrices that represent the operations on the individual bits. 
The idea here is that the tensor product of $A$ and $B$, written $A
⊗ B$, is the matrix $C$ with $C_{ij,k\ell} = A_{ik} B_{j\ell}$.
We're cheating a little bit by allowing the $C$ matrix to have indices
consisting of pairs of indices, one for each of $A$ and $B$; there are
more formal definitions that justify this at the cost of being harder
to understand.

In this particular case, we have
\begin{align*}
    \begin{bmatrix}
        1/3 & 2/3 & 0 & 0 \\
        2/3 & 1/3 & 0 & 0 \\
          0 & 0 & 1/3 & 2/3 \\
          0 & 0 & 2/3 & 1/3
    \end{bmatrix}
    &=
    \begin{bmatrix}
        1 & 0 \\
        0 & 1
    \end{bmatrix}
    ⊗
    \begin{bmatrix}
        1/3 & 2/3 \\
        2/3 & 1/3
    \end{bmatrix}.
\end{align*}
The first matrix in the tensor product gives the update rule for the first
bit (the identity matrix—do nothing), while the second gives the
update rule for the second.

Some operations are not decomposable in this way.  If we swap the
values of the two bits, we get the matrix
\begin{align*}
    S &=
    \begin{bmatrix}
        1 & 0 & 0 & 0 \\
        0 & 0 & 1 & 0 \\
        0 & 1 & 0 & 0 \\
        0 & 0 & 0 & 1
    \end{bmatrix},
\end{align*}
which maps $00$ and $11$ to themselves but maps $01$ to $10$ and vice
versa.

The requirement for all of these matrices is that they be
\index{matrix!stochastic}\indexConcept{stochastic matrix}{stochastic}.
This means that each column has to sum to $1$, or equivalently that
$1A = 1$, where $1$ is the all-ones vector.  This just says that our
operations map probability distributions to probability distributions;
we don't apply an operation and find that the sum of our probabilities
is now $3/2$ or something.  (Proof: If $1A = 1$, then $\abs*{Ax}_1 =
1(Ax) = (1A)x = 1x = \abs*{x}_1$.)

A \index{computation!randomized}\concept{randomized computation} in
this model now consists of a sequence of these stochastic updates to
our random bits, and at the end performing a \concept{measurement} by
looking at what the values of the bits actually are.  If we want to be
mystical about it, we could claim that this measurement collapses 
a probability distribution over states 
into a single unique state, but really we are just opening the
box to see what we got.

For example, we could generate two bias-$2/3$ coin-flips by starting
with $00$ and using the
algorithm flip second, swap bits, flip second, or in matrix terms:
\begin{align*}
    x_{\text{out}}
    &= ASA x_{\text{in}}
    \\&= 
    \begin{bmatrix}
        1/3 & 2/3 & 0 & 0 \\
        2/3 & 1/3 & 0 & 0 \\
          0 & 0 & 1/3 & 2/3 \\
          0 & 0 & 2/3 & 1/3
    \end{bmatrix}
    \begin{bmatrix}
        1 & 0 & 0 & 0 \\
        0 & 0 & 1 & 0 \\
        0 & 1 & 0 & 0 \\
        0 & 0 & 0 & 1
    \end{bmatrix}
    \begin{bmatrix}
        1/3 & 2/3 & 0 & 0 \\
        2/3 & 1/3 & 0 & 0 \\
          0 & 0 & 1/3 & 2/3 \\
          0 & 0 & 2/3 & 1/3
    \end{bmatrix}
    \begin{bmatrix} 1 \\ 0 \\ 0 \\ 0 \end{bmatrix}
    \\&= 
    \begin{bmatrix} 1/9 \\ 2/9 \\ 2/9 \\ 4/9 \end{bmatrix}.
\end{align*}

When we look at the output, we find $11$ with probability $4/9$, as we'd
expect, and similarly for the other values.

\section{Bra-ket notation}

A notational oddity that scares many people away from quantum
mechanics in general and quantum computing in particular is the habit
of practitioners in these fields of writing basis vectors and their
duals using \index{notation!bra-ket}\concept{bra-ket notation}, a kind
of typographical pun invented by the physicist
Paul Dirac~\cite{Dirac1939}.  

This is
based on a traditional way of writing an 
\index{product!inner}\concept{inner product} of two vectors $x$ and
$y$ in \indexConcept{bracket}{bracket form} as $\braket{x}{y}$.  The interpretation of
this is $\braket{x}{y} = x^*y$, where $x^*$ is the
\index{transpose!conjugate}\concept{conjugate transpose} of
$x$.  

For
real-valued $x$ the conjugate transpose is the same as the transpose.  For complex-valued
$x$, each coordinate $x_i = a + bi$ is replaced by its
\index{conjugate!complex}\concept{complex conjugate} $\bar{x}_i = a -
bi$.) Using the conjugate transpose makes $\braket{x}{x}$ equal
$\abs*{x}_2^2$ when $x$ is complex-valued. 

For example, for our vector $x_{\text{in}}$ above that puts all
of its probability on $00$, we have
\begin{align}
    \braket{x_{\text{in}}}{x_{\text{in}}}
    &=
    \begin{bmatrix}
        1 & 0 & 0 & 0
    \end{bmatrix}
    \begin{bmatrix}
        1 \\ 0 \\ 0 \\ 0
    \end{bmatrix}
    = 1.
    \label{eq-braket}
\end{align}

The typographic trick is to split in half both $\braket{x}{y}$ and its
expansion.  For example, we could split \eqref{eq-braket} as
\begin{align*}
    \bra{x_{\text{in}}}
    &=
    \begin{bmatrix}
        1 & 0 & 0 & 0
    \end{bmatrix}
    &
    \ket{x_{\text{in}} }
    &=
    \begin{bmatrix}
        1 \\ 0 \\ 0 \\ 0
    \end{bmatrix}.
\end{align*}

In general, wherever we used to have a bracket $\braket{x}{y}$, we now
have a \concept{bra} $\bra{x}$ and a \concept{ket} $\ket{y}$.  These
are just row vectors and column vectors, and
$\bra{x}$ is always the conjugate transpose of $\ket{x}$.

The second trick in bra-ket notation is to make the contents of the
bra or ket an arbitrary name.  For kets, this will usually describe
some state.  As an example,
we might write $x_{\text{in}}$ as $\ket{00}$ to indicate
that it's the basis vector that puts all of its weight on the state
$00$.  For bras, this is the linear operator that returns
$1$ when applied
to the given state and $0$ when applied to any orthogonal state.  So
$\bra{00} \ket{00} = \braket{00}{00} = 1$ but $\bra{00} \ket{01} =
\braket{00}{01} = 0$.

\subsection{States as kets}

This notation is useful for the same reason that variable names are
useful.  It's much easier to remember that $\ket{01}$ refers to the
distribution assigning probability $1$ to state $01$ than it is to
remember that
$\transpose{\begin{bmatrix}
    0 & 1 & 0 & 0
\end{bmatrix}}$ does.

Other vectors can be expressed using a linear combination of kets.
For example, we can write
\begin{align*}
    x_{\text{out}}
    &= 
    \frac{1}{9} \ket{00}
    + \frac{2}{9} \ket{01}
    + \frac{2}{9} \ket{10}
    + \frac{4}{9} \ket{11}.
\end{align*}
This is not as compact as just writing out the vector as a matrix,
but it has the advantage of clearly
labeling what states the probabilities apply to.

\subsection{Composition of kets}

We'll interpret multiplication of kets as tensor product.
For example:
\begin{align*}
    \ket{0}\ket{1}
    &=
    \begin{bmatrix}
        1 \\ 0
    \end{bmatrix}
    ⊗
    \begin{bmatrix}
        0 \\ 1
    \end{bmatrix}
    =
    \begin{bmatrix}
        0 \\ 1 \\ 0 \\ 0
    \end{bmatrix}
    = \ket{01}.
\end{align*}

This gives a way of composing states together. The general rule is
$\ket{x} \ket{y} = \ket{xy}$. Because this is tensor product
underneath, this is a linear operation, and it associates with other
products in the obvious way.

\subsection{Operators as sums of kets times bras}

A similar trick can be used to express operators, like the swap
operator $S$.  We can represent $S$ as a combination of maps from
specific states to specific other states.  For example, the operator
\begin{align*}
    \ket{01}\bra{10}
    &=
    \begin{bmatrix}
        0 \\ 1 \\ 0 \\ 0
    \end{bmatrix}
    \begin{bmatrix}
        0 & 0 & 1 & 0
    \end{bmatrix}
    =
    \begin{bmatrix}
        0 & 0 & 0 & 0 \\
        0 & 0 & 1 & 0 \\
        0 & 0 & 0 & 0 \\
        0 & 0 & 0 & 0
    \end{bmatrix}
\end{align*}
maps $\ket{10}$ to $\ket{01}$ (Proof: $\ket{01} \bra{10} \ket{10} =
\ket{01} \braket{10}{10} = \ket{01}$) and sends all other states to
$0$.  Add up four of these mappings to get
\begin{align*}
    S &=
    \ket{00}\bra{00}
    + \ket{10}\bra{01}
    + \ket{01}\bra{10}
    + \ket{11}\bra{11}
    =
    \begin{bmatrix}
        1 & 0 & 0 & 0 \\
        0 & 0 & 1 & 0 \\
        0 & 1 & 0 & 0 \\
        0 & 0 & 0 & 1
    \end{bmatrix}.
\end{align*}
Here the bra-ket notation both labels what we are doing and
saves writing a lot of zeros. 

The intuition is that just like a ket
represents a state, a bra represents a test for being in that state.
So something like $\ket{01} \bra{10}$ tests if we are in the $10$
state, and if so, sends us to the $01$ state.

\section{Quantum circuits}

So how do we turn our random circuits into 
\index{circuit!quantum}\index{quantum circuit}\conceptFormat{quantum circuits}?

The first step is to replace our random bits with quantum bits
(qubits).

For a single random bit, the state vector represents a probability
distribution
\begin{align*}
    p_0 \ket{0} + p_1 \ket{1},
\end{align*}
where $p_0$ and $p_1$ are non-negative real numbers with $p_0+p_1=1$.

For a single qubit, the state vector represents amplitudes
\begin{align*}
    a_0 \ket{0} + a_1 \ket{1},
\end{align*}
where $a_0$ and $a_1$ are complex numbers with $\abs*{a_0}^2 +
\abs*{a_1}^2 = 1$.\footnote{The \concept{absolute value},
    \concept{norm}, or \concept{magnitude} $\abs*{a+bi}$ of a
    complex number is given by $√{a^2 + b^2}$.  When $b = 0$, this
    is the same as the absolute value for the corresponding real number.  For any
complex number $x$, the norm can also be written as $√{\bar{x}x}$,
where $\bar{x}$ is the complex conjugate of $x$.  This is because
$√{(a+bi)(a-bi)} = √{a^2 - (bi)^2} = √{a^2 +
b^2}$.  The appearance of the complex conjugate here explains why we
define $\braket{x}{y} = x^{*} y$; the conjugate transpose means that
for $\braket{x}{x}$, when we multiply $x^{*}_i$ by $x_i$ we are
computing a squared norm.}
The reason for this restriction on amplitudes is that if we measure a
qubit, we will see state $0$ with probability $\abs*{a_0}^2$ and state
$1$ with probability $\abs*{a_1}^2$.  Unlike with random bits, these
probabilities are not mere expressions of our ignorance but arise
through a still somewhat mysterious process from the more
fundamental amplitudes.\footnote{In the old days
of ``shut up and calculate,'' this process was thought to involve the unexplained
power of a conscious observer to collapse a superposition into a classical
state.  Nowadays the most favored explanation involves
\concept{decoherence}, the difficulty of maintaining superpositions in
systems that are interacting with large, warm objects with lots of
thermodynamic degrees of freedom, a category that includes most measuring
instruments and brains. The
decoherence explanation is particularly useful for explaining why
real-world quantum computers have a hard time keeping their qubits
mixed even when nobody is looking at them.  Decoherence by itself does
not explain \emph{which} basis states a system collapses to.  Since
bases in linear algebra are pretty much arbitrary, it would seem that
we could end up running into a physics version of Goodman's grue-bleen
paradox~\cite{Goodman1983}, but there are apparently ways of dealing with this too
using a mechanism called \concept{einselection}~\cite{Zurek2003} that
favors classical states over weird ones. Since all of this is (a) well beyond my
own
limited comprehension of quantum mechanics and (b) irrelevant to the
theoretical model we are using, these issues will not be discussed
further.}

With multiple bits, we get amplitudes for all combinations of the
bits, e.g.
\begin{align*}
    \frac{1}{2} \left(\ket{00} + \ket{01} + \ket{10} + \ket{11}\right)
\end{align*}
gives a state vector in which each possible measurement will be
observed with equal probability $\left(\frac{1}{2}\right)^2 =
\frac{1}{4}$.  We could also write this state vector as
\begin{align*}
    \begin{bmatrix}
        1/2 \\
        1/2 \\
        1/2 \\
        1/2
    \end{bmatrix}
\end{align*}
if we aren't worried about forgetting which coordinate corresponds to
which state.

\subsection{Quantum operations}

In the random circuit model, at each step we pick a small number of
random bits, apply a stochastic transformation to them, and replace
their values with the results.  In the quantum circuit model, we do
the same thing, but now our transformations must have the property of
being \index{transformation!unitary}\indexConcept{unitary
transformation}{unitary}.  Just as a stochastic matrix preserves the
property that the probabilities in a state vector sum to $1$, a 
\index{matrix!unitary}
\concept{unitary matrix} preserves the property that the squared norms
of the amplitudes in a state vector sum to $1$.

Formally, a square, complex matrix $A$ is unitary if it preserves
inner products: $\braket{Ax}{Ay} = \braket{x}{y}$ for all $x$ and $y$.
Alternatively, $A$ is unitary if
$A^{*}A = AA^{*}
= I$, where $A^{*}$ is the conjugate transpose of $A$,
because if this holds, $\braket{Ax}{Ay} = (Ax)^*(Ay) = x^*A^*Ay = x^*Iy = x^*y =
\braket{x}{y}$.
Yet another way to state this is that the columns of $A$ form an
orthonormal basis: this means that
$\braket{A_i}{A_j} = 0$ if $i ≠ j$ and $1$ if $i=j$, which is just a
more verbose way to say
$A^{*}A = I$.  The same thing also works if we
consider rows instead of columns.

The rule then is: at each step, we can operate on some constant number
of qubits by applying a unitary transformation to them.  In principle,
this could be \emph{any} unitary transformation, but some particular
transformations show up often in actual quantum
algorithms.\footnote{Deutsch's original paper~\cite{Deutsch1989} shows
    that repeated applications of 
    single-qubit rotations and the CNOT operation (described in
    §\ref{section-quantum-implementations-of-classical-operations})
are enough to approximate any unitary transformation.}

The simplest unitary transformations are permutations on states (like
the operator that swaps two qubits), and rotations of a single state.
One particularly important rotation is the 
\index{operator!Hadamard}\concept{Hadamard operator}
\begin{align*}
    H &=
    √{\frac{1}{2}}
    \begin{bmatrix}
        1 & 1 \\
        1 & -1
    \end{bmatrix}.
\end{align*}
This maps $\ket{0}$ to the superposition
$√{\frac{1}{2}} \ket{0} + √{\frac{1}{2}} \ket{1}$; since this
superposition collapses to either $\ket{0}$ or $\ket{1}$ with
probability $1/2$, the state resulting from $H \ket{0}$ is the
quantum-computing equivalent of a fair coin-flip.  Note that $H\ket{1}
= √{\frac{1}{2}} \ket{0} - √{\frac{1}{2}} \ket{1} ≠
H\ket{0}$.  Even though both yield the same probabilities, these two
superpositions have different \indexConcept{phase}{phases} and may
behave differently when operated on further.  That $H \ket{0}$ and $H
\ket{1}$ are different is necessary, and indeed a similar outcome
occurs for any quantum operation: all
quantum operations are \concept{reversible}, because any unitary
matrix $U$ has an inverse $U^{*}$.

If we apply $H$ in parallel to all the qubits in our system, we get
the $n$-fold tensor product $H^{⊗ n}$, which (if we take our
bit-vector indices as integers $0 \dots N-1 = 2^n-1$ represented in binary)
maps $\ket{0}$ to $√{\frac{1}{N}} ∑_{i=0}^{N-1} \ket{i}$.  So
$n$ applications of $H$ effectively scramble a deterministic initial
state into a uniform distribution across all states.  We'll see this
scrambling operation again when we look at Grover's algorithm in
§\ref{section-Grovers-algorithm}.

\subsection{Quantum implementations of classical operations}
\label{section-quantum-implementations-of-classical-operations}

One issue that comes up with trying to implement classical algorithms
in the quantum-circuit model is that classical operation are generally
not reversible: if I execution $x \leftarrow x ∧ y$, it may not
be possible to reconstruct the old state of $x$.  So I can't implement
AND directly as a quantum operation.

The solution is to use more sophisticated reversible operations from
which standard classical operations can be extracted as a special
case.  A simple example is 
the \concept{controlled NOT} or \concept{CNOT} operator, which
computes the mapping $(x,y) \mapsto (x,x⊕ y)$.  This
corresponds to the matrix (over the basis $\ket{00}, \ket{01},
\ket{10}, \ket{11}$)
\begin{align*}
    \begin{bmatrix}
        1 & 0 & 0 & 0 \\
        0 & 1 & 0 & 0 \\
        0 & 0 & 0 & 1 \\
        0 & 0 & 1 & 0
    \end{bmatrix},
\end{align*}
which is clearly unitary (the rows are just the standard basis
vectors).
We could also write this more compactly as 
$\ket{00} \bra{00}
+\ket{01} \bra{01}
+\ket{11} \bra{10}
+\ket{10} \bra{11}$.

The CNOT operator gives us XOR, but for more destructive operations we
need to use more qubits, possibly including junk qubits that we won't look
at again but that are necessary to preserve reversibility.
The \index{gate!Toffoli}\concept{Toffoli gate} or \concept{controlled
controlled NOT} gate (\concept{CCNOT}) is a 3-qubit gate that was
originally designed to show that classical computation could be
performed reversibly~\cite{Toffoli1980}.  It implements the mapping 
$(x,y,z) \mapsto (x,y, (x ∧ y) ⊕ z)$, which corresponds to
the $8× 8$ matrix
\begin{align*}
    \begin{bmatrix}
        1 & 0 & 0 & 0 & 0 & 0 & 0 & 0 \\
        0 & 1 & 0 & 0 & 0 & 0 & 0 & 0 \\
        0 & 0 & 1 & 0 & 0 & 0 & 0 & 0 \\
        0 & 0 & 0 & 1 & 0 & 0 & 0 & 0 \\
        0 & 0 & 0 & 0 & 1 & 0 & 0 & 0 \\
        0 & 0 & 0 & 0 & 0 & 1 & 0 & 0 \\
        0 & 0 & 0 & 0 & 0 & 0 & 0 & 1 \\
        0 & 0 & 0 & 0 & 0 & 0 & 1 & 0
    \end{bmatrix}.
\end{align*}

By throwing in some extra qubits we don't care about, Toffoli gates
can implement basic operations like NAND ($(x,y,1) \mapsto
(x,y,\neg(x ∧ y))$), NOT ($(x,1,1) \mapsto (x,1,\neg x)$), and
fan-out ($(x,1,0) \mapsto (x,1,x)$).\footnote{In the case of fan-out,
    this only works with perfect accuracy for classical bits and not superpositions, which
run into something called the 
\index{theorem!no-cloning}
\concept{no-cloning theorem}.
For example, applying CCNOT to 
$\frac{1}{√{2}}\ket{010} + \frac{1}{√{2}}\ket{110}$
yields
$\frac{1}{√{2}}\ket{010} + \frac{1}{√{2}}\ket{111}$.
This works, sort of, but the problem is that the first and last bits
are still entangled, meaning we can't operate on them
independently.  This is actually not all that different from what
happens in the probabilistic case (if I make a copy of a random
variable $X$, it's correlated with the original $X$), but it has good or bad consequences depending on whether you
want to prevent people from stealing your information undetected or
run two copies of a quantum circuit on independent replicas of a
single superposition.}
This
gives a sufficient basis for implementing all classical circuits.

What we get at the end of applying this process to a classical circuit
implementing some Boolean function $f$ is a quantum circuit
implementing a unitary operator $f$ that maps $\ket{x,y}$ to
$\ket{x,y⊕f(x)}$ (possibly with some additional auxiliary qubits
getting passed through as well). If we want to treat this operator as
an input to a later computation (say, one that tests some condition on
all of $f$), there is a further simplification step that we can do
that will make $f$ much easier to work with. This replaces the XOR
$y⊕f(x)$ with a change in phase on $\ket{x}$ itself.

\subsection{Phase representation of a function}

The idea of a phase representation of a Boolean function $f$ is to use
an operator $U_f = ∑_x (-1)^{f(x)} \ket{x}\bra{x}$. This has the
effect of mapping each $\ket{x}$ to $-\ket{x}$ when $f(x) = 1$, and
passing it through intact when $f(x)=0$. The result is a diagonal
matrix whose diagonal looks like a truth table for $f$ expressed as
$±1$ values. This has the effect of mapping each $\ket{x}$ to
$-\ket{x}$ when $f(x) = 1$, and passing it through intact when
$f(x)=0$. The result is a diagonal matrix $U_f$ whose diagonal looks like a
truth table for $f$ expressed as $±1$ values, as in this matrix for
the XOR function $f(x_0, x_1) = x_0 ⊕ x_1$:
        \begin{align*}
            \begin{bmatrix}
                1 & 0 & 0 & 0 \\
                0 & -1 & 0 & 0 \\
                0 & 0 & -1 & 0 \\
                0 & 0 & 0  & 1 \\
            \end{bmatrix}.
        \end{align*}
        
        Because the amplitude of each $\ket{x}$ is unchanged, this
        representation is not so useful for
        observing the value of $f(x)$ directly.
        So we will generally be using this representation of $f$ in
        the context of some larger quantum algorithm that exploits the
        phase changes to get useful cancellations for some values of $x$.

The question remains how we get the phase representation $U_f$, given
that our standard construction of a classical circuit gives the
controlled-not representation $f: \ket{xy} ↦ \ket{x,y⊕f(x)}$.
The trick here is to feed this $f$ 
the particular value $y = H \ket{1} = \frac{1}{√{2}}\parens*{\ket{0}-\ket{1}}$
This gives
\begin{align*}
    f \ket{xy}
    &= \frac{1}{√{2}}\parens*{f\ket{x0} - f\ket{x1}}
    \\&= \frac{1}{√{2}}\parens*{\ket{x,f(x)} - \ket{x,¬f(x)}}
    \\&= \frac{(-1)^{f(x)}}{√{2}}\parens*{\ket{x0} - \ket{x1}}.
    \\&= (-1)^{f(x)} \ket{x} \frac{1}{√{2}}\parens*{\ket{0} - \ket{1}}.
    \\&= (U_f \ket{x}) \ket{y}.
\end{align*}

By leaving the auxiliary $y$ alone in subsequent computation, we
effectively end up with just $U_f \ket{x}$.

\subsection{Practical issues (which we will ignore)}

The big practical question is whether any of these operations—or
even non-trivial numbers of independently manipulable qubits—can be
implemented in a real, physical system.  As theorists, we can ignore
these issues, but in real life they are what would make quantum computing
science instead of science fiction.\footnote{Current technology, sadly,
still puts quantum computing mostly in the science fiction category.}

\subsection{Quantum computations}

Putting everything together, a quantum computation consists of three
stages:
\begin{enumerate}
    \item We start with a collection of qubits in some known state
        $x_0$
        (e.g., $\ket{000\dots 0}$).
    \item We apply a sequence of unitary operators $A_1, A_2, \dots
        A_m$ to our qubits. Some of these unitary operators may be
        oracles $U_f$ or $U_w$ representing the input we actually care
        about.
    \item We take a measurement of the final superposition
        $A_m A_{m-1} \dots A_1 x_0$ that collapses it into a single
        state, with probability equal to the square of the amplitude
        of that state.
\end{enumerate}

Our goal is for this final state to tell us what we want to know, with
reasonably high probability.

\section{Deutsch's algorithm}

We now have enough machinery to describe a real quantum algorithm.
Known as \index{algorithm!Deutsch's}{Deutsch's algorithm}, this
computes $f(0) ⊕ f(1)$ while evaluating $f$
once~\cite{Deutsch1989}.
The trick, of course, is that $f$ is applied to a superposition.

Assumption: $f$ is implemented as a unitary operator $U_f$ that
that maps $\ket{x}$ to $(-1)^{f(x)} \ket{x}$.  To compute $f(0) ⊕
f(1)$, evaluate
\begin{align*}
    H U_f H \ket{0}
    &= √{\frac{1}{2}} H U_f \left(\ket{0} + \ket{1}\right)
  \\&= √{\frac{1}{2}} H \left((-1)^{f(0)} \ket{0} +
    (-1)^{f(1)} \ket{1}\right)
  \\&= \frac{1}{2} \left(
    \left((-1)^{f(0)} + (-1)^{f(1)}\right) \ket{0}
   + \left((-1)^{f(0)} - (-1)^{f(1)}\right) \ket{1}\right).
\end{align*}

Suppose now that $f(0) = f(1) = b$.  Then the $\ket{1}$ terms cancel
out and we are left with
\begin{align*}
    \frac{1}{2} \left(2⋅(-1)^b \ket{0}\right)
    &= (-1)^b \ket{0}.
\end{align*}
This puts all the weight on $\ket{0}$, so when we take our measurement
at the end, we'll see $0$.

Alternatively, if $f(0) = b ≠ f(1)$, it's the $\ket{0}$ terms that
cancel out, leaving $(-1)^b \ket{1}$. The phase depends on $b$,
but we don't care about the phase. The important thing is that if we
measure the qubit, we always see $1$.

The result in either case is that with probability $1$, we determine
the value of $f(0) ⊕ f(1)$, after evaluating $f$ once (albeit on
a superposition of quantum states).

This is kind of a silly example, because the huge costs involved in
building our quantum computer almost certainly swamp the factor-of-2
improvement we got in the number of calls to $f$.  But a
generalization of this trick, known as the Deutsch-Josza
algorithm~\cite{DeutschJ1992}, solves the much harder (although still
a bit contrived-looking) problem of
distinguishing a constant Boolean function on $n$ bits from a function
that outputs one for
exactly half of its inputs.  No
deterministic algorithm can solve this problem without computing at
least $2^n/2 + 1$ values of $f$, giving an exponential
speed-up.

The speed-up compared to a randomized algorithm that works with
probability $1-ε$ is less impressive.  With randomization,
we only need to look at
$O(\log 1/ε)$ values of $f$ to see both a $0$ and a $1$ in the
non-constant case.
But even here,
the Deutsch-Josza
algorithm does have the advantage of giving the correct answer with
probability $1$. If we make the same demand of a randomized algorithm,
it does no better than a deterministic algorithm, at least in the
constant-function case.

\section{Grover's algorithm}
\label{section-Grovers-algorithm}

Grover's algorithm~\cite{Grover1996} is one of two main exemplars for
the astonishing power of quantum computers.\footnote{The other is
    Shor's algorithm~\cite{Shor1997}, which allows a quantum computer
to factor $n$-bit integers in time polynomial in $n$.  Sadly, Shor's
algorithm is a bit too complicated to talk about here.}
The idea of Grover's algorithm is that if we have a function $f$ on
$N = 2^n$ possible inputs whose
value is $1$ for exactly one possible input $w$, we can find this $w$
with high probability using $O(√{N})$ quantum evaluations of $f$.
As with Deutsch's algorithm, we assume that $f$ is encoded as an
operator (conventionally written $U_w$)
that maps each $\ket{x}$ to $(-1)^{f(x)} \ket{x}$.

The basic outline of the algorithm:
\begin{enumerate}
    \item Start in the superposition $\ket{s} = √{\frac{1}{N}} ∑_{x}
        \ket{x} = H^{⊗ n} \ket{0}$.
    \item Alternate between applying the 
        \index{operator!Grover diffusion}\concept{Grover diffusion
        operator} $D = 2 \ket{s} \bra{s} - I$ and the $f$ operator
        $U_w = 2 \ket{w} \bra{w} - I$.  Do this $O(√{n})$ times (the exact number of
        iterations is important and will be calculated below).
    \item Take a measurement of the state.  It will be $w$ with high
        probability.
\end{enumerate}

Making this work requires showing that (a) we can generate the original
superposition $\ket{s}$, (b) we can implement $D$ efficiently
using unitary operations on a constant number of qubits each,
and (c) we actually get $w$ at the end of this process.

\subsection{Initial superposition}

To get the initial superposition, start with $\ket{0^n}$ and apply the
Hadamard transform to each bit individually; this gives
$√{\frac{1}{N}} ∑_x \ket{x}$ as claimed.

\subsection{The Grover diffusion operator}

We have the definition $D = 2 \ket{s} \bra{s} - I$.  

Before we try to
implement this, let's start by checking that it is
in fact unitary.  Compute
\begin{align*}
DD^{*} 
&= (2 \ket{s} \bra{s} - I)^2 
\\&= 
4 \ket{s} \bra{s} \ket{s} \bra{s} 
- 4 \ket{s} \bra{s}
+ I^2
\\&= 
4 \ket{s} \bra{s} 
- 4 \ket{s} \bra{s}
+ I
\\&= I.
\end{align*}
Here we use the fact that 
$\ket{s} \bra{s} \ket{s} \bra{s} = 
\ket{s} \braket{s}{s} \bra{s} = 
\ket{s} (1) \bra{s} = \ket{s} \bra{s}$.

Now let's look at implementation.
Recall that $\ket{s} = H^{⊗ n} \ket{0^n}$, where $H^{⊗ n}$
is the result of applying $H$ to each of the $n$ bits individually.
We also have that $H^{*} = H$ and $HH^{*} = I$, from which $H^{⊗
n} H^{⊗ n} = I$ as well.

So we can expand
\begin{align*}
    D
    &=
    2 \ket{s} \bra{s} - I
    \\&= 2 H^{⊗ n} \ket{0^n} (H^{⊗ n} \ket{0^n})^* - I
  \\&= 2 H^{⊗ n} \ket{0^n} \bra{0^n} H^{⊗ n} - I
  \\&= H^{⊗ n} \left(2 \ket{0^n} \bra{0^n} - I\right) H^{⊗
n}.
\end{align*}

The two copies of $H^{⊗ n}$ involve applying $H$ to each of the
$n$ bits, which we can do.  The operator in the middle, $2 \ket{0^n}
\bra{0^n} - I$, maps $\ket{0^n}$ to $\ket{0^n}$ and maps all other
basis vectors $\ket{x}$ to $-\ket{x}$. By our usual output convention
$\ket{x} ↦ (-1)^{f(x)} \ket{x}$, this is
an OR of all the qubits, which we can implement using our tools for classical
computations. So the entire operator $D$ can be implemented using
$O(n)$ qubit operations, most of which can be done in parallel.

\subsection{Effect of the iteration}

To see what happens when we apply $U_w D$, it helps to represent the
state in terms of a particular two-dimensional basis.  The idea here
is that the initial state $\ket{s}$ and the operation $U_w D$ are
symmetric with respect to any basis vectors $\ket{x}$ that aren't $\ket{w}$, so
instead of tracking all of these non-$\ket{w}$ vectors separately, we will
represent all of them by a single composite vector
\begin{align*}
    \ket{u} &= √{\frac{1}{N-1}} ∑_{x ≠ w} \ket{x}.
\end{align*}

The coefficient $√{\frac{1}{N-1}}$ is chosen to make
$\braket{u}{u} = 1$.  As always, we like our vectors to have length
$1$.

Using $\ket{u}$, we can represent
\begin{align}
    \ket{s} = √{\frac{1}{N}} \ket{w} + √{\frac{N-1}{N}} \ket{u}.
    \label{eq-expand-s-into-w-and-u}
\end{align}

A straightforward calculation shows that this indeed puts
$√{\frac{1}{N}}$ amplitude on each $\ket{x}$.

Now we're going to bring in some trigonometry.  Let $\theta =
\sin^{-1} √{\frac{1}{N}}$, so that $\sin \theta =
√{\frac{1}{N}}$ and
$\cos \theta = √{1 - \sin^2 \theta} = √{\frac{N-1}{N}}$.
We can then rewrite \eqref{eq-expand-s-into-w-and-u} as
\begin{align}
    \ket{s} = (\sin \theta) \ket{w} + (\cos \theta) \ket{u}.
    \label{eq-expand-s-into-w-and-u-with-theta}
\end{align}

Let's look at what happens if we expand $D$ using
\eqref{eq-expand-s-into-w-and-u-with-theta}:
\begin{align*}
D
&=  2 \ket{s}\bra{s} - I
\\&= 
     2\left((\sin \theta) \ket{w} + (\cos \theta) \ket{u}\right)
    \left((\sin \theta) \bra{w} + (\cos \theta) \bra{u}\right) - I
    \\&=
    (2 \sin^2 \theta - 1) \ket{w} \bra{w}
    + (2 \sin \theta \cos \theta) \ket{w} \bra{u}
    + (2 \sin \theta \cos \theta) \ket{u} \bra{w}
    + (2 \cos^2 \theta -1) \ket{u} \bra{u}
    \\&=
    (-\cos 2\theta) \ket{w} \bra{w}
    + (\sin 2\theta) \ket{w} \bra{u}
    + (\sin 2\theta) \ket{u} \bra{w}
    + (\cos 2\theta) \ket{u} \bra{u}
    \\&=
    \begin{bmatrix}
        -\cos 2\theta  &  \sin 2\theta \\
        \sin 2\theta   &  \cos 2\theta
    \end{bmatrix},
\end{align*}
where the matrix is over the basis $(\ket{w}, \ket{u})$.

Multiplying by $U_w$ negates all the $\ket{w}$ coefficients.  So we
get
\begin{align}
U_w D
&=
\begin{bmatrix}
    -1 & 0 \\
     0 & 1 
\end{bmatrix}
\begin{bmatrix}
        -\cos 2\theta  &  \sin 2\theta \\
        \sin 2\theta   &  \cos 2\theta
\end{bmatrix}
\nonumber
\\&=
\begin{bmatrix}
        \cos 2\theta  &  -\sin 2\theta \\
        \sin 2\theta   &  \cos 2\theta
\end{bmatrix}.
\label{eq-Grovers-algorithm-rotation-matrix}
\end{align}

Aficionados of computer graphics, robotics, or just matrix algebra in
general may recognize \eqref{eq-Grovers-algorithm-rotation-matrix} as
the matrix that rotates two-dimensional vectors by $2\theta$.  Since
we started with $\ket{s}$ at an angle of $\theta$, after $t$
applications of this matrix we will be at an angle of $(2t+1) \theta$,
or in state
\begin{align*}
    \left(\sin (2t+1) \theta\right) \ket{w}
    + \left(\cos (2t+1) \theta\right) \ket{u}.
\end{align*}

Ideally, we pick $t$ so that $(2t+1) \theta = π/2$, which would put
all of the amplitude on $\ket{w}$.  Because $t$ is an integer, we
can't do this exactly, but setting $t = \floor{\frac{π/2\theta -
1}{2}}$ will get us somewhere between $π/2 - 2\theta$ and $π/2$.
Since $\theta \approx √{\frac{1}{N}}$, this gives us a probability
of seeing $\ket{w}$ in our final measurement of $1-O(√{1/N})$
after $O(√{N})$ iterations of $U_w D$.

Sadly, this is as good as it gets.  A lower bound of
Bennet~\etal~\cite{BennettBBV1997} shows that \emph{any} quantum
algorithm using $U_w$ as the representation for $f$ must apply $U_w$
at least $Ω(√{N})$ times to find $w$.  So we get a quadratic speedup
but not the exponential speedup we'd need to solve $\classNP$-complete
problems directly.

\myChapter{Randomized distributed algorithms}{2025}{}
\label{chapter-randomized-distributed-algorithms}

A \index{algorithm!distributed}\concept{distributed algorithm} is one
that runs on multiple machines that communicate with each other in
some way, typically via \concept{message-passing} (an abstraction of
packet-based computer networks) or \concept{shared memory} (an
abstraction of multi-core CPUs and systems with a common memory bus).
What generally distinguishes \concept{distributed computing} from the
closely-related idea of \concept{parallel computing} is that we expect
a lot of nondeterminism in a distributed algorithm: events take place
at unpredictable times, processes may crash, and in particular bad
cases we may have \concept{Byzantine processes} that work deliberately
against the algorithm. We are not going to assume Byzantine processes;
instead, we'll look at a particular model called \concept{wait-free
shared memory}.

This hostile nondeterminism is modeled by an \concept{adversary} that
controls scheduling and failures.  For a shared-memory model, we have
a collection of processes that each have a \concept{pending operation}
that is either a read or write to some \concept{register}.  The
adversary chooses which of these pending operations happens next (so
that concurrency between processes is modeled by interleaving their
operations).  Unlike the adversary that supplies the worst-case input
to a traditional algorithm before it executes, an adversary scheduler
might be able to observe the execution of a distributed algorithm in
progress and adjust its choices in response.  If it has full
knowledge of the system (including internal states of processes), we
call it an \index{adversary!adaptive}\concept{adaptive adversary}; at
the other extreme is an
\index{adversary!oblivious}\concept{oblivious adversary} 
that chooses the schedule of which
processes execute operations at which times in advance.
This is similar to the distinction between an adaptive and oblivious
adversary required for the analysis of randomized data structures (see
§\ref{section-treaps-oblivious}.

For either adversary, what makes the system \concept{wait-free} is
that any process that gets to take an unbounded number of steps has to
finish whatever it is doing no matter how the other processes are
scheduled. In particular, this means that no process can wait for any
of the others, because they might never be scheduled. The processes
that do run all run the algorithm correctly, so the model is
equivalent to assuming up to $n-1$ of the processes may
\concept{crash} but that none are Byzantine.

As with traditional algorithms, distributed algorithms can use
randomization to make themselves less predictable.  
The assumption is that even if the adversary is adaptive, it can't
predict the outcome of future coin-flips.  For some problems, avoiding
such predictability is provably necessary.

Distributed computing is a huge field, and we aren't going to be able
to cover much of it in the limited space we have here.  So we are
going to concentrate on a few simple problems that give some of the
flavor or randomized distributed algorithms (and that lead to problems
that are still open).

\section{Consensus}
\label{section-consensus}

The \concept{consensus} problem is to get a collection of $n$
processes to agree on a value.  The requirements are that all the
processes that don't crash finish the protocol (with probability $1$
for randomized protocols)
(\concept{termination}), that they all output the same value
(\concept{agreement}), and that this value was an input to one of the
processes (\concept{validity})—this last condition excludes
protocols that always output the same value no matter what happens
during the execution, and makes consensus useful for choosing among
values generated by some other process.

There are many versions of consensus. The original problem as proposed
by Pease, Shostak, and Lamport~\cite{PeaseSL1980} assumes Byzantine
processes in a synchronous message-passing system. Here scheduling is
entirely predictable, and the 
obstacle to agreement is dissension sown by the lying Byzantine
processes. We will instead consider wait-free shared-memory consensus,
where scheduling is unpredictable but the
the processes and memory are trustworthy. Even in this case, the
unpredictable scheduling makes solving the problem deterministically
impossible.

\subsection{Impossibility of deterministic algorithms}

The FLP impossibility result~\cite{FischerLP1985} and its extension to
shared memory~\cite{LouiA1987} show that consensus is impossible to
solve in an asynchronus model with even one crash failure.
The general proof is a bit involved, but 
there is a simple intuitive
proof of this result for the wait-free case where up to $n-1$ of the processes may crash.

Crash all but two processes, one with input $0$ and one with input
$1$.  Define the \concept{preference} of a process as the value it
will decided in a solo execution (this is well-defined because the
processes are deterministic).  In the initial state, the process with
input $b$ has preference $b$ because of validity—it doesn't know the
other process exists.  But before it returns $b$, it has to cause the
other process to change its preference (once it leaves the other
process is running alone).  It can do so only by sending a message or
writing a register.  When it is about to do this, stop it and run the
other process until it does the same thing. 

Either the other process
somehow neutralizes the effect of the delayed message/write during
this time, or it doesn't. In the first case, restart the first process
(which still has its original preference and now must do something
else to make the other process change).  In the second case, deliver
both operations and let the processes exchange preferences.  The
resulting execution looks very much like when two people are trying to
pass each other in a hallway and oscillate back-and-forth—but since
our process's have their timing controlled by an adversary, the
natural damping or randomness that occurs in humans doesn't ever
resolve the situation.

We can avoid this impossibility result with randomness. If either of
the process decides to choose a new preference at random, then there
is a 50\% chance it will end up agreeing with the other one, no matter
what the other one is doing. Assuming we can detect this agreement,
this solves consensus in constant expected time for two processes.

Unfortunately,
this flip-if-confused approach does not generalize very well to $n$ processes. The
first known randomized shared-memory algorithm for consensus, due to
Abrahamson~\cite{Abrahamson1988}, used essentially this idea.
But because it required waiting for a long run of identical coin-flips
across all processes, it took $O\parens*{2^{n^2}}$ steps on average to
finish. In a little while, we will see how to do better.

\section{Leader election}

In a shared-memory system, \concept{leader election} is a very similar
problem to consensus. Here we want one of the processes to decide that
it is the leader, and the rest to decide that they are followers. The
difference from consensus is that we don't require that the followers
learn the identity of the leader. This means that followers can drop
out before the algorithm determines the leader, which may allow a faster algorithm.

Note however that if there are only two processes, the losing process
knows who the winner is, because there is only one possibility. So the
same argument that shows that we can't do consensus deterministically
with two processes also works for leader election.

\section{How randomness helps}

We've already hinted at the idea of using randomness to shake
processes into agreement. In the current literature on shared-memory consensus and
leader election, this idea shows up in two main forms:
\begin{enumerate}
    \item We can replace the $n$ separate coin-flips of stalled
        processes with a single 
        \index{shared coin!weak}
        \index{coin!weak shared}\concept{weak shared coin}
        protocol.
        This is a protocol that has for each outcome $b∈\Set{0,1}$ a
        minimum probability $δ>0$ that every processes receives $b$ as
        the value of the coin. Because $δ$ will typically be less than
        $1/2$, it is possible that we get disagreements, or that the
        adversary can bias the coin toward a particular value that it
        likes. But with a bit of extra machinery that we will not
        describe here, we can get agreement after 
        $O(1/δ)$ calls to the shared coin on
        average~\cite{AspnesH1990consensus}.
        An example of a weak shared coin protocol that assumes an
        adaptive adversary is given in §\ref{section-Attiya-Censor}.
    \item We can eliminate processes quickly using a \concept{sifter},
        which solves a weak version of leader election that allows for
        multiple winners but guarantees that the expected number of
        winners is small relative to $n$. Repeated application of a
        sifter can quickly knock the number of repeated winners down
        to $1$, which solves leader election (§\ref{section-sifter-leader-election}. With some tinkering, it
        is possible to use some sifters to also solve consensus
        (§\ref{section-sifter-consensus}. In both cases we assume an
        oblivious adversary.
\end{enumerate}

\section{Building a weak shared coin}
\label{section-building-a-weak-shared-coin}
\label{section-Attiya-Censor}

The goal of a weak shared coin is to minimize the influence of the
adversary scheduler. An adaptive adversary can bias the outcome of the
coin by looking at the random values generated by the individual
processes, and withholding values that it doesn't like by delaying any
write operations by those processes. So we would like to find a way to
combine the local coins at the processes together into a single global
coin that minimizes the influence of any particular local coin.
The easiest way to do this is by applying the majority function: each
process repeatedly generates $±1$ values with equal probability and
adds them to a common pile. When the pile includes enough local coins,
we take the sign of the total to get the return value of the global
coin.

The nice thing about this approach is that the behavior of the
processes is symmetric. It doesn't matter what order the adversary
runs them in, the next coin-flip is always another fair coin-flip. So
we can analyze the sequence of partial sums of \emph{generated}
coin-flips using the tools we have developed for sums of independent
variables, random walks, and martingales. Unfortunately things get a
little more complicated when we look at the totals actually observed
by any particular processes.

The problem is that we need some mechanism for gathering up the
local coin values. Using read-write registers, we can have each
process write the count and sum of its own local coins to a register
writable by it alone (this avoids processes overwriting each other).
Because the adversary can only delay one coin from each process, both
total count and the total sum are always within $n$ of the correct
value. If we could read all the registers instantaneously and stop as
soon as the count reached some threshold $T$, then for $T=Θ(n^2)$ the
distribution on the generated coins would be spread out enough that
the total would be at least $n+1$ away from $0$ with constant
probability. But the actual situation is more complicated.

The problem is that each process must do $n$ sequential reads to read
all the registers. This is not only expensive—it's not something we
want to do after every local flip—but it also means that more coins
may be generated while I'm reading the registers, meaning any sum I
compute at the end of the algorithm might have only a tenuous
connection with the actual sum of the generated coins at any point in
the execution.

The solution to the first part of this problem was proposed by Bracha
and Rachman~\cite{BrachaR1991}: only check the total after writing out
$Θ(n/\log n)$ local coin-flips, giving an amortized cost per coin-flip of
$Θ(\log n)$ register operations. Unfortunately this makes the second
problem worse: a process $p$ might continue generating many local coins
after the total count crosses the threshold, and each other process
might see a different prefix of these coins depending on when it reads
$p$'s register.

Bracha and Rachman showed that this wasn't too much of a problem using
Azuma's inequality (this is where the $O(\log n)$ factor comes in).
But later work by Attiya and Censor~\cite{AttiyaC2008jacm} allowed for
a simpler analysis of a slightly different algorithm, which we will
describe here.

Pseudocode for the Attiya-Censor coin is given in
Algorithm~\ref{alg-Attiya-Censor}. The algorithm only checks the total
count once for every $n$ coin-flips, giving an amortized cost of $1$
read per coin-flip. But to keep the processes for running too long,
each process checks a multi-writer $\DataSty{stop}$ bit after ever
coin-flip.

\begin{algorithm}
    \While{$∑_{j=1}^{n} r_j.\DataSty{count} < T$}{
        \For{$i ← 1 \dots n$}{
            $\Tuple{r_i.\DataSty{count}, r_i.\DataSty{sum}} ← 
            \Tuple{r_i.\DataSty{count} + 1, r_i.\DataSty{sum} ± 1}$\;
            \If{\DataSty{stop}}{
                \Break;
            }
        }
    $\DataSty{stop} ← \True$\;
    \Return $\sgn\parens*{∑_{j=1}^{n} r_j.\DataSty{sum}}$\;
    }
    \caption{Attiya-Censor weak shared coin~\cite{AttiyaC2008jacm}}
    \label{alg-Attiya-Censor}
\end{algorithm}

Each process may see a different total sum at the end, but our hope is that
if $T$ is large enough, there is at least a $δ$ probability that all
these total sums are positive (or, by symmetry, negative). We can
represent the total sum seen by any particular process $i$ 
as a sum $S+D+X_i-H_i$, where:
\begin{enumerate}
    \item $S$ is the sum of the first $T$ coin-flips. This has a
        binomial distribution, equal to the sum of $T$ independent
        $±1$ random variables. Letting $T=cn^2$ for some reasonably
        large $c$ gives us a constant probability that $\abs{S} > 4n$.
    \item $D$ is the sum of all coin-flips after the first $T$ that
        are generated before some process sets the $\DataSty{stop}$
        bit. There are at most $n^2+n$ such coin-flips, and they form
        a martingale with bounded increments. So Azuma's inequality
        gives us that $\abs{D} ≤ 2n$ with at least constant
        probability, independent of $S$.
    \item $X_i = ∑_{j=1}^n Y_{ij}$, where $Y_{ij}$ is the sum of all
        coin-flips generated by process $j$ between some process
        setting the stop bit and process $i$ reading $r_j$.
        Since each process can generate at most one extra coin-flip
        before checking $\DataSty{stop}$, $\abs{X_i} ≤ n$ always.
    \item $H_i = ∑_{j=1}^n Z_{ij}$, where $Z_{ij}$ is the sum of all
        votes that are generated by process $j$ before $i$ reads
        $r_j$, but that are not included in $r_j.\DataSty{sum}$
        because they haven't been written yet.
        Again, each process can contribute only one coin-flip to
        $H_i$. So $\abs{H_i} ≤ n$ always.
\end{enumerate}

Adding up the error terms $D+X_i-H_i$ gives a total that is bounded by
$4n$ with at least constant probability. This probability is
independent of the constant probability that $\abs{S} > 4n$. So if
both of these events occur, we win.

The total cost of the coin is $O(T+n^2) = O(n^2)$. This also gives a
cost of $O(n^2)$ for consensus.

This turns out to be optimal, because
of a lower bound on the expected total steps for consensus appearing in the same paper~\cite{AttiyaC2008jacm}.
This is a bit disappointing, because an $Ω(n^2)$ lower bound
translates to $Ω(n)$ steps for each process even if we can distribute
the load evenly across all processes (which
Algorithm~\ref{alg-Attiya-Censor} might not).
We'd like to get something that scales better, but to get around the
lower bound we will need to switch to an oblivious adversary.

\section{Leader election with sifters}
\label{section-sifter-leader-election}

The idea of a \concept{sifter}~\cite{AlistarhA2011} is to use
randomization to quickly knock out processes while leaving at least one process always.
The current best known sifter for standard read-write registers, due
to Giakkoupis and Woelfel~\cite{GiakkoupisW2012}, is shown in
Algorithm~\ref{alg-Giakkoupis-Woelfel}.

\begin{algorithm}
    Choose $r∈ℤ^{+}$ such that $\Prob{r=i} = 2^{-i}$\;
    $A[r] ← 1$\;
    \eIf{$A[r+1] = 0$}{
        stay\;
    }{
        leave\;
    }
    \caption{Giakkoupis-Woelfel sifter~\protect{\cite{GiakkoupisW2012}}}
    \label{alg-Giakkoupis-Woelfel}
\end{algorithm}

The cost of executing the sifter is two operations. Each process
chooses an index $r$ according to a geometric distribution, writes
$A[r]$, and then checks if any other process has written $A[r+1]$. The
process stays if it sees nothing.

Because any process with the maximum value of $r$ always stays, at
least one process stays. To show that not too many processes stay, let 
$X_i$ be the number of survivors with
$r=i$. This will be bounded by the number of processes that write to
$A[i]$ before any process writes to $A[i+1]$. We can immediately see
that $\Exp{X_i} ≤ n⋅2^{-i}$, since each process has probability
$2^{-i}$ of writing to $A[i]$. 
But we can also argue that $\Exp{X_i} ≤ 2$ for any value of $n$.

The reason is that because the adversary is oblivious, the choice of
which process writes next is independent of the location it writes to.
If we condition on some particular write being to either $A[i]$ or
$A[i+1]$, there is a $1/3$ chance that it writes to $A[i+1]$. So we
can think of the subsequence of writes that either land in $A[i+1]$ or
$A[i]$ as a sequence of biased coin-flips, and we are counting how
many probability-$2/3$ tails we get before the first probability-$1/3$
heads. This will be $2$ on average, or at most $2$ if we take into
account that we will stop after $n$ writes.

We thus have $\Exp{X_i} ≤ \min(2, n⋅2^{-i})$. So the expected total number of
winners is bounded by $∑_{i=1}^{∞} \Exp{X_i} \min(2, n⋅2^{-i}) = 2 \lg
n + O(1)$.

Now comes the fun part. Take all the survivors of our sifter, and run
them through more sifters. Let $S_i$ be the number of survivors of the
first $i$ sifters. We've shown that $\ExpCond{S_{i+1}}{S_i} = O(\log
S_i)$. Since $\log$ is concave, Jensen's inequality then gives
$\Exp{S_{i+1}} = O(\log \Exp{S_i})$. Iterating this gives $\Exp{S_i} =
O(\log^{(i)} S_0) = O(\log^{(i)} n)$. So there is some $i = O(\log^*
n$ at which $\Exp{S_i}$ is a constant.

This doesn't quite give us leader election because the constant might
not be $1$. But there are known leader election algorithms that run in
$O(1)$ time with $O(1)$ expected participants~\cite{AlistarhAGGG2010},
so we can use these to clean up any excess processes that make it
through all the sifters. The total cost is $O(\log^* n)$ operations
for each process.

\section{Consensus with sifters}
\label{section-sifter-consensus}

We've seen that we can speed up leader election by discarding
potential leaders quickly. For consensus, we want to discard potential
output values quickly. This is a little more complicated, because
processes carrying losing output values can't just drop out if we
haven't determined the winning values yet.

We are going to look at an algorithm by Aspnes and
Er~\cite{AspnesE2019} that solves consensus in $O(\log^* n)$
operations per process, using sifters built on top of a stronger
primitive than normal registers. (This algorithm uses some ideas from an
earlier paper of Aspnes~\cite{Aspnes2015}, which gives a more
complicated $O(\log \log n)$-time algorithm for standard registers.)
As with the Giakkoupis-Woelfel leader election algorithm, we assume an
oblivious adversary scheduler.

The stronger primitive we need is a \index{register!max}\concept{max
register}~\cite{AspnesAC2012}. Unlike a standard register, reading a
max register returns the largest value ever written to it;
equivalently, trying to write a smaller value to a max register than
it already contains has no effect.

What this means for randomized algorithms is that we can knock out
values using a max register by a simple priority scheme. If each
process writes a tuple $\Tuple{r,v}$ to the max register, where $r$ is
a unique random priority and $v$ is the process's preferred value,
then the $i$-th value written appears in the max
register only if it is a left-to-right maximum of the sequence of
priorities, which occurs with probability roughly $1/i$. (This follows from symmetry and the fact that the
oblivious adversary can't selectively schedule smaller priorities
first.) So on average only $H_n = O(\log n)$ of these values will ever
appear in the max register. By having each process read and return a
value from the max register, we've gone from $n$ possible values to
$O(\log n)$ possible values on average.

We now run into two technical obstacles. The first is that we can't
necessarily guarantee that all the random priorities are unique, which
we would need to get an exact $1/i$ probability for survival. This
can be dealt with by choosing priorities over a sufficiently large
range that the rare collisions don't add much to the expected
survivors (say, $O(n^3)$). The second issue is that if we try to
iterate the sifter, in the second round we have many copies of each
input value. So even if one copy drops out because of a low priority,
some other copy might get lucky and survive. To go from $O(\log n)$
expected values to $O(\log \log n)$ expected values, we need to make
sure that no value gets more than one chance at surviving.

Here is where we use an idea that first appeared in~\cite{Aspnes2015}.
Instead of generating a new priority for each value in each round, we
generate a sequence $r_1, r_2, \dots, r_\ell$ of priorities for all
$\ell$ rounds all at once at the beginning. We can then store
$\Tuple{r_i,\dots,r_\ell,v}$ in the max register for round $i$, and
the leading priority $r_i$ controls survival. Now we don't care about
having multiple copies of a value, because they will all have the same
initial $r_i$, and if the first copy to be written doesn't survive,
none of them will.

The result of applying this idea is
Algorithm~\ref{alg-Aspnes-Er}.
Because each iteration of the loop reduces $S_i$ survivors to $S_{i+1}
= O(\log S_i)$ survivors on average, the same application of Jensen's
inequality that we used for Algorithm~\ref{alg-Giakkoupis-Woelfel}
works here as well, so for an appropriate $\ell = O(\log^* n)$ we can
get down to a single surviving value with reasonably high probability.

\begin{algorithm}
    \Procedure{$\FuncSty{sifter}(v)$}{
        Choose random ranks $r_1 \dots r_\ell$\;
        \For{$i ← 1 \dots \ell$}{
            $\WriteMax(M_i, \Tuple{r_i,\dots,r_\ell,v})$\;
            $\Tuple{r_i,\dots,r_\ell,v} ← \ReadMax(M_i)$\;
            \label{line-max-register-conciliator-read}
        }
        \Return $v$\;
    }
    \caption{Sifter using max registers~\cite{AspnesE2019}}
    \label{alg-Aspnes-Er}
\end{algorithm}

This is not quite enough to get consensus, which requires agreement
always, but with some additional machinery it is possible to detect
when the protocol fails and re-run it if necessary. The final result
is consensus in expected $O(\log^* n)$ max register operations.

\appendix

\chapter{Sample assignments from Fall 2025}

\section{Assignment 1, due Thursday 2025-09-11 at 23:59}

    \subsection{RGB search}

    Suppose you have $3n$ closed boxes in front of you, divided into three
    groups: $n$ red boxes, $n$ green boxes, and $n$ blue boxes. In one
    of the boxes is a prize, and opening a box will reveal if the
    prize is inside that box.

    Your goal is to determine the color of the box that contains the
    prize. An optimal randomized algorithm can do this by opening
    $an+O(1)$ boxes on average, where $a$ is a constant.
    Find $a$ and prove that it is both achievable and optimal.

        \subsubsection*{Solution}

        The value of $a$ is $4/3$.
        
        Here is an algorithm that achieves this bound: pick one of the
        three colors uniformly at random. Call this color $C$.

        Line up the $2n$ boxes that do not have color
        $C$, and flip a coin to choose whether to open these boxes
        from left to right or right to left.
        If the prize is in one of these boxes, we know from the
        analysis in §\ref{section-searching-an-array} that it takes
        $\frac{2n+1}{2} = n + 1/2$ probes on average to find the
        prize, which tells us the color of the box containing the
        prize. If the prize is not in one of these boxes, after
        opening all $2n$ of them we can deduce that the color of the
        box containing the prize is $C$.

        Because we chose $C$ at random, the probability of the first
        case is $2/3$ and of the second is $1/3$. So the expected
        total cost is $\frac{2}{3}\parens*{n + 1/2} +
        \frac{1}{3}⋅2n = \frac{4}{3}⋅n + \frac{1}{3} = \frac{4}{3}⋅n +
        O(1)$. This gives us our upper bound on $a$.

        For the lower bound, use Yao's Lemma. Observe that for any
        deterministic algorithm, there is a fixed order in which it
        opens boxes (as long as it doesn't find the prize). If the
        algorithm stops before finding the prize or opening all boxes
        of at least two colors, then for any color it outputs there is
        a box it didn't open of a different color that could have
        contained the prize. So a correct deterministic algorithm can
        be described by a sequence $π$ of $k≥2n$ boxes that it opens
        in sequence, stopping only if it finds the prize or exhausts
        two of the colors. We will assume that $π$ contains no
        duplicates and that the algorithm does indeed stop once it
        finds the prize or exhausts two of the colors, because for any
        deterministic algorithm that doesn't have these properties
        there is a faster deterministic algorithm that does.

        Now suppose the adversary places the prize in a box chosen
        uniformly at random. Let $k'≥2n$ be the smallest value such
        that $π_1 \dots π_{k'}$ includes all boxes of at least two
        colors. For each $i$, box $π_i$ contains the prize with
        probability $\frac{1}{3n}$, so the expected number of boxes opened is
        at least
        \begin{equation*}
            ∑_{i=1}^{k'} \frac{1}{3n}⋅i 
            + ∑_{i=k'+1}^{3n} \frac{1}{3n}⋅2n
        \end{equation*}
        The total is minimized by setting $k'=2n$, giving a lower
        bound of
        \begin{align*}
            ∑_{i=1}^{2n} \frac{1}{3n}⋅i 
            + ∑_{i=2n+1}^{3n} \frac{1}{3n}⋅2n
            &=
            \frac{1}{3n} \parens*{\frac{(2n+1)(2n)}{2} + n⋅2n}
            \\&=
            \frac{2}{3} \parens*{\frac{2n+1}{2} + n}
            \\&=
            \frac{4}{3}⋅n + \frac{1}{3},
        \end{align*}
        exactly matching the upper bound.
    
    \subsection{A randomized vending machine}

    A small child is feeding dollar coins to a capsule vending
    machine. In the vending machine are $n$ capsules, each of which
    contains a cheap plastic toy. The capsules can be ranked $1\dots
    n$ by desirability. Each time the child buys a toy, one of the
    toys is delivered uniformly at random from the remaining toys.
    This removes this particular toy from the stock, meaning we are
    doing sampling without replacement.

    The child keeps buying capsules as long as the last delivered toy
    is ranked higher than all previously delivered toys, assuming any
    capsules are left. If the machine delivers a toy that is ranked
    lower than the highest-ranked toy so far, the child experiences
    disappointment and leaves. For example, with toys ranked $1\dots
    10$, the child might collect $4,5,9,7$ but will stop buying
    capsules after $7$ because it's worse than $9$.

    \begin{enumerate}
        \item Give an asymptotic big-$Θ$ expression for the expected
            number of capsules the child buys as a function of $n$.
        \item Give an asymptotic big-$Θ$ expression for the
            probability that the child obtains the highest-ranked toy
            before giving up.
    \end{enumerate}

        \subsubsection*{Solution}

        This problem gets easier if we pick the right probability
        space. Let's imagine that the child keeps buying until the
        machine is empty. We can then describe the sequence of
        delivered toys as one of of the $n!$ possible permutations of
        the ranks, and by symmetry $n!$ permutations have the same
        probability $1/n!$. The probability of getting any particular
        prefix before quitting will then just be the sum of the
        probabilities of all permutations that have this prefix.

        \begin{enumerate}
            \item There are many ways to calculate this, but a
                particularly sneaky approach is to use linearity of
                expectation. Let $X_i$ be the rank of the $i$-th toy
                in the permutation and  $A_i$ be the indicator variable for
                the event that the child buys the $i$-th toy. Then
                $A_i = 1$ if and only if $X_1 < X_2 < X_{i-1}$.

                We can calculate the probability of this event exactly by
                counting the number of permutations $X$ that have this
                property. The math is a little cleaner if we drop the
                $-1$ and count permutations that are increasing for
                the first $k$ values $X_1 < X_2 < \dots < X_k$.
                Generating such a permutation consists of picking $k$
                initial values that we will sort in increasing order,
                plus an ordering for the remaining $n-k$ values. This
                gives
                \begin{align*}
                    \Prob{X_1 < X_2 < \dots < X_k}
                    &= \frac{\binom{n}{k} ⋅ (n-k)!}{n!}\
                    \\&= \frac{\frac{n!}{k! (n-k)!} ⋅ (n-k)!}{n!}
                    \\&= \frac{1}{k!}.
                \end{align*}

                This is pretty much what we'd expect, and we could
                probably get away with an argument that a uniform
                random permutation makes any ordering of its first $k$
                elements equally likely, but sometimes it's nice to do
                the calculation just to be sure.

                Now observe that the total number of toys bought is $A
                = ∑_{i=1}^n A_i$, and by linearity of expectation
                \begin{align*}
                    \Exp{A}
                    &= ∑_{i=1}^{n} \Exp{A_i}
                    \\&= ∑_{i=1}^{n} \Prob{A_i=1}
                    \\&= ∑_{i=1}^{n} \frac{1}{(i-1)!}
                    \\&≤ ∑_{k=0}^{∞} \frac{1}{k!}
                    \\&= e = O(1).
                \end{align*}

                Since the child always buys at
                least two toys, we have $\Exp{A} ≥ 2 = Ω(1)$ as well,
                giving an asymptotic bound of $Θ(1)$.

            \item For this case, we want to count up all the
                permutations $X$ with the property that $X_1 < X_2 <
                \dots X_i = n$ for some $i$. 
                Given any fixed $i$ with $X_i = n$,
                we can pick $X_1 \dots X_{i-1}$ to be any of the $\binom{n-1}{i-1}$
                possible subsets of $\Set{1,\dots,n-1}$, and pick $X_{i+1}
                \dots X_n$ to be any of the $(n-i)!$ orderings of the
                remaining $n-i$ ranks. Summing over all possible
                choices of $i$ gives
                \begin{align*}
                    \Prob{\text{child buys capsule $n$}}
                    &= ∑_{i=1}^{n} \frac{\binom{n-1}{i-1}⋅(n-i)!}{n!}
                    \\&= ∑_{i=1}^{n} \frac{\frac{(n-1)!}{(i-1)! (n-i)!} ⋅ (n-i)!}{n!}
                    \\&= ∑_{i=1}^{n} \frac{1}{(i-1)!}⋅\frac{1}{n}
                    \\&≤ \frac{1}{n} ∑_{k=0}^{∞} \frac{1}{k!}
                    \\&≤ \frac{e}{n} = O(1/n).
                \end{align*}

                To get the two-sided bound, observe that the child has
                a $1/n = Ω(1/n)$ chance of collecting $n$ on the first try,
                so the probability of winning is $Θ(1/n)$ (and in fact
                very close to $e/n$ for reasonable large $n$).

                There may be less brute-force ways to prove this. One
                idea is to argue that whether the first $(i-1)!$
                values are increasing is independent of whether
                $X_i=n$, which would give the
                $\frac{1}{(i-1)!}⋅\frac{1}{n}$ probability of getting
                $X_1 < X_2 < \dots < X_i = n$ without having to cancel
                out all the factorials. But independence of these
                events is a little slippery, so it might require some
                work to prove it if we want to be really careful.
        \end{enumerate}

\section{Assignment 2, due Thursday 2025-09-25 at 23:59}

    \subsection{A regular community of hipsters}

    One day I wake up and decide to start wearing a particular obscure
    style of hat because I think it is cool. But later one of my
    friends starts wearing one too, and I realize that I need to stop
    wearing mine because this kind of hat is obviously not cool any
    more now that it's gone mainstream.

    Now imagine a community of $n$ such hipsters, each of whom is
    friends with $d$ other hipsters.\footnote{More abstractly, the
    hipsters form $d$-regular graph on $n$ vertices.} Pick a random
    permutation $π$ of the hipsters and run a process where at step
    $i$, hipster $π(i)$ starts wearing a hat and $π(i)$'s $d$
    neighbors each stop wearing a hat if they were wearing one before.
    Let $X$ be the number of hipsters who are still wearing a hat when
    this process finishes.

    \begin{enumerate}
        \item Compute an exact value for $\Exp{X}$ as a function of $n$
            and $d$.
        \item Show that the variance of $X$ is $O(nd)$.
        \item Show that when $d$ is fixed,
            $\Prob{X < \frac{1}{2} \Exp{X}}$ is $O(n^{-c})$
            for every constant $c>0$.
    \end{enumerate}

        \subsubsection*{Solution}

        \begin{enumerate}
            \item For each node $v$ in the graph, let $X_v$ be the
                indicator variable for the event that $V$ has a hat 
                at the end. This occurs if and only if $v$ is the last
                node chosen in among itself and its $d$ neighbors. By
                symmetry, $\Exp{X_v} = \frac{1}{d+1}$.

                Now sum over all $n$ choices of $v$ to get $\Exp{X} =
                \frac{n}{d+1}$.
            \item Define $X_v$ as before. We'd like to compute
                $\Var{X} = ∑_v \Var{v} + ∑_{u≠v} \Cov{X_u}{X_v}$.
                Computing this exactly is going to be a nuisance,
                especially because we don't know the structure of the
                graph. So we will settle for an upper bound that gives
                the desired asymptotic bound. This will involve a lot
                of sloppiness—we are more interested in getting an
                acceptable result quickly than getting a tighter
                result slowly.

                Let $p = \Exp{X_v} = \frac{1}{d+1}$ and let $q=1-p$ as
                usual. For a single node $v$, $\Var{X_v} = pq$.
                The total variance contributed by the $n$ single nodes is
                $npq$.

                If $u$ and $v$ are neighbors,
                whichever one arrives second knocks the hat off the
                first. So $\Cov{X_u}{X_v} = \Exp{X_u X_v} - \Exp{X_u}
                \Exp{X_v} = 0 - p^2 = -p^2$.
                There are $nd$ ordered pairs of neighbors, so between
                them they contribute $-ndp^2$ to the variance.

                Curiously, this exactly cancels out the variance of
                the individual nodes:
                \begin{align*}
                    npq - ndp^2
                    &= np\parens*{q - dp}
                    \\&= np\parens*{1 - p - dp}
                    \\&= np\parens*{1 - (d+1)p}
                    \\&= np\parens*{1 - \frac{d+1}{d+1}}
                    \\&= 0.
                \end{align*}

                This is not entirely surprising: if the hipsters form
                a clique, where every node is adjacent to every other
                node, then only the last node in the permutation gets
                a hat at the end, and the variance is $0$. So if we
                want more variance we are going to need to look at
                nodes at greater distance.

                If $u$ and $v$ share at least one neighbor,
                then $X_u$ and $X_v$ may be positively correlated.
                Calculating the exact covariance as a function of how
                many neighbors $u$ and $v$ share looks painful. But we
                can get a crude bound by observing that 
                $\Cov{X_u}{X_v} ≤ \Exp{X_u X_v} ≤ \Exp{X_u} =
                \frac{1}{d+1}$. For each node $u$ there are at most $d^2$
                possible nodes $v$ at distance $2$, so these pairs
                contribute at most $\frac{nd^2}{d+1} < nd$ to the
                variance.
                
                If $u$ and $v$ share no common neighbors, then $X_u$
                and $X_v$ are independent, giving $\Cov{X_u}{X_v} =
                0$. This means that any pairs nodes at distance more than
                $2$ add nothing to the total variance.

                Adding up all the contributions gives $\Var{X} < 2nd =
                O(nd)$.
            \item This is a job for McDiarmid's inequality.
                To use this inequality, we need to express $X$ as a
                function $f(Y_1,\dots,Y_n)$ where the $Y_i$ are
                independent. The trick is to make each $Y_i$ an
                independent sample from $[0,1]$, and define the
                permutation $π$ as the order that makes the sequence
                $Y_{π(i)}$ increasing.

                Changing $Y_i$ affects at most $d+1$ nodes: $i$ and
                its $d$ neighbors. So let's let $c_i = d+1$ (this
                avoids thinking about the exact details of which nodes
                switch). Then McDiarmid says
                \begin{align*}
                    \Prob{X - \Exp{X} ≤ -\frac{1}{2} \Exp{X}}
                    &≤ \exp\parens*{-\frac{2(n/(d+1))^2}{n(d+1)^2}}
                    \\& = \exp\parens*{-\frac{2n}{(d+1)^3}}
                    \\& = e^{-Ω(n)},
                \end{align*}
                using the assumption that $d$ is a constant. This is
                exponentially small and so is $O(n^{-c})$ for any
                fixed exponent $c$.
        \end{enumerate}

    \subsection{Building a random network}

    Suppose we are trying to build a communication network on
    $n$ nodes $\Set{1,\dots,n}$. Our goal is to produce a network that
    is connected in all but an exponentially small number of cases.
    Since higher-numbered nodes happen to have
    more capacity for routing traffic, we consider two strategies for
    biasing connections in favor of these nodes. Both strategies
    generate roughly the same expected number of edges, but they
    differ in how they assign them to popular and unpopular nodes.

    \begin{enumerate}
        \item In the first strategy, each edge $ij$ appears with
            independent probability $p_{ij} = \frac{i+j}{4n}$.
        \item In the second strategy, each edge $ij$ appears with
            independent probability $p_{ij} = \frac{ij}{n^2}$.
    \end{enumerate}

    For each strategy, prove or disprove the claim that there exists a
    constant $c<1$ such that the given strategy produces a
    connected graph with probability $1-O(c^n)$.

        \subsubsection*{Solution}

        For each strategy, let $G$ be the generated graph and write
        $ij∈G$ if the edge $ij$ appears in $G$.

        \begin{enumerate}
            \item We'll show that $p_{ij} = \frac{i+j}{4n}$ works.

                Call a node $i$ popular if $i ≥ \frac{3}{4}n$.
                Our strategy will be to show that, with all but
                exponentially small error probability, (a) all the popular
                nodes are connected to each other by paths of length
                at most $2$, and (b) every other node is connected to
                some popular node. This shows that there is a path
                between any two nodes, implying that the graph is
                connected.

                \begin{lemma}
                    \label{lemma-hw-random-network-popular-connected}
                    Let $i$ be a popular node and $j≠i$ be any other node.
                    Then $\Prob{ij∈G} ≥ \frac{3}{16}$.
                \end{lemma}
                \begin{proof}
                    Compute 
                    $\Prob{ij∈G}
                    = \frac{i+j}{4n}
                    ≥ \frac{\frac{3}{4}n+1}{4n} 
                    = \frac{3}{16} + \frac{1}{4n}
                    > \frac{3}{16}.$
                \end{proof}

                Using this fact, we can show
                \begin{lemma}
                    \label{lemma-hw-random-network-popular-connected-to-each-other}
                    Let $i$ and $i'$ be distinct popular nodes.
                    Then 
                    \begin{align*}
                        \Prob{¬∃j: ij∈G ∧ i'j∈G} 
                        &≤ \parens*{\frac{247}{256}}^{n-2}.
                    \end{align*}
                \end{lemma}
                \begin{proof}
                    From
                    Lemma~\ref{lemma-hw-random-network-popular-connected},
                    $\Prob{ij ∉ G} ≥ \frac{3}{16}$.
                    and the same bound holds on $\Prob{i'j∉G}$. Since
                    these are independent events, we can compute
                    \begin{align*}
                        \Prob{ij ∉ G ∨ i'j ∉ G}
                        &= 1 - \Prob{ij ∈ G ∧ i'j ∈ G}
                        \\&= 1 - \Prob{ij ∈ G}⋅\Prob{i'j ∈ G}
                        \\&≤ 1 - \parens*{\frac{3}{16}}^2
                        \\&= \frac{247}{256}.
                        \intertext{Different $j$ also give independent
                        events, so}
                        \Prob{∀j∉\Set{i,i'}: ij ∉ G ∨ i'j ∉ G}
                        &= ∏_{j∉\Set{i,i'}} \Prob{ij ∉ G ∨ i'j ∉ G}
                        \\&≤ \parens*{\frac{247}{256}}^{n-2}.
                    \end{align*}
                \end{proof}

                This shows that the popular nodes are all connected.
                For the unpopular nodes, let's show
                \begin{lemma}
                    \label{lemma-hw-random-network-unpopular-connected}
                    Let $i$ be an unpopular node. Then the probability
                    that $i$ is not connected to any popular node is
                    at most $\parens*{\frac{13}{16}}^{\floor{n/4}}$.
                \end{lemma}
                \begin{proof}
                    Immediate from multiplying out the probabilities
                    given by
                    Lemma~\ref{lemma-hw-random-network-popular-connected}.
                \end{proof}

                Combining these bounds gives
                \begin{theorem}
                    \label{theorem-hw-random-network-connected}
                    There is a constant $c<1$ such that probability
                    that a random graph with $p_{ij} = \frac{i+j}{4n}$
                    is non connected is $O(n^{-c})$.
                \end{theorem}
                \begin{proof}
                    This is just a matter of summing up all bounds on bad
                    events from
                    Lemmas~\ref{lemma-hw-random-network-popular-connected-to-each-other}
                    and~\ref{lemma-hw-random-network-unpopular-connected}.

                    The $O(n)$ popular nodes contribute error
                    $O\parens*{n⋅\frac{247}{256}^n}$.

                    The $O(n)$ unpopular nodes contribute error
                    $O\parens*{n⋅\parens*{\parens*{\frac{13}{16}}^{1/4}}^n}$.

                    Choose any $c$ with $\max\parens*{\frac{247}{256},
                    \parens*{\frac{13}{16}}^{1/4}} < c < 1$ and we're
                    done.
                \end{proof}

            \item 
                Dealing with the second strategy is easier because it
                fails in a way that doesn't require checking too many
                details.
                
                In order for the graph
                to be connected, node $1$ needs at least one edge. But
                the union bound tells us that
                \begin{align*}
                    \Prob{∃j: 1j∈G}
                    &≤ ∑_{j=2}^n \Prob{1j∈G}
                    \\&= ∑_{j=2}^n \frac{j}{n^2}
                    \\&= \frac{1}{n^2}⋅\frac{(n-1)(n+2)}{2}
                    \\&= \frac{n^2 + n - 2}{2n^2}
                    \\&= \frac{1}{2} + O(1/n).
                \end{align*}
                So for sufficiently large $n$, there is at least a
                constant probability that node $1$ is isolated, which
                makes the graph not connected no matter what the other
                nodes are doing.
        \end{enumerate}

\section{Assignment 3, due Thursday 2025-10-09 at 23:59}

    \subsection{A fractal load balancing mechanism}

    We would like to construct an extensible load balancing system
    where jobs are assigned to machines, each machine's load is the
    number of jobs assigned to it, and we periodically increase
    the number of machines to keep the average load low.

    Consider the following approach. We have a sequence
    of generations $k=0,1,2,\dots,g$. In each generation $k$:
    \begin{enumerate}
        \item We add $2^{k-1}$ new machines. For $k=0$, this
            will be the single machine $0$.
            For $k>0$, we add machines
            $2^{k-1} \dots 2^k-1$ to
            the old machines $0 \dots 2^{k-1}-1$ left over
            from the previous generations.
            The new machines start with no jobs, while any old machines
            retain whatever jobs they already had.
        \item We then assign $2^k$ new jobs to all $2^k$ machines
            according to a randomized policy described below.
            This gives a total of
            $n=2^{k+1}-1$ jobs at the end of generation $k$.
    \end{enumerate}

    To assign a job in generation $k$ to a machine,
    generate a sequence of $k$ independent random bits where each has a constant
    probability $p$ of being $1$ and $q=1-p$ of being $0$. Interpret
    the resulting sequence as a binary number in the range $0\dots
    2^k-1 = m-1$: this gives the index of the machine that gets the
    job.

    If $p=1/2$, this is equivalent to uniform assignment. By tuning
    $p$ up or down we can favor newer or older machines when
    assigning new jobs. This might be helpful to avoid overloading the
    oldest machines with jobs from many generations.

    Let $X_i$ be the total number of jobs assigned to machine $i$
    after generation $g$. Define the expected maximum load as
    $\Exp{\max_i X_i}$. Our goal is to choose a constant $p$ that
    minimizes the asymptotic expected maximum load as a function of
    $g$.

    Find the best value of $p$ by this metric, compute the
    corresponding asymptotic expected maximum load,
    and show that any other choice of $p$ is asymptotically worse.

        \subsubsection*{Solution}

        First let's compute $\Exp{X_i}$
        and see what choice of $i$ gives the best value for
        $\max_i \Exp{i}$.
        This is not quite the same thing as $\Exp{\max_i X_i}$,
        but maybe we will
        learn something along the way.

        Let $w_i$ be the number of
        one bits in $i$ when expressed in binary. Let $g_i$ be the
        generation in which $i$ is first used.

        Let $Y_{ij}$ be the indicator variable for the
        event that job $j$ is assigned to machine $i$, so that
        $X_i = ∑_j Y_{ij}$ is a sum of independent $0$-$1$
        variables. Let $g_j$ be the generation of job $j$.

        If $g_j < g_i$, then $Y_{ij} = 0$ always.
        If $g_j ≥ g_i$, then $\Exp{Y_{ij}}=\Prob{Y_{ij} = 1} =
        p^{w_i} q^{g_j - w_i} = q^{g_j} (p/q)^{w_i}$.
        Summing over all jobs with $g_j ≥ g_i$ gives 
        \begin{align*}
            \Exp{X_i}
            &= ∑_{k=g_i}^{g} 2^k ⋅ q^k (p/q)^{w_i}
            \\&= ∑_{k=g_i}^{g} (2q)^k (p/q)^{w_i}
            \\&= (p/q)^{w_i} \frac{(2q)^{g_i}-(2q)^{g+1}}{1-2q}.
        \end{align*}

        For fixed $g_i$, we should set $w_i$ to maximize
        $(p/q)^{w_i}$. There are three cases depending on how $p$
        compares to $q$:
        \begin{itemize}
            \item 
                If $p = q = 1/2$, any choice of $w_i$ gives $(p/q)^{w_i}=1$.
                Unfortunately the geometric series formula also collapses
                since $2q$ is now $1$ and we end up dividing zero by
                zero. Instead, we can compute the sum
                directly: $\Exp{X_i} = ∑_{k=g_i}^{g} 2^k⋅(1/2)^k = g - g_i +
                1$. Setting $g_i = 0$ gives this the maximum possible value of
                $g+1 = Θ(g)$.
            \item 
                If $p < q$, we should make $w_i$ as small as possible. For
                $i=0$ we have $w_i = 0$ and $\Exp{X_0} =
                \frac{1-(2q)^{g+1}}{1-2q} = \frac{(2q)^{g+1}}{2q-1} =
                Θ\parens*{(2q)^{g}}$.

                For $i>0$ the minimum possible value of $w_i$ is $1$.
                In this case we get
                $\Exp{X_i} 
                = (p/q) \frac{(2q)^{g_i} - (2q)^{g+1}}{1-2q} 
                = (p/q) \frac{(2q)^{g+1} - (2q)^{g_i}}{1-2q}
                = Θ\parens*{(2q)^{g}}$ again.
            \item 
                If $p > q$, we should make $w_i$ as large as possible,
                which for fixed $g_i$ means setting $w_i = g_i$.
                This gives
                $\Exp{X_i}
                = (p/q)^{g_i} \frac{(2q)^{g_i} - (2q)^{g+1}}{1-2q}
                = \frac{(2p)^{g_i} - (2q)^{g-g_i+1} - (2p)^{g-g_i+1}}{1-2q}$.
                The best choice here for $g_i$ is going to be $g_i=g$,
                which optimizes all three terms in the denominator.
                This corresponds to setting $i=m-1$, giving $\max_i
                \Exp{X_i} = \Exp{X_{m-1}} = Θ\parens*{(2p)^{g}}$.
        \end{itemize}

        Our choices for the asymptotic behavior of $\max_i \Exp{X_i}$
        are either $Θ(g)$ when $p=1/2$ or $Θ(c^g)$ for a
        constant $c>1$ when $p≠1/2$. This suggests we should try to
        bound $\Exp{\max_i X_i}$ for the $p=1/2$ case, then show that
        the other cases are worse.

        Let $p=1/2$.
        Since $\max_i X_i ≥ X_0$, we have $\Exp{\max_i X_i} ≥
        \Exp{X_0} = g+1$. This gives a linear lower bound on
        $\Exp{\max_i X_i}$.
        For an upper bound, we'll use a Chernoff bound argument,
        specifically using the variant $\Prob{S ≥ R} ≤ 2^{-R}$ when
        $R≥2eμ$ (see \eqref{eq-Chernoff-bound-R}).

        Let $R = 6(g+1) ≥ 2e \Exp{X_0} ≥ 2e \Exp{X_i}$ for all $i$.
        Since $X_i$ is a sum of independent $0$-$1$ variables
        $Y_{ij}$, \eqref{eq-Chernoff-bound-R} gives
        \begin{align*}
            \Prob{X_i ≥ 6(g+1)} 
            &≤ 2^{-6(g+1)}
            \intertext{and by the union bound}
            \Prob{\max X_i ≥ 6(g+1)}
            &≤ m⋅2^{-6(g+1)}
            \\&= 2^g ⋅ 2^{-6(g+1)}
            \\&= 2^{-5g+6}.
        \end{align*}
        This is not quite a bound on $\Exp{\max_i X_i}$, since it
        doesn't rule out a rare event where $\max_i X_i$ is
        absurdly large. But we know $\max_i X_i ≤ 2^{g+1}
        -1$ always. So we can split into cases:
        \begin{align*}
            \Exp{\max_i X_i}
            &= \ExpCond{\max_i X_i}{\max_i X_i < 12(g+1)} \Prob{\max_i X_i < 12(g+1)}
            \\&\quad + \ExpCond{\max_i X_i}{\max_i X_i ≥ 12(g+1)} \Prob{\max_i X_i ≥ 12(g+1)}
            \\&≤ 12(g+1) ⋅ 1 + 2^{g+1}⋅2^{-5g+6}
            \\&= 12(g+1) + 2^{-4g+7}
            \\&= O(g).
        \end{align*}

        Combined with the previous lower bound, this gives
        $\Exp{\max_i X_i} = Θ(g)$ when $p=1/2$.

        The good news is that we don't need to use concentration
        bounds to show that the $p≠1/2$ cases are worse. We've already
        shown that $\max_i \Exp{X_i} = Ω(c^g)$.
        But then $\Exp{\max_i X_i} ≥ \max_i \Exp{X_i} = Ω(c^g)$ as well,
        which is asymptotically worse than the $Θ(g)$ bound
        for $p=1/2$.

    \subsection{A randomly balanced binary tree}
    \label{section-hw-random-binary-tree-triangle}

    Here is a simple rules for building a randomly balanced binary
    tree consisting of $n$ layers of nodes, with each layer
    $i∈\Set{1,\dots,n}$ containing exactly $i$ nodes $\Set{
        v_{i1}, v_{i2}, \dots v_{ii}}$. Each node $v_{ij}$ with $1<i$ and $1<j<i$
    flips an independent fair coin to choose between $v_{i-i,j-1}$ and
    $v_{i-1,j}$ as its parent. For the nodes at the ends of each row,
    $v_{i1}$ always has parent $v{i-1,1}$ and $v_{ii}$ always has
    parent $v_{i-1,i-1}$.

    An example of a tree constructed in this way is shown in
    Figure~\ref{fig-hw-random-binary-tree-triangle}.

    \begin{figure}
        \centering
        \begin{tikzpicture}
            \pgfmathsetseed{1371239}
            \newcommand{\n}{9}
\foreach \i  in {1,...,\n} {
                \foreach \j in {1,...,\i} {
                    \node (v\i\j) at (\j-\i/2,-\i) {$v_{\i\j}$};
                }

            }
\foreach \i [evaluate=\i as \ii using {int(\i-1)}] in {2,...,\n} {
                \draw[->] (v\i1) edge (v\ii1);
                \draw[->] (v\i\i) edge (v\ii\ii);
            }
\foreach \i [evaluate=\i as \ii using {int(\i-1)}] in {3,...,\n} {
                \foreach \j in {2,...,\ii} {
\pgfmathsetmacro{\c}{random(0,1)}
                    \pgfmathsetmacro{\jj}{int(\j-\c)}
                    \draw[->] (v\i\j) edge (v\ii\jj);
                }
            }
        \end{tikzpicture}
        \caption{A random tree constructed as described in
        §\ref{section-hw-random-binary-tree-triangle}.}
        \label{fig-hw-random-binary-tree-triangle}
    \end{figure}

    Define the \conceptFormat{weight} $W_{ij}$ of a node $v_{ij}$ as the total
    number of nodes in the subtree rooted at $v_{ij}$. This will be
    equal to the number of descendants of $v_{ij}$ (including itself).

    \begin{enumerate}
        \item Compute an exact value for $\Exp{W_{ij}}$ as a function
            of $i$, $j$, and $n$.
        \item Prove or disprove: For any constants $ε>0$ and $c>0$,
            \begin{align*}
                \Prob{W_{21} ≥ (1+ε) \Exp{W_{21}}} &= O(n^{-c}).
            \end{align*}
    \end{enumerate}

        \subsubsection*{Solution}

        \begin{enumerate}
            \item The easiest way to do this might be to write out a
                recurrence. The size of the subtree rooted at $v_{ij}$
                will be one (for $v_{ij})$ plus the sum of the sizes of the
                subtrees rooted at its children (if any).
                If $1<j<i$, then each potential child of $v_{ij}$ has
                a $1/2$ chance of picking $v_{ij}$ as its parent. For
                $j=1$ or $j=i$, one of the two potential children will always
                pick $v_{ij}$ as its parent. So for $i < n$ we have:
                \begin{align*}
                    \Exp{W_{ij}}
                    &=
                    \begin{cases}
                        1 + \Exp{W_{i+1,j}} + \frac{1}{2} \Exp{W_{i+1,j+1}}
                        & \text{when $j=1$,}
                        \\
                        1 + \frac{1}{2} \Exp{W_{i+1,j}} + \frac{1}{2} \Exp{W_{i+1,j+1}}
                        & \text{when $1 < j < i$, and}
                        \\
                        1 + \frac{1}{2} \Exp{W_{i+1,j}} +  \Exp{W_{i+1,j+1}}
                        & \text{when $j=i$.}
                    \end{cases}
                    \intertext{And for $i=n$:}
                    \Exp{W_{ij}} &= 1.
                \end{align*}

                This recurrence looks ugly but it has a cleaner
                recurrence hiding inside it. When $1<j<i$, the
                equivalent condition $1<j<i+1$ holds for both of
                $v_{ij}$'s potential children. So if we consider only
                the interior nodes with $1<j<i$, we have
                \begin{align*}
                    \Exp{W_{ij}}
                    &= \frac{1}{2} \Exp{W_{i+1,j}} + \frac{1}{2} \Exp{W_{i+1,j+1}}
                    \intertext{with the base case}
                    \Exp{W_{nj}} &= 1.
                    \intertext{This has the easily-checked solution}
                    \Exp{W_{ij}} &= n-i+1.
                \end{align*}

                We could use this to solve a recurrence for
                $\Exp{W_{i1}}$ and $\Exp{W_{ii}}$, but it's possible
                to save a bit of work by approaching the problem
                indirectly. We know that (a) every node in layers $i$
                through $n$ is in exactly one subtree rooted at layer
                $i$, and (b) the total number of such nodes is
                exactly $\frac{i+n}{2}⋅(n-i+1)$.

                There is a special case when $i=1$ where this formula
                gives $\Exp{W_{11}} = W_{11} = \frac{n(n+1)}{2}$
                always. For $i>1$, we can compute
                \begin{align*}
                    \frac{i+n}{2}⋅(n-i+1)
                    &= ∑_{j=1}^{i} W_{ij}
                    \\&= \Exp{∑_{j=1}^{i} W_{ij}}
                    \\&= ∑_{j=1}^{i} \Exp{W_{ij}}
                    \\&= \Exp{W_{i1}} + \Exp{W_{ii}} + (i-2)⋅(n-i+1).
                    \intertext{Solve this to get}
                    \Exp{W_{i1}} + \Exp{W_{ii}}
                    &= \parens*{\frac{i+n}{2} - (i-2)}(n-i+1)
                    \\&= \frac{n-i+4}{2}⋅(n-i+1),
                    \intertext{which by symmetry gives}
                    \Exp{W_{i1}} =  \Exp{W_{ii}}
                    &= \frac{(n-i+4)(n-i+1)}{4}.
                \end{align*}

                We can summarize these results as
                \begin{align*}
                    \Exp{W_{ij}}
                    &=
                    \begin{cases}
                        \frac{n(n+1)}{2} & \text{when $i=j=1$},
                        \\
                        \frac{(n-i+4)(n-i+1)}{4} & \text{when $i>1$
                        and $j∈\Set{1,i}$, and}
                        \\
                        n-i+1 & \text{when $1<j<i$.}
                    \end{cases}
                \end{align*}

            \item 
                We'll prove the claim using Azuma's inequality.

                The idea is that we will construct a martingale by
                exposing the decisions of one layer of nodes at a
                time, starting at the top, and argue that the
                conditional expectation of the final value of $W_{21}$
                changes slowly enough that Azuma works.

                Let $ℱ_i$ be the $σ$-algebra generated by the choices
                of all nodes in rows $1$ through $i$. The key
                observation is that for $i'≥i$, $\ExpCond{W_{i'j}}{ℱ_i} =
                \Exp{W_{i'j}}$ because the value of $W_{i'j}$ depends
                only on the choices of nodes in rows $i+1$ or higher.
                We can also notice that $ℱ_i$ determines which nodes
                in row $i$ contribute to $W_{21}$, since all the edges
                at row $i$ or above are fixed by $ℱ_i$.

                Because $ℱ_1 ⊆ ℱ_2 ⊆ \dots ⊆ ℱ_n$, the sequence $S_i =
                \ExpCond{W_{21}}{ℱ_i}$ is a Doob martingale.

                Write $Z_i$ for the number of nodes in rows $2$
                through $i-1$ that appear in the subtree rooted at
                $v_{21}$, and let $R_i$ be the rightmost node in row
                $i$ that appears in the subtree rooted at $v_{21}$.
                Then
                \begin{align*}
                    W_{21}
                    &= Z_i + ∑_{j=1}^{R_i} W_{ij}
                    \intertext{from which it follows that}
                    S_i 
                    &= \ExpCond{W_{21}}{ℱ_i}
                    \\&= \ExpCond{Z_i + ∑_{j=1}^{R_i} W_{ij}}{ℱ_i}
                    \\&= \ExpCond{Z_i}{ℱ_i} + ∑_{j=1}^{R_i} \ExpCond{W_{ij}}{ℱ_i}
                    \\&= Z_i + ∑_{j=1}^{R_i} \Exp{W_{ij}}.
                \end{align*}

                Let's think a bit about what happens in that sum 
                in the last line. All
                the nodes $v_{i1}$ through $v_{i,R_i}$ are included,
                as are the subtrees rooted at $v_{i+1,1}$ through
                $v_{i+1,R_i}$. The only question is whether we include
                the subtree rooted at $v_{i+1,R_i+1}$ or not, which
                depends on which way $v_{i+1,R_i+1}$ flips its coin.
                So we can expand
                \begin{align*}
                    S_i &= \ExpCond{W_{21}}{ℱ_i}
                    \\&= Z_i + R_i + ∑_{j=1}^{R_i}
                    \ExpCond{W_{i+1,j}}{ℱ_i}
                    \\&\quad
                    + \ExpCond{W_{i+1,j+1}}{ℱ_i}
                    \ProbCond{\text{$v_{i+1,j+1}$ picks $v_{i,j}$}}{ℱ_i}
                    \\&= 
                    Z_{i+1} + ∑_{j=1}^{R_i} \Exp{W_{i+1,j}} +
                    \frac{1}{2} \Exp{W_{i+1,j+1}}.
                    \intertext{Comparing this to}
                    S_{i+1} &= \ExpCond{W_{21}}{ℱ_{i+1}}
                    \\&= Z_{i+1} + ∑_{j=1}^{R_{i+1}} \Exp{W_{i+1,j}},
                    \intertext{we see that}
                    S_{i+1} - S_i
                    &=
                    ± \frac{1}{2} \Exp{W_{i+1,R_i+1}},
                \end{align*}
                depending on whether $R_{i+1}$ is $R_i$ or $R_i+1$.

                This means we can apply Azuma's inequality
                \eqref{eq-Azumas-inequality} to $\Set{S_2,\dots,S_n}$
                with $c_i = \frac{1}{2} \Exp{W_{i,R_i+1}} =
                \frac{1}{2}(n-i+1)$ to get
                \begin{align*}
                    \Prob{W_{21} ≥ \Exp{W_{2i}} + t}
                    &≤ \exp\parens*{\frac{-t^2}{2∑_{i=2}^{n} c_i^2}}
                    \\&=
                    \exp\parens*{\frac{-t^2}{2 ∑_{i=2}^{n}
                    \parens*{\frac{n-i+1}{2}}^2}}
                    \\&=
                    \exp\parens*{\frac{-2t^2}{∑_{k=1}^{n-1} k^2}}
                    \\&=
                    e^{-Θ(t^2/n^3)}.
                \end{align*}
                Since $\Exp{W_{21}} = Θ(n^2)$, we can exceed the
                expectation by $t = ε \Exp{W_{21}} = Θ(n^2)$ with
                probability at most
                $e^{-Θ(n)}$, which is 
                $O(n^{-c})$ for any fixed $c$.

                If we look carefully at the preceding argument, we
                might notice that there is a simpler argument hiding
                inside it that avoids dealing with conditional
                probabilities entirely. The key idea is that $v_{ij}$ is in the
                $v_{21}$ subtree if and only if $j ≤ R_i$, 
                giving $W_{21} = ∑_{i=2}^{n} R_i$,
                and since each
                $R_{i+1}$ is either $R_i$ or $R_{i+1}$ with equal
                probability independent of previous choices, we can
                write $R_i = 1 + ∑_{k=3}^i X_k$ where $X_k$ are fair
                $0$-$1$ coin-flips. So Chernoff bounds apply (with $μ_i =
                \Exp{R_i} = i/2$), giving
                \begin{align*}
                    \Prob{R_i ≥ (1+ε) \Exp{R_i}}
                    &≤
                    e^{-μ_i δ^2/3}
                    \\&=
                    e^{-δ^2 i / 6}.
                \end{align*}

                Now the idea is to pick $δ = ε/2$, and argue
                that even if we take a union bound $∑_{i=2}^{n}
                e^{-δ^2 i / 6}$ on the probability that any of the
                $R_i$ might be big enough to make $W_{21} > (1+ε)\Exp{W_{21}}$, 
                we still get something smaller than any polynomial for
                large $n$. Unfortunately this is not true if we
                include the terms with small $i$. But we can split
                off all $i<√{n}$, and argue that these
                satisfy $R_i ≤ 2\Exp{R_i}$ always and that their total
                is thus $O(n)$. Using Chernoff for the rest of the
                $R_i$ gives $W_{21} ≥ (1+ε/2)
                \Exp{W_{21}} + O(n) = (1+ε/2+o(1)) \Exp{W_{21}}$ with
                probability at most $ne^{-Ω(√{n})}$, which is 
                $O(n^{-c})$ for any fixed $c$.
        \end{enumerate}

\section{Assignment 4, due Thursday 2025-10-30 at 23:59}

    \subsection{A shifty sketch}

    Consider an algorithm running in the streaming model 
    where each of $n$ incoming data points $x_1,\dots,x_n$ is of the form
    $x_t = \Tuple{b_t,i_t,c_t}$, where $b_t∈\Set{0,1}$, $i_t ∈
    \Set{0,\dots,2n}$, and $c_t∈\Set{-1,+1}$.
    As usual, the algorithm maintains a small sketch $s$,
    which it updates by some rule $s←f(s,x_t)$ for each data point.

    Our goal is to determine if the vector represented by the points
    with $b_t = 1$ is a shift of the vector represented by the points
    with $b_t = 0$. Formally, let
    \begin{align*}
        a_{bi} &= ∑_{\SetWhere{t}{b_t = b, i_t = i}} c_t.
        \intertext{and say that $a_1$ is a shift of $a_0$ if there is
        some offset $0≤k≤n$ such that $a_{0i}=0$ for $i>2n-k$ and}
        a_{1i} &=
        \begin{cases}
            0 & \text{for $i < k$, and}
            \\
            a_{0,i-k} & \text{otherwise}.
        \end{cases}
    \end{align*}

    For example, when $n=4$, the vector $a_1 =
    \Tuple{0,0,1,2,-3,4,0,0,0}$ is a shift of $a_0 =
    \Tuple{1,2,-3,4,0,0,0,0,0}$ with offset $k=2$.

    Show that for any $δ>0$, it is possible to construct a
    sketch $s$ of $O\parens*{\log n \log(1/δ)}$ bits, such that:
    \begin{enumerate}
        \item Initializing the sketch, updating the sketch, and
            querying the sketch can all be done in time polynomial in
            $n$ and $\log(1/δ)$.
        \item If $a_1$ is a shift of $a_0$, then querying the sketch
            produces the answer yes with probability $1$.
        \item If $a_1$ is not a shift of $a_0$, then querying the
            sketch produces the answer yes with probability at most
            $δ$.
    \end{enumerate}

        \subsubsection*{Solution}

        First we'll show a solution for fixed $δ$, then use
        amplification to get arbitrary $δ$.

        Pick a prime $p$ such that $8n^2 ≤ p ≤ 16n^2$. We don't even
        have to be particularly clever about this: Bertrand's
        Postulate says that some such $p$ exists, and we can find it
        in $O(n^3)$ time by testing all $O(n^2)$ values $x$ in
        this range by trial division up to $√{x} = O(n)$.

        The sketch will consist of $p$, a value $1≤r≤p-1$ chosen
        uniformly from $Z^*_p$, and two hashes $h_0$ and $h_1$
        given by
        \begin{align}
            h_b &= ∑_{i=1}^{2n} a_{bi} r^i \pmod{p}.
            \label{eq-hw-shifty-sketch-hash}
        \end{align}
        Since all of these values are $O(n^2)$, they take $O(\log n^2)
        = O(\log n)$ bits to represent.

        Initially, $h_0=h_1=0$.

        For the update rule, given $\Tuple{i_t,b_t,c_t}$ set
        \begin{align*}
            h_{b_t} &← h_{b_t} + c_t r^{i_t} \pmod{p}.
        \end{align*}
        This rule takes $O(\log n)$ time per update
        (mostly to compute $r^{i_t}$),
        and a straightforward induction on $t$ shows that it preserves
        \eqref{eq-hw-shifty-sketch-hash}.

        If $a_1$ is a shift of $a_0$ by $k$ positions, then
        \begin{align*}
            h_1
            &=
            ∑_{i=0}^{2n} a_{1i} r^i
            \\&=
            ∑_{i=k}^{2n} a_{1i} r^i
            \\&=
            ∑_{i=k}^{2n} a_{0,i-k} r^i
            \\&=
            ∑_{i=0}^{2n-k} a_{0i} r^{i+k}
            \\&=
            r^k 
            ∑_{i=0}^{2n} a_{0i} r^{i}
            \\&=
            r^k h_0,
        \end{align*}
        with all computation in the finite field $ℤ_p$.

        So we can test if $a_1$ is a shift of $a_0$ by computing $h_1
        - r^k h_0$ for all $k∈\Set{0,\dots,n}$ and returning yes if
        we find a value of $k$ that yields $0$.

        Suppose instead $a_1$ is not a shift of $a_0$.
        Viewing each $h_1 - r^k h_0$ as a polynomial in $r$, its
        degree is at most $3n$, giving at most $3n$ choices $r$
        that make $h_1(r) - r^k h_0(r) = 0$, for a total of at most
        $3n(n+1)$ choices of $r$ that make \emph{any} of the $(n+1)$ polynomials
        $h_1(r) - r^k h_0(r) = 0$. Since we are choosing $r$ from a
        range of at least $8n^2$, this gives a probability strictly
        less than $1/2$ of hitting a bad $r$ that gives a false
        positive.

        To reduce this probability to $δ$, run $\ceil{\lg(1/δ)}$
        copies of the sketch in parallel with independent choices for
        $r$, and return yes to any query only if all copies of the
        sketch answer yes.

    \subsection{A radio network}

    Imagine we are building a relay station for a radio network that accepts
    incoming packets and buffers them for retransmission. The station
    has only one antenna, so it can either receive or transmit at any
    particular moment, but not both. Because the relay station
    receives from relatively weak stations and transmits using more
    power, during any one time interval it can either receive one
    packet or transmit up to two, with received packets having
    priority. Assume that each interval has an independent $1/2$
    probability of receiving a packet. We'd like to get an asymptotic
    estimate of the expected maximum buffer size as a function of the
    number of time intervals $n$.

    More precisely, write $X_t$ for the number of packets in the
    buffer after $t$ intervals.
    Then $X_0=0$ and $X_{t+1}$ satisfies
    \begin{align*}
        X_{t+1}
        &=
        \begin{cases}
            X_t + 1 & \text{with probability $1/2$}
            \\
            \max(0,X_t - 2) & \text{with probability $1/2$}
        \end{cases}
    \end{align*}

    Show that $\Exp{\max_{0≤t≤n} X_t} = Θ(\log n)$.

        \subsubsection*{Solution}

        We'll construct a martingale similar to the exponential one
        from §\ref{section-martingales-for-biased-random-walks}.

        Let $Z_t = a^{X_t}$ where $a$ is a constant to be determined.
        Let $ℱ_t = \Tuple{X_0,\dots,X_t}$.
        Because of the lower bound on $X_t$, in general we do not expect $\Set{Z_t}$
        to satisfy the martingale property. But when $X_t≥2$, we have
        \begin{align*}
            \Exp{Z_{t+1} - Z_t}{ℱ_t}
            &= \frac{1}{2} a^{X_t+1} + \frac{1}{2} a^{X_t - 2} - a^{X_t}.
            \\&= \frac{1}{2} a^{X_t-2} \parens*{a^3 + 1 - 2a^2}.
        \end{align*}

        For $\Set{Z_t}$ to be a martingale we need this expression
        to be $0$, which happens when $a$ is a root of $a^3-2a^2+1$.
        One not very interesting root is $a=1$, which makes $Z_t
        = 1$ always. Dividing out $a-1$ leaves the polynomial
        $a^2-a-1$, which has roots $\frac{1±√{5}}{2}$. We'll go
        with the positive root, which happens to be the golden ratio
        $φ = \frac{1+√{5}}{2}$.

        Start at some time $s$ with $X_s = 2$ and $Z_s = φ^2$. As
        long as $Z_t$ doesn't drop below $2$, the martingale property
        holds. Let $τ$ be the first time at which $X_τ < 2$ or $X_τ =
        b$. To avoid some annoying special cases, imagine that both
        the $\Set{X_t}$ and $\Set{Z_t}$ processes continue forever; we
        will deal with $n$ later.

        Using the usual argument, we can show that $τ$ is finite with
        probability $1$. We also have that $Z_t$ has bounded range.
        So Theorem~\ref{theorem-optional-stopping} applies and
        $\Exp{Z_τ} = \Exp{Z_s} = φ^2$.

        For each $x$, let $p_x$ be the probability that $X_τ=x$. Compute
        \begin{align*}
            φ^2 
            &= \Exp{Z_2}
            \\&= \Exp{Z_τ}
            \\&= p_b⋅φ^b + p_0⋅φ^0 + p_1⋅φ^1
            \intertext{which gives}
            p_b &≤ φ^{2-b}.
            \intertext{Set $b = 2 \log_φ n + 2$ to get}
            p_b &≤ Φ^{-2 \log_φ n} = n^{-2}.
        \end{align*}

        This covers the case of one sequence of values starting at
        some particular time $s$ with $Z_s=2$. In the event that $Z_τ
        < 2$, then eventually $Z_t$ reaches $2$ again and we can
        re-run this argument starting at that time.
        There are at most $n$ possible times
        $s_1,s_2,\dots$ at which $Z_s$ might be $2$; so the union
        bound tells us that the probability that there exists a
        time $s<n$ with $Z_s = 2$ after which $Z_t$ reaches $2 \log_φ
        n + 2$ before reaching $0$ or $1$ is at most $n⋅n^{-2} =
        n^{-1}$. Call the event that this occurs $A$.

        Since in all executions, $Z_t ≤ t$, we have
        \begin{align*}
            \Exp{\max_{0≤t≤n} Z_t}
            &=
            \ExpCond{\max_{0≤t≤n} Z_t}{A} \Prob{A}
            + \ExpCond{\max_{0≤t≤n} Z_t}{\overline{A}}\Prob{\overline{A}}
            \\&≤ n⋅n^{-1} + (2 \log_φ n + 1)⋅1
            \\&= O(\log n).
        \end{align*}

        This gives us an asymptotic upper bound. For the matching
        lower bound,
        Let $k = \floor{\lg√{n}}$, and consider the sequence of
        times $ki$ for $i∈\Set{0,\dots,\floor{n/k} - 1}$.
        Starting at each $ki$, there is an independent $2^{-k} ≥
        n^{-1/2}$ chance that the next $k$ steps all increase $X_t$ by
        $1$, which would leave $X_{ki+k} ≥ k = Ω(\log n)$. These
        events are independent, so the chance that any of them occur
        is at least
        \begin{align*}
            1-\parens*{1-n^{-1/2}}^{\floor{n/k}}
            &≥ 1 - e^{-n^{-1/2}\floor{n/k}}
            \\&=
            1-o(1),
            \intertext{which gives}
            \Exp{\max_{0≤t≤n} X_t}
            &≥ (1-o(1))⋅Ω(\log n)
            \\&= Ω(\log n).
        \end{align*}

\section{Assignment 5, due Thursday 2025-11-13 at 23:59}

    \subsection{Exits}

    Suppose you have a building consisting of an $n×n$ grid of rooms,
    each of which has four doors leading to its north, east, south,
    and west neighbors, or outside in the case of rooms on the edge of
    the grid.

    An exit labeling assigns a direction to each room. The
    labeling is legal if no two rooms point to each other. 
    Figure~\ref{fig-hw-exit-labeling-legal}. 
    gives an example of a legal exit labeling on a $10×10$ grid, where
    the direction of the exit of each room is indicated by an arrow.
    Figure~\ref{fig-hw-exit-labeling-illegal}
    gives an example of a similar exit labeling that is not legal
    because it has two arrows pointing at each other (marked in red).

    \begin{figure}
        \begin{displaymath}
            \begin{array}{cccccccccc}
                ↓&↓&←&→&↑&←&→&↑&↑&→\\
                ←&→&→&↑&←&↓&←&↓&→&→\\
                →&→&→&↑&→&↓&↑&→&↓&↑\\
                ↑&←&↓&↑&↓&←&↓&↓&→&→\\
                ←&↑&←&↓&←&↓&↓&←&←&←\\
                →&↓&←&↓&→&→&→&↓&↑&↑\\
                ←&→&→&↓&←&←&←&←&↓&↓\\
                ↑&←&↑&↓&←&↓&↑&←&↓&←\\
                ↑&↓&↑&←&↑&→&↑&↓&↓&←\\
                ←&↓&↓&↑&↑&↑&↑&→&→&↑
            \end{array}
        \end{displaymath}
        \caption{A legal exit labeling for a $10×10$ grid}
        \label{fig-hw-exit-labeling-legal}
    \end{figure}

    \begin{figure}
        \begin{displaymath}
            \begin{array}{cccccccccc}
                ↓&↓&←&→&↑&←&→&↑&↑&→\\
                ←&→&→&↑&←&↓&←&↓&→&→\\
                →&→&→&↑&→&↓&↑&→&↓&↑\\
                ↑&←&↓&↑&↓&←&↓&↓&→&→\\
                ←&↑&←&↓&←&↓&↓&←&←&←\\
                →&↓&←&↓&→&\textcolor{red}{→}&\textcolor{red}{←}&↓&↑&↑\\
                ←&→&→&↓&←&←&←&←&↓&↓\\
                ↑&←&↑&↓&←&↓&↑&←&↓&←\\
                ↑&↓&↑&←&↑&→&↑&↓&↓&←\\
                ←&↓&↓&↑&↑&↑&↑&→&→&↑
            \end{array}
        \end{displaymath}
        \caption{An illegal exit labeling for a $10×10$ grid}
        \label{fig-hw-exit-labeling-illegal}
    \end{figure}

    Consider the following MCMC algorithm for sampling legal exit
    labelings. Start from any legal labeling $X^0$.
    To choose $X^{t+1}$ given $X^t$, sample a
    uniform random location $\Tuple{i,j}$ and a uniform random
    direction $d∈\Set{↑,→,←,↓}$. 
    Update $X^{t+1}_{ij}$ to $d$ if that
    gives a legal exit labeling $X^{t+1}$; if not, let $X^{t+1} =
    X^t$.

    Show that this process converges to a uniform distribution on
    legal exit labelings with a mixing time polynomial in $n$.

        \subsubsection*{Solution}

        That some stationary distribution of this process is uniform is
        immediate from the fact that for any two labelings $x$ and $y$
        that differ in exactly one position, $p_{xy} = p_{yx} =
        \frac{1}{4n^2}$, and for any two labelings that differ in more
        than one position, $p_{xy} = p_{yx} = 0$. So the detailed
        balance equations hold for $π$ when $π_x = π_y$ for all $x$
        and $y$. Showing that a uniform $π$ is in fact the unique stationary
        distribution will follow from the coupling argument that we
        will use to show mixing.

        As usual, consider two coupled copies of the Markov chain,
        where $X^0$ starts in an arbitrary distribution and $Y^0$
        starts in a uniform distribution. Update $X$ and $Y$ according
        to the rule for the Markov chain,
        choosing the same location $v=\Tuple{i,j}$ and same direction $d$ for
        both. (We'll write $v$ for the location because we are going
        to think of these as vertices in a graph.)

        Given $X^t$ and $Y^t$, let $Z^t = \card*{\SetWhere{v}{X^t_v ≠
        Y^t_v}}$. The Coupling Lemma (\ref{lemma-coupling}) 
        says $d_{TV}(X^t,Y^t) ≤ \Prob{X^t ≠ Y^t} = \Prob{Z^t ≠ 0}$. 
        So we'd like to show $Z^t$ reaches $0$
        in time polynomial in $n$ on average.

        What makes $Z^t$ change?

        Let's write $v+d$ for the vertex in direction $d$ from $v$, or
        $⊥$ if there is no such vertex. Let's also write $-d$ for the
        direction opposite to $d$. Then we are allowed to set
        $X^{t+1}_v$ to $d$ if and only if $v+d = ⊥$ or $X^t_{v+d} ≠
        -d$.

        After choosing $v$ and $d$, $Z_t$
        can change in two ways:
        \begin{itemize}
            \item $Z_{t+1} = Z_t - 1$. This happens when $X^t_v ≠
                Y^t_v$, and either $v+d = ⊥$ 
                or neither of $X^t_{v+d}$ or $Y^t_{v+d}$ is $-d$.
            \item $Z_{t+1} = Z_t + 1$. This happens when $X^t_v =
                Y^t_v$, $v+d≠⊥$, and exactly of of $X^t_{v+d}$ and
                $Y^t_{v+d}$ is $-d$.
        \end{itemize}

        The thing to note is that for every instance of the second
        case there is a corresponding instance of the first. 

        Let's suppose we pick $v$ and $d$ that trigger the second
        case, where $Z^t$ rises. Let $v'=v+d$ (which exists because
        $v+d ≠ ⊥$) and $d' = -d$. We want
        to claim that had we picked $v'$ and $d'$ instead, $Z^t$ would
        have dropped.

        Because the second case holds for $v,d$, we have $X^t_v =
        Y^t_v = d'' ≠ d$, since $d''=d$ would create an illegal
        labeling with whichever of $X^t_{v+d}$ or $Y^t_{v+d}$ is $-d$.
        So from the point of view of $v'$, $X^t_{v'+d'} = Y^t_{v'+d'}
        ≠ -d'$: the first case holds.

        Since these cases occur with the same probability
        $\frac{1}{4n^2}$, summing over all of them gives 
        $\Exp{Z^{t+1} - Z^t} ≤ 0$.

        If we knew this was strictly negative, we'd be done. But it
        might be that the bad and good cases exactly balance each
        other out. So instead we will argue that $Z^t$ behaves no
        worse than a delayed unbiased random walk with a reflecting
        barrier at $n^2$, meaning that it eventually hits $0$ after
        which $X^t=Y^t$ forever.

        Unfortunately, the original version of this argument I put
        here contained a bug: it claimed that for any pair of
        configurations $X^t$ and $Y^t$, there existed a transition
        that gave $Z^{t+1}≠Z^t$. This is not the case, as demonstrated by
        the counterexample in Figure~\ref{fig-hw-exits-locked}
        provided by Shuchen Li.
        \begin{figure}
            \begin{displaymath}
                \begin{array}{cc}
                    \begin{array}{cccc}
                        ↓&↓&↓&↓\\
                        →&→&↓&←\\
                        →&↑&←&←\\
                        ↑&↑&↑&↑
                    \end{array}
                    &
                    \begin{array}{cccc}
                        ↓&↓&↓&↓\\
                        →&↓&←&←\\
                        →&→&↑&←\\
                        ↑&↑&↑&↑
                    \end{array}
                    \bigskip
                    \\
                    X & Y
                \end{array}
            \end{displaymath}
            \caption{A pair of locked configurations for the exits
            process}
            \label{fig-hw-exits-locked}
        \end{figure}

        Instead, we will show that whenever $Z^t>0$, there exists a
        sequence of steps of constant length (and thus polynomial
        probability) that reduces $Z$ by one.

        Construct a directed graph $G^t$ with the grid as vertices and
        an edge from $u$ to $v$ if $u$ points to $v$ in either $X^t$
        or $Y^t$.

        Suppose $v$ has $X^t_v≠Y^t_v$. If the indegree of $v$ in $G^t$
        is not $4$, then there is a direction $d$ we can point $v$
        that works in both $X^t$ and $Y^t$, giving a $\frac{1}{4n^2}$
        chance of reducing $Z^t$ by one. So we can only get a locked
        pair of configurations where $Z^t$ can't drop if there is a
        node in $G^t$ with indegree $4$.

        Alternatively, if $v$ has $X^t_v = Y^t_v$, then its indegree
        can't be higher than $3$, since it can't have an incoming edge
        from the node it is pointing to. If its indegree is $2$ or
        less, then there is a different direction it can be switched
        to in both configurations.

        Now consider a $k×k$ subgrid and ask what the total indegree
        can be in this subgrid. There are at most $2k^2$ edges from
        the subgrid nodes themselves and at most $4k$ edges coming in
        from outside the subgrid, for a total of $2k^2 + 4k$. Dividing
        by the $k^2$ nodes in the subgrid gives an average indegree of
        $2 + 4/k$. For $k=4$ this is $3$, meaning that if there is at
        least one node with indegree $4$ then there must also be at
        least one node with indegree $2$.

        Given a node $v$ with indegree $4$, choose a $4×4$ subgrid that
        includes $v$ as one of its four interior nodes. Somewhere in
        this subgrid is a node $u$ with degree $2$, and its distance
        to $v$ is at most $4$. Consider a sequence of moves along the
        path $u=u_0 \dots u_k = v$ where at each step we point
        $u_i$ away from $u_{i+1}$ in both $X$ and $Y$. This sets
        $X^{t+k+1}_v = Y^{t+k+1}_v$ and occurs with probability
        at least $\frac{1}{(4n^2)^{k+1}} = Ω(n^{-10})$.

        We have shown that (a) $\ExpCond{Z^{t+1}}{Z^t} ≤ Z^t$ always,
        (b) there is always some sequence of $k≤5$ steps that set $Z^{t+k}
        ≠ Z^t$ with probability $Ω(n^{-10})$,
        and (c) if $Z^{t+1} ≠ Z^t$, then $Z^{t+1} = Z^t ± 1$.
        Construct a new process $Q^s$ where $Q^s$ takes a random $±1$
        step whenever $Z^t$ changes, with a reflecting barrier at
        $n^2$, an absorbing barrier at $0$, and correlated with $Z^t$
        so that $Q^s ≥ Z^{t_s}$ where $t_s$ is the step corresponding
        to $s$. Then $Q$ reaches $0$ in at most $n^4$ steps on
        average, where each step of $Q$ corresponds to $O(n^{10})$
        steps of $Z$ on average. This gives an expected waiting time for $Z^t$ to
        reach $0$ of $O(n^{14})$, giving $t_{\mix} =
        t_{\mix}(1/4) ≤ O(n^{14})$ by Markov's inequality.

    \subsection{Trouble moving sideways}

    Consider a random walk on an $n×n$ grid with positions labeled by pairs
    $xy∈ℕ×ℕ$ with $1≤x,y≤n$, where a particle at $xy$ moves on its next
    step to
    \begin{displaymath}
\setlength{\extrarowheight}{4pt}
        \begin{array}{l@{\quad \text{with probability }}l}
        x,\min(y+1,n) & \frac{1}{8}, \\
        x,\max(y-1,1) & \frac{1}{8}, \\
        \min(x+1,n),y & \frac{1}{8}(y/n)^c, \\
        \max(x-1,1),y & \frac{1}{8}(y/n)^c,
        \end{array}
    \end{displaymath}
    and otherwise stays put. This is a symmetric Markov chain, so it
    has a uniform stationary distribution with $π_{xy} = n^{-2}$.

    For which constants $c≥0$ does this walk mix in
    $\widetilde{O}(n^2)$ steps?

    (Assume that $\widetilde{O}$ hides any factors that are polylogarithmic
    in $n$ or $1/ε$.)

        \subsubsection*{Solution}

        For all constants $c≥0$.

        To prove this, we'll use the method of canonical paths. To
        minimize congestion we will want to avoid any edges that are used
        only with low probability in the stationary distribution,
        which will be the horizontal edges where $(y/n)^c$ is small.
        But for $y≥n/2$ we have $(y/n)^c ≥ 2^{-c} = Ω(1)$ since $c$ is
        a constant. We will take advantage of this by packing
        horizontal traffic into the top half of the grid.

        Given a source $s=s_x s_y$ and target $t=t_x t_y$,
        let $y_{\max} = \max(s_y, t_y)$, and let
        \begin{align*}
            z &=
            \begin{cases}
                y_{\max} & \text{if $y_{\max} ≥ n/2$} \\
                y_{\max} + \ceil{n/2} & \text{otherwise}
            \end{cases}
        \end{align*}
        Let $γ_{st}$ be the path that goes from $s_x s_y$ vertically to $s_x
        z$, horizontally to $t_x z$, then vertically to $t_x t_y$.

        For each vertical edge $uv$, $γ_{st}$ crosses $uv$ only if
        $u_x = s_x$ or $u_x = t_x$. In either case, there are at
        most $n^3$ ways to choose the three unspecified
        coordinates, giving at most $2n^3$ paths crossing any
        vertical edge.

        For each horizontal edge $uv$, $γ_{st}$ crosses $uv$ only if
        $z$ as computed above is equal to $u_y = v_y$. There are at
        most two possible choices of $y_{\max}$ yielding this $z$, and
        each could be either $s_y$ or $t_y$. This yields four choices
        altogether, and for each choice there are three remaining
        unspecified coordinates, giving at most $4n^3$ paths
        crossing any horizontal edge.

        Recall that the congestion of an edge $uv$ is given by
        \begin{align*}
            ρ(uv)
            &= \frac{1}{π_u p_{uv}} ∑_{γ_{st}∋uv} π_s π_t.
        \end{align*}

        For a vertical edge, this is
        \begin{align*}
            \frac{1}{n^{-2}⋅\frac{1}{8}} ∑_{γ_{st}∋uv} n^{-4}
            &≤ 8n^2 ⋅ 2n^3 ⋅ n^{-4} ≤ 8n.
            \intertext{For a horizontal edge, we instead get}
            \frac{1}{n^{-2}⋅\frac{1}{8}⋅(y/n)^c} ∑_{γ_{st}∋uv} n^{-4}
            &≤ 8n^2⋅2^{c}⋅4n^3⋅n^{-4} ≤ 2^{c+3} n.
        \end{align*}

        In either case, the congestion is $O(n)$, which gives $τ_2 =
        O(n^2)$ and $t_{\mix}(ε) = O(n^2 (\log n + \log(1/ε))) =
        \widetilde{O}(n^2)$.

\section{Assignment 6, due Thursday 2025-12-04 at 23:59}

    \subsection{Locally incompressible strings}

    It's well-known that long random strings are likely to contain
    substrings that don't look particularly random. For example, the
    allegedly random string
    \begin{displaymath}
        01000001010100100100111011001101010011011011010111100000011111010
    \end{displaymath}
    contains several long runs of $0$ and $1$ bits, as well as a few
    suspiciously repetitive substrings. A natural question is whether
    there exist random-looking strings all of whose sufficiently large
    substrings are also random-looking. But this requires that we
    define what it means for a string to be random-looking.

    In principle the strings $0000000000$ and $0101110010$ both have
    exactly the same $2^{-10}$ chance of being generated randomly, but
    the second one somehow looks a lot more random to the casual
    observer. One way to characterize such ``random-looking'' strings
    is via Kolmogorov complexity, which roughly speaking defines the
    complexity of a string as the length of the shortest program that
    generates it and defines a string as random if its complexity is 
    within an additive constant of its length, meaning it is
    impossible to compress using any program (see~\cite{LiV2019} for
    more details). Since computing the Kolmogorov complexity of a
    string is in general undecidable, we can sometimes approximate
    this notion of randomness by fixing a particular compression
    algorithm and looking for strings that can't be compressed using
    this algorithm.

    Define a block compression algorithm as a function
    $C:\Set{0,1}^*→\Set{0,1}^*$ where for each $k$, the set
    $C(\Set{0,1}^k)$ is prefix-free, meaning that $C(x)$ is not a
    prefix of $C(y)$ if $|x|=|y|=k$ and $x≠y$.

    Let's consider two versions of trying to find a string that is
    locally incompressible by such a $C$, depending on whether or not
    we are looking only at substrings of a fixed size or substrings of
    all sizes greater than some minimum.

    \begin{enumerate}
        \item Prove or disprove: There are constants $a>0$ and
            $k_{\min}$ such that for
            any polynomial-time block compression algorithm $C$, there
            is a randomized algorithm that takes as input $n$
            and $k≥k_{\min}$, uses $C$ as a subroutine, and outputs in expected
            time polynomial in $n$ a string $x$ with $\card{x} = n$
            such that for any substring $y$ of $x$ with $\card{y} =
            k$, $\card{C(y)} > k - a \lg k$.
        \item Prove or disprove: There are constants $a>0$ and
            $k_{\min}$ such that for
            any polynomial-time block compression algorithm $C$, there
            is a randomized algorithm that takes as input $n$,
            uses $C$ as a subroutine, and outputs in expected
            time polynomial in $n$ a string $x$ with $\card{x} = n$
            such that for any substring $y$ of $x$ with $\card{y} ≥
            k_{\min}$, $\card{C(y)} > \card{y} - a √{\card{y}}$.
    \end{enumerate}

        \subsubsection*{Solution}

        \begin{enumerate}
            \item 
                The proof I originally put here used the constructive
                version
                of the Lovász Local Lemma from
                Moser-Tardos~\cite{MoserT2010}, as described in
                Corollary~\ref{corollary-symmetric-lovasz-local-lemma-constructive}.
                I'll keep it below in case somebody tried doing
                something similar and wants to see what it looks like. 
                But it turns out there is a much
                easier proof that was suggested to me by Ruben Carpenter.

                For both proofs a necessary lemma is that any
                prefix-free set of strings over $\Set{0,1}$ includes
                at most $2^\ell$ strings of length $\ell$ or less. So
                the set $C(\Set{0,1}^k)$ contains at most $2^{k-a \lg
                k} = 2^k k^{-a}$ strings of length $k-a \lg k$ or
                less, giving a probability of at most $k^{-a}$ that a
                uniform random string of length $k$ compresses to this
                amount.

                It is also helpful to observe that we can assume $k ≤
                n$, because otherwise a string of length $n$ has no
                length-$k$ substrings and the desired property holds
                trivially. This in particular means that any operation
                that is polynomial in $k$ is also polynomial in $n$.

                The problem we still have to deal with is that a long
                enough random string will have many substrings of
                length $k$, so it becomes very likely that one of them
                is compressible. We can deal with this in two ways:
                \begin{enumerate}
                    \item Ruben Carpenter's argument, as I understand
                        it, is that if we
                        make $x$ be periodic, so that $x_{i+k} = x_i$
                        for all $i$, then every length-$k$ substring
                        $y$ of $x$ is a rotation of $x_1 \dots x_k$.
                        If we choose $x_1 \dots x_k$ uniformly at
                        random, then each rotation has a (not
                        independent) probability of at most $k^{-a}$ of being
                        compressible, giving a union bound of
                        $k^{1-a}$ on the probability that any one of the
                        $k$ rotations is compressible. For any $a>1$ and
                        $k≥k_{\min}=2$ this is at most $2^{1-a} < 1$
                        meaning that we
                        can sample a constant number of strings
                        $x_1,\dots,x_k$ on average until we find one
                        that works. Testing all rotations of 
                        each string requires $k$
                        calls to $C$, so the entire process of finding
                        a good sample
                        is polynomial in $k$. We then
                        have to spend time linear in $n$ to print out
                        the rest of $x$.
                    \item 
                        My original proof, which is a bit more work,
                        uses Moser-Tardos.

                        Let $a=2$ and $k_{\min} = 6$.

                        Let $A_i$ be the event that $\card{C(x_i \dots
                        x_{i+k-1})} ≤ k - a \lg k$ if we choose all the bits
                        in $x$ uniformly at random. We have previously
                        shown that $\Prob{A_i} ≤ p = k^{-a}$.

                        For $d$, $A_i$ shares variables with the events
                        $A_{i-k+1}$ through $A_{i+k-1}$; there are at most
                        $2k-2$ such events if we don't can't $A_i$, giving $d
                        = 2k-2$.

                        So 
                        \begin{align*}
                            ep(d+1)
                            &=
                            e k^{-a} (2k-1)
                            \\&=
                            e \frac{2k-1}{k^2}
                            \\&<
                            \frac{2e}{k}
                            \\&≤
                            \frac{2e}{k_{\min}}
                            \\&=
                            2e/6 < 1.
                        \end{align*}

                        So from
                        Corollary~\ref{corollary-symmetric-lovasz-local-lemma-constructive},
                        not only does there exist an outcome in which no $A_i$
                        occurs, but we can find it in at most $m/d$
                        resampling steps on average, where $m=n-k+1$ is the
                        number of $A_i$ events.

                        The only remaining question is how much it costs to
                        resample. After generating the initial sample of $n$
                        bits, we need to make $Θ(n)$ calls to $C$ on
                        substrings of length $k$ to check for $A_i$ events.
                        We don't know exactly how much this costs, but we do
                        know that each call to $C$ takes time $O(k^c)$ for
                        some constant $c$.
                        Each resampling forces us to re-examine $d+1$ events $A_i$, which
                        gives an expected $Θ(m)$ calls to $C$ until reaching a
                        solution.
                        The total expected cost is thus $O\parens*{(n+m) k^c} =
                        O(n^{c+1})$.
                    \end{enumerate}
            \item Disproof: We'll show that for any choice of $a$ and
                $k_{\min}$, there is a polynomial-time block
                compression algorithm $C$ such that for sufficiently
                large $n$, every string $x$ with $\card{x} = n$ has a
                substring $y$ such that $\card{C(y)} ≤ k - a
                √{\card{y}}$.

                Given $a$ and $k_{\min}$, choose $k$ such that
                $k≥k_{\min}$ and $k - a √{k} > 1$. When $\card{y} =
                k$, let
                \begin{align}
                    C(y) &=
                    \begin{cases}
                        0 & \text{if $y = 0^k$} \\
                        1y & \text{otherwise}
                    \end{cases}
                    \label{eq-hw-locally-incompressible-zeros}
                \end{align}

                Let us call $y$ compressible if $\card{y} ≥
                k_{\min}$ and $\card{C(y)}
                ≤ \card{y} - a √{\card{y}}$.
                If this does not hold, call $y$ incompressible.
                Then for any $C$ satisfying
                \eqref{eq-hw-locally-incompressible-zeros}, there are
                exactly $2^k-1$ incompressible strings of length $k$.

                Now consider a string $x$ consisting of $m$ blocks of
                length $k$ each, where $m$ is a constant to be
                determined. If $x$ has no incompressible
                substrings of length $k$, each block is incompressible, and 
                the number of possible choices for $x$ is bounded by
                \begin{align*}
                    (2^k-1)^m
                    &= 2^{mk} ⋅(1-2^{-k})^m
                    \\&≤ 2^{mk} e^{-2^{-k}m},
                    \intertext{and the number of bits needed to encode
                    one of these choices is}
                    \ceil*{\lg \parens*{2^{mk} e^{-2^{-k}m}}}
                    &= mk - \floor*{2^{-k} m \lg e}.
                \end{align*}

                When $\card{x} = mk$ and none of the $m$ blocks in $x$
                is $0^k$, let $C(x) = 1r_x$ where $r_x$ is the binary
                representation of the rank of $x$ among all such
                strings.\footnote{It's not too hard to show that $r_x$
                can be computed in time polynomial in $mk$, but since
                $mk$ is a constant, we can just be lazy and imagine
                that we compute it in constant time using a table
                lookup.} For all other $x$ with $\card{x} = mk$, let
                $C(x) = 1x$.

                We now claim that for sufficiently large (but
                constant!) $m$, any $x$ that has no compressible
                substrings of length $k$ is itself compressible. For
                this to hold, we need $\floor*{2^{-k} m \lg e} ≥
                a √{mk}$. But since the left-hand side of this
                inequality scales as $Ω(m)$
                and the right-hand side as $√{m}$, no matter what
                value of $k$ and $a$ we picked, there is some $m$
                large enough to make it hold.

                So far we have only defined $C(x)$ when $\card{x} = k$
                or $\card{x} = mk$. For all other values of
                $\card{x}$, let $C(x) = x$.

                Using this $C$, there is no randomized polynomial-time
                algorithm that outputs for all inputs $n$ a string $x$
                with $\card{x} = n$ and no compressible substrings 
                of length at least $k_{\min}$, because when $n≥mk$,
                for any possible
                output $x$, either $x$ contains a compressible
                substring of length $k≥k_{\min}$, or
                $y=x_1,\dots,x_{mk}$ is compressible substring of length 
                $mk≥k_{\min}$.
        \end{enumerate}

    \subsection{Censored palindromes}

    You are given an censored string $s ∈ (Σ∪\Set{-})^n$ over some
    constant-sized alphabet $Σ$, where some of the letters have been
    replaced by a blank symbol \texttt{-}. You would like to count how many
    reconstructions of $s$ that can be obtained by replacing blanks
    with elements of $Σ$ contains a palindrome of length $k$ or more,
    where a palindrome is a string that is equal to its reverse.

    For example, given the censored string
    \begin{displaymath}
        \texttt{-t-e-ble-\phantom{}-\phantom{}-ie-\phantom{}-is-w-\phantom{}-\phantom{}-aq-\phantom{}-nw-\phantom{}-zpr-yd-\phantom{}-wy},
    \end{displaymath}
    one of the many possible palindromes that could be hidden here is
    the famous 19-character Napoleonic palindrome
    \begin{displaymath}
        \texttt{ptte\textcolor{red}{ablewasiereisawelba}qhqnwgkzprwydsnwy}.
    \end{displaymath}
    For a less historically significant example, the censored string
    \begin{displaymath}
        \texttt{-ls-\phantom{}-g-\phantom{}-\phantom{}-\phantom{}-rzbr-\phantom{}-\phantom{}-\phantom{}-z-ulg-\phantom{}-e-\phantom{}-\phantom{}-afq-z-q-\phantom{}-gn},
    \end{displaymath}
    can be extended to contain the 20-character palindrome
    \begin{displaymath}
        \texttt{ulsu\textcolor{red}{ygluyzrzbrrbzrzyulgy}uezidafqvzoqwagn}.
    \end{displaymath}

    Show that there is a fully polynomial-time randomized
    approximation scheme for approximating the number of reconstructions of a
    censored string $s$ that contain at least one palindrome of length
    at least $k$.

        \subsubsection*{Solution}

        Karp-Luby (\cite{KarpL1985}, see also
        §\ref{section-Karp-Luby}) can solve this for us, with a few
        adjustments.

        Let $n = \card{s}$ and for each pair $i,j$ with $1≤i<j≤n$ and
        $j-i ≥ k-1$, let $S_{i,j}$ be the set of all reconstructions
        $x$ of $s$ such that $x_i,\dots,x_{i+j-1}$ is a palindrome. Then 
        the set of all reconstructions of $s$ that contain a
        palindrome of length $k$ or more is just $\bigcup S_{ij}$.

        To apply Karp-Luby, we need to be able to count, sample
        uniformly from, and test membership in each $S_{ij}$, all in
        time polynomial in $n$.

        Call two positions $\ell,\ell'$ in $[i,j]$
        \conceptFormat{matched} if they are mapped to each other when
        reversing $x_i,\dots,x_{i+j-1}$; formally, this means that
        $\ell-i = j-1-(\ell'-i)$.

        If there are matched positions $\ell,\ell'$ in $[i,j]$ where
        $s_\ell$ and $s_{\ell'}$ are not blank with
        $s_\ell≠s_{\ell'}$, then $\card{S_{ij}} = 0$ and we can
        discard $S_{ij}$ when computing the union.

        Otherwise, set the value in each blank position matched to a
        non-blank position to its counterpart value. This leaves some
        number $b$ of blank positions not in $[i,j]$ and some number $c$ of
        matched pairs of positions $\ell,\ell'∈[i,j]$ that are both blank,
        possibly including the degenerate case where $\ell=\ell'$.
        There are exactly $\card{Σ}^{b+c}$ ways to fill in all these
        blanks; we can easily sample from one of these reconstructions
        by choosing how to fill them in at random; and we can test for
        membership of $x$ in $S_{ij}$ by checking that $x_\ell =
        s_\ell$ whenever $s_\ell$ is non-blank and that
        $x_i,\dots,x_{i+j-1}$ is a palindrome. Each of these
        procedures takes time linear in $n=\card{s}$.

        So now let's throw Karp-Luby at this. We have $m = O(n^2)$
        sets $S_{ij}$, sampling from each takes $O(n)$ time, and we
        need $N = Θ\parens*{m ε^{-2} \log (1/δ)}$ samples to get
        within $ε$ relative error with probability at least $1-δ$.
        The total cost is thus $O\parens*{n^3 ε^{-2} \log (1/δ)}$,
        which gives us an FPRAS.

\chapter{Sample assignments from Spring 2024}

\section{Assignment 1, due Thursday 2024-02-01 at 23:59}

    \subsection{Matchings}

    A \concept{matching} in a graph is a subgraph where each vertex
    has degree at most $1$, which means that some vertices are paired
    with other vertices. In this problem, we will consider several
    randomized algorithms for quickly constructing a large matching in
    a $d$-regular graph. For each algorithm, your task is to compute
    an exact closed-form expression for the expected number of edges
    $M(n,d)$ in the matching as a function of $n$ and $d$.  

    \begin{enumerate}
        \item In the simplest algorithm, each vertex $u$ chooses one
            of its $d$ neighbors $v$ independently and uniformly at random. Each
            edge $uv$ is included in the matching
            if $u$ chooses $v$ and $v$ chooses
            $u$.
        \item The second algorithm assumes a bipartite $d$-regular graph where $n$ is even and
            the nodes are partitioned into two subsets $S$ and $T$
            of $n/2$ nodes each, with all
            edges going between $S$ and $T$. In this graph, have each
            node $u$ in $S$ choose one of its $d$ neighbors $v$ in $T$
            independently and uniformly at random. Each edge $uv$ is
            included in the matching if $u$ chooses $v$ and no other
            $u'$ chooses $v$.
        \item The third algorithm attempts to extend the
            bipartite-graph algorithm to a general graph. Each node
            $u$ flips an independent fair coin to decide whether it
            will send or receive. Each sender then picks one of its
            $d$ neighbors independently and uniformly at random. An
            edge $uv$ is included in the matching if $u$ is a sender,
            $v$ is a receiver, and $u$ is the only sender that picks
            $v$; or if the same conditions hold with $u$ and $v$ reversed.
    \end{enumerate}

        \subsubsection*{Solution}

        In each case we'll use linearity of expectation.

        \begin{enumerate}
            \item Let $X_{uv}$ be the indicator for the event that
                $uv$ is included in the matching. Then
                \begin{align*}
                    \Exp{X_{uv}} 
                    &= \Prob{\text{$u$ and $v$ are matched}}
                    \\&= \Prob{\text{$u$ chooses $v$ and $v$ chooses $u$}}
                    \\&= \Prob{\text{$u$ chooses $v$}} \Prob{\text{$v$ chooses $u$}}
                    \\&= (1/d)(1/d)
                    \\&= 1/d^2.
                    \intertext{Now sum over all $nd/2$ edges to get}
                    M(n,d)
                    &= (nd/2)(1/d^2) = \frac{n}{2d}.
                \end{align*}
            \item Again let $X_{uv}$ be the indicator variable for the
                event that $uv$ is included. Now
                \begin{align*}
                    \Exp{X_{uv}} 
                    &= \Prob{\text{only $u$ picks $v$}}
                    \\&= (1/d) (1-1/d)^{d-1}.
                    \intertext{Sum over all $nd/2$ edges to get}
                    M(n,d)
                    &= (nd/2)(1/d)(1-1/d)^{d-1} = \frac{n}{2}(1-1/d)^{d-1}.
                \end{align*}

                Using $1-1/d ≤ e^{-1/d}$, we can show that this is at
                least $\frac{n}{2e}$ for any $d$, meaning we match a
                constant fraction of the nodes on average even when
                $d$ is large.
            \item For this version it's convenient to break symmetry
                and make $X_{uv}$ be the indicator variable for the
                event that $u$ is a sender, $v$ is a receiver, and $u$
                is matched with $v$. This means that we will have two
                variables $X_{uv}$ and $X_{vu}$ for each edge, but we
                can deal with this when we need to.

                Compute
                \begin{align*}
                    \Exp{X_{uv}}
                    &= \Prob{\text{$u$ is sender and $v$ is receiver
                    and only $u$ picks $v$}}
                    \\&= (1/2)(1/2)(1/d)\parens*{1-\frac{1}{2d}}^{d-1}
                    \\&= \frac{1}{4d}\parens*{1-\frac{1}{2d}}^{d-1}.
                    \intertext{Sum over all $nd$ directed edges $uv$
                    to get}
                    M(n,d)
                    &= \frac{n}{4}\parens*{1-\frac{1}{2d}}^{d-1}.
                \end{align*}

                It can be shown that this is also $Θ(n)$ regardless of
                $d$, although the constant is not quite as good as in the
                bipartite case.
        \end{enumerate}

    \subsection{Non-volatile memory}

    A manufacturer has designed a non-volatile memory chip based on
    blowing tiny fuses. Each fuse starts out representing a $0$, but
    by running an excess voltage across the fuse it can be permanently
    changed to a new state representing a $1$.

    To allow storing multiple values in the chip, it is arranged as a
    sequence of cells of $k$ bits each, with an extra bit used to mark
    the cell as discarded. Initially a cell will hold all zeros
    ($0000$). When writing a new value $v$, if it is possible to
    overwrite the previous value without having to turn any $1$ into a
    $0$, the same cell will be re-used. Otherwise it will be marked as
    discarded and a new cell will be used for $v$. An example of this
    process is given in Figure~\ref{fig-assignment-nvram}.

    \begin{figure}
        \begin{tabular}{cl}
            Value & New memory contents \\
            \hline
            (Initial) & 0000 \\
            0101 & 0101 \\
            1011 & \st{0101} 1011 \\
            0010 & \st{0101} \st{1011} 0010 \\
            0110 & \st{0101} \st{1011} 0110 \\
            1100 & \st{0101} \st{1011} \st{0110} 1100 \\
            1101 & \st{0101} \st{1011} \st{0110} 1101 \\
        \end{tabular}
        \caption[Non-volatile memory in action]{Non-volatile memory in
        action. The example shows $n=6$ writes with $k=4$ bits per
        cell. A total of 4 cells are used.}
        \label{fig-assignment-nvram}
    \end{figure}

    In the worst case, every new value requires a new cell. But the
    manufacturer hopes that things will go better with enough
    randomness.

    \begin{enumerate}
        \item Suppose that a sequence $v_1,v_2,\dots,v_n$ of $n≥1$
            values is generated uniformly at random, so
            that each $v_i$ is equally likely to be each of the $2^k$
            possible bit-vectors of length $k$. 
            Give an exact
            closed-form expression for the expected number of cells
            used to write these values, as a function of $n$ and $k$.
        \item Since we can't rely on a random input, let's move the
            randomness into the algorithm. Suppose that a fixed
            permutation $π$ on all $k$-bit vectors
            is chosen uniformly at random and each
            $v_i$ (supplied by an adversary) is mapped to $π(v_i)$ before being written. Now what
            is the expected number of cells used in the worst case?
            (As usual, assume that $π$ is chosen after
            $v_1,\dots,v_n$ are fixed.)
        \item The manufacturer complains that storing $π$ is too
            expensive, and suggests storing a random $k$-bit vector
            $x$ instead. Each $v_i$ is now mapped to $v_i ⊕ x$ before
            being stored, where
            $⊕$ represents bitwise XOR. Now what is the expected
            number of cells used in the worst case?
    \end{enumerate}

        \subsubsection*{Solution}

        \begin{enumerate}
            \item Let $v$ and $v'$ be consecutive values. Then $v'$
                requires a new cell if there is some position $j$ such
                that $v'_j = 0$ but $v_j = 1$. For each position $j$,
                this occurs with probability $1/4$ and does not occur
                with probability $3/4$. Because the positions are
                independent, the probability that $v'$ does not
                require a new cell is $(3/4)^k$ and the probability
                that it does require a new cell is $1-(3/4)^k$.

                For each $i>1$, let $X_i$ be the indicator variable
                for the even that $v_i$ requires a new cell. If we let
                $X_1 = 1$, then $S = ∑_{i=1}^{n} X_i$ gives the total
                number of cells needed. We have calculated above that
                $\Exp{X_i} = 1-(3/4)^k$ for $i>1$, so by linearity of
                expectation
                \begin{align*}
                    \Exp{S}
                    &= ∑_{i=1}^{n} \Exp{X_i}
                    \\&= 1 + (n-1)(1-(3/4)^k).
                \end{align*}
            \item Same setup as before but we need to recompute 
                $\Exp{X_i}$. The problem is that $π(v)$ and $π(v')$
                are no longer independent, because the adversary can
                (and should) pick $v≠v'$, which removes one of the
                possible cases for $π(v')$ since $π$ can't send
                unequal $v$ and $v'$ to the same value.

                We can compensate for this using conditional
                expectations. Let $X$ be the indicator for the event
                that $π(v')$ requires a new cell if $π(v)$ and $π(v')$ 
                are chosen independently at random. Then
                \begin{align*}
                    1-(3/4)^k
                    &=\Exp{X}
                    \\&= \ExpCond{X}{π(v)=π(v')} \Prob{π(v)=π(v')}
                       \\& \quad + \ExpCond{X}{π(v)≠π(v')}
                       \Prob{π(v)≠π(v')}
                    \\&= 0⋅2^{-k} + \ExpCond{X}{π(v)≠π(v')}(1-2^{-k}),
                    \intertext{which we can solve to get}
                    \ExpCond{X}{π(v)≠π(v')} &= \frac{1-(3/4)^k}{1-2^{-k}}.
                \end{align*}
                Since the only choice the adversary can make that
                affects $\Exp{X_i}$ or not is to set $v_i ≠ v_{i-1}$,
                this gives a worst case of
                \begin{align*}
                    \Exp{∑ X_i} &= 1 + (n-1)
                    \frac{1-(3/4)^k}{1-2^{-k}},
                \end{align*}
                which is only a little bit worse than the average
                case.

                An alternative approach to computing $\Exp{X}$ is to
                just count the number of pairs $π(v), π(v')$ that
                require a new cell and divide by the total number of
                possibilities $2^k (2^k-1)$. Here there are $3$
                choices for each bit position that don't require a new
                cell, giving $3^k$ choices overall, but $2^k$ of these
                choices make $π(v)=π(v')$. So the number of distinct
                pairs that don't require a new cell is $3^k-2^k$. We
                can get the number of pairs that do require a new cell
                by subtracting from the total $2^k(2^k-1) = 4^k-2^k$.
                This gives $\Exp{X} = \frac{4^k - 3^k}{4^k-2^k} =
                \frac{1 - (3/4)^k}{1-2^{-k}}$ as computed above.
            \item Here things get substantially worse. Suppose that
                $v = 0^k$ and $v'=1^k$. Now $v'⊕x$ requires a new cell
                as long as $x≠0^k$, which occurs with probability
                $1-2^{-k}$. A similar argument holds when $v = 1^k$
                and $v' = 0^k$. So by alternating between all-0 and
                all-1 values, the adversary can force us to use an expected
                $1+(n-1)(1-2^{-k})$ cells. It can't do worse since any
                choice other than switching all bits would give a
                lower expected cost for each write.
        \end{enumerate}

        All of these bounds are terrible, but they are not equally
        terrible and are modestly distinct for small $k$. For $k=2$,
        for example, the coefficient on $(n-1)$ goes from $7/16$ in
        the average case, to $7/12$ in the permutation case, and all
        the way up to $3/4$ in the bitwise XOR case.

\section{Assignment 2, due Thursday 2024-02-15 at 23:59}

    \subsection{Mediocre cuts}

    Recall that a cut in a graph $G$ is a partition of the vertices
    into sets $S$ and $T$, where the size of the cut is the number of
    edges with one endpoint in each set.

    Given a graph $G$ with $n$ vertices and $m$ edges, we'd like to
    construct a cut with size as close as possible to $m/2$.
    We will consider two algorithms that generate $0$-$1$ random variables
    $X_v$ for each vertex $v$, such that $v$ is in $S$ if $X_v=0$ and $v$ is
    in $T$ if $X_v=1$.

    \begin{enumerate}
        \item In the first algorithm, $X_v$ is simply an independent
            fair coin-flip for each $v$.

            Prove or disprove: This algorithm produces a cut of size $m/2 ±
            o(m)$ with constant nonzero probability.
        \item In this second algorithm, the independent fair coins
            are replaced by pairwise independent coins,
            constructed as
            in~§\ref{section-pairwise-independence-construction}:
            Generate $k=\ceil{\lg (n+1)}$ independent fair random bits
            $Y_1,\dots,Y_k$, assign each vertex $v$ a unique non-empty
            subset $A_v$ of these bits, and let $X_v =
            \bigoplus_{i∈A_v} Y_i$.

            Prove or disprove: This algorithm produces a cut of size $m/2 ±
            o(m)$ with constant nonzero probability.
    \end{enumerate}

        \subsubsection*{Solution}

        \begin{enumerate}
            \item We'll give a proof using Chebyshev's inequality.

                For each edge $uv$ let $Z_{uv} = X_u ⊕ X_v$ be the
                indicator variable for the event that $uv$ is in the
                cut. Let $S = ∑_{uv} Z_{uv}$ be the size of the cut.

                We have $\Exp{S} = ∑ \Exp{Z_{uv}} = m/2$.

                To compute $\Var{S}$, we will first show that the
                variables $Z_{uv}$ are pairwise-independent.
                This holds trivially for any variables corresponding
                to non-incident edges. For the case of two incident
                edges $Z_{uv}$ and $Z_{vw}$, 
                observe that $\ExpCond{Z_{vw}}{Z_{uv}} = 1/2$, because
                whatever value $X_v$ has, adding $X_w$ yields $0$ or
                $1$ with equal probability. So $Z_{uv}$ and $Z_{vw}$ are
                independent as well.

                Since the $Z_{uv}$ are pairwise-independent, $\Var{S}
                = ∑ \Var{Z_{uv}} = m/4$.
                Chebyshev then says
                \begin{align*}
                    \Prob{\abs*{S - m/2} ≥ √{m}}
                    &≤ \frac{m/4}{(√{m})^2} = \frac{1}{4}.
                \end{align*}
                
                This gives a $3/4$ chance of getting
                $S$ in the range $m/2 ± √{m} = m/2 ± o(m)$.
            \item 
                For this version we will give a disproof, by
                constructing a family of arbitrarily large graphs
                where the algorithm generates a cut of size $0$ or $m$
                with equal probability.

                Fix some $n$, and create an edge $uv$ for each pair of
                vertices $u$ and $v$ with $A_u = A_v \symdiff
                \Set{1}$. This will give a total of $m=n/2$ edges.
                Each such edge will have $X_u = X_v ⊕ Y_1$; this will
                put $uv$ in the cut precisely when $Y_1 = 1$. So the
                cut will have $0$ edges with probability $1/2$ and $m$
                edges with probability $1/2$.

                In neither case is the size of the cut within $o(m)$
                of $m/2$, so the algorithm does not guarantee this
                with nonzero probability.
        \end{enumerate}

    \subsection{Training costs}

    A total of $n$ workers arrive at a site in blocks of size $n_1,
    n_2, \dots, n_k$, where $∑ n_i = n$. After the $i$-th block
    arrives, one of the $n_1 + n_2 + \dots + n_i$ workers now on site is chosen
    uniformly at random to be the leader. If the leader arrived
    previously (was one of the $n_1+n_2+\dots+n_{i-1}$ workers already
    at the site), they are said to be \emph{experienced} and require
    no training. If the leader was one of the most recent $n_i$
    arrivals, they are \emph{inexperienced} and must be trained. We'd
    like to get an estimate of the worst-case expected cost of training
    inexperienced leaders, assuming that an adversary chooses the
    sizes and number of the blocks $n_1,\dots,n_k$, and show that the
    actual cost is likely to be close to the expected cost.

    We consider two different models for the cost of training an
    inexperienced leader.

    \begin{enumerate}
        \item In the first model, an inexperienced leader remedies
            their ignorance by calling the help line 1-800-LEADER, at
            a cost of one unit.

            Show an upper bound $f(n)$ on the worst-case
            expected total training cost for this model, and show that
            there is are constants $c>0$ and $c' < 1$ such that actual cost in this
            worst case is within $f(n) ± c' f(n)$ for sufficiently
            large $n$ 
            with probability at least $1-O(n^{-c})$.
        \item \textbf{This part of the problem contained an error
            and has been withdrawn. You do not need to supply a
            solution to this part of the problem and will be given
            full credit for it.}
            
\end{enumerate}

        \subsubsection*{Solution}

        \begin{enumerate}
            \item First we need to figure out the worst-case
                choice of $n_1,\dots,n_k$.

                For any particular choice of the block sizes, the
                expected cost is
                \begin{align*}
                    ∑_{i=1}^{k} \frac{n_i}{∑_{j=1}^i n_j}.
                \end{align*}
                It is not hard to see that this quantity is maximized
                when $k=n$ and $n_i = 1$ for all $i$. The proof is to
                observe that if any $n_i > 1$, it can be replaced by
                consecutive blocks of size $n_i - 1$ and $1$, giving
                an expected cost for these two new blocks of
                \begin{align*}
                    \frac{n_i-1}{∑_{j=i}^{} n_i - 1}
                    + \frac{1}{∑_{j=i}^{i} n_i}
                    &> 
                    \frac{n_i-1}{∑_{j=i}^{} n_i}
                    + \frac{1}{∑_{j=i}^{i} n_i}
                    = \frac{n_i}{∑_{j=i}^{} n_i},
                \end{align*}
                which increases the expected total cost.

                Setting each $n_i=1$, the expected total cost is
                \begin{align*}
                    ∑_{i=1}^{n} \frac{1}{i} 
                    &= H_n = Θ(\log n).
                \end{align*}

                Let $X_i$ be the training cost for the $i$-th block.
                Then each $X_i$ is an independent Bernoulli random
                variable and $S=∑ X_i$ has known expectation
                $μ=H_n=Θ(\log n)$,
                so we can use the two-sided Chernoff bound
                \eqref{eq-Chernoff-two-sided-one-third} to get
                \begin{align*}
                    \Prob{\abs*{S-μ} ≥ (1/2)μ}
                    &≤ 2e^{-μ/12} = 2e^{-Θ(\log n)} = O(n^{-c}),
                \end{align*}
                for some $c>0$.

                But then $S$ lies between $(1/2)μ$ and $(3/2)μ$ with
                probability $1-O(n^{-c})$, giving the desired bound.

\end{enumerate}

\section{Assignment 3, due Thursday 2024-02-29 at 23:59} 

    \subsection{A robot rendezvous problem}

    A warehouse for a large online retailer has $n$ packages, each of
    which has a distinct tracking number $t_i ∈ \Set{0,\dots,t-1}$.
    Each package
    starts out in the possession of one of $n$ warehouse robots and
    needs
    to be transferred to one of $n$ corresponding delivery drones to be delivered.
    We will refer to the robots and drones collectively as
    \emph{workers}.

    Unfortunately there are only $m < n$ landing pads for the 
    drones, and the scheduling system that assigns packages to 
    the workers does not know about the landing pads.

    Instead, each worker knows only $n$, $m$, $t$, the unique tracking number
    $t_i$ for its assigned package, whether it is a robot or a drone, and up to
    $b$ bits of randomness that is shared between all the
    workers, where $b$ is polylogarithmic in $t$ (meaning
    that $b = O(\log^c t)$ for some $c$).\footnote{There is a very
    technical issue here involving how the bound on the number of
    bits interpreted.
    One extreme is that we have $b$ bits of randomness and all $2^b$
    possible bit strings are equally likely. The other extreme is that
    some algorithm generates a random variable or collection of random
    variables that collectively take on at most $2^b$ distinct
    possible values (and thus can be encoded in $b$ bits), but there
    is no requirement that each possible value is equally likely.
    While the problem is solvable in either model, there are fewer
    details to worry about in the second model, so you should feel
    free to assume, for example, that you can encode a six-sided die
    in 3 bits by assigning $0$ probability to $000$ and $111$ and
    $1/6$ probability to each of the remaining six bit vectors.}
    Using this information, it must choose one of
    the $m$ landing pads $0,\dots,m-1$ in one of $k$ rounds
    $0,\dots,k-1$.  If exactly one robot and its
    corresponding drone arrive at a particular pad during a round, the
    package is transferred successfully.  If a robot meets a drone
    expecting a different package, or if two robots or two drones
    attempt to use the same pad during the same round, the transfer
    fails and any workers choosing that pad in that round must try again at a later
    round. Each worker has no ability to communicate with
    other workers beyond being able to detect if a transfer it attempted
    failed or not.

    It is clear that $k$ must be at least $\ceil{n/m}$ for it to be
    possible to deliver all the packages even with perfect
    coordination. Show that this is almost sufficient, by giving an
    algorithm that, for any fixed $c>0$, delivers all packages in $k =
    O\parens*{(n/m) \log n}$ rounds with probability at least $1-O(n^{-c})$.

        \subsubsection*{Solution}

        We'll have a sequence of $p = O(\log n)$ phases of
        $\ceil{2n/m}$ rounds each, and in phase use a hash function to
        assign each package to one of the $\ell = m \ceil{2n/m} ≥ 2n$ slots
        corresponding to a particular
        combination of pad and round within the phase.

        A complication is that we have limited randomness, which is
        going to constrain what hash functions we can use. We'll pick
        an independent linear congruential hash function $h_j$ for
        each phase $j$, which requires $O(\log t)$ bits of randomness
        per phase or $O(\log t \log n) = O(\log^2 t)$ bits of
        randomness across all phases. Tabulation hashing also works
        (we want the version that adds table elements mod $\ell$
        rather than using XOR on bit vectors), but the bits per phase
        goes up to $O(\log^2 t)$ giving $O(\log^3 t)$ overall.

        In each phase, a remaining robot or drone assigned tracking
        number $t_i$ computes $h_j(t_i) ∈ [\ell] $ and goes to pad
        $h_j(t_i) \bmod mod m$ during round $\floor{h_j(t_i)/m}$
        (numbering from $0$ within the phase). Each pair of tracking
        numbers $t_i, t_{i'}$ produces a collision with probability
        $\Prob{h_j(t_i) = h_j(t_{i'})} ≤ \frac{1}{\ell}$. Let $X_j$ be
        the number of robots left after $j$ phases; then $X_0 = n$ and
        \begin{align*}
            \ExpCond{X_{j+1}}{X_j}
            &≤ 2 \binom{X_j}{2} \frac{1}{\ell}
            ≤ \frac{X_j^2}{2n}
            ≤ \frac{X_j}{2}.
        \end{align*}
        Here the $2$ at the start accounts for the fact that each
        collision may send up to two robots to the next phase, the
        $\binom{n}{2}$ counts all the pairs of robots, and the last
        step uses the bound $X_j ≤ n$.

        Iterating this inequality gives $\Exp{X_p} ≤ n⋅2^{-p}$.
        Set $p = \lg n$ to get $\Prob{X_p > 0} ≤ \Exp{X_p}
        ≤ 2^{-c \lg n} = n^{-c}$. This gives the desired error
        probability in $c \lg n \ceil{2n/m} = O(n \log n / m)$ rounds.

    \subsection{A linked list}

    It is well-known that linked lists are terrible data structures,
    with only $O(n)$ guarantees on search time in the worst case.
    However, for some distributions on elements, the average case
    might be better.

    Suppose we are given a set of $n$ elements $1,\dots,n$, where at
    each step element $i$ is supplied with probability $p_i$.
    Suppose that we insert each new element we see at the end of the
    list, until eventually all $n$ elements appear in the list. We
    then sample a new element according to the given probability
    distribution and ask what its expected position in the list is.

    For a uniform distribution, this won't help much: all $n!$
    orderings will be equally likely, and the expected position of a
    random element is just $\frac{n+1}{2} = Θ(n)$, which is (up to
    constants) 
    no better than
    the deterministic worst case. But for more skewed distributions we
    may hope that more probable elements get inserted early, meaning
    that the cost of searching for them will be less than for more
    improbable elements.

    For each of the following distributions, compute a tight (big-$Θ$)
    asymptotic bound on the expected cost to search for a random
    element given by the distribution, assuming that we constructed
    the list as described above by repeatedly sampling from the same
    distribution.

    \begin{enumerate}
        \item A geometric distribution with $p_i \propto 2^{-i}$.
\item A \concept{Zipf's law} distribution with $p_i \propto
            \frac{1}{i}$.
    \end{enumerate}

        \subsubsection*{Solution}

        Let's do what we can for a generic distribution before getting
        into the details of each distribution individually.

        Let $A_{ij}$ be the indicator variable for the event that $i$
        appears in the list at or before the same position as $j$ (the
        ``at'' part covers the case $A_{ii}$, which we take to be $1$).
        Let $D_i$ be the position of $i$ in the list. Then $D_i = ∑_j
        A_{ji}$.

        For $i≠j$, we can compute $\Exp{A_{ij}}$ by looking at
        probability of seeing $i$ conditioned on seeing $i$ or $j$ at
        a particular step. This gives 
        \begin{align*}
            \Exp{A_{ij}} &= \frac{p_i}{p_i+p_j},
            \\
            \Exp{D_i} &= 1 + \sum_{j≠i} \frac{p_j}{p_i+p_j}.
            \intertext{and thus}
            \Exp{\text{search cost}}
            &= \sum_i p_i \Exp{D_i}
            = 1 + \sum_i p_i \sum_{j≠i}
            \frac{p_j}{p_i+p_j}.
        \end{align*}

        Let's see what happens to this for each of our given
        distributions.

        \begin{enumerate}
            \item 
                When $p_i \propto 2^{-i}$, we have
                \begin{align*}
                    p_i &= \frac{2^{-i}}{1-2^{-n-1}} = Θ\parens*{2^{-i}}.
                    \intertext{Even better, the denominators cancel
                    while computing}
                    ∑_{j≠i} \frac{p_j}{p_i + p_j}
                    &= ∑_{j≠i} \frac{2^{-j}}{2^{-i} + 2^{-j}}
                    \\&= 
                    ∑_{j<i} \frac{2^{-j}}{2^{-i} + 2^{-j}}
                    +
                    ∑_{j>i} \frac{2^{-j}}{2^{-i} + 2^{-j}}
                    \\&= 
                    ∑_{j<i} \frac{2^{-j}}{Θ\parens*{2^{-j}}}
                    +
                    ∑_{j>i} \frac{2^{-j}}{Θ\parens*{2^{-i}}}
                    \\&= 
                    ∑_{j<i} Θ(1)
                    +
                    ∑_{j>i} Θ(2^{i-j})
                    \\&= Θ(i) + Θ(1)
                    \\&= Θ(i).
                \end{align*}

                We can now compute the expected search cost as
                \begin{align*}
                    ∑_i p_i \Exp{D_i} &= 1 + ∑_i Θ(2^{-i} i) = Θ(1).
                \end{align*}
\item 
                In this case, we have 
                \begin{align*}
                    p_i &= \frac{1/i}{H_n} = \frac{1}{i H_n}.
                    \intertext{But the $H_n$'s cancel out when we compute}
                    ∑_{j≠i} \frac{p_j}{p_i + p_j}
                    &=
                    ∑_{j≠i} \frac{1/j}{1/i + 1/j}
                    \\&=
                    ∑_{j≠i} \frac{i}{i+j}.
                \end{align*}

                We can use this to get an upper bound on the expected
                search cost
                \begin{align*}
                    ∑_{i=1}^{n} p_i \Exp{D_i}
                    &=
                    ∑_{i=1}^{n} 
                    \frac{1}{i H_n} \parens*{1 + ∑_{j≠i} \frac{i}{i+j}}
                    \\&< 1 + \frac{1}{H_n} ∑_{i=1}^{n} ∑_{j=1}^{n}
                    \frac{1}{i+j}
                    \\&< 1 + \frac{1}{H_n} ∑_{k=1}^{2n}
                    \frac{k-1}{k}
                    \\&< 1 + \frac{1}{H_n}⋅2n
                    \\&= O(n / \log n).
                \end{align*}

                To get a matching lower bound, notice that the best
                possible ordering places each $i$ at position $i$,
                giving an expected search cost of exactly $∑_i
                \frac{1}{i H_n} ⋅ i = n / H_n = Ω(n / \log n)$.
                Since whatever random ordering we end up with is at
                least this bad, this gives us the $Θ(n / \log
                n)$ bound we are looking for.

        \end{enumerate}

\section{Assignment 4, due Thursday 2024-03-28 at 23:59} 

    \subsection{Nearly orthogonal vectors}

    Given two vectors $x$ and $y$, the angle $θ_{xy}$ between them can be
    computed by the formula $θ_{xy} = \cos^{-1} \frac{x⋅y}{\norm{x}
    \norm{y}}$. If $x$ and $y$ are orthogonal, then $x⋅y = 0$, and
    $θ_{xy} = π/2$. Let's call two vectors $x$ and $y$ \concept{nearly
    orthogonal}\index{orthogonal!nearly} to within $ε$ if $π/2 - ε ≤ θ_{xy} ≤ π/2 +
    ε$.

    Given an $n×n$ matrix $A$, we can find all the pairs of rows
    $A_i,A_j$ that are nearly orthogonal by computing $θ_{A_i,A_j}$
    for each pair and comparing it to $π/2±ε$.  Unfortunately, this
    takes $Θ(n^3)$ time if done in the obvious way. So let's allow a
    randomized approximation algorithm.

    Claim: For any fixed $ε$ with $0 < ε ≤ π/4$, there is a randomized algorithm that
    takes as input an $n×n$ matrix $A$ and an error parameter $δ > 0$,
    and returns in time $O(n^2 \log^c n \log^c (1/δ))$, for some
    constant $c$, a list of nearly-orthogonal
    pairs of rows of $A$, such that with probability at least $1-δ$,
    this list includes every pair of rows
    that are nearly orthogonal to within $ε$,
    and only pairs of rows that are nearly orthogonal to within $2ε$.

    Prove this claim.
    
        \subsubsection*{Solution}

        This is a job for the Johnson-Lindenstrauss lemma 
        (distributional version).

        First normalize the rows of $A$ so that each $\norm{A_i} = 1$.
        This takes $O(n^2)$ time and doesn't affect the angles between
        rows.

        Next, observe for any unit vectors $x$ and $y$, $\norm{x-y}_2$
        is an increasing function of $θ = θ_{xy}$.  We don't actually
        need anything more than this, but if we want to we can compute
        the distance exactly by constructing an isosceles triangle with $x$
        and $y$ as the unit edges and chopping it in half: this gives
        $\norm{x-y}_2 = 2 \sin \frac{θ}{2}$. So if $θ ≤ ε$,
        $\norm{x-y}_2 ≤ 2 \sin \frac{ε}{2}$, and if $θ ≥ 2ε$,
        $\norm{x-y}_2 > 2 \sin ε$.

        Pick a threshold halfway between these two bounds.
        By choosing a small enough error term in
        Lemma~\ref{lemma-Johnson-Lindenstrauss-distributional}, we can
        guarantee that any pair $x,y$ that is nearly orthogonal to
        within $ε$ has $\norm{f(x-y)}_2$ close enough to
        $\norm{x-y}^2$ to be within this threshold, and any pair
        $x,y$ that is not nearly orthogonal to within $2ε$ doesn't.
        (We could calculate what the exact error bound we need for
        this, but since it's a constant, we don't care.) If we set the
        probability of exceeding the error bound for a single vector
        to $δ/\binom{n}{2}$, this gives a probability that
        $\norm{A_i-A-j}^2$ exceeds the relative error bound for any
        $i,j$ of at most $δ$. To obtain $δ/\binom{n}{2}$ probability
        of error, we need $k = O(\log (δ/n^2)) = O(\log n + \log δ)$.

        So now we need to feed all of our rows to the
        Johnson-Lindenstrauss transform whose existence is given in
        the lemma and then check the threshold for each pair of rows
        in the output. Expanding out the steps, we must:
        \begin{enumerate}
            \item Compute an $n×k$ transform matrix $B$ implementing
                $f$. This takes $O(nk)$ time.
            \item Compute $AB$. This takes $O(n^2 k)$ time using
                ordinary matrix multiplication.
            \item For each pair of rows $i$ and $j$, compute
                $\norm{(AB)_i - (AB)_j}^2$ and check it against the
                thresholds. This takes $O(k)$ time per pair of rows
                for $O(n^2 k)$ time total.
        \end{enumerate}

        Adding up all of these costs gives $O(n^2 k) = O(n (\log n +
        \log (1/δ)))$ time, which is within the claimed bounds.

    \subsection{Boosting a random walk}

    Suppose you are made to participate in a fair $±1$ random walk,
    starting with $X_0 = k ≤ n/2$, that ends if $X_t ≥ n$ or $X_t ≤ 0$. The
    twist is that once and only once during the walk only you can push a
    ``boost'' button that replaces the normal transition rule $X_{t+1}
    = X_t ±1$ with a rule that doubles the current value: $X_{t+1} = 2X_t$.
    Naturally, your choice to use your boost can't depend on knowledge
    of the future: when choosing to set $X_{t+1} = 2X_t$, you only
    know the outcomes $X_1,\dots,X_t$.

    \begin{enumerate}
        \item Suppose that your goal is to maximize the probability
            that you reach a state $X_t≥n$ before you reach $X_t = 0$. What
            strategy should you use and what probability of reaching
            $X_t≥n$ (as a function of $k$ and $n$) does it give you?
        \item Suppose instead that you want to \emph{minimize} the
            probability that you reach a state $X_t≥n$ before you
            reach reach $0$, but you want to do this in a way that
            always uses the boost before you reach $X_t ≥ n$ or $X_t =
            0$,
            so that suspicious onlookers won't think you
            aren't trying. Now what strategy should you use, and what
            probability of reaching $X_t≥n$ does it give you?
    \end{enumerate}

        \subsubsection*{Solution}

        Let's formalize things a bit. Let $\Set{Δ_t}$ be the set of
        independent fair $±1$ increments, so that $X_{t+1} = X_t +
        Δ_{t+1}$ if we don't use the boost. Then we can let $ℱ_t =
        \Tuple{∆_1,\dots,Δ_t}$ and make the time $σ$ at which we use
        the boost a stopping time with respect to $\Set{ℱ_t}$. This
        makes $[σ≤t]$ measurable $ℱ_t$ for each $t$ and gives
        $\ExpCond{Δ_{t+1}}{ℱ_t} = 0$.

        We can now write the transition rule compactly as
        \begin{align*}
            X_{t+1} &= 2^{[σ=t]} X_t + [σ≠t] Δ_{t+1},
        \end{align*}
        where $[A]$ denotes the indicator variable for the event $A$.

        Because of the boost, $\Set{X_t}$ is not a martingale with
        respect to $\Set{ℱ_t}$, but we can define a process that is.

        Let $Y_t = 2^{[σ≥t]} X_t$. The idea is that the first factor
        represents whether we still have a boost available to use or
        not. 

        Note that $[σ≥t]$ starts at $1$ when $t=0$ and drops to
        $0$ when we use the boost. We can write this as $[σ≥t+1] =
        [σ≥t] - [σ=t]$.
        This lets us compute
        \begin{align*}
            \ExpCond{Y_{t+1}}{ℱ_t}
            &= \ExpCond{2^{[σ≥t+1]} X_{t+1}}{ℱ_t}
            \\&= \ExpCond{2^{[σ≥t+1]} \parens*{2^{[σ=t]}X_t +
            [σ≠t] Δ_{t+1}}}{ℱ_t}
            \\&= 2^{[σ≥t]-[σ=t]}⋅2^{[σ=t]} X_t + 2^{[σ≥t+1]} [σ≠t] \ExpCond{Δ_{t+1}}{ℱ_t}
            \\&= 2^{[σ≥t]} X_t
            \\&= Y_t.
        \end{align*}
        So $\Set{Y_t}$ is a martingale with respect to $\Set{ℱ_t}$.

        (Here we could pull $X_t$, $2^{σ≥t}$, $2^{σ≥t+1} = 2^{[σ>t]}$, and $[σ≠t]$ out of
        the conditional expectation because all these variables are
        measurable $ℱ_t$.)

        Let $τ = \min\SetWhere{t}{X_t ≥ n ∨ X_t = 0}$. Then $τ$ is
        finite with probability $1$ (if nothing else, we eventually
        take $n$ positive steps in a row), and $\abs*{Y_i} ≤ 2n$, so
        we can apply the finite probability / bounded range version of
        the Optional Stopping Theorem to get
        \begin{align*}
            \Exp{Y_0} 
            &= \Exp{Y_τ}
            \\&= \ExpCond{Y_τ}{X_τ ≥ n} \Prob{X_τ ≥ n}
            +  \ExpCond{Y_τ}{X_τ = 0} \Prob{X_τ = 0}
            \\&= \ExpCond{Y_τ}{X_τ ≥ n} \Prob{X_τ ≥ n}.
        \end{align*}
        Since we know $\Exp{Y_0} = 2k$, this gives
        \begin{align}
            \Prob{X_τ ≥ n}
            &= \frac{2k}{\ExpCond{Y_τ}{X_τ ≥ n}}.
            \label{eq-boosted-random-walk-probability}
        \end{align}

        We should thus try to minimize or maximize $\ExpCond{Y_τ}{X_τ≥n}$
        depending on whether we want to maximize or minimize
        $\Prob{X_τ≥n}$.

        \begin{enumerate}
            \item To minimize $\ExpCond{Y_t}{X_τ≥n}$, observe that the
                smallest possible value for $Y_τ$ given $X_τ ≥ n$
                is $Y_τ = X_τ = n$. We can get this by using the boost
                immediately ($σ=0$). Then we have a standard fair
                random walk starting from $X_1 = 2k$, which hits $n$
                before $0$ with probability exactly $\frac{2k}{n}$.

                This is not necessarily the only good strategy, since any
                strategy that guarantees $σ < τ$ and $X_σ ≤ n/2$ will
                also get $\frac{2k}{n}$, but
                \eqref{eq-boosted-random-walk-probability} tells us
                that no strategy can do better.
            \item To maximize $\ExpCond{Y_τ}{X_τ≥n}$ subject to the
                requirement that we must use the boost, we want
                $\ExpCond{X_τ}{X_τ≥n}$ to be as large as possible. The
                best we can hope to get is $X_τ = 2(n-1)$, by
                using the boost on when $X_t = n-1$. But what if we
                don't reach $n-1$ and are forced to use the boost
                anyway?

                Analyzing this using the $Y_t$ martingale turns out to
                be tricky, and an easier approach is just to guess the
                worst strategy and show that it is in fact the worst. Since we
                must use the boost before $τ$, we are forced to use
                the boost if we reach $1$ or $n-1$, since otherwise
                there is a possibility the process finishes before we
                use it. Let's suppose we adopt a maximum-delay
                strategy and set $σ = \min\SetWhere{t}{t ∈ \Set{1,n-1}}$.

                Starting from any position $x$, this $σ$ gives
                $\Prob{X_τ = n}$ of $2/n$ if we reach $1$ first and $1$ if we
                reach $n-1$ first. Which we reach first is just the
                probability of hitting on or the other absorbing
                barrier in a simple random walk in $[1,n-1]$.
                So we can compute
                \begin{align*}
                    \ProbCond{X_τ = n}{X_t = x, t < τ}
                    &= \frac{x-1}{n-2} ⋅ 1 + \parens*{1 - \frac{x-1}{n-2}}⋅\frac{2}{n}
                    \\&= \frac{x-1}{n-2} \parens*{1 - \frac{2}{n}} +
                    \frac{2}{n}
                    \\&= \frac{x-1}{n-2} \parens*{\frac{n-2}{n}} + \frac{2}{n}
                    \\&= \frac{x-1}{n} + \frac{2}{n}
                    \\&= \frac{x+1}{n}.
                \end{align*}

                In particular, starting at $X_0 = k$ gives a
                probability of hitting the upper barrier of
                $\frac{k+1}{n}$.

                Now suppose we consider some other strategy $σ'$.
                Like $σ$, $σ'$ must use the boost at $X_t = 1$ or $X_t
                = n-1$.
                If $X_{σ'} = x$ for some $x$ with $1 < x < n-1$, then
                $\ProbCond{X_τ ≥ n}{X_σ = x} = \frac{2x}{n} >
                \frac{x+1}{n}$ since $x>1$. So any other $σ'$ 
                gives a higher probability of $X_τ ≥ n$.
        \end{enumerate}

\section{Assignment 5, due Thursday 2024-04-11 at 23:59}

    \subsection{Relaxation time for Metropolis-Hastings}

    Suppose we have a lazy irreducible reversible Markov chain on $n$
    states with
    transition probabilities $p_{ij}$, uniform stationary distribution
    $π_i = 1/n$, and relaxation time $τ_2$. Fix some function $f$ with
    $1≤f(x)≤r$ for all states $x$ and apply Metropolis-Hastings to get
    a new Markov chain with transition probabilities $p'_{ij} =
    p_{ij} \min\parens*{1,f(j)/f(i)}$ and stationary distribution $π'_i
    \propto f(i)$. Let $τ'_2$ be the relaxation time of this new
    chain.

    Prove or disprove: Under these assumptions, there is an
    upper bound on $τ'_2$ that is 
    polynomial in $τ_2$ and $r$.

        \subsubsection*{Solution}

        We will prove this by showing that $π'_u$ and $p'_{uv}$ are
        both bounded relative to their unmodified versions by
        polynomial functions of $r$, then show this leads to at most
        polynomial blowup
        going from $τ_2$ to the
        conductance $Φ$ of the original chain to the conductance $Φ'$
        of the modified chain and finally to $τ'_2$.

        Let $n$ be the number of states.
        For $π'_u$, we have
        \begin{align*}
            π'_u &= \frac{f(u)}{∑_v f(v)} ≥ \frac{1}{(n-1)r+1} >
            \frac{1}{nr} = \frac{1}{r} π_u
            \intertext{and in the other direction}
            π'_u &= \frac{f(u)}{∑_v f(v)} ≤ \frac{r}{(n-1)+r} <
            \frac{r}{n} = r π_u.
        \end{align*}

        For $p'_{uv} = p_{uv} \min\parens*{1,f(v)/f(u)}$, we
        get $\frac{1}{r} p_{uv} ≤ p'_{uv} ≤ p_{uv}$.

        So now let's look at some set of states $S$ and try to bound
        $Φ'(S)$.
        \begin{align}
            Φ'(S) &= \frac{∑_{u∈S,u∉S} π'_u p'_{uv}}{π'(S)}
            \nonumber
            \\&> \frac{∑_{u∈S,u∉S} \frac{1}{r} π_u ⋅ \frac{1}{r} p_{uv}}{r π(S)}
            \nonumber
            \\&= \frac{1}{r^3} \frac{∑_{u∈S,u∉S} π_u p_{uv}}{π(S)}
            \nonumber
            \\&= \frac{1}{r^3} Φ(S).
            \label{eq-hw-metropolis-conductance-bound}
        \end{align}

        We'd like to use this to bound $Φ' = \min_{0 < π'(S) ≤ 1/2}
        Φ'(S)$ in terms of $r$ and $Φ = \min_{0 < π(S) ≤ 1/2} Φ(S)$,
        but there is a complication: there may be sets $S$ with $π'(S)
        ≤ 1/2$ but $π(S) > 1/2$ that are included in the computation
        of $Φ'$ but not $Φ$. For these sets we will need to take
        advantage of reversibility.

        Suppose $π'(S) ≤ 1/2$, $π(S) > 1/2$. Let $T = \overline{S}$. 
        Then $π(T) = 1-π(S) < 1/2$, so $Φ(T) ≤ Φ$.
        Now we can argue
        \begin{align}
            Φ'(S) &= \frac{∑_{u∈S,v∈T} π'_u p'_{uv}}{π'(S)}
            \nonumber
            \\&= \frac{∑_{v∈T,u∈S} π'_v p'_{vu}}{π'(S)}
            \nonumber
            \\&= Φ'(T) \frac{π'(T)}{π'(S)}.
            \nonumber
            \\&> \frac{1}{r^3} Φ(T) \frac{π'(T)}{π'(S)}.
            \nonumber
            \\&> \frac{1}{r^3} Φ ⋅ \frac{1/2}{1/2}
            \nonumber
            \\&= \frac{1}{r^3} Φ.
            \label{eq-hw-metropolis-conductance-bound-reversibility}
        \end{align}

        So given $S$ with $0 < π'(S) ≤ 1/2$,
        \eqref{eq-hw-metropolis-conductance-bound} shows $Φ'(S) ≥ r^{-3}
        Φ$ when $π(S) ≤ 1/2$ and
        \eqref{eq-hw-metropolis-conductance-bound-reversibility} shows
        $Φ'(S) ≥ r^{-3} Φ$ when $π(S) > 1/2$. In either case we get $Φ' =
        \min_{0 < π'(S) ≤ 1/2} Φ'(S) ≥ r^{-3} Φ$.

        From \eqref{eq-conductance-relaxation-time}, we have $\frac{1}{2Φ}
        ≤ τ_2$, which gives $Φ ≥ \frac{1}{2τ_2}$. But then we can apply
        the other direction of \eqref{eq-conductance-relaxation-time} to
        get
        \begin{align*}
            τ'_2
            &≤ \frac{2}{(Φ')^2}
            < \frac{2}{(r^{-3} Φ)^2}
            ≤ \frac{2r^6}{(1/(2τ_2))^2}
            = 8 r^6 τ_2^2,
        \end{align*}
        which is polynomial in $τ_2$ and $r$.

        Though it doesn't affect the solution to the problem, it is
        probably worth noting here that there may be a much more efficient
        way to sample proportional to $f$, which is to sample a point
        uniformly (in time $O(τ_2 \log n/ε)$) and then apply rejection
        sampling to adjust the probabilities. This yields a sample in
        $O(r τ_2 \log n/ε)$ expected time in the worst case where $f$ is $1$ for most
        points and we only return a sample with probability $1/r$. So
        we may wish to reserve Metropolis-Hastings for situations where
        $f$ has a much larger range (making rejection sampling too costly)
        and we can show a tighter bound on convergence time in some other
        way.

    \subsection{A constrained random walk}

    Let's imagine we want to sample $n$-bit strings with no
    consecutive pairs of
    ones. So $00101$ is acceptable but $01101$ is not.

    It's possible to do this efficiently and exactly by taking
    advantage of the connection of these strings to Fibonacci numbers,
    but for some unexplained reason we decide to do this using Markov
    Chain Monte Carlo instead. Given an initial string $x$,
    pick an index $i$ and bit $b$ uniformly at random, and set $x_i$
    to $b$ unless this would produce two consecutive ones. It is not
    hard to show that this gives an irreducible aperiodic Markov chain
    that converges to a uniform stationary distribution.

    Show that this chain has the rapid mixing property, which means
    that
    $t_{\mix}(ε)$ is polylogarithmic in the number of states $N$ and
    the total variation distance bound $ε$.

        \subsubsection*{Solution}

        First observe that the number of possible states $N$ is at least
        $2^{\ceil{n/2}}$, the number of strings with zeros in all
        odd-numbered positions.\footnote{This is a rather crude
        estimate that is convenient mostly because itit is instantly
        recognizable as exponential in $n$. As hinted at in the problem
        statement, we can compute $N$ exactly using Fibonacci
        numbers. Each acceptable string of length $2$ or more either
        starts with $0$
        followed by an acceptable string of length $n-1$, or starts
        with $10$ followed by an acceptable string of length $n-2$.
        This gives a recurrence $N(n) = N(n-1) + N(n-2$ with $N(0) =
        1$ and $N(1) = 2$. This is exactly the recurrence for
        Fibonacci numbers shifted so that $N(n) = F_{n+2}$, which is
        not surprising when we recall that Pingala
        invented Fibonacci numbers precisely to count sequences of
        syllables of length $1$ or $2$ adding up to a given total.} So if we can get a mixing time that is
        polynomial in $n$, it will be polylogarithmic in $n$.

        Coupling and canonical paths are both options for showing such a
        bound.

        One possible coupling is to apply the same update rule to both
        the $X$ and $Y$ processes: Pick an index $i$ and a bit $i$ and
        update each of $X_i$ and $Y_i$ to equal $b$ if permitted by
        the no consecutive ones rule. We can then track how the
        Hamming distance between $X$ and $Y$ evolves over time. 

        Let $S^t = \SetWhere{i}{X^t_i≠Y^t_i}$ and let $H^t =
        \card*{S^t}$. Then $H^t$ can
        change in two ways:
        \begin{itemize}
            \item If we pick some $i∈S$, then $H$ always drops
                by $1$ if $b=0$, and drops by $1$ if $b=1$ and neither
                $i-1$ nor $i+1$ is in $S$.
            \item If we pick some $i∉S$, then $H$ stays the same if
                $i-1$ and $i+1$ are both not in $S$ and goes up by $1$
                if either is in $S$.
        \end{itemize}

        Partition $S$ into segments $S_1,\dots,S_k$ where each $S_k$
        is a maximal sequence of consecutive indices.

        If $\card{S_i} = 1$, then the index in $S_i$ contributes a
        $\frac{1}{n}$ chance of lowering $H$, and the adjacent indices
        together contribute at most a $\frac{1}{n}$ chance of increasing $H$.
        
        If $\card{S_i} = k > 1$, then each index in $S_i$ contributes a
        $\frac{1}{2n}$ chance of lowering $H$, for a total of
        $\frac{k}{2n} ≥ \frac{1}{n}$, and the adjacent indices
        together contribute at most a $\frac{1}{n}$ chance of increasing $H$.

        In either case we have $\ExpCond{H^{t+1}}{X^t,Y^t} ≤ H^t$, 
        with at least a $\frac{2}{n}$ chance that $H$ changes if $H>0$.
        We also have that $H$ never exceeds $n$ and never changes 
        once it reaches $0$. So we can bound $H$ by a fair random walk
        $H'$ with a reflecting barrier at $n$ and absorbing barrier at
        $0$, that only takes a step when $H$ does. This shows that $H$
        reaches $0$ in $O(n^3)$ steps on average.

        For canonical paths,
        given two states $x$ and $y$ that differ in $k$ places, we'll
        define a path $γ_{xy}$ of length $k$ that fixes these bits one
        at a time. We can't do strict left-to-right bit fixing,
        because this may create an intermediate state that has two
        consecutive ones. But we can adjust the order a little to get
        around this: at each step, change the leftmost unfixed bit
        that can be changed without violating the no-consecutive ones
        rule.

        A step in such a path changes some string of the form
        $y_1 \dots y_\ell rs x_{\ell+3} \dots x_n$
        to
        $y_1 \dots y_\ell r's' x_{\ell+3} \dots x_n$,
        where the $rs→r's'$ change is one of $00→10$, $01→00$, or
        $10→00$. To reconstruct $x$ and $y$ from this information, 
        it is enough to provide $\ell$ ($n-2$ choices), a string $x_1 \dots
        x_{\ell+1} 0 y_{\ell+3} \dots y_n$ with no consecutive ones ($<N$ choices), and the
        values of $x_{\ell+2}$ and $y_{\ell+2}$ (4 choices). 
        This gives $<4nN$ paths crossing each edge, so
        the congestion $ρ$ is
        \begin{align*}
            \max_{uv ∈ E} \frac{1}{π_{u}p_{uv}} ∑_{γ_{xy} \ni uv} π_{x} π_{y}
            &< \frac{1}{N^{-1} (1/(2n))} ⋅ 4nN ⋅ N^{-2} = 8n^2.
        \end{align*}

        We thus get $τ_2 ≤ 8ρ^2 = O(n^4)$, with a hideous constant
        swept behind the big $O$ for decency's sake. This again is
        polylogarithmic in $N$, so we're done.

\section{Assignment 6, due Thursday 2024-04-25 at 23:59} 

    \subsection{The power of intransigence}

    Let $G$ be a connected graph on $n$ vertices labeled $1,\dots,n$,
    and imagine each vertex $v$ in $G$ represents a person who has an
    opinion $X^t_v∈\Set{0,1}$ at each time $t$. This opinion is
    initially arbitrary, but may evolve over time to match the
    opinions of $v$'s neighbors. 

    Specifically, at each step, we pick one of the $m$ edges in the
    graph uniformly at
    random, and then choose a random orientation of that edge to get
    an ordered pair of
    adjacent vertices $ij$, with each possible pair chosen with
    probability $\frac{1}{2m}$. Most of the time, this results in $j$
    adopting $i$'s opinion, so that $X^{t+1}_j = X^t_i$, while
    $X^{t+1}_k = X^t_k$ for all $k≠j$.

    But not if $j=1$. Vertex $1$ is stubborn: headstrong, willful,
    convinced in its bones that it is right, holding
    a graduate degree in obstinacy in one hand and a thesaurus opened
    to ``stubborn'' in the other to back up its
    unyielding pig-headed beliefs.
    If we are unlucky enough to pick $j=1$, then nothing changes and $X^{t+1} = X^t$.

    Running this process long enough will eventually converge to a
    state where every other vertex gives up and agrees with vertex $1$.
    There is even a constant $c$ such that it reaches this state in
    $Θ(n^c)$ steps on average with the worst possible $G$ and the
    worst possible $X^0$.

    Find $c$, and prove that this bound holds.

        \subsubsection*{Solution}

        We'll show $c=4$.

        For the lower bound, we just need one bad family of graphs
        and starting configurations that gives $Ω(n^4)$ expected
        convergence time. One choice is a lollipop graph, consisting
        of a handle that is a path of $n/2$ vertices and a sucker that
        is $K_{n/2}$, plus an edge linking one end of the path to one
        vertex in the clique.

        Let's make the stubborn node be a member of the clique with
        initial value $0$, and start in a configuration where the
        most distant $n/4$ nodes in the path have $1$ and everybody
        else has $0$. Let $X^t_i$ be the value of node $i$ at time $t$
        and let $Z_t = ∑_i X^t$. Then as long as all the clique nodes
        have $X^t_i = 0$ and some path node has $X^t_i = 1$, we have a
        configuration where the $1$ nodes are exactly the $Z_t$ most
        distant nodes on the path. So $Z^{t+1} = Z_t$ only if we pick
        the unique $0-1$ edge, and we have $Z^{t+1} = Z_t ± 1$ with
        probability $\frac{1}{2m}$ each.

        This give an unbiased random walk that takes steps with
        probability $\frac{1}{m}$. It takes $(n/4)^2$ steps for this
        to reach $Z_t = 0$ or $Z_t = n/2$, which is a lower bound on
        the time for the process as a whole to converge. The waiting
        time for each random walk step is $m$, so the total expected
        time to reach $Z_t = 0$ or $Z_t = n/2$ is $m(n/4)^2$ by Wald's
        Equation. The clique makes $m = Ω(n^2)$, so the expected time
        to converge is $Ω(n^4)$.

        In the other direction, we will show that any graph reaches
        an all-equal state from any initial state in time $O(n^4)$.
        For this argument it's convenient to assume that the stubborn
        node starts with $1$, so we finish when $Z_t = n$.

        Fix some configuration $X^t$, and let $\ell$ be the number of
        $0-1$ edges in $X^t$. Let $ℱ_t = \Tuple{X^0,X^1,\dots,X^t}$ be
        the $σ$-algebra generated by the trajectory of the system up
        until time $t$.
        Then $\Prob{Z^{t+1} - Z_t = 1}{ℱ_t} = \frac{\ell}{2m}$ and
        $\Prob{Z^{t+1} - Z_t = -1}{ℱ_t} ≤ \frac{\ell}{2m}$ (because
        edges incident to the stubborn vertex don't contribute). The
        rest of the time $Z^{t+1} = Z_t$.

        This gives a $±1$ random walk that may be biased toward $+1$
        and that takes steps only some of the time. There are two
        mostly equivalent ways to bound the convergence time of this
        process: one uses an explicit potential function, and the
        other reduces to known results about random walks via a
        coupling argument. We'll describe both below, since either
        gives a reasonable solution to this part of the problem.

        \paragraph{Potential function argument}
        Consider the function $Z_t^2$.
        The expected change in this quantity is
        \begin{align*}
            \ExpCond{Z_{t+1}^2 - Z_t^2}{ℱ_t}
            &≥
            (2Z_t+1) \ProbCond{Z_{t+1} = Z_t + 1}{ℱ+t} 
            \\&\quad
            +
            (-2Z_t+1) \ProbCond{Z_{t+1} = Z_t - 1}{ℱ+t} 
            \\&≥
            (2Z_t+1) \frac{\ell}{2m}
            +
            (-2Z_t+1) \frac{\ell}{2m}
            \\&=
            \frac{\ell}{m} 
            \\&≥ \frac{1}{m}.
        \end{align*}
        (The second step uses $Z_t ≥ 1$, as enforced by the stubborn
        vertex.)

        If we define $Y_t = Z_t^2 - \frac{t}{m}$, then $Y_t$ is a
        submartingale, since the expected increase in $Z_t^2$ always pays
        for the increase in $\frac{t}{m}$.
        So at any stopping time $τ$ that satisfies the conditions of
        the Optional Stopping Theorem for $\Set{Y_t}$, $\Exp{Y_τ} ≥ \Exp{Y_0}$.

        Let $τ$ be the first time at which $Z_τ = n$.
        Then $Y_τ = Z_τ^2 - \frac{τ}{m} = n^2 - \frac{τ}{m}$.
        For $t ≤ τ$, $Y_t$ and $τ$ satisfy the bounded increments and
        finite expected time conditions, because $Y_t$ can't change by
        more than $2n + \frac{1}{m}$ in this range and $τ$ has finite
        expectation by the usual argument for Markov chain hitting
        times. So $\Exp{Y_τ} = n^2 - \frac{1}{m} \Exp{τ} ≥ Y_0 =
        Z_0^2 ≥ 0$. Solve to get $\Exp{τ} ≤ mn^2 = O(n^4)$.

        \paragraph{Coupling argument}
        An alternative argument couples $Z^t$ with an unbiased random
        walk $W_t$ with a reflecting barrier at $0$, that takes steps
        at random times $T_1, T_2, \dots$, where $\Exp{T_{i+1} - T_i}
        ≤ m$. The idea is to move $W_t$ only when we pick a $0-1$ edge
        n the $X^t$ process, and argue that we can always keep $Z_t
        ≥ W_t$ by incrementing $W_t$ whenever we increment $Z_t$ and
        decrementing $W_t$ even if $Z_t$ doesn't go down because of
        stubbornness. So when $W_t$ reaches $n$ after at most $n^2 m$
        expected steps (Wald's Equation again!), $Z_t$ must already
        have reached $n$.

        (This is pretty much the same argument as the potential
        function argument, except that we hide the potential function inside
        the known $n^2$ bound for the unbiased random walk. I think
        the coupling argument is more satisfying in some ways because
        it more intuitively describes what the $Z_t$ process is doing.
        However, at least as written above, it skips over justifying some details
        that don't come up with the potential function, so there is
        also more room for error.)
    
    \subsection{Almost Markov}

    Let $G$ be a $d$-regular graph on $n$ vertices. Assign a label
    $\ell(v) ∈ \Set{0,1}$ to each vertex. Let $V_0 V_1 V_2 \dots$ be a
    random walk on $G$ starting at some initial fixed vertex $v_0$,
    where each $V_{t+1}$ is chosen uniformly among the $d$ neighbors
    of $V_t$, and let $\ell(V_0), \ell(V_1), \ell(V_2), \dots$ be the
    sequence of vertex labels observed during this random walk.

    Call the sequence $\ell(V_0), \ell(V_1), \ell(V_2), \dots$
    \concept{almost Markov} with error $ε$ if, for all $t≥0$,
    $1/2 - ε ≤ \ExpCond{\ell(V_{t+1})}{\ell(V_0),\dots,\ell(V_t)} ≤
    1/2 + ε$.

    Show that, for any fixed $ε > 0$, there is some minimum degree
    $d_ε$, such that for any $d$-regular graph $G$ on $n$ vertices, it
    is possible to compute a labeling on $G$ in expected time
    polynomial in $n$ and $d$, such that a random walk on $G$ starting
    from any vertex produces a sequence of labels 
    $\ell(V_0), \ell(V_1), \ell(V_2), \dots$ that is
    almost Markov with error $ε$. 

        \subsubsection*{Solution}

        Suppose we label each vertex independently and uniformly at
        random. For each vertex $v$, let $N(v)$ be the set of all
        nodes adjacent to $v$, let $L_v = \frac{1}{d} ∑_{u∈N(v)}
        \ell(v)$ be the probability that taking a step starting at V
        lands on a node labeled with a 1, and let
        $A_v$ be the event $\abs*{L_v - 1/2} > ε$. 

        Observe that if none of the events $A_v$ occurs, 
        we get the almost-Markov property because the bound on
        $\Exp{\ell(V_{t+1}}$ will hold conditioned on any previous vertex $V_t$.
        We will use the Lovász Local Lemma to show that a labeling
        exists that makes none of these occur, and then use
        Moser-Tardos to find one. 

        With a bit of tinkering, we can write $L_v-1/2$ as the sum of $d$
        independent $±\frac{1}{2d}$ random variables, so
        Hoeffding's inequality says
        \begin{align*}
            p
            &= \Prob{A_v}
            ≤ \Prob{\abs{L_v-1/2} ≥ ε} 
            ≤ 2e^{-ε^2/2d(1/d^2)}
            = 2e^{-dε^2}.
        \end{align*}

        Let $Γ(A_v) = \SetWhere{A_u}{d(v,u)=2}$. Then $A_v$ is independent of
        any event $A_u ∉ Γ(A_v) ∪ \Set{A_v}$, because only events in this
        set depend on vertices that are adjacent to $v$, which are the
        only ones that affect $A_v$. Because the graph is $d$-regular, there
        are at most $d(d-1)$ vertices at distance 2 from $v$, and thus at
        most $d(d-1)$ events in $Γ(A_v)$.

        The Lovász Local Lemma applies if $ep(d(d-1)+1) < 1$, which holds
        if $e ⋅2e^{-dε^2} ⋅ (d(d-1)+1) < 1$. For any fixed $ε$, the
        expression on the left goes to zero in the limit as $d$ increases, so there
        exists some value $d_ε$ such that the bound holds for all $d ≥
        d_ε$. If the bound does hold, LLL says that there exists a
        labeling that makes none of the $A_v$ happen. 

        Using the Moser-Tardos algorithm~\cite{MoserT2010}, we can
        find such a labeling in $\frac{n}{d(d-1)}$ resamplings on
        average. 
        The efficient way to manage the resamplings is to compute all
        the $L_v$ values initially in $O(nd)$ time, and then pay
        $O(d^2)$ time for each resampling to update the values and
        check any that have changed. 
        If we do this, the initial setup cost dominates on average, giving an
        overall expected $O(nd)$ total time, which is about the best
        we could reasonably hope for given the size of the input, and
        which is polynomial in $n$ and $d$ as required. 

\chapter{Sample assignments from Spring 2023}

\section{Assignment 1, due Thursday 2023-02-16 at 23:59}

    \subsection{Hashing without counting}

    Suppose we are building a hash table with separate chaining, which
    is implemented as an array of
    $m$ positions $A[0], A[1], \dots, A[m-1]$,
    each of which may hold zero or more elements in some
    secondary data structure like a variable-length array or linked
    list (see §\ref{section-hash-tables}). 
    We'll abstract away from the actual hash function and assume that
    when a new element is inserted, its position is chosen
    independently and uniformly at random.

    Normally we would keep track of the number of elements $n$
    inserted into the table and grow the table by some constant factor
    when the load factor $α = n/m$ gets too
    big. But this is work and we are lazy. So instead we adopt the
    following heuristic: grow the table as soon as some element is
    inserted into position $0$. We'd like to make sure that this
    heuristic doesn't produce any bad outcomes, by looking at how
    crowded the positions in the table are at the moment after this
    element is placed in $A[0]$, just before expanding the table.

    Specifically, for this configuration:
    \begin{enumerate}
        \item Compute the exact expected load factor $α=n/m$ as a
            function of $m$.
        \item Compute the best high-probability asymptotic bound you
            can for the maximum number of elements in any table position as a function of $m$.
            This means that you should find a function $f(m)$ such
            that for all $c > 0$, the maximum number of elements is at
            most $O(f(m))$ with probability at least $1-m^{-c}$, where
            the constant in the asymptotic bound may depend on $c$.
    \end{enumerate}

        \subsubsection*{Solution}

        \begin{enumerate}
            \item Let $X$ be the number of elements inserted into the
                table. Then $X$ counts the number of insertions up to
                and including the first insertion to position $0$,
                making it a geometric random variable with
                parameter $p = 1/m$ (see
                §\ref{section-geometric-random-variables}).
                This gives $E[X] = m$.

                The expected load factor is $E[X/m] = 1$.
            \item Pick some $i ≠ 0$, and let $X_i$ be the number of
                elements in position $i$.

                Let's calculate $\Prob{X_i ≥ k}$.
                If $k = 0$, this is $1$.
                Otherwise, when we insert the first element, there is a $1/n$
                chance it lands in position $i$, and a $1/n$ chance it
                lands in position $0$. The rest of the time it lands
                in a position $j∉\Set{0,i}$ and has no effect on
                $X_i$.

                So we have, for $k > 0$,
                \begin{align*}
                    \Prob{X_i ≥ k}
                    &=
                    \frac{1}{n} \ProbCond{X_i ≥ k}{\text{insert $0$}}
                    \\&\quad + \frac{1}{n} \ProbCond{X_i ≥ k}{\text{insert
                    $i$}}
                    \\&\quad + \frac{n-2}{n} \ProbCond{X_i ≥ k}{\text{insert
                    elsewhere}}
                    \intertext{which gives}
                    \Prob{X_i ≥ k}
                    &=
                    \frac{1}{2} \ProbCond{X_i ≥ k}{\text{insert $0$}}
                    \\&\quad + \frac{1}{2} \ProbCond{X_i ≥ k}{\text{insert
                    $i$}}.
                \end{align*}

                The first branch is $0$ when $k > 0$. For the second
                branch, observe that $\ProbCond{X_i ≥ k}{\text{insert
                $i$}} = \Prob{X_i ≥ k-1}$, since we already have one
                element and now we are asking if we can get $k-1$ more
                using the same process that generates $X_i$. This
                gives a recurrence:
                \begin{align*}
                    \Prob{X_i ≥ k} &=
                    \begin{cases}
                        1 & k=0 \\
                        \frac{1}{2} \Prob{X_i ≥ k-1} & k > 0\\
                    \end{cases}
                \end{align*}
                which has the solution $\Prob{X_i ≥ k} = 2^{-k}$.

                Now use the union bound to compute, for $k>1$:
                \begin{align*}
                    \Prob{\max X_i ≥ k}
                    &= \Prob{∃i: X_i ≥ k}
                    \\&≤ ∑_{i=1}^{n-1} \Prob{X_i ≥ k}
                    \\&= (n-1) 2^{-k}.
                \end{align*}

                Pick any $c > 0$. Let $k = (c+1) \lg n$. Then
                \begin{align*}
                    \Prob{\max X_i ≥ k}
                    &≤ (n-1) 2^{-(c+1) \lg n}
                    \\&= (n-1) n^{-c-1}
                    \\&< n^{-c},
                    \intertext{from which it follows that}
                    \Prob{\max X_i < (c+1) \lg n)}
                    &≥ 1-n^{-c}.
                \end{align*}

                Since this makes $\max X_i = O(\log n)$ with
                probability $1-n^{-c}$ for any fixed $c > 0$, we get a
                high-probability bound of $O(\log n)$.

                Though this is not required by the problem, we can see
                that it is not possible to do better than $O(\log m)$
                by looking at the distribution for a single bin. We
                have $\Prob{\max X_i ≥ k} ≥ \Prob{X_1 ≥ k} = 2^{-k}$,
                so $\Prob{\max X_i ≥ c \lg m} ≥ m^{-c}$ for any fixed
                $c$.
        \end{enumerate}

    \subsection{Permutation routing on an incomplete network}

    Suppose we have a network of $n = 2d$ senders and $n=2d$
    receivers, where each sender has outgoing links to exactly $d$
    distinct receivers. Other than this information, 
    the structure of the graph is unconstrained.

    Let the senders be labeled $1\dots n$ and let the receivers be
    labeled $π_1, \dots, π_n$ where $π$ is a permutation of $1 \dots
    n$ chosen uniformly at random. Call a sender $i$ ``successful'' if
    there is a link from sender $i$ to the receiver who is assigned
    label $i$. Let $S$ be the
    number of successful senders. Then $S$ is a random variable that
    depends on the choice of $π$. (Figure~\ref{fig-incomplete-network}
    shows an example.)

    \begin{figure}
        \centering
        \begin{tabular}{cc}
            \begin{tikzpicture}
                \node (s1) at (0, 3) {$1$};
                \node (s2) at (0, 2) {$2$};
                \node (s3) at (0, 1) {$3$};
                \node (s4) at (0, 0) {$4$};
                \node (t1) at (2, 3) {$4$};
                \node (t2) at (2, 2) {$1$};
                \node (t3) at (2, 1) {$3$};
                \node (t4) at (2, 0) {$2$};

                \draw
                (s1) edge[-stealth, color=blue] (t1)
                (s1) edge[-stealth, color=red] (t2)
                (s2) edge[-stealth, color=red] (t1)
                (s2) edge[-stealth, color=blue] (t4)
                (s3) edge[-stealth, color=red] (t2)
                (s3) edge[-stealth, color=blue] (t3)
                (s4) edge[-stealth, color=blue] (t1)
                (s4) edge[-stealth, color=red] (t3)
                ;
            \end{tikzpicture}
            & 
            \begin{tikzpicture}
                \node (s1) at (0, 3) {$1$};
                \node (s2) at (0, 2) {$2$};
                \node (s3) at (0, 1) {$3$};
                \node (s4) at (0, 0) {$4$};
                \node (t1) at (2, 3) {$3$};
                \node (t2) at (2, 2) {$1$};
                \node (t3) at (2, 1) {$2$};
                \node (t4) at (2, 0) {$4$};

                \draw
                (s1) edge[-stealth, color=red] (t1)
                (s1) edge[-stealth, color=blue] (t2)
                (s2) edge[-stealth, color=red] (t1)
                (s2) edge[-stealth, color=red] (t4)
                (s3) edge[-stealth, color=red] (t2)
                (s3) edge[-stealth, color=red] (t3)
                (s4) edge[-stealth, color=red] (t1)
                (s4) edge[-stealth, color=red] (t3)
                ;
            \end{tikzpicture}
        \end{tabular}
        \caption[An incomplete network with $n = 2d = 4$]{An
        incomplete network with $n=2d=4$. In the
        left copy, the random permutation makes all four senders successful, giving $S=4$. In
        the right copy, a different random permutation makes only sender
        $1$ successful, giving $S=1$.}
        \label{fig-incomplete-network}
    \end{figure}

    \begin{enumerate}
        \item What is the exact value of $\Exp{S}$ as a function of $n$?
        \item What are the smallest and largest possible exact values of
            $\Var{S}$ as a function of $n$?
        \item Show that for any choice of
            graph that satisfies the constraints above, at least
            $Ω(n)$ senders are successful with probability at least $1-o(1)$.
    \end{enumerate}

        \subsubsection*{Solution}
        \begin{enumerate}
    \item Let $X_i$ be the indicator variable for the event that
        sender $i$ is successful, so that $S = ∑_{i=1}^n X_i$. Each
        sender $i$ is successful if and only if $i$ is the label of one of $i$'s
        $d$ neighbors. This occurs with probability $d/n = 1/2$, so
        $\Exp{X_i} = 1/2$ and $\Exp{S} = n/2$.
    \item 
        We have $\Var{S} = \sum_i \Var{X_i} + \sum_{i≠j}
        \Cov{X_i}{Y_i}$. Since the $X_i$ are fair coin-flips we get
        $\Var{X_i} = 1/4$. 

                To bound $\Cov{X_i}{Y_i}$
        let $δ_i$ and $δ_j$ be the neighborhoods of senders $i$ and
        $j$, and let $c_{ij} = \card*{δ_i ∩ δ_j}$ be the number of neighbors
        these senders have in common. Then

        \begin{align*}
            \Exp{X_i X_j}
            &= \Prob{i ∈ δ_i ∧ j ∈ δ_j}
            \\&= \Prob{i ∈ δ_i ∩ δ_j} \ProbCond{j ∈ δ_j}{i ∈ δ_i ∩ δ_j}
             \\&\quad + \Prob{i ∈ δ_i ∖ δ_j} \ProbCond{j ∈ δ_j}{i ∖ δ_i ∩ δ_j}
            \\&= \frac{c_{ij}}{n} ⋅ \frac{d-1}{n-1}
            + \frac{d-c_{ij}}{n} ⋅ \frac{d}{n-1}
            \\&= \frac{c_{ij}(d-1) + (d-c_{ij})d}{n(n-1)}
            \\&= \frac{c_{ij}d - c_{ij} + d^2 - c_{ij}d}{n(n-1)}
            \\&= \frac{d^2 - c_{ij}}{n(n-1)}.
            \intertext{This gives}
            \Cov{X_i}{X_j}
            &= \Exp{X_i X_j} - \Exp{X_i}\Exp{X_j}
            \\&= \frac{d^2 - c_{ij}}{n(n-1)} - \frac{d^2}{n^2}
            \\&= \frac{d^2 n - c_{ij}n - d^2(n-1)}{n^2(n-1)}
            \\&= \frac{d^2 - c_{ij}n}{n^2(n-1)}
            \\&= \frac{1/4 - c_{ij}/n}{n-1}.
        \end{align*}

                Summing over all $n(n-1)$ pairs $i≠j$ gives
                \begin{align*}
                    \Var{S}
                    &= ∑_i \Var{X_i} + ∑_{i≠j} \Cov{X_i}{X_j}
                    \\&= \frac{n}{4} + \frac{n}{4} - ∑_{i≠j} \frac{c_{ij}}{n(n-1)}.
                    \\&= \frac{n}{2} - \frac{1}{n(n-1)} ∑_{i≠j} c_{ij}.
                \end{align*}

                To minimize the variance, we want to make $∑_{i≠j}
                c_{ij}$ as large as possible. We can do this by
                routing all $n$ senders to the same $d = n/2$
                receivers, giving $c_{ij} = n/2$ for all $i≠j$ and
                thus $\Var{S} = 0$.
                This is not surprising because in this case, all and only
                those senders whose ids appear among the $d$ favored
                receivers will be successful, making $S$ a constant.

                For the upper bound, we can get crude bound by
                observing that $c_{ij} ≥ 0$ gives $\Var{S} ≤ n/2$. But
                in general there is no graph that actually produces
                $c_{ij} = 0$ for all $i≠j$. Instead, we need to look
                at minimizing $∑ c_{ij}$ subject to the constraint
                that there are $nd$ edges.

                Let $e_{ijk}$ be $1$ if senders $i$ and $j$ share an edge to
                receiver $k$ and $0$ otherwise.
                Then $c_{ij} = ∑_k e_{ijk}$. But if we
                write $d_k$ for the degree of receiver $k$, we also
                have that $∑_{i≠j} e_{ijk} = d_k(d_k-1)$, since
                each ordered pair of distinct neighbors of $k$ contributes one
                nonzero $e_{ijk}$. It follows that 
                \begin{align*}
                ∑_{i≠j} c_{ij} 
                    &= ∑_{i≠j} ∑_k e_{ijk}
                    \\&= ∑_k d_k(d_k-1).
                \end{align*}

                We have the constraint $∑_k d_k = dn$; subject to this
                constraint, $∑_k d_k(d_k-1)$ is minimized when all
                $d_k = d$. This can be achieved by routing each sender
                $i$ to receivers $i, i+1, \dots i+(d-1) \pmod{n}$ or
                by routing all of the first $d$ senders to all of
                the first $d$ receivers and similarly for the second
                $d$ senders and receivers.

                This gives
                \begin{align*}
                    \Var{S}
                    &= \frac{n}{2} - \frac{1}{n(n-1)} ∑_{i≠j} c_{ij}
                    \\&≤ \frac{n}{2} - \frac{1}{n(n-1)} ⋅ n d (d-1)
                    \\&= \frac{n}{2} - \frac{d(d-1)}{n-1}
                    \\&= \frac{n}{2} \parens*{1 - \frac{n/2-1}{n-1}}
                    \\&= \frac{n}{2} ⋅ \frac{2n - 2 - n + 2}{2(n-1}
                    \\&= \frac{n^2}{4(n-1)}.
                \end{align*}

                As a check, we can observe that when $n=2$, this
                gives $\Var{S} = 1 = n/2$, matching the crude bound
                because in this case we can in fact guarantee no
                overlap between senders' neighborhoods.
        \item Since $\Var{S} ≤ n/2$, we can use Chebyshev's inequality to compute
        \begin{align*}
            \Prob{S ≤ n/4}
            &= \Prob{S - \Exp{S} ≤ -n/4}
            \\&≤ \Prob{\abs*{S - \Exp{S}} ≥ n/4}
            \\&≤ \frac{n/2}{(n/4)^2}
            \\&= 1/n.
        \end{align*}
        So our probability of getting at least $n/4 = Ω(n)$ successful
        senders is at least $1-1/n = 1-o(1)$.

        A tighter bound can be obtained using Azuma's inequality with
        an appropriate exposure martingale. Let $ℱ_t$ be the
        $σ$-algebra generated by knowing which receiver is assigned
        label $j$ for all $j≤t$. Let $Y_t = \ExpCond{S}{ℱ_t}$.

        To apply Azuma's inequality, we need a bound on
        $\abs*{Y_{t+1}-Y_T}$. Let's look at $\ExpCond{X_i}{ℱ_t}$ and
        how it changes with $t$.
        \begin{enumerate}
            \item If $i ≤ t$, then the target for sender $i$ is
                already known, and $X_i$ is either $0$ or $1$. The
                change $\ExpCond{X_i}{ℱ_{t+1}}-\ExpCond{X_I}{ℱ_{t+1}}$
                is exactly $0$ in this case.
            \item If $i = t+1$, then $ℱ_{t+1}$ determines $x_i$.
                Whatever the previous conditional expectation, the new
                conditional expectation is either $0$ or $1$. This
                gives a change of at most $±1$.
            \item If $i > t+1$, then there are $n-t$ possible places to
                put label $i$, of which $d_i$ are connected to
                sender $i$ for some $0≤d_i≤d$. The value of $d_i$
                depends on how many receivers connected to sender $i$
                have known labels in $ℱ_t$.

                In this case we have
                \begin{align*}
                    \ExpCond{X_i}{ℱ_{t+1}} - \ExpCond{X_i}{ℱ_t}
                    &≤ \frac{d_i}{n-t-1} - \frac{d_i}{n-t}
                    \\&= \frac{d_i}{(n-t)(n-t-1)}
                    \\&≤ \frac{n-t}{(n-t)(n-t-1)}.
                    \\&≤ \frac{1}{n-t-1}.
                    \intertext{In the other direction,}
                    \ExpCond{X_i}{ℱ_{t+1}} - \ExpCond{X_i}{ℱ_t}
                    &≥ \frac{d_i-1}{n-t-1} - \frac{d_i}{n-t}
                    \\&= -\frac{n-t-d_i}{(n-t)(n-t-1)}
                    \\&≥ -\frac{n-t}{(n-t)(n-t-1)},
                    \\&≥ -\frac{1}{n-t-1}.
                \end{align*}

                In either case the absolute value is bounded by
                $\frac{1}{n-t-1}$. There are $n-t-1$ values of $i$ in
                this range, so the sum for all of them is at most $±1$.
        \end{enumerate}

        Combining the last two cases, we have
        $\abs*{\ExpCond{S}{ℱ_{t+1}}-\ExpCond{S}{ℱ_t}} ≤ 2$.
        So Azuma's inequality gives $\Prob{S - n/2 ≥ t} ≤ e^{-t^2/8n}$
        and similarly $\Prob{S - n/2 ≤ t} ≤ e^{-t^2/8n}$. The constant
        in the denominator could possibly be improved a bit with a
        better analysis, but even as is we get that $S = n/2 ±
        Θ(√{n \log n})$ with high probability and $S ≥ n/4 = Ω(n)$
        with all but exponentially small probability.
\end{enumerate}

\section{Assignment 2, due Thursday 2023-03-30 at 23:59}

    \subsection{Some streaming data structures}

    Suppose you are given a stream $x_1, x_2, \dots, x_{2n}$
    of $2n$ integer values in the range
    $0\dots n-1$. You are presented with these values one at a time.
    Upon receiving each value, you may update a data structure of your
    choosing that fits in $O(\log n)$ bits of space. You have no other storage.

    \begin{enumerate}
        \item Give an algorithm for
            sampling a single value with probability proportional to
            the number of times it occurs in the stream, returning
            this value once all elements of the stream have been
            processed.

            For example, given a stream $1,0,1,2,1,1$, your algorithm
            should return $0$ or $2$ with probability $1/6$ each, and
            return $1$ with probability $2/3$.
        \item Give a Las Vegas algorithm that returns a value that occurs at least twice in the
            stream with probability $Ω(1)$. If it doesn't succeed, it
            should return $⊥$ to indicate
            failure. (This algorithm should never return a value that
            occurs less than twice.)
    \end{enumerate}

    In both cases, prove the correctness of your algorithm. You
    may assume that $n$ is known to the algorithm.

        \subsubsection*{Solution}
        \begin{enumerate}
    \item There are two straightforward ways to do this using $O(\log
        n)$ bits:
        \begin{enumerate}
            \item The easiest is to sample the position of the desired
                element ahead of time, then count down until we hit
                it. This requires $2n$ possible states for the
                countdown timer plus $n$ possible states for the
                stored value, for a total of $\ceil{\lg 3n} = O(\log
                n)$ bits.
            \item Alternatively, we could store a value while counting
                how many values we've seen, and replace the stored
                value with the $k$-th value with probability
                $1/k$.\footnote{This is a special case of a more
                general algorithm known as
                \index{sampling!reservoir}\concept{reservoir
                sampling}, due to Vitter~\cite{Vitter1985}.}
                A straightforward induction argument shows that this
                makes the stored value uniform across all the values
                seen so far. We need $\ceil{\lg 2n}$ bits for the
                counter and $\ceil{lg n}$ bits to store the current
                value, but the total is still $O(\log n)$.
        \end{enumerate}
    \item For this part, the intuition is that we will sample a value
        as we go, then record a winner if we happen to see it again.
        How easy this is to analyze depends on which sampling
        algorithm we use:
        \begin{enumerate}
            \item If we pick a particular position $i$ and then sample
                $x_i$, we win as long as $x_i$ is not the last
                occurrence of this particular value in the sequence.
                Since there are at most $n$ positions that are the
                last occurrence of a value, we have at least a $1-n/2n
                = 1/2$ chance of winning. This requires an extra bit
                to record whether or not the sampled value appeared
                again, for a total of $\ceil{\lg 3n} + 1 = O(\log n)$
                bits.
            \item Alternatively, we could run the continuous sampling
                algorithm above and fix the stored value if we see it again.
                The analysis in this case is a bit more messy.
                Here is one possible approach.

                Consider the original sampling process, and
                let $A_i$ be the event that (a) there exists some $j
                > i$ such that $x_j = x_i$, and (b) value $x_i$ is
                stored at time $i$ and never overwritten.
                There are at least $n$ positions for which (a) holds,
                and for each of these positions we have $\Prob{A_i} =
                1/2n$.

                Unfortunately these events are not independent. But
                for any $i$, we have that 
                $\ProbCond{¬A_i}{¬A_1 ∧ \dots ∧ ¬A_{i-1}}$ is either
                $1$ (if $A_i$ never occurs) or at most $1-1/2n$ (since
                previous events $A_j$ not occurring for $j < i$ i
                can only increase the likelihood of $A_i$ occurring).
                The latter case holds for at least $n$ choices of $i$,
                giving
                \begin{align*}
                    \Prob{\bigwedge ¬A_i}
                    &= ∏ \ProbCond{¬A_i}{¬A_1 ∧ \dots ∧ ¬A_{i-1}}
                    \\&≤ (1-1/2n)^n
                    \\&≤ e^{-1/2},
                \end{align*}
                which is a constant strictly less than $1$. So with
                constant probability, at least one event $A_i$ occurs.
                But then the algorithm observes a duplicate and
                returns it.
        \end{enumerate}
\end{enumerate}

    \subsection{A dense network}

    Suppose we construct a network as follows. Initially, we have $n_0
    = n$ nodes, of which one is an inert base station with no incident edges.
    We also have $m_0 = m$ edges between the $n-1$ remaining nodes. At each step, we pick
    one of the $n_t - 1$ non-base-station nodes uniformly at random and
    duplicate it, creating a new node connected to the same nodes as
    the old one. 

    The \concept{density} $D_t$ of this network after $t$ steps is the ratio
    between the number of edges $m_t$ and the number of possible edges 
    $\binom{n_t}{2}$, where $n_t = n_0 + t$ is the number of nodes
    (including the base station, even though it doesn't otherwise
    participate in the network).
    We'd like to examine how the density evolves over time as new
    nodes are added to the network.

    An example of this process is given in
    Figure~\ref{fig-graph-density-problem}.

\begin{figure}
    \centering
    \begin{tabular}{ccc}
    \begin{tikzpicture}
        \node (bs) at (0, 0) {base};
        \node (a) at (-1, 1) {$a$};
        \node (b) at (1,1) {$b$};
        \path
        (a) edge (b)
        ;
    \end{tikzpicture}
        &
    \begin{tikzpicture}
        \node (bs) at (0, 0) {base};
        \node (a) at (-1, 1) {$a$};
        \node (b) at (1,1) {$b$};
        \node (b2) at (1,2) {$b'$};
        \path
        (a) edge (b)
        (a) edge (b2)
        ;
    \end{tikzpicture}
        &
    \begin{tikzpicture}
        \node (bs) at (0, 0) {base};
        \node (a) at (-1, 1) {$a$};
        \node (a2) at (-1, 2) {$a'$};
        \node (b) at (1,1) {$b$};
        \node (b2) at (1,2) {$b'$};
        \path
        (a) edge (b)
        (a) edge (b2) 
        (a2) edge (b) 
        (a2) edge (b2)
        ;
    \end{tikzpicture}
    \end{tabular}
    \caption[Graph density evolution]{Example of graph density evolution. Initial
    graph has density $D_0 = 1/\binom{3}{2} = 1/3$. After duplicating
    $b$, new density is $D_1 = 2/\binom{4}{2} = 1/3$. After
    duplicating $a$ as well,
    final density is $D_2 = 4/\binom{5}{2} = 2/5$.}
    \label{fig-graph-density-problem}
\end{figure}

    \begin{enumerate}
        \item Compute the exact value of $\Exp{D_t}$ as a function of
            $n$, $m$, and $t$.
        \item 
            Show that, for any $t$, $D_t$ is within $O\parens*{√{\log n / n}}$ of $\Exp{D_t}$
            with high probability.
    \end{enumerate}

        \subsubsection*{Solution}

        \begin{enumerate}
            \item We'll show that $\Set{D_t}$ is a martingale, giving
                $\Exp{D_t} = \Exp{D_0} = m / \binom{n}{2}$ for all
                $t$.

                Let $G_t$ be the graph after $t$ steps.
                Let $ℱ_t$ be the $σ$-algebra generated by $\Tuple{G_0,
                \dots, G_t}$.

                Each edge in
                $G_t$ goes between two non-base-station nodes, so
                there is a chance of exactly $\frac{2}{n_t - 1}$ that
                one of its endpoints is duplicated. Summing over all
                edges gives 
                $\ExpCond{m_{t+1}}{m_t} = m_t \parens*{1 -
                \frac{2}{n-1}}$.
                But then
                \begin{align*}
                    \ExpCond{D_{t+1}}{ℱ_t}
                    &= \frac{m_t \parens*{2}{n-1}}{\binom{n_t+1}{2}}
                    \\&= \frac{m_t \frac{n+1}{n-1}}{\binom{n_t+1}{2}\frac{n+1}{n-1}}
                    \\&= \frac{m_t}{\binom{n_t}{2}}
                    \\&= D_t.
                \end{align*}

                It follows that $\Exp{D_t} = \Exp{D_0} =
                m/\binom{n}{2}$.
            \item We will apply Azuma's inequality. This requires
                getting bounds on the martingale differences $X_t =
                D_t - D_{t-1}$.

                Since the maximum possible degree of any node in
                $G_{t-1}$ is $n_{t-1} - 2$, we have $m_{t} ≤ m_{t-1} + n_{t-1} - 2$
                and thus
                \begin{align*}
                    D_{t} - D_{t-1}
                    &≤ \frac{m_{t-1} + n_{t-1} - 2}{\binom{n_{t-1}+1}{2}}
                    - \frac{m_{t-1}}{\binom{n_{t-1}}{2}}
                    \\&≤ \frac{m_{t-1} + n_{t-1} - 2}{\binom{n_{t-1}}{2}}
                    - \frac{m_{t-1}}{\binom{n_{t-1}}{2}}
                    \\&≤ \frac{n_{t-1}}{\binom{n_{t-1}}{2}}
                    \\&≤ \frac{2}{n_{t-1} - 1}.
                    \\&= \frac{2}{n+t-2}.
                \end{align*}

                In the other direction, duplicating a node destroys no
                edges, so $m_{t} ≥ m_{t-1}$, giving
                \begin{align*}
                    D_{t} - D_{t-1}
                    &≥ \frac{m_{t-1}}{\binom{n_{t-1}+1}{2}}
                    - \frac{m_{t-1}}{\binom{n_{t-1}}{2}}
                    \\&= m_{t-1} \frac{\binom{n_{t-1}}{2} -
                    \binom{n_{t-1}+1}{2}}{\binom{n_{t-1}+1}{2}\binom{n_{t-1}}{2}}
                    \\&= m_{t-1} \frac{-n_{t-1}}{\binom{n_{t-1}+1}{2}\binom{n_{t-1}}{2}}
                    \\&≤ \frac{-n_{t-1}}{\binom{n_{t-1}+1}{2}}
                    \\&= - \frac{2}{n+t}.
                \end{align*}

                So in either direction we have $\abs*{X_t} = \abs*{D_{t+1} - D_t} ≤
                \frac{2}{n+t-2}$.

                Now let us apply Azuma's inequality:
                \begin{align*}
                    \Prob{\abs*{D_t - \Exp{D_t}} ≥ α}
                    &≤ 2\exp\parens*{-α^2 / ∑_{i=1}^{t}
                    \parens*{\frac{2}{n+i-2}}^2}
                    \\&≤ 2\exp\parens*{-α^2 / \parens*{4 ∑_{k=n-1}^{∞}
                    \frac{1}{k^2}}}
                    \\&≤ 2\exp\parens*{-α^2 / \parens*{4 ∫_{x=n-2}^{∞}
                    \frac{1}{x^2} \; dx}}
                    \\&≤ 2\exp\parens*{-α^2 (n-2) / 4}.
                \end{align*}

                For any fixed $c$, setting $α = √{4 (c+1) \ln n /
                (n-2)} = O\parens*{√{\log n / n}}$ gives a
                probability of at most $2n^{-c-1} ≤ n^{-c}$ (when
                $n≥2$) that we are more than $α$ away from the expectation.
        \end{enumerate}

\section{Assignment 3, due Thursday 2023-04-27 at 23:59}

    \subsection{Shuffling a graph}

    Consider the following shuffling algorithm:
    Given a connected graph $G$ with $n$ vertices labeled $1$ through $n$,
    and $m$ edges, at each step, with probability $1/2$, choose one of the edges uniformly at
    random and swap the labels of its endpoints. This process yields a
    Markov chain that is aperiodic and
    irreducible,\footnote{Aperiodicity is immediate from having steps
    that do nothing. For irreducibility, if $G$ is a tree, there is at
    least one vertex $v$ with degree $1$; use a sequence of swaps to
    move the desired label to $v$, then recurse on $G ∖ \Set{v}$ to
    fix the rest of the labeling. For general $G$, apply this
    procedure to a spanning tree of $G$.}
    so it has a unique
    stationary distribution.

    \begin{enumerate}
        \item What is the stationary distribution of this process?
        \item Show that this process converges to within a total
            variation distance of $ε$ of its stationary distribution
            in time polynomial in $n$ and $\log \frac{1}{ε}$.
    \end{enumerate}

        \subsubsection*{Solution}

        \begin{enumerate}
            \item Suppose there is a labeling $i$ and a labeling $j$ such 
                that $j$ is reachable from $i$ in one step. Then these
                labelings differ by swapping one edge so $p_{ij} =
                p_{ji} = \frac{1}{2m}$. For each labeling $i$, let $π_i =
                1/n!$. Then $∑_i π_i = 1$ and, for all $i$ and $j$,
                $π_i p_{ij} = π_j p_{ji} = \frac{m}{n!}$. This shows
                that the chain is reversible and that this particular $π$ is its
                stationary distribution.
            \item We'll do a coupling argument, similar to the
                argument for adjacent-swap shuffling on a line. For
                technical reasons that we will see later, it will be
                helpful to assume that $G$ has at least two edges. In
                the case where $G$ has only one edge, we can argue
                that the process reaches the stationary distribution
                in one step, as there are only two labelings and the
                first step has a $1/2$ chance each of switching or staying
                put.

                Let $X$ and $Y$ be two copies of the process, with $X$
                starting in an arbitrary distribution and $Y$ starting
                in the uniform stationary distribution. 
                At each step, let $D$ be the set of edges with the
                same label on at least one endpoint.
                For each edge $uv$:
                \begin{enumerate}
                    \item If $uv ∈ D$, swap $uv$ in both copies with
                        probability $\frac{1}{2m}$.
                    \item If $uv ∉ D$, swap $uv$ in $X$ only with
                        probability $\frac{1}{2m}$ and in $Y$ only
                        with probability $\frac{1}{2m}$.
                \end{enumerate}
                Do nothing with the remaining probability
                $\frac{\card{D}}{2m}$.

                For each label $\ell$, write $X_\ell$ for the vertex
                labeled with $\ell$ in $X$ and $Y_\ell$ for the vertex
                labeled with $\ell$ in $Y$. As in the adjacent-swap
                case, we can argue that once $X_\ell = Y_\ell$, any
                edge $uv$ with an endpoint labeled $\ell$ is in $D$,
                so $X_\ell$ continues to equal $Y_\ell$ in all
                subsequent states of the coupled process. So the
                question then is how long it takes for two unlinked
                copies of a label to become linked.

                Define a new process $Z^t = \Tuple{U^t,V^t}$ where
                $U^t$ represents the position $X^t_\ell$ and $V^t$
                represents $Y^t_\ell$. Unlike the
                $\Tuple{X^t_\ell,Y^t_\ell}$ process, we will assume
                that at each step of $Z^t$ we always move $U^t$ across
                an incident edge with probability $\frac{1}{2n}$ and
                similarly for $V^t$, without regard to whether $U^t =
                V^t$. This gives a Markov chain with the same
                transition probabilities as
                $\Tuple{X^t_\ell,Y^t_\ell}$ when $X^t_\ell ≠
                Y^t_\ell$, but unlike the $\Tuple{X^t_\ell,Y^t_\ell}$
                process, the $Z^t$ process is both irreducible and
                reversible, with a uniform stationary distribution.
                We can also argue that it is aperiodic under our
                assumption that $m > 1$, because there exists at least
                one state where $U^t$ is not adjacent to some
                edge, so there is a nonzero probability that $Z_{t+1}
                = Z^t$ in this state.

                We will now show that $X^t_\ell$ and $Y^t_\ell$
                collide with high probability after polynomially-many
                steps by showing convergence of $Z^t$ to its
                stationary distribution using Cheeger's inequality.
                Observe that the $Z$ chain has exactly $m^2$ states,
                and that each non-empty subset $S$ of the chain has at
                least one transition that leaves it with probability
                $π_i p_{ij} = \frac{1}{m^2}⋅\frac{1}{2m} =
                \frac{1}{2m^3}$. This gives
                $Φ_S ≥ \frac{1/2m^3}{\card{S} / m^2}$ which is at
                least $\frac{1}{4m^3}$ in the worst case. 
                It follows that $\tau_2 ≤ \frac{2}{Φ^2} ≤ 8m^6$ and
                that the total variation distance between $Z^t$ and
                the uniform distribution is at most $\frac{1}{2m}$
                after $O(m^6 \log m)$ steps. In particular this gives
                a probability of at least $\frac{1}{m} - \frac{1}{2m}
                = \frac{1}{2m}$ of being in a state $\Tuple{U^t,V^t}$
                with $U^t = V^t$ at the end of each such interval.
                Repeating for $O(m \log m)$ intervals with a suitable
                constant gives a probability of at most $m^{-c}$ for
                any fixed $c$ that $U^t = V^t$ at the end of one of
                these intervals, which shows that $X_\ell$ and
                $Y_\ell$ collide within $O(m^7 \log^2 m)$ steps with
                probability at least $1-m^{-c}$.

                Since $m ≥ n-1$, we can choose $c$ so that the
                probability that $X_\ell$ and $Y_\ell$ don't collide
                in $O(m^7 \log^2 m)$ steps is $O(n^{-2})$. By the
                union bound, this gives that $X_\ell$ and $Y_\ell$
                collide in time $O(m^7 \log^2 m)$ with probability at
                least $1-O(n^-1)$. Repeat as needed at most $O(\log
                (1/ε))$ times to drive the probability of failure down
                below $ε$, and use $m = O(n^2)$ to express the
                convergence time in terms of $n$ as $O(n^14 \log^2 n
                \log (1/ε))$. This polynomial in $n$ and $\log
                \frac{1}{ε}$ even though the exponents are terrible.
        \end{enumerate}

    \subsection{Counting unbalanced sets}

    Suppose you are given the usual set balancing set-up, where you
    have a collection of sets $S_1, S_2, \dots, S_m$ with
    $\card*{\bigcup S_i} = n$, and you are asked assign a value $±1$
    to each element $x$ of of $S = \bigcup S_i$ with the goal of
    minimizing the discrepancy $d = \max_i \abs*{∑_{x∈S_i} x}$.

    We know that applying a random assignment gives $d = O(√{n
    \log m})$ with high probability. But this bound doesn't take into
    account any structure the $S_i$ might have. Give an FPRAS 
    that takes the $S_i$ as input along with a bound $t$ and computes
    an approximation for $\Prob{d ≥ t}$, assuming each $x$ is assigned
    $±1$ using an independent fair coin-flip.

        \subsubsection*{Solution}

        Let $A$ be the set of all assignments that produce $d ≥ t$.
        For each $i$, let $X_i = ∑_{x∈S_i} x$ be the random variable representing
        the sum of the assignments of elements of $S_i$.
        We can write $A$ as the union of $B_{ij}$ for all $i ∈ [m]$
        and all $j$ with $\abs{j} ≥ t$, where $B_{ij}$ is the event
        that $X_i = j$. Since this union is not disjoint, we
        will use Karp-Luby (see
        §\ref{section-approximating-sharp-dnf}) to approximate its
        size.

        As with \#DNF, the idea is that we can both count and 
        sample triples $\Tuple{i,j,x}$ where $x$
        is an assignment that makes $X_i = j$. Let $n_i =
        \card{S_i}$. Let $Y_i$ be the number of elements of $S_i$ that
        are assigned $+1$; then $X_i = Y_i - (n_i - Y_i) = 2Y_i -
        n_i$, and solving for $Y_i$ gives $Y_i = \frac{n_i + X_i}{2}$,
        assuming the numerator is even. Given $i$ and $j$, there are
        exactly $2^{n-n_i} \binom{n_i}{(n_i + j)/2}$ assignments with
        $X_i = j$ when $n_i + j$ is even, and none when $n_i+j$ is
        odd. For the even case, we can sample these assignments in
        polynomial time by choosing $(n_i+j)/2$ elements without
        replacement from $S_i$ to have value $+1$, making the
        remaining elements of $S_i$ have value $-1$, and assigning
        $±1$ values independently to any $x ∈ S ∖ S_i$.
        We will abuse notation slightly by writing $B_{ij}$ for the
        set of all possible triples $\Tuple{i,j,x}$ resulting from
        this process for a particular pair of values $i$ and $j$.
        Let $B = \bigcup_{i=1}^{n} \bigcup_{\card{j} ≥ d,
        \text{$j+n_i$ even}} B_{ij}$; note that this union is
        disjoint.

        Our full sampling procedure looks like this:
        \begin{enumerate}
            \item Sample $i$ and $j$ with probability $\card{B_{ij}} /
                \card{B}$.
                Assuming that we precompute the sizes of the $B_{ij}$
                sets and that arithmetic operations take time $O(1)$ time each, this takes
                time at most $O(nm)$ if we are naive about the
                sampling and $O(\log nm)$ if we are a little bit more
                clever and use binary search.
            \item Sample an element $\Tuple{i,j,x}$ of $B_ij$ uniformly as described
                above. This takes time $O(n)$.
            \item Return $1$ if $x$ does not appear in $B_{i'j'}$ for
                any $i'j'$ that precedes $ij$ in lexicographic order.
                This requires checking $O(nm)$ previous sets at a cost
                of $O(n)$ each, for $O(n^2 m)$ time total, which
                dominates the cost of generating the sample.
        \end{enumerate}

        Let $ρ$ be the probability that we obtain $1$ in an iteration
        of this process. Then $\card{A} = ρ \card{B}$, and we can
        approximate $\card{A}$ to within relative error $ε$ by
        approximating $ρ$ within the same bound. Because each element
        of $A$ occurs in at least one and at most $O(nm)$ of the
        $B_{ij}$, we have that $ρ = Ω\parens*{\frac{1}{nm}}$. 
        Applying Lemma~\ref{lemma-sampling}, 
        approximating $ρ$ using $N$ samples where 
        $N = \frac{3}{ε^2 ρ} \ln \frac{2}{δ} = O\parens*{\frac{n^2 m^2
        \log δ}{ε^2}}$ gives us our desired error bounds in time
        polynomial in $nm$, $1/ε$, and $\log δ$, as required.

\chapter{Sample assignments from Fall 2019}

\section{Assignment 1: due Thursday, 2019-09-12, at 23:00}

\subsection*{Bureaucratic part}

Send me email!  My address is
\mailto{james.aspnes@gmail.com}.

In your message, include:

\begin{enumerate}
\item Your name.
\item Your status: whether you are an undergraduate, grad student, auditor, etc.
\item Whether you are taking the course as CPSC 469 or CPSC 569.
\item Anything else you'd like to say.
\end{enumerate}

(You will not be graded on the bureaucratic part, but you should do it anyway.)

\subsection{The golden ticket}

An infamous candy company announces it is placing a golden ticket
uniformly at random in one of $n$ candy bars that are distributed to
$n$ children, with unimaginable wealth to be bestowed on the child who
receives the ticket.  Unfortunately, the candy company owner is notorious for
trickery and lies, so the probability that the golden ticket actually
exists is only $1/2$.

Consider the last child to open their candy bar, and what estimate
they should make of their probability of winning after seeing the
first $k$ children open their candy bars to find no ticket.

An optimist reasons: If the ticket does exist, then the last child's
chances went from $1/n$ to $1/(n-k)$.  So the fact that none of the
first $k$ children got the ticket is good news!

A pessimist reasons: The more candy bars are opened without seeing the
ticket, the more likely it is that the ticket doesn't exist.  So the
fact that none of the first $k$ children got the ticket is bad news!

Which of them is right?  Compute the probability that the last child
receives
the ticket given the first $k$ candy bars come up empty, and compare
this to the probability that the last child receives the ticket given
no information other than the initial setup of the problem.

\subsubsection*{Solution}
    Let $W$ be the event that the last child
    wins, $C$ the event that the candy bar exists, and $L_k$ the event
    that the first $k$ children lose.

    Without conditioning, we have
    \begin{align*}
        \Prob{W}
        &= \ProbCond{W}{C} \Prob{C} + \ProbCond{W}{¬C} \Prob{¬C}
        \\&= \frac{1}{n} ⋅ \frac{1}{2} + 0 ⋅ \frac{1}{2}
        \\&= \frac{1}{2n}.
        \intertext{With conditioning, we have}
        \ProbCond{W}{L_k}
        &= \frac{\Prob{W ∧ L_k}}{\Prob{L_k}}
        \\&= \frac{1/2n}{\Prob{L_k}{C} \Prob{C} + \Prob{L_k}{¬C}\Prob{¬C}}
        \\&= \frac{1/2n}{\frac{n-k}{n} ⋅ \frac{1}{2} + 1 ⋅ \frac{1}{2}}
        \\&= \frac{1/2n}{\frac{n-k}{2n} + \frac{1}{2}}
        \\&= \frac{1/2n}{\frac{2n-k}{2n}}
        \\&= \frac{1}{2n-k}.
    \end{align*}

    Since $\frac{1}{2n-k} > \frac{1}{2n}$ when $k > 0$, the optimist is correct.

\subsection{Exploding computers}
\label{section-problem-exploding-computers}

Suppose we have a system of $3n$ machines $m_0, m_2, \dots m_{3n-1}$ in
a ring, and we turn each machine on at random with independent
probability $p$.  Over time, the machines may overheat: if we turn on
three consecutive machines $m_i, m_{i+1},$ and $m_{i+2}$ (wrapping
indices around mod $3n$), the middle
machine $m_{i+1}$ overheats and explodes.  Explosions happen
simultaneously and are purely a function of which machines are
initially turned on; so if $m_1, m_2, m_3, m_4,$ and $m_5$ are all
turned on, $m_2, m_3,$ and $m_4$ will all explode.

Call a machine a \conceptFormat{active} if it is turned on
and does not explode.

We want to choose $p$ to maximize the number of active machines $X$.  But
there are two ways to interpret this goal.

\begin{enumerate}
    \item Suppose our goal is to maximize the \emph{expected} number
        of active machines $\Exp{X}$.
        What value of $p$ should we choose, and what does this give us
        for $\Exp{X}$?
    \item Suppose instead our goal is to maximize the probability of
        having the maximum possible number of active machines.  More
        formally, letting $M$
        be the number of survivors we could get if we chose which machines
        to turn on optimally, we want to choose $p$ to maximize
        $\Prob{X=M}$.  What value of $p$ should
        we choose, and what does this give us for $\Prob{X=M}$?
\end{enumerate}

    \subsubsection*{Solution}

    \begin{enumerate}
        \item Let $A_i$ be the event that machine $m_i$ is active, and 
            let $X_i$ be the indicator variable for $m_i$ being active
            and not exploding.  We are trying to maximize $\Exp{∑
            X_i} = ∑ \Exp{X_i} = n \Exp{X_1}$, where the last equation
            holds by symmetry.

            Observe that $X_1=1$ if $A_1$ occurs and it is not the
            case that both $A_0$ and $A_2$ occur.
            This gives $\Exp{X_1} = \Prob{X_1=1} = p(1-p^2) = p - p^3$.
            By setting the derivative $1 - 3p^2$ to zero we find a
            unique extremum at $p=√{1/3}$; this is a maximum since
            $p(1-p^2) = 0$ at the two corner solutions $p=0$ and
            $p=1$.  Setting $p=√{1/3}$ gives $3n \Exp{X_i} =
            3n⋅√{1/3}(2/3) = n \frac{2}{√{3}} \approx n ⋅
            1.154701\dots$.
        \item 
            First we need to figure out $M$.  

            We can assume that we set up the optimal configuration so that
            none explode, because any initial configuration that
            produces explosions can be replaced by one where the dead
            machines are just turned off to start with, with no
            reduction in the number of active machines.

            Conveniently, we can divide the machines into consecutive
            groups of three, and observe that (a) turning on the first
            two machines in each group causes no explosions, (b)
            turning on three machines in any group causes the middle one
            to explode. This shows $M=2n$ is the maximum achievable
            number of machines, because (a) lets us get $M$ and (b)
            says we can't do more, since this would give us a group
            with three by the Pigeonhole Principle.

            We still need to characterize how many possible patterns
            of $M=2n$ machines produce no explosions.  Since we can
            only have two consecutive active machines, each pair of
            active machines must be followed by an inactive machine.
            Putting two inactive machines in a row would allow a
            partitioning into groups of three with some group getting
            only one active machine, leaving some other group with
            three, which is trouble. So the pattern must consist of
            pairs of active machines alternating with single inactive
            machines. There are exactly three ways to do this.

            Each of these three patterns occurs with probability
            $p^{2n} q^n$ where $q=1-p$.  Maximizing this probability
            can again be done by taking the derivative $2n p^{2n-1}
            q^n - n p^{2n} q^{n-1}$ and setting to $0$.  Removing
            common factors gives $2q - p = 0$, or $2 - 3p = 0$, giving
            $p = 2/3$, which is about what one would expect.

            We then get $\Prob{X=2n} = 3⋅(2/3)^{2n} (1/3)^n$.  This is
            pretty small for reasonably large $n$, which suggests that
            a randomized algorithm may not be the best in this case.
    \end{enumerate}

\section{Assignment 2: due Thursday, 2019-09-26, at 23:00}

    \subsection{A logging problem}

    An operating system generates a new log entry at time $t$ with
    independent probability $p_t$, and the size of each log entry for
    any $t$ is an integer $X$ chosen uniformly in the range $a ≤ X ≤
    b$, where the choice of $X$ at a particular time $t$ is
    independent of all the other log entry sizes and all choices of
    whether to write a log entry.

    Suppose that over some fixed period of time, the operating
    system generates $n$ log entries on average.  Compute the expectation of
    the total size $S$ of all the log entries, and show that the probability that
    $S$ is more than a small constant fraction larger than $\Exp{S}$
    is exponentially small as a function of $n$.

        \subsubsection*{Solution}

        For each $t$, let $X_t$ be the size of the log entry at time
        $t$. Then $\Exp{S} = ∑_t \Exp{X_t} = ∑_t p_t \frac{a+b}{2} =
        \frac{a+b}{2} \Exp{∑_t p_t} = \frac{a+b}{2}⋅n$.

        We can't use Chernoff bounds directly on the $X_t$ because
        they aren't Bernoulli random variables, and we can't use
        Hoeffding's inequality because we don't actually know how
        many of them there are. So we will use Chernoff first
        to bound the number of nonzero $X_t$ and then use Hoeffding to
        bound their sum.

        The easiest way to think about this is to imagine we generate
        the nonzero $X_t$ by generating a sequence of independent
        values $Y_1, Y_2, \dots$, with each $Y_i$ uniform in $[a,b]$,
        then select the $i$-th nonzero $X_t$ to have value
        $Y_i$.

        Let $Z_t$ be the indicator variable for the event that a log
        entry is written at time $t$. Then $N = ∑_t Z_t$ gives the total
        number of log entries, and $\Exp{N} = \Exp{∑_t Z_t} = ∑_t p_t = n$.

        Pick some $δ < 1$; then Chernoff's inequality
        \eqref{eq-Chernoff-bound-one-third} gives
        \begin{align*}
            \Prob{N ≥ (1+δ)n} &≤ e^{-n δ^2 / 3}.
        \end{align*}

        Suppose $N < (1+δ)n$. Then $∑_t X_t = ∑_{i=1}^{N} Y_i ≤
        ∑_{i=1}^{(1+δ)n} Y_i$. This last quantity is a sum of
        independent bounded random variables, so we can
        apply the asymmetric version of Hoeffding's inequality
        \eqref{eq-Hoeffdings-inequality-asymmetric-nonzero-means} to get
        \begin{align*}
            \Prob{∑_{i=1}^{(1+δ)n} Y_i - (1+δ)n⋅\frac{a+b}{2} ≥ s}
            &≤
            \exp\parens*{-\frac{s^2}{2(1+δ)n\parens*{\frac{b-a}{2}}^2}}
        \end{align*}

        For $s = δn⋅\frac{a+b}{2}$ this gives a bound of
        \begin{align*}
            \exp\parens*{-\frac{-δ^2 n^2
            \parens*{\frac{a+b}{2}}^2}{2(1+δ)n\parens*{\frac{b-a}{2}}^2}}
            &≤ 
            \exp\parens*{-\frac{-δ^2 n
            \parens*{\frac{a+b}{2}}^2}{2(1+δ)\parens*{\frac{b+a}{2}}^2}}
            \\&= 
            \exp\parens*{-\frac{-δ^2 n}{2(1+δ)}}.
        \end{align*}

        If we put this all together, we get
        \begin{align*}
            \Prob{S ≥ (1+2δ)n\frac{a+b}{2}}
            &≤ e^{-nδ^2/3} + e^{-n δ^2 / 2(1+δ)}
            = e^{-Θ(n)}.
        \end{align*}

    \subsection{Return of the exploding computers}

    Suppose we are working in the model described in
    §\ref{section-problem-exploding-computers}, where $3n$ machines
    in a ring are each turned on independently at random with probability $p$,
    and a machine remains active only if it is turned on and at least one of its neighbors is
    not turned on.  We have seen that an optimal choice of $p$ gives
    $cn$ active machines on average for a constant $c$ independent of
    $n$.

    Unfortunately, averages are not good enough for our customers.
    Show that there is a choice of $p$ and constants $c' > 0$ and
    $\delta > 0$
    such that at
    least $c'n$ machines are active with probability at least
    $1-e^{-δn}$ for any $n$.

    \subsubsection*{Solution}

    We can use the method of bounded differences.  Let $X_i$ be the
    indicator for the event that machine $i$ is turned on, and let
    $S = f(X_1,\dots,X_{3n})$ compute the number of active machines.  Then 
    changing one input to $f$ from $0$ to $1$ increases $f$ by at most
    $1$, and decreases it by at most $2$ (since it may be that the new
    machine explodes, producing no increase, and it also explodes both
    neighbors, producing a net $-2$ change).  So McDiarmid's
    inequality \eqref{eq-McDiarmids-inequality} applies with $c_i = 2$
    for all $i$, giving
    \begin{align*}
        \Prob{f(X_1,\dots,X_{3n} - \Exp{f(X_1,\dots,X_{3n}} ≤ t}
        &≤ \exp\parens*{-\frac{2t^2}{3n⋅2^2}}
        &= e^{-t^2/6n}.
    \end{align*}

    Choose $p$ to maximize the expected number of active machines, and
    let $\Exp{S} = c'n$.
    Choose $c$ so that $0 < c < c'$.
    Then $\Prob{S ≤ cn}
    = \Prob{S - \Exp{S} ≤ c'n - cn}
    ≤ e^{-(c'-c)^2 n^2 / 6n}
    = e^{-(c'-c)^2 n / 6}
    = e^{-δn}$,
    where $δ = (c'-c)^2 / 6 > 0$.

    The above is my original sample solution, below are some quick
    sketches of alternative approaches that showed up in submissions:

    \begin{itemize}
        \item
    Analyze the dependence and use
    Lemma~\ref{lemma-Chernoff-almost-independent}. This gives similar
    results to McDiarmid's inequality with a bit more work. The basic
            idea is that $\Exp{X_i|X_1\dots X_{i-1}} ≥ p(1-p)$ for $i
            < 3n$ whether
            or not $X_{i-1}$ is $1$ (we don't care about the other
            variables). So we can let $p_i = p(1-p)$ for these values
            of $i$. To avoid wrapping around at the end, either omit $X_{3n}$ entirely or
            let $p_{3n} = 0$.
\item
    Even simpler: Let $Y_i$ be
    the indicator for the event that machine $i$ is active. Then
    $Y_3,Y_6,\dots,Y_{3n}$ are all independent, and we can bound
    $∑_{i=1}^{n} Y_{3i} ≤ ∑_{i=1}^{3n} Y_i$
    using either Chernoff or Hoeffding.
    \end{itemize}

\section{Assignment 3: due Thursday, 2019-10-10, at 23:00}

    \subsection{Two data plans}

    A cellphone company offers a choice of two data plans, both based
    on penalizing the user whenever they use more data than in the
    past. For both plans the assumption is that your data usage in
    each month $i∈\Set{0\dots \infty}$ is an independent continuous random variable that
    is uniform in the range $[0,c]$ for some fixed constant $c$.

    With \conceptFormat{Plan A}, you pay a penalty of \$1 for every
    month $i$ in which your usage $X_i$ exceeds $\max_{j<i} X_j$.
    This excludes month $0$, for which the max of the previous empty
    set is taken to be $+∞$.

    With \conceptFormat{Plan B}, you pay a penalty of \$100 for every
    pair of months $2i$ and $2i+1$ in which $X_{2i}$ and $X_{2i+1}$
    both exceed $\max_{j<2i} X_j$. As in Plan A, you never pay a
    penalty for the pair of months $0$ and $1$.

    \begin{enumerate}
        \item Suppose that you expect to live forever (and keep
            whatever plan you pick forever as well). If your goal is
            to minimize your expected total cost over eternity, which plan
            should you pick?
        \item Which plan is cheaper on average if you don't plan to live for more than, say, the
            next 1000 years?
    \end{enumerate}

    (Assume no discounting of future payments in either case.)
    
        \subsubsection*{Solution}

        Let $A_i$ be the event that we pay a penalty in month $i$ with
        plan $A$, and let $B_i$ for even $i$ be the event that we pay
        a penalty for months $i$ and $i+1$ in plan $B$. Then for
        $i>0$,
        \begin{align*}
            \Prob{A_i}
            &= \Prob{X_i > \max_{j<i} X_j}
            \\&= \Prob{X_i = \max_{j≤i} X_i} 
            \\&= \frac{i!}{(i+1)!}
            \\&= \frac{1}{i+1}.
            \intertext{Similarly,}
            \Prob{B_i}
            &= \Prob{\text{$X_i > \max_{j<i} X_j$ and $X_{i+1} >
            \max_{j<i}$}}
            \\&= \frac{i!⋅2}{(i+2)!}
            \\&= \frac{2}{(i+2)(i+1)}.
        \end{align*}

        (In the first case, we count all orderings of the $i+1$
        elements $0\dots i$ that make $X_i$ largest, and divide by the
        total number of orderings. In the second case, we count all
        the orderings that make $X_i$ and $X_{i+1}$ both larger than
        all the other values.)

        Now let us do some sums.

        \begin{enumerate}
            \item For the infinite-lifetime version, Plan A costs
                \begin{align*}
                    ∑_{i=1}^{∞} \frac{1}{i+1},
                \end{align*}
                which diverges. Plan B costs
                \begin{align*}
                    100 ⋅ ∑_{i=1}^{∞} \frac{2}{(2i)(2i-1)}
                    &≤ 100 ⋅ ∑_{i=1}^{∞} \frac{2}{(2i)^2}
                    \\&= 100 ⋅ ∑_{i=1}^{∞} \frac{1}{2i^2}
                    \\&= \frac{100}{2} ⋅ \frac{π^2}{6}.
                \end{align*}
                which is finite. So Plan B is
                better for immortals.
            \item For the finite-lifetime version, over $n$ months, Plan A costs
                \begin{align*}
                    ∑_{i=1}^{n} \frac{1}{i}
                    &≈ \ln n + γ
                \end{align*}

                Plan B has a $\frac{2}{3⋅4} = \frac{1}{6}$ chance of
                costing \$100 after the first four months, giving an
                expected cost of at least \$16. As long as you keep
                both for at least four months, for $n \ll e^{16} ≈
                8886110$ months $≈ 740509$ years, plan A is better.
        \end{enumerate}

    \subsection{A randomly-indexed list}

    Consider the following data structure: We store elements in a
    sorted linked list, and as each element is inserted, with
    probability $p$ we also insert it in a balanced binary search tree
    data structure with guaranteed $O(\log n)$-time search and insert operations (say, a
    red-black tree). We can now search for an element $x$ by searching for
    the largest element $y$ in the tree less than or equal to $x$, then walking along
    the list starting from $y$ until we find $x$ (or some value bigger
    than $x$).

    Recall that for a skip list with (small) parameter $p$, the expected
    number of pointers stored per element is $1+O(p)$ and the expected
    cost of a search operation is $O\parens*{\frac{1}{p} \log n}$. 
    
    \begin{enumerate}
        \item 
            What are the
    corresponding asymptotic values for the expected pointers per element and
            the expected cost of a search for this new data structure, as a
    function of $p$ and $n$?
        \item Suppose you are given $n$ in advance. Is it possible to choose
            $p$ based on $n$
            for the tree-based data structure to get lower expected space overhead (expressed in
            terms of pointers per element) 
            than with a skip list, while still getting $O(\log n)$
            expected search time?
        \end{enumerate}

        \subsubsection*{Solution}

        \begin{enumerate}
            \item
        This is actually easier than the analysis for a skip list,
        because there is no recursion to deal with.

        A balanced binary search tree stores exactly $2$ pointers per
        element, so with an expected $pn$ elements in the tree, we get
        $n + 2pn$ pointers total, or $1+2p)$ pointers per element.

        For the search cost, work backwards from the target $x$ to
        argue that we traverse an expected $O(1/p)$ edges in the
        linked list before hitting a tree node. Searching the tree
        takes $O(\log n)$ time independent of $p$. This gives a cost
        of $O\parens*{\frac{1}{p} + \log n}$.
    \item Yes. Let $p = Θ(1/\log n)$. Then we use an expected
        $1+O(1/\log n)$ pointers per element, while the search cost is
                $O\parens*{\frac{1}{1/\log n} + \log n} = O(\log n)$.
                In contrast, getting $1+O(1/\log n)$ expected space
                with a skip list would also require setting $p =
                Θ(1/\log n)$, but this would give an expected search
                cost $Θ(\log^2 n)$.
        \end{enumerate}

\section{Assignment 4: due Thursday, 2019-10-31, at 23:00}

    \subsection{A hash tree}

    Suppose we attempt to balance a binary search tree by inserting
    $n$ elements one at a time, using hashes of the values instead of
    the values themselves.

    Specifically,
    let $M = \Set{1 \dots n^2}$ be the set of possible
    hash values; 
    let $U = \Set{1 \dots u}$ be the universe of possible values,
    where $u \gg n^2$;
    and let $H$ be a $2$-universal family of hash
    functions $h:U→M$.
    After the adversary chooses $n$ distinct values $x_1,\dots,x_n$,
    the algorithm chooses a hash function $h∈H$ uniformly at random,
    and inserts $x_1$ through $x_n$ in order into a binary search tree
    with no rebalancing, where
    the key for each value $x_i$ is $h(x_i)$.

    In the case of duplicate keys (which shouldn't happen very often),
    assume that if we try to insert $x$ into a subtree with root $r$,
    and $h(x) = h(r)$, we just adopt some fixed rule like always inserting
    $x$ into the right subtree of $r$.

    For any $n$ and $u \gg n^2$, prove or disprove: for any $2$-universal
    family $H$ of hash functions, and any sequence of values
    $x_1,\dots,x_n$, the expected depth of a
    tree constructed by the above process is $O(\log n)$.

        \subsubsection*{Solution}

        Disproof: 
        We will construct a $2$-universal family $H$ such
        that $h(x) = x$ for all $h∈H$ whenever $1 ≤ x ≤ m$. We can then
        insert the values $x_i = i$ with corresponding hash values
        $h(x_i) = i$ and get a tree of depth exactly $n$.

        Start with a \emph{strongly} $2$-universal family $H'$. For
        each $h'$ in $H$, construct a corresponding $h$ according to
        the rule $h(x) = x$ when $1 ≤ x ≤ m$ and $h(x) = h'(x)$
        otherwise. We claim that this gives a $2$-universal family of
        hash functions.

        Recall that $H$ is $2$-universal if, for any $x≠y$,
        $\Prob{h(x) = h(y)} ≤ 1/m$, when $h$ is chosen uniformly from
        $H$. We consider three cases:
        \begin{enumerate}
            \item If $1 ≤ x ≤ m$ and $1 ≤ y ≤ m$, then $h(x) = x ≠ y =
                h(y)$ always.
            \item If $1 ≤ x ≤ m$ and $m < y$, then $\Prob{h(y) = h(x)}
                = \Prob{h'(y) = x} = 1/m$, since $h(y)$ is equally
                likely to be any value by strong $2$-universality.
                A symmetric argument works if $m < x$ and $1 ≤ y ≤ m$.
            \item If $m < x$ and $m < y$, then $\Prob{h(x) = h(y)} =
                \Prob{h'(x) = h'(y)} = 1/m$.
        \end{enumerate}

        In each case we have $\Prob{h(x) = h(y)} ≤ 1/m$.
        So $H$ gives a $2$-universal hash family that produces trees
        of depth $n$, much greater than $O(\log n)$.

    \subsection{Randomized robot rendezvous on a ring}

    Suppose we have a ring of size $nm$, consisting of positions
    $0,\dots,nm-1$, with $n$ robots starting at positions $X_{i0} =
    im$.

    At each time step $t$, an adversary chooses one of the $n$ robots
    $i$, and this robot moves from position $X_{it}$ to $X_{i,t+1} = (X_{it} ± 1) \bmod
    mn$, moving clockwise or counterclockwise with equal probability. If
    any other robot or robots $i'$ are at the same position $X_{i't} =
    X_{it}$, they also move to the same new position with robot $i$.
    Eventually, the robots slowly coalesces until all robots
    are in the same position. 

    The process finishes at the first time $τ$ where $X_{iτ} =
    X_{jτ}$ for all $i$ and $j$. What is $\Exp{τ}$?

    (Assume that the adversary's choice of
    robot to move at each time $t$ doesn't involve predicting the
    future. Formally, this choice must be 
    measurable $ℱ_t$, where $ℱ_t$ is the $σ$-algebra generated by all
    the positions of the robots at times up to and including $t$.)

        \subsubsection*{Solution}

        We'll use a variant of the $X_t^2 - t$ martingale for random
        walks.

        For each $i$ and $t$, let $Y_{it} = X_{i+1,t} - X_{it}$ be the
        size of the gap between $X_{i+1}$ and $X_i$ (wrapping around
        in the obvious way). 

        Suppose at some step $t$ the adversary chooses robot $j$ to
        move. Let $i$ and $k$ be the smallest and largest robot ids
        such that $X_{it} =
        X_{jt} = X_{kt}$ (again wrapping around in the obvious way).

        Then $Y_{i-1,t}$ and $Y_{kt}$ are the only gaps that change,
        and each rises or drops by 1 with equal probability. So
        $\ExpCond{Y_{i-1,t+1}^2}{ℱ_t} = Y_{i-1,t}^2 + 1$
        and
        $\ExpCond{Y_{j,t+1}^2}{ℱ_t} = Y_{j,t}^2 + 1$.
        If we define $Z_t = ∑_{i=1}^{n} Y_{it}^2 - 2t$, then
        $\Set{Z_t}$ is a martingale with respect to $\Set{ℱ_t}$.

        For any adversary strategy, starting from any configurations,
        there is a sequence of at most
        $n^2m$ coin-flips that causes all robots to coalesce. So
        $\Exp{τ} ≤ n^2m 2^{n^2m} < ∞$. We also have that
        $\abs*{Z_{t+1} - Z_t} ≤ 2$, so we can apply the
        finite-expectation/bounded-increments case of
        Theorem~\ref{theorem-optional-stopping} to show that
        $\Exp{Z_τ} = \Exp{Z_0} = nm^2$.
        But at time $τ$, all but one interval $Y_{it}$ is zero, and
        the remaining interval has length $mn$. This gives
        $\Exp{Z_τ} = \Exp{(mn)^2 - 2τ}$, giving $\Exp{τ} = \frac{1}{2}
        \parens*{n^2m^2 - nm^2} = \binom{n}{2} m^2$.

        The nice thing about this argument is that it instantly shows
        that the adversary's strategy doesn't affect the expected time
        until all robots coalesce, but the price is that the argument
        is somewhat indirect. For specific adversary strategies,
        it may be possible to show the coalescence time more directly.
        For example, consider an adversary that always picks robot
        $0$. Then we can break down the resulting process into a
        sequence of phases delimited by collisions between $0$ and 
        robots that have not yet been added to its entourage. The
        first such collision happens after robot $0$ hits one of the
        two absorbing barriers at $-m$ and $+m$, which occurs in $m^2$
        steps on average. At this point, we now have a new random walk
        with absorbing barriers at $-m$ and $+2m$ relative to the
        position of robot $0$ (possibly after flipping the directions
        to get the signs right). This second random walk takes $2m^2$
        steps on average to hit one of the barriers. Continuing in
        this way gives us a sequence of random walks with absorbing
        barriers at $-m$ and $+km$ for each $k∈\Set{1\dots n-1}$. The
        total time is thus $∑_{k=1}^{n-1} km^2 = \frac{(n-1)n}{2}m^2 =
        \binom{n}{2} m^2$, as shown above for the general case.

\section{Assignment 5: due Thursday, 2019-11-14, at 23:00}

    \subsection{Non-exploding computers}

    Engineers from the Exploding Computer Company have devised a new
    scheme for managing the ring of overheating computers
    from Problem~\ref{section-problem-exploding-computers}. In the new scheme, 
    we start with an $n$-bit vector $S^0 = 0^n$. Each $1$ represents a
    computer that is turned on, and each $0$ represents a computer
    that is turned off.\footnote{The marketing department convinced
    the company to change the number of computers in the ring from
    $3n$ to $n$ in response to consumer complaints. If the engineers'
    scheme works, the next meeting of the marketing department
    will consider choosing a new name for the company.}

    Call a state $S$ \conceptFormat{permissible} if there is no $j$
    such that $S^{t+1}_j = S^{t+1}_{j+1} = S^{t+1}_{j+2} = 1$, where
    the indices wrap around mod $n$.

    At each step, some position $i$ is selected uniformly at random.
    If $S^t_i = 1$, $S^{t+1}_i$ is set to $0$ with probability $p$.
    If $S^t_i = 0$, $S^{t+1}_i$ is set to $1$ if this creates a
    permissible configuration, and stays $0$ otherwise.
    For any positions $j≠i$, $S^{t+1}_j = S^t_j$.
    
    \begin{enumerate}
        \item Show that when $0<p<1$, the sequence of values $S^t$ forms a Markov
            chain with states consisting of all permissible
            sets, and that this chain has
            a unique stationary distribution $π$ in which
            $π(S) = π(T)$ whenever $\card{S} = \card{T}$, where
            $\card{S}$ is the number of $1$ bits in $S$.
        \item Let $τ$ be the first time at which
            $\card{S^τ} ≥ n/2$. Show that for any $n$, there is a
            choice of $p≥0$
            such that $\Exp{τ} = O(n \log n)$.
    \end{enumerate}

        \subsubsection*{Solution}

        \begin{enumerate}
            \item First observe that the update rule for generating
                $S^{t+1}$ depends only on the state $S^t$; this gives
                that $\Set{S^t}$ is a Markov chain. It is irreducible,
                because there is always a nonzero-probability path
                from any permissible configuration $S$ to any
                permissible configuration $T$ that goes through the
                empty configuration $0^n$. It is aperiodic, because for any
                nonempty configuration $S^t$, there is a nonzero probability
                that $S^{t+1} = S^t$, since we can pick a position
                $i$ with $S^t_i = 1$ and then choose not set
                $S^{t+1}_i$ to $0$.
                So a unique stationary distribution $π$ exists.

                We can show that $π(S)$ depends only on $\card{S}$ by
                using the fact that the Markov chain is 
                reversible. Let $S$ and $T$ be reachable states that
                differ only in position $i$. Suppose that $S^t_i = 0$
                and $T^t_i = 1$.
                Then $p_{ST} = 1/n$ and $p_{TS} = p/n$.
                Let $c = ∑_S p^{-\card{S}}$, where $S$ ranges over all
                admissible states, and let $π_S =
                p^{-\card{S}}/c$. Pick some $S$ and $T$ as above and
                let $k = \card{S}$.
                \begin{align*}
                    π_S p_{ST}
                    &= \frac{p^{-k}}{c}\frac{1}{n}
                    \\&= \frac{p^{-k-1}}{c}\frac{p}{n}
                    \\&= π_T p_{TS}.
                \end{align*}

                So this particular choice of $π$ satisfies the
                detailed balance equations and is the stationary
                distribution of the chain. Since $π(S)$ only depends
                on $\card{S}$, it is immediate that $π(S) = π(T)$ when
                $\card{S} = \card{T}$.
            \item 
                For this part, we'll set $p=0$, then walk
                directly up to a configuration with $\card{S^t} ≥ n/2$
                in $O(n \log n)$ steps on average. Setting $p=0$
                means that the chain is no longer irreducible, but we
                won't care about this since we are not worried about
                convergence.

                Let $S$ be a permissible configuration with
                $\card{S} = k$. We want to get a lower bound on the
                number of positions $i$ such that $S[i/1]$ is also
                permissible. There $n-k$ zeroes in $S$; each such $0$
                can be switched to $1$ unless it would create three
                ones in a row. We will show that at most $k$ of the
                $n-k$ zeroes cannot be switched, by constructing an
                injection $f$ from the set of unswitchable $0$ positions
                to the set of $1$ positions.

                Let $i$ be a position such that $S_i=0$ but $S[i/1]$
                is not permissible. Then one of the following cases
                holds:
                \begin{enumerate}
                    \item $S_{i-2} = S_{i-1} = 1$. Let $f(i) = i-1$.
                    \item $S_{i-1} = S_{i+1} = 1$. Let $f(i) = i-1$.
                    \item $S_{i+1} = S_{i+2} = 1$. Let $f(i) = i+1$.
                \end{enumerate}

                To show that this is an injection, observe that the
                only way for $f(i) = f(j)$ to occur is if $f(i) = i+1$
                and $f(j) = j-1$ (or vice versa). But if $f(i) =
                i+1$, then $S_j = S_{i+2} = 1$, and $S_j$ is not an
                unswitchable zero.

                So out of $n-k$ zeroes we have at most $k$
                unswitchable zeroes, giving at least $n-2k$ switchable zeroes.
                The expected waiting time to hit one of these at least $n-2k$ switchable
                zeroes is at most $\frac{n}{n-2k}$, giving a total
                expected waiting time of at most $∑_{k=0}^{\floor{(n-1)/2}}
                \frac{n}{n-2k} ≤ n H_n = O(n \log n)$.

                I suspect the correct expected waiting time is in fact
                $O(n)$, because (a) the pattern $11011$ includes at
                least one $1$ that is not the image of any
                unswitchable zero; (b) after, say, $n/8$ steps it is
                likely (via linearity of expectation and McDiarmid's
                inequality) that there are a linear number $\ell$ of
                such patterns in $S$; so that (c) even as $k$ gets
                close to $n/2$, the number of unswitchable zeros is at
                least $n-2k+\ell = Θ(n)$, giving constant expected waiting
                times to find the next switchable zero.
                But actually proving (b)
                given all the dependencies flying around looks
                at least mildly painful, so we may want to just hope that an $O(n \log
                n)$ bound is good enough.
        \end{enumerate}

        \subsection{A wordy walk}

        Consider the following random walk on strings over some
        alphabet $Σ$. At each step:
        \begin{enumerate}
            \item With probability $p$, if $\card{X^t} > 1$, delete a
                character from a position chosen uniformly at random
                from all positions in $X^t$. If $\card{X^t} = 1$, do
                nothing.
            \item With probability $q$, insert a new character chosen
                uniformly at random from $Σ$ into $X^t$. The new
                character will be inserted after the first $i$
                characters in $X^t$, where $i$ is chosen uniformly at
                random from $\Set{0\dots \card{X^t}}$.
            \item With probability $r$, choose one of the character
                positions in $X^t$ uniformly at random and replace the
                character in that position with a character chosen
                uniformly at random from $Σ$.
        \end{enumerate}

        For example, here is the result of running the above process
        for a few steps
        with $Σ$ the lowercase Latin alphabet, $p = 1/2$, $q = 1/4$, and $r=1/4$, starting from the
        string \texttt{markov}: \texttt{markov markzv marzv carzv caurzv cauzv cauav cauav cavav avav}.

        Assume $p+q+r=1$, $p > q > 0$, and $r > 0$. 

        Suppose that we start with $\card{X^0} = n$. Show that for
        any constants $p,q,r$ satisfying the above constraints, and any
        constant $ε > 0$, there is a time $t = O(n)$ such that
        $d_{TV}(X^t,π) ≤ ε$, where $π$ is a stationary distribution
        of this process.

            \subsubsection*{Solution}

            We can simplify things a bit by tracking $X^t =
            \card{S^t}$. Since changes in the length of $S$ don't
            depend on the characters in $S$, $\Set{X^t}$
            is also a Markov chain, with transition
            probabilities
            \begin{align*}
                p_{ij} &=
                \begin{cases}
                    p & j = i-1 \\
                    q & j = i+1 \\
                    r & \text{otherwise}
                \end{cases}
            \end{align*}

            Applying the detailed balance equations gives a stationary
            distribution $ρ$ for $X^t$ of $ρ_i = (q/p)^{i-1} /
            (1-q/p)$.  By symmetry, this gives a stationary
            distribution $π_S = \frac{(q/p)^{\card{S}-1}
            \card{Σ}^{-\card{S}}}{1-q/p}$. 
            For $X$ with distribution $ρ$, we have $\Exp{X}$ finite
            when $p > q$ since $\Exp{X} = \frac{1}{1-q/p} ∑_{n=1}^{∞}
            n (q/p)^{n-1}$ converges.

            We'll show convergence for $S^t$ starting from a particular $S^0$
            using a coupling with a stationary copy of the chain $\Set{T^t}$.

            Our strategy will go as follows:
            \begin{enumerate}
                \item Let $m = \card{T^0}$. We will argue that a
                    coupling that applies the same operation (delete,
                    insert, change) to both copies of the chain
                    reaches a configuration where $\card{S^t} =
                    \card{T^t} = 1$ in $O\max(n,m)$ expected steps.
                \item From a state with $\card{S^t} = \card{T^t} = 1$,
                    with probability $r>0$, both chains switch to a
                    new character. Make this the same character and
                    get $S^t = Y^t$.
                \item If this doesn't work, repeat the argument from
                    the start. Note that we now have $\card{S^t} =
                    \card{T^t} ≤ 2$, so after $O(1)$ expected steps we
                    are back at $\card{S^t} = \card{T^t} = 1$ again.
                    The waiting time for convergence is now geometric
                    and has expectation $O(1)$.
            \end{enumerate}

            Let us now justify the claim that the waiting time to
            reach $\card{S^t} = \card{T^t} = 1$ is $O(\max(n,m))$.
            This is just the usual argument for a biased random walk
            with one absorbing barrier.
            Let $Z^t = \max\parens*{\card{S^t},\card{Y^t}}$. Let $τ$
            be the stopping time at which $Z^t$ is first equal to $1$.
            Observe that for $t < τ$, $\Exp{Z^{t+1}}{Z^0\dots Z^t} =
            Z^t + q - p$, so $Q^t = Z^t + (p-q)t$ is a martingale with
            bounded increments, and we can apply OST to get that
            $\Exp{Q_τ} = 1 - (p-q)τ = \Exp{Q_0} = \max{n,m}$. This
            gives $\Exp{τ} = \frac{\max{n,m}-1}{p-q}$.

            We now have a coupling that gives an expected
            d coalescence time of $O(\max(n,m))$. If $\max(n,m) = n$,
            we are done. But it could be that $\max(n,m) = m$.
            Conditioning on $\card{T^t} > n$, we are truncating the
            distribution of $\card{T^t}$ to be a geometric
            distribution starting at $n+1$; this has expected value
            $n+O(1)$, giving $\Exp{\max(n,m)} = n + O(1)$ in either
            case. So the coalescence time is $O(n+O(1)) = O(n)$.

            Since $ε$ is a constant, after applying the Coupling Lemma
            and Markov's inequality, the $O$ eats it.

\section{Assignment 6: due Monday, 2019-12-09, at 23:00}

    \subsection{Randomized colorings}

    Suppose we want to use a constant number $k$ of colors
    to color the vertices of a graph $G$ with $n$
    vertices and $m$ edges
    so that we minimize the number of monochromatic edges. A
    particularly simple approach is to color the vertices
    independently and uniformly at random. This gives $m/k$
    monochromatic edges on average. But we'd like to derandomize the
    algorithm to get at most $m/k$ monochromatic edges always.

    There are two standard ways to do this:
    \begin{enumerate}
        \item Replace the independent random colors with pairwise
            independent random colors and then consider all possible
            random choices. 
            Show that we can do this by generating $n$
            pairwise-independent colors from a small number of random
            values, and determine the asymptotic time complexity, as a
            function of $n$, $m$, and $k$, of generating
            and testing all the resulting colorings to find the best
            one.
        \item Apply the method of conditional probabilities and assign
            colors one at a time to minimize the expected number of
            monochromatic edges conditioned on the assignment so far.
            Show that we can do so in this case, and compute the
            asymptotic time complexity as a function of $n$, $m$, and $k$ of
            the resulting deterministic algorithm.
    \end{enumerate}

        \subsubsection*{Solution}

        \begin{enumerate}
            \item The obvious way to generate $n$ pairwise-independent
                random values in $\Set{0,\dots,k-1}$ is to apply the
                subset-sum construction described in
                §\ref{section-pairwise-independence-construction}.
                Let $\ell = \ceil{\lg (n+1)}$, assign a distinct
                nonempty subset $S_v$ of $\Set{1,\dots,\ell}$ to each
                vertex $v$, and let $X_v = \sum_{i∈C_v} r_i$ where
                $r_1,\dots,r_\ell$ are independent random values
                chosen uniformly from $ℤ_k$. Then the $X_v$ are
                pairwise independent, giving a probability of $1/k$
                that any particular edge is monochromatic. Since this
                gives $m/k$ monochromatic edges on average,
                enumerating all choices of $r_1,\dots,r_\ell$ will
                find a particular coloring that gets at least $m/k$
                monochromatic edges.

                There are $k^\ell = k^{\ceil{\lg (n+1)}}$ possible
                choices. This is bounded by $k^{2 + \lg n} =
                O\parens*{k^2 n^{\lg k}} = O\parens*{n^{\lg k}}$, which is
                polynomial in $n$.

                Generating each coloring takes $O(n \log n)$ time, and
                testing the edges takes $O(m)$ time, for a total time
                of $O\parens{n^{\lg k} (m + n \log n)}$.

                With some data structure tinkering we can reduce the
                complexity a bit. We can enumerate all of
                $r_1,\dots,r_k$ with an amortized $O(1)$ values changing at
                each iteration. This still
                requires changing about half the colors, but we can
                update each dynamically by subtracting off the old
                value of $r_i$ and adding the new one, reducing the
                amortized cost per iteration from $O(n \log n)$ to
                $O(n)$. We still end up checking most of the edges, so
                the cost will now be $O\parens*{n^{\lg k} (m+n)}$. I don't
                see an obvious way to do better.
            \item For this part we let $X_1,\dots,X_n$ be the colors
                assigned to the vertices (ordering them arbitrarily)
                and let $f(X_1,\dots,X_n)$ be the number of
                monochromatic edges given a particular coloring.
                Having fixed $X_1,\dots,X_i$, we want to choose
                $X_{i+1}$ to minimized
                $\ExpCond{f}{X_1,\dots,X_{i+1}}$.

                Observe that $f(X_1,\dots,X_n) = ∑_{uv∈E} [X_u=X_v]$,
                so $\Exp{f|X_1,\dots,X_i} = ∑_{uv∈E} \Prob{X_u=X_v}$.
                For each $uv$, if $i+1 ∉ \Set{u,v}$, then
                $[X_u = X_v]$ is independent of $X_{i+1}$, giving
                $\ProbCond{X_u=X_v}{X_1,\dots,X_{i+1}} =
                \ProbCond{X_u=X_v}{X_1,\dots,X_{i+1}}$.
                So we only care about the cases where $i+1 ∈ \Set{u,v}$.
                Assume without loss of generality that $i+1=u$. 
                If $v ≤ i$, then $[X_u=X_v]$ is either $1$ or $0$
                depending on whether we set $X_u = X_v$ or not.
                If $v > i+1$, then $X_v$ is independent of $X_{i+1}$,
                so $\ProbCond{X_u=X_v}{X_1,\dots,X_{i+1}} =
                \ProbCond{X_u=X_v}{X_1,\dots,X_i}$.

                This means that to minimize
                $\ExpCond{f}{X_1,\dots,X_{i+1}}$, the only edges we
                need to consider are those edges $X_{uv}$ where one
                endpoint is $i+1$ and the other endpoint is in
                $\Set{1,\dots,i}$. Minimizing the conditional
                expectation of $f$ turns into the obvious greedy
                algorithm of picking a color for $v_{i+1}$ that is
                least represented among its neighbors, which we can do
                in $O\parens*{1+d(v_{i+1})}$ time. Adding up over all vertices
                $v$ gives $O\parens*{∑_{v∈V} (1+d(v))} = O(n+m)$ time.
        \end{enumerate}

    \subsection{No long paths}

    Given a directed graph $G$ with $n$ vertices and maximum
    out-degree $d$, we would like to find an induced subgraph $G'$
    with at least $(1-δ) n / (d+1)$ vertices, such that $G'$ has no
    simple paths\footnote{For the purposes of this problem, a
    \index{path!simple}\concept{simple path} of length $\ell$ is a
    directed path with $\ell$ edges and no repeated vertices.} of
    length $\ceil{2 d \ln n}$ or greater.

    Find an algorithm that does this in polynomial expected time for
    any constant $d > 0$ and $δ > 0$, and sufficiently large $n$.

        \subsubsection*{Solution}
        We'll put each vertex $v$ to $G'$ with independent probability
        $p = 1/(d+1)$. We now have two possible bad outcomes:
        \begin{enumerate}
            \item We may get a path of length $\ceil{2 d \ln n}$ in
                $G'$.
            \item We may get fewer than $(1-δ) n / (d+1)$ vertices in
                $G'$.
        \end{enumerate}

        We'll start by showing that the sum of the probabilities of
        these bad outcomes is not too big.

        Let $\ell = \ceil{2d \ln n}$.
        Observe that we can describe a path of length $\ell$ in $G$ by
        specifying its starting vertex ($n$ choices) and a sequence of
        $\ell$ edges each leaving the last vertex added so far ($d$ choices each).
        This gives at most $nd^\ell$ paths of length $\ell$ in $G$.
        Each such path appears in $G'$ with probability exactly
        $p^{\ell+1}$. Using the usual probabilistic-method argument,
        we get
        \begin{align*}
            \Prob{\text{$G'$ includes at least one path of length
            $\ell$}}
            &≤ \Exp{\text{number of paths of length $\ell$ in $G'$}}
            \\&≤ nd^{\ell} p^{\ell+1}
            \\&= np (pd)^\ell.
            \\&= \frac{1}{d+1} n \parens*{\frac{d}{d+1}}^\ell
            \\&= \frac{1}{d+1} n \parens*{1-\frac{1}{d+1}}^\ell
            \\&= \frac{1}{d+1} n \parens*{1-\frac{1}{d+1}}^{\ceil{2d \ln n}}
            \\&≤ \frac{1}{d+1} n \parens*{1-\frac{1}{d+1}}^{2d \ln n}
            \\&≤ \frac{1}{d+1} n e^{-\ln n}
            \\&= \frac{1}{d+1}.
            \\&≤ \frac{1}{2}.
        \end{align*}

        For the second-to-last inequality, we use the inequality
        $\parens*{1-\frac{1}{d+1}}^{2d} ≤ e^{-1}$ for $d ≥ 1$.
        This is easiest to demonstrate numerically, but if we had to
        prove it we could argue that it holds when $d=1$ (since $1/4 < 1/e$)
        and that $\parens{1-\frac{1}{d+1}}^{2d}$ is a decreasing function of
        $d$.

        For getting too few vertices, use Chernoff bounds. Let $X$ be
        the number of vertices in $G'$. Let $μ = \Exp{X} =
        \frac{n}{d+1}$.
        
        \begin{align*}
            \Prob{X < (1-δ) n / (d+1)}
            &= \Prob{X < (1-δ)μ}
            \\&≤ e^{-μδ^2/2}
            \\&= e^{-nδ^2/(2(d+1))}
            \\&< 1/4,
        \end{align*}
        for any fixed $d > 0$, $δ > 0$, and sufficiently large $n$.

        Adding these error probabilities together gives a total
        probability of failure of at most $3/4$. We can select the
        vertices in $G'$ in time linear in $n$. So if we can detect
        failure in polynomial time and try again, we only have to run
        at most $4$ poly-time attempts on average until we win, giving
        polynomial expected cost.

        So now let us detect failure. There are at most $nd^\ell$
        paths of length $\ell$ in $G$. We can enumerate all of them
        and check if each is in $G'$ in time 
        \begin{align*}
            O\parens*{nd^\ell \ell}
            &= O\parens*{n\ell d^{2 d \ln n + 1}}
            \\&= O\parens*{n \ell d e^{2 d \ln n \ln d}}
            \\&= O\parens*{n^{2 d \ln d + 1} d^2 \log n},
        \end{align*}
        which is polynomial in $n$ for fixed $d$.

\section{Final exam}

Write your answers in the blue book(s).  Justify your answers.  Work
alone.  Do not use any notes or books.  

There are three problems on this exam, each worth 20
points, for a total of 60 points.
You have approximately three hours to complete this
exam.

\subsection{A uniform ring}

Suppose we have a ring of $n$ nodes, each of which holds a bit.
At each step, we pick a node uniformly at random and have it copy
the bit of its left neighbor. For example, if we start in state
$010$, then picking the rightmost node will send us to state $011$,
and if we then pick the leftmost node we will reach state $111$. From
this last state no further changes are possible.

Suppose $n$ is even and we start with a state where positions $1$
through $n/2$ are all $0$ and positions $n/2+1$ through $n$ are all
$1$.
What is the
expected number of steps, as an asymptotic function of $n$, to reach a
state where all bits are the same?

    \subsubsection*{Solution}
    First observe that any transition preserves the property that our
    state looks like a string of the form $0^k 1^{n-k}$,
    where both the zeroes and ones all appear in consecutive positions
    around the ring.
    
    Let $X_t$ be the number of zeroes after $t$
    steps. We have $X_0 = n/2$, and our possible transitions when $0 <
    X_t < n$ are:
    \begin{enumerate}
        \item With probability $1/n$, we choose the leftmost zero and
            replace it with a one. This makes $X_{t+1} = X_t - 1$.
        \item With probability $1/n$, we choose the leftmost one and
            replace it with a zero. This makes $X_{t+1} = X_t + 1$.
        \item With the remaining probability $1-2/n$, we choose some
            node whose bit is equal to that of its left neighbor. This
            makes $X_{t+1} = X_t$.
    \end{enumerate}

    When $X_t = 0$ or $X_t = n$, we have converged: no further changes
    are possible.

    Looking at the transitions for $X_t$, we see that it is doing a
    very lazy random walk with absorbing barriers at $0$ and $n$. If
    we remember that a non-lazy random walk converges in $O(n^2)$
    steps on average, we can
    argue that the lazy version converges in $O(n^3)$ steps on
    average,
    because the expected waiting time between changes in $X_t$ is
    $n/2 = O(n)$, and we can multiply this by the number of changes because of
    Wald's equation.

    If we want to be very formal about this, we can also calculate the
    exact expected convergence time using the Optional Stopping
    Theorem. Let's guess that $Y_t = X_t^2 - ct$ is a martingale for some
    coefficient $c$. To find $c$, observe that we need
    \begin{align*}
        \ExpCond{\parens*{X_{t+1}^2 - c(t+1)}
        - \parens*{X_t^2 - ct}}{X_t}
        &= \ExpCond{X_{t+1}^2 - X_t^2 - c}{X_t}
        \\&= \frac{1}{n} (2X_t + 1) + \frac{1}{n} (-2X_t+1) - c
        \\&= \frac{2}{n} - c
        \\&= 0.
    \end{align*}

    So $\Set{Y_t}$ is a martingale when $c = \frac{2}{n}$.

    Let $τ$ be the stopping time at which $X_τ$ first equals $0$ or
    $n$. The usual argument shows that $\Exp{τ}$ is finite, and
    $\Set{Y_t}$ has bounded increments because $Y_t$ never changes by
    more than $2X_t + 1 ≤ 2n + 1$. So OST applies and $\Exp{Y_τ} =
    \Exp{Y_0} = (n/2)^2$. But we can expand 
    \begin{align*}
        \Exp{Y_τ} &= \frac{1}{2}⋅0
    + \frac{1}{2}⋅n^2 - \frac{2}{n}⋅\Exp{τ} 
        \\&= \frac{n^2}{2} -
    \frac{2}{n}⋅\Exp{τ}.
    \end{align*}
    This gives $\frac{n^2}{4} = \frac{n^2}{2} -
    \frac{2}{n}⋅τ$. Solving for $\Exp{τ}$ gives $\Exp{τ} =
    \frac{n^3}{8} = O(n^3)$.

\subsection{Forbidden runs}

Given a sequence of coin-flips, a \concept{run} is
a maximal subsequence of either all heads or all tails. The
\concept{length} of a run is the number of flips in the run. For
example, the sequence $\coinFlips{HHHTTHTHH}$ has five runs: a run
$\coinFlips{HHH}$ of length $3$, a run $\coinFlips{TT}$ of length $2$, a run
$\coinFlips{H}$ of length $1$, a run $\coinFlips{T}$ of length $1$, and a run
$\coinFlips{HH}$ of length $2$.

Suppose we are attempting to transmit a sequence of $n$ fair,
independent coin-flips across a communication channel that uses a run
of length $k$ to signal an emergency condition. To avoid false alarms,
we must encode our sequence to include no runs of length $k$. We do
this by encoding any run of length $\ell ≥ k$ as a run of length
$\ell+1$ of the same value. For example, when $k=2$, we would encode the previous
example $\coinFlips{HHHTTHTHH}$ as $\coinFlips{HHHHTTTHTHHH}$.

As a function of $n$ and $k$, what is the expected length of the
encoded sequence?

    \subsubsection*{Solution}

    Let $i$ range from $1$ to $n$, and let $X_i$ be the $i$-th
    coin-flip in the sequence. Let $R_i$ be the indicator variable for
    the event that there is a run of length $k$ or more starting at
    position $i$. Then the expected number of runs of length $k$ or
    more is given by $∑_{i=1}^{n} \Exp{R_i}$ by linearity of
    expectation, and because each such run adds one to the length of
    the encoded sequence, the expected encoded length is $n+∑_{i=1}^n
    \Exp{R_i}$.

    To compute $\Exp{R_i}$, we consider three cases:
    \begin{enumerate}
        \item If $i=1$, then $R_i=1$ when
            $X_1,\dots,X_k$ are all
            $\coinFlips{H}$ or all $\coinFlips{T}$. This gives $\Exp{R_i} =
            2⋅2^{-k} = 2^{-k+1}$.
        \item If $1 < i ≤ n-k+1$, then $R_i=1$ when
            $X_1,\dots,X_k$ are all $\coinFlips{H}$ or all $\coinFlips{T}$,
            and $X_{i-1}$ is not this common value (because otherwise
            the run starts earlier). This gives $\Exp{R_i} =
            2⋅2^{-k-1} = 2^{-k}$.
        \item If $i > n-k+1$, then $R_i = 0$, because there aren't
            enough coins left to get a run of length $k$.
    \end{enumerate}

    Summing over all the cases gives
    \begin{align*}
        ∑_{i=1}^n \Exp{R_i}
        &= 2^{-k+1} + ∑_{i=2}^{n-k+1} 2^{-k}
        \\&= 2⋅2^{-k} + (n-k)⋅2^{-k}
        \\&= (n-k+2)⋅2^{-k}.
    \end{align*}

    For the expected encoded sequence length, we must add back $n$ to
    get $n+(n-k+2)⋅2^{-k}$.

\subsection{A derandomized balancing scheme}

Suppose we construct an $n×n$ matrix $A$ where the elements $A_{ij}$
are $n^2$ independent fair $±1$ random variables.
Then Hoeffding's inequality tells us that 
\begin{equation*}
    \Prob{\abs*{∑_{i=1}^{n} ∑_{j=1}^{n} A_{ij}} ≥ t} ≤ 2\exp\parens*{-\frac{t^2}{2n^2}}.
\end{equation*}
Unfortunately this requires using $n^2$ random bits.

As an alternative, suppose we use $2n$ random bits to
generate two sequences $X$ and $Y$ of
$n$ independent fair $±1$ random variables, and define $B_{ij} =
X_i Y_j$.  

Show that
        \begin{align*}
    \Prob{\abs*{∑_{i=1}^{n} ∑_{j=1}^{n} B_{ij}} ≥ t} ≤ 4\exp\parens*{-\frac{t}{2n}}.
        \end{align*}

    \subsubsection*{Solution}

    Start by factoring
    \begin{align*}
    ∑_{i=1}^n ∑_{j=1}^n B_{ij}
        &= 
    ∑_{i=1}^n ∑_{j=1}^n X_i Y_j
       \\&=
    \parens*{∑_{i=1}^n X_i}\parens*{∑_{j=1}^n Y_j}.
    \end{align*}

    Each factor is a sum of $n$ independent $±1$ random variables with
    expectation $0$, so Hoeffding's inequality applies, giving
    \begin{align*}
    \Prob{\abs*{∑_{i=1}^n X_i} ≥ s} 
        &≤ 2 \exp\parens*{-\frac{s^2}{2n}}
        \intertext{and}
    \Prob{\abs*{∑_{j=1}^n Y_j} ≥ s} 
        &≤ 2 \exp\parens*{-\frac{s^2}{2n}}.
    \end{align*}
    If neither of these events holds, then
    $\abs*{∑_{i=1}^n ∑_{j=1}^n B_{ij}}
    =\abs*{∑_{i=1}^n X_i}⋅\abs*{∑_{j=1}^n Y_j}
    < s^2$.

    So by the union bound,
    \begin{align*}
        \Prob{\abs*{∑_{i=1}^{n} ∑_{j=1}^{n} B_{ij}} ≥ s^2}
        &≤ \Prob{\abs*{∑_{i=1}^n X_i} ≥ s} +
    \Prob{\abs*{∑_{j=1}^n Y_j} ≥ s}
        \\&≤ 4 \exp\parens*{-\frac{s^2}{2n}}.
    \end{align*}

    Now substitute $t$ for $s^2$.

\chapter{Sample assignments from Fall 2016}

\section{Assignment 1: due Sunday, 2016-09-18, at 17:00}

\subsection*{Bureaucratic part}

Send me email!  My address is
\mailto{james.aspnes@gmail.com}.

In your message, include:

\begin{enumerate}
\item Your name.
\item Your status: whether you are an undergraduate, grad student, auditor, etc.
\item Whether you are taking the course as CPSC 469 or CPSC 569.
\item Anything else you'd like to say.
\end{enumerate}

(You will not be graded on the bureaucratic part, but you should do it anyway.)

\subsection{Bubble sort}

Algorithm~\ref{alg-bubble-sort-one-pass} implements one pass of the
(generally pretty slow) \index{sort!bubble}\concept{bubble sort}
algorithm.  This involves comparing each array location with its
successor, and swapping the elements if they are out of order.
The full bubble sort algorithm repeats this loop until
the array is sorted, but here we just do it once.

\begin{algorithm}
    \Procedure{$\FuncSty{BubbleSortOnePass}(A,n)$}{
        \For{$i ← 1$ \KwTo $n-1$}{
            \If{$A[i] > A[i+1]$}{
                Swap $A[i]$ and $A[i+1]$\;
            }
        }
    }
    \caption{One pass of bubble sort.}
    \label{alg-bubble-sort-one-pass}
\end{algorithm}

Suppose that $A$ starts out as a uniform random permutation of
distinct elements.  As a function of $n$, what is the exact expected
number of swaps performed by Algorithm~\ref{alg-bubble-sort-one-pass}?

    \subsubsection*{Solution}

    The answer is $n-H_n$, where $H_n = ∑_{i=1}^{n} \frac{1}{i}$ is
    the $n$-th harmonic number.  There are a couple of ways to prove this.
    Below, we let $A_i$ represent the original contents of $A[i]$,
    before doing any swaps.  In each case, we will use the fact that
    after $i$ iterations of the loop, $A[i]$ contains the largest of
    $A_i,\dots,A_i$; this is easily proved by induction on $i$.

    \begin{itemize}
        \item Let $X_{ij}$ be the indicator variable for the event
            that $A_i$ is eventually swapped with $A_j$.  For this to
            occur, $A_i$ must be bigger than $A_j$, and must be
            present in $A[j-1]$ after $j-1$ passes through the loop.
            This happens
            if and only if $A_i$ is the largest value in
            $A_1,\dots,A_j$.  Because these values are drawn from a
            uniform random permutation, by symmetry $A_i$ is largest
            with probability exactly $1/j$.  So $\Exp{X_{ij}} = 1/j$.

            Now sum $X_{ij}$ over all pairs $i < j$.  It is easiest to
            do this by summing over $j$ first:
            \begin{align*}
                \Exp{∑_{i<j} X_{ij}} &= ∑_{i<j} \Exp{X_{ij}}
                \\&= ∑_{j=2}^{n} ∑_{i=1}^{j-1} \Exp{X_{ij}}
                \\&= ∑_{j=2}^{n} ∑_{i=1}^{j-1} \frac{1}{j}
                \\&= ∑_{j=2}^{n} \frac{j-1}{j}
                \\&= ∑_{j=1}^{n} \frac{j-1}{j}
                \\&= ∑_{j=1}^{n} \parens{1 - \frac{1}{j}}
                \\&= n - ∑_{j=1}^{n} \frac{1}{j}
                \\&= n - H_n.
            \end{align*}
        \item Alternatively, let's count how many values are
            \emph{not}
            swapped from $A[i]$ to $A[i-1]$.  We can then subtract
            from $n$ to get the number that are.

            Let $Y_i$
            be the indicator variable for the event that $A_i$ is
            not
            swapped into $A[i-1]$.  This occurs if, when testing
            $A[i-1]$ against $A[i]$, $A[i]$ is larger.  Since we know
            that at this point $A[i-1]$ is the largest value among
            $A_1,\dots,A_{i-1}$, $Y_i = 1$ if and only if $A_i$ is
            greater than all of $A_1,\dots,A_{i-1}$, or equivalently
            if $A_i$ is the largest value in $A_1,\dots,A_i$.  Again
            by symmetry we have $\Exp{Y_i} = \frac{1}{i}$, and summing
            over all $i$ gives an expected $H_n$ values that are not
            swapped down.  So there are $n-H_n$ values on average that are
            swapped down, which also gives the expected number of swaps.
    \end{itemize}

\subsection{Finding seats}

A huge lecture class with $n$ students meets in a room with $m ≥ n$
seats.  Because the room is dark, loud, and confusing, the students
adopt the following seat-finding algorithm:
\begin{enumerate}
    \item The students enter the room one at a time.  No student
        enters until the previous student has found a seat.
    \item To find a seat, a student chooses one of the $m$ seats
        uniformly at random.  If the seat is occupied, the student
        again picks a seat uniformly at random, continuing until they
        either find a seat or have made $k$ attempts that led them to
        an occupied seat.
    \item If a student fails to find a seat in $k$ attempts, they
        contact an Undergraduate Seating Assistant, who leads them to
        an unoccupied seat.
\end{enumerate}

Give an asymptotic (big-$Θ$) expression for the number of students who
are helped by Undergraduate Seating Assistants, as a function of $n$,
$m$, and $k$, where $k$ is a constant independent of $n$ and $m$.  As
with any asymptotic formula, this expression should work for large
$n$ and $m$; and should be as simple as possible.\footnote{An earlier
version of this problem allowed $k$ to grow with $n$, which causes
some trouble with the asymptotics.}

    \subsubsection*{Solution}

    Define the $i$-th student to be the student who takes their seat
    after $i$ students have already sat (note that we are counting
    from zero here, which makes things a little easier).  Let $X_i$ be
    the indicator variable for the even that the $i$-th student does
    not find a seat on their own and needs help from a USA.  Each
    attempt to find a seat fails with probability $i/m$.  Since each
    attempt is independent of the others, the probability that all $k$
    attempts fail is $(i/m)^k$.

    The number of students who need help is $∑_{i=0}^{n-1} X_i$, so
    the expected number is
    \begin{align*}
        \Exp{∑_{i=0}^{n-1} X_i}
        &= ∑_{i=0}^{n-1} \Exp{X_i}
        \\&= ∑_{i=0}^{n-1} (i/m)^k
        \\&= m^{-k} ∑_{i=0}^{n-1} i^k.
    \end{align*}

    Unfortunately there is no simple expression for $∑ i^k$ that works
    for all $k$.  However, we can approximate it from above and below
    using integrals.  From below, we have
    \begin{align*}
        ∑_{i=0}^{n-1} i^k
        &= ∑_{i=1}^{n-1} i^k
        \\&≥ ∫_{0}^{n-1} x^k dx
        \\&= \left[ \frac{1}{k+1} x^{k+1} \right]_{0}^{n-1}
        \\&= \frac{1}{k+1} (n-1)^{k+1}
        \\&= Ω\parens*{\frac{1}{k+1} n^{k+1}}.
    \end{align*}

    From above, an almost-identical computation gives
    \begin{align*}
        ∑_{i=0}^{n-1} i^k
        &≤ ∫_{0}^{n} x^k dx
        \\&= \left[ \frac{1}{k+1} x^{k+1} \right]_{0}^{n}
        \\&= frac{1}{k+1} n^(k+1)
        \\&= O\parens{\frac{1}{k+1} n^{k+1}}.
    \end{align*}

    Combining the two cases gives that the sum is
    $Θ\parens*{\frac{1}{k+1} n^{k+1}}$.  We must also include $m^{-k}$,
    giving the full answer
    \begin{displaymath}
        Θ\parens*{\frac{1}{k+1} n^{k+1} m^{-k}}.
    \end{displaymath}

    I like writing this as $Θ\parens*{n \frac{(n/m)^k}{k+1} }$ to
    emphasize that the cost per student is
    $Θ\parens*{\frac{(n/m)^k}{k+1}}$, but there are many different
    ways to write the same thing.  Note that since we are dealing with
    asymptotics, we could also replace the $\frac{1}{k+1}$ with just
    $\frac{1}{k}$ and still be correct, but that seems less informative somehow.

\section{Assignment 2: due Thursday, 2016-09-29, at 23:00}

    \subsection{Technical analysis}

    It is well known that many human endeavors—including 
    \concept{sports reporting}, \concept{financial analysis}, and 
    \concept{Dungeons and Dragons}—involve
    building narratives on top of the output of a weighted random
    number generator~\cite{Munroe2011}, a process that is now often
    automated~\cite{Holmes2016}.
    For this problem, we are
    going to consider a simplified model of automated financial analysis.

    Let us model the changes in a stock price over $n$ days using a
    sequence of independent $±1$ random variables
    $X_1,X_2,\dots,X_n$, where each $X_i$ is $±1$ with probability
    $1/2$.  Let $S_i = ∑_{j=1}^{i} X_j$, where $i∈\Set{0,\dots,n}$,
    be the price after $i$ days.

    Our automated financial reporter will declare a \concept{dead cat
    bounce} if a stock falls in price for two days in a row, followed
    by rising in price, followed by falling in price again.  Formally,
    a dead cat bounce occurs on day $i$ if $i ≥ 4$ and $S_{i-4} > S_{i-3} > S_{i-2} <
    S_{i-1} > S_i$.  Let $D$ be the number of dead cat bounces over
    the $n$ days.

    \begin{enumerate}
        \item What is $\Exp{D}$?
        \item What is $\Var{D}$?
        \item Our investors will shut down our system if it doesn't
            declare at least one dead cat bounce during the first $n$
            days.  What upper bound you can get on $\Prob{D = 0}$
            using Chebyshev's inequality?
        \item Show $\Prob{D=0}$ is in fact exponentially small in $n$.
    \end{enumerate}

    Note added 2016-09-28: It's OK if your solutions only work for
    sufficiently large $n$.  This should save some time dealing with
    weird corner cases when $n$ is small.

        \subsubsection*{Solution}

        \begin{enumerate}
            \item Let $D_i$ be the indicator variable for the event
                that a dead cat bounce occurs on day $i$. 
                Let $p = \Exp{D_i} = \Prob{D_i = 1}$.
                Then 
                \begin{equation*}
                    p = \Prob{X_{i-3} = -1 ∧ X_{i-2} = -1 ∧ X_{i-1 = +1} ∧ X_i = -1} = \frac{1}{16},
                \end{equation*}
                since the $X_i$ are
                independent.

                Then
                \begin{align*}
                    \Exp{D}
                    &= \Exp{∑_{i=4}^{n} D_i}
                    \\&= ∑_{i=4}^{n} \Exp{D_i}
                    \\&= ∑_{i=4}^{n} \frac{1}{16}
                    \\&= \frac{n-3}{16},
                \end{align*}
                assuming $n$ is at least $3$.
            \item For variance, we can't just sum up $\Var{D_i} =
                p(1-p) = \frac{15}{256}$, because the $D_i$ are not
                independent.  Instead, we have to look at covariance.

                Each $D_i$ depends on getting a particular sequence of
                four values for $X_{i-3}$, $X_{i-2}$, $X_{i-1}$, and
                $X_i$.  If we consider how $D_i$ overlaps with $D_j$
                for $j>i$, we get these cases:

                \begin{tabular}{lll}
                    Variable & Pattern & Correlation \\\hline 
                    $D_i$     & \texttt{-{}-+-}    & \\
                    $D_{i+1}$ & \texttt{~-{}-+-}   & inconsistent \\
                    $D_{i+2}$ & \texttt{~~-{}-+-}  & inconsistent \\
                    $D_{i+3}$ & \texttt{~~~-{}-+-} & overlap in one place \\
                \end{tabular}

                For larger $j$, there is no overlap, so $D_i$ and
                $D_j$ are independent for $j≥4$, giving
                $\Cov{D_i}{D_j} = 0$ for these cases.

                Because $D_i$ can't occur with $D_{i+1}$ or $D_{i+2}$,
                when $j ∈ \Set{i+1,i+2}$ we have
                \begin{equation*}
                    \Cov{D_i}{D_{j}}
                    = \Exp{D_i D_j} - \Exp{D_i} \Exp{D_j}
                    = 0 - (1/16)^2
                    = -1/256.
                \end{equation*}

                When $j=i+3$, the probability that both
                $D_i$ and $D_j$ occur is $1/2^7$, since we only need
                to get $7$ coin-flips right.  This gives
                \begin{equation*}
                    \Cov{D_i}{D_{i+3}}
                    = \Exp{D_i D_{i+3}} - \Exp{D_i} \Exp{D_{i+3}}
                    = 1/128 - (1/16)^2
                    = 1/256.
                \end{equation*}

                Now apply \eqref{eq-variance-sum-asymmetric}:
                \begin{align*}
                    \Var{D}
                    &= \Var{∑_{i=4}^{n} D_i}
                    \\&=
                       ∑_{i=4}^{n} \Var{D_i} 
                       + 2 ∑_{i=4}^{n-1} \Cov{D_i}{D_{i+1}}
                       + 2 ∑_{i=4}^{n-2} \Cov{D_i}{D_{i+2}}
                       + 2 ∑_{i=4}^{n-3} \Cov{D_i}{D_{i+3}}
                    \\&=
                       (n-3) \frac{15}{256}
                       + 2 (n-4) \frac{-1}{256}
                       + 2 (n-5) \frac{-1}{256}
                       + 2 (n-6) \frac{1}{256}
                    \\&=
                       \frac{13n - 39}{256},
                \end{align*}
                where to avoid special cases, we assume $n$ is at
                least $6$.
            \item At this point we are just doing algebra.  For $n≥6$,
                we have:
                \begin{align*}
                    \Prob{D=0}
                    &≤ \Prob{\abs{D-\Exp{D}} ≥ \Exp{D}}
                    \\&= \Prob{\abs{D-\Exp{D}} ≥ \frac{n-3}{16}}
                    \\&≤ \frac{(13n-39)/256}{((n-3)/16)^2}
                    \\&= \frac{13n-39}{(n-3)^2}
                    \\&= \frac{13}{n-3}.
                \end{align*}
                This is $Θ\parens*{1/n}$, which is not an
                especially strong bound.
            \item Here is an upper bound that avoids the mess of
                computing the exact probability, but is still
                exponentially small in $n$.  Consider the events
                $[D_{4i} = 1]$ for $i ∈ \Set{1,2,\dots,\floor{n/4}}$.
                These events are independent since they are functions
                of non-overlapping sequences of increments.  So we can
                compute
                $\Prob{D = 0} ≤ \Prob{D_{4i} = 0,
                ∀i∈\Set{1,\dots,\floor{n/4}}} =
                (15/16)^{\floor{n/4}}$.

                This expression is a little awkward, so if we want to
                get an asymptotic estimate we can simplify it using
                $1+x ≤ e^x$ to get $(15/16)^{\floor{n/4}} ≤
                \exp\parens*{-(1/16)\floor{n/4}} =
                O\parens*{\exp\parens*{-n/64}}$.

                With a better analysis, we should be able to improve the
                constant in the exponent; or, if we are real fanatics,
                calculate $\Prob{D=0}$ exactly.  But this answer is
                good enough given what the problems asks for.
        \end{enumerate}

    \subsection{Faulty comparisons}

    Suppose we want to do a comparison-based sort of
    an array of $n$ distinct elements, but an
    adversary has sabotaged our comparison subroutine.  Specifically,
    the adversary chooses a sequence of times $t_1, t_2, \dots, t_n$,
    such that for each $i$, the $t_i$-th comparison we do returns the
    wrong answer.  Fortunately, this can only happen $n$ times.
    In addition, the adversary is oblivious: thought it can examine
    the algorithm before choosing the input and the times $t_i$, it
    cannot predict any random choices made by the algorithm.

    A simple deterministic work-around is to replace each comparison
    in our favorite $Θ(n \log n)$-comparison sorting algorithm with
    $2n+1$ comparisons, and take the majority value.  But this gives
    us a $Θ(n^2 \log n)$-comparison algorithm.  We would like to do
    better using randomization.

    Show that in this model, for every fixed $c > 0$ and sufficiently
    large $n$, it is possible to sort
    correctly with probability at least $1-n^{-c}$
    using $O(n \log^2 n)$ comparisons.

    \subsubsection*{Solution}

    We'll adapt QuickSort.  If the bad comparisons occurred randomly,
    we could just take the majority of $Θ(\log n)$ faulty comparisons
    to simulate a non-faulty comparison with high probability.  But
    the adversary is watching, so if we do these comparisons at
    predictable times, it could hit all $Θ(\log n)$ of them and stay
    well within its fault budget.  So we are going to have to be more
    sneaky.

    Here's this idea: Suppose we have a list of $n$ comparisons
    $\Tuple{x_1,y_1}, \Tuple{x_2,y_2}, \dots, \Tuple{x_n,y_n}$ that
    we want to perform.  We will use a subroutine that carries out
    these comparisons with high probability by doing $k n \ln n + n$ possibly-faulty
    comparisons, with $k$ a constant to be chosen below, 
    where each of the possibly-faulty comparisons looks
    at $x_r$ and $y_r$ where each $r$ is chosen independently and uniformly
    at random.  The subroutine collects the results of these comparisons for
    each pair $\Tuple{x_i,y_i}$ and takes the majority value.

    At most $n$ of these comparisons are faulty, and we get an error
    only if some pair $\Tuple{x_i,y_i}$ gets more faulty comparisons than non-faulty
    ones.  
    Let $B_i$ be the number of bad comparisons of the pair
    $\Tuple{x_i,y_i}$ and $G_i$ the number of good comparisons.
    We want to bound the probability of the event $B_i > G_i$.

    The probability that any particular comparison lands on a
    particular pair is exactly $1/n$; so $\Exp{B_i} = 1$ and
    $\Exp{G_i} = k \ln n$.
    Now apply Chernoff bounds.  From \eqref{eq-Chernoff-bound-R},
    $\Prob{B_i ≥ (k/2) \ln n} ≤ 2^{-(k/2) \ln n} = n^{-(k \ln 2)/2)}$
    provided $(k/2) \ln n$ is at least $6$.
    In the other direction, \eqref{eq-Chernoff-bound-negative-one-half}
    says that $\Prob{G_i ≤ (k/2) \log n} ≤ e^{-(1/2)^2 k \log n / 2} =
    n^{-k/8}$.  So we have a probability of at most $n^{-k/2} +
    n^{-k/8}$ that we get the wrong result for this particular pair,
    and from the union bound we have a probability of at most
    $n^{1-k/2} + n^{1-k/8}$ that we get the wrong result for any pair.
    We can simplify this a bit by observe that $n$ must be at least
    $2$ (or we have nothing to sort), so we can put a bound of
    $n^{2-k/8}$ on the probability of any error among our $n$
    simulated comparisons, provided $k ≥ 12 / \ln 2$.
    We'll choose $k = \max(12/ \ln 2, 8(c+3))$ to get a probability of
    error of at most $n^{-c-1}$.

    So now let's implement QuickSort.  The first round of QuickSort
    picks a pivot and performs $n-1$ comparisons.  We perform
    these comparisons using our subroutine
    (note we can always
    throw in extra comparisons to bring the total up to $n$).  Now we
    have two piles on which to run the algorithm recursively.
    Comparing all nodes in each pile to the pivot for that pile
    requires $n-3$ comparisons, which can again be done by our
    subroutine.  At the next state, we have four piles, and we can
    again perform the $n-7$ comparisons we need using the subroutine.
    Continue until all piles have size $1$ or less; this takes O($\log
    n)$ rounds with high probability.  Since each round does $O(n \log
    n)$ comparisons and fails with probability at most $n^{-c-1}$, the
    entire process takes $O(n \log^2 n)$ comparisons and fails with
    probability at most $n^{-c-1} O(\log n) ≤ n^{-c}$ when $n$ is
    sufficiently large.

\section{Assignment 3: due Thursday, 2016-10-13, at 23:00} 

    \subsection{Painting with sprites}
    \label{section-hw-painting-with-sprites}

    In the early days of computing, memory was scarce, and primitive home game
    consoles like the \concept{Atari 2600} did not possess enough RAM to store
    the image they displayed on your television.  Instead, these
    machines used a graphics chip that would composite together fixed
    \indexConcept{sprite}{sprites}, bitmaps stored in ROM that could
    be displayed at locations specified by the CPU.\footnote{In the
    actual Atari 2600 graphics hardware, the CPU did this not by
    writing down the locations in RAM somewhere (which would take
    memory), but by counting instruction cycles since the graphics chip started
    refreshing the screen and screaming ``display this sprite NOW!''
    when the CRT beam got to the right position.  For this problem we
    will assume a more modern approach, where we can just give a list
    of locations to the graphics hardware.}  One common programming
    trick of this era was to repurpose existing data in memory as
    sprites, for example by searching through a program's machine code
    instructions to find sequences of bytes that looked like 
    explosions when displayed.

    For this problem, we will imagine that we have a graphics device
    that can only display one sprite.  This is a bitmap consisting of
    $m$ ones, at distinct relative positions $\Tuple{y_1,x_1}, \Tuple{y_2,x_2},
    \dots \Tuple{y_m,x_m}$.  
    Displaying a sprite at position $y,x$ on our $n×n$ screen sets the
    pixels at positions $\Tuple{(y+y_i) \bmod n,(x+x_i) \bmod n}$ for
    all $i ∈ \Set{1,\dots,m}$.  The screen is initially blank (all
    pixels $0$) and setting a pixel at some position $\Tuple{y,x}$
    changes it to $1$.  Setting the same pixel more than once has no
    effect.

    We would like to use these sprites to
    simulate white noise on the screen, by placing them at independent
    uniform random locations with the goal of setting
    roughly half of the pixels.  Unfortunately, because the
    contents of the screen are not actually stored anywhere, we can't
    detect when this event occurs.  Instead, we want to fix the number
    of sprites $\ell$ so that we get $1/2 ± o(1)$ of the total number
    of pixels set to $1$ with high probability, by which we mean
    that $(1/2±o(1))n^2$ total pixels are set with probability
    $1-n^{-c}$ for any fixed $c>0$ and sufficiently large $n$,
    assuming $m$ is fixed.

\begin{figure}
        \centering
        \begin{tabular}{cc}
        \includegraphics[scale=4.0]{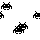}
            &
        \includegraphics[scale=4.0]{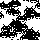}
        \end{tabular}
        \caption[Filling a screen with Space Invaders]{Filling a
        screen with Space Invaders.  The left-hand image places
        four copies of a 46-pixel sprite in random positions on a $40×40$ screen.
        The right-hand image does the same thing with $24$ copies.}
        \label{fig-space-invaders-explosion}
    \end{figure}

    An example of this process, using Taito Corporation's classic
    \indexConcept{Space Invaders}{Space Invader} bitmap, is shown in
    Figure~\ref{fig-space-invaders-explosion}.

    Compute the value of $\ell$ that we need as a function of $n$ and
    $m$, and show that this choice of $\ell$ does in fact get $1/2 ±
    o(1)$ of the pixels set with high probability.

        \subsubsection*{Solution}

        We will apply the following strategy.  First, we'll choose
        $\ell$ so that each individual pixel is set with probability
        close to $1/2$.  Linearity of expectation then gives
        roughly $n^2/2$ total pixels set on average.  To show that we
        get close to this number with high probability, we'll use the
        method of bounded differences.

        The probability that pasting in a single copy of the sprite
        sets a particular pixel is exactly $m/n^2$.  So the
        probability that the pixel is not set after $\ell$ sprites is
        $(1-m/n^2)^\ell$.  This will be exactly $1/2$ if $\ell =
        \log_{1-m/n^2} 1/2 = \frac{\ln (1/2)}{\ln (1-m/n^2)}$.

        Since $\ell$ must be an integer, we can just round this
        quantity to the nearest integer to pick our actual $\ell$.
        For example, when $n=40$ and $m=46$, $\frac{\ln (1/2)}{\ln
        (1-46/40^2)} = 23.7612...$ which rounds to $24$.

        For large $n$, we can use the approximation $\ln (1+x) ≈ x$
        (which holds for small $x$) to
        approximate $\ell$ as $(n^2 \ln 2 / m)+o(1)$.
        We will use this to get a concentration bound using
        \eqref{eq-McDiarmids-inequality}.  

        We have $\ell$ independent random variables corresponding to
        the positions of the $\ell$ sprites.  Moving a single sprite 
        changes the number of set pixels by at most $m$.  So the
        probability that the total number of set pixels deviates from
        its expectation by more than $an √{\ln n}$ is bounded by
        \begin{align*}
            2\exp\parens*{-\frac{\parens*{an √{\ln n}}^2}{\ell m^2}}
            &= 2\exp\parens*{-\frac{a^2 n^2 \ln n}{(n^2 \ln 2 / m + o(1)) m^2}}
            \\&= 2\exp\parens*{-\frac{a^2 \ln n}{m \ln 2 + o(m^2/n^2)}}
            \\&≤ \exp\parens*{-(a^2/m) \ln n}
            \\&= n^{-a^2/m},
        \end{align*}
        where the inequality holds when $n$ is sufficiently large.
        Now choose $a = √{cm}$ to get a probability of deviating
        by more than $n √{cm \ln n}$ of at most $n^{-c}$.  Since
        $n √{cm \ln n} = o(1) n^2$, this give us our desired
        bound.

    \subsection{Dynamic load balancing}

    A low-cost competitor to better-managed data
    center companies offers a scalable dynamic load balancing system
    that supplies a new machine whenever a new job shows up, so that
    there are always $n$ machines available to run $n$ jobs.
    Unfortunately, one of the ways the company cuts cost is by not
    hiring programmers to replace the code from their previous
    randomized load balancing mechanism, so when the $i$-th job
    arrives, it is still assigned uniformly at random to one of the
    $i$ available machines.  This means that job $1$ is always assigned to machine
    $1$; job $2$ is assigned with equal probability to machine $1$
    or $2$; job $3$ is assigned with equal probability to machine $1$,
    $2$, or $3$; and so on.  These choices are all independent.
    The company claims that the maximum load is still not too bad, and
    that the reduced programming costs that
    they pass on to their customers make up for their 
    poor quality control and terrible service.

    Justify the first part of this claim by 
    showing that, with high probability, the most loaded machine after
    $n$ jobs has arrived has load $O(\log n)$.

        \subsubsection*{Solution}

        Let $X_{ij}$ be the indicator variable for the event that
        machine $i$ gets job $j$.  Then $\Exp{X_{ij}} = 1/j$ for all
        $i≤j$, and $\Exp{X_{ij}} = 0$ when $i>j$.

        Let $Y_i = ∑_{j=1}^n X_{ij}$ be the load on machine $i$.

        Then $\Exp{Y_i} = \Exp{\sum_{j=1}^n X_{ij}} = ∑_{j=i}^n 1/j ≤
        ∑_{j=1}^n 1/j = H_n ≤ \ln n + 1$.

        From \eqref{eq-Chernoff-bound-R}, we have $\Prob{Y_i ≥ R} ≤
        2^{-R}$ as long as $R > 2e\Exp{Y_i}$.  So if we let $R = (c+1)
        \lg n$, we get a probability of at most $n^{-c-1}$ that we get
        more than $R$ jobs on machine $i$.  Taking the union bound
        over all $i$ gives a probability of at most $n^{-c}$ that any
        machine gets a load greater than $(c+1) \lg n$.  This works as
        long as $(c+1) \lg n ≥ 2e (\ln n + 1)$, which holds for
        sufficiently large $c$.  For smaller $c$, we can just choose a
        larger value $c'$ that does work,
        and get that $\Prob{\max Y_i ≥ (c'+1) \lg n} ≤ n^{-c'} ≤ n^{-c}$.

        So for any fixed $c$, we get that
        with probability at least $1-n^{-c}$ the maximum load is
        $O(\log n)$.

\section{Assignment 4: due Thursday, 2016-11-03, at 23:00}

    \subsection{Re-rolling a random treap}

    Suppose that we add to a random treap (see §\ref{section-treaps})
    an operation that re-rolls the priority of the node with a given
    key $x$.  This replaces the old priority for $x$ with a new priority generated
    uniformly at random, independently of the old priority any other
    priorities in the treap.  We assume that the range of priorities
    is large enough that the probability that this produces a
    duplicate is negligible, and that the choice of which node to
    re-roll is done obliviously, without regard to the current
    priorities of nodes or the resulting shape of the tree.

    Changing the priority of $x$ may break the heap property.  To
    fix the heap property, we either rotate $x$ up (if
    its priority now exceeds its parent's priority) or down (if its
    priority is now less than that of one or both of its children).  In the
    latter case, we rotate up the child with the higher priority.
    We repeat this process until the heap property is restored.

    Compute the best constant upper bound you can on the expected number of
    rotations resulting from executing a re-roll operation.

        \subsubsection*{Solution}

        The best possible bound is at most $2$ rotations on average in the worst
        case.

        There is an easy but incorrect argument that $2$ is an upper
        bound, which says that if we rotate up, we do at most the same
        number of rotations as when we insert a new element, and if we
        rotate down, we do at most the same number of rotations as
        when we delete an existing element.  This gives the right
        answer, but for the wrong reasons: the cost of deleting $x$,
        conditioned on the event that re-rolling its priority gives a
        lower priority, is likely to be greater than $2$, since the
        conditioning means that $x$ is likely to be higher up in the
        tree than average; the same thing happens in the other direction when $x$
        moves up.  Fortunately, it turns out that the fact that we
        don't necessarily rotate $x$ all the way to or from the bottom
        compensates for this issue.

        This can be formalized using the following argument, for which
        I am indebted to Adit Singha and Stanislaw Swidinski.  Fix
        some element $x$, and suppose its old and new priorities are
        $p$ and $p'$.  If $p < p'$, we rotate up, and the sequence of
        rotations is exactly the same as we get if we remove all
        elements of the original treap with priority $p$ or less and
        then insert a new element $x$ with priority $p'$.  But now if
        we condition on the number $k$ of elements with priority
        greater than $p$, their priorities together with $p'$ are all
        independent and identically distributed, since they are all
        obtained by taking their original distribution and
        conditioning on being greater than $p$.  So all $(k+1)!$
        orderings of these priorities are equally likely, and this
        means that we have the same expected cost as an insertion into
        a treap with $k$ elements, which is at most $2$.  Averaging
        over all $k$ shows
        that the expected cost of rotating up is at most $2$, and,
        since rotating down is just the reverse of this process with a
        reversed distribution on priorities
        (since we get it by choosing $p'$ as our old priority and $p$
        as the new one), the expected cost of rotating down is also at
        most $2$.  Finally, averaging the up and down cases gives that the expected
        number of rotations without conditioning on anything is at most $2$.

        We now give an exact analysis of the expected number of
        rotations, which will show that $2$ is in fact the best bound
        we can hope for.

        The idea is to notice that whenever we do a rotation
        involving $x$, we change the number of ancestors of $x$ by
        exactly one.  This will always be a decrease in the number of
        ancestors if the priority of $x$ went up, or an increase if
        the priority of $x$ went down, so the total number of
        rotations will be equal to the change in the number of
        ancestors.
        
        Letting $A_i$ be the indicator for the event that
        $i$ is an ancestor of $x$ before the re-roll, and $A'_i$ the
        indicator for the even that $i$ is an ancestor of $x$ after
        the re-roll, then the number of rotations is just
        $\abs*{∑_i A_i - ∑_i A'_i}$, which is equal to $∑_i \abs*{A_i
        - A'_i}$ since we know that all changes $A_i - A'_i$ have the
        same sign.  So the expected number of rotations is just 
        $\Exp{∑_i \abs*{A_i - A'_i}} = ∑_i \Exp{\abs*{A_i - A'_i}}$, by
        linearity of expectation.

        So we have to compute $\Exp{\abs*{A_i - A'_i}}$.  
        Using the same argument as in §\ref{section-treaps-analysis},
        we have that $A_i = 1$ if and only if $i$ has the highest
        initial priority of all elements in the range $[\min(i,x),\max(i,x)]$,
        and the same holds for $A'_i$ if we consider updated
        priorities.  So we want to know the probability that changing
        only the priority of $x$ to a new random value changes whether
        $i$ has the highest priority.

        Let $k = \max(i,x) - \min(i,x) + 1$ be the number of elements
        in the range under consideration.  
        To avoid writing a lot of mins and maxes, let's renumber
        these elements as $1$ through $k$, with $i=1$ and
        $x=k$ (this may involve flipping the sequence if $i>x$).
        Let $X_1,\dots,X_k$ be the priorities of these elements, and
        let $X'_k$ be the new priority of $x$.  These $k+1$ random
        variables are independent and identically distributed, so
        conditioning on the even that no two are equal, all $(k+1)!$
        orderings of their values are equally likely.

        So now let us consider how many of these orderings result in
        $\abs*{A_i - A'_i} = 1$.  For $A_i$ to be $1$, $X_1$ must exceed
        all of $X_2,\dots,X_k$.  For $A'_i$ to be $0$, $X_1$ must not
        exceed all of $X_2,\dots,X_{k_1},X'_k$.  The intersection of
        these events is when
        $X'_k > X_1 > \max(X_2,\dots,X_{k})$.  Since
        $X_2,\dots,X_{k}$ can be ordered in any of $(k-1)!$ ways, this gives
        \begin{equation*}
            \Prob{A_i = 1 ∧ A'_i = 0} 
            = \frac{(k-1)!}{(k+1)!}
            = \frac{1}{k(k+1)}.
        \end{equation*}
        In the other direction, for $A_i$ to be $0$ and $A'_i$ to be $1$, we
        must have $X_k > X_1 > \max(X_1,\dots,X_{k-1},X_k)$.  This
        again gives
        \begin{align*}
            \Prob{A_i = 0 ∧ A'_i = 1} &= \frac{1}{k(k+1)}
            \intertext{and combining the two cases gives}
            \Exp{\abs*{A_i - A'_i}} = \frac{2}{k(k+1)}.
        \end{align*}

        Now sum over all $i$:
        \begin{align*}
            \Exp{\text{number of rotations}}
            &= ∑_{i=1}^{n} \Exp{\abs*{A'_i-A_i}}
            \\&= ∑_{i=1}^{x-1} \frac{2}{(x-i+1)(x-i+2)} + ∑_{i=x+1}^{n} \frac{2}{(i-x+1)(i-x+2}
            \\&= ∑_{k=2}^{x} \frac{2}{k(k+1)} + ∑_{k=2}^{n-x+1} \frac{2}{k(k+1)}
            \\&= ∑_{k=2}^{x} \parens*{\frac{2}{k} - \frac{2}{k+1}}
               + ∑_{k=2}^{n-x+1} \parens*{\frac{2}{k} - \frac{2}{k+1}}
            \\&= 2 - \frac{2}{x+1} - \frac{2}{n-x+2}.
        \end{align*}

        In the limit as $n$ goes to infinity, choosing
        $x = \floor{n/2}$ makes both fractional terms converge to $0$,
        making the expected number of rotations
        arbitrarily close to $2$.  So $2$ is the best possible bound
        that doesn't depend on $n$.

    \subsection{A defective hash table}

    A foolish programmer implements a hash table with $m$ buckets,
    numbered $0,\dots,m-1$,
    with the property that any value hashed to an even-numbered bucket
    is stored in a linked list as usual, but any value hashed to an
    odd-numbered bucket is lost.

    \begin{enumerate}
        \item Suppose we insert a set $S$ of $n = \card{S}$ items into
            this hash table, using a
            hash function $h$ chosen at random from a strongly $2$-universal
            hash family $H$.  Show that there is a constant $c$ such
            that, for any $t>0$, the probability that at least $n/2+t$ items are
            lost is at most $c n / t^2$.
        \item Suppose instead that we insert a set $S$ of $n =
            \card{S}$ items into this hash
            table, using a hash function $h$ chosen at random from a
            hash family $H'$ that is $2$-universal, but that may not
            be strongly $2$-universal.
            Show that if $n ≤ m/2$, it is possible for an adversary
            that knows $S$ to design a $2$-universal hash family $H'$
            that ensures that all $n$ items are lost with probability
            $1$.
    \end{enumerate}

        \subsubsection*{Solution}

        \begin{enumerate}
            \item Let $X_i$ be the indicator variable for the event
                that the $i$-th element of $S$ is hashed to an
                odd-numbered bucket.  Since $H$ is strongly
                $2$-universal, $\Exp{X_i} ≤ 1/2$ (with equality when
                $m$ is even), from which it follows that $\Var{X_i} =
                \Exp{X_i} (1-\Exp{X_i}) ≤ 1/4$; 
                and the $X_i$ are pairwise independent.
                Letting $Y=∑_{i=1}^n X_i$ be the total number of items
                lost, we get $\Exp{Y} ≤ n/2$ and
                $\Var{Y} ≤ n/4$.  But then we can apply Chebyshev's
                inequality to show
                \begin{align*}
                    \Prob{Y ≥ n/2 + t} 
                    &≤ \Prob{Y ≥ \Exp{Y} + t}
                    \\&≤ \frac{\Var{Y}}{t^2}
                    \\&≤ \frac{n/4}{t^2}
                    \\&= \frac{1}{4} (n/t^2).
                \end{align*}
                So the desired bound holds with $c = 1/4$.
            \item Write $[m]$ for the set $\Set{0,\dots,m-1}$.
                Let $S=\Set{x_1,x_2,\dots,x_n}$.

                Consider the set $H'$ of all functions $h:U→[m]$ with the
                property that $h(x_i) = 2i+1$ for each $i$.  Choosing
                a function $h$ uniformly from this set corresponds to
                choosing a random function conditioned on this
                constraint; and this conditioning guarantees that every
                element of $S$ is sent to an odd-numbered bucket and
                lost.  This gives us half of what we want.
                
                The other half is that $H'$ is $2$-universal.  Observe
                that for any $y ∉ S$, the conditioning does
                not constrain $h(y)$, and so $h(y)$ is equally
                likely to be any element of $[m]$; in addition,
                $h(y)$ is independent of $h(z)$ for any
                $z≠y$.  So for any $y≠z$, $\Prob{h(y) = h(z)} = 1/m$
                if one or both of $y$ and $z$ is not an element of
                $S$, and $\Prob{h(y) = h(z)} = 0$ if both are elements
                of $S$.  In either case,
                $\Prob{h(y)=h(z)}≤1/m$, and so $H'$ is $2$-universal.
        \end{enumerate}

\section{Assignment 5: due Thursday, 2016-11-17, at 23:00}

    \subsection{A spectre is haunting Halloween}

    An adult in possession of $n$ pieces of leftover Halloween candy
    is distributing them to $n$ children.  The candies
    have known integer values $v_1, v_2, \dots, v_n$, and the adult starts by
    permuting the candies according to a uniform random permutation
    $π$.  They then give candy $π(1)$ to child $1$, candy
    $π(2)$ to child $2$, and so on.  Or at least, that is their plan.

    Unbeknownst to the adult, child $n$ is a very young Karl
    Marx, and after observing the first $k$ candies $π(1), \dots,
    π(k)$, he can launch a communist revolution and seize the means
    of distribution.  This causes the
    remaining $n-k$ candies $π(k+1)$ through $π(n)$ to be distributed evenly
    among the remaining $n-k$ children, so that each receives a value
    equal to exactly $\frac{1}{n-k} ∑_{i=k+1}^{n} v_{π(i)}$.  Karl may
    declare the revolution at any time before the last candy is
    distributed (even at time $0$, when no candies have been
    distributed).  However, his youthful understanding of the
    mechanisms of historical determinism are not detailed enough to
    predict the future, so his decision to declare a revolution after
    seeing $k$ candies can only depend on those $k$ candies and his
    knowledge of the values of all of the candies but not their order.

    Help young Karl optimize his expected take by devising an
    algorithm that takes as input $v_1,\dots,v_n$ and
    $v_{π(1)},\dots,v_{π(k)}$,
    where $0 ≤ k < n$,
    and outputs whether or not to launch a revolution at this time.
    Compute Karl's exact expected return as a function of $v_1,\dots,v_n$
    when running your algorithm, and show that no algorithm can do better.

        \subsubsection*{Solution}

        To paraphrase an often-misquoted line of Trotsky's, young Karl
        Marx may not recognize the Optional Stopping Theorem,
        but the Optional Stopping Theorem does not permit him to escape its
        net.  No strategy, no matter how clever, can produce an
        expected return better or worse than simply waiting for the
        last candy.

        Let $X_t$
        be the expected return if Karl declares the revolution after
        seeing $t$ cards.  We will show that $\Set{X_t,ℱ_t}$ is a
        martingale where each $ℱ_t$ is the $σ$-algebra generated by
        the random variables $v_{π(1)}$ through $v_{π(t)}$.  

        The value of $X_t$ is
        exactly 
        \begin{equation*}
                X_t = \frac{1}{n-t} ∑_{i=t+1}^{n} v_{π(i)} = 
            \frac{1}{n-t} \parens*{∑_{i=t+1}^{n} v_{n} - ∑_{i=1}^{t} v_{π(i)}}.
        \end{equation*}
        This is also $\ExpCond{v_{π(t+1)}}{ℱ_t}$, since 
        the next undistributed candy is equally likely to be any of
        the remaining candies, and $ℱ_t$ determines the value of
        $v_{π(i)}$ for all $i ≤ t$.
        So
        \begin{align*}
            \ExpCond{X_{t+1}}{ℱ_t}
            &= \ExpCond{\frac{1}{n-t-1} 
                \parens*{∑_{i=1}^n v_i - ∑_{i=1}^{t+1} v_{π(i)}}}{ℱ_t}
            \\&= \frac{1}{n-t-1} 
                \parens*{∑_{i=1}^{n} v_i - ∑_{i=1}^{t} v_{π(i)}}
                - \frac{1}{n-t-1} \ExpCond{v_{π(t+1)}}{ℱ_T}
            \\&= \frac{n-t}{n-t-1} X_t
                - \frac{1}{n-t-1} X_t
            \\&= X_t.
        \end{align*}

        Now fix some strategy for Karl, and let $τ$ be the time at which
        he launches the revolution.  Then $τ < n$ is a stopping time with
        respect to the $ℱ_t$, and the Optional Stopping Theorem
        (bounded time version) says that $\Exp{X_τ} = \Exp{X_0}$.
        So any strategy is equivalent (in expectation) to launching
        the revolution immediately.

    \subsection{Colliding robots on a line}

    Suppose we have $k$ robots on a line of $n$ positions, numbered
    $1$ through $n$.  No two robots can pass each other or occupy
    the same position, so we can specify the positions of all the robots
    at time $t$ as an increasing vector $X^t =
    \Tuple{X^t_1,X^t_2,\dots,X^t_k}$.  At each time $t$, we pick one of
    the robots $i$ uniformly at random, and also choose a direction
    $d = ±1$ uniformly and independently at random.  Robot $i$ moves to position
    $X^{t+1} = X^t_i + d$ if (a) $0 ≤ X^t_i + d < n$, and (b) no other
    robot already occupies position $X^t_i + d$.  If these conditions
    do not hold, robot $i$ stays put, and $X^{t+1}_i = X^t_i$.

    \begin{enumerate}
        \item Given $k$ and $n$, determine the stationary distribution
            of this process.
        \item Show that the mixing time $t_{\mix}$ for this
            process is polynomial in $n$.
    \end{enumerate}

        \subsubsection*{Solution}
        \begin{enumerate}
            \item The stationary distribution is a uniform
                distribution on all $\binom{n}{k}$ placements of the
                robots.  To prove this, observe that two vectors
                increasing vectors $x$ and $y$ are adjacent if and
                only if there is some $i$ such that $x_j = y_j$ for
                all $j≠i$ and $x_j + d = y_j$ for some $d ∈
                \Set{-1,+1}$.  In this case, the transition
                probability $p_{xy}$ is $\frac{1}{2n}$, since there is
                a $\frac{1}{n}$ chance that we choose $i$ and a
                $\frac{1}{2}$ chance that we choose $d$.  But this is
                the same as the probability that starting from $y$ we
                choose $i$ and $-d$.  So we have $p_{xy} = p_{yx}$ for
                all adjacent $x$ and $y$, which means that a uniform
                distribution $π$ satisfies $π_x p_{xy} = π_y p_{yx}$
                for all $x$ and $y$.

                To show that this stationary distribution is unique,
                we must show that there is at least one path between any two states
                $x$ and $y$.  One way to do this is to show that there
                is a path from any state $x$ to the state
                $\Tuple{1,\dots,k}$, where at each step we move the
                lowest-index robot $i$ that is not already at position
                $i$.  Since we can reverse this process to get to $y$,
                this gives a path between any $x$ and $y$ that occurs
                with nonzero probability.

            \item 
                This one could be done in a lot of ways.
                Below I'll give sketches of three possible
                approaches, ordered by increasing difficulty.
                The first reduces to card shuffling by adjacent swaps,
                the second uses an explicit coupling, and the third
                uses conductance.  
                
                Of these approaches, I am
                personally only confident of the coupling argument,
                since it's the one I did before handing out the
                problem, and indeed this is the only one I have
                written up in enough detail below to be even remotely
                convincing.  But the reduction to card shuffling is also pretty straightforward
                and was used in several student solutions, so I am
                convinced that it can be made to work as well.  
                The conductance idea I am not sure works at all, but it seems
                like it could be made to work with enough effort.
                
                \begin{enumerate}
                    \item 
                Let's start with the easy method.  Suppose that
                instead of colliding robots, we have a deck of $n$
                cards, of which $k$ are specially marked.
                Now run a shuffling algorithm that swaps adjacent
                cards at each step.  If we place a robot at the
                position of each marked card, the trajectories of the
                robots follow the pretty much the same distribution as in the
                colliding-robots process.  This is trivially the case
                when we swap a marked card with an unmarked card
                (a robot moves), but it also works when we swap two
                marked cards (no robot moves, since the positions of
                        the set of marked cards stays the same; this
                        corresponds to a robot being stuck).

                        Unfortunately we can't use exactly the same
                        process we used in
                        §\ref{section-shuffling-with-adjacent-swaps},
                        because this (a) allows swapping the cards in
                        the first and last positions of the deck, and
                        (b) doesn't include any moves corresponding to
                        a robot at position $1$ or $n$ trying to move
                        off the end of the line.

                        The first objection is easily dealt with, and
                        indeed the cited result of
                        Wilson~\cite{Wilson2004} doesn't allow such
                        swaps either.  The second can be dealt with by
                        adding extra no-op moves to the card-shuffling
                        process that occur with probability $1/n$,
                        scaling the probabilities of the other
                        operations to keep the sum to $1$.  This
                        doesn't affect the card-shuffling convergence
                        argument much, but it is probably a good idea
                        to check that everything still works.

                        Finally, even after fixing the card-shuffling
                        argument, we still have to argue that
                        convergence in the card-shuffling process
                        implies convergence in the corresponding
                        colliding-robot process.  Here is where the
                        definition of total variation distance helps.
                        Let $C^t$ be the permutation of the cards
                        after $t$ steps, and let $f:C^t↦X^t$ map
                        permutations of cards to positions of robots.
                        Let $π$ and $π'$ be the stationary
                        distributions of the card and robot
                        processes, respectively.
                        Then $d_{TV}(f(C^t),f(π)) = \max_A
                        \abs*{\Prob{X^t∈A}-π'(A)} = \max_A
                        \abs*{\Prob{C^t∈f^{-1}(A)}-π(f^{-1}(A))}
                        ≤ \max_B \abs*{\Prob{C^t∈B}-π(B)} =
                        d_{TV}(C^t,π)$.  So convergence of $C$ implies
                        convergence of $X = f(C)$.
            \item Alternatively, 
                we can construct a coupling between two copies of the
                process $X^t$ and $Y^t$,
                where as usual we start $X^0$ in our initial distribution
                (whatever it is) and $Y^0$ in the uniform stationary
                distribution.
                At each step we
                generate a pair $(i,d)$ uniformly at random from
                $\Set{1,\dots,n}×\Set{-1,+1}$.  Having generated this
                pair, we move robot $i$ in direction $d$ in both
                processes if possible.  
                It is clear that once
                $X^t=Y^t$, 
                it will continue to hold that $X^{t'} = Y^{t'}$ for all $t' > t$.

                To show convergence we must show that $X^t$ will
                in fact eventually hit $Y^t$.  To do so, let us track
                $Z^t = \norm{X^t-Y^t}_1 = ∑_{i=1}^{k} \abs{X^t_i - Y^t_i}$.
                We will show that $Z_t$ is a supermartingale with
                respect to the history $ℱ_t$ generated by the random
                variables $\Tuple{X^s,Y^s}$ for $s ≤ t$.
                Specifically, we will show that at each
                step, the expected change in $Z^t$ conditioning
                on $X^t$ and $Y^t$ is non-positive.

                Fix $X^t$, and consider all $2n$ possible moves $(i,d)$.
                Since $(i,d)$ doesn't change $X^t_j$ for any $j≠i$,
                the only change in $Z^t$ will occur if one of $X^t_i$
                and $Y^t_i$ can move and the other can't.  There are
                several cases:
                \begin{enumerate}
                    \item If $i=k$, $d=+1$, and exactly one of $X^t_k$
                        and $Y^t_k$ is $n$, then the copy of the robot
                        not at $n$ moves toward the copy at $n$,
                        giving $Z_{t+1} - Z_t = -1$.
                        The same thing occurs if $i=1$, $d=-1$, and
                        exactly one of $X^t_1$ and $Y^t_1$ is $1$.

                        It is interesting to note that these two cases will account for the entire
                        nonzero part of $\ExpCond{Z_{t+1} -
                        Z_t}{ℱ_t}$, although we will not use this
                        fact.
                    \item If $i<k$, $d=+1$, $X^t_i+1 = X^t_{i+1}$, but
                        $Y^t_i + 1 < X^t_{i+1}$, and $X^t_i ≤ Y^t_i$,
                        then robot $i$ can move right in $Y^t$ but not
                        in $X^t$.  This gives $Z^{t+1} - Z^t = +1$.

                        However, in this case the move $(i+1,-1)$
                        moves robot $i+1$ left in $Y^t$ but not in
                        $X^t$, and since $Y^t_{i+1} > Y^t_i + 1 >
                        X^t_i = X^{t+1}_i$, this move gives $Z^{t+1} -
                        Z^t = -1$.  Since the moves $(i,+1)$ and
                        $(i+1,-1)$ occur with equal probability, these
                        two changes in $Z^t$ cancel out on average.

                        Essentially the same analysis applies if we
                        swap $X^t$ with $Y^t$, and if we consider
                        moves $(i,-1)$ where $i>1$ in which one
                        copy of the robot is blocked but not the
                        other.  If we enumerate all such cases, we
                        find that for every move of these types that
                        increase $Z^t$, there is a corresponding move
                        that decreases $Z^t$; it follows that these
                        changes sum to zero in expectation.
                \end{enumerate}

                Because each $X^t_i$ and $Y^t_i$ lies in the range $[1,n]$, it
                holds trivially that $Z^t ≤ k(n-1) < n^2$.  Consider
                an unbiased
                $±1$ random walk $W^t$ with a reflecting barrier at $n^2$
                (meaning that at $n^2$ we always move down) and an
                absorbing barrier at $0$.  Then the expected time to
                reach $0$ starting from $W^t = x$ is given by
                $x(2n^2-x) ≤ n^4$.  Since this unbiased random walk
                dominates the biased random walk corresponding to
                changes in $Z^t$, this implies that $Z^t$ can change
                at most $n^4$ times on average before reaching $0$.

                Now let us ask how often $Z^t$ changes.  The analysis
                above implies that there is a
                chance of at least $1/n$ that $Z^t$ changes in any
                state where some robot $i$ with $X^t_i ≠ Y^t_i$ 
                is blocked from moving freely
                in one or both directions in either the $X^t$ or $Y^t$
                process.  To show that this occurs after polynomially
                many steps, choose some $i$ with $X^t_i ≠ Y^t_i$ that
                is not blocked in either process.  Then $i$ is
                selected every $\frac{1}{n}$ steps on average, and it
                moves according to a $±1$ random walk as long as it is
                not blocked.  But a $±1$ random walk will reach
                position $1$ or $n$ in $O(n^2)$ steps on average (corresponding
                to $O(n^3)$ steps of the original Markov chain), and
                $i$ will be blocked by the left or right end of the
                line if it is not blocked before then.  It follows
                that we reach a state in which $Z^t$ changes with
                probability at least $\frac{1}{n}$ after $O(n^3)$
                expected steps, which means that $Z^t$ changes every
                $O(n^4)$ on average.  Since we have previously
                established that $Z^t$ reaches $0$ after $O(n^4)$
                expected changes, this gives a total of $O(n^8)$
                expected steps of the Markov process before $Z^t = 0$,
                giving $X^t = Y^t$.  Markov's inequality and the
                Coupling Lemma then give $t_{\mix} = O(n^8)$.

                    \item 
                I believe it may also be possible to show convergence
                using a conductance argument.  The idea is to 
                represent a state $\Tuple{x_1,\dots,x_k}$ as the
                differences $\Tuple{Δ_1,\dots,Δ_n}$, where $Δ_1 = x_1$
                and $Δ_{i+1} = x_{i+1} - x_i$.  This representation
                makes the set of possible states 
                $\SetWhere{Δ}{Δ_i ≥ 1, ∑_{i=1}^k Δ_i ≤ n}$ look like a
                simplex, a pretty well-behaved geometric object.  In
                principle it should be
                possible to show an isoperimetric inequality 
                that the lowest-conductance sets in
                this simplex are the $N$ points closest to any given
                corner, or possibly get a lower bound on conductance
                using canonical paths.  But the $Δ$ version of the
                process is messy (two coordinates change every time a
                robot moves), and there are some complication with
                this not being a lazy random walk, so
                I didn't pursue this myself.
                \end{enumerate}
        \end{enumerate}

\section{Assignment 6: due Thursday, 2016-12-08, at 23:00}

    \subsection{Another colliding robot}

    A warehouse consists of an $n×n$ grid of locations, which we can
    think of as indexed by pairs $(i,j)$ where $1 ≤ i,j ≤ n$.  At
    night, the warehouse is patrolled by a robot executing a lazy
    random walk.  Unfortunately for the robot, there are also $m$
    crates scattered about the warehouse, and if the robot attempts to
    walk onto one of the $m$ grid locations occupied by a crate, it
    fails to do so, and instead emits a loud screeching noise.
    We would like to use the noises coming from inside the warehouse
    to estimate $m$.

    Formally, if the robot is at position $(i,j)$ at time $t$, then
    at time $t+1$ it either (a) stays at $(i,j)$ with probability
    $1/2$; or (b) chooses a position $(i',j')$ 
    uniformly at random from $\Set{(i-1,j),(i+1,j),(i,j-1),(i,j+1)}$ and
    attempts to move to it.  If the robot chooses not to move, it makes no
    noise.  If it chooses to move, but $(i',j')$ is off the grid or occupied
    by one of the $m$ crates, then the robot stays at $(i,j)$ and
    emits the noise.  Otherwise it moves to $(i',j')$ and remains
    silent.  The robot's position at time $0$ can be any unoccupied
    location on the grid.

    To keep the robot from getting walled in somewhere,
    whatever adversary placed the
    crates was kind enough to ensure that if a crate was placed at
    position $(i,j)$, then all of the eight positions
    \begin{displaymath}
        \begin{array}{ccc}
            (i-1,j+1) & (i,j+1) & (i+1,j+1) \\
            (i-1,j  ) &         & (i+1,j  ) \\
            (i-1,j-1) & (i,j-1) & (i+1,j-1) \\
        \end{array}
    \end{displaymath}
    reachable by a king's move from $(i,j)$ are unoccupied grid
    locations.  This means that they are not off the grid and not
    occupied by another crate, so that the robot can move to any of
    these eight positions.

    Your job is to devise an algorithm for estimating $m$ to within
    $ε$ relative error with probability at least $1-δ$, based on the
    noises emitted by the robot.  The input to
    your algorithm is the sequence of bits $x_1,x_2,x_3,\dots$, where
    $x_i$ is $1$ if the robot makes a noise on its $i$-th step.  Your
    algorithm should run in time polynomial in $n$, $1/ε$, and
    $\log(1/δ)$.

        \subsubsection*{Solution}

        It turns out that $ε$ is a bit of a red herring: we can
        in fact compute the \emph{exact} number of crates with probability
        $1-δ$ in time polynomial in $n$ and $\log(1/δ)$.

        The placement restrictions and laziness make this an
        irreducible aperiodic chain, so it has a unique stationary
        distribution $π$.
        It is easy to argue from reversibility that
        this is uniform, so each of the $N=n^2-m$ unoccupied
        positions occurs with probability exactly $1/N$.

        It will be useful to observe that we can assign three unique
        unoccupied positions to each crate, to the east, south, and
        southeast, and this implies $m ≤ n^2/4$.

        The idea now is to run the robot until the distribution on
        its position is close to $π$, and then see if it hits an
        obstacle on the next step.  We can easily count the number of
        possible transitions that hit an obstacle, since there are
        $4m$ incoming edges to the crates, plus $4n$ incoming edges to
        the walls.  Since each edge $uv$ has probability $π_u p_{uv} =
        \frac{1}{8N}$ of being selected in the stationary
        distribution, the probability $q$ that we hit an obstacle starting
        from $π$ is exactly $\frac{n+m}{2N} = \frac{n+m}{n^2-m}$.
        This function is not trivial to invert, but we don't have to
        invert it: if we can compute its value (to some reasonable
        precision), we can compare it to all $n^2/4 + 1$ possible
        values of $m$, and see which it matches.
        But we still have to deal with both getting enough samples and 
        the difference
        between our actual distribution when we take a sample and the
        stationary distribution.
        For the latter, we need a convergence bound.

        We will get this bound by constructing a family of canonical
        paths between each distinct pair of locations $(i,j)$ and
        $(i',j')$.  
        In the absence of crates, we can use the L-shaped
        paths that first change $i$ to $i'$ and then $j$ to $j'$, one step
        at a time in the obvious way.  If we encounter a crate at some
        position along this path, we replace that position with a
        detour that uses up to three locations off the original path
        to get around the obstacle.  The general rule is that if we
        are traveling left-to-right or right-to-left, we shift up one
        row to move past the obstacle, and if we are moving up-to-down
        or down-to-up, we shift left one column.  An annoying special
        case is when the obstacle appears exactly at the corner
        $(i',j)$; here we replace the blocked position with the unique
        diagonally adjacent position that still gives us a path.

        Consider the number of paths $(i,j)→(i',j')$ crossing a particular 
        left-right edge $(x,y)→(x+1,y)$.
        There are two cases we have to consider:
        \begin{enumerate}
            \item We cross the edge as part of the left-to-right
                portion of the path (this includes left-to-right moves
                that are part of detours).  In this case we have
                $\abs{j-y} ≤ 1$.  This gives at most
                $3$ choices for $j$, giving at most $3n^3$ possible
                paths.  (The constants can be improved here.)
            \item We cross the edge as part of a detour on the
                vertical portion of the path.
                Then $\abs{i-x} ≤ 1$, and so we again have at most
                $3n^3$ possible paths.
        \end{enumerate}

        This gives at most $6n^3$ possible paths across each edge,
        giving a congestion
        \begin{align*}
            ρ 
            &≤ \frac{1}{π_{ij} p_{ij,i'j'}} (6n^3) π_{ij} π_{i'j'}
            \\&= \frac{1}{N^{-1} (1/8)} (6n^3) N^{-2}
            \\&= 24 n^3 N^-1
            \\&≤ 24 n^3 (\frac{3}{4}n^2)^{-1}
            \\&= 32 n,
        \end{align*}
        since we can argue from the placement restrictions that $N ≤
        n^2/4$.  This immediately gives a bound $τ_2 ≤ 8ρ^2 ≤ O(n^2)$.
        using Lemma~\ref{lemma-congestion}.

        So let's run for $\ceil{51 τ_2 \ln n} = O(n^2 \ln n)$ 
        steps for each sample.  Starting from any
        initial location, we will reach some distribution $σ$ with
        $d_{TV}{σ,π} = O(n^{-50})$.
        Let $X$ be the number of obstacles (walls or crates)
        adjacent to the current position, then we can apply 
        Lemma~\ref{lemma-total-variation-expectation} to 
        get  
        $\abs{\E_σ(X) - \E_π(X)} ≤ 4 d_{TV}(σ,π) = O(n^{-50})$.
        The same bound (up to constants) also applies to the
        probability $ρ = X/8$ of hitting an obstacle, giving $ρ =
        4(n+m)/(n^2-m) ± O(n^{-50})$.  Note that $ρ$ is $Θ(n^{-1}$ for
        all values of $m$.

        Now take $n^{10} √{\ln (1/δ)}$ samples, with a gap of $\ceil{51 τ_2
        \ln n}$ steps between each.  The expected number of positive
        samples is $μ = (ρ + O(n^{-50})) n^{10} √{\ln (1/δ)} =
        Θ(n^9 √{\ln (1/δ)})$,
        and from Lemma~\ref{lemma-Chernoff-almost-independent}, the
        probability that the number of positive samples exceeds
        $μ(1+n^{-4})$ is at most 
        $\exp(-μn^{-8}/3) = \exp(-Θ(n \ln(1/δ))) = δ^{Θ(n)} ≤ δ/2$ for
        sufficiently large $n$.  A similar bound holds on the other
        side.  So with probability at least $1-δ$ we get an estimate
        $\hat{ρ}$ of $ρ = 4(n+m)/(n^2-m)$ that is accurate to within a
        relative error of $n^{-4}$.  Since $\frac{dρ}{dm}
        = O(n^{-2})$ throughout the interval spanned by $m$, and since
        $ρ$ itself is $Θ(n^{-1})$, any relative error that is
        $o(n^{-1})$ will give us an exact answer for $m$ when we round to the
        nearest value of $ρ$.  We are done.

        We've chosen big exponents because all we care about is
        getting a polynomial bound.
        With a cleaner analysis we could probably get better
        exponents.  It is also worth observing that the problem is
        ambiguous about whether we have random access to the data
        stream, and whether the time taken by the robot to move around
        counts toward our time bound.  If we are processing the stream
        after the fact, and are allow to skip to specific places in
        the stream at no cost, we can get each sample in $O(1)$ time.
        This may further reduce the exponents, and demonstrates why it
        is important to nail down all the details of a model.

    \subsection{Return of the sprites}

\begin{figure}
        \centering
        \begin{tabular}{cc}
        \includegraphics[scale=4.0]{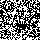}
            &
        \includegraphics[scale=4.0]{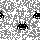}
        \end{tabular}
        \caption[Hidden Space Invaders]{Two hidden Space Invaders.  On
        the left, the Space Invaders hide behind random pixels.  On
        the right, their positions are revealed by turning the other
        pixels gray.}
        \label{fig-hidden-space-invader}
    \end{figure}

    The Space Invaders from
    Problem~\ref{section-hw-painting-with-sprites} are back, and now
    they are trying to hide in random $n×n$ bitmaps, where each pixel
    is set with probability $1/2$.
    Figure~\ref{fig-hidden-space-invader} shows an example of this,
    with two sprites embedded in an otherwise random bitmap.

    The sprites would like your assistance in making their hiding
    places convincing.  Obviously it's impossible to disguise their
    presence completely, but to make the camouflage as realistic as
    possible, they would like you to provide them with an algorithm
    that will generate an $n×n$ bitmap uniformly at random,
    conditioned on containing \emph{at least two} copies of a given $m$-pixel
    sprite at distinct offsets.  (It is OK for the sprites to overlap,
    but they cannot be directly on top of each other, and sprites that
    extend beyond the edges of the bitmap wrap around as in
    Problem~\ref{section-hw-painting-with-sprites}.)  Your algorithm
    should make the probability of each bitmap that contains at least
    two copies of the sprite be exactly equally likely, and should run
    in expected time polynomial in $n$ and $m$.  

        \subsubsection*{Solution}

        Rejection sampling doesn't work here because if the sprites
        are large, the chances of getting two sprites out of a random
        bitmap are exponentially small.  Generating a random bitmap
        and slapping two sprites on top of it also doesn't work,
        because it gives a non-uniform distribution: if the sprites
        overlap in $k$ places, there are $2^{n^2-2m+k}$ choices for
        the remaining bits, which means that we would have to adjust
        the probabilities to account for the effect of the overlap.
        But even if we do this, we still have issues with bitmaps that
        contain more than two sprites: a bitmap with three sprites can
        be generated in three different ways, and it gets worse
        quickly as we generate more.  It may be possible to work
        around these issues, but a simpler approach is to use the
        sampling mechanism from Karp-Luby~\cite{KarpL1985} (see also
        §\ref{section-approximating-sharp-dnf}).

        Order all $\binom{n^2}{2}$ positions $u<v$ for the two planted
        sprites in lexicographic order.  For each pair of positions
        $u<v$, let $A_{uv}$ be the set of all bitmaps with sprites
        at $u$ and $v$.  Then we can easily calculate 
        $\card{A_{uv}} = 2^{n^2-2m+k}$ where $k$
        is the number of positions where sprites at $u$ and $v$
        overlap, and so we can sample a particular $A_{uv}$ with
        probability $\card{A_{uv}}/∑_{st} \card{A_{st}}$ and then
        choose an element of $A_{uv}$ uniformly at random by filling
        in the positions not covered by the sprites with independent
        random bits.  To avoid overcounting, we discard any bitmap
        that contains a sprite at a position less than $\max(u,v)$;
        this corresponds to discarding any bitmap that appears in
        $A_{st}$ for some $st < uv$.  By the same argument as in
        Karp-Luby, we expect to discard at most $\binom{n^2}{2}$
        bitmaps before we get one that works.  This gives an expected
        $O(n^4)$ attempts, each of which requires about $O(n^2)$ work to
        generate a bitmap and $O(n^2 m)$ work to test it, 
        for a nicely polynomial $O(n^6 m)$ expected
        time in total.

        If we don't want to go through all this analysis, we could
        also reduce to Karp-Luby directly by encoding the existence of
        two sprites as a large DNF formula consisting of
        $\binom{n^2}{2}$ clauses, and then argue that the sampling
        mechanism inside Karp-Luby gives us what we want.

\section{Final exam}
\label{appendix-final-MMXVI}

Write your answers in the blue book(s).  Justify your answers.  Work
alone.  Do not use any notes or books.  

There are three problems on this exam, each worth 20
points, for a total of 60 points.
You have approximately three hours to complete this
exam.

\subsection{Virus eradication (20 points)}

Viruses have infected a network structured as an
undirected graph $G = (V,E)$.  At time $0$,
all $n = \card{V}$ nodes are infected.
At each step, as long as there is at least one infected node,
we can choose a node $v$ to disinfect; this causes $v$
to become uninfected, but there is an independent $\frac{1}{2d(v)}$
chance that each of $v$'s neighbors becomes infected if it is not
infected already, where $d(v)$ is the degree of $v$.

Let $X_t$ be the number of nodes that are infected after $t$ disinfection steps.  Let $τ$ be
the first time at which $X_τ = 0$.  Show that, no matter what strategy
is used to select nodes to disinfect, $\Exp{τ} = O(n)$.

\emph{Clarification added during exam:} When choosing a node to
disinfect, the algorithm can only choose nodes that are currently
infected.

    \subsubsection*{Solution}

    Let $Y_t = X_t + \frac{1}{2} \min(t,τ)$.  We will show that
    $\Set{Y_t}$
    is a supermartingale.  Suppose we disinfect node $v$ at time $t$,
    and that $v$ has $d' ≤ d(v)$ uninfected neighbors.  Then
    $\ExpCond{X_{t+1}}{X_t} = X_t - 1 + \frac{d'}{2d(v)} ≤ X_t -
    \frac{1}{2}$, because we always disinfect $v$ and each of its $d'$ uninfected
    neighbors contributes $\frac{1}{2d(v)}$ new infections on average.
    We also have $\min(τ,t)$ increasing by $1$ in this case.  This
    gives $\ExpCond{Y_{t+1}}{Y_t} ≤ Y_t -\frac{1}{2} + \frac{1}{2} =
    Y_t$ when $t < τ$; it also holds trivially that $Y_{t+1} = Y_t$
    when $t ≥ τ$.  So $\ExpCond{Y_{t+1}}{Y_t} ≤ Y_t$ always.

    Now apply the optional stopping theorem.  We have bounded
    increments, because $\card{Y_{t+1}-Y_t} ≤ n + \frac{1}{2}$.  We
    also have bounded expected time, because there is a nonzero
    probability $ε$ that each sequence of $n$ consecutive disinfections
    produces no new infections, giving $\Exp{τ} ≤ 1/ε$.  So
    $n = Y_0 ≥ \Exp{Y_τ} = \frac{1}{2} \Exp{τ}$, giving $\Exp{τ} ≤
    2n = O(n)$.

\subsection{Parallel bubblesort (20 points)}

Suppose we are given a random bit-vector and choose to sort it using
parallel bubblesort.  The initial state consists of $n$ independent
fair random bits
$X^0_1,X^0_2,\dots,X^0_n$, and in each round, for each pair of
positions $X^t_i$ and $X^t_{i+1}$, we set new values $X^{t+1}_i = 0$ and
$X^{t+1}_{i+1} = 1$ if $X^t_i = 1$ and $X^t_{i+1} = 0$.  When it holds
that $X^t_i = 1$ and $X^t_{i+1} = 0$, we say that there is a swap at
position $i$ in round $t$.  Denote the total number of swaps in round
$t$ by $Y_t$.
\begin{enumerate}
    \item What is $\Exp{Y_0}$?
    \item What is $\Exp{Y_1}$?
    \item Show that both $Y_0$ and $Y_1$ are concentrated around their
        expectations, by showing, for any fixed $c > 0$ and each $t ∈
        \Set{0,1}$, that there exists a
        function $f_t(n) = o(n)$ such that
        $\Prob{\abs*{Y_t - \Exp{Y_t}} ≥ f_t(n)} ≤ n^{-c}$.
\end{enumerate}

    \subsubsection*{Solution}

    \begin{enumerate}
        \item This is a straightforward application of linearity of
            expectation.  Let $Z^t_i$, for each $i ∈ \Set{1,\dots,n-1}$, be the indicator for the event that 
            $X^t_i = 1$ and $X^t_{i+1} = 0$.  
            For $t=0$, $X^0_i$ and $X^0_{i+1}$ are independent, so this event occurs with probability
            $\frac{1}{4}$.
            So
            $\Exp{Y_0} = \Exp{∑_{i=1}^{n-1} Z^0_i} = ∑_{i=1}^{n-1}
            \Exp{Z^0_i} = (n-1)\frac{1}{4} = \frac{n-1}{4}$.
        \item Again we want to use linearity of expectation, but we
            have to do a little more work to calculate $\Exp{Z^1_i}$.
            For $Z^1_i$ to be $1$, we need $X^1_i = 1$ and $X^1_{i+1}
            = 0$.  If we ask where that $1$ in $X^1_i$ came from,
            either it started in $X^0_{i-1}$ and moved to position $i$
            because $X^0_i = 0$, or it started in $X^0_i$ and didn't
            move because $X^0_{i+1} = 1$.  Similarly the $0$ in
            $X^1_{i+1}$ either started in $X^0_{i+2}$ and moved down
            to replace $X^0_{i+1} = 1$ or started in $X^1_{i+1}$ and
            didn't move because $X^0_i = 0$.  Enumerating all the
            possible assignments of 
            $X^0_{i-1}$, $X^0_i$, $X^0_{i+1}$, and $X^0_{i+2}$
            consistent with these conditions gives $0110$, $1000$,
            $1001$, $1010$, and $1110$, making $\Exp{Z^1_i} = 5⋅2^{-4} =
            \frac{5}{16}$ when $2 ≤ i ≤ n-2$.  This accounts for $n-3$
            of the positions.

            There are two annoying special cases: when $i=1$ and when
            $i=n-1$.  We can handle these with the above analysis by
            pretending that $X^t_0 = 0$ and $X^t_{n+1} = 1$ for all
            $t$; this leaves the initial patterns $0110$ for $i=1$ and
            $1001$ for $i=n-1$, each of which occurs with probability
            $\frac{1}{8}$.

            Summing over all cases then gives $\Exp{Y_1} = \frac{1}{8} +
            ∑_{i=2}^{n-3} \frac{5}{16} + \frac{1}{8} =
            \frac{5n-11}{16}$.

        \item The last part is just McDiarmid's inequality.  Because
            the algorithm is deterministic aside from the initial
            random choice of input, We can
            express each $Y_t$ as a function $g_t(X^0_1,\dots,X^0_n)$
            of the independent random variables $X^0_1,\dots,X^0_n$.
            Changing some $X^0_i$ can change $Z^0_{i-1}$ and
            $Z^0_{i}$, for a total change to $Y_0$ of at most $2$;
            similarly, changing $X^0_i$ can change at most
            $Z^1_{i-2}$, $Z^1_{i-1}$, $Z^1_i$, and $Z^i_{i+1}$, for a
            total change of at most $4$.  So McDiarmid says that, for
            each $t ∈ \Set{0,1}$, 
            $\Prob{\abs{Y_t-\Exp{Y_t}} ≤ s} ≤ 2\exp{-Θ(s^2/n)}$.
            For any fixed $c>0$, there is some $s = Θ(√{n \log n}) = o(n)$ 
            such that the right-hand side is less than $n^{-c}$.
    \end{enumerate}

\subsection{Rolling a die (20 points)}

Suppose we have a device that generates a sequence of independent fair
random coin-flips, but what we want is a six-sided die that generates
the values $1,2,3,4,5,6$ with equal probability.
\begin{enumerate}
    \item Give an algorithm that does so using $O(1)$ coin-flips on
        average.
    \item Show that any correct algorithm for this problem will use more than
        $n$ coin-flips with probability at least $2^{-O(n)}$.
\end{enumerate}

    \subsubsection*{Solution}

    \begin{enumerate}
        \item Use rejection sampling: generate $3$ bits, and if the
            resulting binary number is not in the range $1,\dots,6$,
            try again.  Each attempt consumes $3$ bits and
            succeeds with probability $3/4$, so we need to generate
            $\frac{4}{3}⋅3 = 4$ bits on average.

            It is possible to improve on this by reusing the last bit
            of a discarded triple of bits as the first bit in the next
            triple.  This requires a more complicated argument to show
            uniformity, but requires only two bits for each attempt
            after the first, for a total of $3 +
            \frac{1}{4}⋅\frac{4}{3}⋅2 = \frac{11}{3}$ bits on average.
            This is still $O(1)$, so unless that $\frac{1}{3}$ bit
            improvement is
            really important, it's probably easiest just to do
            rejection sampling.
        \item Suppose that we have generated $n$ bits so far.  Since
            $6$ does not evenly divide $2^n$ for any $n$, we cannot
            assign an output from $1,\dots,6$ to all possible $2^n$ sequences of bits
            without giving two outputs different probabilities.  So we
            must keep going in at least one case, giving a probability 
            of at least $2^{-n} = 2^{-O(n)}$ that we continue.
    \end{enumerate}

\chapter{Sample assignments from Spring 2014}

\section{Assignment 1: due Wednesday, 2014-09-10, at 17:00} 

\subsection{Bureaucratic part}

Send me email!  My address is
\mailto{james.aspnes@gmail.com}.

In your message, include:

\begin{enumerate}
\item Your name.
\item Your status: whether you are an undergraduate, grad student, auditor, etc.
\item Anything else you'd like to say.
\end{enumerate}

(You will not be graded on the bureaucratic part, but you should do it anyway.)

\subsection{Two terrible data structures}

Consider the following data structures for maintaining a sorted list:
\begin{enumerate}
    \item A sorted array $A$.  Inserting a new element at position
        $A[i]$ into an array that current contains $k$ elements
        requires moving the previous values in $A[i]\dots
        A[k]$ to $A[i+1] \dots A[k+1]$.  Each
        element moved costs one unit.  
        
        For example, if the array currently contains
        \texttt{1 3 5}, and we insert \texttt{2}, the resulting array
        will contain \texttt{1 2 3 5} and the cost of the operation
        will be $2$, because we had to move \texttt{3} and \texttt{5}.
    \item A sorted doubly-linked list.  Inserting a new element at
        a new position requires moving the head pointer from its
        previous position to a neighbor of the new node, and then to
        the new node; each pointer move costs one unit.

        For example, if the linked list currently contains \texttt{1 3
        5}, with the head pointer pointing to \texttt{5}, and we
        insert \texttt{2}, the resulting linked list will contain
        \texttt{1 2 3 5}, with the head pointing to \texttt{2}, and
        the cost will be $2$, because we had to move the pointer from
        \texttt{5} to \texttt{3} and then from \texttt{3} to
        \texttt{2}.  Note that we do not charge for updating the
        pointers between the new element and its neighbors.  We will
        also assume that inserting the first element is free.
\end{enumerate}

Suppose that we insert the elements $1$ through $n$, in random order,
into both data structures.  Give an exact closed-form expression for
the expected cost of each sequence of insertions.

    \subsubsection*{Solution}

    \begin{enumerate}
        \item Suppose we have already inserted $k$ elements.  Then the
            next element is equally likely to land in any of the
            positions $A[1]$ through $A[k+1]$.  The number of displaced
            elements is then uniformly distributed in $0$ through $k$,
            giving an expected cost for this insertion of
            $\frac{k}{2}$.

            Summing over all insertions gives
            \begin{align*}
                ∑_{k=0}^{n-1} \frac{k}{2}
                &= \frac{1}{2} ∑_{k=0}^{n-1}
                \\&= \frac{n(n-1)}{4}.
            \end{align*}

            An alternative proof, which also uses linearity of
            expectation, is to define $X_{ij}$ as the indicator
            variable for the event that element $j$ moves when element
            $i<j$ is inserted.  This is $1$ if an only if $j$ is
            inserted before $i$, which by symmetry occurs with
            probability exactly $1/2$.  So the expected total number
            of moves is
            \begin{align*}
                ∑_{1≤i<j≤n} \Exp{X_{ij}}
                &= ∑_{1≤i<j≤n} \frac{1}{2}
                \\&= \frac{1}{2} \binom{n}{2}
                \\&= \frac{n(n-1)}{4}.
            \end{align*}

            It's always nice when we get the same answer in situations
            like this.

        \item Now we need to count how far the pointer moves between
            any two consecutive elements.  Suppose that we have
            already inserted $k-1 > 0$ elements, and let $X_k$ be the cost
            of inserting the $k$-th element.  Let $i$ and $j$ be the
            indices in the sorted list of the new and old pointer
            positions after the $k$-th insertion.  By symmetry, all
            pairs of distinct positions $i≠j$ are equally likely.  So
            we have
            \begin{align*}
                \Exp{X_k}
                &= \frac{1}{k(k-1)} ∑_{i≠j} \abs*{i-j}
                \\&= \frac{1}{k(k-1)} 
                    \left(
                        ∑_{1≤i<j≤k} (j-i)
                        + ∑_{1≤j<i≤k} (i-j)
                    \right)
                \\&= \frac{2}{k(k-1)} ∑_{1≤i<j≤k} (j-i)
                \\&= \frac{2}{k(k-1)} ∑_{j=1}^{k} ∑_{\ell=1}^{j-1} \ell
                \\&= \frac{2}{k(k-1)} ∑_{j=1}^{k} \frac{j(j-1)}{2}
                \\&= \frac{1}{k(k-1)} ∑_{j=1}^{k} j(j-1).
                \\&= \frac{1}{k(k-1)} ∑_{j=1}^{k} \left(j^2 - j\right)
                \\&= \frac{1}{k(k-1)} ⋅ \left(\frac{(2k+1)k(k+1)}{6} - \frac{k}{k+1}{2}\right)
                \\&= \frac{1}{k(k-1)} ⋅ \frac{(2k-2)k(k+1)}{6}
                \\&= \frac{k+1}{3}.
            \end{align*}

            This is such a simple result that we might reasonably
            expect that there is a faster way to get it, and we'd be
            right.  A standard trick is to observe that we can
            simulate choosing $k$ points uniformly at random
            from a line of $n$ points by
            instead choosing $k+1$ points uniformly at random from a cycle of $n+1$ points,
            and deleting the first point chosen to turn the cycle back into a
            line.  In the cycle, symmetry implies that the expected
            distance between each point and its successor is the same
            as for any other point; there are $k+1$ such distances,
            and they add up to $n+1$, so each expected distance is
            exactly $\frac{n+1}{k+1}$.

            In our particular case, $n$ (in the formula) is $k$ and
            $k$ (in the formula) is $2$, so we get
            $\frac{k+1}{3}$.  Note we are sweeping the whole absolute
            value thing under the carpet here, so maybe the more
            explicit derivation is safer.

            However we arrive at $\Exp{X_k} = \frac{k+1}{3}$ (for
            $k>1$), we can sum these expectations to get our total
            expected cost:
            \begin{align*}
                \Exp{∑_{k=2}^{n} X_k}
                &= ∑_{k=2}^{n} \Exp{X_k}
                \\&= ∑_{k=2}^{n} \frac{k+1}{3}
                \\&= \frac{1}{3} ∑_{\ell=3}^{n+1} \ell
                \\&= \frac{1}{3} \left(\frac{(n+1)(n+2)}{2} - 3\right)
                \\&= \frac{(n+1)(n+2)}{6} - 1.
            \end{align*}

            It's probably worth checking a few small cases to see
            that this answer actually makes sense.
    \end{enumerate}

    For large $n$, this shows that the doubly-linked list
    wins, but not by much: we get roughly $n^2/6$ instead of
    $n^2/4$.  This is a small enough difference that in practice it is
    probably dominated by other constant-factor differences that we have
    neglected.

\subsection{Parallel graph coloring}

Consider the following algorithm for assigning one of $k$ colors to
each node in a graph with $m$ edges:
\begin{enumerate}
    \item\label{item-parallel-graph-coloring-coloring}
        Assign each vertex $u$ a color $c_u$, chosen uniformly at
        random from all $k$ possible colors.
    \item\label{item-parallel-graph-coloring-recoloring} For each vertex $u$, if $u$ has a neighbor $v$ with $c_u =
        c_v$, assign $u$ a new color $c'_u$, again chosen uniformly at
        random from all $k$ possible colors.  Otherwise, let $c'_u = c_u$.
\end{enumerate}

Note that any new colors $c'$ do not affect the test $c_u=c_v$.  A node
changes its color only if it has the same original color as the
original color of one or more of its neighbors.

Suppose that we run this algorithm on an $r$-regular\footnote{Every
vertex has exactly $r$ neighbors.} triangle-free\footnote{There are
    no vertices $u$, $v$, and $w$ such that all are neighbors of each
other.} graph.  As a function of $k$, $r$, and $m$, give an exact
closed-form expression for the expected number of
monochromatic\footnote{Both endpoints have the same color.} edges
after running both steps of the algorithm.

    \subsubsection*{Solution}

    We can use linearity of expectation to compute the probability
    that any particular edge is monochromatic, and then multiply by
    $m$ to get the total.

    Fix some edge $uv$.  If either of $u$ or $v$ is recolored in step
    \ref{item-parallel-graph-coloring-recoloring}, then the
    probability that $c'_u = c'_v$ is exactly $1/k$.  If neither is
    recolored, the probability that $c'_u = c'_v$ is zero (otherwise
    $c_u = c_v$, forcing both to be recolored).  So we can calculate
    the probability that $c'_u = c'_v$ by conditioning on the event
    $A$ that neither vertex is recolored.

    This event occurs if both $u$ and $v$ have no neighbors with the
    same color.  The probability that $c_u = c_v$ is $1/k$.
    The probability that any particular neighbor $w$ of $u$ has $c_w =
    c_u$ is also $1/k$; similarly for any neighbor $w$ of $v$.  These
    events are all independent on the assumption that the graph is
    triangle-free (which implies that no neighbor of $u$ is also a
    neighbor of $v$).  So the probability that none of these $2r-1$ events
    occur is $(1-1/k)^{2r-1}$.

    We then have
    \begin{align*}
        \Prob{c_u = c_v}
        &= \ProbCond{c_u = c_v}{\overline{A}} \Prob{\overline{A}} + \ProbCond{c_u = c_v}{A} \Prob{A}
        \\&= \frac{1}{k} ⋅ \left(1-\left(1-\frac{1}{k}\right)^{2r-1}\right).
    \end{align*}

    Multiply by $m$ to get
    \begin{displaymath}
        \frac{m}{k} ⋅ \left(1-\left(1-\frac{1}{k}\right)^{2r-1}\right).
    \end{displaymath}

    For large $k$, this is approximately $\frac{m}{k} ⋅
    \left(1-e^{-(2r-1)/k}\right)$, which is a little bit better than 
    than the $\frac{m}{k}$ expected monochrome
    edges from just running step
    \ref{item-parallel-graph-coloring-coloring}.

    Repeated application of step
    \ref{item-parallel-graph-coloring-recoloring} may give better
    results, particular if $k$ is large relative to $r$.  We will see
    this technique applied to a more general class of problems in 
    §\ref{section-constructive-lll}.

\section{Assignment 2: due Wednesday, 2014-09-24, at 17:00}

    \subsection{Load balancing}

    Suppose we distribute $n$ processes independently and uniformly at
    random among $m$ machines, and pay a
    communication cost of $1$ for each pair of processes assigned to
    different machines.  Let $C$ be the total communication cost, that
    is, the number of pairs of processes assigned to different
    machines.  What is the expectation and variance of $C$?

    \subsubsection*{Solution}

        Let $C_{ij}$ be the communication cost between machines $i$
        and $j$.  This is just an indicator variable for the event
        that $i$ and $j$ are assigned to different machines, which
        occurs with probability $1-\frac{1}{m}$.  We have $C =
        ∑_{1≤i<j≤n} C_{ij}$.
        \begin{enumerate}
            \item Expectation is a straightforward application of
                linearity of expectation.  There are $\binom{n}{2}$
                pairs of processes, and $\Exp{C_{ij}} = 1-\frac{1}{m}$
                for each pair, so
                \begin{displaymath}
                    \Exp{C} = \binom{n}{2}\left(1-\frac{1}{m}\right).
                \end{displaymath}
            \item Variance is a little trickier because the $C_{ij}$
                are not independent.  But they are pairwise
                independent: even if we fix the location of $i$ and
                $j$, the expectation of $C_{jk}$ is still
                $1-\frac{1}{m}$, so $\Cov{C_{ij}}{C_{jk}} = \Exp{C_{ij}
                    C_{jk}} - \Exp{C_{ij}}⋅\Exp{C_{jk}} = 0$.  So we
                    can compute
                    \begin{displaymath}
                        \Var{C}
                        = ∑_{1≤i<j<n} \Var{C_{ij}}
                        = \binom{n}{2} \frac{1}{m} \left(1-\frac{1}{m}\right).
                    \end{displaymath}
        \end{enumerate}

    \subsection{A missing hash function}

    A clever programmer inserts $X_i$ elements in each of $m$ buckets
    in a hash table, where each bucket $i$ is implemented as a balanced
    binary search tree with search cost at most $\ceil{\lg (X_i+1)}$.  We are
    interested in finding a particular target element $x$, which is
    equally likely to be in any of the buckets, but we don't know what
    the hash function is.

    Suppose that we know that the $X_i$ are independent and
    identically distributed with $\Exp{X_i} = k$, and that the
    location of $x$ is independent of the values of the $X_i$.  What
    is best upper bound we can put on the expected cost of finding $x$?

        \subsubsection*{Solution}

        The expected cost of searching bucket $i$ is
        $\Exp{\ceil{\lg (X_i+1)}}$.  This is the expectation of a
        function of $X_i$, so we would like to bound it using
        Jensen's inequality (§\ref{section-jensens-inequality}).

        Unfortunately the function $f(n) = \ceil{\lg (n+1)}$ is not
        concave (because of the ceiling), but $1+\lg (n+1) >
        \ceil{\lg(n+1)}$ is.  So the expected cost of searching bucket
        $i$ is bounded by $1+\lg(\Exp{X_i} + 1) = 1+\lg(k+1)$.

        Assuming we search the buckets in some fixed order until we find $x$, we
        will search $Y$ buckets where $\Exp{Y} = \frac{n+1}{2}$.
        Because $Y$ is determined by the position of $x$, which is
        independent of the $X_i$, $Y$ is also independent of the
        $X_i$.  So Wald's equation
        \eqref{eq-Walds-equation-simple} applies, and 
        the total cost is bounded by 
        \begin{displaymath}
            \frac{n+1}{2} \left(1+\lg(k+1)\right).
        \end{displaymath}

\section{Assignment 3: due Wednesday, 2014-10-08, at 17:00}

    \subsection{Tree contraction}
    \label{section-hw-tree-contraction}

    Suppose that you have a tree data structure with $n$ nodes, in which each node
    $i$ has a pointer $\Parent(u)$ to its parent (or itself, in the
    case of the root node $\Root$).

    Consider the following randomized algorithm for shrinking paths in
    the tree: in the first phase, each node $u$ first determines its parent $v = \Parent(u)$
    and its grandparent $w=\Parent(\Parent(u))$.  
    In the second phase, it sets $\Parent'(u)$ to $v$ or $w$ according
    to a fair coin-flip independent of the coin-flips of all the other
    nodes.

    Let $T$ be the tree generated by the $\Parent$ pointers and $T'$
    the tree generated by the $\Parent'$ pointers.  An example of two
    such trees is given in Figure~\ref{fig-hw-tree-contraction}.

    \begin{figure}
        \centering
        \begin{tabular}{cc}
            \begin{tikzpicture}
                \node {0}
                    child { 
                        node {\textbf{1}} 
                        child {
                            node {2}
                            child {
                                node {\textbf{3}}
                            }
                            child {
                                node {4}
                            }
                        }
                        child {
                            node {\textbf{5}}
                            child { node{\textbf{6}} }
                        }
                    }
                ;
            \end{tikzpicture}
            &
            \hspace{2cm}
            \begin{tikzpicture}
                \node {0}
                    child { 
                        node {\textbf{1}} 
                        child {
                            node {2}
                            child {
                                node {4}
                            }
                        }
                        child { node {\textbf{3}} }
                        child { node{\textbf{6}} }
                    }
                    child {
                        node {\textbf{5}}
                    }
                ;
            \end{tikzpicture}
    \end{tabular}
    \caption[Example of tree contraction for
    Problem~\ref{section-hw-tree-contraction}]
    {Example of tree contraction for
    Problem~\ref{section-hw-tree-contraction}.
    Tree $T$ is on the left, $T'$ on the right.
    Nodes $1$, $3$, $5$, and $6$ (in boldface) switch to their
grandparents.  The other nodes retain their original parents.}
    \label{fig-hw-tree-contraction}
    \end{figure}

    Recall that the \indexConcept{depth!tree}{depth} 
    of a node is defined by $\depth(\Root) = 0$
    and $\depth(u) = 1+\depth(\Parent(u))$ when $u≠\Root$.
    The depth of a tree is equal to the maximum depth of any node.

    Let $D$ be depth of $T$, and
    $D'$ be the depth of $T'$.  Show that there is a constant $a$,
    such that for any
    fixed $c>0$, with probability at least $1-n^{-c}$ it holds that
    \begin{equation}
        \label{eq-hw-tree-contraction-goal}
        a⋅D - O\left(√{D \log n}\right) ≤ D' ≤ a⋅D + O\left(√{D \log n}\right).
    \end{equation}

        \subsubsection*{Solution}

        For any node $u$,
        let $\depth(u)$ be the depth of $u$ in $T$ and $\depth'(u)$ be
        the depth of $u$ in $T'$.  Note that $\depth'(u)$ is a random
        variable.  We will start by computing $\Exp{\depth'(u)}$ as a
        function of $\depth(u)$, by solving an appropriate recurrence.

        Let $S(k) = \depth'(u)$ when $\depth(u)=k$.  The base cases
        are $S(0)=0$ (the depth of the root never changes) and
        $S(1)=1$ (same for the root's children).  For larger $k$, we
        have 
        \begin{align*}
            \Exp{\depth'(u)} 
            &=   \frac{1}{2}\Exp{1+\depth'(\Parent(u))}
               + \frac{1}{2}\Exp{1+\depth'(\Parent(\Parent(u)))}
            \intertext{or}
            S(k) &= 1 + \frac{1}{2} S(k-1) + \frac{1}{2} S(k-2).
        \end{align*}

        There are various ways to solve this recurrence.  The most
        direct may be to define a generating function
        $F(z) = ∑_{k=0}^{∞} S(k) z^k$.  Then the recurrence becomes
        \begin{equation*}
            \label{eq-hw-tree-contraction-generating-function}
            F = \frac{z}{1-z} + \frac{1}{2} z F + \frac{1}{2} z^2 F.
        \end{equation*}

        Solving for $F$ gives
        \begin{align*}
            F
            &= \frac{\frac{z}{1-z}}{1-\frac{1}{2}z-\frac{1}{2}z^2}
            \\&= \frac{2z}{(1-z)(2 - z - z^2)}.
            \\&= \frac{2z}{(1-z)^2(2+z)}
            \\&= 2z \left(\frac{1/3}{(1-z)^2} + \frac{1/9}{(1-z)} + \frac{1/18}{1+\frac{1}{2}z}\right)
            \\&= 
                \frac{2}{3}⋅\frac{z}{(1-z)^2}
                + \frac{2}{9}⋅\frac{z}{1-z}
                + \frac{1}{9}⋅\frac{z}{1+\frac{1}{2}z},
            \intertext{from which we can read off the exact solution}
            S(k)
            &= \frac{2}{3}⋅k + \frac{2}{9} + \frac{1}{9} \left(-\frac{1}{2}\right)^{k-1}
        \end{align*}
        when $k ≥ 1$.\footnote{A less direct but still effective
            approach is to guess that $S(k)$ grows linearly, and find
            $a$ and $b$ such that $S(k) ≤ ak + b$.  For this we need
            $ak+b ≤ 1 + \frac{1}{2}(a(k-1)+b) +
            \frac{1}{2}(a(k-2)+b)$.  The $b$'s cancel, leaving
            $ak ≤ 1 + ak - \frac{3}{2}a$.  Now the $ak$'s cancel,
            leaving us with $0 ≤ 1 - \frac{3}{2}a$ or $a ≥ 2/3$.
            We then go back and make $b = 1/3$ to get the right
            bound on $S(1)$, giving the bound $S(k) ≤
            \frac{2}{3}⋅k + \frac{1}{3}$.  
            We can then repeat the argument for $S(k) ≥ a'k+b'$ to get
            a full bound $\frac{2}{3}k ≤ S(k) ≤ \frac{2}{3}k +
        \frac{1}{3}$.}

        We can easily show that $\depth'(u)$ is tightly concentrated
        around $\depth(u)$ using McDiarmid's inequality
        \eqref{eq-McDiarmids-inequality}.  Let $X_i$, for $i=2\dots
        \depth{u}$, be the choice
        made by $u$'s depth-$i$ ancestor.  Then changing one $X_i$
        changes $\depth'(u)$ by at most $1$.  So we get
        \begin{align}
            \Prob{\depth'(u) ≥ \Exp{\depth'(u) + t}}
            &≤ e^{-2t^2/(\depth(u)-1)}
            \label{eq-hw-tree-contraction-mcdiarmid-upper-bound}
            \intertext{and similarly}
            \Prob{\depth'(u) ≤ \Exp{\depth'(u) - t}}
            &≤ e^{-2t^2/(\depth(u)-1)}.
            \label{eq-hw-tree-contraction-mcdiarmid-lower-bound}
        \end{align}

        Let $t = √{\frac{1}{2}D \ln \left(1/ε\right)}$.  Then the right-hand
        side of \eqref{eq-hw-tree-contraction-mcdiarmid-upper-bound}
        and \eqref{eq-hw-tree-contraction-mcdiarmid-lower-bound}
        becomes $e^{-D \ln \left(1/ε\right) / \left(\depth\left(u\right)-1\right)} < e^{-\ln \left(1/ε\right)} =
        ε$.  For $ε=\frac{1}{2} n^{-c-1}$, we get $t=√{\frac{1}{2}D
        \ln\left(2n^{c+1}\right)} = √{\frac{c+1}{2}D
        (\ln n + \ln 2)} = O\left(√{D \log
        n}\right)$ when $c$ is constant.

        For the lower bound on $D'$, when can apply
        \eqref{eq-hw-tree-contraction-mcdiarmid-lower-bound} to some
        single node $u$ with $\depth\left(u\right) = D$; this node by itself will
        give $D' ≥ \frac{2}{3} D - O\left(√{D \log n}\right)$ with
        probability at least $1-\frac{1}{2}n^{-c-1}$.  For the upper bound, we need to take the maximum
        over all nodes.  In general, an upper bound on the maximum of a bunch of random
        variables is likely to be larger than an upper bound on any
        one of the random variables individually, because there is a
        lot of room for one of the variables to get unlucky, but we
        can apply the union bound to get around this.  For each
        individual $u$, we have $\Prob{\depth'\left(u\right) ≥ \frac{2}{3}D +
        O\left(√{D \log n}\right)} ≤ \frac{1}{2}n^{-c-1}$, so $\Prob{D' ≥ \frac{2}{3}D +
        O\left(√{D \log n}\right)} ≤ ∑_u \frac{1}{2} n^{-c-1} =
        \frac{1}{2} n^{-c}$.  This
            completes the proof.

    \subsection{Part testing}

    You are running a factory that produces parts for some important
    piece of machinery.  Many of these parts are defective, and must
    be discarded.  There are two levels of tests that can be
    performed:
    \begin{itemize}
        \item A normal test, which passes a part with probability
            $2/3$.
        \item A rigorous test, which passes a part with probability
            $1/3$.
    \end{itemize}

    At each step, the part inspectors apply the following rules to
    decide which test to apply:
    \begin{itemize}
        \item For the first part, a fair coin-flip decides between the
            tests.
        \item For subsequent parts, if the previous part passed, the
            inspectors become suspicious and apply the rigorous test; if
            it failed, they relax and apply the normal test.
    \end{itemize}

    For example, writing 
    \texttt{N+} for a part that passes the normal test, 
    \texttt{N-} for one that fails the normal test, 
    \texttt{R+} for a part that passes the rigorous test, 
    and
    \texttt{R-} for one that fails the rigorous test, 
    a typical execution of the testing procedure might look like
    \texttt{N- N+ R- N- N+ R- N+ R- N- N- N- N+ R+ R- N+ R+}.
    This execution tests $16$ parts and passes $7$ of them.

    Suppose that we test $n$ parts.  Let $S$ be the number that pass.

    \begin{enumerate}
        \item Compute $\Exp{S}$.
        \item Show that there a constant $c > 0$ such that, for any $t > 0$,
            \begin{equation}
                \label{eq-hw-normal-vs-rigorous-testing-bound}
                \Prob{\abs*{S-\Exp{S}} ≥ t} ≤ 2e^{-ct^2/n}.
            \end{equation}
    \end{enumerate}

        \subsubsection*{Solution}

        \paragraph{Using McDiarmid's inequality and some cleverness}

        Let $X_i$ be the indicator variable for the event that part
        $i$ passes, so that $S = ∑_{i=1}^{n} X_i$.

        \begin{enumerate}
            \item We can show by induction that $\Exp{X_i} = 1/2$ for
                all $i$.  The base case is $X_1$, where
                $\Prob{\text{part 1 passes}} =
                \frac{1}{2}\Prob{\text{part 1 passes rigorous test}}
                +\frac{1}{2}\Prob{\text{part 1 passes normal test}}
                = \frac{1}{2}\left(\frac{1}{3}+\frac{2}{3}\right)
                = \frac{1}{2}$.  For $i > 1$, $\Exp{X_{i-1}} = 1/2$
                implies that part $i$ is tested with the normal and
                rigorous tests with equal probability, so the analysis
                for $X_1$ carries through and gives $\Exp{X_i} = 1/2$
                as well.  Summing over all $X_i$ gives $\Exp{S} =
                n/2$.
            \item We can't use Chernoff, Hoeffding, or Azuma here,
                because the $X_i$ are not independent, and do not
                form a martingale difference sequence even after
                centralizing them by subtracting off their
                expectations.  So we are left with McDiarmid's
                inequality unless we want to do something clever and
                new (we don't).  Applying McDiarmid to the $X_i$
                directly doesn't work so well, but we can make it work
                with a different set of variables that generate the
                same outcomes.

                Let $Y_i ∈ \Set{A,B,C}$ be the grade of part $i$, where $A$
                means that it passes both the rigorous and the normal test,
                $B$ means that fails the rigorous test but passes the normal
                test, and $C$ means that it fails both tests.
                In terms of the $X_i$, $Y_i = A$ means $X_i = 1$, $Y_i
                = C$ means $X_i = 0$, and $Y_i = B$ means $X_i =
                1-X_{i-1}$ (when $i > 1$).  We get the right
                probabilities for passing each test by assigning equal
                probabilities.

                We can either handle the coin-flip at the beginning by
                including an extra variable $Y_0$, or we can combine
                the coin-flip with $Y_1$ by assuming that $Y_1$ is either $A$
                or $C$ with equal probability.  The latter approach improves
                our bound a little bit since then we only have $n$
                variables and not $n+1$.

                Now
                suppose that we fix all $Y_j$ for $j≠i$ and ask what happens
                if $Y_i$ changes.

                \begin{enumerate}
                    \item If $j<i$, then $X_j$ is not affected by $Y_i$.
                    \item Let $k>i$ be such that $Y_k ∈ \Set{A,C}$.
                        Then $X_k$ is not affected by $Y_i$, and
                        neither is $X_j$ for any $j > k$.
                \end{enumerate}

                It follows that changing $Y_i$ can only change $X_i,
                \dots, X_{i+\ell}$, where $\ell$ is the number of $B$
                grades that follow position $i$.

                There are two cases for the sequence $X_0\dots
                X_\ell$:
                \begin{enumerate}
                    \item If $X_i = 0$, then $X_1 = 1$, $X_2 = 0$,
                        etc.
                    \item If $X_i = 1$, then $X_1 = 0$, $X_2 = 1$,
                        etc.
                \end{enumerate}

                If $\ell$ is odd, changing $Y_i$ thus has no effect on
                $∑_{j=i}^{i+\ell} X_i$, while if $\ell$ is even,
                changing $Y_i$ changes the sum by $1$.  In either
                case, the effect of changing $Y_i$ is bounded by $1$,
                and McDiarmid's inequality applies with $c_i = 1$,
                giving
                \begin{displaymath}
                    \Prob{\abs*{S-\Exp{S}} ≥ t} ≤ 2e^{-2t^2/n}.
                \end{displaymath}
            \end{enumerate}

\section{Assignment 4: due Wednesday, 2014-10-29, at 17:00}

    \subsection{A doubling strategy}

    A common strategy for keeping the load factor of a hash table down
    is to double its size whenever it gets too full.  Suppose that we
    start with a hash table of size $1$ and double its size whenever
    two items in a row hash to the same location.

    Effectively, this means that we attempt to insert all elements
    into a table of size $1$; if two consecutive items hash to the
    same location, we start over and try to do the same to a table of
    size $2$, and in general move to a table of size $2^{k+1}$
    whenever any two consecutive elements hash to the same location in
    our table of size $2^k$.
    
    Assume that
    we are using an independently chosen $2$-universal hash function
    for each table size.
    Show that the expected final table size is $O(n)$.

        \subsubsection*{Solution}

        Let $X$ be the random variable representing the final table
        size.  Our goal is to bound $\Exp{X}$.

        First let's look at the probability that we get at least one
        collision between consecutive elements when inserting $n$
        elements into a table with $m$ locations.  
        Because the pairs
        of consecutive elements overlap, computing the exact
        probability that we get a collision is complicated, but we 
        only need an upper bound. 
        
        We have $n-1$ consecutive pairs, and each produces a
        collision with probability at most $1/m$.  This gives a total
        probability of a collision of at most $(n-1)/m$.

        Let $k = \ceil{\lg n}$, so that $2^k ≥ n$.  Then the
        probability of a consecutive collision in a table with
        $2^{k+\ell}$ locations is at most $(n-1)/2^{k+\ell} <
        2^k/2^{k+\ell} = 2^{-\ell}$.  Since the events that collisions
        occur at each table size are independent, we can compute, for
        $\ell > 0$,
        \begin{align*}
            \Prob{X = 2^{k+\ell}}
            &≤ \Prob{X ≥ 2^{k+\ell}}
            \\&≤ ∏_{i=0}^{\ell-1} 2^{-i}
            \\&= 2^{-\ell(\ell-1)/2}.
            \intertext{From this it follows that}
            \Exp{X}
            &= ∑_{i=0}^{∞} 2^i \Prob{X = 2^i}
            \\&< 2^k \Prob{X ≤ 2^k} + ∑_{l=1}^{∞} 2^{k+\ell} \Prob{X = 2^{k+\ell}}
            \\&≤ 2^k + ∑_{l=1}^{∞} 2^{k+\ell} ⋅ 2^{-\ell(\ell-1)/2}
            \\&≤ 2^k + 2^k ∑_{l=1}^{∞} 2^{\ell - \ell(\ell-1)/2}
            \\&≤ 2^k + 2^k ∑_{l=1}^{∞} 2^{-(\ell^2 - 2\ell)/2}
            \\&= O(2^k),
        \end{align*}
        since the series converges to some constant that does not
        depend on $k$.  But we chose $k$ so that $2^k =
        O(n)$, so this gives us our desired bound.

    \subsection{Hash treaps}

    Consider the following variant on a treap: instead of choosing the
    heap keys randomly, we choose a hash function $h:U→[1\dots m]$ from some
    strongly $2$-universal hash family, and use $h(x)$ as the heap key
    for tree key $x$.  Otherwise the treap operates as
    usual.\footnote{If $m$ is small, we may get collisions in the heap
        key values.  Assume in this case that a node will stop rising
    if its parent has the same key.}

    Suppose that $\card*{U} ≥ n$.
    Show that there is a sequence of $n$
    distinct tree keys such that the total expected time to insert
    them into an initially empty hash treap is
    $Ω(n^2/m)$.

        \subsubsection*{Solution}

        Insert the sequence $1\dots n$.  

        Let us first argue by induction on $i$ that all elements $x$
        with $h(x) = m$ appear as the uppermost elements of the right
        spine of the treap.  Suppose that this holds for $i-1$.  If
        $h(i) = m$, then after insertion $i$ is rotated up until it
        has a parent that also has heap key $m$; this extends the
        sequence of elements with heap key $m$ in the spine by $1$.
        Alternatively, if $h(i) < m$, then $i$ is never rotated above
        an element $x$ with $h(x) = m$, so the sequence of elements
        with heap key $m$ is unaffected.

        Because each new element has a larger tree key than all
        previous elements, inserting a new element $i$ requires moving
        past any elements in the right spine, and in particular
        requires moving past any elements $j < i$ with $h(j) = m$.  So
        the expected cost of inserting $i$ is at least the expected
        number of such elements $j$.  Because $h$ is chosen from a
        strongly $2$-universal hash family, $\Prob{h(j)=m} = 1/m$ for
        any $j$, and by linearity of expectation,
        $\Exp{\card*{\SetWhere{j<i}{h(j)=m}}} = (i-1)/m$.  Summing this
        quantity over
        all $i$ gives a total expected insertion cost of at least
        $n(n-1)/2m = Ω(n^2/m)$.

\section{Assignment 5: due Wednesday, 2014-11-12, at 17:00}

    \subsection{Agreement on a ring}

    Consider a ring of $n$ processes at positions $0, 1, \dots n-1$.
    Each process $i$ has a value $A[i]$ that is initially $0$ or $1$.
    At each step, we choose a process $r$ uniformly at random, and
    copy $A[r]$ to $A[(r+1) \bmod n]$.  Eventually, all $A[i]$ will have
    the same value.

    Let $A_t[i]$ be the value of $A[i]$ at time $t$, and let $X_t =
    ∑_{i=0}^{n-1} A_t[i]$ be the total number of ones in $A_t$.
    Let $τ$ be the first time at which $X_τ ∈ \Set{0,n}$.

    \begin{enumerate}
        \item Suppose that we start with $X_0 = k$.
            What is the probability that we eventually reach a state
            that is all ones?
        \item What is $\Exp{τ}$, assuming we start from the worst
            possible initial state?
    \end{enumerate}

        \subsubsection*{Solution}

        This is a job for the optional stopping theorem.  Essentially
        we are going to follow the same analysis from
        §\ref{section-stopping-times-and-random-walks} for a random
        walk with two absorbing barriers, applied to $X_t$.

        Let $ℱ_t$ be the $σ$-algebra generated by
        $A_0,\dots,A_t$.  Then $\Set{ℱ_t}$ forms a
        filtration, and each $X_t$ is $ℱ_t$-measurable.

        \begin{enumerate}
            \item For the first part, we show that $(X_t,
                ℱ_t)$ is a martingale.  The intuition is
                that however the bits in $A_t$ are arranged, there are
                always exactly the same number of positions where a
                zero can be replaced by a one as there are where a one
                can be replaced by a zero.

                Let $R_t$ be the random location chosen in state
                $A_t$.
                Observe that
                \begin{equation*}
                    \label{eq-hw-ring-agreement-step}
                    \ExpCond{X_{t+1}}{ℱ_t,R_t=i} = X_t + A[i] - A[(i+1) \bmod n].
                \end{equation*}

                But then
                \begin{align*}
                    \ExpCond{X_{t+1}}{ℱ_t} 
                    &= ∑_{i=0}^{n-1} \frac{1}{n} \left(X_t + A[i] - A[(i+1) \bmod n]\right)
                    \\&= X_t + \frac{1}{n} X_t - \frac{1}{n} X_t
                    \\&= X_t,
                \end{align*}
                which establishes the martingale property.

                We also have that (a) $τ$ is a stopping time with respect to
                the $ℱ_i$; (b) $\Prob{τ<∞} = 1$, because
                from any state there is a nonzero chance that all
                $A[i]$ are equal $n$ steps later; and (c) $X_t$ is
                bounded between $0$ and $n$.  So the optional
                stopping theorem applies, giving
                \begin{equation*}
                    \Exp{X_τ} = \Exp{X_0} = k.
                \end{equation*}

                But then
                \begin{align*}
                    \Exp{X_τ} = n ⋅ \Prob{X_τ = n} + 0 ⋅ \Prob{X_τ= 0} = k,
                \end{align*}
                so $\Exp{X_τ} = k/n$.

            \item For the second part, we use a variant on the $X_t^2 - t$ martingale.

                Let $Y_t$ count the number of positions $i$ for which
                $A_t[i] = 1$ and $A_t[(i+1) \bmod n] = 0$.  Then,
                conditioning on $ℱ_t$, we have
                \begin{equation*}
                    X_{t+1}^2 = 
                    \begin{cases}
                        X_t^2 + 2X_t + 1 & \text{with probability $Y_t/n$,} \\
                        X_t^2 - 2X_t + 1 & \text{with probability $Y_t/n$, and} \\
                        X_t^2 & \text{otherwise.}
                    \end{cases}
                \end{equation*}

                The conditional expectation sums to
                \begin{equation*}
                    \ExpCond{X_{t+1}^2}{ℱ_t} 
                    = X_t^2 + 2Y_t/n.
                \end{equation*}

                Let $Z_t = X_t^2 - 2t/n$ when $t≤τ$ and $X_τ^2 - 2τ/n$
                otherwise.  Then, for $t<τ$, we have
                \begin{align*}
                    \ExpCond{Z_{t+1}}{ℱ_t} 
                    &= \ExpCond{X_{t+1}^2 - 2(t+1)/n}{ℱ_t} 
                    \\&= X_t^2 + 2Y_t/n - 2(t+1)/n
                    \\&= X_t^2 - 2t/n + 2Y_t/n - 2/n
                    \\&= Z_t + 2Y_t/n - 2/n
                    \\&≤ Z_t.
                \end{align*}

                For $t≥τ$, $Z_{t+1} = Z_t$, so in either case 
                $Z_t$ has the submartingale property.

                We have previously established that $τ$ has bounded
                expectation, and it's easy to see that $Z_t$ has
                bounded step size.  So the optional stopping theorem
                applies to $Z_t$, and $\Exp{Z_0} ≤ \Exp{Z_τ}$.

                Let $X_0 = k$.  Then $\Exp{Z_0} = k^2$, 
                and $\Exp{Z_τ} = (k/n)⋅n^2 - 2\Exp{τ}/n = kn - 2\Exp{τ}/n$.
                But then
                \begin{align*}
                    k^2 &≤ kn - 2\Exp{τ}/n
                    \intertext{which gives}
                    \Exp{τ}
                    &≤ \frac{kn - k^2}{2/n}
                    \\&= \frac{kn^2 - k^2n}{2}.
                    \intertext{This is maximized at $k=\floor{n/2}$,
                    giving}
                    \Exp{τ}
                    &≤ \frac{\floor{n/2}⋅n^2 - (\floor{n/2})^2 ⋅ n}{2}.
                \end{align*}

                For even $n$, this is just $n^3/4 - n^3/8 = n^3/8$.

                For odd $n$, this is $(n-1)n^2/4 - (n-1)^2 n/8 = n^3/8
                - n/8$.  So there is a slight win for odd $n$ from not
                being able to start with an exact half-and-half split.

                To show that this bound applies in the worst case,
                observe that if we start with have contiguous regions
                of $k$ ones and $n-k$ zeros in $A_t$, then (a) $Y_t =
                1$, and (b) the two-region property is preserved in
                $A_{t+1}$.  In this case, for $t < τ$ it holds that
                $\ExpCond{Z_{t+1}}{ℱ_t} = Z_t + 2Y_t/n - 2/n
                = Z_t$, so $Z_t$ is a martingale, and thus
                $\Exp{τ} = \frac{kn^2-k^2n}{2}$.  This shows that the
                initial state with $\floor{n/2}$ consecutive zeros and
                $\ceil{n/2}$ consecutive ones (or vice versa) gives
                the claimed worst-case time.
        \end{enumerate}

    \subsection{Shuffling a two-dimensional array}

    A programmer working under tight time constraints needs to write a
    procedure to shuffle the elements of an $n×n$ array, so
    that all $(n^2)!$ permutations are equally likely.  Some quick
    Googling suggests that this can be reduced to shuffling a
    one-dimensional array, for which the programmer's favorite
    language provides a convenient library routine that runs in time
    linear in the size of the array.  Unfortunately,
    the programmer doesn't read the next paragraph about converting
    the 2-d array to a 1-d array first, and instead decides to
    pick one of the $2n$ rows or columns uniformly at random at each
    step, and call the 1-d shuffler on this row or column.

    Let $A^t$ be the state of the array after $t$ steps, where each
    step shuffles one row or column, and let $B$
    be a random variable over permutations of the original array
    state that has a uniform
    distribution.  Show that the 2-d shuffling procedure above is
    asympotically worse than the direct approach,
    by showing that
    there is some $f(n) = ω(n)$ such that
    after $f(n)$ steps, $d_{TV}(A^t, B) = 1-o(1)$.\footnote{I've been
        getting some questions about what this means, so here is an
        attempt to translate it into English.

        Recall that $f(n)$ is $ω(g(n))$ if $\lim_{n→∞}
        \frac{f(n)}{g(n)}$ goes to infinity, and
        $f(n)$ is $o(g(n))$ if 
        $\lim_{n→∞} \frac{f(n)}{g(n)}$ goes to zero.

        The problem is asking you to show that there is some $f(n)$
        that is more than a constant times $n$, such that the total
        variation distance between $A^{f(n)}$ becomes arbitrarily
        close to $1$ for sufficiently large $n$.

        So for example, if you showed that at $t=n^4$, $d_{TV}(A^t,B)
        ≥ 1-\frac{1}{\log^2 n}$, that would demonstrate the claim,
        because $\lim_{n→∞} \frac{n^2}{n}$ goes to infinity and 
        $\lim_{n→∞} \frac{1/\log n}{1} = 0$.  These functions are, of
        course, for
        illustration only.  The actual process might or might
    not converge by time $n^4$.)}

        \subsubsection*{Solution}

             Consider the $n$ diagonal elements in
                positions $A_{ii}$.  For each such element, there is
                a $1/n$ chance at each step that its row or column is
                chosen.  The time until every diagonal node is picked
                at least once maps to the coupon collector
                problem, which means that it is $Ω(n \log n)$ with
                high probability using standard concentration bounds.

                Let $C$ be the event that there is at least one
                diagonal element that is in its original position.
                If there is some diagonal node that has not moved, $C$
                holds; so with probability $1-o(1)$, $A^t$ holds at some
                time $t$ that is $Θ(n \log n) = ω(n)$.  But by the union bound,
                $C$ holds in $B$ with probability at most
                $n⋅n^{-2} = 1/n$.  So the difference between the
                probability of $C$ in $A^t$ and $B$ is at least
                $1-o(1)-1/n = 1-o(1)$.

\section{Assignment 6: due Wednesday, 2014-12-03, at 17:00}

    \subsection{Sampling colorings on a cycle}

    Devise a Las Vegas algorithm for sampling $3$-colorings of a cycle.

    Given $n>1$ nodes in a cycle,
    your algorithm should return a random coloring $X$ of the nodes
    with the usual constraint that no edge should have the same color
    on both endpoints.  Your algorithm should generate all possible
    colorings with equal probability,
    and run in time polynomial in $n$ on average.

        \subsubsection*{Solution}

        Since it's a Las Vegas algorithm, Markov chain Monte Carlo is
        not going to help us here.  So we can set aside couplings and
        conductance and just go straight for generating a solution.

        First, let's show that we can generate uniform colorings of an
        $n$-node line in linear time.  Let $X_i$ be the color of node
        $i$, where $0 ≤ i < n$.  Choose $X_0$ uniformly at random from
        all three colors;
        then for each $i > 0$, choose $X_i$ uniformly at random from
        the two colors not chosen for $X_{i-1}$.  Given a coloring,
        we can show by induction on $i$ that it can be generated by
        this process, and because each choice is uniform and each
        coloring is generated only once, we get all $3⋅2^{n-1}$
        colorings of the line with equal probability.

        Now we try hooking $X_{n-1}$ to $X_0$.  If $X_{n-1} = X_0$,
        then we don't have a cycle coloring, and have to start over.
        The probability that this event occurs is at most $1/2$,
        because for every path coloring with $X_{n-1} = X_0$, there is
        another coloring where we replace $X_{n-1}$ with a color not
        equal to $X_{n-2}$ or $X_0$.  So after at most $2$ attempts on
        average we get a good cycle coloring.  This gives a total
        expected cost of $O(n)$.

    \subsection{A hedging problem}

    Suppose that you own a portfolio of $n$ investment vehicles
    numbered $1,
    \dots n$, where investment $i$ pays out $a_{ij} ∈
    \Set{-1,+1}$ at time $j$, where $0 < j ≤ m$.  You have carefully
    chosen these investments so that your total payout $∑_{i=1}^n
    a_{ij}$ for any time $j$ is zero, eliminating all risk.
    
    Unfortunately, in doing so you
    have run afoul of securities regulators, who demand that you sell
    off half of your holdings—a demand that, fortunately, you
    anticipated by making $n$ even.

    This will leave you with a subset $S$ consisting of $n/2$ of the
    $a_i$, and your net worth at time $t$ will now be 
    $w_t = ∑_{j=1}^{t} ∑_{i ∈ S} a_{ij}$.  If your net worth drops too
    low at any time $t$, all your worldly goods will be repossessed,
    and you will have nothing but your talent for randomized
    algorithms to fall back on.

    \textbf{Note: an earlier version of this problem demanded a
    tighter bound.}

    Show that when $n$ and $m$ are sufficiently large, it is always
    possible to choose a subset $S$ of size
    $n/2$ so that $w_t ≥ -m √{n \ln nm}$
    for all $0 < t ≤ m$, and
    give an algorithm that finds such a subset in time polynomial in
    $n$ and $m$ on average.

        \subsubsection*{Solution}

        Suppose that we flip a coin independently to choose whether to
        include each investment $i$.  There are two bad things that
        can happen:
        \begin{enumerate}
            \item We lose too much at some time $t$ from the
                investments the coin chooses.
            \item We don't get exactly $n/2$ heads.
        \end{enumerate}

        If we can show that the sum of the probabilities of these bad
        events is less than $1$, we get the existence proof we need.
        If we can show that it is enough less than $1$, we also get an
        algorithm, because we can test in time $O(nm)$ if a particular
        choice works.

        Let $X_i$ be the indicator variable for the event that we
        include investment $i$.  Then

        \begin{align*}
            w_t
            &= ∑_{i=1}^{n} \parens*{ X_i ∑_{j=1}^{t} a_{ij} }
            \\&= ∑_{i=1}^{n} \parens*{ \parens*{\frac{1}{2} +
    \parens*{X_i - \frac{1}{2}}} ∑_{j=1}^{t} a_{ij} }
            \\&= \frac{1}{2} ∑_{j=1}^{t} ∑_{i=1}^{n} a_{ij}
                 + ∑_{i=1}^{n} \parens*{X_i - \frac{1}{2}}
                 \parens*{∑_{i=1}^{t} a_{ij}}
            \\&= ∑_{i=1}^{n} \parens*{X_i - \frac{1}{2}} \parens*{∑_{i=1}^{t} a_{ij}}.
        \end{align*}

        Because $\abs*{X_i - \frac{1}{2}}$ is always $\frac{1}{2}$,
        $\Exp{X_i - \frac{1}{2}} = 0$,
        and each $a_{ij}$ is $±1$, each term in the outermost sum
        is a zero-mean random variable that
        satisfies $\abs*{\parens*{X_i-\frac{1}{2}} ∑_{j=1}^{t} a_{ij}}
        ≤ \frac{t}{2} ≤ \frac{m}{2}$.  So Hoeffding's inequality
        says
        \begin{align*}
            \Prob{w_t - \Exp{w_t} < m √{n \ln n m}}
            &≤ e^{-m^2 n \ln n m / (2 n (m/2)^2)}
            \\&= e^{- 2 \ln n m}
            \\&= (nm)^{-2}.
        \end{align*}

        Summing over all $t$, the probability that this bound is
        violated for any $t$ is at most $\frac{1}{n^2m}$.

        For the second source of error, we have $\Prob{∑_{i=1}^{n} X_i
        ≠ n/2} = 1 - \binom{n}{2} / 2^n = 1 - Θ(1/√{n})$.  So the
        total probability that the random assignment fails is bounded
        by $1-Θ(1/√{n})+1/n$, giving a probability that it
        succeeds of at least $Θ(1/√{n}) - 1/(n^2m) = Θ(√{n})$.  It
        follows that generating and testing random assignments gives
        an assignment with the desired characteristics after
        $Θ(√{n})$ trials on average, giving a total expected cost
        of $Θ(n^{3/2} m)$.

\section{Final exam}

Write your answers in the blue book(s).  Justify your answers.  Work
alone.  Do not use any notes or books.  

There are four problems on this exam, each worth 20
points, for a total of 80 points.
You have approximately three hours to complete this
exam.

\subsection{Double records (20 points)}

Let $A[1\dots n]$ be an array holding the integers $1 \dots n$ in random
order, with all $n!$ permutations of the elements of $A$ equally likely.

Call $A[k]$, where $1 ≤ k ≤ n$, a \concept{record} if $A[j] < A[k]$ for all $j < k$.

Call $A[k]$, where $2 ≤ k ≤ n$, a \index{record!double}\concept{double record} if both
$A[k-1]$ and $A[k]$ are records.

Give an asymptotic (big-$Θ$) bound on the expected number of double
records in $A$.

    \subsubsection*{Solution}

    Suppose we condition on a particular set of $k$ elements appearing
    in positions $A[1]$ through $A[k]$.  By symmetry, all $k!$
    permutations of these elements are equally likely.  Putting the
    largest two elements in $A[k-1]$ and $A[k]$ leaves $(k-2)!$
    choices for the remaining elements, giving a probability of a
    double record at $k$ of exactly $\frac{(k-2)!}{k!} =
    \frac{1}{k(k-1)}$.

    Applying linearity of expectation gives a total expected number of
    double records of
    \begin{align*}
        ∑_{k=2}^{n} \frac{1}{k(k-1)}
        &≤ ∑_{k=2}^{n} \frac{1}{(k-1)^2}
        \\&≤ ∑_{k=2}^{∞} \frac{1}{(k-1)^2}
        \\&= ∑_{k=1}^{∞} \frac{1}{k^2}
        \\&= \frac{π^2}{6}
        \\&= O(1).
    \end{align*}

    Since the expected number of double records is at least $1/2 =
    Ω(1)$ for $n≥2$, this gives a tight asymptotic bound of $Θ(1)$.

    I liked seeing our old friend $π^2/6$ so much that I didn't notice
    an easier exact bound, which several people supplied in their
    solutions:
    \begin{align*}
        ∑_{k=2}^{n} \frac{1}{k(k-1)}
        &= ∑_{k=2}^{n} \parens*{\frac{1}{k-1} - \frac{1}{k}}
        \\&= ∑_{k=1}^{n-1} \frac{1}{k} - ∑_{k=2}^{n} \frac{1}{k}
        \\&= 1 - \frac{1}{n}.
        \\&= O(1).
    \end{align*}

\subsection{Hipster graphs (20 points)}

Consider the problem of labeling the vertices of a $3$-regular graph with labels
from $\Set{0,1}$ to maximize the number of \emph{happy} nodes, where a node
is \emph{happy} if its label is the opposite of the majority of its
three neighbors.

Give a \emph{deterministic} algorithm that takes a $3$-regular graph
as input and computes, in time
polynomial in the size of the graph, a labeling that makes at least
half of the nodes happy.

    \subsubsection*{Solution}

    This problem produced the widest range of solutions, including
    several very clever deterministic algorithms.  
    Here are some examples.

    \paragraph{Using the method of conditional probabilities}

    If we are allowed a randomized algorithm, it's easy to get at
    exactly half of the nodes on average: simply label each node
    independently and uniformly at random, observe that each node individually has a
    $1/2$ chance at happiness, and apply linearity of expectation.

    To turn this into a deterministic algorithm, we'll apply the
    method of conditional expectations.  Start with an unlabeled
    graph.  At each step, pick a node and assign it a label that
    maximizes the expected number of happy nodes conditioned on the
    labeling so far, and assuming that all later nodes are labeled
    independently and uniformly at random.  We can compute this
    conditional expectation in linear time by computing the value for
    each node (there are only $3^4$ possible partial labelings of the node and
    its immediate neighbors, so computing the expected happiness of
    any particular node can be done in constant time by table lookup;
    to compute the table, we just enumerate all possible assignments
    to the unlabeled nodes).
    So in $O(n^2)$ time we get a labeling in which the number of happy
    nodes is at least the $n/2$ expected happy nodes we started with.

    With a bit of care, the cost can be reduced to linear: because
    each new labeled node only affects its own probability of
    happiness and those of its three neighbors, we can update the
    conditional expectation by just updating the values for those four
    nodes.  This gives $O(1)$ cost per step or $O(n)$ total.

    \paragraph{Using hill climbing}

    The following algorithm is an adaptation of the solution of Rose
    Sloan, and demonstrates that it is in fact possible to make
    \emph{all} of the nodes happy in linear time.

    Start with an arbitrary labeling (say, all $0$).  At each step,
    choose an unhappy node and flip its label.  This reduces the
    number of monochromatic edges by at least $1$.  Because we have
    only $3n/2$ edges, we can repeat this process at most $3n/2$
    times before it terminates.
    But it only terminates when there are no unhappy nodes.

    To implement this in linear time, maintain a queue of all unhappy
    nodes.

\subsection{Storage allocation (20 points)}

Suppose that you are implementing a stack in an array, and you need to
figure out how much space to allocate for the array.  At time $0$, the
size of the stack is $X_0 = 0$.  At each subsequent time $t$, the user
flips an independent fair coin to choose whether to push onto the
stack ($X_t = X_{t-1} + 1$) or pop from the stack ($X_t = X_{t-1} -
1$).  The exception is that when the stack is empty, the user always
pushes.

You are told in advance that the stack will only be used for $n$ time
units.  Your job is to choose a size $s$ for the array so that it will
overflow at most half the time: $\Prob{\max_t X_t > s} < 1/2$.  As an asymptotic (big-$O$) function
of $n$, what is the smallest value of $s$ you can choose?

    \subsection*{Solution}

    We need two ideas here.  First, we'll show that $X_t - t^2$ is a
    martingale, despite the fact that $X_t$ by itself isn't.  Second,
    we'll use the idea of stopping $X_t$ when it hits $s+1$, creating a
    new martingale $Y_t = X_{t∧τ}^2 - (t∧τ)$ where $τ$ is the first
    time where $X_t = s$ (and $t∧τ$ is shorthand for $\min(t,τ)$).
    We can then apply Markov's inequality to $X_{n∧τ}$.

    To save time, we'll skip directly to showing that $Y_t$ is a
    martingale.  There are two cases:
    \begin{enumerate}
        \item If $t < τ$, then
            \begin{align*}
                \ExpCond{Y_{t+1}}{Y_t,t<τ}
                &= \frac{1}{2} \parens*{(X_t+1)^2 - (t+1)} +
                   \frac{1}{2} \parens*{(X_t-1)^2 - (t+1)}
                \\&= X_t^2 + 1 - (t+1)
                \\&= X_t^2 - t
                \\&= Y_t.
            \end{align*}
        \item If $t ≥ τ$, then $Y_{t+1} = Y_t$, so $\ExpCond{Y_{t+1}}{Y_t,t≥τ} = Y_t$.
    \end{enumerate}
    In either case the martingale property holds.

    It follows that $\Exp{Y_n} = \Exp{Y_0} = 0$, or $\Exp{X_{n∧τ}^2 -
    n} = 0$, giving $\Exp{X_{n∧τ}^2} = n$.  Now apply Markov's
    inequality:
    \begin{align*}
        \Prob{\max X_t > s}
        &= \Prob{X_{n∧τ} ≥ s+1}
        \\&= \Prob{X_{n∧τ}^2 ≥ (s+1)^2}
        \\&≤ \frac{\Exp{X_{n∧τ}^2}}{(s+1)^2}
        \\&= \frac{n}{(√{2n} + 1)^2}
        \\&< \frac{n}{2n}
        \\&= 1/2.
    \end{align*}

    So $s = √{2n} = O(√{n})$ is enough.

\subsection{Fault detectors in a grid (20 points)}

A processing plant for rendering discarded final exam problems
harmless consists of $n^2$ processing units arranged in an $n×n$ grid
with coordinates $(i,j)$ each in the range $1$ through $n$.
We would like to monitor these processing units to make sure that they
are working correctly, and have access to a supply of monitors that
will detect failures.  Each monitor is placed at some position
$(i,j)$, and will detect failures at any of the five positions $(i,j)$,
$(i-1,j)$, $(i+1,j)$, $(i,j-1)$, and $(i,j+1)$ that are within the
bounds of the grid.  This plus-shaped detection range is awkward
enough that the engineer designing the system has given up on figuring
out how to pack the detectors properly, and instead places a detector
at each grid location with independent probability $1/4$.  

The engineer reasons that since a typical monitor covers $5$ grid
locations, using $n^2/4$ monitors on average should cover $(5/4)n^2$
locations, with the extra monitors adding a little bit of safety to
deal with bad random choices.  So few if any processing units should
escape.

\begin{enumerate}
    \item Compute the exact expected number of processing units that are not
        within range of any monitor, as a function of $n$.  You may
        assume $n > 0$.
    \item Show that for any fixed $c$, 
        the actual number of unmonitored processing units
        is within $O(n √{\log n})$ of the expected number with
        probability at least $1-n^{-c}$.
\end{enumerate}

    \subsubsection*{Solution}

    \begin{enumerate}
        \item This part is actually more annoying, because we have to
            deal with nodes on the edges.  There are three classes of nodes:
            \begin{enumerate}
                \item The four corner nodes.  To be unmonitored, each
                    corner node needs to have no monitor in its own
                    location or either of the two adjacent locations.
                    This event occurs with probability $(3/4)^3$.
                \item The $4(n-2)$ edge nodes.  These have three
                    neighbors in the grid, so they are unmonitored
                    with probability $(3/4)^4$.
                \item The $(n-2)^2$ interior nodes.  These are each
                    unmonitored with probability $(3/4)^5$.
            \end{enumerate}

            Adding these cases up gives a total expected number of
            unmonitored nodes of
            \begin{equation}
                \label{eq-final-exam-problem-unmonitored-nodes}
                4⋅(3/4)^3 + 4(n-2)⋅(3/4)^4 + (n-2)^2⋅(3/4)^5 
                = \frac{243}{1024} n^2 - \frac{81}{32} n + \frac{189}{64}.
            \end{equation}

            For $n=1$, this analysis breaks down; instead, we can
            calculate directly that the sole node in the grid has a
            $3/4$ chance of being unmonitored.

            Leaving the left-hand side of \eqref{eq-final-exam-problem-unmonitored-nodes} 
            in the original form is probably a
            good idea for understanding how the result works, but the
            right-hand side demonstrates that this
            strategy leaves slightly less than a quarter of the
            processing units uncovered on average.

        \item 
            Let $X_{ij}$ be the indicator for a monitor
            at position $(i,j)$.  
            Recall that we have assumed that these variables are
            independent.

            Let $S = f(X_{11}, \dots, X_{nn}$)
            compute the number of uncovered processing units.  Then
            changing a single $X_{ij}$ changes the value of $f$ by at
            most $5$.  So we can apply McDiarmid's inequality to get
            \begin{align*}
                \Prob{\abs*{S-\Exp{S}} ≥ t} 
                &≤ 2e^{-2t^2/\parens*{∑c_{ij}^2}}
                &= 2e^{-2t^2/\parens*{n^2⋅5^2}}
                &= 2e^{-2t^2/(25n^2)}.
            \end{align*}

            (The actual bound is slightly better, because we are
            overestimating $c_{ij}$ for boundary nodes.)

            Set this to $n^{-c}$ and solve for $t = n √{(25/2) (c \ln n + \ln 2)} = O(n √{\log n})$.  Then the bound
            becomes $2 e^{- c \ln n - \ln 2} = n^{-c}$, as desired.
    \end{enumerate}

\section{Alternative final exam}

Write your answers in the blue book(s).  Justify your answers.  Work
alone.  Do not use any notes or books.  

There are four problems on this exam, each worth 20
points, for a total of 80 points.
You have approximately three hours to complete this
exam.

\section{An approximate counter (20 points)}

Consider the following approximate counting algorithm.
Start with $n$ people standing up, where $n$ is unknown.  In each round, each
person flips an independent fair coin and sits down if it comes up
heads.  Let $R$ be the number of rounds until no one is standing up;
for example, if everybody sits down in the first round, $R=1$.

Show that for any fixed $n > 0$, $\Prob{2^{R-2} > n} ≤ 1/2$; in other words, if we estimate
that $n$ is approximately $2^{R-2}$, this will be an overestimate 
at most half the time.

    \subsection*{Solution}

    The expected number of people standing after $k$ rounds is
    $n⋅2^{-k}$, so after $\ceil{\lg n} + 1$ rounds, the expected
    number drops to at most $1/2$ and the probability that someone is
    still standing is at most $1/2$.  Suppose that $2^{R-2} > n$.
    Then $R-2 > \lg n$ and someone is still standing after $2+\lg n$
    rounds.  But $2+\lg n > 1 + \ceil{\lg n}$ so someone is still
    standing after $1+\ceil{\lg n}$ rounds, which occurs with
    probability at most $1/2$.

    Check: For $n=3$, we get an overestimate if $R≥4$, which occurs
    with probability bounded by $3/8$.

\section{Balls in bins}

Suppose we throw $n$ balls in $n$ bins, independently and uniformly at
random.  As a function of $n$, what is the variance of the number of empty bins?
Give a closed-form expression if possible.

    \subsection*{Solution}

    This is just calculation.  Let $X_i$ be the indicator for the
    event that the $i$-th bin is empty, and let $S = ∑ X_i$ be the
    total number of empty bins.  Then:
    \begin{align*}
        \Var{S}
        &= \Var{∑_i X_i}
        \\&= ∑_i \Var{X_i} + 2 ∑_{i<j} \Cov{X_i}{X_j}
        \\&= n \parens*{(1-1/n)^n - (1-1/n)^{2n}} 
        + 2 \binom{n}{2} \parens*{(1-2/n)^n - (1-1/n)^{2n}}.
    \end{align*}

    I don't think there is much more that can be done to simplify
    this, although for large $n$ this will tend to $n(e^{-1}
    - e^{-2})$.

\section{Finding a maximum}

Here is a randomized algorithm for finding the maximum element of an
array $A$ with $n$ elements.  At each step, choose one of the elements
of the array uniformly at random with replacement.  After $t$ such steps, report the
largest element seen.

Let $c>0$ be a fixed constant. 
In asymptotic terms, how large should $t$ be to give a probability of
at most $n^{-c}$ that this algorithm fails to return the correct
value?
(Assume that all the elements of the array are distinct.)

    \subsection*{Solution}

    If we observe the maximum value, we'll return it.  The probability
    that we fail to observe it after $t$ steps is $(1-1/n)^t ≤
    e^{-t/n}$.  So solve $e^{-t/n} ≤ n^{-c}$ for $t$ to get $t ≥ cn
    \ln n$ or $t = Θ(n \log n)$.

\section{Flipping coins}

Suppose that we flip independent fair coins until a coin comes up on the same side as
the coin flipped two steps earlier.  So, for example, if we flip
\coinFlips{HTH} we stop after three flips because the last heads is equal to the
coin-flip two steps before, and if we flip \coinFlips{HHTTT} we stop
after five flips because the last tails is equal to the tails two
steps before.

How many coins do we flip on average using this rule?

    \subsection*{Solution}

    We never stop on the first two flips.  After this, each coin has
    an independent $1/2$ probability of being equal to the coin two
    steps before.  So the expected number of additional flips is
    geometrically distributed with mean $2$.  This gives a total of
    $4$ flips on average before we stop.

\chapter{Sample assignments from Spring 2013}

\section{Assignment 1: due Wednesday, 2013-01-30, at 17:00}

\subsection{Bureaucratic part}

Send me email!  My address is
\mailto{james.aspnes@gmail.com}.

In your message, include:

\begin{enumerate}
\item Your name.
\item Your status: whether you are an undergraduate, grad student, auditor, etc.
\item Anything else you'd like to say.
\end{enumerate}

(You will not be graded on the bureaucratic part, but you should do it anyway.)

\subsection{Balls in bins}

Throw $m$ balls independently and uniformly at random into $n$ bins
labeled $1\dots n$.  What is the expected number of positions $i < n$ such that
bin $i$ and bin $i+1$ are both empty?

\subsubsection*{Solution}

If we can figure out the probability that bins $i$ and $i+1$ are both
empty for some particular $i$, then by symmetry and 
linearity of expectation we can
just multiply by $n-1$ to get the full answer.

For bins $i$ and $i+1$ to be empty, every ball must choose another
bin.  This occurs with probability $(1-2/n)^m$.  The full answer is
thus $n (1-2/n)^m$, or approximately $n e^{-2m/n}$ when $n$ is large.

\subsection{A labeled graph}

Suppose you are given a graph $G=(V,E)$, where $\card*{V} = n$, and you want
to assign labels to each vertex such that the sum of the labels of
each vertex and its neighbors modulo $n+1$ is nonzero.  
Consider the
na\"ive randomized algorithm that assigns a label in the range $0\dots
n$ to each vertex independently and uniformly at random and then tries again if the
resulting labeling doesn't work.  Show that this algorithm
will find a correct labeling in
time polynomial in $n$ on average.

\subsubsection*{Solution}

We'll use the law of total probability.  First observe that the
probability that a random labeling yields a zero sum for any single
vertex and its neighbors is exactly $1/(n+1)$; the easiest way to see
this is that after conditioning on the values of the neighbors, there
is only one value in $n+1$ that can be assigned to the vertex itself
to cause a failure.  Now sum this probability over all $n$ vertices to
get a probability of failure of at most $n/(n+1)$.  It follows that
after $n+1$ attempts on average (each of which takes $O(n^2)$ time to
check all the neighborhood sums), the algorithm will find a good
labeling, giving a total expected time of $O(n^3)$.

\subsection{Negative progress}

An algorithm has the property that if it has already run for $n$
steps, it runs for an additional $n+1$ steps on average.  Formally,
let $T ≥ 0$ be the random variable representing the running time of the
algorithm, then
\begin{align}
    \ExpCond{T}{T≥ n} &=
    2n+1.\label{eq-negative-progress-conditional-expectation}
\end{align}

For each $n ≥ 0$, what is the conditional probability $\ProbCond{T=n}{T≥ n}$ that the
algorithm stops just after its $n$-th step?

\subsubsection*{Solution}

Expand \eqref{eq-negative-progress-conditional-expectation} using the
definition of conditional expectation to get

\begin{align}
    2n+1
    &= ∑_{x=0}^{∞} x \ProbCond{T=x}{T≥ n} \nonumber\\
    &= ∑_{x=0}^{∞} x \frac{\Prob{T=x ∧ T≥ n}}{\Prob{T≥
    n}} \nonumber\\
    &= \frac{1}{\Prob{T≥ n}} ∑_{x=n}^{∞} x \Prob{T=x},
    \nonumber
    \intertext{which we can rearrange to get}
    ∑_{x=n}^{∞} x \Prob{T = x} &= (2n+1) \Prob{T ≥ n},
    \label{eq-negative-progress-sum}
\end{align}
provided $\Prob{T ≥ n}$ is nonzero.  We can justify this assumption
by observing that (a) it holds for $n=0$, because $T ≥ 0$ always;
and (b) if there is some $n > 0$ such that $\Prob{T≥ n}= 0$,
then $\ExpCond{T}{T≥ n-1} = n-1$, contradicting
\eqref{eq-negative-progress-conditional-expectation}.

Substituting $n+1$ into \eqref{eq-negative-progress-sum} and
subtracting from the original gives the equation
\begin{align*}
    n \Prob{T=n} &=
    ∑_{x=n}^{∞} x \Prob{T = x} 
    - ∑_{x=n+1}^{∞} x \Prob{T = x} \\
    &= (2n+1) \Prob{T ≥ n} - (2n+3) \Prob{T ≥ n+1} \\
    &= (2n+1) \Prob{T = n} + (2n+1) \Prob{T ≥ n+1} - (2n+3) \Prob{T ≥ n+1} \\
    &= (2n+1) \Prob{T = n} - 2 \Prob{T ≥ n+1}.
\end{align*}

Since we are looking for $\ProbCond{T=n}{T≥ n} = \Prob{T=n}/\Prob{T≥ n}$,
having an equation involving $\Prob{T≥ n+1}$ is a bit annoying.  But
we can borrow a bit of $\Prob{T=n}$ from the other terms to make it
work:

\begin{align*}
    n \Prob{T=n}
    &= (2n+1) \Prob{T=n} - 2 \Prob{T ≥ n+1} \\
    &= (2n+3) \Prob{T=n} - 2 \Prob{T=n} - 2 \Prob{T ≥ n+1} \\
    &= (2n+3) \Prob{T=n} - 2 \Prob{T≥ n}.
\end{align*}

A little bit of algebra turns this into
\begin{align*}
    \ProbCond{T=n}{T≥ n}
    &= \frac{\Prob{T=n}}{\Prob{T ≥ n}} 
    = \frac{2}{n+3}.
\end{align*}

\section{Assignment 2: due Thursday, 2013-02-14, at 17:00}

\subsection{A local load-balancing algorithm}

Suppose that we are trying to balance $n$ jobs evenly between two
machines.  Job 1 chooses the left or right machine with equal
probability.
For $i > 1$, job $i$ chooses the same machine as job $i-1$ with probability $p$, and
chooses the other machine with probability $1-p$.  This process
continues until every job chooses a machine.  We want to estimate how
imbalanced the jobs are at the end of this process.

Let $X_i$ be $+1$ if the $i$-th job chooses the left machine and $-1$
if it chooses the right.  Then $S = ∑_{i=1}^{n} X_i$ is the
difference between the number of jobs that choose the left machine and
the number that choose the right.  By symmetry, the expectation of $S$
is zero.  What is the variance of $S$ as a function of $p$ and $n$?

\subsubsection*{Solution}

To compute the variance, we'll use \eqref{eq-variance-sum-asymmetric},
which says that $\Var{∑_i X_i} = ∑_i \Var{X_i} + 2 ∑_{i < j}
\Cov{X_i}{X_j}$.

Recall that $\Cov{X_i}{X_j} = \Exp{X_i X_j} - \Exp{X_i} \Exp{X_j}$.  Since
the last term is $0$ (symmetry again), we just need to figure out
$\Exp{X_i X_j}$ for all $i ≤ j$ (the $i=j$ case gets us $\Var{X_i}$).

First, let's compute $\ExpCond{X_j = 1}{X_i = 1}$.
It's easiest to do this starting with the $j=i$ case:
$\ExpCond{X_{i} }{ X_i = 1} = 1$.
For larger $j$, compute
\begin{align}
    \ExpCond{X_{j} }{ X_{j-1}}
    &= p X_{j-1} + (1-p) (-X_{j-1}) \nonumber \\
    &= (2p-1) X_{j-1}.\nonumber
    \intertext{It follows that}
    \ExpCond{X_j}{X_i=1}
    &= \ExpCond{(2p-1)X_{j-1}}{X_i=1} \nonumber \\
    &= (2p-1) \ExpCond{X_{j-1}}{X_i=1}.
    \label{eq-local-load-balancing-assignment-conditional-expectation}
\end{align}
The solution to this recurrence is $\ExpCond{X_j}{X_i=1} = (2p-1)^{j-i}$.

We next have
\begin{align*}
    \Cov{X_i}{X_j}
&= \Exp{X_i X_j} \\
&= \ExpCond{X_i X_j }{ X_i = 1} \Prob{X_i = 1}
+ \ExpCond{X_i X_j}{X_i = -1} \Prob{X_i = -1} \\
&= \frac{1}{2} \ExpCond{X_j }{ X_i = 1} + \frac{1}{2} \ExpCond{-X_j }{ X_i = -1} \\
&= \frac{1}{2} \ExpCond{X_j }{ X_i = 1} + \frac{1}{2} \ExpCond{X_j }{ X_i = 1} \\
&= \ExpCond{X_j }{ X_i = 1} \\
&= (2p-1)^{j-i}
\end{align*}
as calculated in
\eqref{eq-local-load-balancing-assignment-conditional-expectation}.

So now we just need to evaluate the horrible sum.
\begin{align}
    ∑_{i} \Var{X_i} + 2 ∑_{i<j} \Cov{X_i}{X_j}
    &= n + 2 ∑_{i=1}^{n} ∑_{j=i+1}^{n} (2p-1)^{j-i} \nonumber \\
    &= n + 2 ∑_{i=1}^{n} ∑_{k=1}^{n-i} (2p-1)^{k} \nonumber \\
    &= n + 2 ∑_{i=1}^{n} \frac{(2p-1) - (2p-1)^{n-i+1}}{1-(2p-1)}
    \nonumber \\
    &= n + \frac{n (2p-1)}{1-p} - \frac{1}{1-p} ∑_{m=1}^{n} (2p-1)^{m}
    \nonumber \\
    &= n + \frac{n (2p-1)}{1-p} - \frac{(2p-1)
- (2p-1)^{n+1}}{2(1-p)^2}.
\label{eq-local-load-balancing-assignment-variance}
\end{align}

This covers all but the $p=1$ case, for which the geometric series
formula fails.  Here we can compute directly that $\Var{S} = n^2$,
since $S$ will be $\pm n$ with equal probability.  

For smaller values
of $p$, plotting
\eqref{eq-local-load-balancing-assignment-variance} shows the
variance increasing smoothly starting at $0$ (for even $n$) or $1$
(for odd $n$) at $p=0$ to $n^2$ in the limit as $p$ goes to $1$, with
an interesting intermediate case of $n$ at $p=1/2$, where all terms
but the first vanish.  This makes a certain intuitive sense: when
$p=0$, the processes alternative which machine they take, which gives
an even split for even $n$ and a discrepancy of $\pm 1$ for odd $n$;
when $p=1/2$, the processes choose machines independently, giving
variance $n$; and for
$p=1$, the processes all choose the same machine, giving $n^2$.

\subsection{An assignment problem}

Here is an algorithm for placing $n$ balls into $n$ bins.  For each
ball, we first select two bins uniformly at random without
replacement.  If at least one of the chosen bins is unoccupied, the
ball is placed in the empty bin at no cost.  If both chosen bins are
occupied, we execute an expensive parallel scan operation to find an
empty bin and place the ball there.

\begin{enumerate}
    \item Compute the exact value of the expected number of scan
        operations.
    \item Let $c >0$.  Show that the absolute value of the
        difference between the actual number of scan operations
        and the expected number is at most
        $O(√{cn \log n})$ with probability at least
        $1-n^{-c}$.
\end{enumerate}

\subsubsection*{Solution}

\begin{enumerate}
    \item Number the balls $1$ to $n$.  For ball $i$, there are $i-1$
        bins already occupied, 
        giving a probability of 
        $\left(\frac{i-1}{n}\right)\left(\frac{i-2}{n-1}\right)$
        that we choose an occupied bin on both attempts and incur a scan.  Summing over
        all $i$ gives us that the expected number of scans is:
        \begin{align*}
            ∑_{i=1}^{n} \left(\frac{i-1}{n}\right)
            \left(\frac{i-2}{n-1}\right)
            &= \frac{1}{n(n-1)} ∑_{i=1}^{n-1} (i^2-i) \\
            &= \frac{1}{n(n-1)} \left(\frac{(n-1)n(2n-1)}{6} -
            \frac{(n-1)n)}{2}\right) \\
            &= \frac{(2n-1)}{6} -
            \frac{1}{2} \\
            &\frac{n-2}{3},
        \end{align*}
        provided $n ≥ 2$.  For $n < 2$, we incur no scans.
    \item It's tempting to go after this using Chernoff's inequality,
        but in this case Hoeffding gives a better bound.  Let $S$ be
        the number of scans.  Then $S$ is the 
        sum of $n$ independent Bernoulli random
        variables, so~\ref{eq-Hoeffdings-inequality-asymmetric} says
        that $\Prob{\abs*{S-\Exp{S}} ≥ t} ≤ 2e^{-2t^2/n}$.
        Now let $t=√{c n \ln n} = O(√{c n \log n})$
        to make the right-hand side
        $2 n^{-2c} ≤ n^{-c}$ for sufficiently large $n$.
\end{enumerate}

\subsection{Detecting excessive collusion}

Suppose you have $n$ students in some unnamed Ivy League university's
\emph{Introduction to Congress} class,
and each generates a random $\pm 1$ vote, with both outcomes
having equal probability.  It is expected that members of the same
final club will vote together, so it may be that many groups of up to $k$
students each will all vote the same way (flipping a single coin to
determine the vote of all students in the group, with the coins for
different groups being independent).  However, there may also be a
much larger conspiracy of exactly 
$m$ students who all vote the same way (again
based on a single independent fair coin),
in violation of academic honesty regulations.

Let $c > 0$.
How large must $m$ be in asymptotic terms as a function of $n$, $k$, and $c$
so that the existence of a conspiracy can be
detected solely by looking at the total vote,
where the probability of error (either incorrectly claiming a conspiracy
when none exists or incorrectly claiming no conspiracy when one exists)
is at most $n^{-c}$?

    \subsubsection*{Solution}

    Let $S$ be the total vote.
    The intuition here is that if there is no conspiracy, $S$ is
    concentrated around $0$, and if there is a conspiracy, $S$ is
    concentrated around $\pm m$.  So if $m$ is sufficiently large and
    $\abs*{S} ≥ m/2$, we can
    reasonably guess that there is a conspiracy.

    We need to prove two bounds: first, that the probability that we
    see $\abs*{S} ≥ m/2$ when there is no conspiracy is small, and
    second, that the probability that we see $\abs*{S} < m/2$ when
    there is a conspiracy is large.

    For the first case, let $X_i$ be the total vote cast by the $i$-th
    group.  This will be $\pm n_i$ with equal probability, where $n_i
    ≤ k$ is the size of the group.  This gives $\Exp{X_i} = 0$.
    We also have that $∑ n_i = n$.

    Because the $X_i$ are all bounded, we can use Hoeffding's
    inequality \eqref{eq-Hoeffdings-inequality}, so long as we can
    compute an upper bound on $∑ n_i^2$.  Here we use the fact that
    $∑ n_i^2$ is maximized subject to $0 ≤ n_i ≤ k$ and $∑
    n_i = 0$ by setting as many $n_i$ as possible to $k$; this follows
    from convexity of $x \mapsto x^2$.\footnote{The easy way to see
        this is that if $f$ is strictly convex, then $f'$ is increasing.  So if
        $0 < n_i ≤ n_j < k$, increasing $n_j$ by $ε$ while
        decreasing $n_i$ by $ε$ leaves $∑ n_i$ unchanged
        while increasing $∑ f(n_i)$ by 
        $ε(f'(n_j)-f'(n_i)) + O(ε^2)$, which will be
        positive when
        $ε$ is small enough.  So at any global maximum, we
        must have that at least one of $n_i$ or $n_j$ equals $0$ or
    $k$ for any $i ≠ j$.}
    We thus have
    \begin{align*}
        ∑ n_i^2 
        &≤ \floor{n/k} k^2 + (n \bmod k)^2 \\
        &≤ \floor{n/k} k^2 + (n \bmod k) k \\
        &≤ (n/k) k^2 \\
        &= nk.
    \end{align*}

    Now apply Hoeffding's inequality to get
    \begin{align*}
        \Prob{\abs*{S} ≥ m/2} ≤ 2 e^{-(m/2)^2/2nk}.
    \end{align*}

    We want to set $m$ so that the right-hand side is less than
    $n^{-c}$.  Taking logs as usual gives
    \begin{align*}
        \ln 2 - m^2/8nk &≤ -c \ln n,
        \intertext{so the desired bound holds when}
        m &≥ √{8nk(c\ln n + \ln 2)} \\
          &= Ω(√{c k n \log n}).
    \end{align*}

    For the second case, repeat the above analysis on the $n-m$ votes except
    the $\pm m$ from the conspiracy.  Again we get that if
    $m=Ω(√{ckn \log n})$, the probability that these votes
    exceed $m/2$ is bounded by $n^{-c}$.  So in both cases $m =
    Ω(√{ckn \log n})$ is enough.

\section{Assignment 3: due Wednesday, 2013-02-27, at 17:00} 

    \subsection{Going bowling}

    For your crimes, you are sentenced to play $n$ frames of
    \concept{bowling}, a
    game that involves knocking down pins with a heavy ball, which we
    are mostly interested in because of its complicated scoring
    system.
    
    \newcommand{\BowlingSpare}{\mbox{\texttt{/}}}
    \newcommand{\BowlingStrike}{\mbox{\texttt{X}}}

    In each frame, your result may be any of the integers 
    $0$ through $9$, a
    \concept{spare} (marked as \BowlingSpare), or a \concept{strike}
    (marked as \BowlingStrike).  We can think of your result on the $i$-th
    frame as a random variable $X_i$.  Each result gives a base score
    $B_i$, which is equal to $X_i$ when $X_i ∈ \Set{0 \dots 9}$ and
    $10$ when $X_i ∈ \Set{\BowlingSpare, \BowlingStrike}$.  The actual
    score $Y_i$ for frame $i$ is the sum of the base scores for the frame and
    up to two subsequent frames, according to the rule:
    \begin{align*}
        Y_i &=
        \begin{cases}
            B_i & \text{when $X_i ∈ \Set{0 \dots 9}$,} \\
            B_i + B_{i+1} & \text{when $X_i = \BowlingSpare$, and} \\
  B_i + B_{i+1} + B_{i+2} & \text{when $X_i = \BowlingStrike$.}
        \end{cases}
    \end{align*}

    To ensure that $B_{i+2}$ makes sense even for $i=n$, assume that
    there exist random variables $X_{n+1}$ and $X_{n+2}$ and the
    corresponding $B_{n+1}$ and $B_{n+2}$.

    Suppose that the $X_i$ are independent (but not necessarily
    identically distributed).  Show that your final score
    $S = ∑_{i=1}^{n} Y_i$ is exponentially
    concentrated\footnote{This means ``more concentrated than you can
    show using Chebyshev's inequality.''} around its expected value.

        \subsubsection*{Solution}

        This is a job for McDiarmid's inequality
        \eqref{eq-McDiarmids-inequality}.  Observe that
        $S$ is a function of $X_1 \dots X_{n+2}$.
        We need to show that changing any one of the $X_i$ won't
        change this function by too much.

        From the description of the $Y_i$, we have that $X_i$ can
        affect any of $Y_{i-2}$ (if $X_{i-2} = \BowlingStrike$),
        $Y_{i-1}$ (if $X_{i-1} = \BowlingSpare$) and $Y_i$.
        We can get a crude bound by observing that each $Y_i$ ranges
        from $0$ to $30$, so changing $X_i$ can change $∑ Y_i$ by
        at most $\pm 90$, giving $c_i ≤ 90$. 
        A better bound can be obtained by observing
        that $X_i$ contributes only $B_i$ to each of $Y_{i-2}$ and
        $Y_{i-1}$, so changing $X_i$ can only change these values by
        up to $10$; this gives $c_i ≤ 50$. 
        An even more pedantic bound can be obtained by observing that
        $X_1$, $X_2$, $X_{n+1}$, and $X_{n+2}$ are all special cases,
        with $c_1 = 30$, $c_2 = 40$, $c_{n+1} = 20$, and $c_{n+2} =
        10$, respectively; these values can be obtained by detailed
        meditation on the rules above.

        We thus have 
        $∑_{i=1}^{n+2} c_i^2 = (n-2)50^2 + 30^2 + 40^2 + 20^2 +
        10^2 = 2500(n-2) + 3000 = 2500n - 2000$, assuming $n ≥ 2$.
        This gives $\Prob{\abs*{S - \Exp{S}} ≥ t} ≤
        \exp(-2t^2/(2500n - 2000))$, with the symmetric bound holding on
        the other side as well.

        For the standard game of bowling, with $n=10$, this bound
        starts to bite at $t = √{11500} \approx 107$, which is
        more than a third of
        the range between the minimum and maximum possible
        scores.  There's a lot of variance in bowling, but this looks
        like a pretty loose bound for players who don't
        throw a lot of strikes.  For large $n$, we get the usual bound
        of $O\left(√{n \log n}\right)$ with high probability: the averaging
        power of endless repetition eventually overcomes any slop in the
        constants.

    \subsection{Unbalanced treaps}

    Recall that a \concept{treap}
    (§\ref{section-treaps}) is only likely to be balanced if the
    sequence of insert and delete operations applied to it is
    independent of the priorities chosen by the algorithm.

    Suppose that we insert the keys $1$ through $n$ into a treap with random
    priorities as usual, but then allow the adversary to selectively
    delete whichever keys it wants to after observing the priorities
    assigned to each key.

        Show that there is an adversary strategy that produces a
            path in the treap after deletions that has expected length $Ω\left(√{n}\right)$.

    \subsubsection*{Solution}

        An easy way to do this is
            to produce a tree that consists of a single
            path, which we can do by arranging that the
            remaining keys have priorities that are ordered the same
            as their key values.
            
            Here's a simple strategy that works.  Divide the keys
            into $√{n}$ ranges of $√{n}$ keys each
            ($1\dots √{n}$, $√{n+1}\dots 2 √{n}$,
            etc.).\footnote{To make our life easer, we'll assume that
                $n$ is a square.  This doesn't affect the asymptotic
            result.}
            Rank the priorities from $1$ to $n$.
            From each range $(i-1)√{n} \dots i √{n}$, choose a
            key to keep whose priority is also ranked in the range
            $(i-1)√{n} \dots i √{n}$ (if there is one), or
            choose no key (if there isn't).  Delete all the other
            keys.

            For a particular range, we are drawing $√{n}$ samples
            without replacement from the $n$ priorities, and there are
            $√{n}$ possible choices that cause us to keep a key in
            that range.  The probability that every draw misses is
            $\prod_{i=1}^{√{n}} (1-√{n} / (n-i+1))
        ≤ (1-1/√{n})^{√{n}} ≤ e^{-1}$.
             So each range contributes at least $1-e^{-1}$ keys on
             average.  Summing over all $√{n}$ ranges gives a
             sequence of keys with increasing priorities with expected
             length at least $(1-e^{-1})√{n} =
             Ω\left(√{n}\right)$.

             An alternative solution is to apply the
             \index{theorem!Erd\H{o}s-Szekeres}
             \concept{Erd\H{o}s-Szekeres theorem}~\cite{ErdosS1935},
             which says that every sequence of length $k^2+1$ has either
             an increasing subsequence of length $k+1$ or a decreasing
             sequence of $k+1$.  Consider the sequence of priorities
             corresponding to the keys $1 \dots n$; letting
             $k=\floor{√{n-1}}$ gives a subsequence of length
             at least $√{n-1}$ that is either increasing or decreasing.
             If we delete all other elements of the treap, the
             elements corresponding to this subsequence will form a
             path, giving the desired bound.  Note that this does not
             require any probabilistic reasoning at all.

             Though not required for the problem, it's possible to
             show that $\Theta(√{n})$ is the best possible bound
             here.  The idea is that the number of
             possible sequences of keys that correspond to a path of
             length $k$ in a binary search tree is exactly
             $\binom{n}{k}2^{k-1}$; the $\binom{n}{k}$ corresponds to
             choosing the keys in the path, and the $2^{k-1}$ is
             because for each node except the last, it must contain
             either the smallest or the largest of the remaining keys
             because of the binary search tree property.

             Since each such
             sequence will be a treap path only if the priorities are
             decreasing (with probability $1/k!$), the union bound
             says that the probability of having any 
             length-$k$ paths is at most
             $\binom{n}{k}2^{k-1} /k!$.  
             But
             \begin{align*}
             \binom{n}{k}2^{k-1} /k!
             &≤ \frac{(2n)^k}{2(k!)^2} \\
             &≥ \frac{(2n)^k}{2 (k/e)^{2k}} \\
             &= \frac{1}{2} (2e^2 n/k^2)^k.
             \end{align*}
             This is exponentially small for $k \gg √{2e^2 n}$,
             showing that with high probability all possible paths
             have length $O(√{n})$.

\subsection{Random radix trees}

A \index{tree!radix}\concept{radix tree} over an alphabet of size $m$
is a tree data structure where each node has up to $m$ children, each
corresponding to one of the letters in the alphabet.  A string is
represented by a node at the end of a path whose edges are labeled
with the letters in the string in order.  For example, in
Figure~\ref{figure-radix-tree}, the string \texttt{ab} is stored at
the node reached by following the \texttt{a} edge out of the root,
then the \texttt{b} edge out of this child.

\begin{figure}
\centering
    \begin{forest}
        [$⋅$
            [$⋅$,edge label={node[midway,left]{\texttt{a}}}
                [\texttt{aa},edge label={node[midway,left]{\texttt{a}}}]
                [\texttt{ab},edge label={node[midway,right]{\texttt{b}}}]
            ]
            [$⋅$,edge label={node[midway,right]{\texttt{b}}}
                [,no edge]
                [\texttt{bb},edge label={node[midway,right]{\texttt{b}}}]
            ]
        ]
    \end{forest}
\caption[Radix tree]{A radix tree, storing the strings \texttt{aa},
\texttt{ab}, and \texttt{ba}.}
\label{figure-radix-tree}
\end{figure}

The only nodes created in the radix tree are those corresponding to
stored keys or ancestors of stored keys.

Suppose you have a radix tree into which you 
have already inserted $n$ strings of length $k$ from an
alphabet of size $m$, generated uniformly at random with replacement.
What is the expected number of new nodes you need to create to insert
a new string of length $k$?

\subsubsection*{Solution}

We need to create a new node for each prefix of the new string that is
not already represented in the tree.

For a prefix of length $\ell$, the chance that none of the $n$ strings
have this prefix is exactly $\left(1-m^{-\ell}\right)^n$.  Summing
over all $\ell$ gives that the expected number of new nodes is
$∑_{\ell=0}^{k} \left(1-m^{-\ell}\right)^n$.

There is no particularly clean expression for this, but we can observe
that $(1-m^{-\ell})^n ≤ \exp(-nm^{-\ell})$ is close to zero for
$\ell < \log_m n$ and close to $1$ for $\ell > \log_m n$.  This
suggests that the expected value is $k - \log_m n + O(1)$.

\section{Assignment 4: due Wednesday, 2013-03-27, at 17:00}

    \subsection{Flajolet-Martin sketches with deletion}

    A \index{sketch!Flajolet-Martin}\concept{Flajolet-Martin
    sketch}~\cite{FlajoletM1985}
    is a streaming data structure for approximately counting the
    number of distinct items $n$ in a large data
    stream using only $m = O(\log n)$ bits of storage.\footnote{If you actually need to do this,
        there exist better data structures for this
    problem.  See \cite{KaneNW2010}.}
    The idea is to
    use a hash function $h$ that generates each value $i ∈ \Set{1,\dots
    m}$ with probability $2^{-i}$, and for each element $x$ that
    arrives in the data stream, we write a $1$ to
    $A[h(x)]$.  (With probability $2^{-m}$ we get a value outside this 
    range and write nowhere.)  After
    inserting $n$ distinct elements, we estimate $n$ as $\hat{n} = 2^{\hat{k}}$, where
$\hat{k} = \max \SetWhere{ k }{ A[k] = 1}$, and argue that this is likely to be reasonably close to $n$.

    Suppose that we modify the Flajolet-Martin sketch to allow an
    element $x$ to be deleted by writing $0$ to
    $A[h(x)]$.  After $n$ insertions and $d$ deletions (of distinct
    elements in both cases), 
    we estimate the number of remaining elements $n-d$ as before by
    $\widehat{n-d} = 2^{\hat{k}}$, where $\hat{k} = \max
\SetWhere{ k }{ A[k] = 1 }$.  
    
    Assume that we never delete an element that has not previously
    been inserted, and that the values of $h$ are for different inputs
    are independent of each other and
    of the sequence of insertions and deletions.

    Show that there exist constants $c > 1$ and $ε >
    0$, such that for $n$ sufficiently large,
    after inserting $n$ distinct elements then 
    deleting $d ≤ ε n$
    of them, $\Prob{(n-d)/c ≤ \widehat{n-d} ≤ (n-d)c} ≥ 2/3$.

        \subsubsection*{Solution}

        We'll apply the usual error budget approach and show that the
        probability that $\widehat{n-d}$ is too big and the
        probability that $\widehat{n-d}$ is too small 
        are both small.  For the moment, we will leave $c$ and
        $ε$ as variables, and find values that work at the end.

        Let's start with the too-big side.  To get $A[k]=1$, we need $h(x_i)
        = k$ for some $x_i$ that is inserted but not subsequently
        deleted.  There are $n-d$ such $x_i$, and each gives $h(x_i) =
        k$ with probability $2^{-k}$.
        So $\Prob{A[k] = 1} ≤ (n-d) 2^{-k}$.
        This gives 
        \begin{align*}
            \Prob{\widehat{n-d} ≥ (n-d)c} 
            &= \Prob{\widehat{k} ≥ \ceil{\lg\left((n-d)c \right)}}
          \\&≤ ∑_{k=\ceil{\lg\left((n-d)c\right)}}^{∞} (n-d) 2^{-k}
          \\&= 2 (n-d) 2^{-\ceil{\lg\left((n-d)c\right)}}
          \\&≤ \frac{2}{c}.
        \end{align*}

        On the too-small side, fix $k = \ceil{\lg\left((n-d)/c)\right)}$.
        Since $A[k] = 1$ gives $\hat{k} ≥ k ≥ \ceil{\lg\left((n-d)/c)\right)}$,
        we have $\Prob{\widehat{n-d} < (n-d)/c} = \Prob{\hat{k} < \lg
        (n-d)/c} ≤ \Prob{A[k]=0}$.
        (We might be able to get a
        better bound by looking at larger indices, but to solve
        the problem this one $k$ will turn out to be enough.)

        Let $x_1 \dots x_{n-d}$ be the values that are inserted and
        not later deleted, and $x_{n-d+1} \dots x_n$ the values that
        are inserted and then deleted.  For $A[k]$ to be zero, either
        (a) no $x_i$ for $i$ in $1 \dots x_{n-d}$ has $h(x_i) = k$; or
        (b) some $x_i$ for $i$ in $n-d+1 \dots x_n$ has $h(x_i) = k$.
        The probability of the first event is $\left(1-2^{-k}\right)^{n-d}$; the
        probability of the second is $1-\left(1-2^{-k}\right)^d$.  So we have
        \begin{align*}
            \Prob{A[k] = 0}
            &≤ \left(1-2^{-k}\right)^{n-d} + \left(1 -
            \left(1-2^{-k}\right)^d\right)
          \\&≤ \exp\left(-2^{-k}(n-d)\right) +
            \left(1-\exp\left(-\left(2^{-k}+2^{-2k}\right)d\right)\right)
            \\&≤ \exp\left(-2^{-\ceil{\lg\left((n-d)/c)\right)}}(n-d)\right) +
            \left(1-\exp\left(-2⋅ 2^{-\ceil{\lg\left((n-d)/c)\right)}}d\right)\right)
            \\&≤ \exp\left(-2^{-\lg\left((n-d)/c)\right)}(n-d)\right) +
            \left(1-\exp\left(-2⋅ 2^{-\lg\left((n-d)/c)\right)+1}d\right)\right)
            \\&= e^{-c} + 
            \left(1-\exp\left(-\frac{4cd}{n-d}\right)\right)
            \\&≤ e^{-c} + \frac{4cd}{n-d}
            \\&≤ e^{-c} + \frac{4cε}{1-ε}.
        \end{align*}

        So our total probability of error is bounded by $\frac{2}{c} +
        e^{-c} + \frac{4cε}{1-ε}$.  Let $c = 8$ and
        $ε = 1/128$ to make this less than $\frac{1}{4} +
        e^{-8} + \frac{128}{127} ⋅ \frac{1}{16} \approx 0.313328 <
        1/3$, giving the desired bound.

    \subsection{An adaptive hash table}

    Suppose we want to build a hash table, but we don't know how many
    elements we are going to put in it, and because we allow
    undisciplined C programmers to obtain pointers directly to our
    hash table entries, we can't move an element once we assign it a
    position to it.  Here we will consider a data structure that
    attempts to solve this problem.

    Construct a sequence of tables $T_0, T_1, \dots$, where
    each $T_i$ has $m_i = 2^{2^i}$ slots.  For each table
    $T_i$, choose $k$ independent strongly $2$-universal hash
    functions $h_{i1}, h_{i2}, \dots h_{ik}$.

    The insertion procedure is given in
    Algorithm~\ref{alg-adaptive-hash-table-problem-insertion}.  The
    essentially idea is that we make $k$ attempts (each with a
    different hash function) to fit $x$ into
    $T_0$, then $k$ attempts to fit it in $T_1$, and so on.

    \begin{algorithm}
        \Procedure{$\Insert(x)$}{
            \For{$i \leftarrow 0$ \KwTo $∞$}{
                \For{$j \leftarrow 1$ \KwTo $k$}{
                    \If{$T_i[h_{ij}(x)] = \bot$}{
                        $T_i[h_{ij}(x)] \leftarrow x$\\
                        \Return
                    }
                }
            }
        }
        \caption{Adaptive hash table insertion}
        \label{alg-adaptive-hash-table-problem-insertion}
    \end{algorithm}

    If the tables $T_i$ are allocated only when needed, the space
    complexity of this data structure is given by the sum of $m_i$ for
    all tables that have at least one element.

    Show that for any fixed $ε > 0$, there is 
    a constant $k$ such that after inserting $n$ elements:
    \begin{enumerate}
        \item The expected cost of
    an additional insertion is $O(\log \log n)$, and
\item The expected space complexity
    is $O\left(n^{2+ε}\right)$.
    \end{enumerate}

        \subsubsection*{Solution}

        The idea is that we use $T_{i+1}$ only if we get a collision
        in $T_i$. 
        Let $X_i$ be the indicator for the event that there
        is a collision in $T_i$.  Then
        \begin{align}
            \label{eq-problem-adaptive-hash-table-rounds}
            \Exp{\text{steps}} &≤ 1 + ∑_{i=0}^{∞} \Exp{X_i}
            \intertext{and}
            \label{eq-problem-adaptive-hash-table-space}
            \Exp{\text{space}} &≤ m_0 + ∑_{i=0}^{∞} \Exp{X_i}
            m_{i+1}.
        \end{align}

        To bound $\Exp{X_i}$,
        let's calculate an upper bound on the probability that
        a newly-inserted element $x_{n+1}$ collides with any of the previous $n$
        elements $x_1 \dots x_n$ in table $T_i$.  This occurs if, for
        every location $h_{ij}(x_{n+1})$, there is some $x_r$ and some
        $j'$ such that $h_{ij'}(x_r) = h_{ij}(x_{n+1})$.  The chance
        that this occurs for any particular $j$, $j'$, and $r$ is at
        most $1/m_i$ (if $j=j'$, use $2$-universality of $h_{ij}$, and
        if $j ≠ j'$, use independence and uniformity), giving a
        chance that it occurs for fixed $j$ that is at most $n/m_i$.
        The chance that it occurs for all $j$ is at most $(n/m_i)^k$,
        and the expected number of such collisions summed over any
        $n+1$ elements that we insert is bounded by
        $n(n/m_i)^k$ (the first element can't collide with any
        previous elements).  So we have 
        $\Exp{X_i} ≤ \min\left(1, n(n/m_i)^k\right)$.

        Let $\ell$ be the largest value such that $m_\ell ≤
        n^{2+ε}$.  We will show that, for an appropriate choice
        of $k$, we are sufficiently unlikely to get a collision in
        round $\ell$ that
        the right-hand sides of
        \eqref{eq-problem-adaptive-hash-table-space} and
        \eqref{eq-problem-adaptive-hash-table-space} end up being not
        much more than the corresponding sums up to $\ell-1$.
        
        From our choice of $\ell$,
        it follows that (a) $\ell ≤ \lg \lg
        n^{2+ε} = \lg \lg n + \lg (2+ε) = O(\log \log
        n)$; and (b) $m_{\ell+1} >
        n^{2+ε}$, giving $m_\ell = √{m_{\ell+1}} >
        n^{1+ε/2}$.  From this we get 
        $\Exp{X_\ell} ≤ n(n/m_\ell)^k < n^{1-kε/2}$.

        By choosing $k$ large enough, we can make this an arbitrarily
        small polynomial in $n$.  Our goal is to wipe out the
        $\Exp{X_\ell} m_{\ell+1}$ and subsequent terms in
        \eqref{eq-problem-adaptive-hash-table-space}.

        Observe that 
        $\Exp{X_i} m_{i+1} ≤ n(n/m_i)^k m_i^2 = n^{k+1} m_i^{2-k}$.
        Let's choose $k$ so that this is at most $1/m_i$, when $i ≥
        \ell$, so we get a
        nice convergent series.\footnote{We could pick a smaller $k$,
        but why not make things easier for ourselves?}
        This requires $n^{k+1} m_i^{3-k} ≤ 1$ or
        $k+1 + (3-k) \log_n m_i ≤ 0$.
        If $i ≥ \ell$, we have $\log_n m_i > 1+ε/2$,
        so we win if $k+1 + (3-k) (1+ε/2) ≤ 0$.
        Setting $k ≥ 8/ε + 3$ works.
        (Since we want $k$ to be an integer, we probably want
        $k=\ceil{8/ε}+3$ instead.)

        So now we have
        \begin{align*}
            \Exp{\text{space}} 
            &≤ m_0 + ∑_{i=0}^{∞} \Exp{X_i} m_{i+1}
          \\&≤ ∑_{i=0}^{\ell} m_{i} + ∑_{i=\ell}^∞
            \frac{1}{m_i}
            \\&≤ 2 m_{\ell} + \frac{2}{m_\ell}
            \\&= O(n^{2+ε}).
        \end{align*}

        For $\Exp{\text{steps}}$, compute the same sum without all
        the $m_{i+1}$ factors.  This makes the tail terms even
        smaller, so they
        is still bounded by a constant, and the head becomes just
        $∑_{i=0}^{\ell} 1 = O(\log \log n)$.

    \subsection{An odd locality-sensitive hash function}

    A deranged computer scientist decides that if taking
    one bit from a random index in a bit vector is a good way to do
    locality-sensitive hashing (see
    §\ref{section-locality-sensitive-hashing-for-hamming-distance}),
    then taking the exclusive OR of $k$ independently chosen indices
    must be even better.

    Formally, given a bit-vector $x_1 x_2 \dots x_n$,
    and a sequence of
    indices $i_1 i_2 \dots i_k$, define 
    $h_i(x) = \bigoplus_{j=1}^{k} x_{i_j}$.
    For example, if $x = 00101$ and $i = 3,5,2$, $h_i(x) = 1 ⊕ 1
    ⊕ 0 = 0$.

    Suppose $x$ and $y$ are bit-vectors of length $n$ that
    differ in $m$ places.  

    \begin{enumerate}
        \item 
    Give a \emph{closed-form expression} for the
    probability that $h_i(x) ≠ h_i(y)$, assuming $i$ consists of $k$
    indices chosen uniformly and independently at random from $1\dots
    n$
\item Use this to compute the exact probability that $h_i(x) ≠
    h_i(y)$ when $m=0$,
    $m=n/2$, and $m=n$.
    \end{enumerate}

    Hint: You may find it helpful to use the identity $(a \bmod 2) =
    \frac{1}{2}(1-(-1)^a)$.

        \subsubsection*{Solution}

        \begin{enumerate}
            \item 
        Observe that $h_i(x) ≠ h_i(y)$ if an only if $i$ chooses an
        odd number of indices where $x$ and $y$ differ.  
        Let $p = m/n$ be the probability that each index in $i$ hits a
        position where $x$ and $y$ differ, and let $q = 1-p$.  Then
        the event that we get an odd number of differences is
        \begin{align*}
            ∑_{j=0}^{k} (j \bmod 2) \binom{k}{j} p^j q^{k-j}
            &= ∑_{j=0}^{k} \frac{1}{2}\left(1-(-1)^j\right) \binom{k}{j} p^j q^{k-j}
          \\&=
            \frac{1}{2} ∑_{j=0}^{k} \binom{k}{j} p^j q^{k-j}
            - 
            \frac{1}{2} ∑_{j=0}^{k} \binom{k}{j} (-p)^j q^{k-j}
            \\&=
            \frac{1}{2} (p+q)^k - \frac{1}{2} (-p+q)^k
            \\&= \frac{1 - \left(1-2(m/n)\right)^k}{2}.
        \end{align*}
            \item 
        \begin{itemize}
            \item 
        For $m=0$, this is $\frac{1-1^k}{2} = 0$.
    \item 
        For $m=n/2$, it's $\frac{1-0^k}{2} = \frac{1}{2}$ (assuming
        $k > 0$).
    \item 
        For $m=n$, it's $\frac{1-(-1)^k}{2} = (k \bmod 2)$.
        \end{itemize}
\end{enumerate}

In fact, the chances of not colliding as a function of $m$ are symmetric
around $m = n/2$ if $k$ is even and increasing if $k$ is odd.  So we
can only hope to use this as locality-sensitive hash function in the
odd case.

\section{Assignment 5: due Friday, 2013-04-12, at 17:00}

    \subsection{Choosing a random direction}

    Consider the following algorithm for choosing a random direction
    in three dimensions.  Start at the point $(X_0, Y_0, Z_0) = (0,0,0)$.  At each step,
    pick one of the three coordinates uniformly at random and add 
    $\pm 1$ to it with equal probability.  Continue until the
    resulting vector has length at least $k$, i.e., until 
    $X_t^2 + Y_t^2 + Z_t^2 ≥ k^2$.  Return this vector.

    What is the expected running time of this algorithm, as an
    asymptotic function of $k$?

        \subsubsection*{Solution}

        The trick is to observe that $X_t^2 + Y_t^2 + Z_t^2 - t$ is a
        martingale, essentially 
        following the same analysis as for $X_t^2 - t$ for
        a one-dimensional random walk.  Suppose we pick $X_t$ to
        change.  Then 
        \begin{align*}
            \ExpCond{X_{t+1}^2}{X_t}
            &= \frac{1}{2} \left((X_t+1)^2 + (X_t-1)^2\right)
          \\&= X_t^2 + 1.
        \end{align*}
        So
        \begin{align*}
            \ExpCond{X_{t+1}^2 + Y_{t+1}^2 + Z_{t+1}^2 - (t+1)}{X_t, Y_t,
            Z_t, \text{$X$ changes}}
        &= X_t^2 + Y_t^2 + Z_t^2 - t.
        \end{align*}
        But by symmetry, the same equation holds if we condition on
        $Y$ or $Z$ changing.  It follows that
        $\ExpCond{X_{t+1}^2 + Y_{t+1}^2 + Z_{t+1}^2 - (t+1)}{X_t, Y_t,
        Z_t}
        = X_t^2 + Y_t^2 + Z_t^2 - t$, and that we have a martingale as
        claimed.

        Let $τ$ be the first time at which $X_t^2 + Y_t^2 + Z_t^2
        ≤ k^2$.  From the optional stopping theorem
        (specifically, the bounded-increments case of Theorem~\ref{theorem-optional-stopping}), $\Exp{X_τ^2 + Y_τ^2 + Z_τ^2 - τ} =
        0$, or equivalently $\Exp{τ} = \Exp{X_τ^2 + Y_τ^2 +
        Z_τ^2}$.  This immediately gives $\Exp{τ} ≥ k^2$.  
        
        To
        get an upper bound, observe that
        $X_{τ-1}^2+Y_{τ-1}^2+Z_{τ-1}^2 < k^2$, and that
        exactly one of these three term increases between $τ-1$ and
        $τ$.  Suppose it's $X$ (the other cases are symmetric).
        Increasing $X$ by $1$ sets $X_τ^2 = X_{τ-1}^2 +
        2X_{τ-1} + 1$.  So we get
        \begin{align*}
            X_τ^2 + Y_τ^2 + Z_τ^2 
            &= \left(X_{τ-1}^2+Y_{τ-1}^2+Z_{τ-1}^2\right) +
            2X_{τ-1} + 1
            \\&< k^2 + 2k + 1.
        \end{align*}

        So we have $k^2 ≤ \Exp{τ} < k^2 + 2 k + 1$,
        giving $\Exp{τ} = \Theta(k^2)$.  (Or $k^2 + O(k)$ if
        we are feeling really precise.)

    \subsection{Random walk on a tree}

    Consider the following random walk on a (possibly unbalanced)
    binary search tree:
    At each step, with probability $1/3$ each, move to the
    current node's parent, left child, or right child.  If the target
    node does not exist, stay put.

    Suppose we adapt this random walk using Metropolis-Hastings (see
    §\ref{section-Metropolis-Hastings}) so that the probability of
    each node at depth $d$ in the stationary
    distribution is proportional to $α^{-d}$.  
    
    Use a coupling argument to show that, for any constant
    $α > 2$,
    this adapted random walk converges in $O(D)$ steps, where $D$ is
    the depth of the tree.

        \subsubsection*{Solution}
        As usual, let $X_t$ be a copy
        of the chain starting in an arbitrary initial state and $Y_t$
        be a copy starting in the stationary distribution.

        From the Metropolis-Hastings algorithm, the probability that
        the walk moves to a particular child is $1/3α$, so the
        probability that the depth increases after one step is
        at most 
        $2/3α$.  The probability that the walk moves to the
        parent (if we are not already at the root) is $1/3$.  

        We'll use the same choice (left, right, or parent) in both the
        $X$ and $Y$ processes, but it may be that only one of the
        particles moves (because the target node doesn't exist).  To
        show convergence, we'll track
        $Z_t = \max(\depth(X_t), \depth(Y_t))$.
        When $Z_t = 0$, both $X_t$ and $Y_t$ are the root node.

        There are two ways that $Z_t$ can change:
        \begin{enumerate}
            \item Both processes choose ``parent''; if $Z_t$ is not
                already $0$, $Z_{t+1} = Z_t - 1$.  This case occurs
                with probability $1/3$.
            \item Both processes choose one of the child directions.
                If the appropriate child exists for the deeper
                process (or for either process if they are at the same
                depth), we get $Z_{t+1} = Z_t + 1$.  This even occurs
                with probability at most $2/3α < 1/3$.
        \end{enumerate}

        So the expected change in $Z_{t+1}$ conditioned on $Z_t > 0$
        is at most $-1/3 + 2/3α = -(1/3)(2/\alpha-1)$.  Let
        $τ$ be the first time at which $Z_t = 0$.  Then the
        process $Z'_t = Z_t - (1/3)(2/α - 1)t$ for $t ≤ τ$
        and $0$ for $t > τ$ is a supermartingale, so $\Exp{Z_τ}
        = \Exp{Z_0} = \Exp{\max(\depth(X_0), \depth(Y_0))} ≤ D$.
        This gives $\Exp{τ} ≤ \frac{3D}{2/α - 1}$.

    \subsection{Sampling from a tree}

    Suppose we want to sample from the stationary distribution of the
    Metropolis-Hastings walk in the previous problem, but we don't
    want to do an actual random walk.  Assuming $α >
    2$ is a constant, give an algorithm for sampling
    \emph{exactly} from the stationary distribution that runs in constant
    expected time.
    
    Your algorithm should not require knowledge of
    the structure of the tree.  Its only input should be a pointer to
    the root node.

    \emph{Clarification added 2013-04-09:} Your algorithm can
    determine the children of any node that it has already found.  The
    idea of not knowing the structure of the tree is that it can't,
    for example, assume a known bound on the depth, counts of the
    number of nodes in subtrees, etc., without searching through the
    tree to find this information directly.

        \subsubsection*{Solution}
        We'll use rejection sampling.  The idea is
        to choose a node
        in the infinite binary tree with probability proportional to
        $α^{-d}$, and then repeat the process if we picked a node
        that doesn't actually exist.  Conditioned on finding a node
        $i$
        that exists, its probability will be
        $\frac{α^{\depth(x)}}{∑_{j} α^{-\depth(j)}}$.

        If we think of a node in the infinite tree as indexed by a
        binary strength of length equal to its depth, we can generate
        it by first choosing the length $X$ and then choosing the bits in
        the string.  We want $\Prob{X = n}$ to be proportional
        to $2^n α^{-n} = (2/α)^{n}$.  Summing the geometric series gives
        \begin{align*}
            \Prob{X = n} &=
            \frac{(2/α)^n}{1-(2/α)}.
        \end{align*}
        This is a geometric distribution, so we can sample it by
        repeatedly flipping a biased coin.  The number of
        such coin-flips is $O(X)$, as is the number of random bits we
        need to generate and the number of tree nodes we need to
        check.  The \emph{expected} value of $X$ is given by the
        infinite series
        \begin{align*}
            ∑_{n=0}^{∞} 
            \frac{(2/α)^n n }{1-(2/α)};
        \end{align*}
        this series converges to some constant by the ratio test.
        
        So each probe costs $O(1)$ time on average, and has at least a
        constant probability of success, since we choose the root with
        $\Prob{X=0} = \frac{1}{1-2/α}$.  Using Wald's equation
        \eqref{eq-walds-equation}, the total expected time to run the
        algorithm is $O(1)$.

        This is a little surprising, since the output of the algorithm
        may have more than constant length.  But we are interested in
        expectation, and when $α > 2$ most of the weight lands
        near the top of the tree.

\section{Assignment 6: due Friday, 2013-04-26, at 17:00} 

    \subsection{Increasing subsequences}

    Let $S_1, \dots, S_m$ be sets of indices in the range $1\dots n$.
    Say that a permutation $π$ of $1 \dots n$ is increasing on $S_j$
    if $π(i_1) < π(i_2) < \dots < π(i_{k_j})$ where
    $i_1 < i_2 < \dots < i_{k_j}$ are the elements of $S_j$.

    Given a fully polynomial-time randomized approximation scheme that
    takes as input $n$ and a sequence of sets $S_1, \dots, S_m$, and
    approximates the
    number of permutations $π$ that are increasing on at least one
    of the $S_j$.

        \subsubsection*{Solution}

        This can be solved using a fairly straightforward application
        of Karp-Luby~\cite{KarpL1985} (see
        §\ref{section-approximating-sharp-dnf}).  Recall that 
        for Karp-Luby we need to be able to express our target set $U$ as
        the union of a polynomial number of covering sets $U_j$, where
        we can both compute the size of each $U_j$ and sample
        uniformly from it.  We can then estimate $\card*{U} =
        ∑_{j, x ∈ U_j} f(j, x) = \left(∑_j
        \card*{U_j}\right)\Prob{f(j, x) = 1}$ where $f(j,x)$ is the
        indicator for the event that $x \not∈ U_{j'}$ for any $j' <
        j$ and in the probability, the pair $(j,x)$ is chosen
        uniformly at random from $\SetWhere{(j, x)}{x ∈ U_j}$.

        In this case, let $U_j$ be the set of all permutations that
        are increasing on $S_j$.  We can specify each such permutation
        by specifying the choice of which $k_j = \card*{S_j}$ elements
        are in positions $i_1 \dots i_{k_j}$ (the order of these
        elements is determined by the requirement that the permutation
        be increasing on $S_j$) and specifying the order of the
        remaining $n-k_j$ elements.  
        This gives $\binom{n}{k_j} (n-k_j)! = (n)_{n-k_j}$ such
        permutations.  Begin by computing these counts for all $S_j$,
        as well as their sum.

        We now wish to sample uniformly from pairs $(j, π)$ where
        each $π$ is an element of $S_j$.  First sample each $j$ with
        probability $\card*{S_j} / ∑_\ell \card*{S_\ell}$, using the
        counts we've already computed.
        Sampling a permutation uniformly from $S_j$
        mirrors the counting argument: choose a $k_j$-subset
        for the positions in $S_j$, then order the remaining elements
        randomly.  The entire sampling step can easily be done in $O(n+m)$ time.

        Computing $f(j, π)$ requires testing $π$ to see if it is
        increasing for any $S_{j'}$ for $j < j'$; without doing anything particularly
        intelligent, this takes $O(nm)$ time.  So we can construct and
        test one sample in $O(nm)$ time.  Since each sample has at
        least a $ρ = 1/m$ chance of having $f(j, π) = 1$, from
        Lemma~\ref{lemma-sampling} we need
        $O\left(\frac{1}{ε^2 ρ} \log
        \frac{1}{δ}\right) = O\left(m ε^{-2}
        \log\frac{1}{δ}\right)$ samples to get relative error
        with probability at least $1-δ$, for a total cost of
        $O\left(m^2 n ε^{-2} \log\frac{1}{δ}\right)$.

    \subsection{Futile word searches}
    \label{section-hw-futile-word-searches}

    A \index{puzzle!word search}\concept{word search puzzle} consists
    of an $n×n$ grid of letters from some alphabet $Σ$,
    where the goal is to find
    contiguous sequences of letters in one of the eight orthogonal or
    diagonal directions that form words from some lexicon.  For example,
    in Figure~\ref{figure-word-search}, the left grid contains an
    instance of \texttt{aabc} (running up and left from the rightmost
    \texttt{a} character on the last line), while the right grid
    contains no instances of this word.

    \begin{figure}
\begin{alltt}
bacab\quad{}bacab
\textcolor{red}{c}caac\quad{}ccaac
b\textcolor{red}{b}bac\quad{}babac
bb\textcolor{red}{a}aa\quad{}bbaaa
acb\textcolor{red}{a}b\quad{}acbab
\end{alltt}
\caption[Word searches]{Non-futile (left) and futile (right) word
    search grids for the
lexicon $\Set{\mbox{\texttt{aabc}}, \mbox{\texttt{ccca}}}$}
\label{figure-word-search}
    \end{figure}

    For this problem, you are asked to build an algorithm for
    constructing word search puzzles
    with no solution for a given lexicon.  That is, given a set 
    of words $S$ over some alphabet and a grid size $n$, the output
    should be an $n × n$ grid of letters such that no word in $S$
    appears as a contiguous sequence of letters in one of the eight
    directions anywhere in the grid.  We will refer to such puzzles as
    \index{puzzle!word search!futile}
    \index{word search puzzle!futile}
    \index{futile word search puzzle}
    \conceptFormat{futile word search puzzles}.
    
    \begin{enumerate}
        \item Suppose the maximum length of any word in $S$ is $k$.
            Let $p_S$ be the probability that some word in $S$ is a
            prefix of an infinite string generated by picking letters
            uniformly and independently from $Σ$.  Show that
            there is a constant $c > 0$ such that for any $k$,
            $Σ$, and $S$,
            $p_S < ck^{-2}$ implies
            that there exists, for all $n$, 
            an $n× n$ futile word search puzzle
            for $S$ using only letters from $Σ$.
        \item Give an algorithm that
            constructs a futile word search puzzle given $S$ and $n$
            in expected time
            polynomial in $\card*{S}$, $k$, and $n$, provided $p_S <
            ck^{-2}$ as above.
    \end{enumerate}

        \subsubsection*{Solution}

        \begin{enumerate}
            \item We'll apply the symmetric version of the Lovász
                local lemma.  Suppose the grid is filled in
                independently and uniformly at random with characters
                from $Σ$.  Given a position $ij$ in the grid, let
                $A_{ij}$ be the event that there exists a word in $S$
                whose first character is at position $ij$; observe
                that $\Prob{A_{ij}} ≤ 8p_S$ by the union bound (this
                may be an overestimate, both because we might run off
                the grid in some directions and because the choice of
                initial character is not independent).  Observe also
                that $A_{ij}$ is independent of any event $A_{i'j'}$
                where $\abs*{i-i'} ≥ 2k-1$ or $\abs*{j-j'} ≥ 2k-1$,
                because no two words starting at these positions can
                overlap.  So we can build a dependency graph
                with $p ≤ 8p_S$ and $d ≤ (4k-3)^2$.  The Lovász
                local lemma shows that there exists an assignment
                where no $A_{ij}$ occurs provided $ep(d+1) < 1$ or
                $8e p_S ((4k-3)^2+1) < 1$.  This easily holds if $p_S
                < \frac{1}{8e(4k^2)} = \frac{1}{128e} k^{-2}$.
            \item For this part, we can just use
                Moser-Tardos~\cite{MoserT2010}, particularly the
                symmetric version described in
                Corollary~\ref{corollary-symmetric-lovasz-local-lemma-constructive}.
                We have a collection of $m = O(n^2)$ bad events,
                with $d = \Theta(k^2)$, so the expected number of
                resamplings is bounded by $m/d = O(n^2/k^2)$.  Each
                resampling requires checking every position in the new
                grid for an occurrence of some string in $S$; this
                takes $O(n^2 k ⋅ \card*{S})$ time per resampling
                even if we are not very clever about it.  So the total
                expected cost is $O(n^4 ⋅ \card*{S} / k)$.

                With some more intelligence, this can be improved.  We
                don't need to recheck any position at distance greater
                than $k$ from any of the at most $k$ letters we
                resample, and if we are sensible, we can store $S$
                using a radix tree or some similar data structure that
                allows us to look up all words that appear as a prefix
                of a given length-$k$ string in time $O(k)$.  This
                reduces the cost of each resampling to $O(k^3)$, with
                an additive cost of $O(k ⋅ \card*{S})$ to
                initialize the data structure.  So the total expected 
                cost is now $O(n^2 k + \card*{S})$.
    \end{enumerate}

\subsection{Balance of power}

Suppose you are given a set of $n$ MMORPG players, and a sequence of
subsets $S_1, S_2, \dots, S_m$ of this set, where each subset $S_i$ gives the
players who will participate in some raid.  Before any of
the raids take place, you are to assign each player permanently
to one of three
factions.  If for any $i$, $\card*{S_i}/2$ or more of the players are
in the same faction, then instead of carrying out the raid they
will overwhelm and rob the other participants.

Give a randomized algorithm for computing a faction assignment that
prevents this tragedy from occurring (for all $i$) and thus allows all 
$m$ raids
to be completed without incident, assuming that $m > 1$ and 
$\min_i \card*{S_i} ≥ c \ln m$ for some constant $c > 0$ that does not depend on
$n$ or $m$.
Your algorithm should run in expected time
polynomial in $n$ and $m$.

    \subsubsection*{Solution}

    Assign each player randomly to a faction.  Let $X_{ij}$ be the
    number of players in $S_i$ that are assigned to faction $j$.
    Then $\Exp{X_{ij}} = \card*{S_i}/3$.
    Applying the Chernoff bound \eqref{eq-Chernoff-bound-one-third}, we have
    \begin{align*}
        \Prob{X_{ij} ≥ \card*{S_i}/2}
        &= \Prob{X_{ij} ≥ \left(1+\frac{1}{2}\right) \Exp{X_{ij}}}
      \\&≤
        \exp\left(-\left(\card*{S_i}/3\right)\left(\frac{1}{2}\right)^2/3\right)
        \\&= e^{-\card*{S_i}/36}.
    \end{align*}

    Let $c = 3⋅ 36= 108$.  Then if $\min_i \card*{S_i} ≥ c \ln m$, for each
    $i$, $j$, it holds that $\Prob{X_{ij} ≥ \card*{S_i}/2} ≤ e^{-3
    \ln m} = m^{-3}$.  So the probability that this bound is exceeded
    for any $i$ and $j$ is at most $(3m)m^{-2} = 3/m^2$.  So a random
    assignment works with at least $1/4$ probability for $m > 1$.

    We can generate and test each assignment in $O(nm)$ time.  So our expected time
    is $O(nm)$.

\section{Final exam}

Write your answers in the blue book(s).  Justify your answers.  Work
alone.  Do not use any notes or books.  

There are four problems on this exam, each worth 20
points, for a total of 80 points.
You have approximately three hours to complete this
exam.

\subsection{Dominating sets}

A \concept{dominating set} in a graph is a subset $S$ of the vertices
for which every vertex $v$ is either in $S$ or adjacent to a vertex in
$S$.

Show that for any graph $G$, there is an aperiodic, irreducible Markov
chain on the dominating sets of $G$, such that (a) the transition rule
for the chain can be implemented in polynomial time; and (b) the
stationary distribution of the chain is uniform.  (You do not need to
say anything about the convergence rate of this chain.)

    \subsection*{Solution}

        Suppose we are have state $S_t$ at time $t$.
            We will use a random walk where 
            we choose a vertex
            uniformly at random to add to or remove from $S_t$, and
            carry out the action only if the resulting set is still a
            dominating set.

            In more detail: For each
            vertex $v$, with
            probability $1/n$, $S_{t+1} = S_t \cup \Set{v}$ if $v
            \not∈ S_t$, $S_{t+1} = S_t ∖ \Set{v}$ if $v ∈
            S_t$ and $S_t ∖ \Set{v}$ is a dominating set, and
            $S_{t+1} = S_t$ otherwise.  To implement this transition
            rule, we need to be able to choose a vertex $v$ uniformly at
            random (easy) and test in the case where $v ∈ S_t$ if
            $S_t ∖ \Set{v}$ is a dominating set (also
            polynomial: for each vertex, check if it or one of its
            neighbors is in $S_t ∖ \Set{v}$, which takes time
            $O(\card*{V} + \card*{E})$).
            Note that we do not need to
            check if $S_t \cup \Set{v}$ is a dominating set.

            For any pair of adjacent states $S$ and $S' = S ∖
            \Set{v}$ the probability of moving from $S$ to $S'$ and
            the probability of moving from $S'$ to $S$ are both
            $1/n$.  So the Markov chain is reversible with a uniform
            stationary distribution.  

            This is an aperiodic chain, because there exist minimal
            dominating sets for which there is a nonzero chance that
            $S_{t+1} = S_t$.

            It is irreducible, because for any dominating set $S$,
            there is a path to the complete set of vertices $V$ 
            by adding each vertex in $V ∖ S$ one at a time.
            Conversely, removing these vertices from $V$ gives a path
            to $S$.  This gives a path $S \leadsto V \leadsto T$ 
            between any two dominating
            sets $S$ and $T$.

\subsection{Tricolor triangles}

Suppose you wish to count the number of assignments of colors 
$\Set{r,g,b}$ to nodes of a graph $G$ that have the property that some 
triangle in $G$ contains all three colors.  For
example, the four-vertex graph shown below is labeled with one of $18$ such
colorings ($6$ permutations of the colors of the triangle nodes
times $3$ unconstrained choices for the degree-$1$ node).

\begin{center}
\begin{tikzpicture}
    \node[labeled] (r1) {$r$};
    \node[labeled] (g) [below right of=r1] {$g$};
    \node[labeled] (b) [below left of=g] {$b$};
    \node[labeled] (r2) [right of=g] {$r$};
    \path
        (r1) edge (g) edge (b)
        (g) edge (b) edge (r2)
        ;
\end{tikzpicture}
\end{center}

Give a fully polynomial-time randomized approximation scheme for this
problem.

    \subsection*{Solution}

    Though it is possible, and tempting, to go after this using Karp-Luby (see
    §\ref{section-approximating-sharp-dnf}),
    naive sampling is enough.  

    If a graph has at least one triangle (which can be checked in
    $O(n^3)$ time just by enumerating all possible triangles),
    then the probability that that particular triangle is tricolored
    when colors are chosen uniformly and independently at random is
    $6/27 = 2/9$.  This gives a constant hit rate $ρ$, so by
    Lemma~\ref{lemma-sampling}, we can get $ε$ relative
    error with $1-δ$ probability using
    $O\left(\frac{1}{ε^2} \log \frac{1}{δ}\right)$
    samples.  Each sample costs $O(n^3)$ time to evaluate (again,
    brute-force checking of all possible triangles), for a total
    cost of 
    $O\left(n^3ε^{-2} \log
    \frac{1}{δ}\right)$.

\subsection{The $n$ rooks problem}

The $n$ rooks problem requires marking as large a subset as possible
of the squares in an $n× n$ grid, so that no two squares in
the same row or column are marked.\footnote{This is not actually a
hard problem.}

Consider the following randomized algorithm that attempts to solve
this problem:
\begin{enumerate}
\item Give each of the $n^2$ squares a distinct label
using a uniformly chosen random
permutation of the integers $1 \dots n^2$.
\item Mark any square whose label is larger than any other label in
its row and column.
\end{enumerate}

What is the expected number of marked squares?

    \subsection*{Solution}
Each square is marked if it is the largest of the $2n-1$ total
squares in its row and column.  By symmetry, each of these $2n-1$
squares is equally likely to be the largest, so the probability that a
particular square is marked is exactly $\frac{1}{2n-1}$.  By linearity
of expectation, the total expected number of marked squares is then
$\frac{n^2}{2n-1}$.

\subsection{Pursuing an invisible target on a ring}

Suppose that you start at position $0$ on a ring of size $2n$, while a
target particle starts at position $n$.  At each step, starting at
position $i$, you can choose whether to move to any of positions
$i-1$, $i$, or $i+1$.  At the same time, the target moves from its
position $j$ to either $j-1$ or $j+1$ with equal probability,
independent of its previous moves or your moves.  Aside
from knowing that the target starts at $n$ at time $0$, you cannot
tell where the target moves.

Your goal is to end up on the same node as the target after some
step.\footnote{Note that you must be in the same place at the end of
    the step: if you move from $1$ to $2$ while the target moves from
$2$ to $1$, that doesn't count.}
Give an algorithm for choosing your moves such that, for any $c > 0$,
you encounter the target in at most $2n$
steps with probability at least $1-n^{-c}$ for
sufficiently large $n$.

    \subsection*{Solution}

    The basic idea is to just go through positions $0, 1, 2, \dots$
    until we encounter the target, but we have to be a little careful about
    parity to make sure we don't pass it by accident.\footnote{This is
    not the only possible algorithm, but there are a lot of
    plausible-looking algorithms that turn out not to work.  One
    particularly tempting approach is to run to position $n$ using the
    first $n$ steps and then spend the next $n$ steps trying to hit
    the target in the immediate neighborhood of $n$, either by staying
    put (a sensible strategy when lost in the woods in real life,
    assuming somebody is looking for you), or moving in a random walk
    of some sort starting at $n$.  This doesn't work if we want a
    high-probability bound.  To see this, observe that the target has
    a small but nonzero 
    constant probability in the limit of begin at some position
    greater than or equal to $n + 4 √{n}$ after exactly $n/2$ steps.
    Conditioned on starting at $n+4√{n}$ or above,
    its chances of moving below $n + 4√{n} - 2√{n} = n +
    2√{n}$ at any
    time in the
    next $3n/2$ steps is bounded by $e^{-4n/2(3n/2)} = e^{-4/3}$ (Azuma), 
    and a similar
    bound
    holds independently for our chances of getting up to $n+2√{n}$
    or above.  Multiplying out all these constants gives a constant
    probability of failure.  A similar but bigger disaster occurs if
    we don't rush to $n$ first.}

    Let $X_i = \pm 1$ be the increment of the target's $i$-th move, 
    and let $S_i = ∑_{j=1}^{i} X_j$, so
    that its position after $i$ steps is $n + S_i \bmod{2n}$.

    Let $Y_i$ be the position of the pursuer after $i$ steps.

    First move: stay at $0$ if $n$ is odd, move to $1$ if $n$ is
    even.  The purpose of this is to establish the invariant that
    $n+S_i - Y_i$ is even starting at $i = 1$.  For subsequent moves,
    let $Y_{i+1} = Y_i + 1$.  Observe that this maintains the
    invariant.

    We assume that $n$ is at least $2$.  This is necessary to ensure
    that at time $1$, $Y_1 ≤ n + S_1$.

    Claim: if at time $2n$, $Y_{2n} ≥ n + S_{2n}$, then at some time $i ≤
    2n$, $Y_i = n + S_i$.  Proof: Let $i$ be the first time at which
    $Y_i ≥ n + S_i$; under the assumption that $n ≥ 2$, $i ≥ 1$.
    So from the invariant, we can't have $Y_i = n +
    S_i + 1$, and if $Y_i ≥ n + S_i + 2$, we have
    $Y_{i-1} ≥ Y_i-1 ≥ n+S_i+1 ≥ n+S_{i-1}$, contradicting our
    assumption that $i$ is minimal.  The remaining alternative is that
    $Y_i = n + S_i$, giving a collision at time $i$.

    We now argue that $Y_{2n} ≥ n-1$ is very likely to be at least
    $n + S_{2n}$.  Since $S_{2n}$ is a sum of $2n$ independent $\pm 1$
    variables, from Hoeffding's inequality we have $\Prob{Y_n < n +
    S_{2n}} ≤ \Prob{S_{2n} ≥ n} ≤ e^{-n^2/4n} = e^{-n/4}$.
    For sufficiently large $n$, this is much smaller than $n^{-c}$ for
    any fixed $c$.

\chapter{Sample assignments from Spring 2011}

\section{Assignment 1: due Wednesday, 2011-01-26, at 17:00}

\subsection{Bureaucratic part}

Send me email!  My address is
\mailto{james.aspnes@gmail.com}.

In your message, include:

\begin{enumerate}
\item Your name.
\item Your status: whether you are an undergraduate, grad student, auditor, etc.
\item Anything else you'd like to say.
\end{enumerate}

(You will not be graded on the bureaucratic part, but you should do it anyway.)

\subsection{Rolling a die}

The usual model of a randomized algorithm assumes a source of fair,
independent random bits.  This makes it easy to generate uniform
numbers in the range $0\dots 2^n-1$, but not so easy for other ranges.
Here are two algorithms for generating a uniform random integer $0 ≤
s < n$:
\begin{itemize}
\item \index{sampling!rejection}\indexConcept{rejection sampling}{Rejection sampling} generates
a uniform random integer $0 ≤ s < 2^{\ceil{\lg n}}$.  If $s < n$,
return $s$; otherwise keep trying until you get some $s < n$.
\item \indexConcept{arithmetic coding}{Arithmetic coding} or 
\concept{range coding} 
generates a sequence of bits $r_1, r_2, \dots r_k$ until the half-open
interval
$[∑_{i=1}^{k} 2^{-i} r_i, ∑_{i=1}^{k} 2^{-i} r_i + 2^{-k-1})$ 
is a subset of $[s/n, (s+1)/n)$ for some $s$; it then returns $s$.
\end{itemize}

\begin{enumerate}
\item Show that both rejection sampling and range coding produce a
uniform value $0 ≤ s < n$ using an expected $O(\log n)$ random bits.
\item Which algorithm has a better constant?
\item Does there exist a function $f$ and 
an algorithm that produces a uniform value $0
≤ s < n$ for any $n$ using $f(n)$ random bits with probability
$1$?
\end{enumerate}

\subsubsection*{Solution}

\begin{enumerate}
\item For rejection sampling, each sample requires $\ceil{\lg n}$ bits
and is accepted with probability $n/2^{\ceil{\lg n}} ≥ 1/2$.
So rejection sampling returns a value after at most $2$ samples on
average, using no more than an expected $2 \ceil{\lg n} < 2 (\lg n + 1)$ expected
bits for the worst $n$.

For range coding, we keep going as long as one of the $n-1$ nonzero
endpoints $s/n$ lies inside the current interval.  After
$k$ bits, the probability that one of the $2^k$
intervals contains an endpoint is at most $(n-1)2^{-k}$; in
particular, it drops below $1$ as soon as $k = 2^{\ceil{\lg n}}$ and
continues to drop by $1/2$ for each additional bit, requiring $2$ more
bits on average.  So the expected cost of range coding is at most
$\ceil{\lg n} + 2 < \lg n + 3$ bits.
\item We've just shown that range coding beats rejection sampling by a
factor of $2$ in the limit, for worst-case $n$.  It's worth noting that other
factors might be more important if random bits are cheap: rejection
sampling is much easier to code and avoids the need for division.
\item There is no algorithm that produces a uniform value $0 ≤ s <
n$ for all $n$ using any fixed number of bits.  Suppose such an algorithm
existed.  Fix some $n$.  For all $n$ values $s$ to be equally
likely, the sets of random bits $M^{-1}(s) = \SetWhere{r}{M(r) = s}$ must
have the same size.  But this can only happen if $n$ divides
$2^{f(n)}$, which works only for $n$ a power of $2$.
\end{enumerate}

\subsection{Rolling many dice}

Suppose you repeatedly roll an $n$-sided die.  Give an asymptotic
(big-$\Theta$) bound on the expected
number of rolls until you roll some number you have already rolled
before.

\subsubsection*{Solution}
In principle, it is possible to compute this value exactly, but we are
lazy.

For a lower bound, observe that after $m$ rolls, each of the 
$\binom{m}{2}$ pairs of rolls has probability $1/n$ of being equal, for
an expected total of $\binom{m}{2}/n$ duplicates.  For $m =
√{n}/2$, this is less than $1/8$, which shows that the expected
number of rolls is $Ω(√{n})$.

For the upper bound, suppose we have already rolled the die $√{n}$
times.  If we haven't gotten a duplicate already, each new roll has
probability at least $√{n}/n = 1/√{n}$ of matching a previous
roll.  So after an additional $√{n}$ rolls on average, we get a
repeat.  This shows that the expected number of rolls is
$O(√{n})$.

Combining these bounds shows that we need $\Theta(√{n})$ rolls on
average.

\subsection{All must have candy}
\label{problem-all-must-have-candy}

A set of $n_0$ children each reach for one of $n_0$ candies, with each
child choosing a candy independently and uniformly at random.  If 
a candy is chosen by exactly one child, the candy and child drop out.
The remaining $n_1$ children and candies then repeat the process for
another round, leaving $n_2$ remaining children and candies, etc.  The
process continues until ever child has a candy.

Give the best bound you can on the expected number of rounds until
every child has a candy.

\subsubsection*{Solution}

Let $T(n)$ be the expected number of rounds remaining given we are
starting with $n$ candies.  We can set up a probabilistic recurrence
relation $T(n) = 1 + T(n-X_n)$ where $X_n$ is the number of candies 
chosen by eactly one child.  It is easy to compute $\Exp{X_n}$, since the
probability that any candy gets chosen exactly once is
$n(1/n)(1-1/n)^{n-1} = (1-1/n)^{n-1}$.  
Summing over all
candies gives $\Exp{X_n} = n(1-1/n)^{n-1}$.

The term $(1-1/n)^{n-1}$ approaches $e^{-1}$ in
the limit, so for any fixed $ε > 0$, we have $n(1-1/n)^{n-1} ≥
n(e^{-1} - ε)$ for sufficiently large $n$.  We can get a quick
bound by choosing $ε$ so that $e^{-1}-ε ≥ 1/4$ (for
example) and then applying the Karp-Upfal-Wigderson inequality
\eqref{eq-karp-upfal-wigderson} with $μ(n) = n/4$ to get
\begin{align*}
    \Exp{T(n)} &≤ \int_1^n \frac{1}{t/4} \,dt
\\ &= 4 \ln n.
\end{align*}
There is a sneaky trick here, which is that we stop if we get down to
$1$ candy instead of $0$.  This avoids the usual problem with KUW and
$\ln 0$, by observing that we can't ever get down to exactly one
candy: if there were exactly one candy that gets grabbed twice or not
at all, then there must be some other candy that also gets grabbed
twice or not at all.

This analysis is sufficient for an asymptotic estimate: the last candy
gets grabbed in $O(\log n)$ rounds on average.  For most
computer-sciency purposes, we'd be done here.

We can improve the constant slightly by observing that $(1-1/n)^{n-1}$
is in fact always greater than or equal to $e^{-1}$.  The easiest way
to see this is to plot the function, but if we want to prove it
formally we can show that $(1-1/n)^{n-1}$ is a decreasing function by
taking the derivative of its logarithm:
\begin{align*}
\frac{d}{dn} \ln (1-1/n)^{n-1} 
&= \frac{d}{dn} (n-1) \ln (1-1/n)
\\ &= \ln (1-1/n) + \frac{n-1}{1-1/n} ⋅ \frac{-1}{n^2}.
\end{align*}
and observing that it is negative for $n > 1$ (we could also take the
derivative of the original function, but it's less obvious that it's
negative).  So if it approaches
$e^{-1}$ in the limit, it must do so from above, implying
$(1-1/n)^{n-1} ≥ e^{-1}$.

This lets us apply \eqref{eq-karp-upfal-wigderson} with $μ(n) = n/e$,
giving $\Exp{T(n)} ≤ e \ln n$.

If we skip the KUW bound and use the analysis in
§\ref{section-probabilistic-recurrences-detailed-analysis}
instead, we get that 
$\Prob{T(n) ≥ \ln n + \ln(1/ε)} ≤ ε$.
This suggests that the actual expected value should be 
$(1+o(1)) \ln n$.

\section{Assignment 2: due Wednesday, 2011-02-09, at 17:00}

\subsection{Randomized dominating set}

A \concept{dominating set} in a graph $G=(V,E)$ is a set of vertices
$D$ such that each of the $n$ vertices in $V$ is either in $D$ or adjacent to a
vertex in $D$.

Suppose we have a $d$-regular graph, in which every vertex has exactly
$d$ neighbors.  Let $D_1$ be a random subset of $V$ in which each
vertex appears with independent probability $p$.  Let $D$ be the union
of $D_1$ and the set of all vertices that are not adjacent to any
vertex in $D_1$.  
(This construction is related to a classic maximal
independent set algorithm of Luby~\cite{Luby1985}, and has the
desirable property in a distributed system of finding a dominating set
in only one round of communication.)

\begin{enumerate}
\item What would be a good value of $p$ if our goal is to minimize
    $\Exp{\card*{D}}$, and what bound on $\Exp{\card*{D}}$ does this value give?
\item For your choice of $p$ above, what bound can you get on
    $\Prob{\card*{D} - \Exp{\card*{D}} ≥ t}$?
\end{enumerate}

\subsubsection*{Solution}

\begin{enumerate}
    \item First let's compute $\Exp{\card*{D}}$.  Let $X_v$ be the indicator
for the event that $v∈ D$.  Then $X_v = 1$ if either (a) $v$ is in
$D_1$, which occurs with probability $p$; or (b) $v$ and all $d$ of
its neighbors are not in $D_1$, which occurs with probability
$(1-p)^{d+1}$.  Adding these two cases gives 
$\Exp{X_v} = p + (1-p)^{d+1}$ and thus
\begin{align}
\label{eq-2011-assignment-2-dominating-set-size}
\Exp{\card*{D}} = ∑_v \Exp{X_v} = n\left(p+(1-p)^{d+1}\right).
\end{align}

We optimize $\Exp{\card*{D}}$ in the usual way, by seeking a minimum for
$\Exp{X_v}$.  Differentiating with respect to $p$ and setting to $0$ gives
$1 - (d+1)(1-p)^d = 0$, which we can solve to get
$p = 1 - (d+1)^{-1/d}$.  (We can easily observe that this must be a
minimum because setting $p$ to either $0$ or $1$ gives $\Exp{X_v} = 1$.)

The value of $\Exp{\card*{D}}$ for this value of $p$ is
the rather nasty expression
$n(1-(d+1)^{-1/d}+(d+1)^{-1-1/d})$.

Plotting the $d$ factor up suggests that it goes to $\ln d / d$ in the
limit, and both Maxima and \url{www.wolframalpha.com} agree with this.
Knowing the answer, we can prove it by showing
\begin{align*}
\lim_{d\rightarrow ∞}
\frac{1-(d+1)^{-1/d}+(d+1)^{-1-1/d}}{\ln d / d}
&=
\lim_{d\rightarrow ∞}
\frac{1-d^{-1/d}+d^{-1-1/d}}{\ln d / d}
\\
&=
\lim_{d\rightarrow ∞}
\frac{1-d^{-1/d}}{\ln d / d}
\\
&=
\lim_{d\rightarrow ∞}
\frac{1-e^{-\ln d/d}}{\ln d / d}
\\
&=
\lim_{d\rightarrow ∞}
\frac{1-\left(1 - \ln d / d + O(\ln^2 d / d^2)\right)}{\ln d / d}
\\
&=
\lim_{d\rightarrow ∞}
\frac{\ln d / d + O(\ln^2 d / d^2)}{\ln d / d}
\\
&=
\lim_{d\rightarrow ∞}
\left(1 + O(\ln d / d)\right)
\\
&=
1.
\end{align*}

This lets us write $\Exp{\card*{D}} = (1+o(1)) n \ln d / d$, where we
bear in mind that the $o(1)$ term depends on $d$ but not $n$.

\item Suppose we fix $X_v$ for all but one vertex $u$.  Changing $X_u$
from $0$ to $1$ can increase $\card*{D}$ by at most one (if $u$ wasn't
already in $D$) and can decrease it by at most $d-1$ (if $u$ wasn't
already in $D$ and adding $u$ to $D_1$ lets all $d$ of $u$'s neighbors
drop out).  So we can apply the method of bounded differences with
$c_i = d-1$ to get 
\begin{align*}
\Prob{\card*{D} - \Exp{\card*{D}} ≥ t}
&≤ \exp\left(-\frac{t^2}{2n(d-1)^2}\right).
\end{align*}

A curious feature of this bound is that it doesn't depend on $p$ at
all.  It may be possible to get a tighter bound using a better
analysis, which might pay off for very large $d$ (say, $d \gg
√{n})$.
\end{enumerate}

\subsection{Chernoff bounds with variable probabilities}
\label{problem-variable-probability-chernoff-bounds}

Let $X_1 \dots X_n$ be a sequence of Bernoulli random variables, where
for all $i$, $\ExpCond{X_i }{ X_1 \dots X_{i-1}} ≤ p_i$.
Let $S = ∑_{i=1}^{n} X_i$ and $μ = ∑_{i=1}^{n} p_i$.  Show
that, for all $δ ≥ 0$,
\begin{align*}
\Prob{S ≥ (1+δ)μ} 
&≤
\left(\frac{e^δ}{(1+δ)^{1+δ}}\right)^μ.
\end{align*}

\subsubsection*{Solution}

Let $S_t = ∑_{i=1}^{t} X_i$, so that $S = S_n$, and let $μ_t =
∑_{i=1}^{t} p_i$.  We'll show by
induction on $t$ that 
$\Exp{e^{α S_t}} ≤ \exp(e^{α - 1}μ_t)$, when $α > 0$.

Compute
\begin{align*}
\Exp{e^{α S}}
&= \Exp{e^{α S_{n-1}} e^{α X_n}}
\\
&= \Exp{e^{α S_{n-1}} \ExpCond{e^{α X_n}}{X_1,\dots,X_{n-1}}}
\\
&= 
\Exp{e^{α S_{n-1}}
  \left(
    \ProbCond{X_n = 0 }{ X_1, \dots, X_{n-1}}
   + e^α \ProbCond{X_n = 1 }{ X_1,\dots,X_{n-1} }
  \right)
  }
\\
&=
\Exp{e^{α S_{n-1}}
  \left(
    1 
   + (e^α - 1) \ProbCond{X_n = 1 }{ X_1,\dots,X_{n-1} }
  \right)
  }
\\
&≤
\Exp{e^{α S_{n-1}}
  \left(
    1
   + (e^α - 1) p_n
  \right)
  }
\\
&≤
\Exp{e^{α S_{n-1}}
  \exp\left(
   (e^α - 1) p_n
  \right)
  }
\\
&≤
\Exp{e^{α S_{n-1}}}
  \exp\left(
   (e^α - 1) p_n
  \right)
\\
&≤
\exp\left(e^α - 1)μ_{n-1}\right)
  \exp\left(
   (e^α - 1) p_n
  \right)
\\
&=
\exp\left(e^α - 1)μ_{n}\right).
\end{align*}

Now apply the rest of the proof of \eqref{eq-Chernoff-bound} to get
the full result.

\subsection{Long runs}

Let $W$ be a binary string of length $n$, generated uniformly at
random.  Define a \concept{run} of ones as a maximal sequence of
contiguous ones; for example, the string $11100110011111101011$
contains 5 runs of ones, of length 3, 2, 6, 1, and 2.

Let $X_k$ be the number of runs in $W$ of length $k$ or more.

\begin{enumerate}
\item Compute the exact value of $\Exp{X_k}$ as a function of $n$ and
$k$.
\item Give the best concentration bound you can for $\abs*{X_k - \Exp{X_k}}$.
\end{enumerate}

\subsubsection*{Solution}

\begin{enumerate}
\item We'll compute the probability that any particular position
$i=1\dots n$ is the start of a run of length $k$ or more, then sum
over all $i$.  For a run of length $k$ to start at position $i$,
either (a) $i=1$ and $W_i\dots W_{i+k-1}$ are all $1$, or (b) $i>1$,
$W_{i-1} = 0$, and $W_i\dots W_{i+k-1}$ are all $1$.  Assuming $n ≥
k$, case (a) adds $2^{-k}$ to $\Exp{X_k}$ and case (b) adds
$(n-k)2^{-k-1}$, for a total of $2^{-k} + (n-k)2^{-k-1} =
(n-k+2)2^{-k-1}$.
\item 
We can get an easy bound without too much cleverness
using McDiarmid's inequality \eqref{eq-McDiarmids-inequality}.  
Observe that $X_k$ is a function of
the independent random variables $W_1\dots W_n$ and that changing one
of these bits changes $X_k$ by at most $1$ (this can happen in several
ways: a previous run of length $k-1$ can become a run of length $k$ or
vice versa, or two runs of length $k$ or more separated by a single
zero may become a single run, or vice versa).  So
\eqref{eq-McDiarmids-inequality} gives
$\Prob{\abs*{X-\Exp{X}} ≥ t} ≤ 2\exp\left(-\frac{t^2}{2n}\right)$.

We can improve on this a bit by grouping the $W_i$ together into
blocks of length $\ell$.  If we are given control over a block of $\ell$
consecutive bits and want to minimize the number of runs, we can
either (a) make all the bits zero, causing no runs to appear within
the block and preventing adjacent runs from extending to length $k$
using bits from the block, or (b) make all the bits one, possibly
creating a new run but possibly also causing two existing runs on
either side of the block to merge into one.  In the first case,
changing all the bits to one except for a zero after every $k$
consecutive ones creates at most $\floor{\frac{\ell+2k-1}{k+1}}$ new
runs.  
Treating each of the $\ceil{n}{\ell}$ blocks as a single
variable then gives
$\Prob{\abs*{X-\Exp{X}} ≥ t} ≤
2\exp\left(-\frac{t^2}{2\ceil{n/\ell}\left(\floor{(\ell+2k-1)/(k+1)}\right)^2}\right)$.
Staring at plots of the denominator for a while suggests that it is
minimized at $\ell=k+3$, the largest value with
$\floor{(\ell+2k-1)/(k+1)} ≤ 2$.
This gives
$\Prob{\abs*{X-\Exp{X}} ≥ t} ≤
2\exp\left(-\frac{t^2}{8\ceil{n/(k+3)}}\right)$, improving the bound
on $t$ from
$\Theta(√{n \log(1/ε)})$ to
$\Theta(√{(n/k) \log(1/ε)})$.

For large $k$, the expectation of any individual $X_k$ becomes small,
so we'd expect that Chernoff bounds would work better on the upper
bound side than the method
of bounded differences.  
Unfortunately, we don't have independence.  But
from
Problem~\ref{problem-variable-probability-chernoff-bounds}, we know
that the usual Chernoff bound works as long as we can show
$\ExpCond{X_i}{X_1,\dots,X_{i-1}} ≤ p_i$ for some sequence of fixed bounds $p_i$.

For $X_1$, there are no previous $X_i$, and we have $\Exp{X_1} = 2^{-k}$
exactly.

For $X_i$ with $i > 1$, fix $X_1,\dots,X_{i-1}$; that is, condition on
the event $X_j = x_j$ for all $j < i$ 
with some fixed sequence $x_1,\dots,x_{i-1}$.
Let's call this event $A$.  Depending on the particular values of the
$x_j$, it's not clear how conditioning on $A$ will affect $X_i$; but
we can split on the value of $W_{i-1}$ to show that either it has no
effect or $X_i = 0$:
\begin{align*}
\ExpCond{X_i}{A}
&=
 \ExpCond{X_i}{A, W_{i-1} = 0} \ProbCond{W_{i-1} = 0}{A}
+\ExpCond{X_i}{A, W_{i-1} = 1} \ProbCond{W_{i-1} = 1}{A}
\\
&≤
 2^{-k} \ProbCond{W_{i-1} = 0}{A} + 0
\\
&≤ 2^{-k}.
\end{align*}

So we have $p_i ≤ 2^{-k}$ for all $1 ≤ i ≤ n-k+1$.  This gives
$μ = 2^{-k}(n-k+1)$, and
$\Prob{X ≥ (1+δ)μ} ≤
\left(\frac{e^δ}{(1+δ)^{(1+δ)}}\right)^μ$.

If we want a two-sided bound, we can set $δ = 1$ (since $X$ can't
drop below $0$ anyway, and get
$\Prob{\abs*{X - \Exp{X}} > 2^{-k}(n-k+1)} ≤
\left(\frac{e}{4}\right)^{2^{-k}(n-k+1)}$.
This is exponentially small for $k = o(\lg n)$.  If $k$ is much bigger
than $\lg n$,
then we have $\Exp{X} \ll 1$, so Markov's inequality alone gives us a
strong concentration bound.

However, in both cases, the bounds are competitive with the previous
bounds from McDiarmid's inequality only if $\Exp{X} = O(√{n \log
(1/ε)})$.  So McDiarmid's inequality wins for $k = o(\log n)$,
Markov's inequality wins for $k = ω(\log n)$, and Chernoff bounds
may be useful for a small interval in the middle.
\end{enumerate}

\section{Assignment 3: due Wednesday, 2011-02-23, at 17:00} 

\subsection{Longest common subsequence}

A \concept{common subsequence} of two sequences $v$ and $w$ is
a sequence $u$ of length $k$ such that there exist indices
$i_1 < i_2 < \dots < i_k$ and
$j_1 < j_2 < \dots < j_k$ with $u_\ell = v_{i_\ell} = w_{j_\ell}$ for
all $\ell$.  For example, \texttt{ardab} is a common subsequence of
\texttt{abracadabra} and \texttt{cardtable}.

Let $v$ and $w$ be words of length $n$ over an alphabet of size
$n$ drawn independently and uniformly at random.
Give the best upper bound you can on the expected length of the
longest common subsequence of $v$ and $w$.

\subsubsection*{Solution}

Let's count the expectation of the number $X_k$ 
of common subsequences of length $k$.
We have $\binom{n}{k}$ choices of positions in $v$, and $\binom{n}{k}$
choices of positions in $w$; for each such choices, there is a
probability of exactly $n^{-k}$ that the corresponding positions
match.  This gives
\begin{align*}
\Exp{X_k} 
&= {\binom{n}{k}}^2 n^{-k}
\\
&< \frac{n^k}{(k!)^2}.
\\
&< \frac{n^k}{(k/e)^{2k}}
\\
&= \left(\frac{ne^2}{k^2}\right)^k.
\end{align*}

We'd like this bound to be substantially less than $1$.  We can't
reasonably expect this to happen unless the base of the exponent is
less than $1$, so we need $k > e√{n}$.  

If $k = (1+ε)e√{n}$ for any $ε > 0$, then 
$\Exp{X_k} < (1+ε)^{-2e√{n}} < \frac{1}{n}$ for sufficiently
large $n$.  It follows that the expected length of the longest common
subsequent is at most $(1+ε)e√{n}$ for sufficiently large
$n$ (because if there are no length-$k$ subsequences, the longest
subsequence has length at most $k-1$, and if there is at least one,
the longest has length at most $n$; this gives a bound of at most 
$(1-1/n)(k-1) + (1/n)n < k$).  So in general we have the length of the
longest common subsequence is at most $(1+o(1))e√{n}$.

Though it is not required by the problem, here is a quick argument
that the expected length of the longest common subsequence is
$Ω(√{n})$, based on the \concept{Erd\H{o}s-Szekeres
theorem}~\cite{ErdosS1935}.\footnote{As suggested by Benjamin Kunsberg.}
The Erd\H{o}s-Szekeres theorem says that any permutation of $n^2+1$
elements contains either an increasing sequence of $n+1$ elements
or a decreasing sequence of $n+1$ elements.  Given two random
sequences of length $n$, let $S$ be the set of all elements that
appear in both, and consider two permutations $ρ$ and $σ$ of $S$
corresponding to the order in which the elements appear in $v$ and
$w$, respectively (if an element appears multiple
times, pick one of the occurrences at random).  Then the
Erd\H{o}s-Szekeres theorem says that $ρ$ contains a sequence of
length at least $\floor{√{\card*{ρ}-1}}$ that is either increasing or decreasing
with respect to the order given by $σ$; by symmetry, the
probability that it is increasing is at least $1/2$.  This gives an
expected value for the longest common subsequence that is at least
$\Exp{√{\card*{ρ}-1}}/2$.

Let $X = \card*{ρ}$.  We can compute a lower bound $\Exp{X}$ easily; each possible
element fails to occur in $v$ with probability $(1-1/n)^n ≤ e^{-1}$,
and similarly for $w$.  So the chance that an element appears in both
sequences is at least $(1-e^{-1})^2$, and thus $\Exp{X} ≥ n(1-e^{-1})^2$.
What we want is $\Exp{√{X-1}}/2$; but here the fact that $√{x}$ is
concave means that $\Exp{√{X-1}} ≥ √{\Exp{X-1}}$ by
Jensen's inequality \eqref{eq-jensens-inequality}.  So we have
$\Exp{√{\card*{ρ}-1}}/2 ≥ √{n(1-e^{-1})^2-1}/2 \approx
\frac{1-e^{-1}}{2}√{n}$.

This is not a very good bound (empirically, the real bound seems to be
in the neighborhood of $1.9 √{n}$ when $n=10000$), but
it shows that the upper bound of $(1+o(1))e √{n}$ is tight up to
constant factors.

\subsection{A strange error-correcting code}
\label{section-a-strange-error-correcting-code}

Let $Σ$ be an alphabet of size $m+1$ that includes $m$ non-blank
symbols and a special blank symbol.  
Let $S$ be a set of
$\binom{n}{k}$ strings of length $n$
with non-blank symbols in exactly $k$ positions each, such that no two
strings in $S$ have non-blank symbols in the same $k$ positions.

For what value of $m$ can you show $S$ exists such that no two strings
in $S$ have the same non-blank symbols in $k-1$ positions?

\subsubsection*{Solution}

This is a job for the Lovász Local Lemma.  And it's even symmetric,
so we can use the symmetric version
(Corollary~\ref{corollary-symmetric-lovasz-local-lemma}).

Suppose we assign the non-blank symbols to each string uniformly and
independently at random.  For each $A⊆ S$ with $\card*{A} = k$, 
let $X_A$ be the string that has non-blank symbols in all positions in
$A$.  For each pair of subsets $A,B$ with $\card*{A} = \card*{B} = k$
and $\card*{A \cap B} = k-1$,
let $C_{A,B}$ be the event that $X_A$ and $X_B$ are identical on
all positions in $A \cap B$.  Then $\Prob{C_{A,B}} = m^{-k+1}$.

We now now need to figure out how many events are in each
neighborhood $Γ(C_{A,B})$.  Since $C_{A,B}$ depends only on the
choices of values for $A$ and $B$, it is independent of any events
$C_{A',B'}$ where neither of $A'$ or $B'$ is equal to $A$ or $B$.  So
we can make $Γ(C_{A,B})$ consist of all events $C_{A,B'}$ and
$C_{A',B}$ where $B' ≠ B$ and $A' ≠ A$.

For each fixed $A$, there are exactly $(n-k)k$ events $B$ that overlap it
in $k-1$ places, because we can specify $B$ by choosing the elements
in $B ∖ A$ and $A ∖ B$.  This gives $(n-k)k - 1$
events $C_{A,B'}$ where $B' ≠ B$.  Applying the same argument for
$A'$ gives a total of $d=2(n-k)k-2$ events in $Γ(C_{A,B})$.
Corollary~\ref{corollary-symmetric-lovasz-local-lemma} applies if
$ep(d+1) ≤ 1$, which in this case means $em^{-(k-1)}(2(n-k)k-1) ≤
1$.  Solving for $m$ gives
\begin{align}
\label{eq-strange-ecc-bound}
m &≥ \left(2e(n-k)k-1\right)^{1/(k-1)}.
\end{align}

For $k \ll n$, the $(n-k)^{1/(k-1)} \approx n^{1/(k-1)}$ 
term dominates the shape of the
right-hand side asymptotically as $k$ gets large, since everything
else goes to $1$.  This suggests we
need $k = Ω(\log n)$ to get $m$ down to a constant.

Note that \eqref{eq-strange-ecc-bound}
doesn't work very well when $k=1$.\footnote{Thanks to
Brad Hayes for pointing this out.}  For the $k=1$ case, there is no
overlap in non-blank positions between different strings, so $m=1$ is
enough.

\subsection{A multiway cut}

Given a graph $G=(V,E)$, a \concept{$3$-way cut} is a set of
edges whose endpoints lie in different parts of a partition of the
vertices $V$ into three disjoint parts $S\cup T \cup U = V$.

\begin{enumerate}
\item Show that any graph with $m$ edges has a $3$-way cut with at
least $2m/3$ edges.
\item Give an efficient deterministic algorithm for finding such a cut.
\end{enumerate}

\subsubsection*{Solution}

\begin{enumerate}
\item Assign each vertex independently to $S$, $T$, or $U$ with 
probability $1/3$ each.  Then the probability that any edge $uv$ is
contained in the cut is exactly $2/3$.  Summing over all edges gives
an expected $2m/3$ edges.
\item We'll derandomize the random vertex assignment using the method
of conditional probabilities.  Given a partial assignment of the
vertices, we can compute the conditional expectation of the size of
the cut assuming all other vertices are assigned randomly: each 
edge with matching assigned endpoints contributes $0$ to the total,
each edge with non-matching assigned endpoints contributes $1$, and
each edge with zero or one assigned endpoints contributes $2/3$.  
We'll pick values for the vertices in some arbitrary order to maximize
this conditional expectation (since our goal is to get a large cut).
At each step, we need only consider the effect on edges incident to
the vertex we are assigning whose other endpoints are already
assigned, because the contribution of any other edge is not changed by
the assignment.  Then maximizing the conditional probability is done by
choosing an assignment that matches the assignment of the fewest
previously-assigned neighbors: in other words, the natural greedy
algorithm works.  The cost of this algorithm is $O(n+m)$, since we
loop over all vertices and have to check each edge at most once for
each of its endpoints.
\end{enumerate}

\section{Assignment 4: due Wednesday, 2011-03-23, at 17:00}
\label{appendix-2011-hw4}

\subsection{Sometimes successful betting strategies are possible}

You enter a casino with $X_0 = a$ dollars, and leave if you reach $0$
dollars or $b$ or more dollars, where $a,b ∈ ℕ$.
The casino is unusual in that it
offers arbitrary fair games subject to the requirements that:
\begin{itemize}
\item Any payoff resulting from a bet must be a nonzero integer in
the range $-X_t$ to $X_t$, inclusive, where $X_t$ is your current wealth.
\item The expected payoff must be exactly $0$.  (In other words, your
assets $X_t$ should form a martingale sequence.)
\end{itemize}
For example, if you have $2$ dollars, you may make a bet that pays off
$-2$ with probability $2/5$, $+1$ with probability $2/5$ and $+2$ with
probability $1/5$; but you may not make a bet that pays off $-3$,
$+3/2$, or $+4$ under any circumstances, or a bet that pays off $-1$
with probability $2/3$ and $+1$ with probability $1/3$.

\begin{enumerate}
\item What strategy should you use to maximize your
chances of leaving with at least $b$ dollars?
\item What strategy should you use to maximize your
changes of leaving with nothing?
\item What strategy should you use to maximize the
number of bets you make before leaving?
\end{enumerate}

\subsubsection*{Solution}

\begin{enumerate}
\item Let $X_t$ be your wealth at time $t$, and let $τ$ be the
stopping time when you leave.  Because $\Set{X_t}$ is a martingale,
$\Exp{X_0} = a = \Exp{X_τ} = \Prob{X_τ ≥ b}
\ExpCond{X_τ}{X_τ ≥ b}$.  So $\Prob{X_τ ≥ b}$ is
maximized by making $\ExpCond{X_τ }{ X_τ ≥ b}$ as small as possible.  It can't be any smaller than
$b$, which can be obtained exactly by making only $\pm 1$ bets.
The gives a probability of leaving with $b$ of exactly $a/b$.
\item Here our goal is to minimize $\Prob{X_τ ≥ b}$, so we want to
make $\ExpCond{X_τ}{X_τ ≥ b}$ as large as possible.
The largest value of $X_τ$ we can possibly reach is $2(b-1)$; we
can obtain this value by betting $\pm 1$ until we reach $b-1$, then
making any fair bet with positive payoff $b-1$ (for example, $\pm
(b-1)$ with equal probability works, as does a bet that pays off $b-1$
with probability $1/b$ and $-1$ with probability $(b-1)/b$).
In this case we get a probability of leaving with $0$ of
$1-\frac{a}{2(b-1)}$.
\item 
For each $t$,
let $δ_t = X_t - X_{t-1}$ and
$V_t = \VarCond{δ_t }{ ℱ_t}$.
We have previously shown (see the footnote to
§\ref{section-stopping-times-and-random-walks})
that $\Exp{X_τ^2} = \Exp{X_0^2} + \Exp{∑_{t=1}^{τ} V_t}$ where
$τ$ is an appropriate stopping time.
When we stop, we know that $X_τ^2 ≤ (2(b-1))^2$, which puts an upper
bound on $\Exp{∑_{i=1}^{τ} V_i}$.  We can spend this bound most
parsimoniously by minimizing $V_i$ as much as possible.  If we make
each $δ_t = \pm 1$, we get the smallest possible value for $V_t$
(since any change contributes at least $1$ to the variance).  However,
in this case we don't get all the way out to $2(b-1)$ at the high
end; instead, we stop at $b$,
giving an expected number of steps equal to $a(b-a)$.

We can do a bit better than this by changing our strategy at $b-1$.
Instead of betting $\pm 1$, let's pick some $x$ and 
place a bet that pays off $b-1$ with probability $\frac{1}{b}$
and $-1$ with probability $\frac{b-1}{b} = 1 - \frac{1}{b}$.  (The idea here is to
minimize the conditional variance while still allowing ourselves to
reach $2(b-1)$.)  Each ordinary random walk step has $V_t = 1$; a
``big'' bet starting at $b-1$ has $V_t = 1 - \frac{1}{b} +
\frac{(b-1)^2}{b} = \frac{b - 1 + b^2 - 2b + 1}{b} = b - \frac{1}{b}$.

To analyze this process, observe that starting from $a$, we first
spending $a(b-a-1)$ steps on average to reach either $0$ (with
probability $1-\frac{a}{b-1}$ or $b-1$ (with probability
$\frac{a}{b-1}$.  In the first case, we are done.  Otherwise, we take
one more step, then with probability $\frac{1}{b}$ we lose and with
probability $\frac{b-1}{b}$ we continue starting from $b-2$.  We can
write a recurrence for our expected number of steps $T(a)$ starting
from $a$, as:
\begin{align}
T(a)
&= a(b-a-1) + \frac{a}{b-1} \left(1 + \frac{b-1}{b} T(b-2)\right).
\label{eq-maximum-steps-recurrence}
\intertext{When $a = b-2$, we get}
T(b-2)
&= (b-2) + \frac{b-2}{b-1} \left(1 + \frac{b-1}{b} T(b-2)\right)
\nonumber
\\
&= (b-2) \left(1 + \frac{1}{b-1}\right) + \frac{b-2}{b} T(b-2),
\nonumber
\intertext{which gives}
T(b-2) &= \frac{(b-2)\frac{2b-1}{b-1}}{2/b} \nonumber \\
&= \frac{b(b-2)(2b-1)}{2(b-1)}.
\nonumber
\intertext{Plugging this back into \eqref{eq-maximum-steps-recurrence}
gives}
T(a)
&= a(b-a-1) + \frac{a}{b-1} \left(1 + \frac{b-1}{b} \frac{b(b-2)(2b-1)}{2(b-1)}\right)
\nonumber
\\
&= ab - a^2 + a + \frac{a}{b-1} + \frac{a(b-2)(2b-1)}{2(b-1)}
\nonumber
\\
&= \frac{3}{2} ab + O(b).
\end{align}
This is much better than the $a(b-a)$
value for the straight $\pm 1$ strategy, especially when $a$ is also
large.

I don't know if this particular strategy is in fact optimal, but
that's what I'd be tempted to bet.
\end{enumerate}

\subsection{Random walk with reset}

Consider a random walk on $ℕ$ that goes up with probability
$1/2$, down with probability $3/8$, and resets to $0$ with probability
$1/8$.  When $X_t > 0$, this gives:
\begin{align*}
X_{t+1} &=
\begin{cases}
X_t + 1 & \text{with probability $1/2$,} \\
X_t - 1 & \text{with probability $3/8$, and} \\
0       & \text{with probability $1/8$.}
\end{cases}
\end{align*}
When $X_t = 0$, we let $X_{t+1} = 1$ with probability $1/2$ and $0$
with probability $1/2$.

\begin{enumerate}
\item What is the stationary distribution of this process?
\item What is the mean recurrence time $μ_n$ for some state $n$?
\item Use $μ_n$ to get a tight asymptotic (i.e., big-$\Theta$) bound on $μ_{0,n}$, the
expected time to reach $n$ starting from $0$.
\end{enumerate}

\subsubsection*{Solution}

\begin{enumerate}
\item For $n>0$, we have $π_n = \frac{1}{2} π_{n-1} + 
\frac{3}{8} π_{n+1}$, with a base case $π_0 = \frac{1}{8} +
\frac{3}{8} π_0 + \frac{3}{8} π_1$.

The $π_n$ expression is a linear homogeneous recurrence, so its
solution consists of linear combinations of terms $b^n$, where $b$
satisfies $1 = \frac{1}{2} b^{-1} + \frac{3}{8} b$.  The solutions to
this equation are $b=2/3$ and $b=2$; we can exclude the $b=2$ case
because it would make our probabilities blow up for large $n$.  So we
can reasonably guess $π_n = a(2/3)^n$ when $n > 0$.

For $n=0$, substitute
$π_0 = \frac{1}{8} + \frac{3}{8} π_0 + \frac{3}{8} a (2/3)$ to get
$π_0 = \frac{1}{5} + \frac{2}{5} a$.
Now substitute
\begin{align*}
π_1 
&= (2/3) a 
\\
&= \frac{1}{2} π_0 + \frac{3}{8} a (2/3)^2
\\
&= \frac{1}{2} \left(\frac{1}{5} + \frac{2}{5}a\right) + \frac{3}{8} a (2/3)^2
\\
&=
\frac{1}{10} + \frac{11}{30} a,
\end{align*}
which we can solve to get $a = 1/3$.

So our candidate $π$ is $π_0 = 1/3$, $π_n = (1/3)(2/3)^n$, and
in fact we can drop the special case for $π_0$.

As a check, $∑_{i=0}^{n} π_n = (1/3) ∑_{i=0}^{n} (2/3)^n =
\frac{1/3}{1-2/3} = 1$.

\item Since $μ_n = 1/π_n$, we have $μ_n = 3(3/2)^n$.

\item In general, 
let $μ_{k,n}$ be the expected time to reach $n$ starting at
$k$.  Then $μ_n = μ_{n,n} = 1 + \frac{1}{8} μ_{0,n} + \frac{1}{2}
μ_{n+1,n} + \frac{3}{8} μ_{n-1,n} ≥ 1 + μ_{0,n}/8$.  It
follows that $μ_{0,n} ≤ 8μ_n + 1 = 24(3/2)^n + 1 = O((3/2)^n)$.

For the lower bound, observe that $μ_n ≤ μ_{n,0} + μ_{0,n}$.
Since there is a $1/8$ chance of reaching $0$ from any state, we have
$μ_{n,0} ≤ 8$.  It follows that $μ_n ≤ 8 + μ_{0,n}$ or
$μ_{0,n} ≥ μ_n - 8 = Ω((3/2)^n)$.
\end{enumerate}

\subsection{Yet another shuffling algorithm}

Suppose we attempt to shuffle a deck of $n$ cards by picking a card
uniformly at random, and swapping it with the top card.  Give the best
bound you can on the mixing time for this process to reach a total
variation distance of $ε$ from the uniform distribution.

\subsubsection*{Solution}

It's tempting to use the same coupling as for move-to-top (see
§\ref{section-shuffling-algorithms}).  This would be that at
each step we choose the same card to swap to the top position, which
increases by at least one the number of cards that are in the same position in
both decks.  The problem is that at the next step, these two cards are
most likely separated again, by being swapped with other cards in two
different positions.

Instead, we will do something slightly more clever.  Let $Z_t$ be the
number of cards in the same position at time $t$.  If the top cards
of both decks are equal, we swap both to the same position chosen
uniformly at random.  This has no effect on $Z_t$.
If the top cards of both decks are not
equal, we pick a card uniformly at random and swap it to the top in
both decks.  This increases $Z_t$ by at least $1$, unless we happen to pick
cards that are already in the same position; so $Z_t$ increases by at
least $1$ with probability $1-Z_t/n$.

Let's summarize a state by an ordered pair $(k,b)$ where $k = Z_t$ and
$b$ is $0$ if the top cards are equal and $1$ if they are not equal.
Then we have a Markov chain where $(k,0)$ goes to $(k,1)$ with
probability $\frac{n-k}{n}$ (and otherwise stays put); and $(k,1)$
goes to $(k+1,0)$ (or higher) with probability $\frac{n-k}{n}$ and to $(k,0)$ 
with probability $\frac{k}{n}$.

Starting from $(k,0)$, we expect to wait $\frac{n}{n-k}$ steps on
average to reach $(k,1)$, at which point we move to $(k+1,0)$ or back
to $(k,0)$ in one more step; we iterate through this process
$\frac{n}{n-k}$ times on average before we are successful.  This gives
an expected number of steps to get from $(k,0)$ to $(k+1,0)$ (or
possibly a higher value) of
$\frac{n}{n-k}\left(\frac{n}{n-k}+1)\right)$.  Summing over $k$ up to
$n-2$ (since once $k > n-2$, we will in fact have $k=n$, since $k$
can't be $n-1$), we get
\begin{align*}
\Exp{τ}
&≤ ∑_{k=0}^{n-2} \frac{n}{n-k} \left(\frac{n}{n-k} + 1\right)
\\
&= ∑_{m=2}^{n} \left(\frac{n^2}{m^2} + \frac{n}{m}\right)
\\
&≤ n^2 \left(\frac{π^2}{6} - 1\right) + n \ln n.
\\
&= O(n^2).
\end{align*}

So we expect the deck to mix in $O(n^2 \log (1/ε))$ steps.  (I don't know if this
is the real bound; my guess is that it should be closer to $O(n \log
n)$ as in all the other shuffling procedures.)

\section{Assignment 5: due Thursday, 2011-04-07, at 23:59}

\subsection{A reversible chain}

Consider a random walk on $ℤ_m$, where
$p_{i,i+1} = 2/3$ for all $i$ and $p_{i,i-1} = 1/3$ for all $i$ except
$i=0$.  Is it possible to assign values to $p_{0,m-1}$ and $p_{0,0}$
to make this
chain reversible, and if so, what stationary distribution do you get?

\subsubsection*{Solution}

Suppose we can make this chain reversible, and let $π$ be the
resulting
stationary distribution.  From the
detailed balance equations, we have $(2/3) π_i = (1/3) π_{i+1}$ or
$π_{i+1} = 2 π_i$ for $i = 0\dots m-2$.  The solution to this
recurrence is $π_i = 2^i π_0$, which gives $π_i =
\frac{2^i}{2^m-1}$ when we set $π_0$ to get $∑_i π_i = 1$.

Now solve $π_0 p_{0,m-1} = π_{m-1} p_{m-1,0}$ to get
\begin{align*}
p_{0,m-1}
&= \frac{π_{m-1} p_{m-1,0}}{π_0}
\\
&= 2^{m-1} (2/3)
\\
&= 2^{m}/3.
\end{align*}

This is greater than $1$ for $m > 1$, so except for the degenerate
cases of $m=1$ and $m=2$, it's not possible to make the chain reversible.

\subsection{Toggling bits}

Consider the following Markov chain on an array of $n$ bits $a[1],
a[2], \dots a[n]$.  At each step, we choose a position $i$ uniformly
at random.  We then change $A[i]$ to $\neg A[i]$ with probability
$1/2$, provided $i = 1$ or $A[i-1] = 1$ (if neither condition holds
hold, do nothing).\footnote{Motivation: Imagine each bit represents
whether a node in some distributed system is inactive (0) or active
(1), and you can only change your state if you have an active left neighbor
to notify.  Also imagine that there is an always-active \emph{base
station} at $-1$ (alternatively, imagine that
this assumption makes the problem easier than
the other natural arrangement where we put all the nodes in a ring).}

\begin{enumerate}
\item What is the
stationary distribution?
\item How quickly does it converge?
\end{enumerate}

\subsubsection*{Solution}

\begin{enumerate}
\item 
First let's show irreducibility.  Starting from an arbitrary
configuration, repeatedly switch the leftmost $0$ to a $1$ (this is
always permitted by the transition rules); after at most $n$ steps, we
reach the all-$1$ configuration.  Since we can repeat this process in
reverse to get to any other configuration, we get that every
configuration is reachable from every other configuration in at most
$2n$ steps ($2n-1$ if we are careful about handling the all-$0$
configuration separately).

We also have that for any two adjacent configurations $x$ and $y$,
$p_{xy} = p_{yx} = \frac{1}{2n}$.  So we have a reversible,
irreducible, aperiodic (because there exists at least one self-loop)
chain with a uniform stationary distribution $π_x = 2^{-n}$.

\item Here is a bound using the obvious coupling, where we choose the
same position in $X$ and $Y$ and attempt to set it to the same value.
To show this coalesces, given $X_t$ and $Y_t$ define $Z_t$ to be the
position of the rightmost $1$ in the common prefix of $X_t$ and $Y_t$,
or $0$ if there is no $1$ in the common prefix of $X_t$ and $Y_t$.
Then $Z_t$ increases by at least $1$ 
if we attempt to set position $Z_t+1$ to $1$, which occurs
with probability $\frac{1}{2n}$, and decreases by at most $1$ if we
attempt to set $Z_t$ to $0$, again with probability $\frac{1}{2n}$.

It follows that $Z_t$ reaches $n$ no later than a $\pm 1$ random walk
on $0\dots n$ with reflecting barriers that takes a step every $1/n$
time units on average.  The expected number of steps to reach $n$ from
the worst-case starting position of $0$ is exactly $n^2$.  (Proof: 
model the random walk with a reflecting barrier at $0$ by 
folding a random walk with absorbing barriers at $\pm n$ in half, then
use the bound from
§\ref{section-stopping-times-and-random-walks}.)
We must then multiply this by $n$ to get an expected $n^3$ steps in the
original process.  So the two copies coalesce in at most $n^3$ expected
steps.  My suspicion is one could 
improve this bound with a better analysis by
using the bias toward increasing $Z_t$ to get the expected time to
coalesce down to $O(n^2)$, but I don't know any clean way to do this.

The path coupling version of this is that we look at two adjacent
configurations $X_t$ and $Y_t$, use the obvious coupling again, and
see what happens to $\ExpCond{d(X_{t+1},Y_{t+1})}{X_t,Y_t}$, where the
distance is the number of transitions needed to convert $X_t$ to $Y_t$
or vice versa.  If we pick the position $i$ where $X_t$ and $Y_t$ differ,
then we coalesce; this occurs with probability $1/n$.  If we change
the $1$ to the left of $i$ to a $0$, then $d(X_{t+1},Y_{t+1})$ rises
to $3$ (because to get from $X_{t+1}$ to $Y_{t+1}$, we have to change
position $i-1$ to $1$, change position $i$, and then change position
$i-1$ back to $0$); this occurs with probability $1/2n$ if $i > 1$.
But we can also get into trouble if we try to change position $i+1$;
we can only make the change in one of $X_t$ and $Y_t$, so we get
$d(X_{t+1},Y_{t+1}) = 2$ in this case, which occurs with probability
$1/2n$ when $i < n$.  Adding up all three cases gives a
worst-case expected change of $-1/n + 2/2n + 1/2n = 1/2n > 0$.  So unless
we can do something more clever, path coupling won't help us here.

However, it is possible to get a bound using canonical paths, but the best
bound I could get was not as good as the coupling
bound.  The basic idea is that we will change $x$ into $y$ one bit
at a time (from left to right, say), so that we will go through a
sequence of intermediate states of the form $y[1] y[2] \dots y[i]
x[i+1] x[i+2] \dots x[n]$.  But to change $x[i+1]$ to $y[i+1]$, we may
also need to reach out with a tentacle of $1$ bits from from the last
$1$ in the current prefix of $y$
(and then retract it afterwards).  Given a particular
transition where we change a $0$ to a $1$, we can reconstruct the
original $x$ and $y$ by specifying (a) which bit $i$ at or after our
current position we are trying to
change; (b) which $1$ bit before our current position is the last
``real'' $1$ bit in $y$ as opposed to something we are creating to
reach out to position $i$; and (c) the values of $x[1]\dots x[i-1]$
and $y[i+1] \dots y[i]$.  A similar argument applies to $1→0$
transitions.  So we are routing at most $n^2
2^{n-1}$ paths across each transition, giving a bound on the
congestion
\begin{align*}
ρ &≤
\left(\frac{1}{2^{-n}/2n}\right) n^2 2^{n-1} 2^{-2n}
\\
&= n^3.
\end{align*}
The bound on $τ_2$ that follows from this is $8n^6$, which is
pretty bad (although the constant could be improved by counting the
(a) and (b) bits more carefully).
As with the coupling argument, it may be that there is a less
congested set of canonical paths that gives a better bound.
\end{enumerate}

\subsection{Spanning trees}

Suppose you have a connected graph $G = (V,E)$ with $n$ nodes and $m$
edges.  Consider the following Markov process.  Each state $H_t$ is a
subgraph of $G$ that is either a \index{spanning tree}spanning tree
or a spanning tree plus an additional edge.  At each step, flip a
fair coin.  If it comes up heads, choose an edge $e$ uniformly at
random from $E$ and let $H_{t+1} = H_t \cup \Set{e}$ if $H_t$ is a
spanning tree and let $H_{t+1} = H_t ∖ \Set{e}$ if $H_t$ is not
a spanning tree and $H_t ∖ \Set{e}$ is connected.  If it comes
up tails and $H_t$ is a spanning tree, let $H_{t+1}$ be some other
spanning tree, sampled uniformly at random.  In all other cases, let
$H_{t+1} = H_t$.

Let $N$ be the number of states in this Markov chain.
\begin{enumerate}
\item What is the stationary distribution?
\item How quickly does it converge?
\end{enumerate}

\subsubsection*{Solution}

\begin{enumerate}
\item Since every transition has a matching reverse transition 
with the same transition probability, the chain is reversible with a
uniform stationary distribution $π_H = 1/N$.
\item Here's a coupling that coalesces in at most $4m/3 + 2$ expected steps:
\begin{enumerate}
\item If $X_t$ and $Y_t$ are both trees, then send them to the same
tree with probability $1/2$; else let them both add edges
independently (or we could have them add the same edge—it doesn't
make any difference to the final result).
\item If only one of $X_t$ and $Y_t$ is a tree, with probability $1/2$
scramble the tree while attempting to remove an edge from the
non-tree, and the rest of the time scramble the non-tree (which has no
effect) while attempting to add an edge to the tree.  Since the
non-tree has at least three edges that can be removed, this puts
$(X_{t+1},Y_{t+1}$ in the two-tree case with probability at least
$3/2m$.
\item If neither $X_t$ nor $Y_t$ is a tree, attempt to remove an edge
from both.  Let $S$ and $T$ be the sets of edges that we can remove
from $X_t$ and $Y_t$, respectively, and let $k =
\min(\card*{S},\card*{T}) ≥ 3$.  Choose $k$ edges from each of $S$ and
$T$ and match them, so that if we remove one edge from each pair, we
also remove the other edge.  As in the previous case, this puts
$(X_{t+1},Y_{t+1}$ in the two-tree case with probability at least
$3/2m$.
\end{enumerate}
To show this coalesces, starting from an arbitrary state, we reach a
two-tree state in at most $2m/3$ expected steps.  After one more step,
we either coalesce (with probability $1/2$) or restart from a new
arbitrary state.  This gives an expected coupling time of at most
$2(2m/3+1) = 4m/3 + 2$ as claimed.
\end{enumerate}

\section{Assignment 6: due Monday, 2011-04-25, at 17:00}

\subsection{Sparse satisfying assignments to DNFs}

Given a formula in disjunctive normal form, we'd like to estimate the
number of satisfying assignments in which exactly $w$ of the variables
are true.  Give a fully polynomial-time randomized approximation
scheme for this problem.

\subsubsection*{Solution}

Essentially, we're going to do the Karp-Luby covering
trick~\cite{KarpL1985} described in
§\ref{section-approximating-sharp-dnf}, but will tweak the
probability distribution when we generate our samples so that we only
get samples with weight $w$.

Let $U$ be the set of assignment with weight $w$ (there are exactly
$\binom{n}{w}$ such assignments, where $n$ is the number of
variables).  For each clause $C_i$, 
let $U_i = \SetWhere{ x ∈ U }{ C_i(x) = 1 }$.  Now observe that:
\begin{enumerate}
\item We can compute $\card*{U_i}$.  Let $k_i = \card*{C_i}$ be the
number of variables in $C_i$ and
$k^+_i = \card*{C^+_i}$ the number of
variables that appear in positive form in $C_i$.  Then $\card*{U_i} =
\binom{n-k_i}{w-k^+_i}$ is the number of ways to make a total of $w$
variables true using the remaining $n-k_i$ variables.
\item We can sample uniformly from $U_i$, by sampling a set of $w-k^+_i$
true variables not in $C_i$ uniformly from all variables not in $C_i$.
\item We can use the values computed for $\card*{U_i}$ to sample $i$
proportionally to the size of $\card*{U_i}$.
\end{enumerate}

So now we sample pairs $(i,x)$ with $x ∈ U_i$ uniformly at random by
sampling $i$ first, then sampling $x ∈ U_i$.  As in the original
algorithm, we then count $(i,x)$ if and only if $C_i$ is the leftmost
clause for which $C_i(x) = 1$.  The same argument that at least $1/m$
of the $(i,x)$ pairs count applies, and so we get the same bounds as
in the original algorithm.

\subsection{Detecting duplicates}

Algorithm~\ref{algorithm-dubious-duplicate-detector} attempts to
detect duplicate values in an input array $S$ of length $n$.

\begin{algorithm}
\caption{Dubious duplicate detector}
\label{algorithm-dubious-duplicate-detector}
Initialize $A[1\dots n]$ to $\bot$\\
Choose a hash function $h$\\
\For{$i \leftarrow 1 \dots n$}{
    $x \leftarrow S[i]$ \\
    \eIf{$A[h(x)] = x$}{
        \Return \True
    }{
        $A[h(x)] \leftarrow x$
    }
}
\Return \False
\end{algorithm}

It's easy to see that
Algorithm~\ref{algorithm-dubious-duplicate-detector} never returns
\True\ unless some value appears twice in $S$.  But maybe it misses
some duplicates it should find.

\begin{enumerate}
\item Suppose $h$ is a random function.  What is the worst-case
probability that Algorithm~\ref{algorithm-dubious-duplicate-detector}
returns $\False$ if $S$ contains two copies of some value?
\item Is this worst-case probability affected if $h$ is drawn instead
from a $2$-universal family of hash functions?
\end{enumerate}

\subsubsection*{Solution}

\begin{enumerate}
\item Suppose that $S[i] = S[j] = x$ for $i < j$.  Then the algorithm will
see $x$ in $A[h(x)]$ on iteration $j$ and return $\True$, unless it is
overwritten by some value $S[k]$ with $i < k < j$.  This occurs if
$h(S[k]) = h(x)$, which occurs with probability exactly
$1-(1-1/n)^{j-i-1}$ if we consider all possible $k$.  This quantity is
maximized at $1-(1-1/n)^{n-2} \approx 1-(1-1/n)^2/e \approx 1 -
(1-1/2n)/e$ when $i=1$ and $j=n$.
\item As it happens, the algorithm can fail pretty badly if all we
know is that $h$ is $2$-universal.  What we can show is that the
probability that some $S[k]$ with $i < j < k$ gets hashed to the same
place as $x = S[i] = S[j]$ in the analysis above is at most
$(j-i-1)/n ≤ (n-2)/n = (1-2/n)$, since each $S[k]$ has at most a $1/n$ chance of colliding
with $x$ and the union bound applies.  But it is possible to construct
a $2$-universal family for which we get exactly this probability in
the worst case.

Let $U = \Set{0 \dots n}$, and define for each $a$ in $\Set{0\dots n-1}$
$h_a: U \rightarrow n$ by $h_a(n) = 0$ and $h_a(x) = (x + a) \bmod
n$ for $x ≠ n$.  Then $H = \Set{ h_a }$ is $2$-universal, since
if $x ≠ y$ and neither $x$ nor $y$ is $n$, $\Prob{h_a(x)=h_a(y)} = 0$,
and if one of $x$ or $y$ is $n$, $\Prob{h_a(x) = h_a(y)} = 1/n$.
But if we use this family in
Algorithm~\ref{algorithm-dubious-duplicate-detector} with $S[1]=S[n] =
n$ and $S[k] = k$ for $1 < k < n$, then there are
$n-2$ choices of $a$ that put one of the middle values in $A[0]$.
\end{enumerate}

\subsection{Balanced Bloom filters}

A clever algorithmist decides to solve the problem of \index{Bloom
filter}Bloom filters
filling up with ones by capping the number of ones at $m/2$.
As in a standard Bloom filter, an 
element is inserted by writing ones to $A[h_1(x)], A[h_2(x)],
\dots, A[h_k(x)]$; but
after writing each one, if the number of one bits in the
bit-vector is more than $m/2$, 
one of the ones in the vector (chosen uniformly at random) is changed back to a
zero.

Because some of the ones associated with a particular element might be
deleted, the membership test answers yes if at least $3/4$ of the bits
$A[h_1(x)] \dots A[h_k(x)]$ are ones.

To simplify the analysis, you may assume that the $h_i$ are
independent random functions.  You may also assume that $(3/4)k$ is an
integer.

\begin{enumerate}
\item Give an upper bound on
the probability of a false positive when testing for a
value $x$ that has never been inserted.
\item Suppose that we insert $x$ at some point, and then follow this
insertion with a sequence of insertions of new, distinct values $y_1,
y_2, \dots$.  Assuming a worst-case state before inserting $x$, give
asymptotic upper and lower bounds on the expected number of insertions
until a test for $x$ fails.
\end{enumerate}

\subsubsection*{Solution}

\begin{enumerate}
\item The probability of a false positive is maximized when exactly half the bits in $A$ are one.
If
$x$ has never been inserted, each $A[h_i(x)]$ is equally likely to be
zero or one.  So $\Prob{\text{false positive for $x$}} = \Prob{S_k ≥
(3/4)k}$ when $S_k$ is a binomial random variable with parameters
$1/2$ and $k$.  Chernoff bounds give 
\begin{align*}
\Prob{S_k ≥ (3/4)k}
&= \Prob{S_k ≥ (3/2)\Exp{S_k}}
\\
&≤ \left(\frac{e^{1/2}}{(3/2)^{3/2}}\right)^{k/2}
\\
&≤ (0.94734)^k.
\end{align*}

We can make this less than any fixed $ε$ by setting $k ≥
20 \ln(1/ε)$ or thereabouts.
\item For false negatives, we need to look at how quickly the bits for
$x$ are eroded away.  A minor complication is
that the erosion may start even as we are setting $A[h_1(x)] \dots
A[h_k(x)]$.

Let's consider a single bit $A[i]$ and look at how it changes after
(a) setting $A[i] = 1$, and (b) setting some random $A[r] = 1$.

In the first case, $A[i]$ will be $1$ after the assignment unless it
is set back to zero, which occurs with probability $\frac{1}{m/2+1}$.  This
distribution does not depend on the prior value of $A[i]$.

In the second case, if $A[i]$ was previously $0$, it becomes $1$ with
probability 
\begin{align*}
\frac{1}{m}  \left(1-\frac{1}{m/2+1}\right)
&= \frac{1}{m} ⋅ \frac{m/2}{m/2+1}
\\
&= \frac{1}{m+2}.
\end{align*}

If
it was previously $1$, it becomes $0$ with probability 
\begin{align*}
\frac{1}{2} ⋅ \frac{1}{m/2+1}
&= \frac{1}{m+2}.
\end{align*}

So after the initial assignment, $A[i]$ just flips its value with
probability $\frac{1}{m+2}$.  

It is convenient to represent $A[i]$ as $\pm 1$; let $X^t_i = -1$ if
$A[i] = 0$ at time $t$, and $1$ otherwise.  Then $X^t_i$ satisfies the
recurrence
\begin{align*}
\ExpCond{X^{t+1}_i}{X^t_i}
&= \frac{m+1}{m+2} X^t_i - \frac{1}{m+2} X^t_i
\\
&= \frac{m}{m+2} X^t_i.
\\
&= \left(1-\frac{2}{m+2}\right) X^t_i.
\end{align*}

We can extend this to $\ExpCond{X^t_i}{X^0_i} = \left(1-\frac{2}{m+2}\right)^t
X^0_i \approx e^{-2t/(m+2)} X^0_i$.

Similarly, after setting $A[i] = 1$, we get $\Exp{X_i} = 1 - 2
\frac{1}{m/2+1} = 1 - \frac{4}{2m+1} = 1-o(1)$.

Let $S^t = ∑_{i=1}^{k} X^t_{h_i(x)}$.  Let $0$ be the time at which
we finish inserting $x$.  Then each for each $i$ we have 
\begin{displaymath}
1-o(1) e^{-2k/(m+2)}
≤ \Exp{X^0_{h_i(x)}}
≤ 1-o(1),
\end{displaymath}
from which it follows that
\begin{displaymath}
k (1-o(1)) e^{-2k/(m+2)}
≤ \Exp{S^0}
≤ 1-o(1)
\end{displaymath}
and in general that
\begin{displaymath}
k (1-o(1)) e^{-2(k+t)/(m+2)}
≤ \Exp{S^t}
≤ 1-o(1) e^{-2t/(m+2)}.
\end{displaymath}

So for any fixed $0 < ε < 1$ and sufficiently large $m$, we
will have $\Exp{S^t} = ε k$ for some $t'$ where $t ≤ t' ≤ k+t$
and $t = \Theta(m \ln (1/ε))$.

We are now looking for the time at which $S^t$ drops below $k/2$ (the
$k/2$ is because we are working with $\pm 1$ variables).  We will
bound when this time occurs using Markov's inequality.

Let's look for the largest time $t$ with
$\Exp{S^t} ≥ (3/4)k$.  Then $\Exp{k - S^t} ≤ k/4$ and
$\Prob{k - S^t ≥ k/2} ≤ 1/2$.  It follows that after $\Theta(m) - k$
operations, $x$ is still visible with probability $1/2$, which implies
that the expected time at which it stops being visible is at least
$(Ω(m) - k)/2$.  To get the expected number of insert operations,
we divide by $k$, to get $Ω(m/k)$.

For the upper bound, apply the same reasoning to the first time at
which $\Exp{S^t} ≤ k/4$.  This occurs at time $O(m)$ at the latest
(with a different constant), so after $O(m)$ steps there is at most a
$1/2$ probability that $S^t ≥ k/2$.  If $S^t$ is still greater than
$k/2$ at this point, try again using the same analysis; this gives us
the usual geometric series argument that $\Exp{t} = O(m)$.  Again, we
have to divide by $k$ to get the number of insert operations, so we
get $O(m/k)$ in this case.

Combining these bounds, we have that $x$ disappears after
$\Theta(m/k)$ insertions on average.
This seems like about what we
would expect.
\end{enumerate}

\section{Final exam}
\label{appendix-2011-final-exam}

Write your answers in the blue book(s).  Justify your answers.  Work
alone.  Do not use any notes or books.  

There are four problems on this exam, each worth 20
points, for a total of 80 points.
You have approximately three hours to complete this
exam.

\subsection{Leader election}

Suppose we have $n$ processes and we want to elect a leader.
At each round, each process flips a coin, and drops out if the coin
comes up tails.  We win if in some round there is exactly one
process left.

Let $T(n)$ be the probability that this event eventually occurs
starting with $n$ processes.  For small $n$, we have $T(0)=0$ and
$T(1)=1$.  Show that there is a constant $c > 0$ such that $T(n) ≥
c$ for all $n > 1$.

\subsubsection*{Solution}

Let's suppose that there is some such $c$.  We will necessarily have
$c ≤ 1 = T(1)$, so the induction hypothesis will hold in the base
case $n=1$.

For $n ≥ 2$, compute
\begin{align*}
T(n)
&= ∑_{k=0}^{n} 2^{-n} \binom{n}{k} T(k)
\\&= 2^{-n} T(n) + 2^{-n} n T(1) + ∑_{k=2}^{n-1} 2^{-n}
\binom{n}{k} T(k)
\\&≥ 2^{-n} T(n) + 2^{-n} n + 2^{-n}(2^n - n - 2) c.
\end{align*}

Solve for $T(n)$ to get
\begin{align*}
T(n)
&≥ \frac{n + (2^n-n-2)c}{2^n-1}
\\&= c \left(\frac{2^n - n - 2 + n/c}{2^n - 1}\right).
\end{align*}

This will be greater than or equal to $c$ if $2^n-n-2+n/c ≥ 2^n-1$
or $n/c ≥ n+1$, which holds if $c ≤ \frac{n}{n+1}$.  The worst
case is $n=2$ giving $c = 2/3$.

Valiant and Vazirani~\cite{ValiantV1986} used this approach to reduce
solving general instances of \index{SAT}{SAT} to solving instances of
SAT with unique solutions; they prove essentially the result given
above (which shows that fixing variables in a SAT formula is likely to
produce a SAT formula with a unique solution at some point) with a
slightly worse constant.

\subsection{Two-coloring an even cycle}

Here is a not-very-efficient algorithm for $2$-coloring an even cycle.
Every node starts out red or blue.  At each step, we pick one of the
$n$ nodes uniformly at random, and change its color if it has the same
color as at least one of its neighbors.  We continue until no node has
the same color as either of its neighbors.

Suppose that in the initial state there are exactly two monochromatic
edges.  What is the worst-case expected number of steps until
there are no monochromatic edges?

\subsubsection*{Solution}

Suppose we have a monochromatic edge surrounded by non-monochrome
edges, e.g. $RBRRBR$.  If we pick one of the endpoints of the edge
(say the left endpoint in this case), then the monochromatic edge
shifts in the direction of that endpoint: $RBBRBRB$.  Picking any node
not incident to a monochromatic edge has no effect, so in this case
there is no way to increase the number of monochromatic edges.

It may also be that we have two adjacent monochromatic edges:
$BRRRB$.  Now if we happen to pick the node in the middle, we end up
with no monochromatic edges ($BRBRB$) and the process terminates.
If on the other hand we pick one of the nodes on the outside, then the
monochromatic edges move away from each other.

We can thus model the process with $2$ monochromatic edges as a random
walk, where the difference between the leftmost nodes of the edges
(mod $n$) increases or decreases with equal probability $2/n$ except
if the distance is $1$ or $-1$; in this last case, the distance
increases (going to $2$ or $-2$) with probability $2/n$, but decreases
only with probability $1/n$.  We want to know when this process hits
$0$ (or $n$).

Imagine a random walk starting from position $k$ with absorbing
barriers at $1$ and $n-1$.  This reaches $1$ or $n-1$ after
$(k-1)(n-1-k)$ steps on average, which translates into
$(n/4)(k-1)(n-1-k)$ steps of our original process if we take into
account that we only move with probability $4/n$ per time unit.
This time is maximized by setting $k=n/2$, which gives
$(n/4)(n/2-1)^2 = n^3/16 - n^2/4 + n/4$ expected time units to reach
$1$ or $n-1$ for the first time.

At $1$ or $n-1$, we wait an addition $n/3$ steps on average; then with
probability $1/3$ the process finishes and with probability $2/3$ we
start over from position $2$ or $n-2$; in the latter case, we run
$(n/4)(n-3) + n/3$ time units on average before we may finish again.
On average, it takes $3$ attempts to finish.  Each attempt incurs the
expected $n/3$ cost before taking a step, and all but the last attempt incur
the expected $(n/4)(n-3)$ additional steps for the random walk.  So the last phase of the process
the process adds $(1/2)n(n-3) + n = (1/2)n^2 - (5/4)n$ steps.

Adding up all of the costs gives $n^3/16 - n^2/4 + n/4 + n/3 +
(1/2)n^2 - (5/4)n = \frac{1}{16} n^3 + \frac{1}{4} n^2 -
\frac{2}{3}n$ steps.

\subsection{Finding the maximum}

\begin{algorithm}
\caption{Randomized max-finding algorithm}
\label{alg-max-finder}
Randomly permute $A$\\
$m \gets -∞$\\
\For{$i \gets 1 \dots n$}{
\If{$A[i] > m$}{
   $m \gets A[i]$
   \nllabel{line-max-finder-max-update}
}
}
\Return $m$
\end{algorithm}

Suppose that we run
Algorithm~\ref{alg-max-finder}
on an array with $n$ elements, all of which are distinct.
What is the expected number of
times Line~\ref{line-max-finder-max-update} is executed as a function
of $n$?

\subsubsection*{Solution}

Let $X_i$ be the indicator variable for the event that
Line~\ref{line-max-finder-max-update} is executed on the $i$-th pass
through the loop.  This will occur if $A[i]$ is greater than $A[j]$
for all $j < i$, which occurs with probability exactly $1/i$ (given
that $A$ has been permuted randomly).  So the expected number of calls
to Line~\ref{line-max-finder-max-update} is $∑{i=1}^{n} \Exp{X_i} =
∑_{i=1}^{n} \frac{1}{i} = H_n$.

\subsection{Random graph coloring}

Let $G$ be a random $d$-regular graph on $n$ vertices, that is, a
graph drawn uniformly from the family of all $n$-vertex graph in which
each vertex has exactly $d$ neighbors.  Color the vertices of $G$ red
or blue independently at random.

\begin{enumerate}
\item What is the expected number of monochromatic edges in $G$?
\item Show that the actual number of monochromatic edges is tightly
concentrated around its expectation.
\end{enumerate}

\subsubsection*{Solution}

The fact that $G$ is itself a random graph is a red herring; all we
really need to know is that it's $d$-regular.

\begin{enumerate}
\item Because $G$ has exactly $dn/2$ edges, and each edge has
probability $1/2$ of being monochromatic, the expected number of
monochromatic edges is $dn/4$.
\item This is a job for Azuma's inequality.  Consider the vertex
exposure martingale.  Changing the color of any one vertex changes the
number of monochromatic edges by at most $d$.  So we have
$\Prob{\abs*{X - \Exp{X}} ≥ t}
≤ 2\exp\left(-t^2/2∑_{i=1}^{n} d^2\right) = 2e^{-t^2/2nd^2}$,
which tells us that the deviation is likely to be not much more than
$O(d√{n})$.
\end{enumerate}

\chapter{Sample assignments from Spring 2009}

\section{Final exam, Spring 2009}

\newcommand{\finalMMIXproblem}[1]{\subsection{#1 (20 points)}}
\newenvironment{finalMMIXsolution}{\subsubsection*{Solution}}{}

Write your answers in the blue book(s).  Justify your answers.  Work
alone.  Do not use any notes or books.  

There are four problems on this exam, each worth 20
points, for a total of 80 points.
You have approximately three hours to complete this
exam.

\finalMMIXproblem{Randomized mergesort}

Consider the following randomized version of the mergesort algorithm.
We take an unsorted list of $n$ elements and split it into two lists
by flipping an independent fair coin for each element to decide which
list to put it in.  We then recursively sort the two lists, and merge
the resulting sorted lists.  The merge procedure involves repeatedly
comparing the smallest element in each of the two lists and removing
the smaller element found, until one of
the lists is empty.

Compute the expected number of comparisons needed to perform
this final merge.  (You do not need to consider the cost of performing
the recursive sorts.)

\begin{finalMMIXsolution}
Color the elements in the final merged list red or blue based on which
sublist they came from.  The only elements that do not require a
comparison to insert into the main list are those that are followed
only by elements of the same color; the expected number of such
elements is equal to the expected length of the longest monochromatic
suffix.  By symmetry, this is the same as the expected longest
monochromatic prefix, which is equal to the expected length of the
longest sequence of identical coin-flips.

The probability of getting $k$ identical coin-flips in a row followed
by a different coin-flip is exactly $2^{-k}$; the first coin-flip sets
the color, the next $k-1$ must follow it (giving a factor of
$2^{-k+1}$, and the last must be the opposite color (giving an
additional factor of $2^{-1}$).  For $n$ identical coin-flips, there
is a probability of $2^{-n+1}$, since we don't need an extra coin-flip
of the opposite color.  So the expected length is $∑_{k=1}^{n-1}
k 2^{-k} + n 2^{-n+1} = ∑_{k=0}^{n} k 2^{-k} + n 2^{-n}$.  

We can simplify the sum using generating functions.  The sum
$∑_{k=0}^{n} 2^{-k} z^k$ is given by $\frac{1 -
(z/2)^{n+1}}{1-z/2}$.  Taking the derivative with respect to $z$ gives
$∑_{k=0}^{n} 2^{-k} k z^{k-1} = (1/2) \frac{1-(z/2)^{n+1}}{1-z/2}^2
+ (1/2) \frac{(n+1) (z/2)^n}{1-z/2}$.  At $z=1$ this is $2(1-2^{-n-1}) -
2(n+1)2^{-n} = 2 - (n+2)2^{-n}$.  Adding the second term gives $\Exp{X}
= 2 - 2 ⋅ 2^{-n} = 2 - 2^{-n+1}$.

Note that this counts the expected number of elements for which we do
not have to do a comparison; with $n$ elements total, this leaves $n-2
+ 2^{-n+1}$ comparisons on average.
\end{finalMMIXsolution}

\finalMMIXproblem{A search problem}

Suppose you are searching a space by generating new instances of some
problem from old ones.  Each instance is either good or bad; if you
generate a new instance from a good instance, the new instance is also
good, and if you generate a new instance from a bad instance, the new
instance is also bad.

Suppose that your start with $X_0$ good instances and $Y_0$ bad
instances, and that at each step you choose one of the instances you
already have uniformly at random to generate a new instance.  What is
the expected number of good instances you have after $n$ steps?

Hint: Consider the sequence of values $\Set{ X_t / (X_t+Y_t) }$.

\begin{finalMMIXsolution}
We can show that the suggested sequence is a martingale, by computing
\begin{align*}
    \ExpCond{\frac{X_{t+1}}{X_{t+1} + Y_{t+1}} }{ X_t, Y_t}
&= \frac{X_t}{X_t+Y_t} ⋅ \frac{X_t+1}{X_t+Y_t+1} 
   + \frac{Y_t}{X_t+Y_t} ⋅ \frac{X_t}{X_t+Y_t+1} 
\\
&= \frac{X_t(X_t+1) + Y_t X_t}{(X_t+Y_t)(X_t+Y_t+1)}
\\
&= \frac{X_t(X_t+Y_t+1)}{(X_t+Y_t)(X_t+Y_t+1)}
\\
&= \frac{X_t}{X_t+Y_t}.
\end{align*}

From the martingale property we have
$\Exp{\frac{X_n}{X_n+Y_n}} = \frac{X_0}{X_0+Y_0}$.  But
$X_n+Y_n = X_0+Y_0+n$, a constant, so we can multiply both sides by
this value to get $\Exp{X_n} = X_0
\left(\frac{X_0+Y_0+n}{X_0+Y_0}\right)$.
\end{finalMMIXsolution}

\finalMMIXproblem{Support your local police}

At one point I lived in a city whose local police department
supported themselves in part by collecting fines for speeding tickets.
A speeding ticket would cost 1 unit (approximately \$100), and it was
unpredictable how often one would get a speeding ticket.  For a price
of 2 units, it was possible to purchase a metal placard to go on
your vehicle identifying
yourself as a supporter of the police union,
which (at least according to local legend) would
eliminate any fines for subsequent speeding tickets, but which would
not eliminate the cost of any previous speeding tickets.

Let us consider the question of when to purchase a placard as a
problem in on-line algorithms.  It is possible to achieve a
strict\footnote{I.e., with no additive constant.}
competitive ratio of 2 by purchasing a placard after the second
ticket.  If one receives fewer than 2 tickets, both the on-line and
off-line algorithms pay the same amount, and at 2 or more tickets the
on-line algorithm pays 4 while the off-line pays 2 (the off-line
algorithm purchased
the placard before receiving any tickets at all).

\begin{enumerate}
\item Show that no deterministic algorithm can achieve a lower
(strict) competitive ratio.
\item Show that a randomized algorithm can do so, against an oblivious
adversary.
\end{enumerate}

\begin{finalMMIXsolution}
\begin{enumerate}
\item Any deterministic algorithm essentially just chooses some fixed 
number $m$ of tickets to collect before buying the placard.  Let $n$
be the actual number of tickets issued.  For
$m=0$, the competitive ratio is infinite when $n=0$.
For $m=1$, the competitive ratio is 3 when $n=1$.
For $m>2$, the competitive ratio is $(m+2)/2 > 2$ when $n=m$.  So
$m=2$ is the optimal choice.
\item Consider the following algorithm: with probability $p$, we
purchase a placard after 1 ticket, and with probability $q = 1-p$, we
purchase a placard after 2 tickets.  This gives a competitive ratio of
$1$ for $n=0$, $1+2p$ for $n=1$, and $(3p+4q)/2 = (4-p)/2 = 2-p/2$ for
$n≥ 2$.  There is a clearly a trade-off between the two ratios
$1+2p$ and $2-p/2$.  The break-even point is when they are equal, at
$p=2/5$.  This gives a competitive ratio of $1+2p = 9/5$, which is
less than $2$.
\end{enumerate}
\end{finalMMIXsolution}

\finalMMIXproblem{Overloaded machines}

Suppose $n^2$ jobs are assigned to $n$ machines with each job choosing
a machine independently and uniformly at random.  Let the load on a
machine be the number of jobs assigned to it.  Show that for any fixed
$δ > 0$ and sufficiently large $n$, there is a constant $c < 1$ such
that the maximum load exceeds $(1+δ)n$ with probability at
most $nc^n$.

\begin{finalMMIXsolution}
This is a job for Chernoff bounds.  For any particular machine, the
load $S$ is a sum of independent indicator variables and the mean load is
$μ = n$.  So we have
\begin{align*}
\Prob{S ≥ (1+δ)μ} &≤
\left(\frac{e^δ}{(1+δ)^{1+δ}}\right)^n.
\end{align*}

Observe that $e^δ/(1+δ)^{1+δ} < 1$ for $δ >
0$.  One proof of this fact is to take the log to get $δ -
(1+δ) \log (1+δ)$, which equals $0$ at $δ = 0$, and
then show that the logarithm is decreasing by showing that
$\frac{d}{dδ} \dots = 1 - \frac{1+δ}{1+δ} -
\log(1+δ) = - \log(1+δ) < 0$ for all $δ > 0$.

So we can
let $c = e^δ/(1+δ)^{1+δ}$ 
to get a bound of $c^n$ on the probability
that any particular machine is overloaded and a bound of $nc^n$ (from the
union bound) on the probability that any of the machines is
overloaded.
\end{finalMMIXsolution}

\backmatter

\clearpage
\phantomsection
\addcontentsline{toc}{chapter}{Bibliography}
\bibliography{notes}

\clearpage
\phantomsection
\addcontentsline{toc}{chapter}{Index}
\printindex

\end{document}